%% file: ms.tex
\documentclass[pdftex,colorlinks]{sty/panda_book}
\pdfoutput=1
\usepackage[english]{babel}
\usepackage[T1]{fontenc}
\usepackage[utf8]{inputenc}
\usepackage{polski}
\usepackage{array}
\newcolumntype{P}[1]{>{\centering\arraybackslash}p{#1}}
\newcolumntype{M}[1]{>{\centering\arraybackslash}m{#1}}

\def\PANDA{\texorpdfstring{$\overline{\mbox{\sf P}}${\sf ANDA}}{Panda}\xspace}%
\def\Panda{\texorpdfstring{$\overline{\mbox{\sf P}}${\sf ANDA}}{Panda}\xspace}%

\usepackage{graphicx}
\usepackage{subfig}
\usepackage{colortbl}
\definecolor{orange}{RGB}{255,158,62}
\definecolor{yellow}{RGB}{250,250,10}
\usepackage{todo}
\usepackage{multicol}
\usepackage{enumitem}

\usepackage{lipsum}

\usepackage{sty/widetext}

\usepackage{authblk}

\usepackage[square,numbers]{natbib}
\usepackage[sectionbib]{bibunits}
\usepackage[toc,page]{appendix}

\usepackage{wrapfig}
\usepackage{comment}

\RequirePackage{multirow}
\RequirePackage{bibunits}
\input{sty/panda_symbols}
\input{sty/panda_detector}

\bibliographyunit[\chapter]
\title{\PANDA Barrel DIRC\\\vspace{0.5cm}Technical Design Report}

\author{A.~N.~Author and A.~N.~Other~Author}

\date{}
\begin{document}

\pagenumbering{roman}

\input{titlepage/titlepages.tex}
\cleardoublepage
\newpage

\pagenumbering{arabic}
\setcounter{page}{1}

\cleardoublepage
\input{executive/executive_summary.tex}
\cleardoublepage
\input{panda/panda.tex}
\cleardoublepage
\input{design/design.tex}
\cleardoublepage
\input{simulation/simulation-reconstruction.tex}
\cleardoublepage
\input{components/components.tex}
\cleardoublepage
\input{performance-validation/performance-validation.tex}

\cleardoublepage
\input{mechanics/mechanics.tex}

\cleardoublepage
\input{organization/organization.tex}
\cleardoublepage

\end{document}

%% file: titlepage/titlepages.tex
\pagenumbering{roman}
\thispagestyle{empty}
\onecolumn
%
%
\begin{center}
{\bfseries \sffamily \huge Technical Design Report for the

\panda{}  Barrel DIRC Detector
\vskip 0.2cm
{\sffamily \small (Anti\underline{P}roton \underline{An}nihilations at \underline{Da}rmstadt)}
\vskip .2cm
{\sffamily \small  Strong Interaction Studies with Antiprotons}}
\vskip 1cm
{\large \sffamily \panda{} Collaboration}
%
\vskip .5cm
%
\vskip 0.3cm
\fbox{August 2017}
\end{center}
%
%
\vspace*{20mm }
\begin{center}
\includegraphics[width=0.75\textwidth]{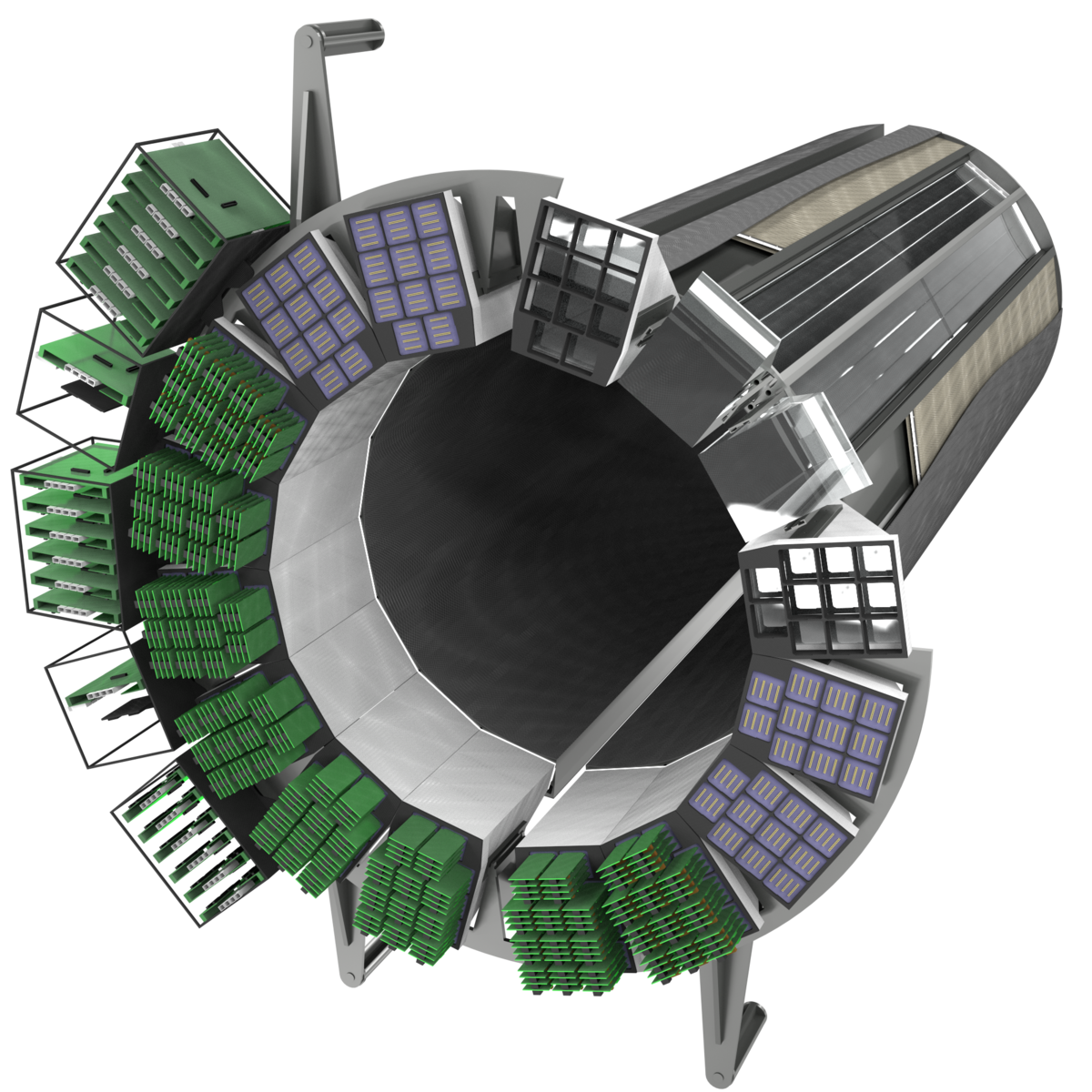}
\end{center}
\vfill
%
%
\newpage
\begin{center}
\vspace*{3mm }
{\LARGE \bfseries \sffamily The \panda{} Collaboration}
\vskip 7mm
\input{./collaboration/authors.tex}
\end{center}
\newpage
%
%
\vfill
\hrulefill
\begin{tabbing}

Editors:  \hspace{25mm} \= Anastastios Belias   \hspace{6mm}  \= Email: \verb$a.belias@gsi.de$ \\                        
                        \> Roman Dzhygadlo \> Email: \verb$r.dzhygadlo@gsi.de$ \\
                        \> Erik Etzelm\"uller \> Email: \verb$Erik.Etzelmueller@physik.uni-giessen.de$\\ 
                        \> Andreas Gerhardt \> Email: \verb$a.gerhardt@gsi.de$ \\
                        \> Klaus G\"otzen \> Email: \verb$k.goetzen@gsi.de$ \\ 
                        \> Matthias Hoek \>  Email: \verb$hoek@uni-mainz.de$ \\                         
                        \> Marvin Krebs \> Email: \verb$m.krebs@gsi.de$ \\
                        \> Albert Lehmann \>  Email: \verb$albert.lehmann@physik.uni-erlangen.de$ \\
                        \> Frank Nerling \> Email: \verb$f.nerling@gsi.de$ \\                           
                        \> Georg Schepers \> Email: \verb$g.schepers@gsi.de$ \\                         
                        \> Carsten Schwarz \>  Email: \verb$c.schwarz@gsi.de$ \\
                        \> Jochen Schwiening \>  Email: \verb$j.schwiening@gsi.de$ \\[2mm]
                         
Technical Coordinator: \> Lars Schmitt  \> Email: \verb$l.schmitt@gsi.de$\\[2mm]

Spokesperson:  \> Klaus Peters  \> Email:  \verb$k.peters@gsi.de$\\[2mm]
Deputy: \> Tord Johansson \> Email:  \verb$tord.johansson@physics.uu.se$
\end{tabbing}
\hrulefill










%
\vfill
%
%
\newpage
\input{./preface/preface.tex}
\cleardoublepage
%
%
\tableofcontents
\twocolumn

\newpage
%
%

%% file: collaboration/authors.tex
%
%
\institem{Aligarth Muslim University, Physics Department,{ \bf Aligarth}, India}
\authitem{B.~Singh}
\lastitem
\institem{Universität Basel,{ \bf Basel}, Switzerland}
\authitem{W.~Erni},
\authitem{B.~Krusche},
\authitem{M.~Steinacher},
\authitem{N.~Walford}
\lastitem
\institem{Institute of High Energy Physics, Chinese Academy of Sciences,{ \bf Beijing}, China}
\authitem{B.~Liu},
\authitem{H.~Liu},
\authitem{Z.~Liu},
\authitem{X.~Shen},
\authitem{C.~Wang},
\authitem{J.~Zhao}
% ...
\lastitem

%% file: preface/preface.tex
\begin{center}
\vspace*{2cm}
{\Large \bfseries \sffamily Abstract}\addcontentsline{toc}{chapter}{Abstract}
\vskip 2cm
\begin{minipage}[t]{12cm}
\sloppy\large
This documents describes the technical design and the expected performance of 
the Barrel DIRC detector for the \panda experiment. 
The Barrel DIRC will provide hadronic charged particle identification  
in the polar angle range of $22^\circ$ to $140^\circ$ for particle
momenta between 0.5~GeV/c and 3.5~GeV/c.

The design is based on the successful BaBar DIRC with several key
improvements.
The performance and system cost were optimized in detailed detector simulations
and validated with full system prototypes using particle beams at GSI and 
CERN.
The final design meets or exceeds the PID goal of clean $\pi/K$ separation 
with at least 3 standard deviations over the entire phase space of charged
kaons in the Barrel DIRC.
\end{minipage}
\end{center}

\clearpage
\vspace*{18cm}
\hrulefill\\
\vspace*{2cm}
\begin{minipage}[t]{10cm}
\sloppy
The use of registered names, trademarks, \etc in this publication does not
imply, even in the absence of specific statement, that such names are exempt
from the relevant laws and regulations and therefore free for general use.
\end{minipage}
\vfill
%

%% file: executive/executive_summary.tex
\chapter{Executive Summary}
\label{cha:executivesummary}
\begin{bibunit}[unsrt]

\subsubsection*{The \panda Experiment}

The \panda experiment~\cite{panda1} will be one of the four flagship experiments at the new 
international accelerator complex FAIR (Facility for Antiproton and Ion Research) in Darmstadt, Germany.
\panda will perform unique experiments using the high-quality antiproton beam with momenta in the 
range of 1.5~GeV/c to 15~GeV/c, stored in the HESR (High Energy Storage Ring)~\cite{talk:prasuhn2014}, 
to explore fundamental questions of hadron physics in the charmed and multi-strange hadron sector 
and deliver decisive contributions to the open questions of QCD.
The scientific program of \panda~\cite{panda-physics} includes hadron spectroscopy, properties of hadrons 
in matter, nucleon structure, and hypernuclei.
The cooled antiproton beam colliding with a fixed proton or nuclear target will allow hadron production 
and formation experiments with a luminosity of up to $2\times10^{32}cm^{-2}s^{-1}$. 
Excellent Particle Identification (PID) is crucial to the success of the \panda physics program.

\subsubsection*{Particle Identification in \panda}
The \panda PID system comprises a range of detectors using different technologies.
Dedicated PID devices, such as several Time-of-Flight and Cherenkov counters and a Muon 
detection system~\cite{muontdr}, are combined with PID information delivered by the 
Micro Vertex Detector~\cite{mvdtdr} and the Straw Tube Tracker~\cite{trktdr} 
as well as by the Electromagnetic Calorimeter~\cite{emctdr}.

While the specific energy loss measurements from the \panda tracking detectors, in combination with the
Time-of-Flight information, provide $\pi$/K separation at low momentum, dedicated hadronic PID systems 
are required for the positive identification of kaons with higher momentum ($p > 1$~GeV/c) and for the suppression of large pionic backgrounds.
Two Ring Imaging Cherenkov (RICH) counters using the DIRC (Detection of Internally Reflected Cherenkov 
light) principle~\cite{ratcliff:dircprinciple1,ratcliff:dircprinciple2,ratcliff:dircprinciple3} 
in the Target Spectrometer (TS) and an aerogel RICH counter in the Forward Spectrometer (FS)
will provide this charged hadron PID.

The DIRC concept was introduced and successfully used by the BaBar experiment~\cite{babar:barreldirc} 
where it provided excellent $\pi$/K separation up to 4.2~GeV/c and proved to 
be robust and easy to operate.
In \panda the Barrel DIRC, modeled after the BaBar DIRC, will surround the interaction point at a 
distance of about 50~cm and cover the central region of $22^\circ < \theta < 140^\circ$
while the novel Endcap Disc DIRC~\cite{eddreport} will cover the smaller forward angles, 
$5^\circ < \theta < 22^\circ$ and $10^\circ < \theta < 22^\circ$ in the vertical and horizontal 
direction, respectively.

\subsubsection*{The \panda Barrel DIRC}

The Barrel DIRC design described in this report will provide a clean separation of charged pions and 
kaons with 3 standard deviations (s.d.) or more in the range of 0.5 -- 3.5~GeV/c.
The scientific merit of the Barrel DIRC is that the particle identification performance enables 
a wide range of physics measurements in \panda with kaons in the final state, i.e. the study of light 
hadron reactions, charmed baryons, charmonium spectroscopy, and open charm events.

The design concept is based on the successful BaBar DIRC~\cite{babar:barreldirc} and key results from 
the R\&D for the SuperB FDIRC~\cite{superb:dirc1}.
The main design difference compared to the BaBar DIRC, the replacement of the large water tank 
expansion volume (EV) by 16 compact prisms, is due to the fact that the plans for the magnet and 
the upstream endcap of the \panda detector did not allow the DIRC bars to penetrate the iron, 
requiring a small EV that can be placed inside the already crowded \panda detector volume.
This compact EV in turn meant that focusing optics and smaller sensor pixels are needed to keep 
the Cherenkov angle resolution similar to the performance obtained by the BaBar DIRC.

In the \panda Barrel DIRC baseline design the circular cross section of the barrel part is 
approximated by a hexadecagon. 
Each of the $16$ flat sections contains three fused silica radiator bars 
($17 \times 53 \times 2400$~mm$^3$). 
Cherenkov photons, produced along the charged particle track in the bar, are guided inside 
the radiator via total internal reflection. 
A flat mirror is attached to the forward end of the bar to reflect photons towards the read out 
end, where they are focused by a multi-component spherical lens on the back of a 
$30$~cm-deep solid fused silica prism, serving as expansion volume. 

An array of lifetime-enhanced Microchannel Plate Photomultiplier Tubes 
(MCP-PMTs)~\cite{fred02}, 
each with 8~$\times$~8 pixels of about 6.5~$\times$~6.5~mm$^2$ size, is used to 
detect the photons and measure their arrival time on a total of about 11,300 pixels 
with a precision of $100$~ps or better in the magnetic field of approximately 1~T.

The sensors are read out by an updated version of the Trigger and Readout Board 
(TRB)~\cite{traxler:trb3}, developed for the HADES experiment~\cite{hades}, 
in combination with the PADIWA front-end amplification and discrimination 
card~\cite{ugur:trb3}, mounted directly on the MCP-PMTs. 
This FPGA-based system provides measurements of both the photon arrival time and Time-over-Threshold 
(TOT), which is related to the pulse height of the analog signal and can be used to monitor 
the sensor performance and to perform time-walk corrections to achieve the required 
precision of the photon timing.

The focusing optics has to produce a flat image to match the shape of the back surface 
of the fused silica prism.
This is achieved by a combination of focusing and defocusing elements in a spherical triplet 
lens made from one layer of lanthanum crown glass (NLaK33, refractive index n=1.786 for 
$\lambda$=380~nm) between two layers of synthetic fused silica (n=1.473 for $\lambda$=380~nm).
Such a 3-layer lens works without any air gaps, minimizing the photon loss that would 
otherwise occur at the transition from the lens to the expansion
volume.

The mechanical system is modular with components made of aluminum alloy and 
Carbon-Fiber–Reinforced Polymer (CFRP).
The optical components are placed in light-tight CFRP containers that are installed in
the \panda  detector by sliding them on rails into slots in two rings which are 
attached to the main central support beams.
Boil-off dry nitrogen flows through the CFRP containers to remove moisture and
residue from outgassing. 
The entire readout unit, comprising the prisms, sensors, and electronics, can be 
detached from the \panda detector to facilitate access to the tracking systems 
during scheduled extended shutdowns.

Industrial fabrication of the fused silica radiators remains a significant technological 
challenge, just as it was during the construction of the BaBar DIRC and the 
Belle~II~TOP~\cite{Inami:2014nra}.
An excellent surface polish with an RMS roughness of 10~\AA\ or better is needed for
efficient photon transport since Cherenkov photons are internally reflected up to 400 
times before exiting the bar.
The radiator surfaces have to be perpendicular to each other within 0.25~mrad to preserve the magnitude 
of the Cherenkov angle during these reflections. 
Due to the tight optical and mechanical tolerances the price of radiator fabrication 
is, together with the price of the photon detectors, the dominant contribution to the 
Barrel DIRC construction cost.

A substantial reduction of the radiator fabrication cost is achieved by increasing the 
width of the fused silica bars by 50\% compared to the BaBar DIRC.

A further significant cost reduction may be possible if the three radiator bars per section 
are replaced by one 16~cm-wide plate since even fewer pieces would have to be produced.
However, although a design with an even wider plate (with a width of 45~cm) 
is being built for the Belle~II experiment, so far the PID performance of a Barrel 
DIRC design with wide plates has not been validated experimentally.
Therefore, until such an experimental validation is achieved, the wide plates remain only 
as an alternative cost-saving design option to the narrow bar geometry baseline design.

The use of legacy detector components as a cost-saving measure was investigated in 
2013 when the SLAC National Accelerator Laboratory issued a call for proposals for 
reuse of the BaBar DIRC bar boxes.
The \panda collaboration submitted a letter of interest and detailed proposal
for using three BaBar DIRC bar boxes, which, after disassembly, could have 
yielded all the narrow radiator bars needed for the \panda Barrel DIRC.
The formal review by SLAC and U.S. Department of Energy (DOE), Office of High 
Energy Physics (OHEP), decided in 2014 that the reuse of the BaBar DIRC bar boxes
would only be granted to experiments that keep the bar boxes intact.
This was not an option for \panda since the length of the BaBar DIRC bar boxes 
is about 490~cm, twice the length of the \panda Barrel DIRC.

\subsubsection*{Simulation and Prototyping}

A detailed physical simulation of the \panda Barrel DIRC was developed in the PANDARoot 
framework~\cite{pandaroot-1,pandaroot-2}, which uses the Virtual Monte Carlo (VMC) 
approach to easily switch between Geant3 and Geant4~\cite{Exc-Geant4} for systematic studies.
The simulation is tuned to include measured values for the sensor quantum and collection
efficiency and the timing resolution~\cite{fred02}.
It includes the coefficient of total internal reflection of DIRC radiator bars as a 
function of photon energy~\cite{babar:dirc1}, the bulk transmission of bars, glue, and 
lenses, the wavelength-dependent refractive indices of fused silica, NLaK33, and the 
photocathode, as well as the reflectivity of the forward mirrors.
Background from hadronic interaction and delta electrons is simulated as well as 
contributions from MCP-PMT dark noise and charge sharing between anode pads. 
Additional simulation tools employed during the R\&D phase include Zemax~\cite{Zemax}, 
used primarily in the design of the focusing optics, and \mbox{DircProp}, a stand-alone ray-tracing 
package designed at GSI, for the development of prototype configurations.

For the purpose of design evaluation the single photon Cherenkov angle resolution (SPR) 
and the photon yield were selected as figures of merit.
These two quantities allow a comparison of different design options to the performance
of prototypes in test beams and to published results for the BaBar DIRC and other
RICH counters.
A fast reconstruction method based on lookup-tables, similar to the approach used for 
the BaBar DIRC, was utilized to determine the SPR and photon yield for a wide range 
of particle angles and momenta for each simulated design with narrow bars.
For the evaluation of the design option with wide plates an alternative reconstruction 
algorithm was developed~\cite{MZ-MZuehlsdorf-PHD-THESIS}, the so-called time-based imaging 
method, similar to the approach used by the Belle~II~TOP~\cite{belle2-software}.

In the process of optimizing the design of the \panda Barrel DIRC for cost and 
performance many different design aspects were tested in simulations.
These include the thickness and width of the radiators, the number of bars per sector, 
the material and shape of the focusing lenses, the material, shape, and size of the 
expansion volume, and the sensor layout on the focal plane~\cite{MP-MPatsyuk-PHD-THESIS}.
The SPR and photon yield were determined for each configuration and evaluated 
as a function of momentum, polar and azimuthal angle for the entire \panda phase
space.
The simulation effort identified several designs that meet or exceed the 
\panda PID performance requirements for the entire kaon phase space.

\subsubsection*{Design Validation}

A number of the most promising design elements were implemented in prototypes and 
tested under controlled conditions in a dedicated optics laboratory or with particle beams.

A total of more than 30 radiator prototypes, narrow bars as well as wide 
plates, were produced by eight manufacturers using different materials and 
fabrication processes.
The goal was to identify companies capable of producing high-quality radiators 
for the full-scale \panda Barrel DIRC production.
The radiator surface properties were measured by internally reflecting laser beams of 
different wavelengths to determine the coefficient of internal reflection and to 
study subsurface damage effects.
The bar angles were measured using an autocollimator.
The results show that several of the prototype manufactures are able to produce 
high-quality bars or plates that meet the specifications.

A series of increasingly complex \panda Barrel DIRC system prototypes 
were tested in particle beams at GSI and CERN from 2011--2016 to determine 
the PID performance and to validate the simulation results. 
The prototypes all featured a dark box containing a radiator bar or plate coupled 
via optional focusing to an expansion volume equipped with a photon detector array on the image plane.
The sensors were read out by a TRB in combination with 
an amplification and discrimination card mounted directly on the MCP-PMTs.

During the two most recent prototype tests at the CERN PS, in 2015 and 2016, 
the experimental data obtained with the narrow bar and a 3-layer spherical 
lens and with the wide plate and a 2-layer cylindrical lens both showed 
good agreement of the Cherenkov hit patterns with simulation in the 
pixel space and in the photon hit time space.

A single photon Cherenkov angle resolution of 10--12~mrad and a yield of 15--80 
detected Cherenkov photons per particle, depending on the polar angle, were obtained 
for the narrow bar.
These values are comparable to the performance of the BaBar DIRC, are consistent 
with the simulation of the experimental setup, and demonstrate that this 
design is technically feasible.

The observed $\pi/p$ separation power for a momentum of 7~GeV/c and a polar angle
of $25^{\circ}$, corresponding to the most demanding region in \panda, was 
3.6~s.d. for the narrow bar and 3.1~s.d. for the wide plate.
Simulation was used to extrapolate the prototype results to the 
expected PID performance of the \panda Barrel DIRC.
Both radiator geometries meet or exceed the PID goal for the entire final
state kaon phase space, validating both as possible designs for \panda.

The PID performance of the narrow bar geometry was found to be superior to
the design with the wide plate and to be significantly less sensitive to 
a deterioration of the timing precision and to provide a larger 
margin for error during the early phase of \panda operation.

Because of these advantages the geometry with the narrow bars and the 
3-layer spherical lens was selected as the baseline design for the \panda Barrel 
DIRC.

\putbib[./literature/lit_executive]

\end{bibunit}

%% file: panda/panda.tex
\chapter{The \panda Experiment} \label{chap:panda} 

\begin{bibunit}[unsrt]

 \input{panda/pandaexp.tex}

 \input{panda/pandadet.tex}

 \putbib[./literature/lit_panda]

\end{bibunit}

%% file: panda/pandaexp.tex
\section{The \panda Experiment} \label{sec:pandaexp2}

\subsection{The Scientific Program}
The \panda (anti-Proton ANnihiliation at DArmstadt) collaboration 
\cite{panda} 
envisages a physics core program~\cite{panda-physics2} that comprises 
\begin{itemize}
\item charmonium spectroscopy
with precision measurements of mass, width, and decay branches;
\item the investigation of states that are assumed to have 
more exotic configurations like multiquark states,
charmed hybrids, and glueballs;
\item spectroscopy of (multi-)strange and charmed baryons;
\item the search for medium modifications of charmed hadrons in nuclear matter;
\item the $\gamma$-ray spectroscopy of hypernuclei,
in particular double $\Lambda$ states.
\end{itemize}

In the charmonium and open-charm regions, many new states have been observed in the
last years, that do not match the patterns predicted in those regimes \cite{LuiXYZ_2014}. 
There are even several states unambiguously being of exotic nature, raising the question about
the underlying mechanism to form such kind of states \cite{Kalashnikova_2010}.  

The production of charmonium and open-charm states in $e^+e^-$ interactions
is to first order restricted to initial spin-parities of $J^{PC} = 1^{--}$. This limits the possibility
to precisely scan and investigate these resonances in formation reactions. The use of $\bar{p}p$ annihilation does not suffer from this limitation. Combined with the excellent
energy resolution of down to about 25~keV, these kind of reactions offer a unique 
opportunity to perform hadron and charmonium spectroscopy in that energy range. 

Since the decay of charm quarks predominantly proceeds via strangeness production, the 
identification of kaons in the final state is mandatory to separate the signal events
from the huge pionic background.

\subsection{High Energy Storage Ring}\label{sec:hesr}

\begin{figure}[htb]
\begin{center}
\resizebox{0.99\columnwidth}{!}{%
\includegraphics{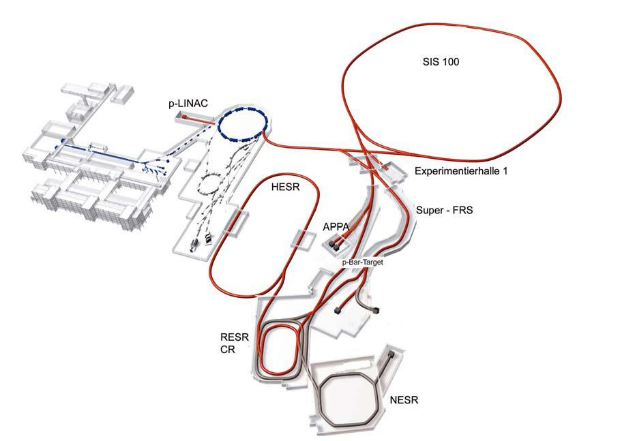}}
\caption{
Schematic of the future FAIR layout 
incorporating the current GSI installations on the left;
on the right the future installations, the SIS~100 synchrotron
the storage and cooler ring complex including CR and HESR
and the Super FRS experiment being some of the new parts.}
\label{fig:FAIR-schematic}
\end{center}
\end{figure}

The combination of HESR and \panda aims 
at both high reaction rates and high resolution
to be able to study rare production processes and small
branching ratios. With a design value of $10^{11}$ stored antiprotons for
beam momenta from 1.5~GeV/c to 15~GeV/c and high density targets
   the anticipated antiproton production rate of 2$\cdot$10$^7$~s$^{-1}$
   governs the experiment interaction rate in the order of
   cycle-averaged 1$\cdot$10$^7$~s$^{-1}$.
The stored antiprotons do not have a bunch structure, and
with 10\% to 20\% allocated to a barrier bucket, the antiprotons
are continuously spread over about 80\% of the HESR circumference.

\begin{figure}[htb]
\begin{center}
\resizebox{0.95\columnwidth}{!}{%
\includegraphics{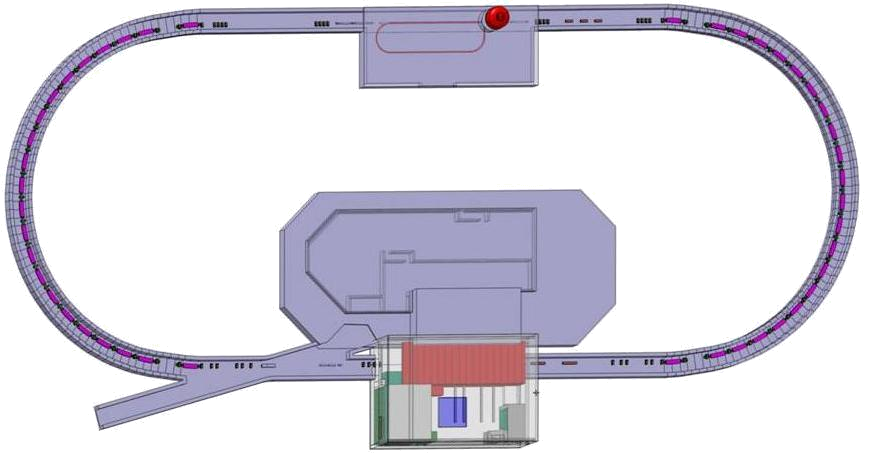}}
\caption{
The HESR ring with the \panda experimental area at the bottom
and the electron cooler installation at the top. 
Standard operation has the antiproton injection from RESR
(during the modularized startup phase from CR) 
from the left, or protons at reversed field polarities.
}
\label{fig:HESR-topview}
\end{center}
\end{figure}

Two complementary operating modes are planned, named
{\it high luminosity} mode and {\it high resolution} mode. 
The high luminosity mode with $\Delta p/p=10^{-4}$,
stochastic cooling and a pellet target density of
$4\cdot 10^{15}$~cm$^{-2}$ 
will have an average luminosity of up to
 $L=1.6\cdot 10^{32}$~cm$^{-2}$s$^{-1}$. 
For the high resolution mode 
$\Delta p/p=5\cdot 10^{-5}$ will be  
achieved with stochastic cooling and will operate in conjunction with a cluster jet target to limit
the energy broadening caused by the target.
 The cycle-averaged luminosity is expected to be 
$L=1.6\cdot 10^{31}$~cm$^{-2}$s$^{-1}$.

The values described here are the design values for the HESR and the \panda experiment.

In the modularized start version the Recycled Experimental Storage Ring (RESR) 
will not be available to accumulate the anti-protons. 
Instead, the accumulation process has to be done with the HESR itself. 
The absence of the dedicated RESR has the implication that, on one hand, the maximum number 
of anti-protons is reduced by one order of magnitude to $N_{max} = 10^{10}$ compared to 
the high luminosity mode. 
On the other hand the accumulation process, which takes a finite time, cannot be performed 
in parallel but further worsens the duty cycle (for more detail see~\cite{Lumi-Goetzen}).
However, since the full version of FAIR is decided to be built, the requirements for detectors of the \panda experiment have to be set up regarding the original design values.

\subsection{Targets}

The \panda Target Spectrometer
is designed to allow the installation of different targets.
For hydrogen as target material both Cluster Jet Targets
and Pellet Targets are being prepared. One main technical challenge
is the distance of 2~m between the target injection point 
and the dumping region.

The cluster jet target has a constant thickness as a function of time
whereas a pellet target with average velocities of around 50~m/s
and average pellet spacing of 3~mm 
  has pellet target density variations 
on the 10-100~$\mu$s timescale. 

An extension of the targets to heavier gases
such as deuterium, nitrogen, or argon is planned
for complementary studies with nuclear targets. 
In addition wire or foil targets are used in a dedicated setup for the 
production of hypernuclei. 

\subsection{Luminosity Considerations}

\begin{figure}[htb]
\begin{center}
\resizebox{0.95\columnwidth}{!}{%
\includegraphics{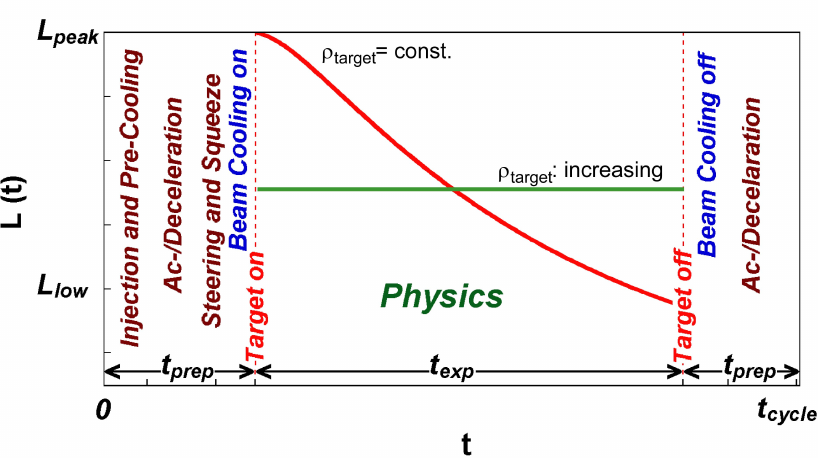}}
\caption{ 
Time dependent macroscopic luminosity profile $L(t)$
in one operation cycle for constant (solid red) and increasing
(green dotted) target density $\rho_{target}$~\cite{trk-tdr}. Different measures for
beam preparation are indicated. Pre-cooling is performed at
3.8~GeV/$c$. A maximum ramp of 25~mT/s is specified for acceleration and deceleration of the beam.                                          
}
\label{fig:macroscopic-luminosity-profile}
\vspace*{-2mm}
\end{center}
\end{figure}

The luminosity is linked to the number of stored antiprotons
and the maximum luminosity depends on the antiproton production
rate. In first approximation the cycle-averaged antiproton production
and reaction rates should be equal. Due to injection time and possible
dumping of beam particles at the end of a cycle the time-averaged
reaction rate will be lower.
In Fig.~\ref{fig:macroscopic-luminosity-profile} the 
beam preparation periods with target off and data taking periods
with target on are depicted. The red curve showing the luminosity at constant
target thickness is proportional to the decreasing number of
antiprotons during data taking. In order to provide a constant
luminosity, measures to implement a target density increasing with time
are studied in order to achieve a constant luminosity. 

In the case of a pellet target, variations of the instantaneous
luminosity will occur. These are depending on antiproton beam profile,
pellet size, pellet trajectories and the spacing between pellets.
In the case of an uncontrolled pellet sequence 
(the variation of pellet velocities can be at most in the order of 10\%)
target density fluctuations with up to 2-3 pellets in beam do occur
during a timescale of 10-100~$\mu$s, the pellet transit time.
Even if only one pellet was present in the beam at any given time, the maximum interaction rate of 32~MHz \cite{targettdr}
is still a factor of 3 above the average interaction rate of about 10~MHz.
The pellet high luminosity mode (PHL mode) features smaller droplet sizes, lower spreads in pellet relative velocity and average pellet distances.
 The latter being much smaller than the beam size. Here the high intensity fluctuations are reduced a lot.

%% file: panda/pandadet.tex
\section{The \panda Detector} \label{sec:pandadet}

\begin{figure*}[htb]
\begin{center}
\resizebox{1.8\columnwidth}{!}{%
\includegraphics{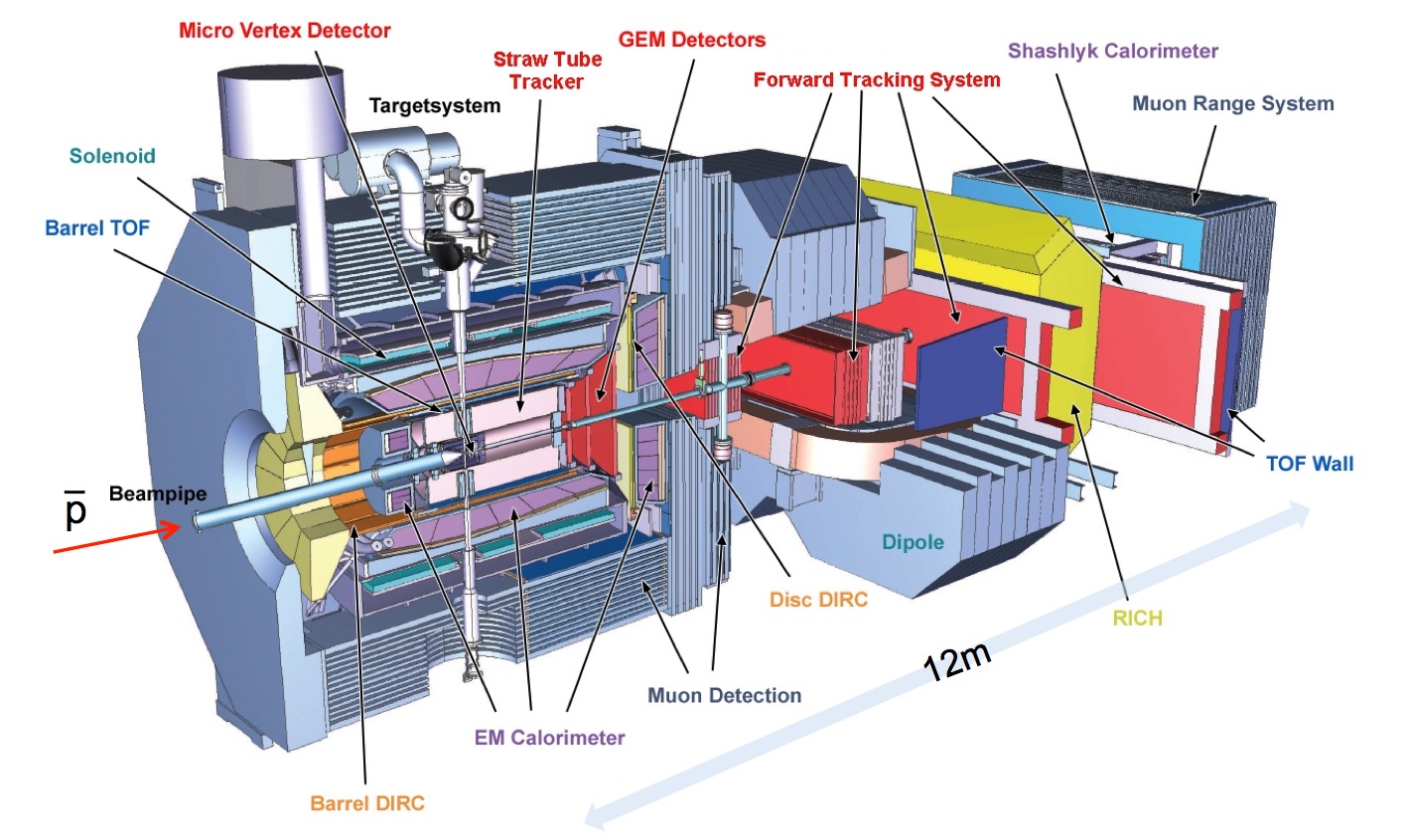}}
\caption{
Side view of \panda with the Target Spectrometer (TS) on the left side,
and the Forward Spectrometer (FS) starting with the dipole magnet
center on the right.
The antiproton beam enters from the left.
}
\label{fig:PANDA-sideview}
\end{center}
\end{figure*}

Figure~\ref{fig:PANDA-sideview} shows the \panda detector as a partial cut-out.
   As a fixed target experiment, it is asymmetric having two parts, 
the Target Spectrometer~(TS) and the Forward Spectrometer~(FS).
The antiproton beam is scattered off a pellet or cluster jet target
(left side in Fig.~\ref{fig:PANDA-sideview}).
\panda will measure $\bar{p}p$ reactions
comprehensively and exclusively, which
 requires simultaneous measurements of leptons and photons as well as
charged and neutral hadrons, with high multiplicities.

The physics requirements for the detectors are:
\begin{itemize}
\item to cover the full solid angle of the final state particles, 
\item to detect momenta of the reaction products, and
\item to identify  particle types  over the full range of momenta of the reaction products.
\end{itemize}

\subsection{Target Spectrometer}

The TS, which is almost hermetically sealed
to avoid solid angle gaps and which provides little spare space inside, 
consists of a solenoid magnet with a field of~2~T and a set of detectors for 
the energy determination of neutral and charged particles as well as for the 
tracking and PID for charged tracks housed within the superconducting solenoid magnet:
The silicon microvertex detector (MVD) closely abuts the beam pipe surrounding 
the target area and provides secondary vertex sensitivity for particles 
with decay lengths on the order of 100~$\mu m$.

Surrounding the MVD the main tracker is a straw tube tracker (STT).
There will be several tracking stations in the forward direction 
based on Gaseous Electron Multiplier foils (GEM) as gas amplification 
stages in order to stand the high forward particle rates.
The tracking detectors like MVD and STT also provide 
information on the specific energy loss in their data stream.

Two Detectors for Internally Reflected Cherenkov light (DIRC) are to be 
located within the TS. Compared to other types of Ring Imaging Cherenkov 
(RICH) counters the possibility of using thin radiators
and placing the readout elements outside the acceptance
favors the use of DIRC designs as Cherenkov imaging detectors for PID.
The Barrel DIRC, which is the topic of this document, covers the polar angles
$\theta$ from 22$^\circ$ to 140$^\circ$ inside the \panda TS with at least a 
3~s.d. $\pi$-K~separation up to 3.5~GeV/c.
The Endcap Disc DIRC covers the polar angles $\theta$
from  10$^\circ$ to 22$^\circ$ in the horizontal plane and 5$^\circ$ to 
22$^\circ$ in the horizontal plane.
For the analysis of the DIRC data the tracking information is needed,
as the Cherenkov angle is measured between the Cherenkov photon direction
and the momentum vector of the radiating particle. 
The track error of the measurement of the polar angle from the tracking system 
is expected to be 2-3~mrad.

The Scintillation Tile (SciTil) detector consisting 
of small scintillator tiles (3~cm~$\times$~3~cm), read out by Silicon 
PhotoMultipliers (SiPMs), and situated in the support frame outside the 
Barrel DIRC will have a time precision of 100~ps or less. 
In the absence of a start detector the SciTil will provide in combination with 
the forward time of flight system a good relative timing and event start time.

The lead tungstate (PWO) crystals of the electromagnetic calorimeters (EMC) 
are read out with Avalanche Photo Diodes (APD) or vacuum tetrodes. 
Both the light output and the APD performance improve with lower 
temperature. Thus the plan is, to operate the EMC detectors at 
\mbox{T=--25$^\circ$C}.
The EMC is subdivided into backward endcap, barrel and forward endcap,
all housed within the solenoid magnet return yoke.

Besides the detection of photons, the EMC is also the most powerful detector 
for the identification of electrons. 
The identification and measurement
of this particle species will play an essential role 
for the physics program of \panda.

The return yoke for the solenoid magnet in the \panda TS
is laminated to accommodate layers of muon detectors.
They form a range stack,
with the inner muon layer being able to detect low energy muons
and the cumulated iron layer thickness in front of the
outer layers providing enough hadronic material to stop
the high energy pions produced in \panda.

\subsection{Forward Spectrometer}

The FS angular acceptance has an ellipsoidal form
with a maximum value of $\pm$10 degrees horizontally
and $\pm$5 degrees vertically w.r.t. the beam direction.

The FS starts with a dipole magnet
to provide bending power with a B-field perpendicular to the forward tracks.
Most of the detector systems (except parts of the tracking sensors)
are located downstream outside the dipole magnet.

An aerogel RICH detector will be located between the dipole magnet and the 
Forward EMC.
A Time-of-Flight wall covers the identification of slow particles below the Cherenkov light threshold.

In the FS,
a Shashlyk-type electromagnetic calorimeter, consisting of 1512 channels of 
55~$\times$~55~mm$^2$ cell size, covers an area of 4.9~$\times$~2.2~m$^2$.
For the determination of the luminosity a detector based on four layers of 
monolithic active pixel sensors close to the beam pipe detects hits from the tracks of elastically scattered antiprotons. 

\subsection{The Particle Identification System} \label{sec:pandapid}

The charged particle identification (PID) will combine the information from 
the time-of-flight, tracking, dE/dx, and calorimetry with the output from the 
Cherenkov detectors. The latter focus on positive identification of kaons.

The individual \panda subsystems contributing to a global PID information 
have been reviewed in the report of a \panda study group on 
PID~\cite{schepers2} and are desribed in Sec.~\ref{cha:design}.

\subsection{Data Acquisition} \label{sec:daqbrief}
The data flow and processing is spatially separated 
into the Front End Electronics (FEE) part located on 
the actual detector subsystems and the Data Acquisition (DAQ), 
located off-detector in the counting room.

The FEE comprises analog electronics, digitization, low level pre-processing and optical data transmission to the DAQ system.

While each sub-detector implements detector specific FEE systems
 the DAQ features a common architecture and hardware for the complete \panda detector.

Operating the \panda detector at interaction rates of $2\times10^7$/s, typical event sizes of 4-20~kB lead to mean data rates of $\sim$\,200~GB/s.

The \panda DAQ design does not use fixed hardware based triggers but 
features a continuously sampling system where the various subsystems
are synchronized with a precision time stamp distribution system.

Event selection is based on real time feature extraction, filtering and high level correlations.

The main elements of the \panda DAQ are the data concentrators, the compute nodes, and high speed interconnecting networks.
The data concentrators aggregate data via point-to-point links from the FEE 
and the compute nodes provide feature extraction, event building and physics driven event selection.

A data rate reduction of about 1000 is envisaged in order to write event data of interest to permanent storage. 
 
Peak rates above the mean data rate of $\sim$200~GB/s and increased pile-up 
may occur due to antiproton beam time structure, target density fluctuations
(in case of pellet target) and luminosity variations during the HESR operation cycle.
 
FPGA based compute nodes serve as basic building blocks for the \panda DAQ system
 exploiting parallel and pipelined processing to implement the various real-time tasks,
 while multiple high speed interconnects provide flexible scalability to meet the rate demands.

\subsection{Infrastructure}

The \panda detector is located in an experimental hall,
encased in smaller tunnel-like concrete structure for radiation protection. 
Most subsystems connect their FEE-components via
cables and tubes placed in movable cable ducts to the installations
in the counting house, where three levels are foreseen to accommodate
cooling, gas supplies, power supplies, electronics,
and worker places. Only subcomponents, where cables must be as short as 
possible, will place racks or crates directly on the outside of the TS.

%% file: design/design.tex
\chapter{Design of the Barrel DIRC}
\label{cha:design}
\begin{bibunit}[unsrt]

The main objectives of the design of a Barrel DIRC counter for the
\panda experiment were to achieve clean separation of pions and kaons
for momenta up to 3.5~GeV/c, to follow a conservative approach,
inspired by the successful BaBar DIRC and optimized for the smaller
\panda experiment, and to minimize the production cost.

\section{Goals and Requirements}
\label{cha:design-goals}
The many different topics of the \panda physics program and the large 
investigated center-of-mass energy range between 2.2~GeV and 5.5~GeV require 
a rather wide phase space coverage with particle identification systems. 
Although a fixed target experiment tends to produce tracks with rather 
low $p_t$, pointing preferentially forward, many particles are emitted 
into the barrel region of the target spectrometer, defined as the
polar angle range between $22^{\circ}$ and $140^{\circ}$. 

Since signal reactions, e.g. from open charm and charmonium decays, 
predominantly proceed via strangeness production from weak decays of 
the charm quarks, the fraction of kaons going into the barrel part of the 
TS is of particular interest. 
In order to quantify this fraction, the following 16 event types (M1--M16) 
with kaons in the final state, comprising light hadron reactions, charmed 
baryons, charmonium and open charm events, were investigated:
\begin{multicols}{2}
\begin{enumerate}[label=(M\arabic*)]
\item $D^0\bar{D}^0$
\item $D^0\bar{D}^0\gamma$
\item $D^{\ast 0} \bar{D}^{\ast 0}$
\item $D^{0\ast} \bar{D}^{0\ast}\gamma$
\item $\Lambda_c^+ \Lambda_c^-$
\item $D^+ D^-$
\item $D^+ D^- \gamma$
\item $D_s^+ D_s^-$
\item $D_s^+ D_s^-\gamma$
\item $D^{\ast +} D^{\ast -}$
\item $\phi\phi$
\item $K^+ K^- \gamma$
\item $\eta_c \pi^+\pi^-$
\item $\eta_c \gamma$
\item $K^+ K^- 2\pi^+2\pi^-$
\item $K^+ K^- \pi^+\pi^-$
\end{enumerate}
\end{multicols}
All the reactions were generated with the EvtGen~\cite{EvtGen} event generator 
for anti-proton beam momenta between 4~GeV/$c$ and 15~GeV/$c$ to study 
the kinematic distributions of the final state kaons.
All possible decay channels were allowed for the generated particles.

\begin{figure}[h]
\begin{center}
\includegraphics[width=0.9\columnwidth]{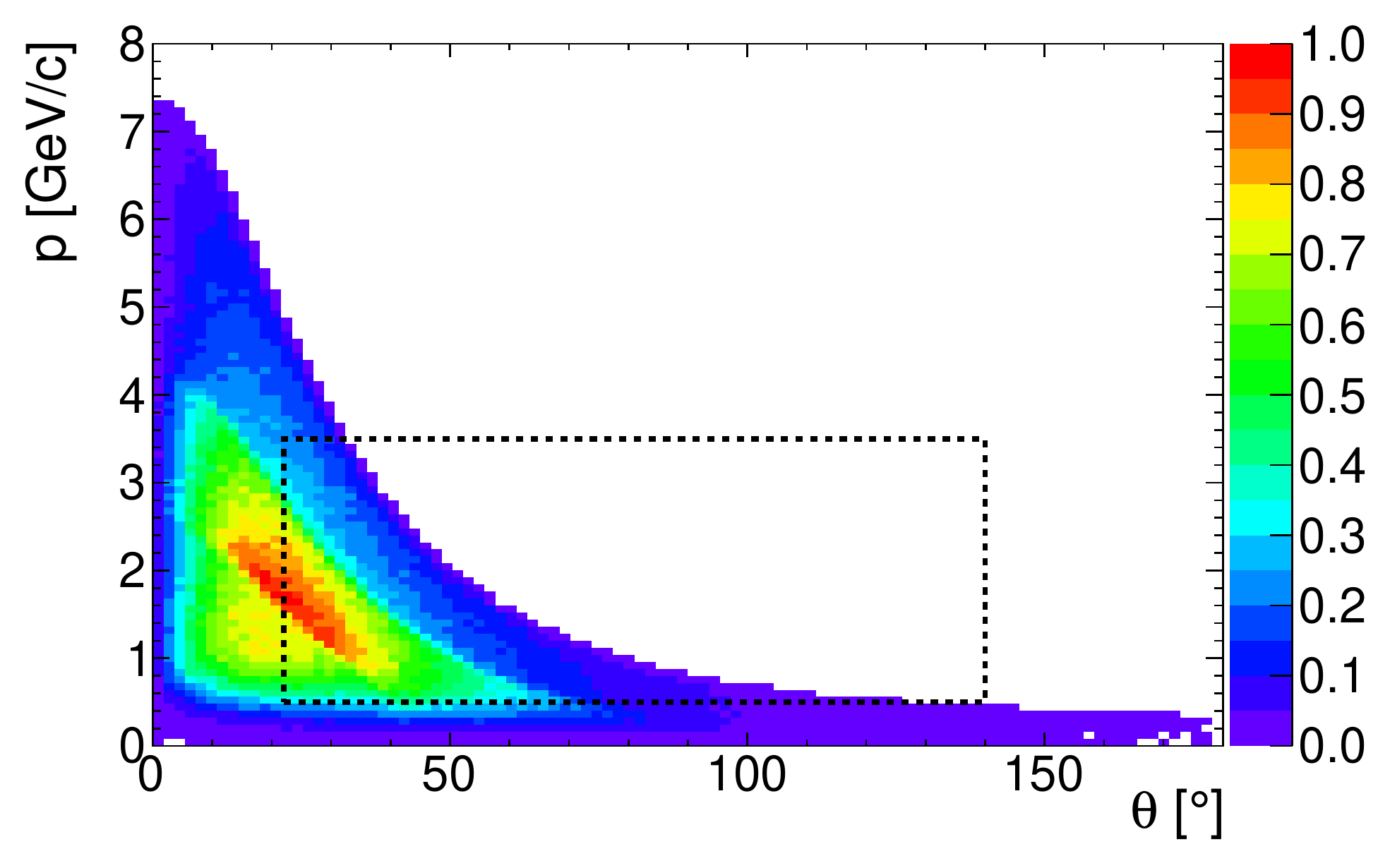}
\includegraphics[width=0.9\columnwidth]{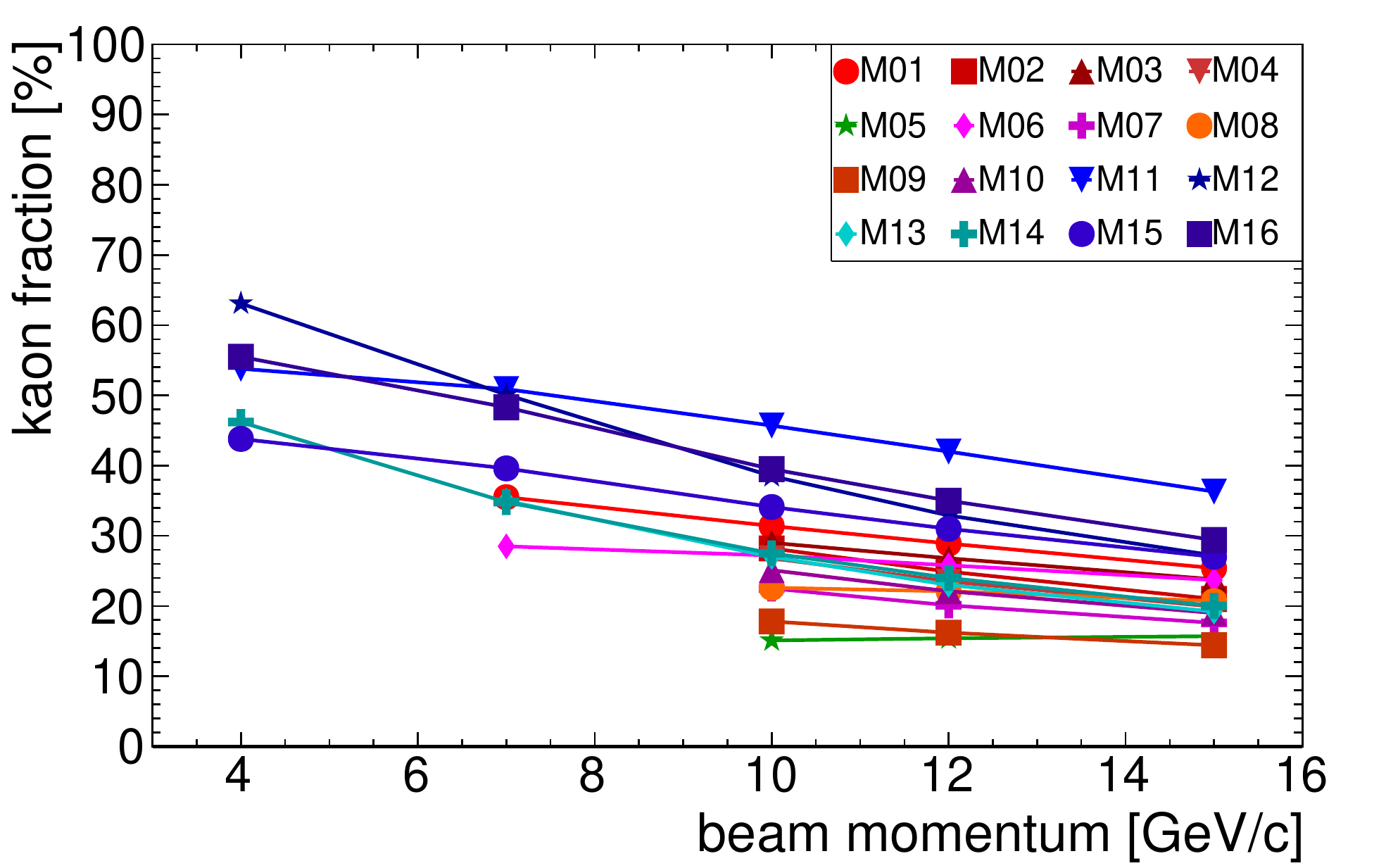}
\includegraphics[width=0.9\columnwidth]{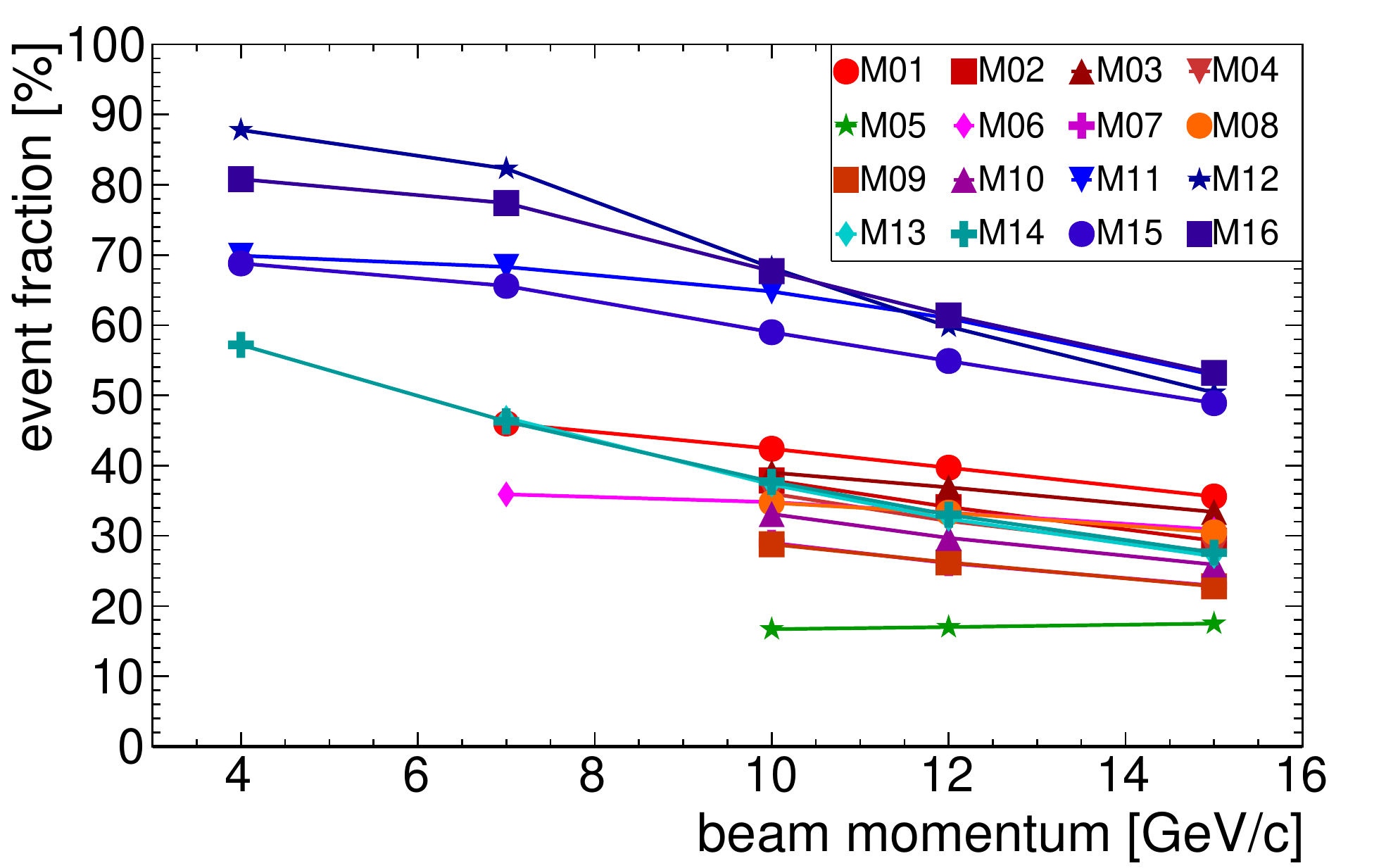}
\caption{\textit{Top:} Phase space distributions of kaons emitted for 
$p_{\bar{p}}$=~7~GeV/$c$ for eight benchmark channels (for details, see text). 
The Barrel DIRC coverage is marked with the dashed rectangle. 
\textit{Center:} Fractions of kaons within the Barrel DIRC phase space 
for 16 different reactions (see text for details) and beam momenta between 
4~GeV/$c$ and 15~GeV/$c$. 
\textit{Bottom:} Fractions of events producing at least one kaon in the 
Barrel DIRC phase space for 16 different reactions and beam momenta between 
4~GeV/$c$ and 15~GeV/$c$. 
}
\label{fig:kaon-phasespace}
\end{center}
\end{figure}

As an example, the top plot of Fig.~\ref{fig:kaon-phasespace} shows a 
superposition of the two-dimensional distributions of track momentum $p$ vs. 
track polar angle $\theta$ for the relevant channels at 
an antiproton beam momentum of 7~GeV/$c$, namely M1, M6, M11, M12, M13, M14, M15 and M16. 
In the polar angle region between $22^{\circ}$ and $140^{\circ}$, corresponding 
to the \panda TS barrel region, a large fraction of kaons have momenta below 
3.5 GeV/$c$. 
Summed over these eight equally weighted channels, 43\% of the kaons from 63\% of the 
reactions with final state kaons fall into that region of the TS. 

Investigations for all 16 event types (M1--M16) across the full beam momentum 
range are summarized in the two others plots of Fig.~\ref{fig:kaon-phasespace}. 
The center plot shows the fraction of kaons from the individual benchmark 
reactions within the Barrel DIRC phase space, defined as a momentum in the range
0.5--3.5~GeV/c and a polar angle of $22^{\circ}$ to $140^{\circ}$.
For small and intermediate beam momenta below 7~GeV/$c$, about 30\%--65\% 
of the kaons have to be detected by the Barrel DIRC. 
While the fraction of kaons in the barrel is reduced for higher beam momenta, 
even at the highest beam momentum of $p_{\bar{p}} = 15$~GeV/$c$ up 
to 40\% of the kaons are emitted into the barrel part of the phase space, 
depending on the reaction.

The bottom plot of Fig.~\ref{fig:kaon-phasespace} shows the fraction of events 
from the individual benchmark reactions with kaons in the final state 
producing at least one kaon in the Barrel DIRC phase space. 
Between 50\% and 90\% of the light hadron reactions (M11, M12, M15, M16) 
are affected over the full beam momentum range. 
Furthermore, at least one third of various open charm (e.g. M1, M3) and charmonium
reactions (M13, M14) require kaon identification in that region.

Given the fact that most of the hadrons produced in $\bar{p}p$ annihilations 
are pions, the hadronic charged PID in the TS has to be able to cleanly separate 
pions from kaons for momenta up to 3.5~GeV/$c$. 
The figure of merit in that respect is chosen to be the \textit{separation power} $N_{\rm sep}$. 
For Gaussian likelihood distributions it is defined as the absolute value of the 
difference of the two mean values ($\mu_1, \mu_2$) in units of the average of
the two standard deviations ($\sigma_1, \sigma_2$):
\begin{equation}
  N_{\rm sep} =\frac{|\mu_1-\mu_2|}{0.5(\sigma_1+\sigma_2)}\quad
  \label{eq:seppower}
\end{equation}
To ensure clean kaon identification, this quantity is required to be 
$N_{\rm sep}\geq 3$~s.d. over the full phase space 
$22^{\circ} < \theta< 140^{\circ}$ with $0.5\mbox{ GeV}/c < p < 3.5$~GeV/$c$. 
This corresponds to a mis-identification level of less than 4.3\% at 90\% efficiency. 

\begin{figure}[h]
\begin{center}
\includegraphics[width=0.9\columnwidth]{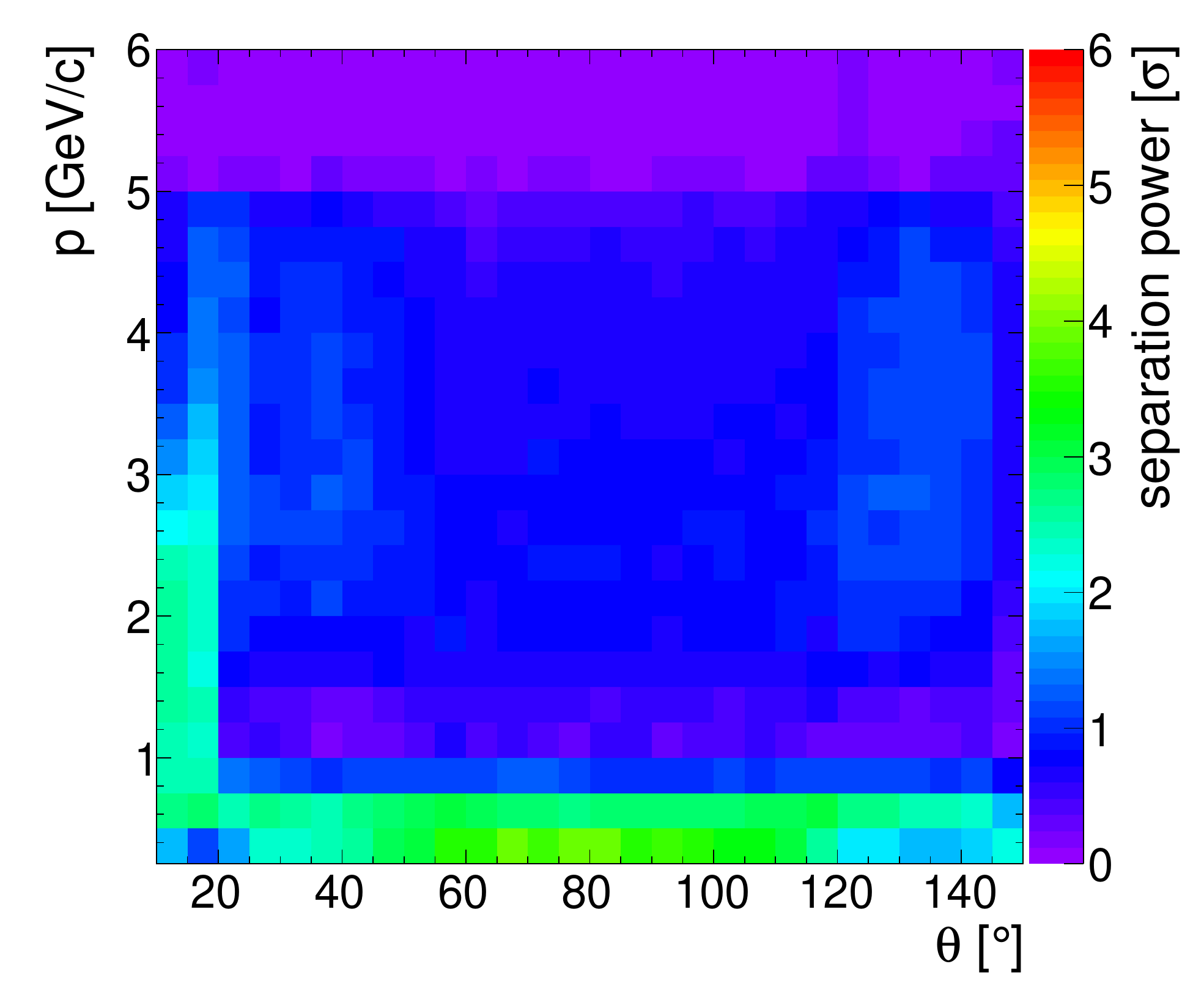}
\caption{Phase space map of the achievable $\pi$/$K$ separation power in standard 
deviations without a dedicated Cherenkov detector in the TS region. 
The map is based on $5\cdot 10^6$ single track kaon/pion events simulated and 
reconstructed with the PandaRoot framework. 
}
\label{fig:sep_wo_dirc}

\end{center}
\end{figure}

Figure~\ref{fig:sep_wo_dirc} shows the PID quality in terms of $\pi$/$K$ separation 
power for a \panda TS design without a dedicated PID system in the barrel region. 
For most of the phase space, the $\pi$/$K$ separation is at the level of 1~s.d.
or less.
While the tracking detectors provide a reasonable kaon identification via ${\rm d}E/{\rm d}x$
measurements for momenta $p<0.5$~GeV/$c$, this is not the case for higher momenta. 
The only detector providing PID in that region is the electro-magnetic calorimeter. 
It delivers a rather low hadron PID quality in the order of $N_{\rm sep}<1$.

The planned time-of-flight detector in the barrel region,
a scintillator tile hodoscope (SciTil) with a radius of $R=0.5$~m, 
is not yet fully implemented in the \panda software. 
However, assuming the time resolution to be $\sigma_t\approx 100$~ps on both the 
start and stop time, such a TOF system would only be able to contribute significantly 
to the identification of charged particles below 1~GeV/c.

A RICH counter using the DIRC
principle~\cite{ratcliff:dircprinciple01,ratcliff:dircprinciple02,ratcliff:dircprinciple03}
meets all the requirements for PID in the barrel region of the Target Spectrometer.
The first, and so far only, DIRC counter for a large high-energy physics experiment was 
used successfully in the BaBar 
experiment~\cite{design-babar:barreldirc} where it achieved more than 3~s.d. 
$\pi$/$K$ separation up to a momentum of 4.2~GeV/c.
A DIRC counter has many attractive features. 
It is thin in comparison to other PID systems, both in radius and radiation 
length, making it possible to decrease the size of the solenoid and the 
outer detectors, in particular the electromagnetic calorimeter, leading to
substantial overall cost savings.
Due to the dual nature of the DIRC fused silica bars, serving both as 
Cherenkov radiators and as light guides, the photon detection and readout
can be moved outside the densely populated active area of the central
region of the \panda detector.
Modern sensors and electronics make it possible to detect single
photons even in the magnetic field of about 1~T and at average interaction 
rates of about 10--20~MHz, making it possible to place the DIRC photon sensors
and readout electronics inside the magnetic yoke. 

\section{DIRC Principle}
\label{cha:dirc-principle}

\begin{figure}[htb]
\begin{center}
\includegraphics[width=0.95\columnwidth]{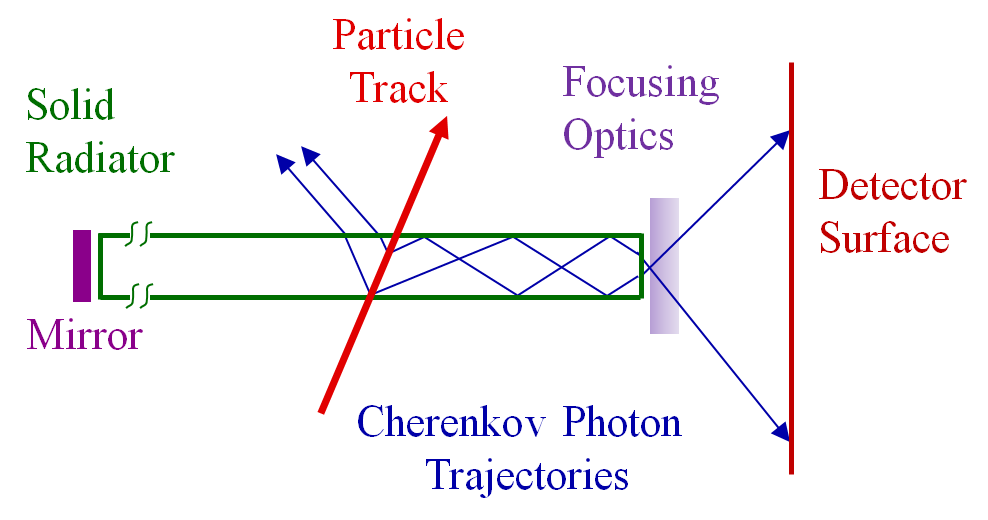}
\caption{Schematic of the basic DIRC principle.}
\label{fig:dirc-priciple}
\end{center}
\end{figure}

The basic principle of a DIRC counter is illustrated in Fig.~\ref{fig:dirc-priciple}.
Cherenkov photons are produced by a charged particle passing through a 
solid radiator with the refractive index $n$ if the velocity $v$ is larger than the 
speed of light in that medium \mbox{$(v > c/n)$}.
The photons are emitted on a cone with a half opening angle of 
$\cos \theta_C = 1/\beta n(\lambda)$, where, in a dispersive medium, 
$\theta_C$, the so-called Cherenkov angle, is a function of the 
photon wavelength $\lambda$.

\begin{figure}[htb]
\begin{center}
\includegraphics[width=0.90\columnwidth]{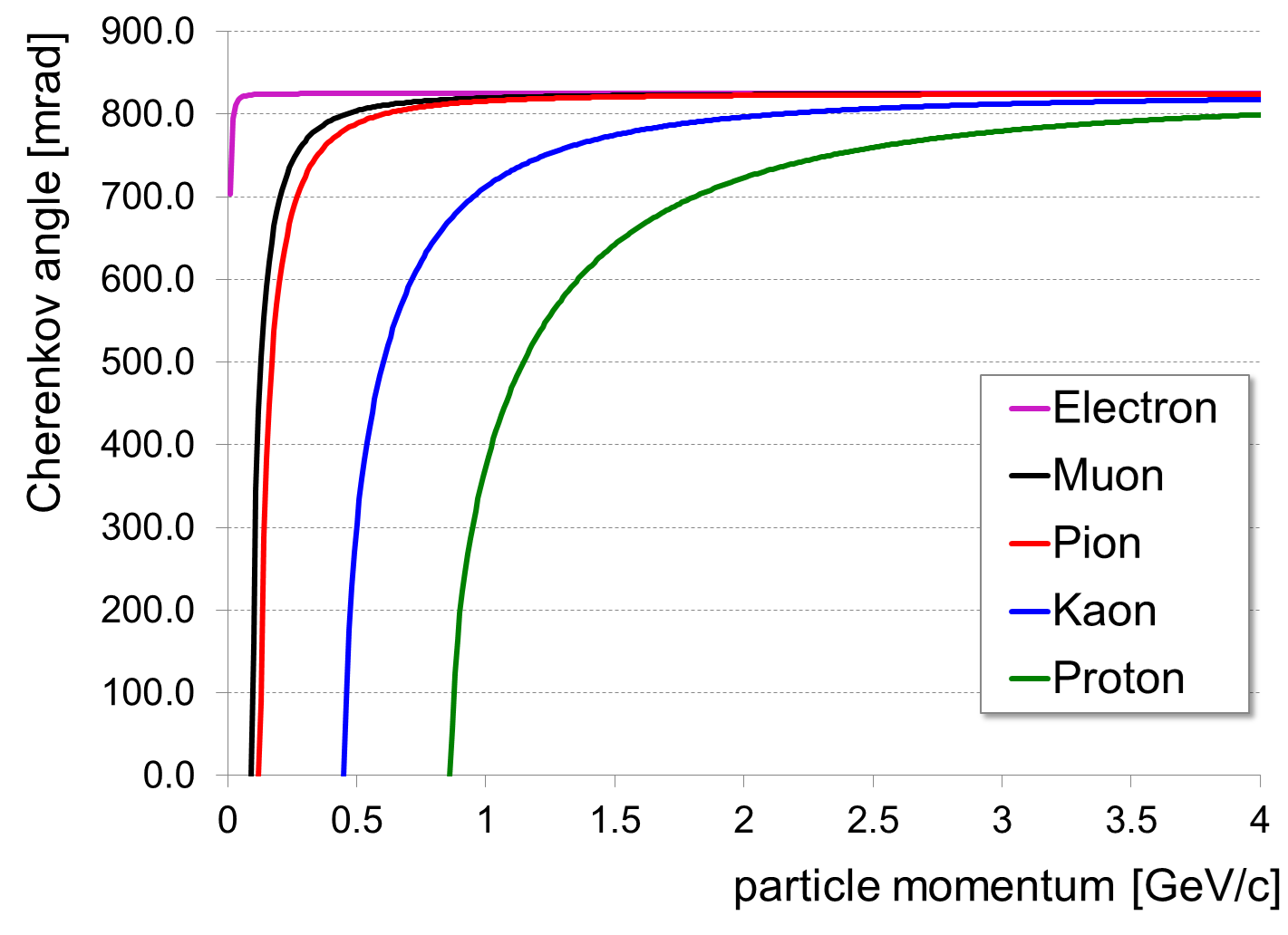}

 \vspace{5mm}

 \includegraphics[width=0.90\columnwidth]{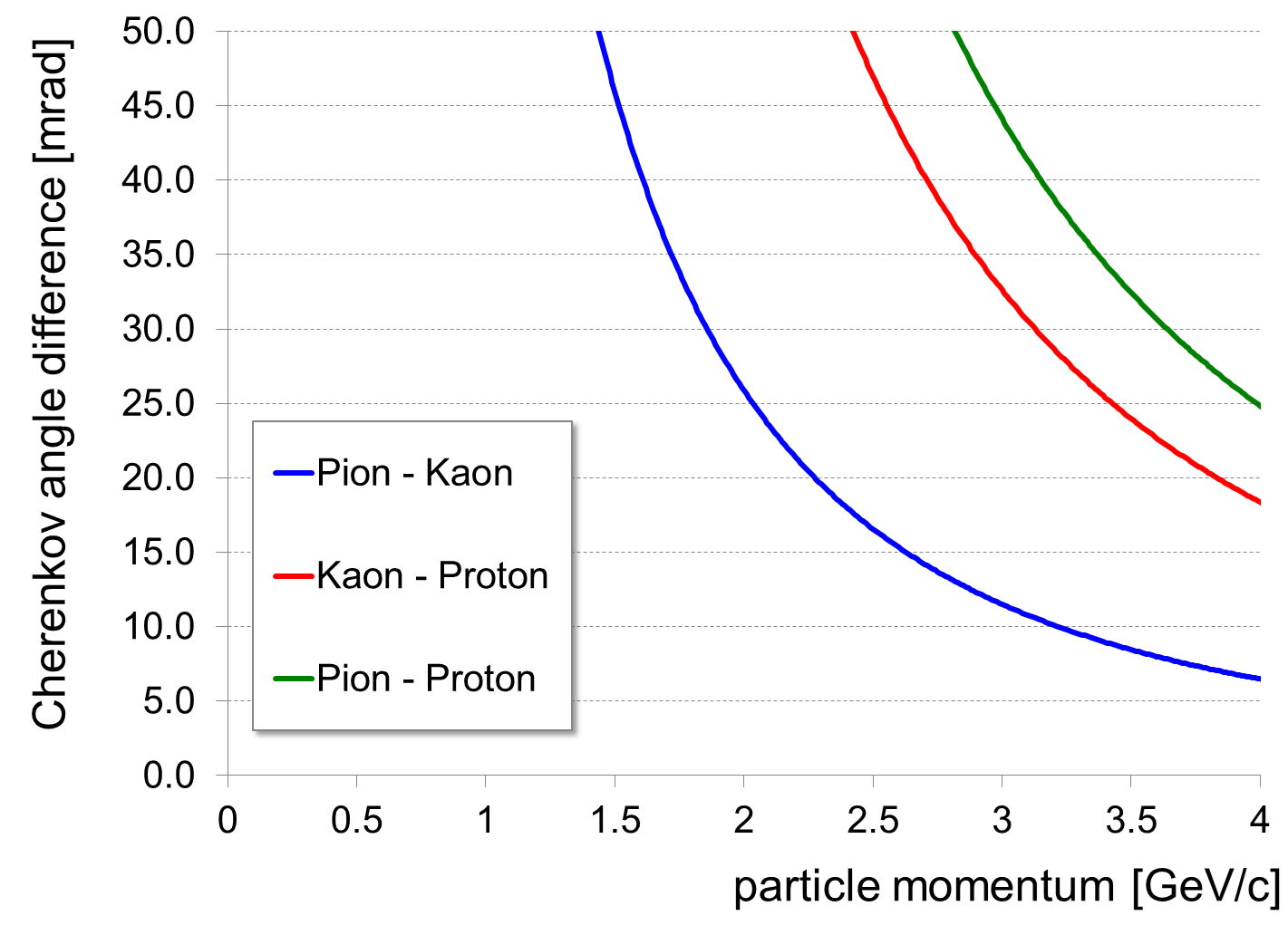}
\caption{\textit{Top:} Cherenkov angle as function of the particle momentum 
for charged particles in synthetic fused silica. 
\textit{Bottom:} Cherenkov angle difference in synthetic fused silica for pions and kaons, kaons and protons, and for pions and protons.}
\label{fig:pid-cherenkov}
\end{center}
\end{figure}

The radiator for a DIRC counter is typically a highly-polished bar made of 
synthetic fused silica.
The average Cherenkov angle for synthetic fused silica ($n=1.473$ at 380~nm) 
is shown as a function of the particle momentum in Fig.~\ref{fig:pid-cherenkov}~(top). 
For a particle with $\beta \approx 1$ some of the photons will always be trapped 
inside the radiator due to total internal reflection and propagate towards 
the ends of the bar.
A mirror is attached to the forward end of the bar to redirect the photons to 
the backward (readout) end.
If the bar is rectangular and highly polished the magnitude of the Cherenkov 
angle will be conserved during the reflections until the photon
exits the radiator via optional focusing optics into the expansion volume (EV).
The Cherenkov ring expands in the EV to transform the position information 
of the photon at the end of the bar into a direction measurement by determining
the positions on the detector plane. 
By combining the particle momentum measurements, provided by the tracking
detectors, with the photon direction and propagation time obtained by the 
photon sensor pixel, the Cherenkov angle and the corresponding PID likelihoods
are determined.

\section{DIRC PID Performance}
\label{cha:design-PIDperformance}

The PID performance of a DIRC counter is driven by the Cherenkov track angle 
resolution $\sigma_C$, which can be written as
\begin{equation}
\label{equ:sigma-track}
\sigma_C^2=\sigma_{C,\gamma}^2/N_\gamma+\sigma_{\mathrm{track}}^2 ,
\end{equation}
where $N_\gamma$ is the number of detected photons and $\sigma_{C,\gamma}$ is the
resolution of the Cherenkov angle measurement per photon (single photon resolution, 
SPR). 
$\sigma_{\mathrm{track}}$ is the uncertainty of the track direction within the
DIRC, which is dominated by multiple scattering and the resolution of the 
\panda tracking detectors, which is expected to be about 2~mrad for high-momentum
particles in the barrel region.

The SPR is defined by a number of contributions, 
\begin{equation}
\label{equ:spr}
\sigma_{C,\gamma}^2=\sigma_{\mathrm{det}}^2+\sigma_{\mathrm{bar}}^2+\sigma_{\mathrm{trans}}^2+\sigma_{\mathrm{chrom}}^2 ,
\end{equation}
where $\sigma_{\mathrm{det}}$ is the error due to the detector pixel size, 
$\sigma_{\mathrm{bar}}$ is the contribution from the size of the image of the bar,
including optical aberration and imaging errors, 
$\sigma_{\mathrm{trans}}$ is the error due to plate imperfections, such as non-squareness, 
and $\sigma_{\mathrm{chrom}}$ is the uncertainty in the photon production 
angle due to the chromatic dispersion $n(\lambda$) of the fused silica material.

The track Cherenkov angle resolution $\sigma_C$ required to cleanly separate charged 
pions and kaons in the DIRC can be extracted from 
Fig.~\ref{fig:pid-cherenkov}~(bottom) where the Cherenkov angle difference 
between pions, kaons, and protons is shown as a function of the particle momentum.
For the momentum of 3.5~GeV/c the $\pi/K$ separation is only
$\Delta(\theta_C)=8.5$~mrad.
Therefore, the design goal for the \panda Barrel DIRC is $\sigma_C < 2.8$~mrad
for the highest-momentum forward-going particles.

\section{The \panda Barrel DIRC}
\label{sec:barrel}

The concept of a DIRC counter as barrel PID system was proven by 
BaBar, further advanced by the R\&D for the SuperB FDIRC~\cite{superb:dirc,superb:dirc-optics},
and has been selected for the Belle~II experiment~\cite{belleII:TOP}.
The BaBar DIRC achieved a single photon Cherenkov angle resolution of 
$\sigma_{C,\gamma} \approx 10$~mrad, a photon yield of $N_\gamma=15-60$ 
photons per particle, depending on the polar angle, a track Cherenkov 
angle resolution of $\sigma_C = 2.4$~mrad, and clean $\pi$/$K$ separation
of 3~s.d. or more for momenta up to 4.2~GeV/c.

Since the BaBar DIRC performance meets the \panda PID requirements, the
conservative approach was to follow the BaBar DIRC design when possible 
and to modify and optimize it for \panda when necessary.

\subsection{The \panda Barrel DIRC Baseline Design}
\label{subsec:designbarrel}

\begin{figure}[h]
\begin{center}
\includegraphics[width=0.95\columnwidth]{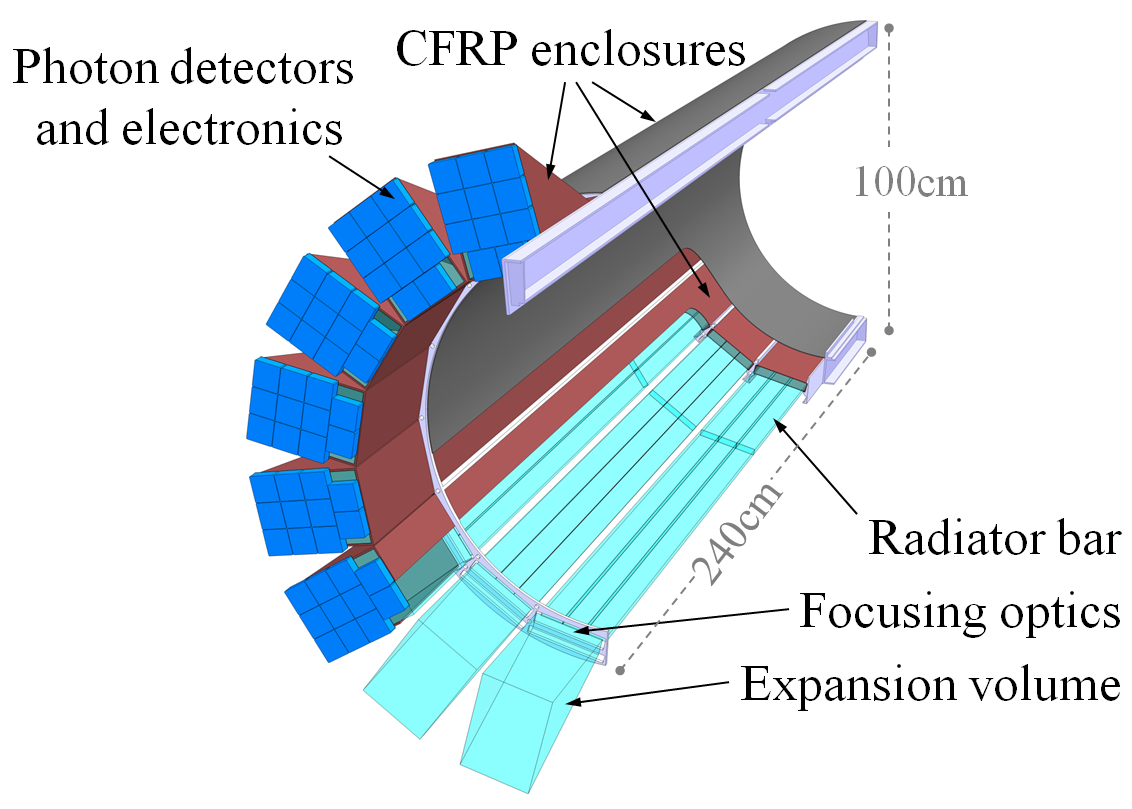}

\caption{
Schematic of the Barrel DIRC baseline design. Only one half of the detector is shown.
}
\label{fig:barrel-dirc-design}
\end{center}
\end{figure}

\begin{figure}[h]
\begin{center}
\includegraphics[width=0.95\columnwidth]{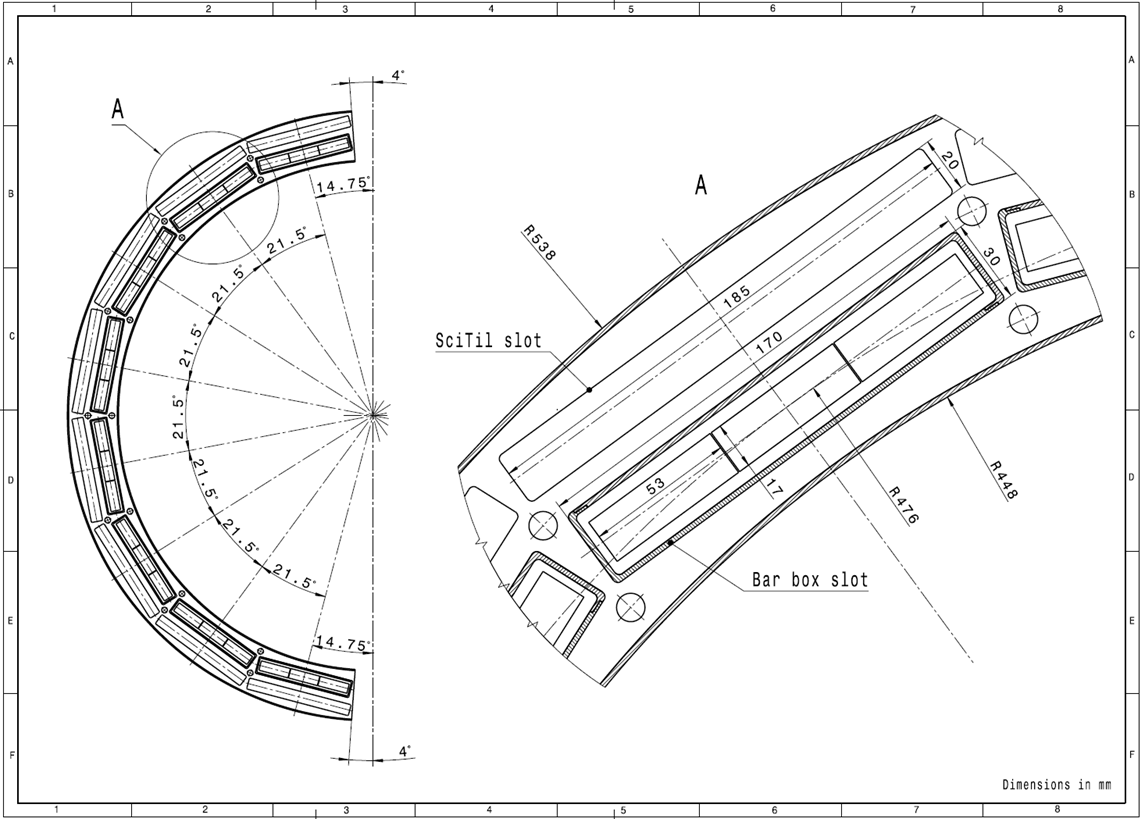}
\caption{
Central cross section view of the nominal Barrel DIRC geometry, including 
the space for the SciTil detector. 
}
\label{fig:barrel-dirc-design-cross-section}
\end{center}
\end{figure}

The baseline design of the \panda Barrel DIRC detector is
shown in Fig.~\ref{fig:barrel-dirc-design} and Fig.~\ref{fig:barrel-dirc-design-cross-section}.
16 optically isolated sectors, each comprising a bar box and a solid 
fused silica prism, surround the beam line in a 16-sided polygonal barrel 
with a radius of 476~mm and cover the polar angle range of  
22$^\circ$ to 140$^\circ$. 

Each bar box contains three bars of 17~mm thickness, 53~mm width, 
and 2400~mm length (produced by gluing two 1200~mm-long bars back-to-back
using Epotek 301-2~\cite{lit-design-epotek}), 
placed side-by-side, separated by a small air gap.
A flat mirror is attached to the forward end of each bar to reflect 
photons towards the read-out end, where they are focused by a three-component 
spherical compound lens on the back of a $30$~cm-deep solid prism, 
made of synthetic fused silica, serving as expansion volume. 
The location and arrival time of the photons are measured by an array 
of 11 lifetime-enhanced Microchannel Plate PhotoMultiplier Tubes 
(MCP-PMTs) with a precision of about 2~mm and $100$~ps, respectively.
The MCP-PMTs are read out by an updated version of the HADES trigger and 
readout board (TRB)~\cite{trb3-jinst} in combination with a front-end
amplification and discrimination card mounted directly on the 
MCP-PMTs~\cite{cardinali:padiwa}.
The sensors and readout electronics are located in the region close to 
the backward end-cap of the solenoid where the magnetic field strength is 
\mbox{$B\approx 1$~T}.

Since the image plane is located on the back surface of the prism, 
a complex multi-layer spherical compound lens is required to match 
the focal plane to this 
geometry using a combination of focusing and defocusing elements.
A layer of lanthanum crown glass (NLaK33, refractive index n=1.786 
for $\lambda$=380~nm) between two layers of synthetic fused silica 
(n=1.473 for $\lambda$=380~nm), creates two refracting surfaces.
The transition from fused silica to NLaK33 is defocusing while the  
transition into fused silica focuses the photons.
Due to the smaller refractive index differences the use of a 
high-refractive index material avoids the total internal
reflection losses at the lens transitions that are associated with 
air gaps.
The lens is glued to the bar with Epotek 301-2 and serves also as exit window 
of the bar box.
The optical coupling between the bar box and the prism will be provided by a 
silicone cookie, made, for example, from Momentive TSE3032~\cite{momentive}
material. 

The components of the modular mechanical system are made of 
Carbon--Fiber–-Reinforced Polymer (CFRP).
The light-tight CFRP containers for the bars (bar boxes) slide into the 
\panda detector on rails that connect slots in two rings which are 
attached to the main support beams (see Fig.~\ref{fig:barrel-dirc-design}).
A cross section of the CFRP structure and a bar box can be seen
in Fig.~\ref{fig:barrel-dirc-design-cross-section}.
Similar CFRP containers house the prisms and front-end cards so that
each sector is joined into one light-tight unit and optically isolated 
from all other sectors.
To remove moisture and residue from outgassing of the bar box components
as well as glue and silicone materials, the CFRP containers are constantly 
flushed by boil-off dry nitrogen.
To facilitate access to the inner detectors of \panda, the modular design 
allows the entire frame holding the prisms, sensors, and electronics to be 
detached from the barrel structure that holds the bar boxes during extended
shutdowns periods.
An additional advantage of the modular design is that the installation of bar
boxes could be staged, in case of fabrication delays, with minimal impact
on the neighboring \panda subsystems.

\begin{figure}[htb]
\begin{center}
\vspace*{-3mm}
\includegraphics[width=0.8\columnwidth]{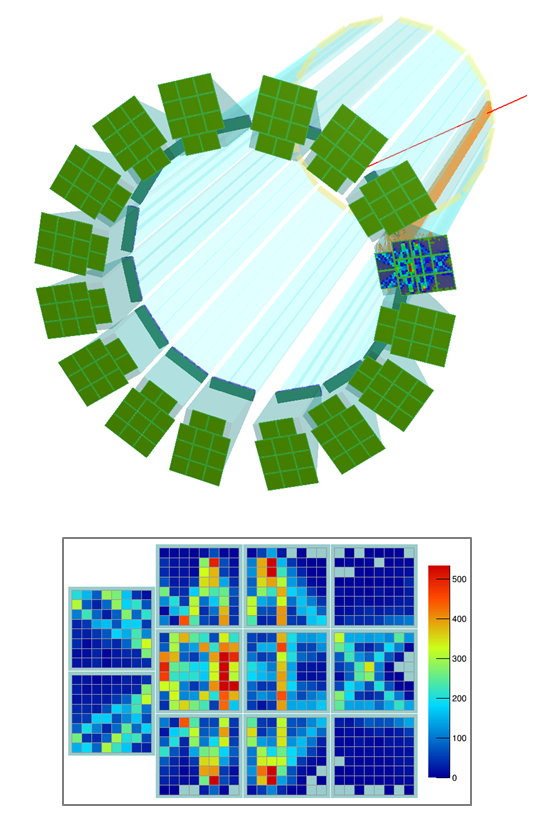}
\caption{
Geant simulation of the \panda Barrel DIRC baseline design. 
The colored histogram at the bottom shows the accumulated hit pattern from 
1000~$K^+$ at 3.5~GeV/c and $25^\circ$ polar angle.
}
\label{fig:barrel-dirc-baseline-geant}
\end{center}
\end{figure}

A Geant simulation of the baseline design is shown in
Fig.~\ref{fig:barrel-dirc-baseline-geant}. 
A kaon track (red line) produces Cherenkov photons (orange lines), which
are detected on the MCP-PMT array. 
The accumulated histogram shows the distinctive hit pattern, typical for
DIRC counters, where the conic section of the Cherenkov ring is projected on 
the flat detector plane after many internal reflections in the bar and prism.

\begin{table*}[htb!]
\caption{Comparison of Barrel DIRC design parameters.}
\label{Tab:dirc-designs}
\vspace*{3mm}
\begin{center}
\begingroup
\setlength{\tabcolsep}{6pt} 
\renewcommand{\arraystretch}{1.5} 

{\small\begin{tabular*}{0.95\textwidth}[]{llll}
\hline  
& BaBar & Belle~II TOP & \panda \\ 
\hline 
Radiator geometry   &   Narrow bars (35~mm)	   &  Wide plates (450~mm)	    &   Wide bars (53~mm)\\
Barrel radius 	    &   845~mm			   &  1150~mm			     &  476~mm\\
Bar length 	    &   4900~mm (4$\times$1225~mm) &   2500~mm (2$\times$1250~mm)	     &  2400~mm (2$\times$1200~mm)\\
Number of long bars  &  144 (12$\times$12 bars)  &   16 (16$\times$1 plate)	&       48 (16 $\times$3 bars)\\
EV material    &   Ultrapure water	   &  Fused silica 	     &  Fused silica\\
EV depth    &   1100~mm	   &  100~mm 	     &  300~mm\\
Focusing 	    &   None (pinhole)			&	Mirror			     &  Lens system\\
Photon detector 	    &   $\approx$~11k PMTs		  &   $\approx$~8k MCP-PMT pixels 	     &  $\approx$~11k MCP-PMT pixels\\
Timing resolution   &   $\approx$~1.7~ns			   &  $\approx$~0.1~ns			     &  $\approx$~0.1~ns\\
Pixel size	     &  25~mm diameter		   &  5.6~mm $\times$ 5.6~mm	     &  6.5~mm $\times$ 6.5~mm\\
PID goal 	     &  3 s.d. $\pi$/$K$ to 4~GeV/c  &   3 s.d. $\pi$/$K$ to 4~GeV/c	&       3 s.d. $\pi$/$K$ to 3.5~GeV/c\\
Time line	     &  Operation 1999 -- 2008		   &  Installation 2016		     &  Installation 2023\\[1mm]
\hline
\end{tabular*} 
}
\endgroup
\end{center}
\end{table*}

\subsubsection*{Key Design Improvements} 

Since \panda is smaller than the BaBar detector, several design modifications 
were required compared to the BaBar DIRC.
Additional changes were the result of the optimization of cost vs. performance.
The main parameters of the DIRC counters for BaBar, Belle~II, and \panda
are summarized in Table.~\ref{Tab:dirc-designs}.

\begin{itemize}
\item \textit{Radiator bar size}

Due to the tight optical and mechanical spe\-ci\-fi\-cations the fabrication 
of the radiator bars remains one of the dominant cost drivers for DIRC counters.
A significant cost reduction is only possible if fewer pieces have to be polished.
Detailed physical simulation studies (see Sec.~\ref{ch:performance} and 
Ref.~\cite{MP-MPatsyuk-PHD-THESIS-D}) demonstrated that reducing the number of
bars per bar box from 5 bars (32~mm width) to 3 bars (53~mm width) does not
affect the PID performance since the lens system is able to correct for 
the increase in bar size.

\item \textit{Compact fused silica prism as expansion volume}

The overall design of the \panda experiment required that the large water
tank used by the BaBar DIRC is replaced with a compact expansion volume,
placed inside the detector.
Initial tests with a 30~cm-deep tank filled with mineral oil showed a good
single photon Cherenkov angle resolution.
However, the use of mineral oil inside the detector caused concern for possible
spills and the optical quality of the oil led to a loss of some 20-30\% of
photons inside of the tank.

Fused silica as material and separated smaller units as expansion volume 
were already favored by the SuperB FDIRC and the Belle~II TOP.
The superior optical quality increases the photon yield and the direct match
of a bar box to a prism EV simplifies the alignment.

The prism also allows a smaller EV opening angle compared to a larger tank
since the image is folded within the EV after reflections off much 
higher-quality optical surfaces than a tank would provide.
This reduces the photon detection area and, thus, the number of required
MCP-PMTs, the other main cost driver for the Barrel DIRC.
The additional reflections inside the prism are taken into account in the
reconstruction software and do not cause any PID performance degradation.

\item \textit{Focusing optics}

The larger bar size and the smaller expansion volume make the use of 
focusing elements necessary.
Initial tests, performed with a traditional spherical fused silica lens 
with an air gap showed sharp ring images but an almost complete loss of 
photon yield for track polar angles near $\theta = 90^\circ$ due to 
total internal reflection at the air gap.
This photon loss is avoided by using the high-refractive index material
in the compound lens.
Several iterations of 2-layer and 3-layer cylindrical and spherical lens
designs were tested in prototypes in the optical lab and with particle beams.
The latest 3-layer spherical lens achieves a flat focal surface, which 
is an excellent match to the prism geometry, as well as a consistently high 
photon yield for all polar angles.
The radiation hardness of the NLaK33 material is a possible concern for
\panda and measurements of the radiation hardness in an X-ray source are
currently ongoing.

\item \textit{Compact multi-anode photon detectors}

The smaller expansion volume requires not only focusing optics to reduce the
contribution from the bar size to the angular resolution but also smaller 
photodetector pixels.
With a pixel size of 6.5~mm$\times$6.5~mm MCP-PMTs meet the requirements for 
spatial resolution and provide a single photon timing resolution of 30--40~ps
for a gain of about 10$^6$.
They work in the magnetic field of 1~T and tolerate the expected photon hit rates of 200~kHz/pixel.

For many years the main challenge for the use of MCP-PMTs in \panda was the 
photon flux, expressed as the integrated anode charge. 
Recent improvements in the fabrication technique have increased the lifetime 
of MCP-PMTs to significantly more than the 5~C/cm$^2$ integrated anode charge
expected during 10 years of operating the Barrel DIRC at design luminosity.

The excellent photon timing provided by the MCP-PMTs, in combination with 
fast readout electronics, make it possible to measure the photon time of
propagation with about 100~ps resolution. 
This fast timing is essential in the use of the time-based imaging, required
for the wide plate design and helps in the reconstruction of the Cherenkov 
angle for the baseline design by suppressing ambiguities due to reflections 
in the bar and prism.
Ultimately it may even make it possible to mitigate the influence of the
chromatic dispersion of the Cherenkov angle (see Eqn.~\ref{equ:spr}) and
to further improve the PID performance~\cite{Benitez08}.

\end{itemize}

\subsection{The \panda Barrel DIRC Design Option: Wide Radiator Plates}

\label{subsec:platedesign}

A significant additional reduction of the cost of radiator fabrication
would be possible if one wide plate per bar box would be used instead of 3 bars.
The Belle~II TOP counter demonstrated that high-quality wide plates can be 
fabricated by optical industry~\cite{belleII:TOP-Krizan}.
During the \panda Barrel R\&D phase two 160~mm-wide prototype plates 
were produced by industry and found to meet the specifications.

Geant simulation and the implementation of a time-based likelihood reconstruction 
approach~\cite{MZ-MZuehlsdorf-PHD-THESIS-d}, inspired by the Belle~II TOP, 
demonstrated that two designs with a wide plate, either with a cylindrical 
3-layer lens or without any focusing, meet the PID requirements for \panda.

\begin{figure}[h]
\begin{center}

\includegraphics[width=0.95\columnwidth]{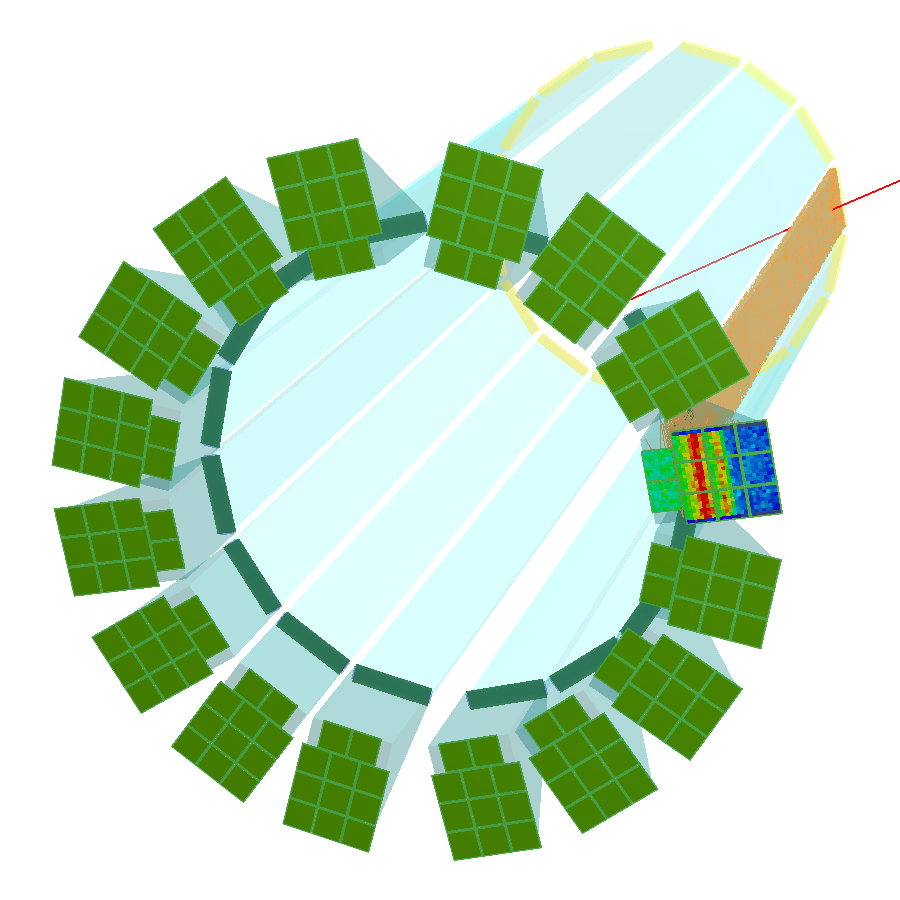}
\caption{
Geant simulation of the \panda Barrel DIRC design option with wide plates. 
The colored histogram shows the accumulated hit pattern from  1000~$\pi^+$ at 3.5~GeV/c 
momentum and $25^\circ$ polar angle.
}
\label{fig:barrel-dirc-plate-option}
\end{center}
\end{figure}

A Geant simulation of the design option with wide plates is shown in
Fig.~\ref{fig:barrel-dirc-plate-option}. 
A pion track (red line) produces Cherenkov photons (orange lines), which
are detected on the MCP-PMT array. 
The accumulated histogram shows the plate hit pattern which no longer 
exhibits the typical DIRC ring segments that were visible for the narrow 
bar design (Fig.~\ref{fig:barrel-dirc-baseline-geant}).
The time-based reconstruction approach, however, is able to process this
hit pattern to cleanly separate
pions from kaons for the entire \panda phase space.

It is important to note that the choice of the radiator width has little or 
no impact on the mechanical design of the barrel or expansion volume 
components.
Aside from the choice of focusing the only difference is that one wide 
mirror would replace 3 narrow mirrors. 
The construction of the bar boxes would stay the same and assembly would 
be simplified since the careful separation of the bars in the bar box
is no longer required.

\putbib[./literature/lit_design]

\end{bibunit}

%% file: simulation/simulation-reconstruction.tex
\chapter{Simulation and Reconstruction}
\label{cha:simulation}
\begin{bibunit}[unsrt]

A detailed physical simulation of the \panda Barrel DIRC was developed 
in Geant~\cite{Geant4} to design the detector, optimize the performance, 
and to reduce the system cost. 
Two reconstruction algorithms were used to perform the performance evaluation 
for a number of different Barrel DIRC designs with narrow bars and wide plates.

\section{Input to the Simulation}
\label{sec:sim-input}

\begin{figure}[htb]
  \centering

    \includegraphics*[width=0.45\textwidth]{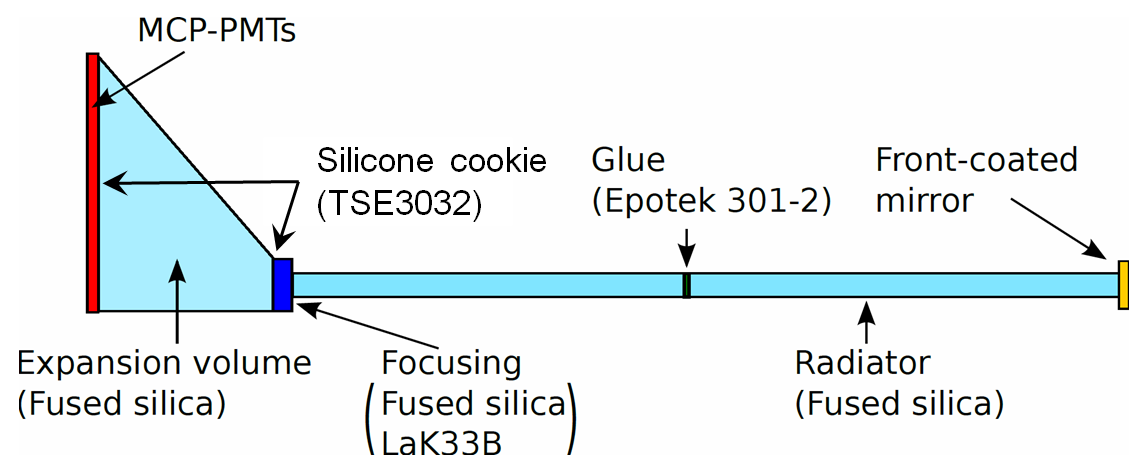} 
  \caption{Simplified side view of one section of the \panda Barrel DIRC with the main components and their materials.
    Not to scale.}
  \label{lSimpleDirc}
\end{figure}

All detector components are assembled as individual volumes in Geant and used as media for 
particle transport.  
Figure~\ref{lSimpleDirc} shows a schematic side view of one of the 16 sections of the \panda Barrel 
DIRC with the main components and materials used in the simulation. 
These include the synthetic fused silica prism and radiators, lenses made from fused 
silica and NLaK33B~\cite{nlak} material, front-coated mirrors, as well as Epotek 301-2~\cite{lit-sim-epotek}
glue, Momentive TSE3032~\cite{momentive2} silcone cookies, and Eljen EJ-550~\cite{eljen} optical 
grease for connecting different components. 
The MCP-PMTs are constructed from a fused silica entrance window and 
a bialkali photocathode. 
All mechanical structures are made of Carbon--Fiber–-Reinforced Polymer (CFRP).

\begin{figure}[htb]
\begin{center}
\vspace*{-3mm}

\includegraphics[width=0.8\columnwidth]{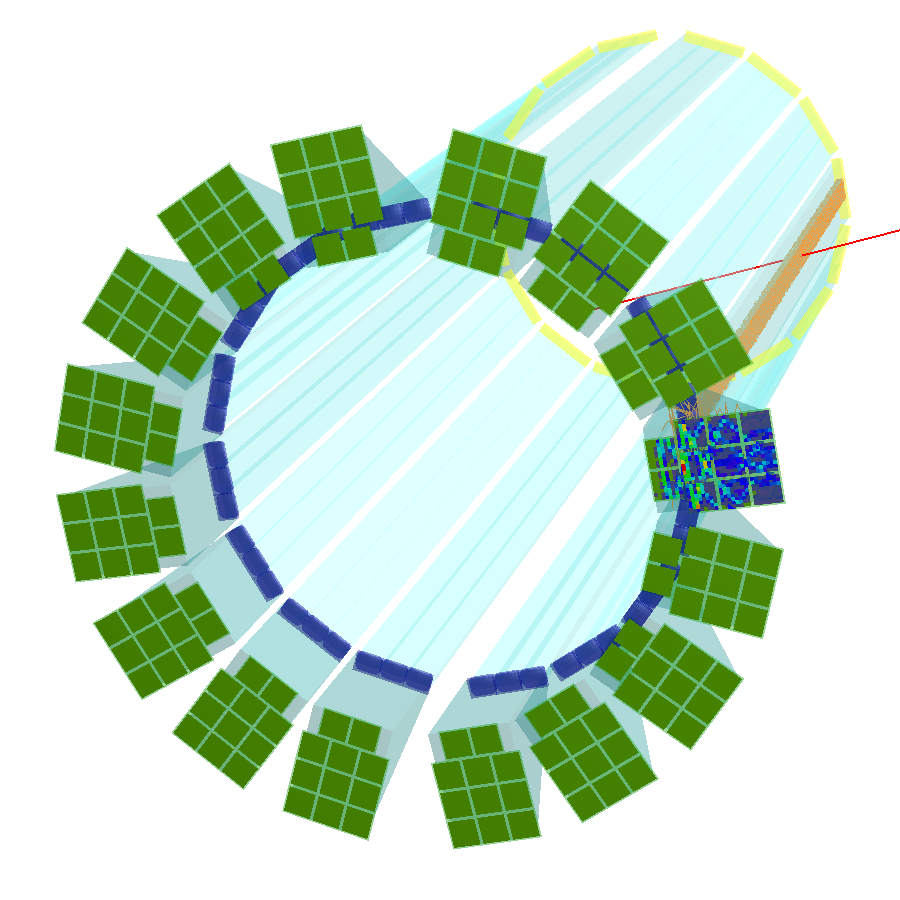}
\includegraphics[width=0.8\columnwidth]{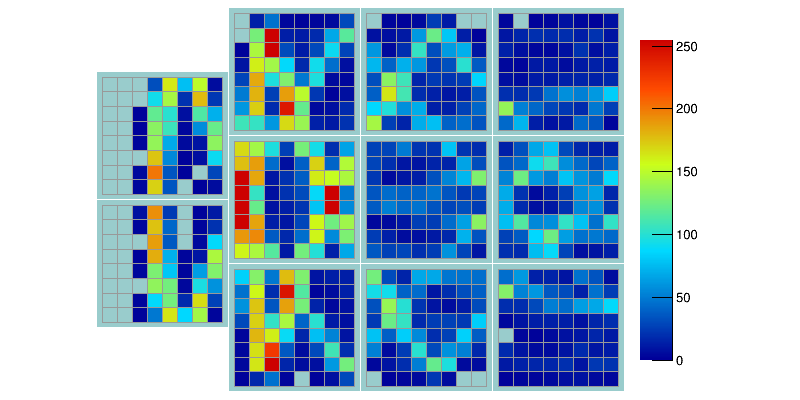}
\caption{
  Geant simulation of the baseline geometry of the \panda Barrel DIRC.
  The colored histogram at the bottom shows the accumulated hit pattern from 1000~$K^{+}$
  at 3.5~GeV/c momentum and $55^{\circ}$ polar angle.}
\label{lSimExample-sr}
\end{center}
\end{figure}

The simulation is performed within the PandaRoot framework~\cite{pandaroot} and
includes event generation, particle transport, digitization, hit finding, and reconstruction. 
The particle transport uses the Virtual Monte Carlo approach, which allows for
easy switching between Geant3 and Geant4 for systematic studies.
All Geant simulation results shown in the report were obtained using Geant4.

Figure~\ref{lSimExample-sr} shows the Geant representation of the \panda Barrel DIRC
baseline design together with an example of the accumulated hit pattern produces by 
Cherenkov photons (orange lines) from 1000 $K^{+}$ tracks (red line).

\begin{figure}[htb]
  \includegraphics*[width=0.48\textwidth]{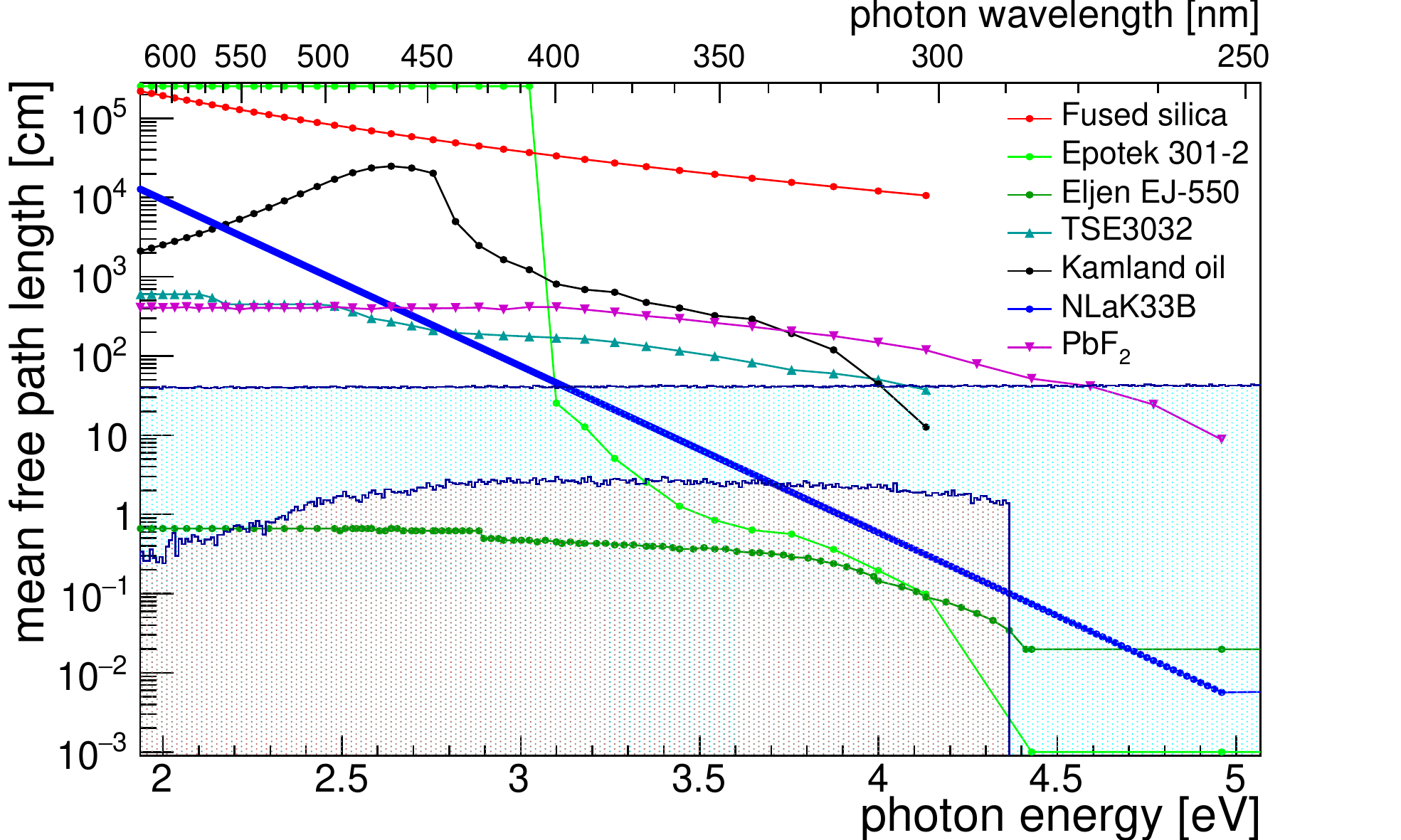}
  \caption{Wavelength dependent attenuation length for the optical materials used.
  The shaded areas indicate the energy spectra of generated and detected Cherenkov 
  photons, respectively.}
  \label{fig:atenuationlength}
\end{figure}

The simulation of the transport of Cherenkov photons includes the wavelength-dependent
properties of all optical materials, such as the index of refraction of fused silica,
NLaK33B, the bialkali photocathode, the coefficient of total internal reflection for 
a surface roughness of polished fused silica bars, and the attenuation length, 
shown in Fig.~\ref{fig:atenuationlength}

The digitization stage simulates the realistic detector response of the
photon detectors. 
This includes charge sharing, dark noise, collection efficiency, quantum 
efficiency, and the single photon timing resolution measured for the 
MCP-PMTs studied for the Barrel DIRC (see Sec.~\ref{sec:sensors} for details).

Background from hadronic interaction and delta electrons is simulated as well as 
backsplash particles from the electromagnetic calorimeter.

\begin{figure}[htb]
     \includegraphics[width=0.5\textwidth]{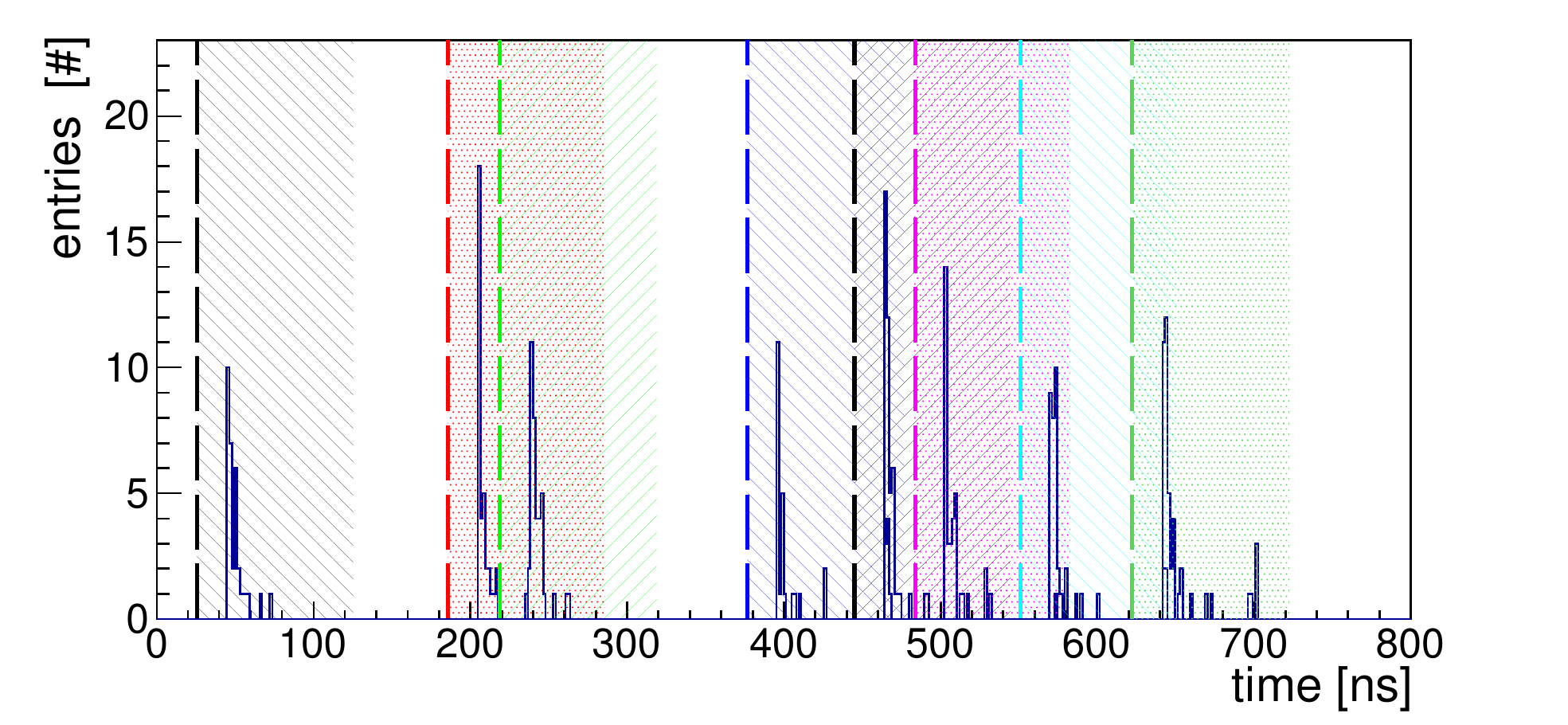}
  \vspace*{-3mm}
  \caption{Time spectrum of the hits in the Barrel DIRC at 20~MHz event rate after the hit finder. 
  Different colors represent different events. 
  The vertical lines indicate the start time of the events. 
  Shaded areas indicate the time window of reconstructed events.}
  \label{lStime}
\end{figure}

Most of the simulation results presented in this report were obtained using
event-based simulation with either an event generator, like the Dual Parton 
Model (DPM)~\cite{dpm}, or with single particles generated at the nominal 
interaction point.
To evaluate the possible impact from the dead time of the readout 
electronics or the time structure of \panda data events at high 
luminosities on the Barrel DIRC reconstruction, a time-based 
simulation was implemented as well.

One challenge of the time-based structure of the data is the ambiguous
assignment of hits to the events due to the pile-up effect. 
Cherenkov photons, generated by a track in a radiator, may propagate by
total internal reflection for up to 30~ns before they reach the sensor.
At high luminosities, when the time between subsequent interactions 
approaches 100~ns or less, a reconstructed event may therefore obtain 
DIRC hits from neighboring events.
Figure~\ref{lStime} shows an example of eight events with at least one 
track in the Barrel DIRC, produced with time-based simulation for 
a 20~MHz interaction rate.
Some of these events show overlapping event reconstruction time windows
and several of the Cherenkov photon signals are located in more than one 
event time window. 
However, the tracks from different events usually hit Barrel DIRC radiators 
in different sectors and are thus detected by sensors attached to
different prisms.
If two tracks hit the same bar box they are still usually well-separated 
in space and time and can be successfully assigned to the correct event 
without loss of photons. 
The most challenging case is two particles hitting the same radiator bar, 
causing overlapping photon hit patterns, which happens in about 2\% of the 
events, according to the DPM event generator.
Even for those events, simulation shows that the hits will be correctly 
assigned in 80--90\% of the cases by assigning hits based on the calculated 
difference between the measured and reconstructed photon propagation 
time in the radiator and prism.

The combination of the proposed Barrel DIRC photon detectors and readout 
electronics is expected to have a dead time of up to 40~ns.
The probability for a second Cherenkov photon to hit a pixel within
this dead time is driven by the number of photons produced in one event, 
the size of the sensor area covered by the ring images, and the event rate.
The compact expansion volume of the \panda Barrel DIRC baseline design
causes the Cherenkov hit pattern to consist of overlapping ring segments, 
spread over a small area.
Time-based simulation, using the DPM event generator with an event rate of 
20~MHz, predicts that about 11\% of the Cherenkov photons are lost on 
average due to pile-up and dead time.

The impact of these photon loss processes on the Barrel DIRC PID performance 
is rather small. 
A worst case estimate can be made for the region with the lowest photon yield,
polar angles around 80$^\circ{}$, where the 10~\% loss of photons due to dead 
time and pile-up would reduce the yield from about 20 photons per track to 18.
According to Eqn.~\ref{equ:sigma-track} this would worsen 
the track Cherenkov angle resolution by only 0.1~mrad for typical values 
of the Cherenkov angle resolution per photon around 10-12~mrad.
For the 2\% of events with tracks in the same bar the additional photon loss 
would deteriorate the track Cherenkov angle resolution by another 0.2~mrad,
which is still not a significant effect.

\section{Reconstruction Methods}
\label{sec:reco-bars}
Two different reconstruction approaches were developed to evaluate
the detector resolution and PID performance of the various designs.

\subsection{Geometrical Reconstruction}
\label{sec:reco-geo-bars}
\begin{figure}[htb]
  \centering
  \includegraphics*[width=0.45\textwidth]{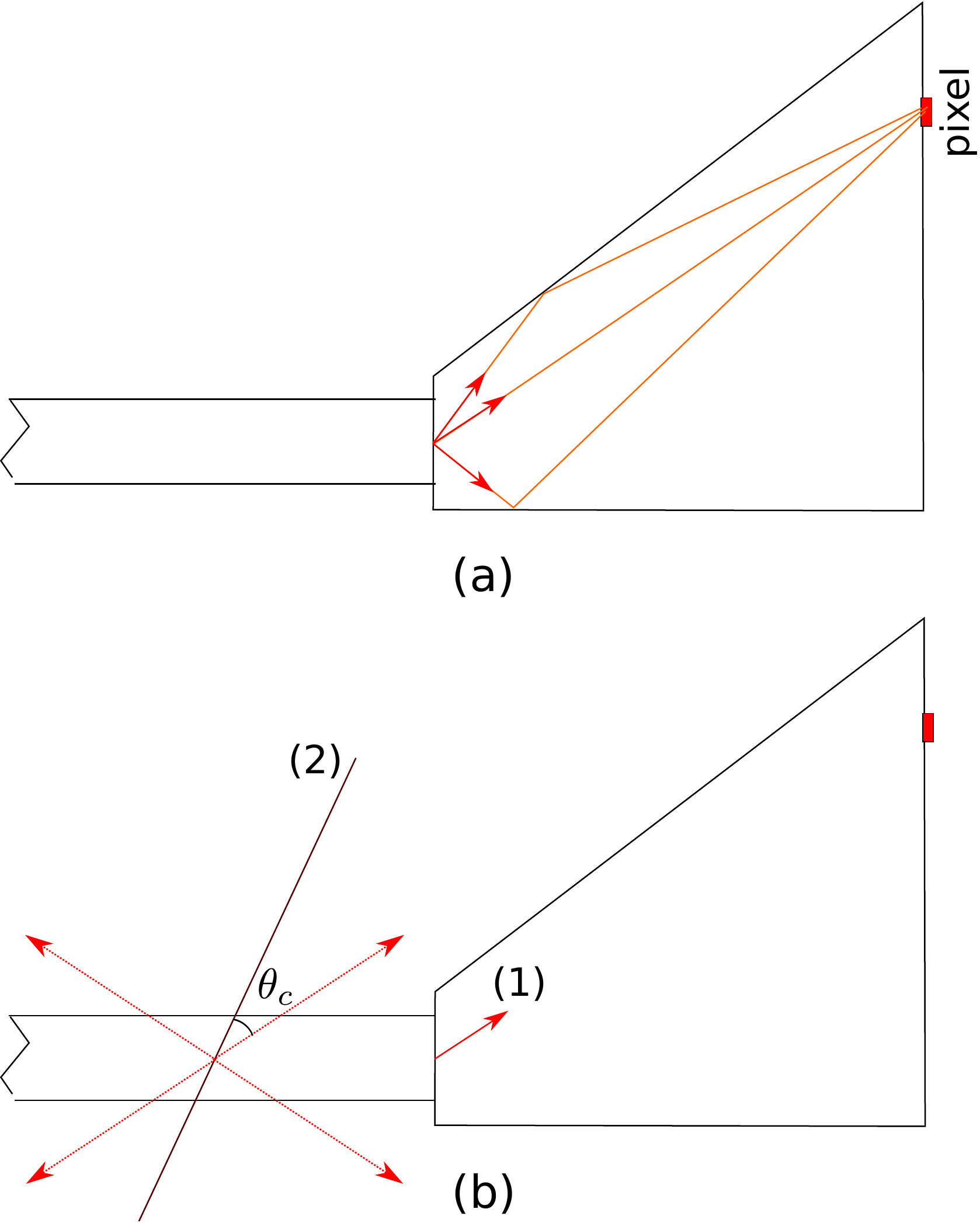}
  \caption{Schematic of the geometrical reconstruction method. 
   (a) Different photon paths in the prism expansion volume (EV) hitting 
   the same pixel are stored in look-up tables (LUT). 
   (b) Determining the Cherenkov angle by calculating the angle between 
   the photon direction from the LUT (1) and the charged track direction (2). 
   Eight different combinations are possible (four are shown), leading 
   to combinatorial background.}
  \label{lGeomReco}
\end{figure}

\begin{figure}[htb]

  \vspace*{3mm}\includegraphics[width=0.5\textwidth]{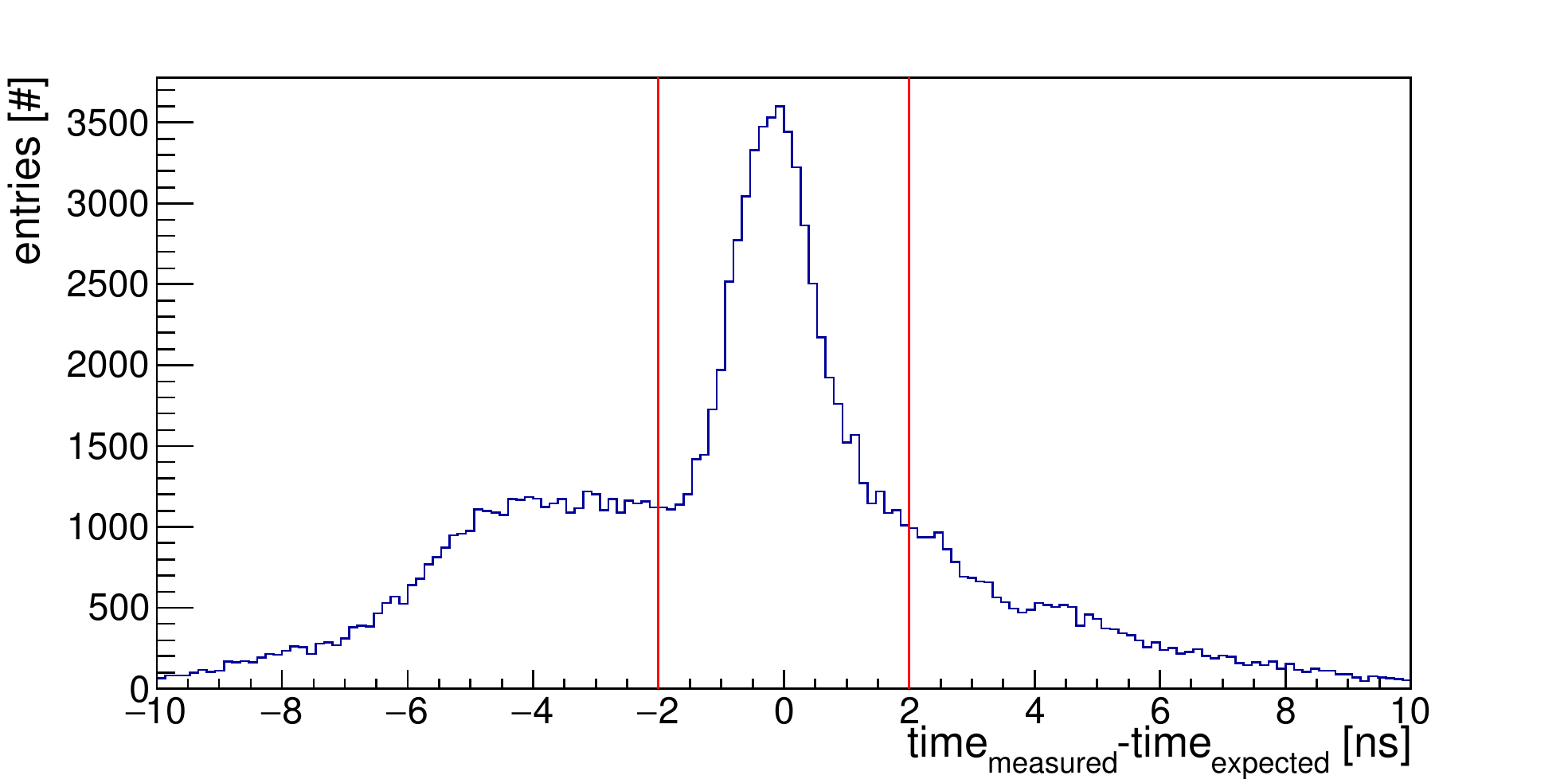}
  \includegraphics[width=0.5\textwidth]{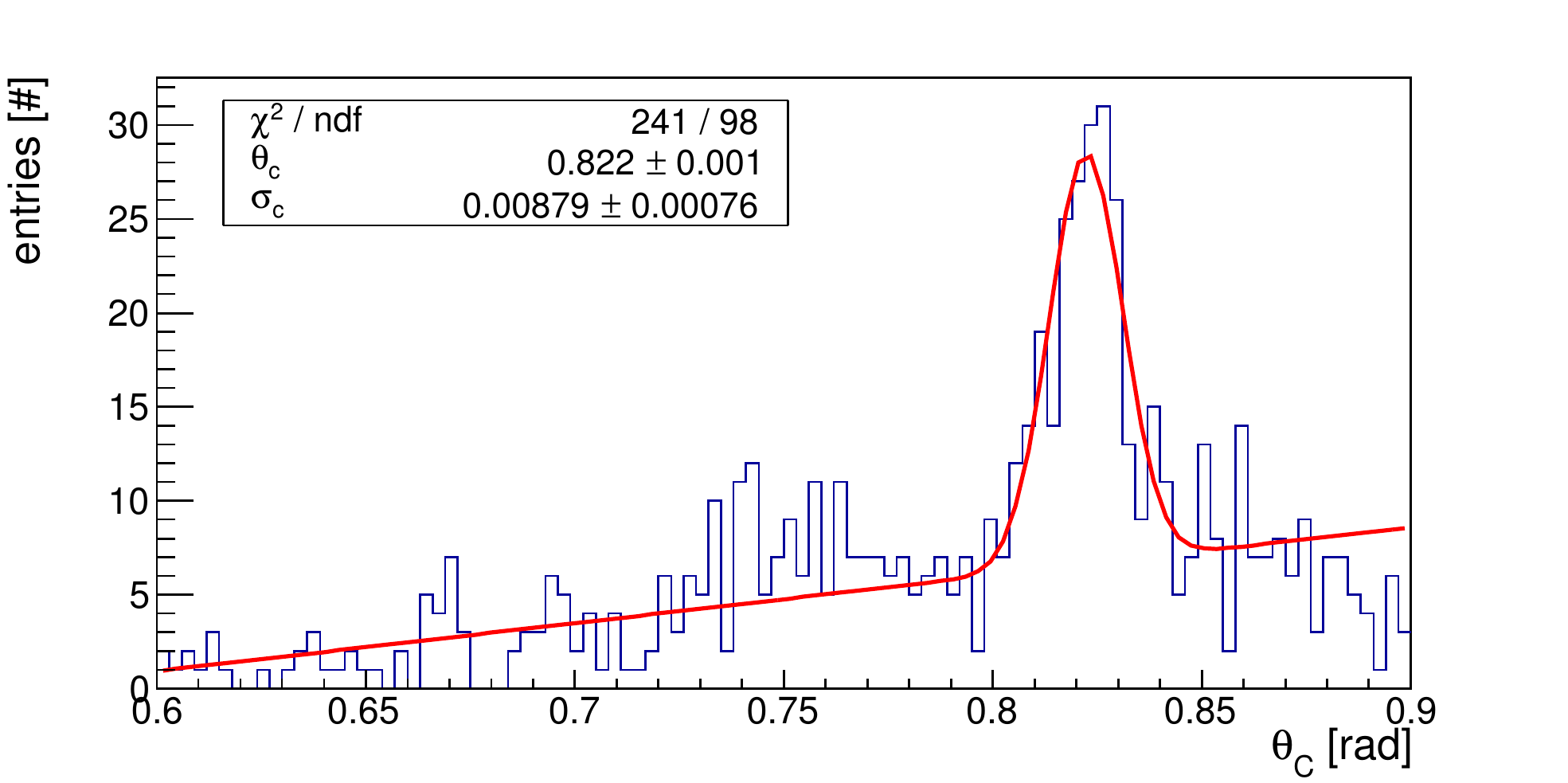}
    
  \caption{(a) Time difference between 
    measured and expected arrival time of the Cherenkov photons from 100 charged kaons.
    The vertical lines indicate the selection region.
    (b) Example of the single photon Cherenkov angle resolution (SPR)
        for a single $K^{+}$ track with 3.5~GeV/c momentum emitted at $25^{\circ}$ polar angle.
        The fit results in an SPR value of $\approx$9~mrad.
        }
  \label{lGeomRecoCh}
\end{figure}

The geometrical reconstruction method, developed for the BaBar DIRC~\cite{Adam5},
transforms the known spatial positions of the bar through which the track passed
and the pixel with a detected photon into the Cherenkov coordinate system.
The direction of a detected photon is approximated by the three-dimensional 
vector between the center of the bar and the center of the pixel, taking 
refraction at all material interfaces into account.
The full simulation is used to calculate these photon direction vectors
for every possible bar-pixel combination.
This is done by simulating the production of optical photons at the end of the 
bar and tracking them through the lens and prism to the sensor pixels.
Photons are produced for polar angles between 90$^{\circ}$ and 270$^{\circ}$ and azimuthal 
angles between 0$^{\circ}$ and 360$^{\circ}$ and for every pixel the average direction 
vector between the bar and \ pixel is stored in a look-up table (LUT) 
(see Fig.~\ref{lGeomReco}a).
 
In the reconstruction those direction vectors are combined with the particle 
momentum vector, provided by the tracking system, to determine the Cherenkov 
angle $\theta_{C}$ for each photon (see  Fig.~\ref{lGeomReco}b). 
Since the exact path of the photon during the many internal reflections 
in the bar is unknown, the reconstructed photon direction is ambiguous. 
Eight different direction combinations are possible inside the bar 
(forward/backward, top/bottom, and left/right).
They are taken into account by combining the direction from the LUT in eight 
different ways with the particle direction, leading to up to eight values 
for the reconstructed photon Cherenkov angle.
Additional reconstruction ambiguities arise from the various possible 
reflections inside the prism so that for some angles a total of 50 
possible photon paths and more are considered in the reconstruction.
This number is reduced by considering only angles that are internally 
reflected in fused silica and by requiring the photon Cherenkov angle 
to be smaller than 1000~mrad.

Most of the reconstructed photon paths correspond to Cherenkov angles far away 
from the expected value and form a combinatorial background under the peak 
associated with the correct photon path.
A further reduction of the combinatorial background is achieved by applying 
a selection cut on the difference between the measured arrival time of the photon 
and the expected arrival time. 
The latter is calculated from the reconstructed photon path in the bar and 
the prism assuming a group velocity corresponding to a photon with the 
wavelength of 380~nm, which is the average wavelength of detected photons 
determined from simulation.
Figure~\ref{lGeomRecoCh}a shows the 
time difference distribution for 100 charged kaons at 3.5~GeV/c momentum and 25 degree 
polar angle.

Figure~\ref{lGeomRecoCh}b shows the resulting reconstructed Cherenkov angles per photon,
including all reconstruction ambiguities, for one 3.5~GeV/c $K^{+}$ at $25^{\circ}$
polar angle which produced 52 detected Cherenkov photons. 
A clear peak at the correct value of the Cherenkov angle can be seen.
The width of the peak corresponds to the single photon Cherenkov angle 
resolution (SPR) and is found to be SPR$\approx$9~mrad for this track.

In the final step the distribution of Cherenkov angle per photon is fit with
a Gaussian plus a linear background to calculate the likelihood for 
the distribution to originate from a $e$, $\mu$, $\pi$, $K$, or $p$ and to 
determine the mean Cherenkov angle for the track.

The main advantage of this reconstruction method
is that it delivers a measurement of the SPR and the Cherenkov angle of 
the track as well as the yield of signal and background photons,
which are all important variables for the detector design.
Furthermore, the algorithm is very fast since the LUTs depend only on the 
detector geometry and not on the particle properties, and thus can be 
created prior to event reconstruction.

\subsection{Time-based Imaging}
\label{sec:reco-plate}

The geometrical reconstruction approach is not suitable for wide plates since
the fundamental assumption that the photon exits from the center of the radiator 
is no longer valid.
An alternative algorithm was developed for the wide plates but can also be
used for narrow bars. 
This time-based imaging method is based on the approach used by the Belle~II 
time-of-propagation (TOP) counter~\cite{belle-pdfs}.
The basic concept is that the measured arrival time of Cherenkov photons
in each single event is compared to the expected photon arrival time for 
every pixel and for every particle hypothesis, yielding the PID likelihoods.

The expected photon arrival times can be calculated either from an analytical 
function or from the simulation, and the latter approach was used to evaluate 
the time-based imaging method for \panda. 

\begin{figure}[htb]
\includegraphics[width=0.5\textwidth]{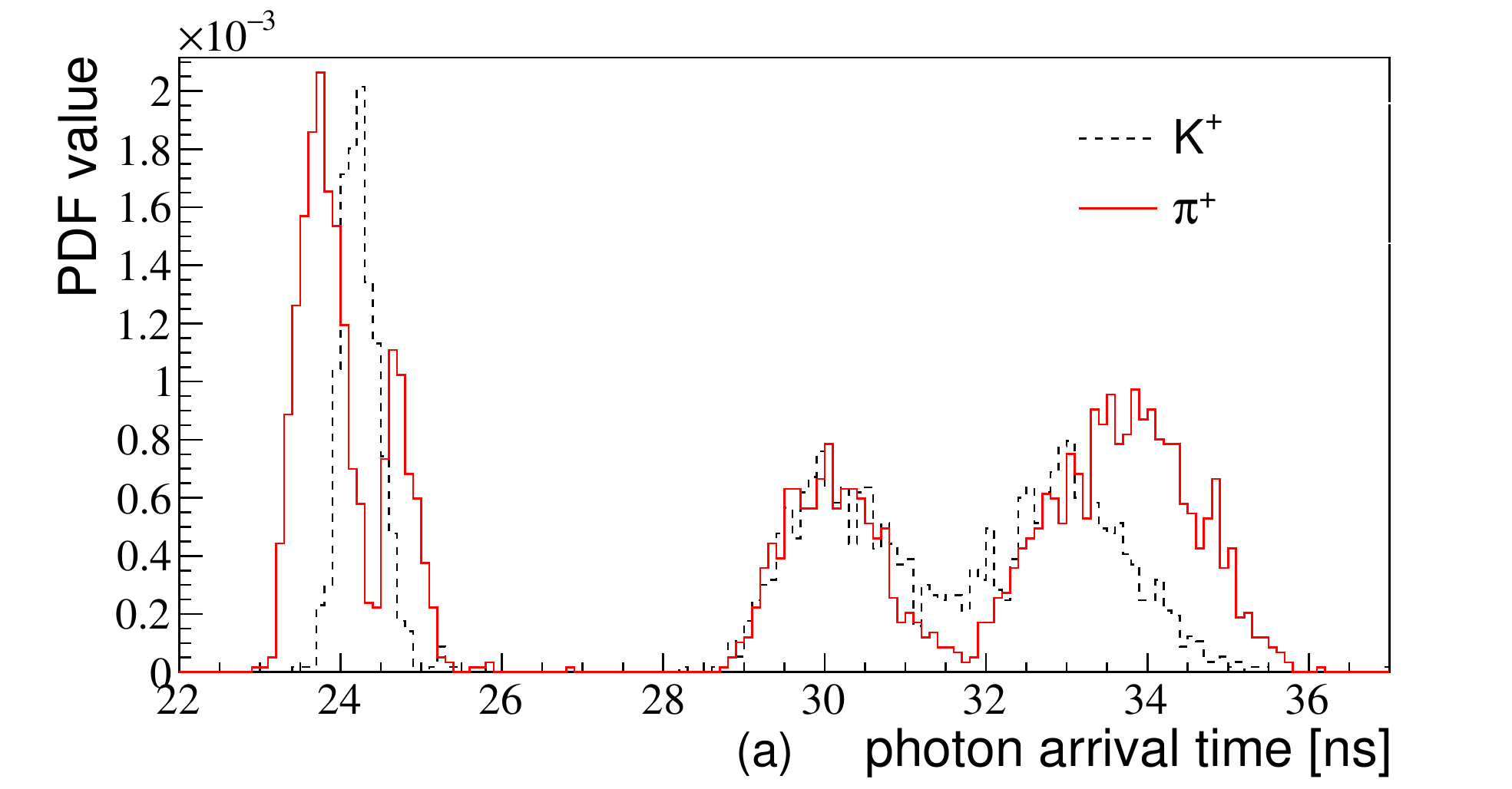}
\includegraphics[width=0.5\textwidth]{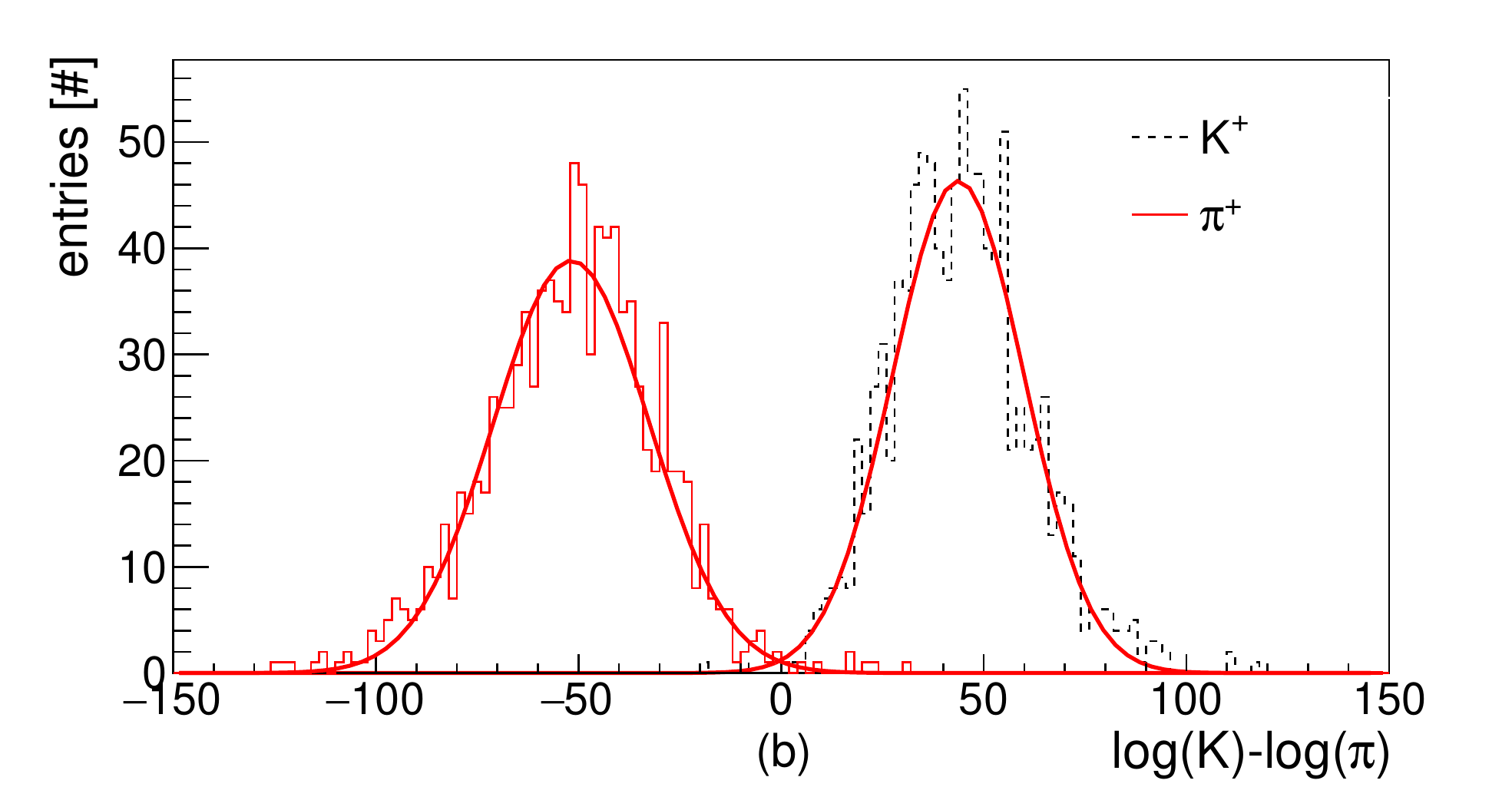} 
\vspace*{-2mm}
\caption{Examples for the time-based reconstruction of the plate geometry with
  a prism EV but without focusing optics: photon arrival time for charged pions and
  kaons for a selected MCP-PMT pixel (a) and log-likelihood difference for
  kaon and pion hypotheses for a sample of 3.5~GeV/c pions and kaons
  at 22$^{\circ}$ polar angle (b).
}
\label{time}
\end{figure}

The full detector simulation is used to generate a large number of tracks 
with the observed momentum and charge of the particle.
The arrival time of the Cherenkov photons produced by $e$, $\mu$, $\pi$, $K$, 
and $p$ is recorded for every pixel and stored in an array of normalized 
histograms to produce probability density functions (PDF).
An example for one MCP-PMT pixel is shown in Fig.~\ref{time}a.
The resolution of the detected time was chosen to be 100~ps, the expected
single photon timing resolution for the Barrel DIRC.

For a given track the observed photon arrival time for each hit pixel
is compared to the histogram array to calculate the time-based likelihood for
the photons to originate from a given particle hypothesis.
Combining this likelihood with the Poissonian PDF of the number of observed
photons creates the full likelihood.
Figure~\ref{time}b shows the log-likelihood difference for kaon and pion hypotheses
for a sample of 3.5~GeV/c pions and kaons at 22$^{\circ}$ polar angle.
The $\pi$/$K$ separation of this design, calculated as the difference of the two 
mean values of the fitted Gaussians divided by the average width, corresponds 
to more than 5.1~s.d. in this case.

This time-based imaging method works well, not only for wide plates
but also for narrow bars, where the performance of the time-based imaging is 
found to be superior to geometric reconstruction results.

It should be noted that the current implementation, which is based on generating a 
large number of simulated events for every possible particle direction, momentum, 
charge, and type, and storing all photon timing information in histogram arrays, 
is not practical for use in \panda since the corresponding 
time histogram arrays would require large storage capacities and slow down 
reconstruction. 
The Belle~II TOP group has shown in Ref.~\cite{belle-pdfs} that the timing PDFs 
can be calculated analytically instead.
They found that these analytical PDFs deliver a performance similar to PDFs from
the full simulation at a much faster reconstruction speed.
A first version of this algorithm was implemented for the \panda Barrel DIRC in 
Ref.~\cite{MZ-MZuehlsdorf-PHD-THESIS_S}.
Initial results were promising but additional work is required to extend 
the method to describe the \panda Barrel DIRC data in more detail.

\section{Evaluation of Design Options}
\label{sec:sim-design-options}

\subsection{Baseline Design}

A figure of merit is needed to quantify important aspects of the design
and to compare the performance of different \panda Barrel DIRC designs 
to each other and to other DIRC counters.
It is important that this figure of merit can be measured with 
DIRC prototypes in different types of particle beams since each critical design 
element needs to be validated with experimental data.
Since the Cherenkov angle resolution can be seen as the critical driver of the DIRC PID 
performance, the photon yield $N_\gamma$ and the single photon Cherenkov angle 
resolution (SPR) are selected as figures of merit because those 
two quantities are closely related to the PID performance 
(see Eqn.~\ref{equ:sigma-track}).
They can be reliably determined in test beams and were previously used for
qualifying the performance of the BaBar DIRC and the SuperB FDIRC.

The initial simulation studies were focused on finding at least one  
Barrel DIRC design that matches the figures of merit reported by
the BaBar DIRC and, thus, meets the PID requirements for \panda.
After the geometry with 5 narrow bars per bar box, a large oil tank, and a 
2-layer spherical lens for each bar was found to meet or exceed the 
required figures of merit~\cite{RICH13_sim,DIRC13_sim},
additional studies were performed to optimize the performance, while simultaneously 
minimizing the overall Barrel DIRC costs~\cite{MP-MPatsyuk-PHD-THESIS_S}.

A wide range of design options was investigated, including 
\begin{itemize}
\item the material, type, shape, and size of the expansion volume, 
\item the material, type, and shape of the focusing lenses, 
\item the number of bars per bar box, 
\item the thickness and width of the radiators, 
\item the offset between the bottom of the bar and the bottom of the EV, and
\item the sensor layout on the focal plane.
\end{itemize}

Each design was evaluated in terms of photon yield and SPR for the
entire range of polar angles and momenta in the Barrel DIRC.
The results of the most important studies, summarized in 
Tab.~\ref{Tab:simres-ev},\,\ref{Tab:simres-radiatorwidth} and \,\ref{Tab:simres-MCPlayout}, are discussed in some detail below.
In all cases, unless specified differently, the geometry used 5 narrow bars per 
bar box, a 2-layer spherical focusing lens, a large oil tank filled with mineral 
oil, and 5 rows of MCP-PMTs.\\

\begin{figure}[]
  \begin{center}
\vspace*{-2mm} \includegraphics[width=0.33\textwidth]{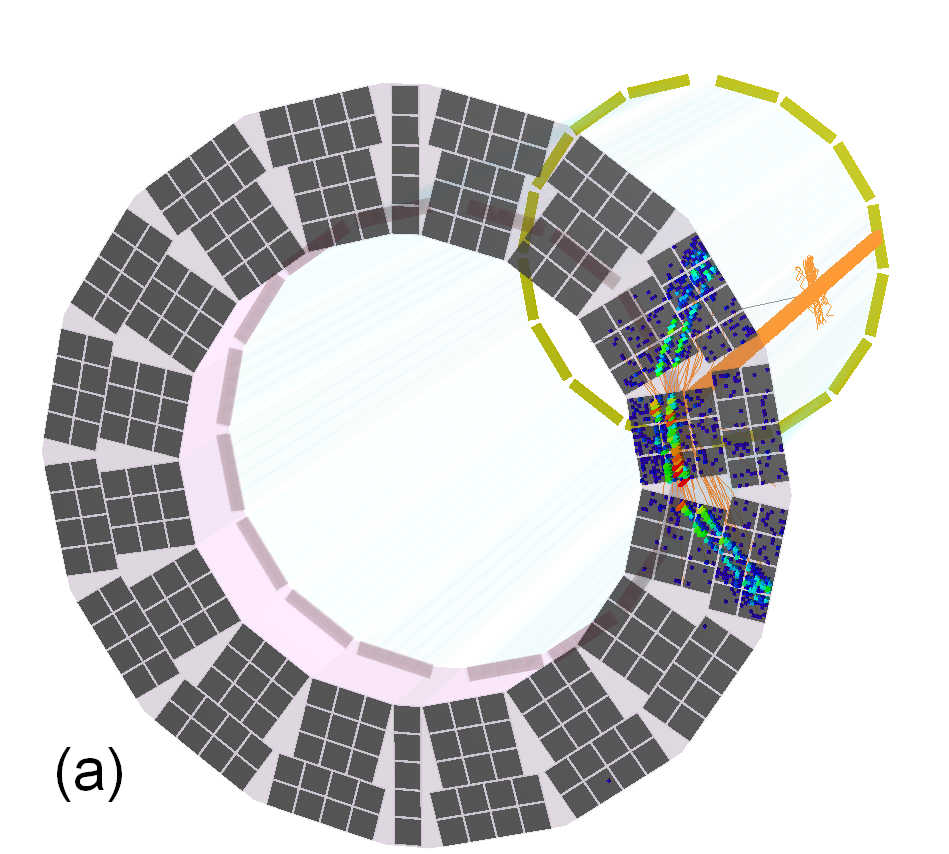}
\vspace*{-2mm} \includegraphics[width=0.33\textwidth]{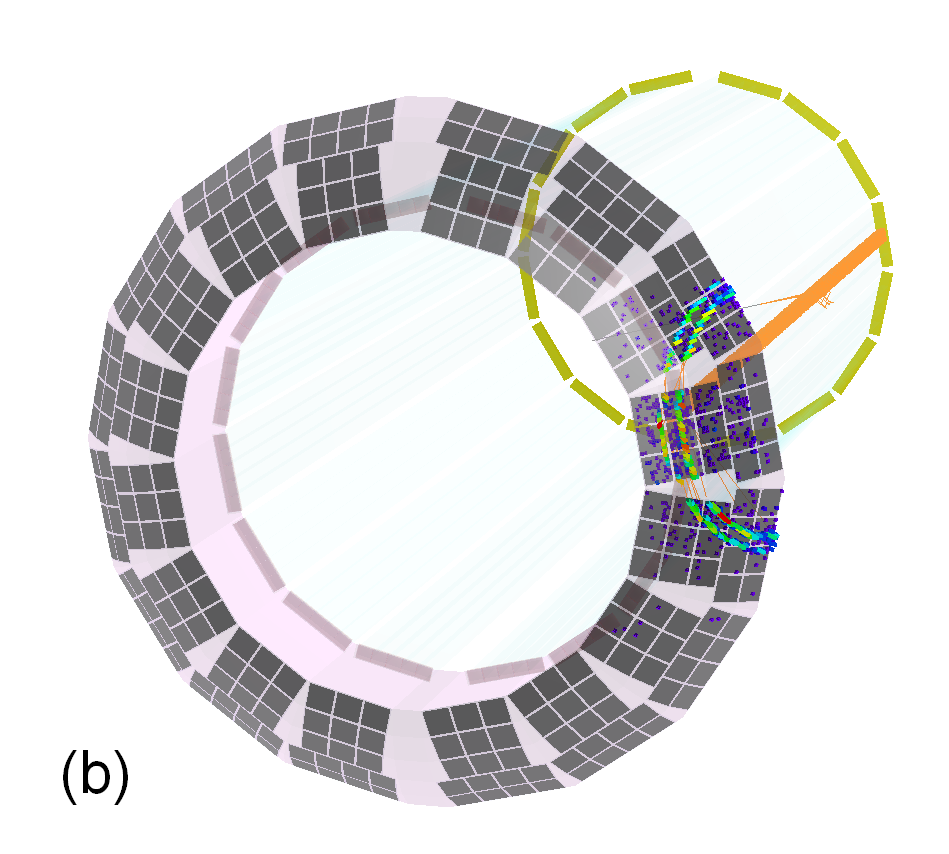}
\vspace*{-2mm} \includegraphics[width=0.33\textwidth]{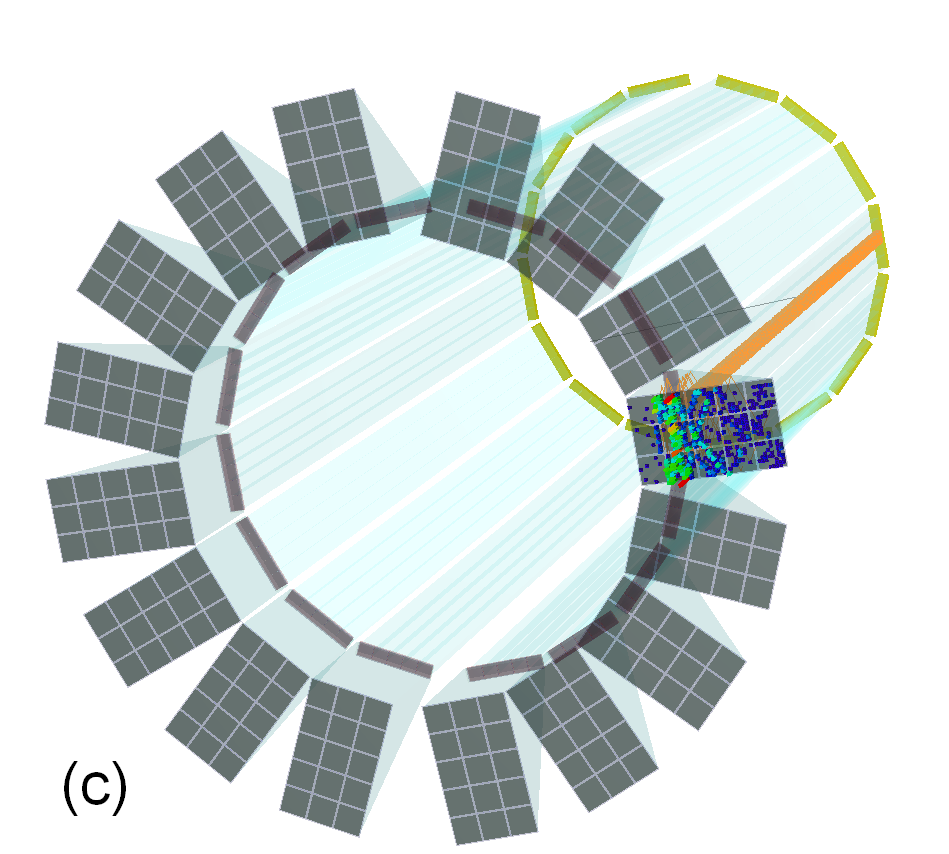}
\vspace*{-2mm} \includegraphics[width=0.33\textwidth]{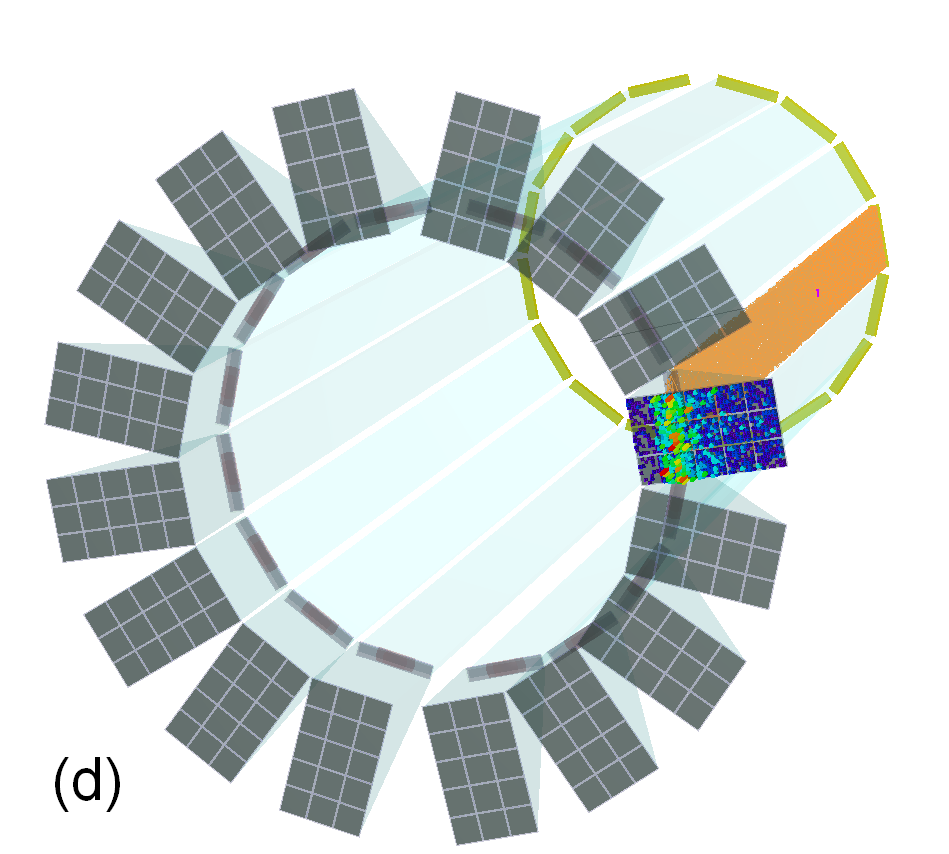}
 \end{center}
\vspace*{-2mm}
\caption{Geant simulation of the \panda Barrel DIRC using different design options.
  An oil tank with a straight (a) or curved (b) imaging plane is shown at the top.
  Solid fused silica prisms are used as EV in the design on the bottom (c), (d).
  Narrow bars (a), (b), (c) or wide plates (d) are used as radiator.
  Cherenkov photon trajectories from a 3~GeV/c kaon are shown in orange.
  The colored histogram shows the accumulated hit pattern from 100 kaons of
  the same momentum.
}
\label{dircBarrel}
\end{figure}

\textbf{Expansion Volume Shape}

The two options considered were the large oil tank, filled with mineral oil,
and separate fused silica prisms.
Figure~\ref{dircBarrel} shows four examples of EV geometry options simulated 
in Geant and the corresponding accumulated hit patterns from 100 charged kaons.
The readout side of the EV could be perpendicular to the bottom surface,
tilted at an oblique angle or toroidal in shape.

\begin{figure}[h]
	\vspace*{-3mm}\includegraphics[width=0.5\textwidth]{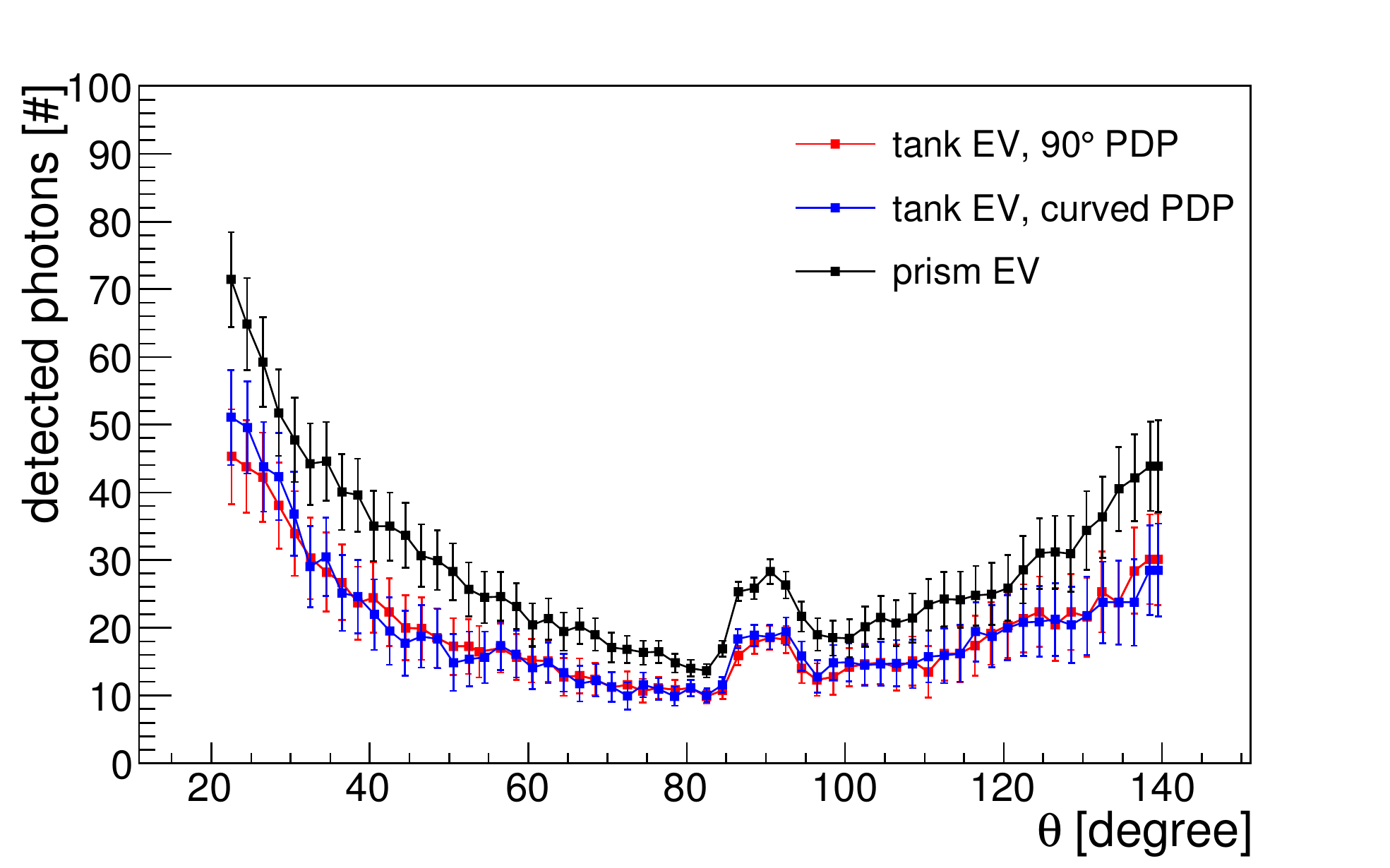}
	\includegraphics[width=0.5\textwidth]{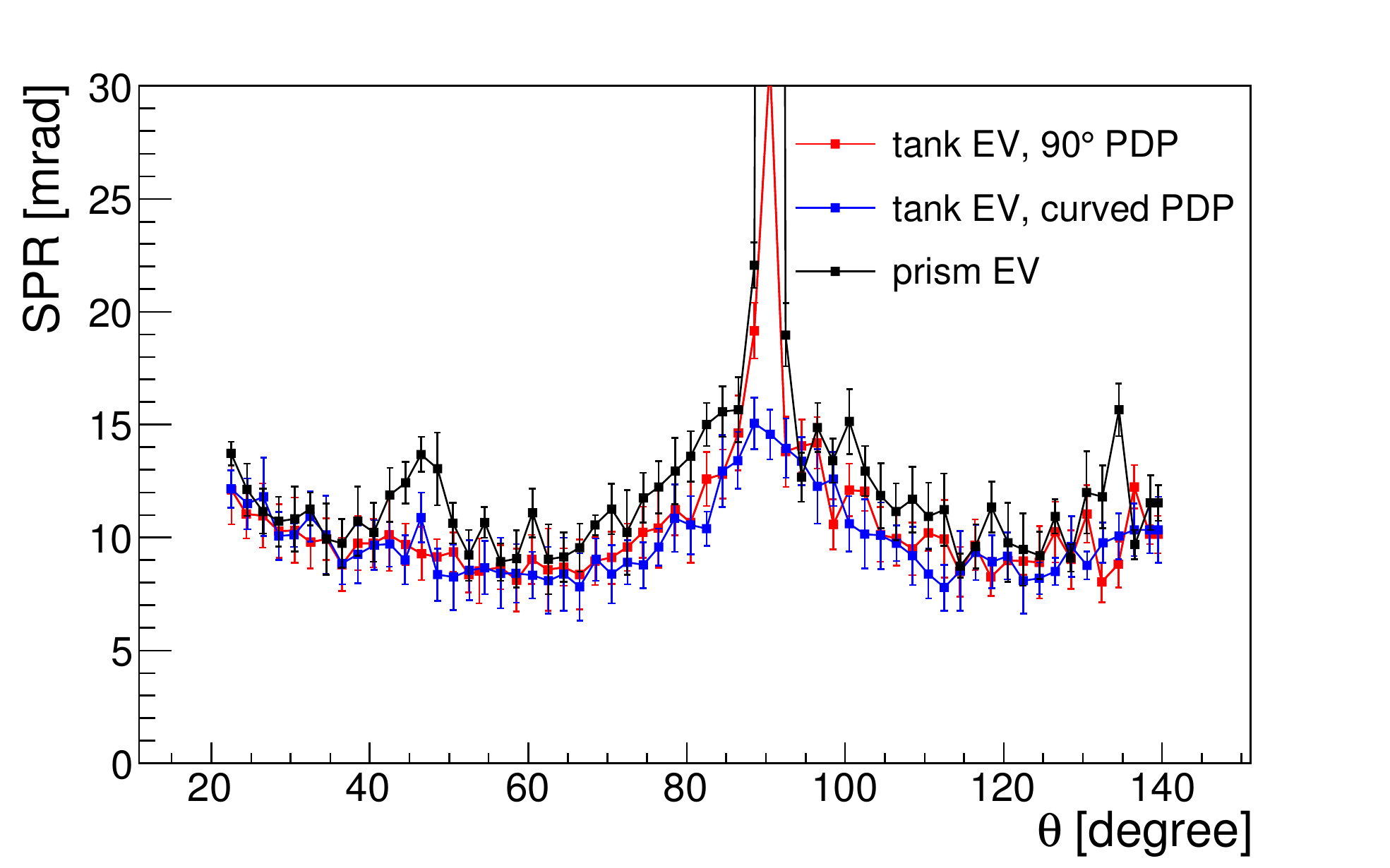}
	\caption{
		Geant simulation study of the impact of the shape and type of EV and 
		photon detector plane (PDP) on the photon yield (top) and SPR (bottom).
		The distributions are shown for the geometry with 5 bars per bar box, 
		2-layer spherical lenses and pions emitted at 3.5~GeV/c momentum.
		The error bands correspond to the RMS of the distributions in each bin.
	}
	\label{fig:opt_ev}
\end{figure}

Figure~\ref{fig:opt_ev} shows the performance summary for three different 
EV types. 
The photon yield and the single photon Cherenkov angle resolution
are shown as a function of the polar angle for a sample of pions 
with 3.5~GeV/c momentum, generated uniformly in azimuthal angle.

The different oil tank geometries with the flat and curved focal 
planes show a similar performance while the fused silica prism 
performs slightly better in the important region of steep forward angles.
Due to the much better optical properties, the photon yield with 
the prism exceeds the yield of the tank EV by about 40\% with about 
70 photons per track at 22$^\circ$ and about 22 at 60$^\circ$.
In this region the SPR for the prism is only slightly worse than
for the tank EV, resulting in an overall significantly better performance
for the prism EV, even for the not yet fully optimized 2-layer spherical lens.
Furthermore, the total area to be covered with MCP-PMTs is considerably
smaller for the prism EV than the tank EV, which leads to a significant
cost reduction.

Other prism parameters studied include the depth and the opening angle.
The SPR was calculated for a prism depth between 250~mm and 400~mm, the 
maximum depth possible within the space available for the Barrel DIRC readout.
While a larger prism size improves the angular resolution, it also increases
the cost of the prism and can be the source of more combinatorial background
since additional reflections inside the prism become possible.
Furthermore, for a given prism opening angle, a larger prism depth creates
a larger area that has to be equipped with more photon sensors, further increasing
the cost.
Since the SPR was found to depend only weakly on the EV depth for values over
250~mm, a 300~mm prism depth was selected as best compromise between the cost 
and performance while still keeping sufficient space for the readout electronics
and cabling.
The prism opening angle was varied between 30$^\circ$ and 48$^\circ$.
While a smaller opening angle means lower fabrication cost and fewer required
sensors, larger angles may reduce the number of ambiguous photon paths in the
prism, leading to less background and a more stable reconstruction.
A study of the SPR as function of the prism opening angle favored smaller 
angle values, in the range of 35$^\circ$ and below.
The value of 33$^\circ$ was selected to match the size of commercially 
available MCP-PMTs.\\

\textbf{Focusing System}

Due to the compact expansion volume the design of the focusing system is 
particularly important.
Spherical and cylindrical lenses, with and without air gaps between the
lens and EV, were simulated, as well as a design without any focusing optics.
For lenses without air gap, versions with two and three layers of optical 
material (fused silica in combination with either NLaK33B or PbF$_2$
as high-refractive index material, see Sec.~\ref{subsec:lenses}) were considered.

\begin{figure}[h]
	\vspace*{-3mm}\includegraphics[width=0.5\textwidth]{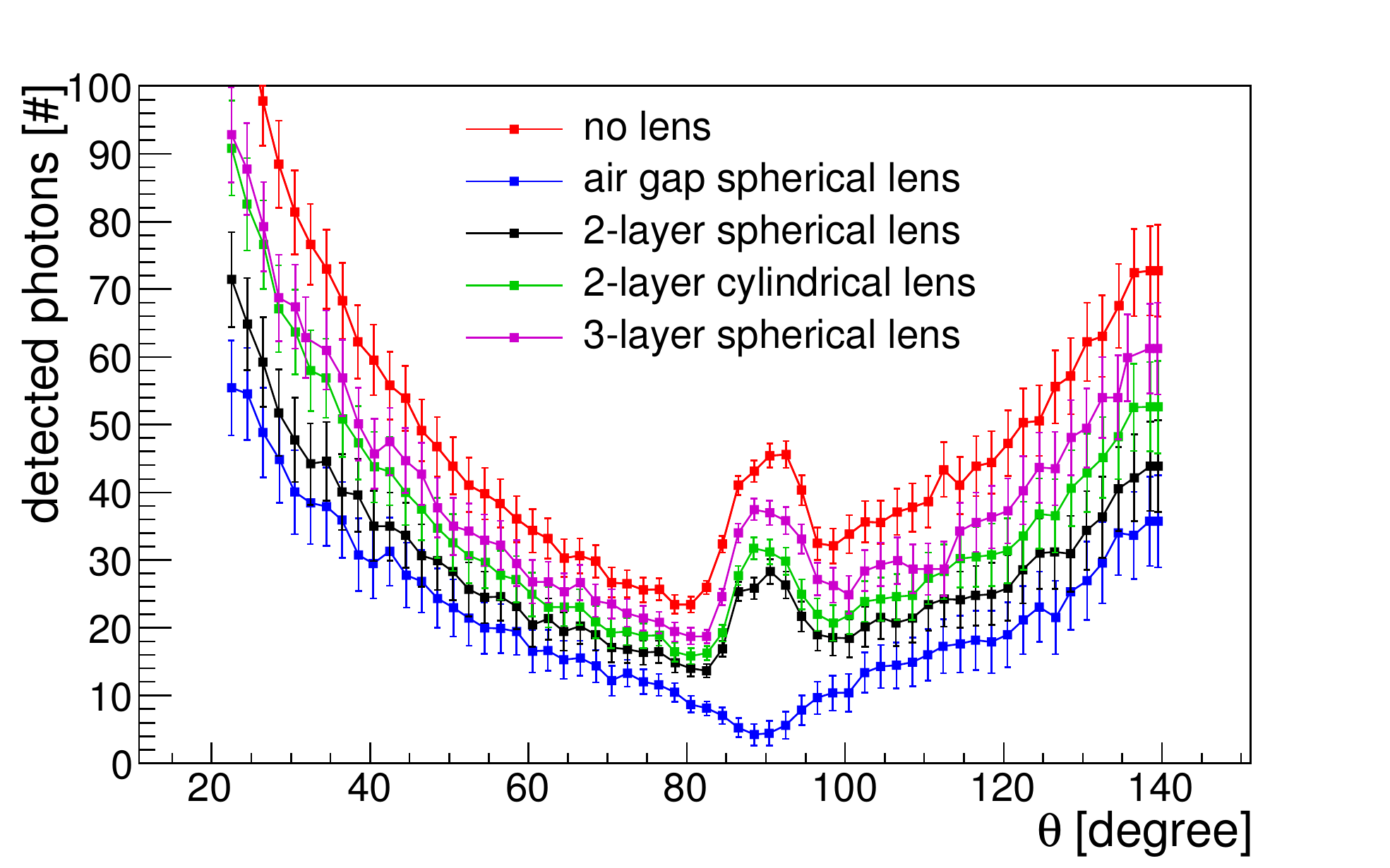}
	\includegraphics[width=0.5\textwidth]{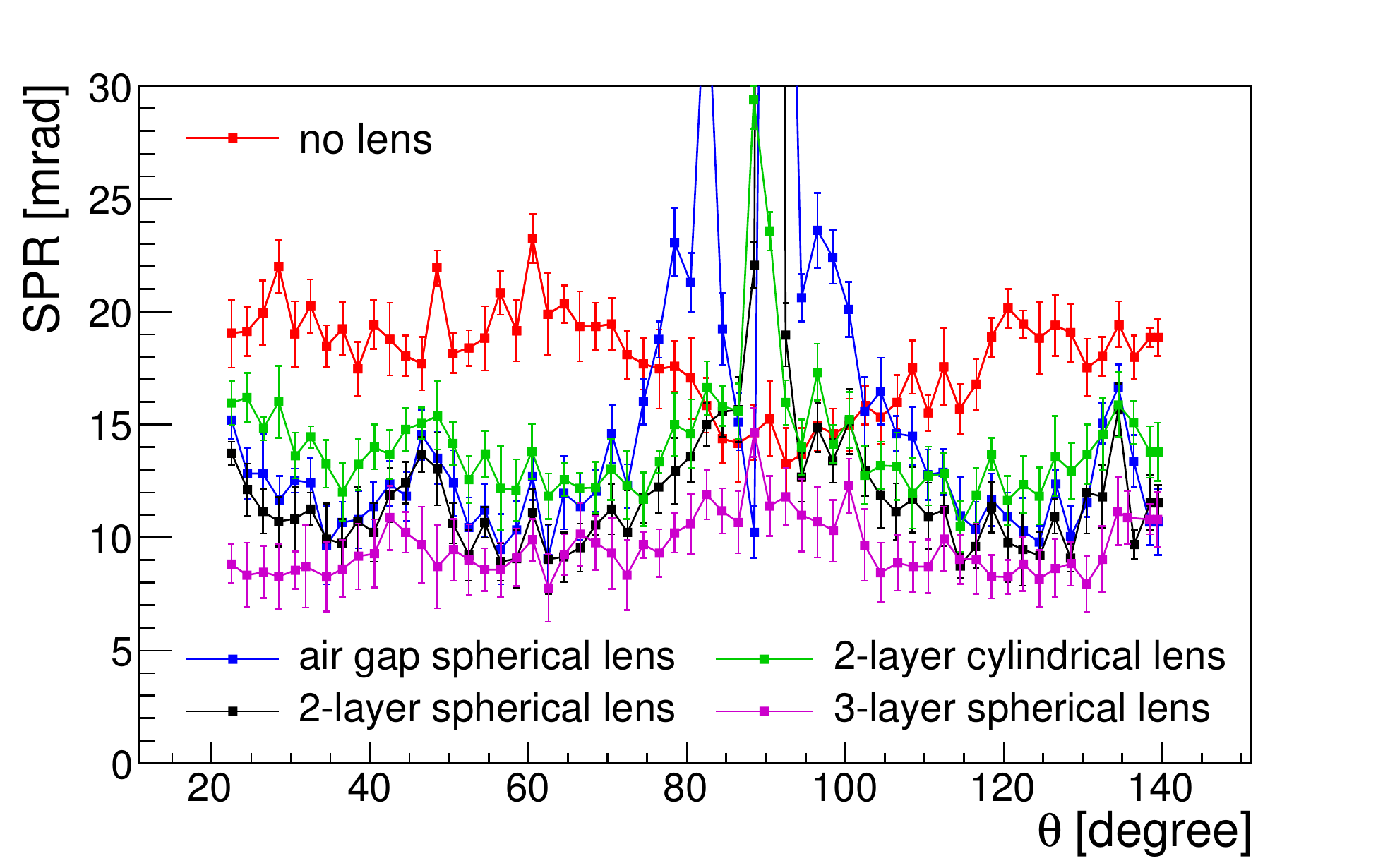}
	\caption{
		Geant simulation study of the impact of different focusing options on the 
		photon yield (top) and SPR (bottom).
		The distributions are shown for the geometry with 5 bars per bar box,
		a fused silica prism EV and pions emitted at 3.5~GeV/c momentum.
		The error bands correspond to the RMS of the distributions in each bin.
	}
	\label{opt_lens}
\end{figure}

Figure~\ref{opt_lens} compares the performance of five such design options:
No lenses between the 5 bars and the prism, with spherical lenses and air 
gaps, with 2-layer spherical lenses, 2-layer cylindrical lenses,
and with the 3-layer spherical lenses. 

While the design without focusing shows a very high photon yield, the poor 
SPR values lead to a track Cherenkov angle 
resolution at 3.5~GeV/c significantly worse than the 2.8~mrad required for
the 3~s.d. $\pi/K$ separation.
The fused silica spherical lens with an air gap shows a better single photon
resolution for most polar angles.
However, the lens suffers from unacceptable photon yield losses near 90$^\circ$
polar angles due to total internal reflection of the photons at the lens-air 
interface.
This not only leads to a poor track Cherenkov angle resolution but also makes
the design very sensitive to track- and event-related backgrounds, 
including Cherenkov photons from $\delta$ electrons and nearby tracks, 
possible accelerator-induced background from $\gamma$ and neutrons,
as well as backsplash particles from the electromagnetic calorimeter, which,
according to the simulation, may produce up to 10 background photons per event.

The focusing with a 3-layer spherical lens is superior to all other 
lens solutions.
The single photon resolution is in the range of 8--10~mrad, except for
angles of 80-100$^\circ$, where the combinatorial background from ambiguous 
photon paths between bar and pixel is most severe.
However, even for those angles the SPR is still significantly better 
than required.

Thus, the prism EV with the 3-layer high-refractive index spherical lens 
reaches the design goals in both the photon yield and the SPR. 
The track Cherenkov angle resolution is below 2.5~mrad at forward angles
and considerably better than the 3~s.d. requirement for the entire angle range.\\

\textbf{Number of Bars per Bar Box}

A significant cost reduction can be achieved if the total number of bars can be
reduced by increasing the bar width without performance loss.

\begin{figure}[h]
\centering
	\vspace*{-3mm}\includegraphics[width=0.5\textwidth]{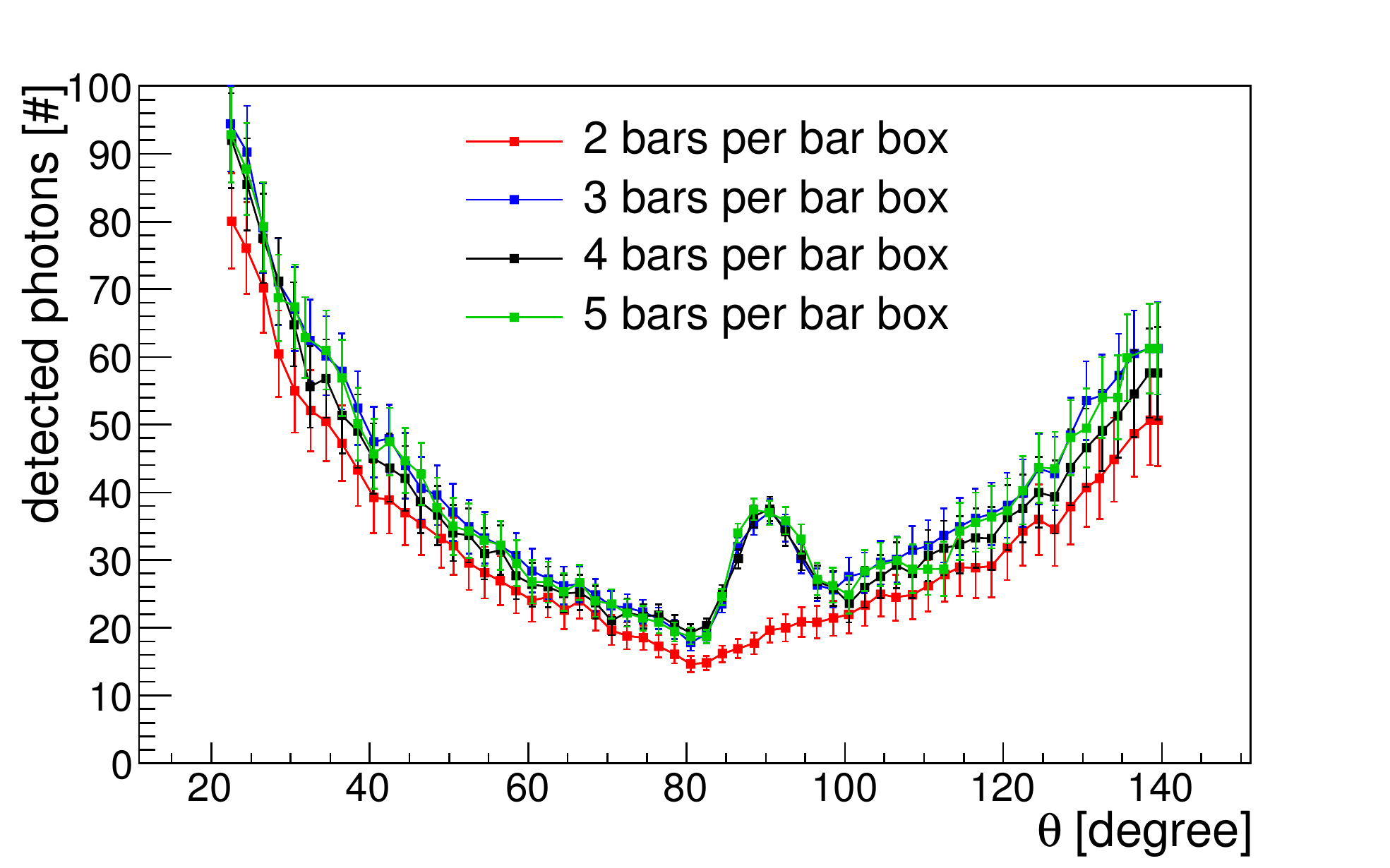} 
	\includegraphics[width=0.5\textwidth]{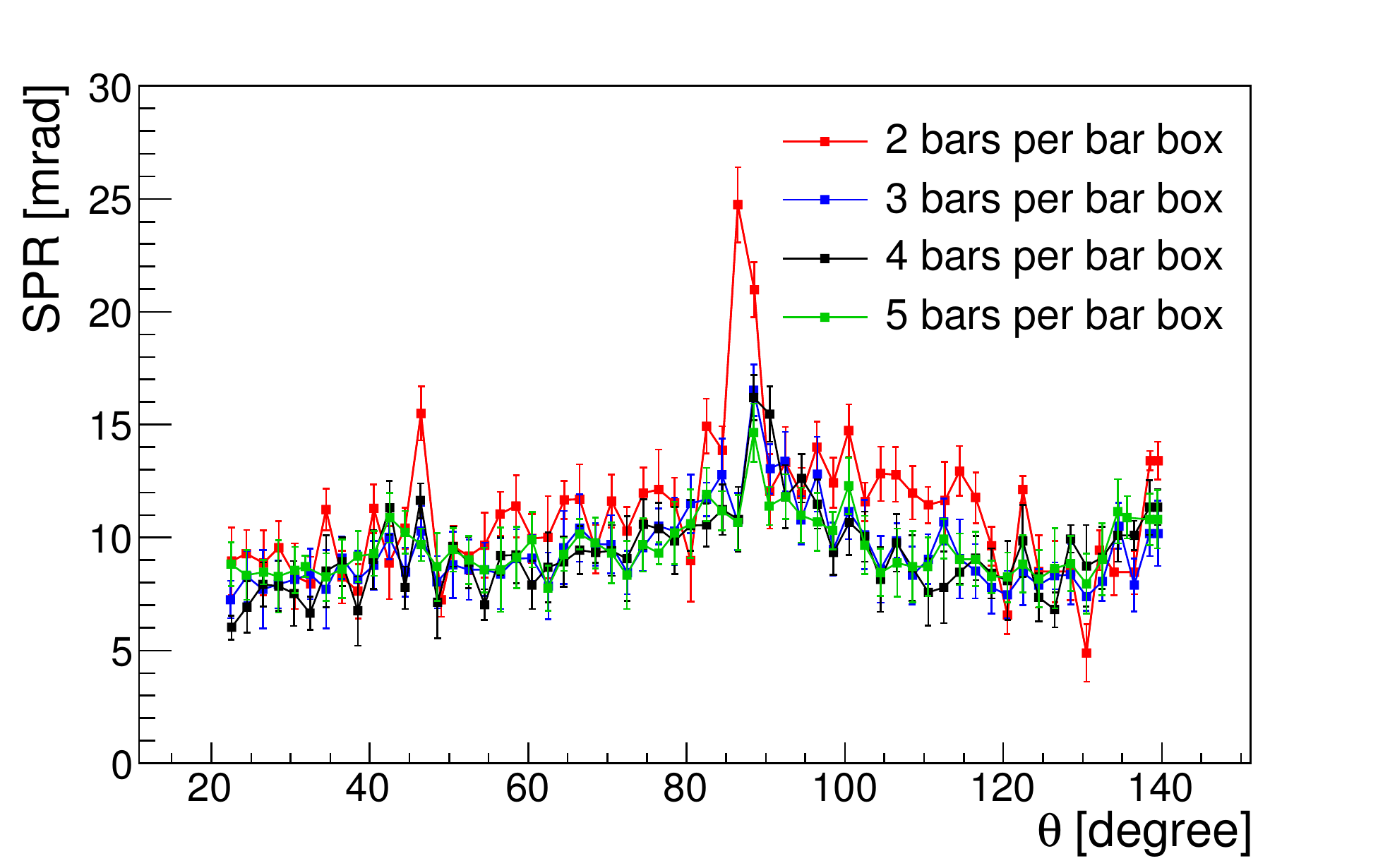}
	\caption{
		Geant simulation study of the impact of the number of bars per bar box on the 
		photon yield (top) and SPR (bottom).
		The distributions are shown for the geometry with a fused silica prism EV, 
		3-layer spherical lenses and pions emitted at 3.5~GeV/c momentum.
		The error bands correspond to the RMS of the distributions in each bin.
	}
	\label{opt_bar}
\end{figure}

Figure~\ref{opt_bar} shows the comparison of the photon yield and the SPR for 
2--5 bars per bar box.
The azimuthal coverage of the DIRC is kept constant by defining the bar 
width to be the 160~mm-width of the bar box, divided by the number of bars.

The larger bar width requires a thicker 3-layer spherical lens design as
well as different curvatures of the focusing layers.
The thicker lens leads to an additional loss of photon yield inside
the lens, especially visible for two bars per bar box and polar angles near 
90$^\circ{}$, since the sides of the lenses are assumed to be unpolished and 
non-reflecting.
Furthermore, the multi-layer lens no longer succeeds in creating a flat focal plane.
The geometry with two bars per bar box, therefore, does not meet the requirements
for the Barrel DIRC.

The SPR for 3--5 bars per bar box is about the same and lies in the 
8--11~mrad range, depending on the polar angle of the track.
In combination with the yield between 20 and 90 photons per track this means
that those designs exceed the $\theta_C$ resolution requirements for the entire 
kaon phase space.

As the cost of the geometry with 3 bars per bar box is the lowest,
this width is selected as baseline geometry.\\

\textbf{MCP-PMT Coverage of the Prism}

\begin{figure}[htb]
	\centering
     \vspace*{-3mm}\includegraphics[width=0.5\textwidth]{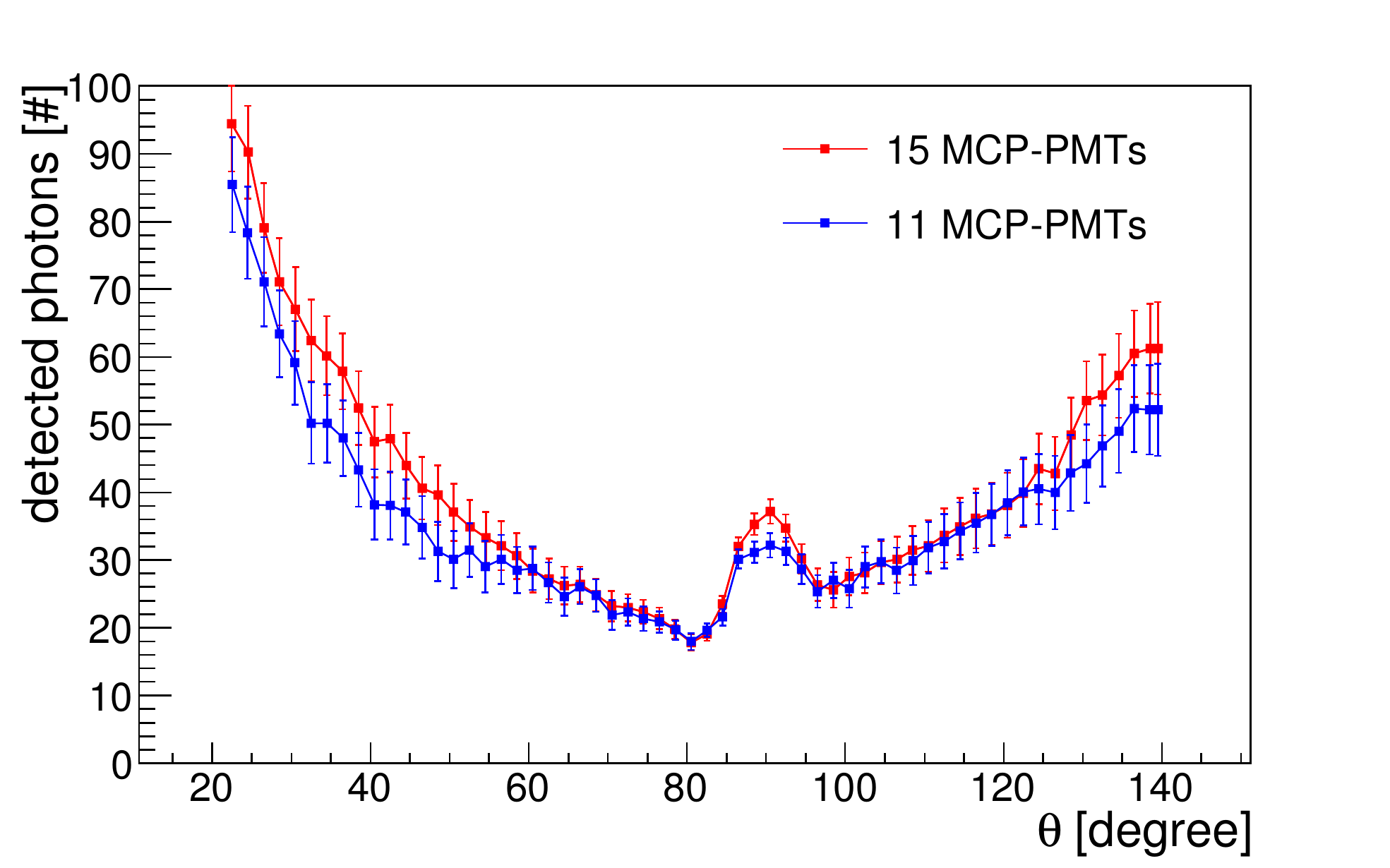}
  \includegraphics[width=0.5\textwidth]{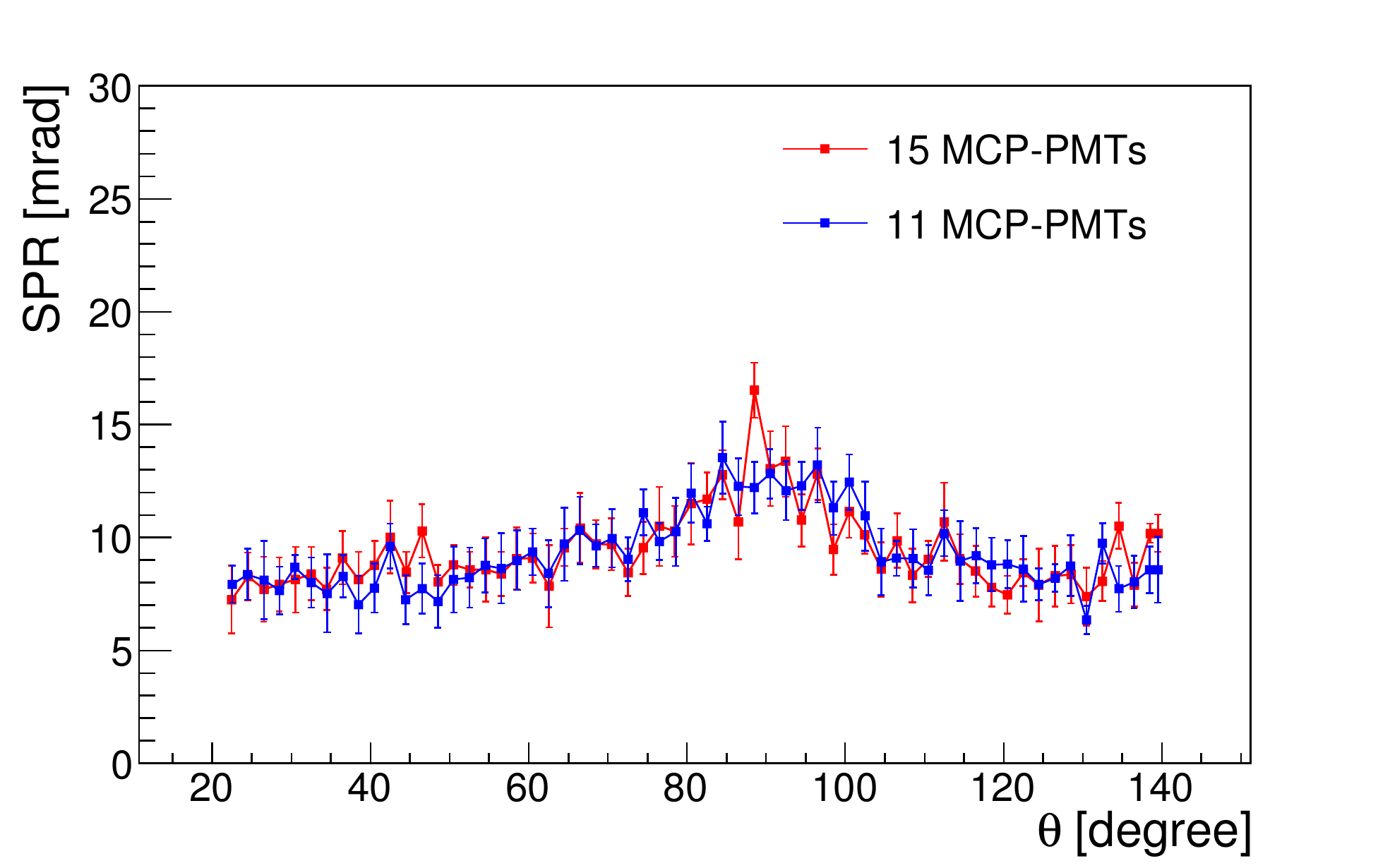}
  \caption{
      Geant simulation study of the impact of the number of MCP-PMTs covering the
      fused silica prism EV on the photon yield (top) and SPR (bottom).
      The distributions are shown for the geometry with 3 bars per bar box,
      3-layer spherical lenses and pions emitted at 3.5~GeV/c momentum.
      The error bands correspond to the RMS of the distributions in each bin.  
  }
  \label{opt_mcp}
\end{figure}

Since the production of the photon detectors is one of the two main
cost drivers for the Barrel DIRC, designs with different numbers of
MCP-PMTs per prism were studied.
Figure~\ref{opt_mcp} compares the photon yield and SPR for a prism
with a 40$^\circ{}$ top angle and 5 rows of MCP-PMTs (for a total of 
15 MCP-PMTs per prism) to a prism with 33$^\circ{}$ top angle 
and 4 rows of MCP-PMTs.
For the smaller prism the number of MCP-PMTs per prism is further reduced
to account for the fact that the size of commercially available MCP-PMTs 
is such that only two MCP-PMTs will fit side-by-side at the inner 
radius of the prism for a total of 11 MCP-PMTs per prism.

The SPR is nearly identical for the two prism sizes and the photon yield
drops only by 10-15\% and remains always above 20 photons per track,
making the cost-saving smaller prism with 11 MCP-PMTs the preferred option.


\subsubsection*{Evaluation of the Baseline Design}
\label{sec:eval-baselinedesign}
\begin{figure}[htb]
	\centering
      \vspace*{-3mm}\includegraphics[width=0.5\textwidth]{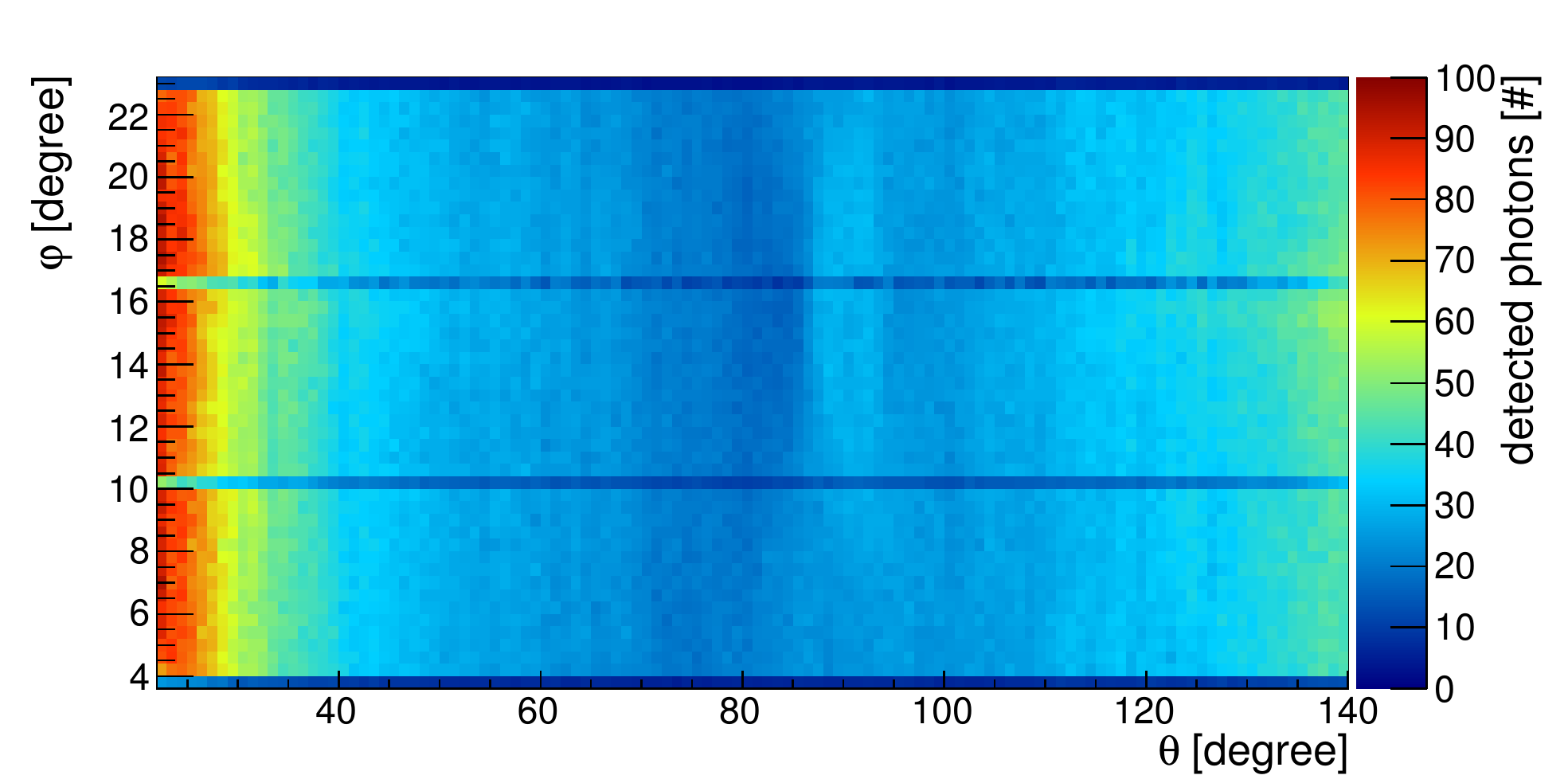}
   \includegraphics[width=0.5\textwidth]{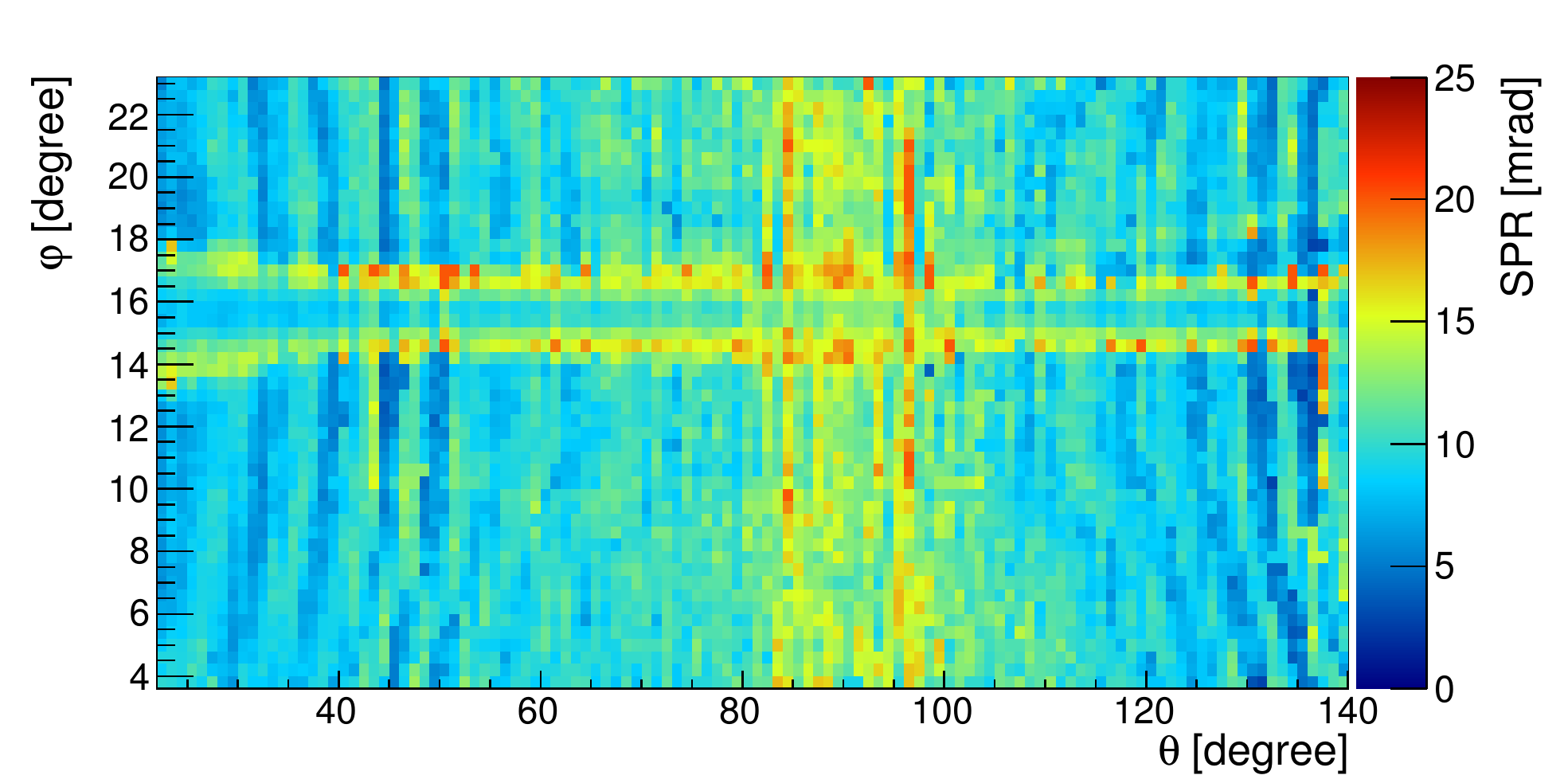}
   \caption{
   Maps of photon yield (top) and SPR (bottom) vs. azimuthal and polar angle 
   from Geant simulation for a geometry with three bars per bar box, a fused 
   silica prism EV, and 3-layer spherical lenses for kaons with 3.5~GeV/c 
   momentum.
   The color scale corresponds to the number of detected photons (top) and 
   the SPR (bottom).}
  \label{phimap}
\end{figure}

For the final baseline design, three bars per bar box, 3-layer spherical lenses,
and a prism with 11 MCP-PMTs, Fig.~\ref{phimap} shows the detailed analysis of the
photon yield and SPR as a function of the charged kaon polar and azimuth angle 
across one bar box.
The photon yield at the top shows the familiar shape of increasing yield for steeper angles,
due to longer track length within the bar, and the bump near 90$^\circ{}$ polar
angle due to both sides of the Cherenkov ring being totally internally reflected.
The structures in the SPR, shown in Fig.~\ref{phimap} (bottom), are dominated by 
reconstruction ambiguities and gaps 
between rows of MPC-PMTs, both of which are strongly correlated with the polar angle.
The horizontal structure in the azimuthal angle is caused by the track hitting the bar
at perpendicular incidence for 16$^\circ{}$ azimuthal angle in the \panda magnetic field,
causing several of the reconstruction ambiguities to overlap.

For the same design the track Cherenkov angle resolution $\sigma_{\theta_C}$ 
is calculated from the photon yield N$_{\gamma}$ and the SPR via
\begin{equation}
\sigma_{\theta_C}^2=\mathrm{SPR}^2/N_{\gamma}+\sigma_{\mathrm{track}}^2 .
\end{equation}
$\sigma_{\mathrm{track}}$ is the uncertainty of the track direction 
in the DIRC, dominated by multiple scattering and the resolution of the 
\panda tracking detectors, and was determined from detector simulation 
to be $\sigma_{track} \approx$1.7--2.3~mrad, depending on the polar angle, in 
the latest \panda design.

\begin{figure}[h]
	\centering
	\vspace*{-3mm}\includegraphics[width=0.5\textwidth]{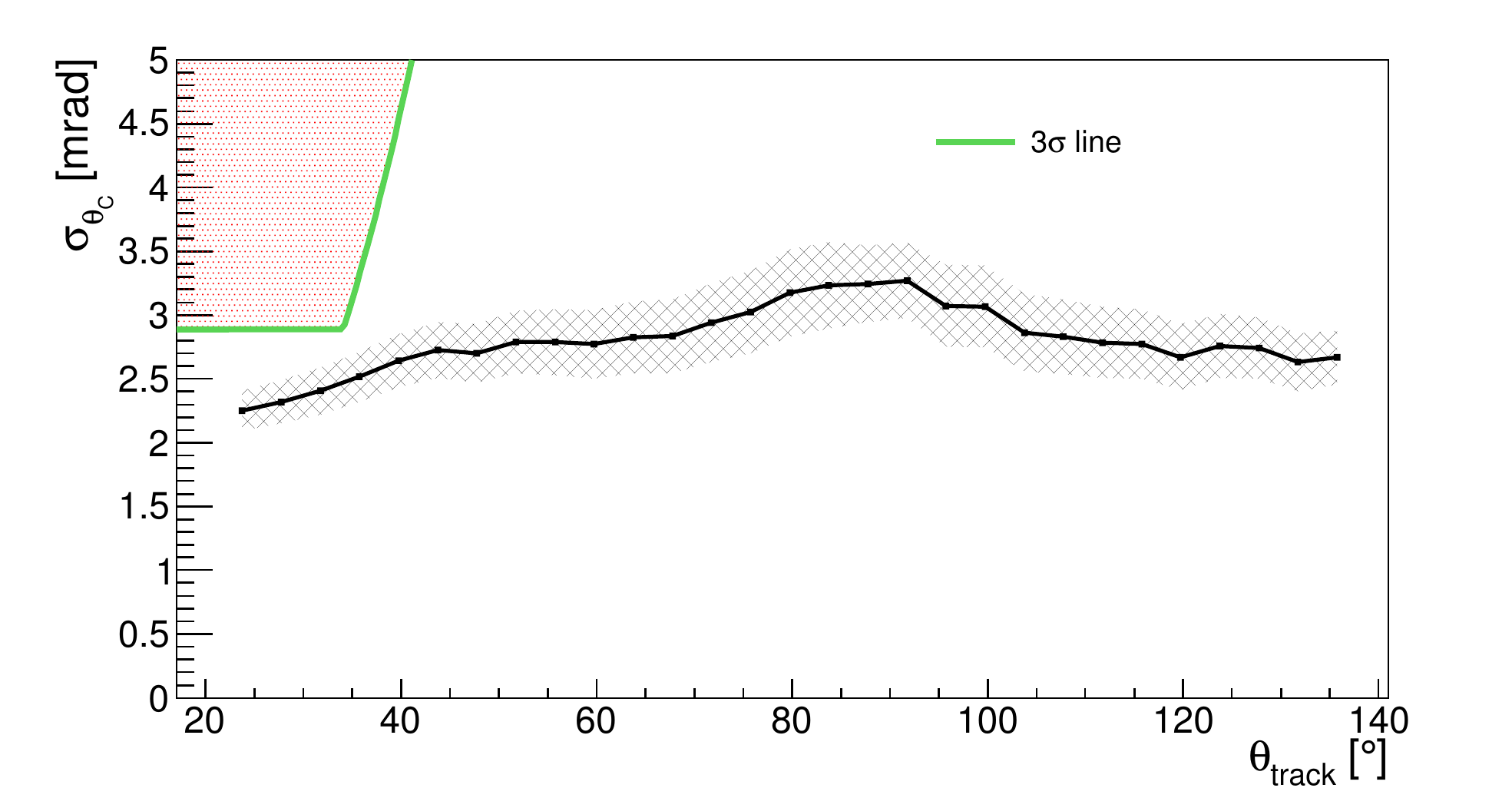}
	\caption{
		Track Cherenkov angle resolution calculated from the SPR and the photon yield
		for 3.5~GeV/c kaons in Geant simulation as a function
		of the polar angle for a design with three bars per bar box, a fused silica prism EV,
		and 3-layer spherical lenses.
	} 
	\label{trackresol}
\end{figure}

Figure~\ref{trackresol} shows $\sigma_{\theta_C}$ as a function of the polar
angle. 
The green curve corresponds to the 3~s.d. $\pi$/$K$ separation goal for the 
Barrel DIRC, which strongly depends on the polar angle, as discussed in 
Sec.~\ref{cha:design}, and is most demanding for the forward region.
In this representation of the track Cherenkov angle resolution
all points outside the red area meet the Barrel DIRC PID goals.
The obtained track Cherenkov angle resolution values of 2.2--3.3~mrad are
better than the required resolution for $\pi$/$K$ separation of at least 3~s.d.
for the entire polar angle range.


\begin{figure}[h]
\centering
\vspace*{-3mm}	\raggedright{\small{Narrow bar, spherical lens, geometrical reconstruction}}
	\includegraphics[width=0.5\textwidth]{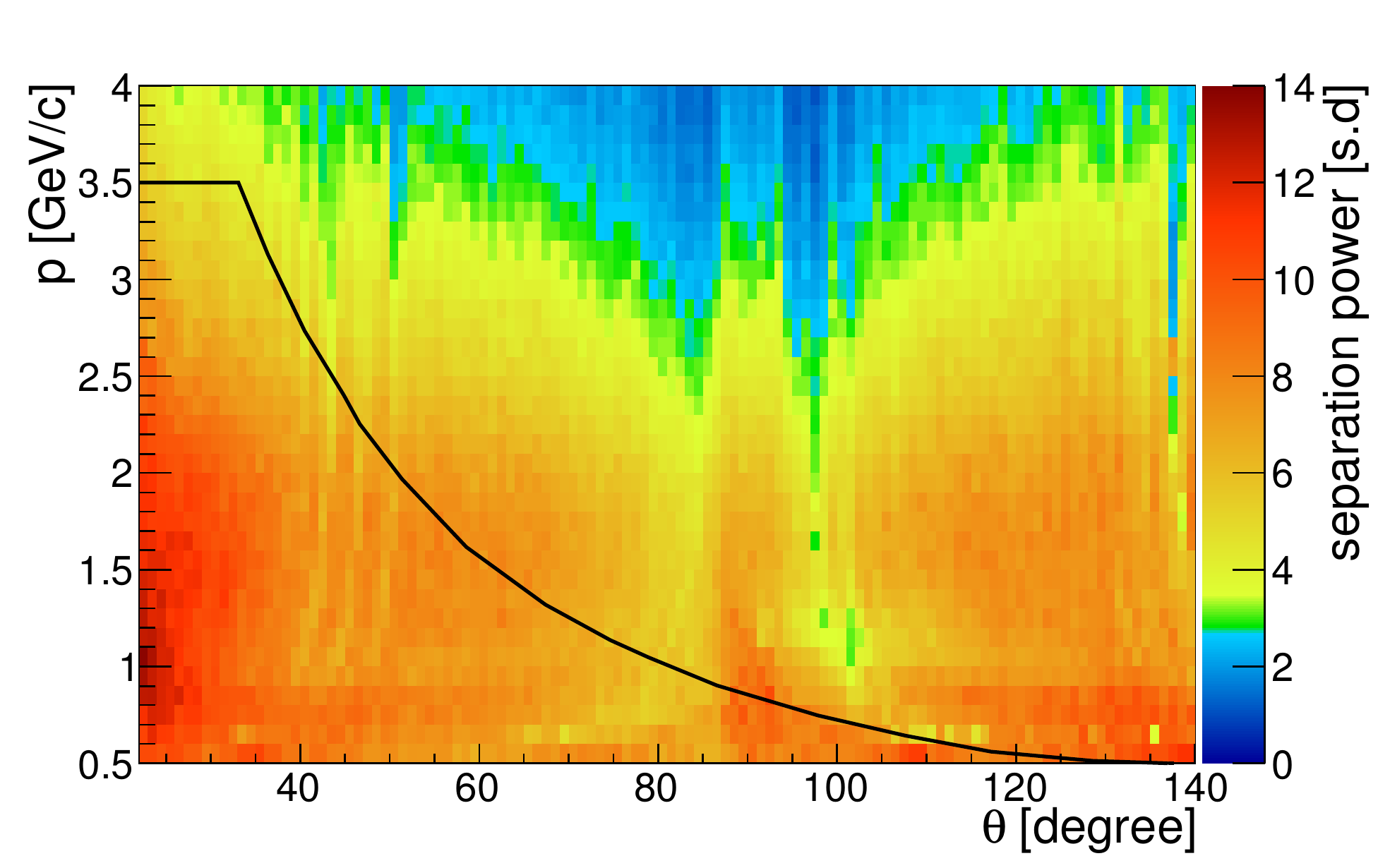}
	\raggedright{\small{Narrow bar, spherical lens, time-based imaging}}
	\includegraphics[width=0.5\textwidth]{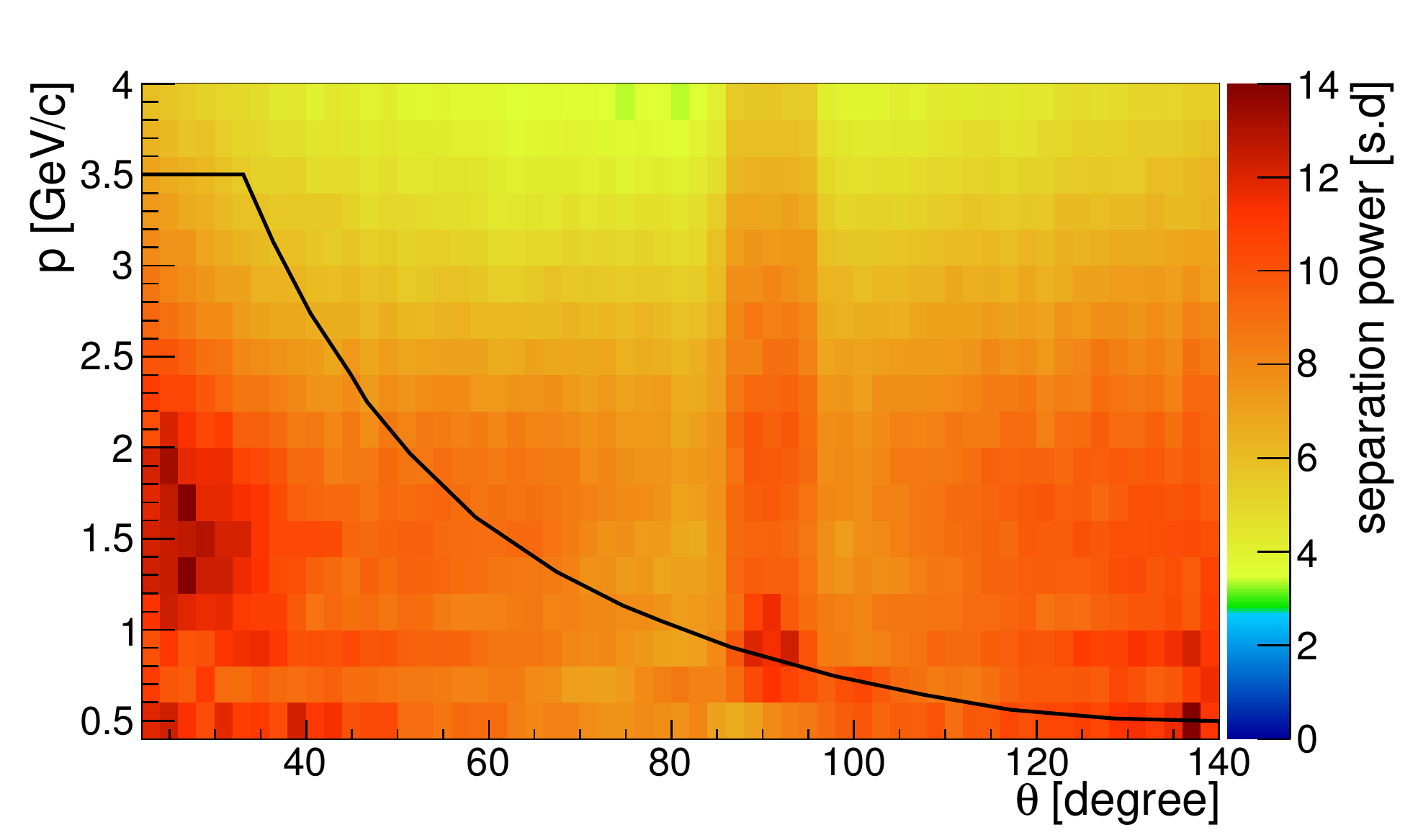}
	\vspace*{-2mm}
\caption{
	$\pi$/$K$ separation power as a function of particle momentum and
	polar angle in Geant simulation for the geometry with 
	three bars per bar box, a fused silica prism EV, and 3-layer spherical lenses.
  The separation power was determined by the geometrical reconstruction using 
  track-by-track maximum likelihood fits (top) or by the time-based imaging 
  method (bottom). 
  The area below the black line corresponds to the final-state phase space
  for charged kaons from various benchmark channels.
}
\label{lh_bar}
\end{figure}

The $\pi/K$ separation power of the baseline design with three bars per bar box, 
3-layer spherical lenses, and a prism with 11 MCP-PMTs is shown as a function of the 
particle momentum and polar angle in Fig.~\ref{lh_bar} for two different PID algorithms.
With a separation power of 4--16 s.d. the baseline design exceeds the \panda PID 
requirement for the entire charged kaon phase space, indicated by the area below 
the black line.
The performance of the time-based imaging method (bottom) is even better than the 
result of the track-by-track maximum likelihood fit (top) due to the optimized use
of the high-precision photon timing information, but both algorithms provide 
excellent $\pi/K$ separation for \panda.

\subsection{Design Option with Wide Plates}

The optimization process for the geometry with narrow bars identified prisms 
with a top angle of 33$^\circ{}$ and 11 MCP-PMTs as the optimum EV design.
The geometry with two bars per bar box showed that the thickness required for 
such wide spherical lenses creates an unacceptable photon loss due to reflections 
inside the lens.
Therefore, the main remaining optimization for the plate design was either 
no focusing or a cylindrical lens.
While it is expected that focusing improves the PID performance of the 
plate, the geometry without lens is attractive because it avoids possible
issues with the radiation hardness of multi-layer lenses, simplifies 
the assembly, and will have a slightly lower cost. 

\begin{figure}[htb]
\centering
   \raggedright{\small{Wide plate, no focusing, time-based imaging}}
	\includegraphics[width=0.5\textwidth]{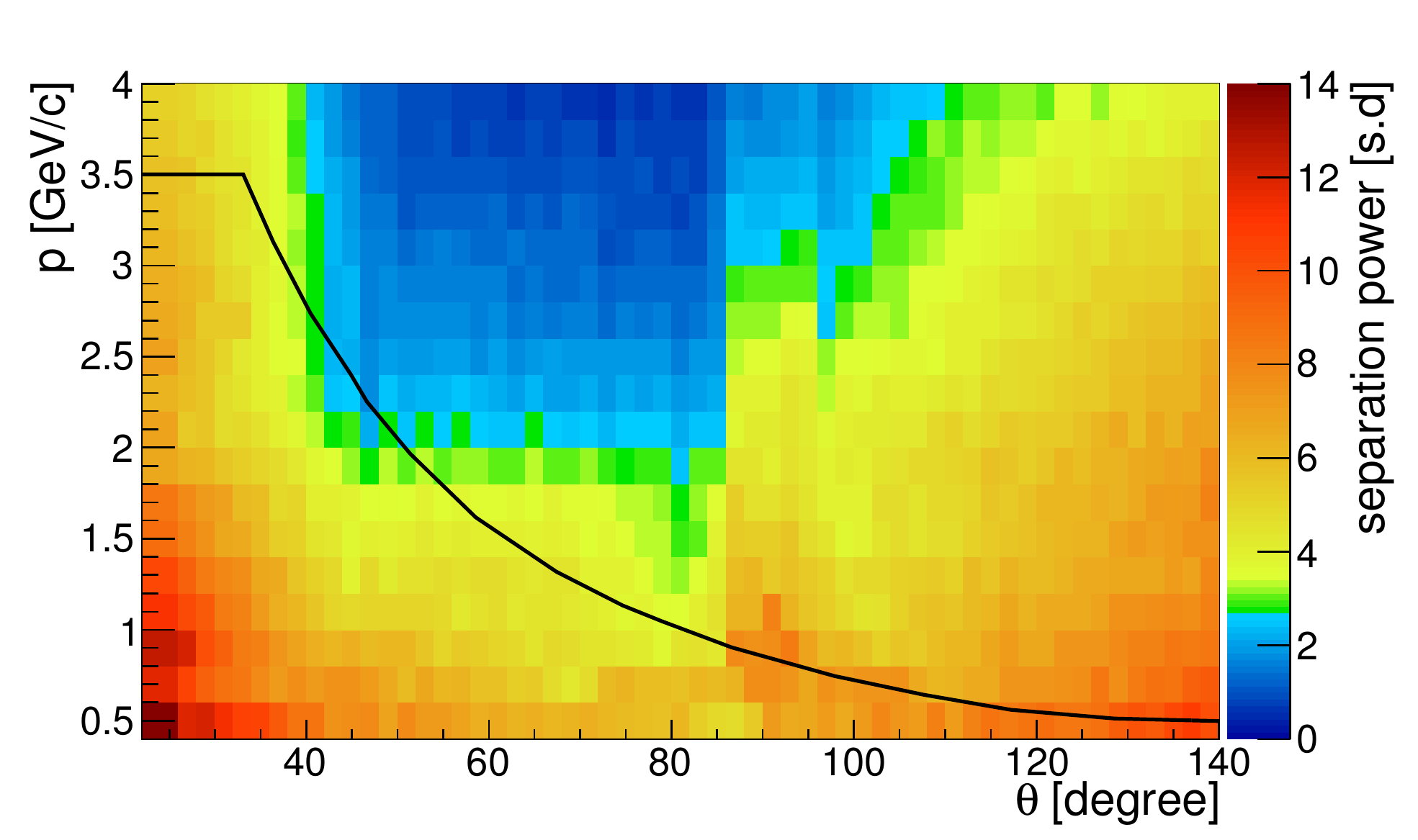}
   \raggedright{\small{Wide plate, cylindrical lens, time-based imaging}}
   \includegraphics[width=0.5\textwidth]{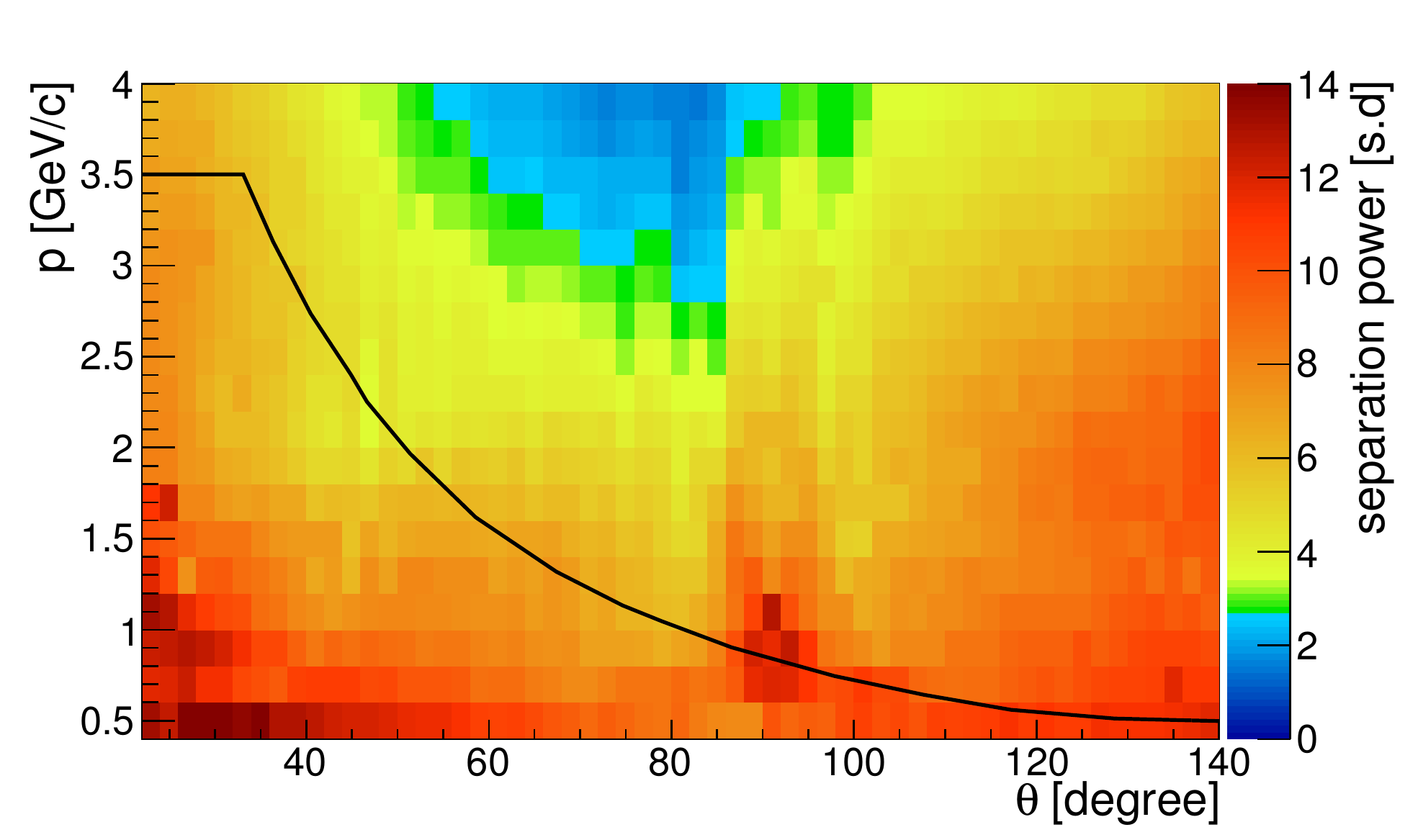}
\vspace*{-2mm}
\caption{
	$\pi$/$K$ separation power as a function of particle momentum and
	polar angle in Geant simulation, determined by the time-based imaging 
	method. 
	The area below the black line corresponds to the final-state phase space
	for charged kaons from various benchmark channels.
	\textit{Top:} Geometry with a wide plate and a fused silica prism EV without focusing optics.
	\textit{Bottom:} Geometry with a wide plate, a fused silica prism EV, and a 3-layer 
	cylindrical lens.
}
\label{sp_barrel}
\end{figure}

The time-based imaging reconstruction method was used to evaluate the
$\pi/K$ separation power for many points of the Barrel DIRC phase space 
acceptance region.
Figure~\ref{sp_barrel} shows the results for the two plate design options,
without focusing optics (top) or with a 3-layer cylindrical lens (bottom).
For both designs the $\pi$/$K$ separation power exceeds the \panda Barrel DIRC
PID requirements for the entire final-state phase space distribution of the kaons,
corresponding to the area below the black line (see Sec.~\ref{cha:design} 
for details).
The plate design with the 3-layer cylindrical lens shows the better performance
with 4--14 s.d. $\pi/K$ separation, second only to the performance of
the narrow bar design with the 3-layer spherical lens.

\begin{table*}[htb]
	\centering
	\caption{
		Performance summary from Geant simulation for the design parameters: 
		focusing type and EV type. The geometry used 5 bars per bar box.
	}
	\label{Tab:simres-ev}
	\vspace*{3mm}
	\begingroup
	\setlength{\tabcolsep}{6pt} 
	\renewcommand{\arraystretch}{1.5} 
	{\small
		\begin{tabular*}{0.7\textwidth}[]{@{\extracolsep{\fill}}llllll}
			\hline EV  type & Lens type  & N$\gamma $  & SPR [mrad] \\
			\hline\multirow{6}{*}{Oil tank} & none & 17 -- 65 & 10 -- 35  &\\
			& 2-layer spherical  & 13 -- 48 & 9 -- 30 &\\
			& 3-layer spherical  & 12 -- 47 & 8 -- 11  &\\
			& 2-layer cylindrical  & 16 -- 62 & 8 -- 17  &\\
			& 3-layer cylindrical  & 15 -- 56 & 9 -- 16  &\\
			& spherical with air gap & 4 -- 34  & 9 -- 19  &\\
			
			\hline \multirow{6}{*}{Prism} & none & 25 -- 108 & 13 -- 24  &\\
			& 2-layer spherical  & 18 -- 70 & 10 -- 27  &\\
			& 3-layer spherical  & 20 -- 94 & 8 -- 14   &\\
			& 2-layer cylindrical  & 22 -- 98 & 13 -- 23  &\\
			& 3-layer cylindrical  & 18 -- 85 & 9 -- 16   &\\
			& spherical with air gap & 4 -- 55  & 10 -- 35  &\\
			
			\hline
		\end{tabular*} 
	}
	\endgroup
\end{table*}

\begin{table*}[htb]
	\centering
	\caption{
		Performance summary from Geant simulation for the design parameters: 
		number of bars per sector, bar width (W) and thickness (T), focusing type and EV type. 
	}
	\label{Tab:simres-radiatorwidth}
	\vspace*{3mm}
	\begingroup
	\setlength{\tabcolsep}{6pt} 
	\renewcommand{\arraystretch}{1.5} 
	{\small\begin{tabular*}{0.9\textwidth}[]{@{\extracolsep{\fill}}lllllll}
			\hline EV type & Number of bars & Bar size W$\times$T [mm$^2$] & Lens type & N$\gamma $ &  SPR [mrad] & \\
			\hline \multirow{6}{*}{Oil tank} & 5 &  32$\times$10 & none & 9 -- 37  & 8 -- 17  &\\
			& 5 &  32$\times$10 & 3-layer spherical  & 5 -- 27  & 7 -- 9   &\\
			& 5 &  32$\times$17 & none & 17 -- 65 & 10 -- 23 &\\
			& 5 &  32$\times$17 & 3-layer spherical  & 12 -- 47 & 8 -- 11  &\\
			& 5 &  32$\times$20 & none & 19 -- 75 & 10 -- 26 &\\
			& 5 &  32$\times$20 & 3-layer spherical  & 12 -- 54 & 8 -- 12  &\\
			\hline \multirow{4}{*}{Prism} & 5 &  32$\times$17 & 3-layer spherical  & 20 -- 95 & 8 -- 14  &\\
			& 4 &  40$\times$17 & 3-layer spherical  & 20 -- 95 & 7 -- 15  &\\
			& 3 &  53$\times$17 & 3-layer spherical  & 18 -- 92 & 8 -- 15  &\\
			& 2 &  80$\times$17 & 3-layer spherical  & 15 -- 80 & 8 -- 25  &\\
			\hline
		\end{tabular*} 
	}
	\endgroup
\end{table*}

\begin{table*}[htb!]
	\centering
	\caption{
		Performance summary from Geant simulation for the design parameter: 
		number of MCP-PMTs per prism. The geometry used 3 bars per bar box and 
		3-layer spherical lenses.
	}
	\label{Tab:simres-MCPlayout}
	\vspace*{3mm}
	\begingroup
	\setlength{\tabcolsep}{6pt} 
	\renewcommand{\arraystretch}{1.5} 
	{\small\begin{tabular*}{0.7\textwidth}[]{@{\extracolsep{\fill}}lllll}
			\hline EV type & Number of MCP-PMTs & N$\gamma $   & SPR [mrad] \\
			\hline\multirow{2}{*}{Prism} &  11 & 20 -- 85 & 8 -- 13 &\\
			&  15 & 20 -- 95 & 8 -- 16 &\\
			\hline
		\end{tabular*} 
	}
	\endgroup
\end{table*}

\putbib[./literature/lit_simulation]
\end{bibunit}

%% file: components/components.tex
\chapter{Components}
\label{cha:components}
\begin{bibunit}[unsrt]

  The \panda Barrel DIRC detector consists of three main parts, in particular:

  \begin{enumerate}
  \item Optical Elements
    \begin{itemize}
    \item Radiator and Lightguide
    \item Focusing Lens
    \item Expansion Volume
    \end{itemize}
  \item Photon Sensors
  \item Front-end Electronics
  \end{enumerate}
 
  While the design is based on the successful BaBar DIRC, several key
  aspects of the \panda Barrel DIRC were optimized to reduce the
  total detector cost, while keeping the required performance for the
  \panda PID, described in Sec.~\ref{cha:design-goals}. 
  The detector cost drivers are the number of photon sensors, 
  which depends on the size and shape of the expansion volume, and 
  the fabrication of radiators, in particular the total number of surfaces to be
  polished.
The cost of the fabrication would be reduced significantly, if the 48 narrow bars foreseen 
in the \panda Barrel DIRC baseline design were replaced by only 16 wide plates. 
It was shown by the Belle~II~TOP counter collaboration that wide plates can be produced 
by optical industry~\cite{plate:industry} with the necessary high quality. 
This choice, however, implies the use of cylindrical instead of spherical 
lenses for the focusing (see Sec.~\ref{cha:simulation}).

\section{Optical Elements}
\label{sec:optical_elements} 
The optical elements of the \panda Barrel DIRC are the radiator, which also serves as 
lightguide, the flat mirror, the focusing lens, and the expansion volume (EV). 
These components have been optimized to collect the maximum
number of the produced Cherenkov photons and to focus them on a flat focal
plane, designed with a shape to be easily equipped with the optimal number
of photon detectors. 

\subsection{Radiator and Lightguide}
\label{sec:radiator}

The Cherenkov radiators are the largest non-mechanical parts of the
\panda Barrel DIRC. 
In contrast to a radiator in a RICH detector, a DIRC radiator also serves as 
a light guide, as the emitted Cherenkov light propagates inside the radiator 
towards its upstream end, where it enters focusing optics and is detected 
by the array of photon sensors.
Despite their large size, the radiators are precision optical components and 
have very strict requirements regarding mechanical and optical tolerances 
and the choice of the material.

\subsubsection*{Requirements}

In order to conserve the angle of the propagating photons and to
avoid light loss, high demands are placed on the squareness and
parallelism, as well as on the surface quality of the
radiators. 
The very valuable results of the BaBar DIRC regarding the appropriate 
surface specifications for their fused silica bars~\cite{babar:dirc} 
have been adapted to the requirements and geometry of the \panda Barrel DIRC. 
Our own simulations (see Sec.~\ref{cha:simulation}) and laboratory tests (described 
in this section) resulted in a set of specifications tailored to the 
needs of \panda. 
The procedures developed and utilized for the Barrel DIRC R\&D will also 
be part of the quality assurance process to cross-check the radiator 
properties after delivery (see Sec.~\ref{subsec:QA-Radiators}).

  In the \panda Barrel DIRC baseline design photons may have several hundred internal
  reflections inside the radiator. Scalar scattering theory for smooth
  surfaces predicts that the light loss due to surface scattering is
  proportional to the square of the surface roughness. This leads to the 
  requirement to have a maximum surface roughness of 10~\AA~RMS for the 
  large surfaces and 25~\AA~RMS for the ends of the bar.

To limit angular smearing, the parallelism and squareness of
the long radiator sides and faces are an important part of the specification. 
Due to the large number of internal reflections the squareness must not exceed a
value of 0.25~mrad for side-to-face angles and the total thickness
variation is required to be 25~$\mu$m or less. 
The demands for the ends can be less restrictive, i.e. the squareness 
of the side-to-end and face-to-end angles must not exceed 0.5~mrad.  
The length of each fabricated radiator piece is 1200$^{\mathrm{+0}}_{\mathrm{-1}}$~mm 
and two radiator pieces are glued end-to-end to form a long bar, covering the 
full length of the Barrel DIRC. 
In the baseline design the thickness and width are 17$^{\mathrm{+0}}_\mathrm{{-0.5}}$~mm 
and 53$^{\mathrm{+0}}_{\mathrm{-0.5}}$~mm, respectively. 
In the optional design with a wide plate the radiator has a width of 
160$^{\mathrm{+0}}_\mathrm{{-0.5}}$~mm and a thickness of 17$^{\mathrm{+0}}_\mathrm{{-0.5}}$~mm.

  \subsubsection*{Choice of Material}

  The material for the optical components in the \panda
  Barrel DIRC has to fulfill the following requirements:
  \begin{itemize}
  \item Excellent optical properties,
  \item Radiation hardness,
  \item Excellent polishability.
  \end{itemize}
  Quartz (chemically $\mathrm{SiO_2}$), which meets the
  above mentioned criteria, exists in three different
  compositions. The crystalline form of quartz (natural quartz) is
  birefringent, contains a high level of impurities and hence cannot
  be used for the DIRC optics. An amorphous form of quartz (natural
  fused silica) is produced by crushing and melting natural
  quartz. Although the optical properties would fit the requirements,
  a considerably large amount of impurities remains, which reduces
  radiation hardness.

  A third form of quartz (synthetic fused silica) is made of different
  feedstock, such as silicon tetra-chloride ($\mathrm{SiCl_4}$). This
  material is burned in an oxygen atmosphere at around $2000^\circ
  \mathrm{C}$ and forms a large ingot, which is then processed 
  further~\cite{babar:dirc}. This process results in a very pure
  material, which is widely used in optical applications. 
  Depending on the level of interstitial hydrogen, the radiation hardness 
  can be tailored to the application. 
  The optical homogeneity was a concern during the selection of fused silica 
  material for the BaBar DIRC. 
  Several candidates materials showed significant striae and/or inclusions, 
  which would have led to unacceptable photon yield or resolution losses.
  Since then, improvements to the material production process resulted in 
  much better optical homogeneity.
  
  Available materials (amongst others) are
  Spectrosil~2000 and Suprasil~1 and 2
  by Heraeus~\cite{Heraeus}, HPFS~7980
  by Corning~\cite{corning}, NIFS-S
  by Nikon~\cite{nikon} and Lithosil~Q0
  by Schott~\cite{schott}.

  \subsubsection*{Radiation Hardness of Radiator Material}
\label{sec:RadHardnessMatthias}

\newcommand*{\cary}{Cary~300}

The optical properties of the radiator material for DIRC-type Cherenkov detectors 
are crucial for the overall performance. 
The generated Cherenkov photons travel a substantial distance inside the radiator 
material, unlike traditional RICH counters, and undergo many reflections off 
the surfaces. 
Thus its optical properties must remain unchanged in the radiation fields as 
encountered in \panda. 

\begin{figure*}[tb]
  \begin{center}	
    \resizebox{1.8\columnwidth}{!}{\includegraphics{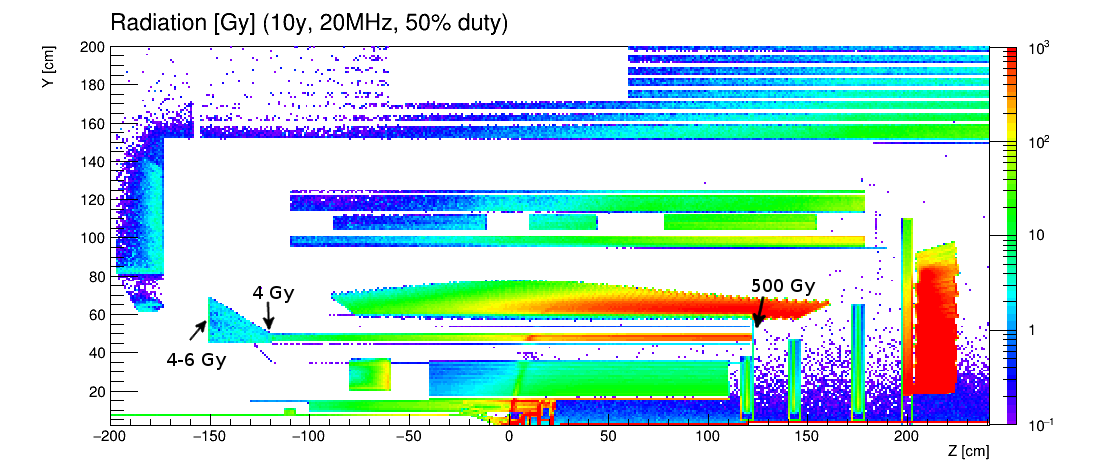}}
    \caption{Simulated radiation map of the \panda target spectrometer for 10 years of operation 
      at an average interaction rate of $20$~MHz and beam on a proton target for 50\% of the year. 
      Radiation levels within the figures are indicated at prominent positions: 
      backside of the expansion volume where the photon detectors and electronics are attached (left),
      the position of the focusing lenses (middle), 
      and the downstream side of the radiator bars, equipped with mirrors.}  
    \label{fig:radmap_dirc}
  \end{center}
\end{figure*}

The PandaRoot simulation framework was used to estimate the dose level expected for
the \panda experiment.
A sample of $10^8$ events of antiproton-proton collisions at a momentum of 
\mbox{$p = 15$~GeV/c} were generated with the DPM event generator.
The results were scaled to 10 years of \panda operation, assuming an average interaction 
rate of $20$~MHz and a detector operation during 50\% of the year. 
The radiation map is shown in Fig.~\ref{fig:radmap_dirc} together with selected values at 
specific positions, showing that the expected doses for the optical elements is between
4~Gy and 500~Gy. 
For the photon detectors and the front-end electronic, the flux of particles is of interest. 
The simulated flux at the upstream side of the expansion volume is 
\mbox{$2\times10^{11}$ cm$^{-2}$}, half of it due to neutrons.

Synthetic fused silica has already been identified as the most suitable material for radiators in DIRC-type RICH detectors by the BaBar DIRC group \cite{babar:dirc}. 

In general, synthetic fused silica is produced by flame hydrolysis of the raw materials. However, different processing of the raw materials and conditions during production lead to different categories. Most noteworthy, regarding the radiation hardness, is the OH-content which distinguishes ``dry'' and ``wet'' types. Dry types contain typically a few hundred ppm OH while wet types are between 800~ppm and 1300~ppm.

Several types of wet synthetic fused silica were investigated using a proton irradiation facility at KVI, Groningen, The Netherlands, in order to test the effects of radiation up to 10~Mrad, well beyond the expected lifetime dose for \panda. 
Further tests, using $\gamma$-ray irradiation at the University of Giessen, Germany, were performed on dry and wet synthetic fused silica samples. The induced radiation damage in the UV region was studied to reveal possible damage mechanisms.
The various types of synthetic fused silica studied for the \panda Barrel DIRC are listed in Tab.~\ref{Tab:fused_silica-types}.

\begin{table}[hbt]
\centering
\setlength{\tabcolsep}{6pt} 
\renewcommand{\arraystretch}{1.5} 

\caption{List of synthetic fused silica types investigated for the \panda Barrel DIRC
(see text).}
\ \

\label{Tab:fused_silica-types}
{\begin{tabular}[]{@{\extracolsep{\fill}}llll}
\hline  
Vendor & Type & Irradiation & OH-level \\ 
\hline 
Corning & HPFS 7980 & Proton & wet\\
Heraeus & Suprasil 1 & Proton & wet \\
Schott & Lithosil Q0 & Proton & wet \\
\hline
Heraeus & Suprasil 2A & $\gamma$-ray & wet\\
Heraeus & Suprasil 311 & $\gamma$-ray & dry \\
Nikon & NIFS-S & $\gamma$-ray & wet \\
Nikon & NIFS-U & $\gamma$-ray & wet \\
Nikon & NIFS-A & $\gamma$-ray & wet \\
Nikon & NIFS-V & $\gamma$-ray & dry \\
\hline\end{tabular} 
}
\end{table}

\begin{figure}[h]
	\centering
		\includegraphics[width=0.9\linewidth]{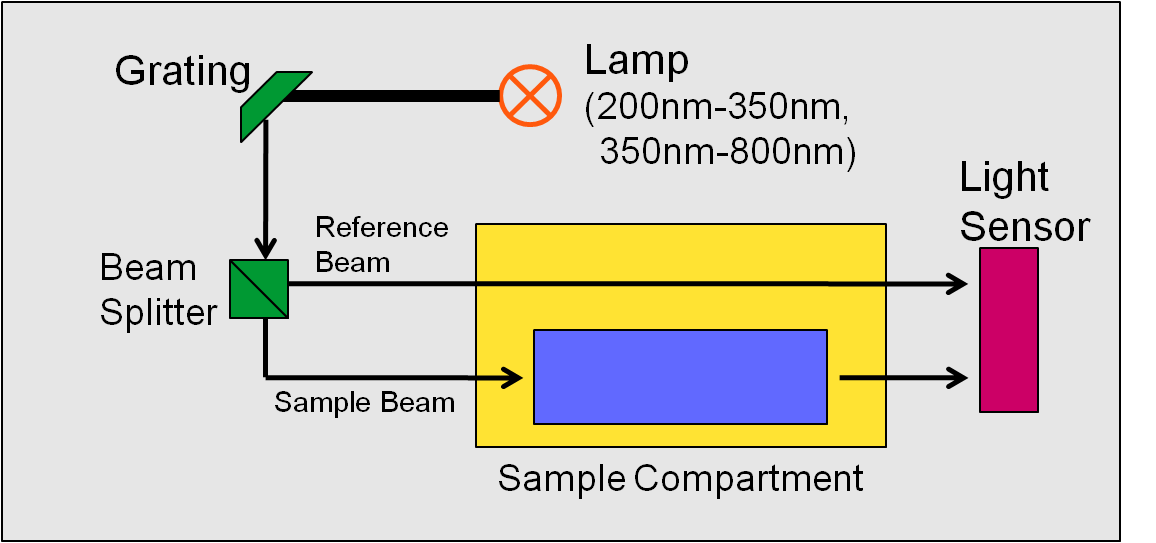}
	\caption{Schematic drawing of the light paths in a Varian Cary~300 spectrophotometer. The light source consists of two lamps covering the spectral range from 200~nm to 350~nm and 350~nm to 800~nm, respectively.}
	\label{fig:Cary300_schematic}
\end{figure}

The optical transmission of the fused silica samples was measured with commercial spectrophotometers covering a spectral range of 200~nm to 800~nm.
The spectrophotometers feature a dual-beam set-up (see Fig.~\ref{fig:Cary300_schematic}) allowing for very precise measurements. However, absolute measurements are not possible due to design-inherent beam properties so that the recorded transmission values are influenced by the sample length and position within the sample compartment. Comparative measurements are not affected, provided certain parameters, e.g. sample size and position, are identical.

\begin{figure}[!h]
\begin{center}
	\includegraphics[width=0.8\linewidth]{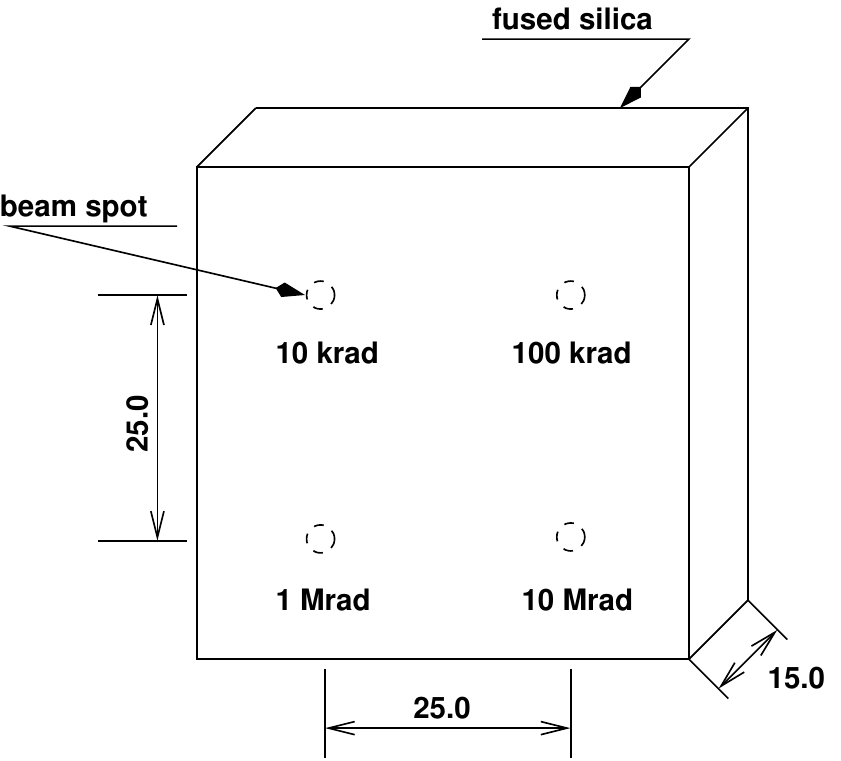}
\end{center}
\caption{Schematic drawing of expected dose distribution across a sample tile for KVI proton irradiation.}
\label{fig:irradiation}
\end{figure}

\begin{figure}[!h]
\begin{center}
	\includegraphics[width=0.9\linewidth]{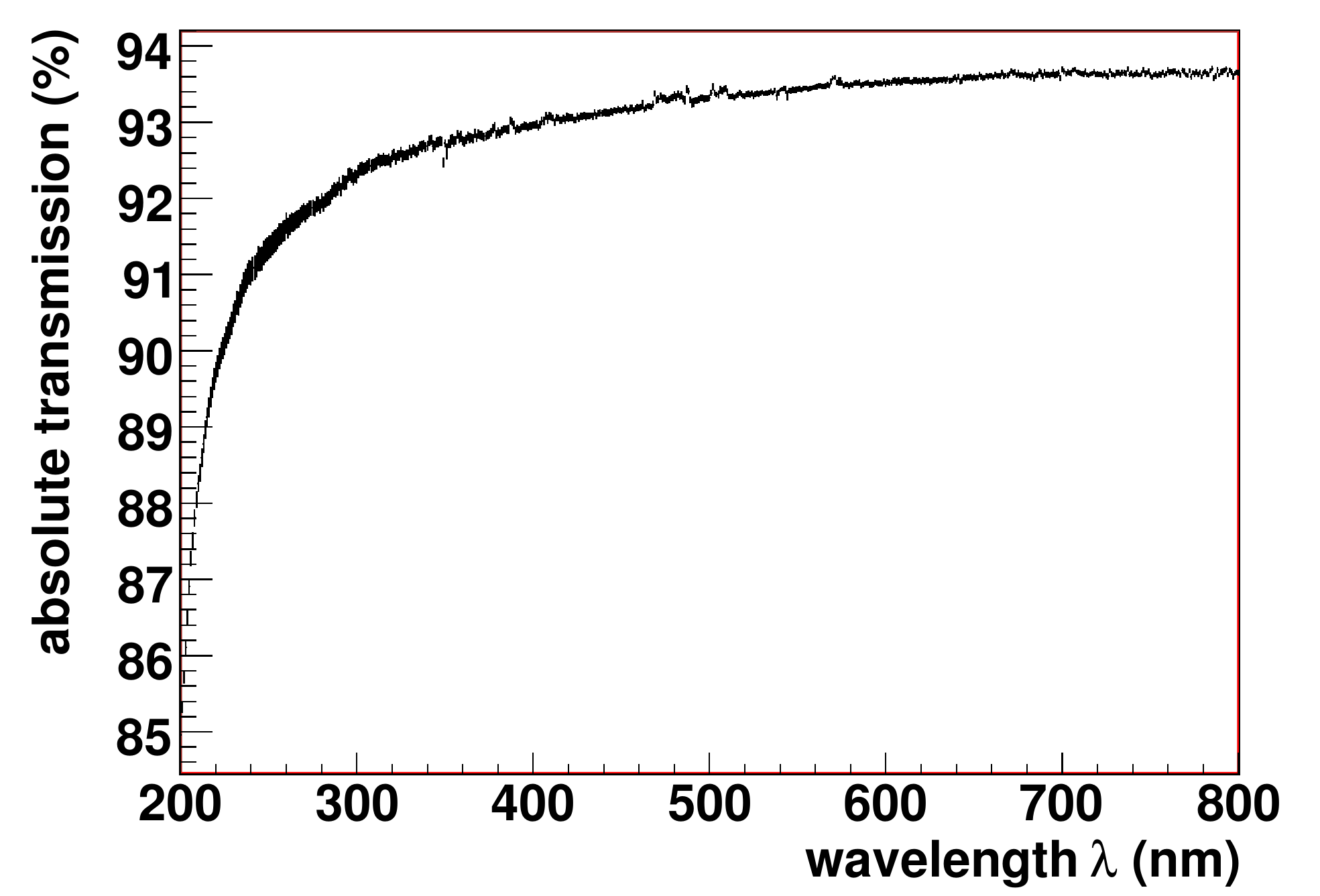}
\end{center}
\caption{Absolute transmission as function of wavelength for Suprasil~1. Transmission values were not corrected for Fresnel losses. The error bars display the statistical fluctuations only, no systematic effects were included.}
\label{fig:transmission}
\end{figure}

The proton irradiation was carried out at KVI in Groningen using a proton beam extracted from KVI's cyclotron with an energy of 150~MeV. 
The beam passed through a 0.4~mm scattering foil and, after traveling 450~mm, through a collimator with an aperture of 5~mm and a length of 45~mm. The dimensions of the Suprasil and Lithosil samples were 50$\times$50$\times$15~mm$^3$, whereas for the Corning sample they were 80$\times$80$\times$20~mm$^3$.
The samples were placed 130~mm downstream of the exit point of the collimator. On each sample four spots, each separated by about 25~mm from its adjacent spot (Fig.~\ref{fig:irradiation}), were irradiated with different dose levels (10~krad, 100~krad, 1~Mrad and 10~Mrad) to cover the range of the expected total irradiation dose.
The repeatability of the results across the sample surface was estimated by measuring at 4 different spots prior to irradiation. The obtainable precision is estimated to be $\pm$\,0.4\% of the absolute transmission. The spots were chosen to closely match the planned irradiation spots. Figure~\ref{fig:transmission} shows the averaged transmission of the Suprasil~1 sample prior to irradiation.
Problems with the beam position during the 10~krad run were discovered only after the samples 
were returned from KVI examination.
Instead of a disc-shaped irradiation spot a broad, elongated band towards the edge of the tiles was visible for a reference crown glass sample. An estimate for the true accumulated dose for this run was not possible and the results were discarded. However, the other runs at higher dose levels were not affected.

The transmission measured after irradiation was compared to the reference measurements prior to irradiation. The result is given as normalized transmission loss $\Delta T'$:
\begin{equation}
	\label{eq:trans_loss}
	\Delta T' = \frac{T_{before} - T_{after}}{T_{before}}\quad,
\end{equation}
which is used to account for Fresnel losses occurring at the surfaces of the samples. $\Delta T'$ thus describes the change of transmission due to absorption effects inside the bulk material. The uncertainty of $\Delta T'$ is better than $\pm$\,1\% absolute transmission.
\begin{figure}[!h]
\begin{center}
\begin{minipage}{0.49\textwidth}
	\includegraphics[width=0.93\textwidth]{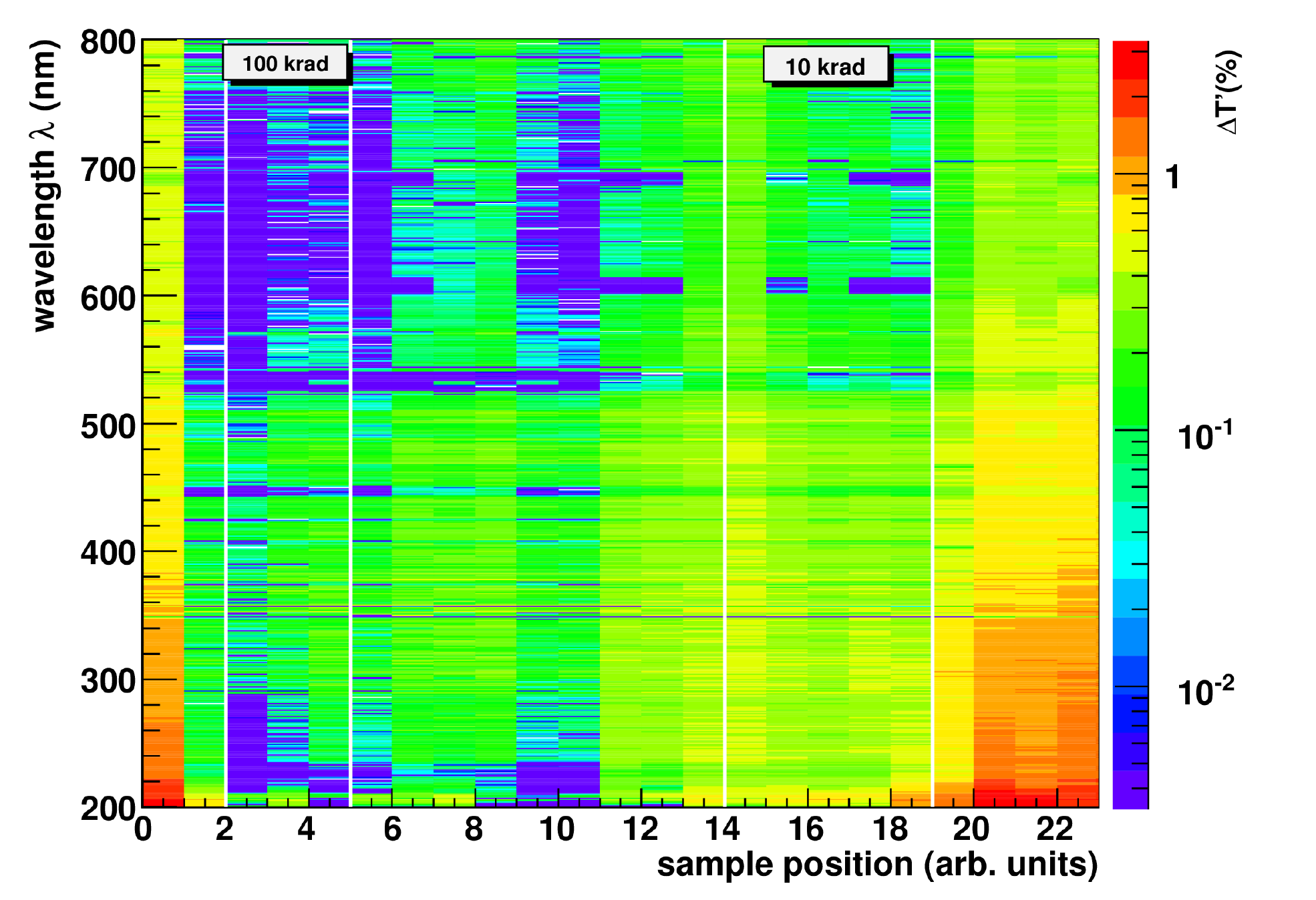}
\end{minipage}
\hfill
\begin{minipage}{0.49\textwidth}
	\includegraphics[width=0.93\textwidth]{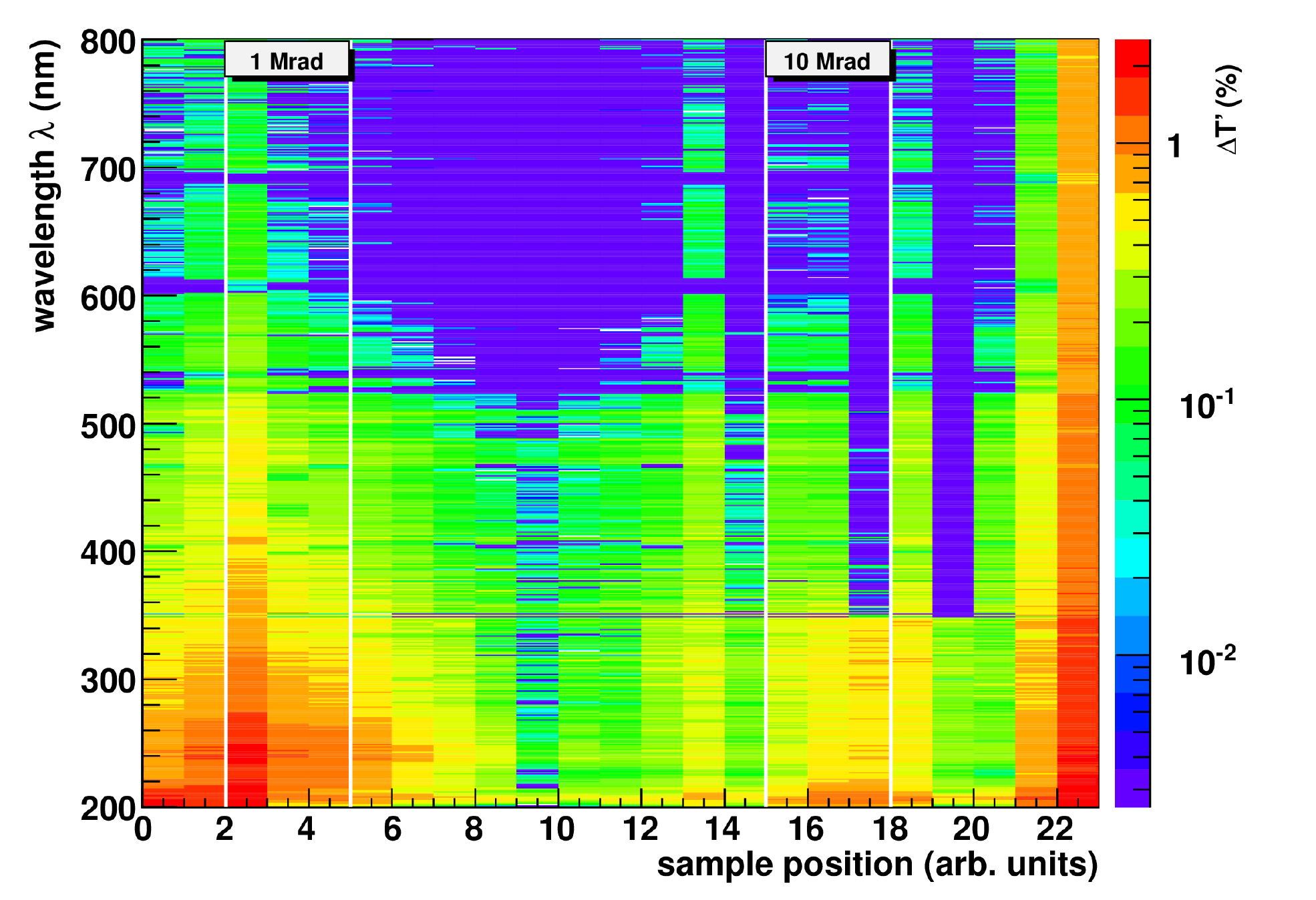}
\end{minipage}
\end{center}
\caption{Normalized transmission difference $\Delta T'$ for Heraeus Suprasil~1. The vertical lines indicate the expected position of the radiation spots. No distinct features corresponding to the irradiation spots are observed within the obtained precision. Large deviations at the corners are attributed to edge effects.}
\label{fig:supra}
\end{figure}
No degradation of transmittance was found for all three fused silica pieces. The results for Suprasil~1 are shown, as example in Fig.~\ref{fig:supra}, with the other two samples showing similar results \cite{panda:radiation_hardness_2008}. None of the samples exhibit any significant radiation damage. Large deviations at the corners are attributed to edge effects of the measurement. The peculiar difference at the 1~Mrad spot in Fig.~\ref{fig:supra} is thought to be caused by surface contamination from previous cleaning using Propanol and Methanol, especially since the 10~Mrad spot does not show any degradation.
Previous studies on Suprasil Standard \cite{babar:dirc}, by contrast, found a significant transmission reduction in the UV region after irradiation with a dose of 280~krad. Despite the fact that a different sample geometry was used, the results from \cite{babar:dirc} suggest that a significant deterioration in the sample should have occurred, however such a deterioration has not been observed in our sample.

In the following, more irradiation studies using a $^{60}$Co source at the University of Giessen, Germany, included also dry synthetic fused silica types. Furthermore, the sample geometry was changed to a cylindrical shape with a length of $L = $100~mm to improve the sensitivity to radiation induced damages in the UV region. Synthetic fused silica defect mechanisms are well-studied within UV laser applications \cite{fused_silica:defects}. The defect models developed in this area suggest two absorption lines at wavelengths of 210~nm and 260~nm, respectively. Furthermore, these models predict a dependency of the radiation damage induced by laser light on the amount of interstitial hydrogen present in the sample. 

\begin{figure}[!h]
 \centering
 \includegraphics[width=0.49\textwidth]{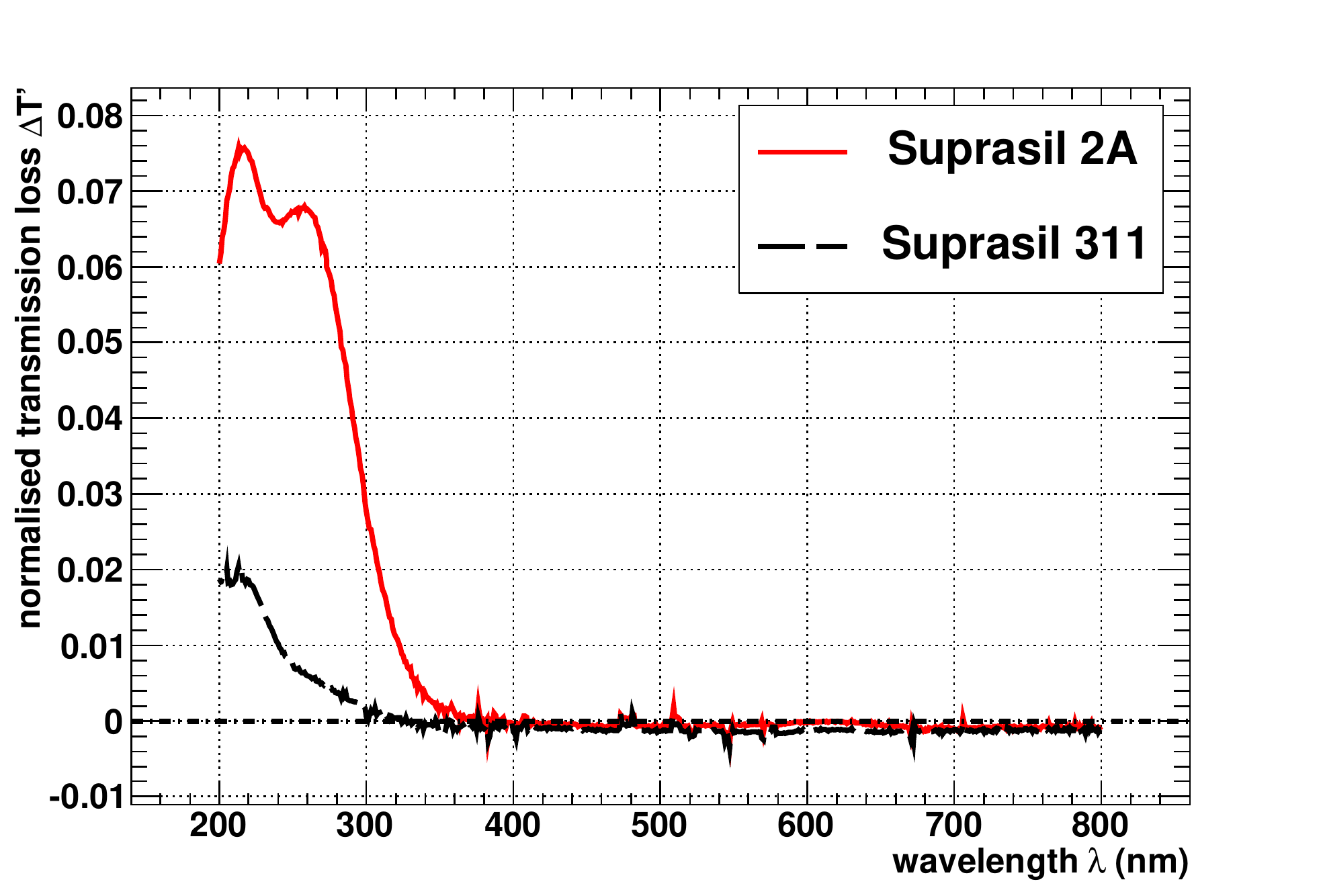}
 \caption{Radiation induced change of transmission for long samples ($L=$100~mm) of Suprasil~2 (shown in red) and 311 (shown in black). The error bars include systematic uncertainties from day-to-day variations of the calibration of the \cary{} spectrophotometer.}
 \label{fig:suprasil_damage_comp}
\end{figure}

A first comparison of readily available samples of Heraeus Suprasil~2A and Suprasil~311 revealed significant differences with respect to radiation hardness (see Fig.~\ref{fig:suprasil_damage_comp}). Both materials show damage after irradiation with a dose of 100~krad, which was achieved in approximately 4~h. 

The level of damage, however, varies between Suprasil~2A and 311. Suprasil~2A is affected much more, the magnitude of the computed transmission change $\Delta T'$ is 6-7 times larger compared to Suprasil~311 and the damage extends to wavelengths well above 300~nm, which is the critical wavelength for the Barrel DIRC.

Two distinct absorption bands became visible in the irradiated Suprasil~2A specimen at 217$\pm$3~nm and 255$\pm$8~nm. Suprasil~311 was affected much less and with only one absorption band at 214$\pm$17~nm and the damage limited to wavelengths below 300~nm. The reported absorption bands most probably correspond to well-known absorption bands at 214~nm and 242~nm, respectively \cite{Cohen1957,Nelson1960,Cohen1955,Lell1960}. The former is related to the silica network and does not require any impurity atoms whereas the latter is attributed to either metallic impurities (e.g. Ge) or interstitial silicon atoms. According to information available from the manufacturer \cite{Heraeus2008}, Suprasil~2A and 311 only differ in the OH-content (see also Tab.~\ref{Tab:fused_silica-types}).

Based on the initial results Heraeus specially prepared four samples each of Suprasil~2A and Suprasil~311 with varying interstitial hydrogen content to test the defect models \cite{panda:radiation_hardness_2010}. The hydrogen content of each sample was measured by Heraeus using Raman spectroscopy (see Tab.~\ref{tab:SuprasilSamplesAndhydrogenContent}). Three samples (090BP, 090BF and 090BG) have hydrogen levels below the sensitivity limit which differs from sample to sample due to the experimental set-up \cite{heraeus:bkuehn:raman}. Of these the two Suprasil~311 samples (090BF and 090BG) do have, however, a different hydrogen content which is known from production parameters \cite{heraeus:bkuehn:raman}.

\begin{table}[h]
	\centering
\setlength{\tabcolsep}{6pt} 
\renewcommand{\arraystretch}{1.5} 
	\caption{Hydrogen content as determined by Raman spectroscopy of all Suprasil samples prepared by Heraeus Quartzglas. Three samples (090BP, 090BF and 090BG) have hydrogen levels below the sensitivity limit.}
	\ \
	
	\label{tab:SuprasilSamplesAndhydrogenContent}
		\begin{tabular}{ccr@{\,$\times$\,}l}
		\hline
		Type & Sample ID & \multicolumn{2}{c}{hydrogen content} \\
		\hline
		\multirow{4}{*}{2A} & 090BP & $<$ 1.0 & 10$^{15}$ mol/cm$^3$ \\
		                    & 090BL &     1.3 & 10$^{16}$ mol/cm$^3$ \\
		                    & 090BN &     1.4 & 10$^{17}$ mol/cm$^3$ \\
		                    & 090BK &     1.7 & 10$^{18}$ mol/cm$^3$ \\
		\hline
		\multirow{4}{*}{311} & 090BF & $<$ 0.9 & 10$^{15}$ mol/cm$^3$ \\
		                     & 090BG & $<$ 1.2 & 10$^{15}$ mol/cm$^3$ \\
		                     & 090BH &     1.6 & 10$^{16}$ mol/cm$^3$ \\
		                     & 090BJ &     2.3 & 10$^{17}$ mol/cm$^3$ \\
		\hline
		\end{tabular}
\end{table}

\begin{figure}[!h]
	\centering
		\includegraphics[width=0.95\linewidth]{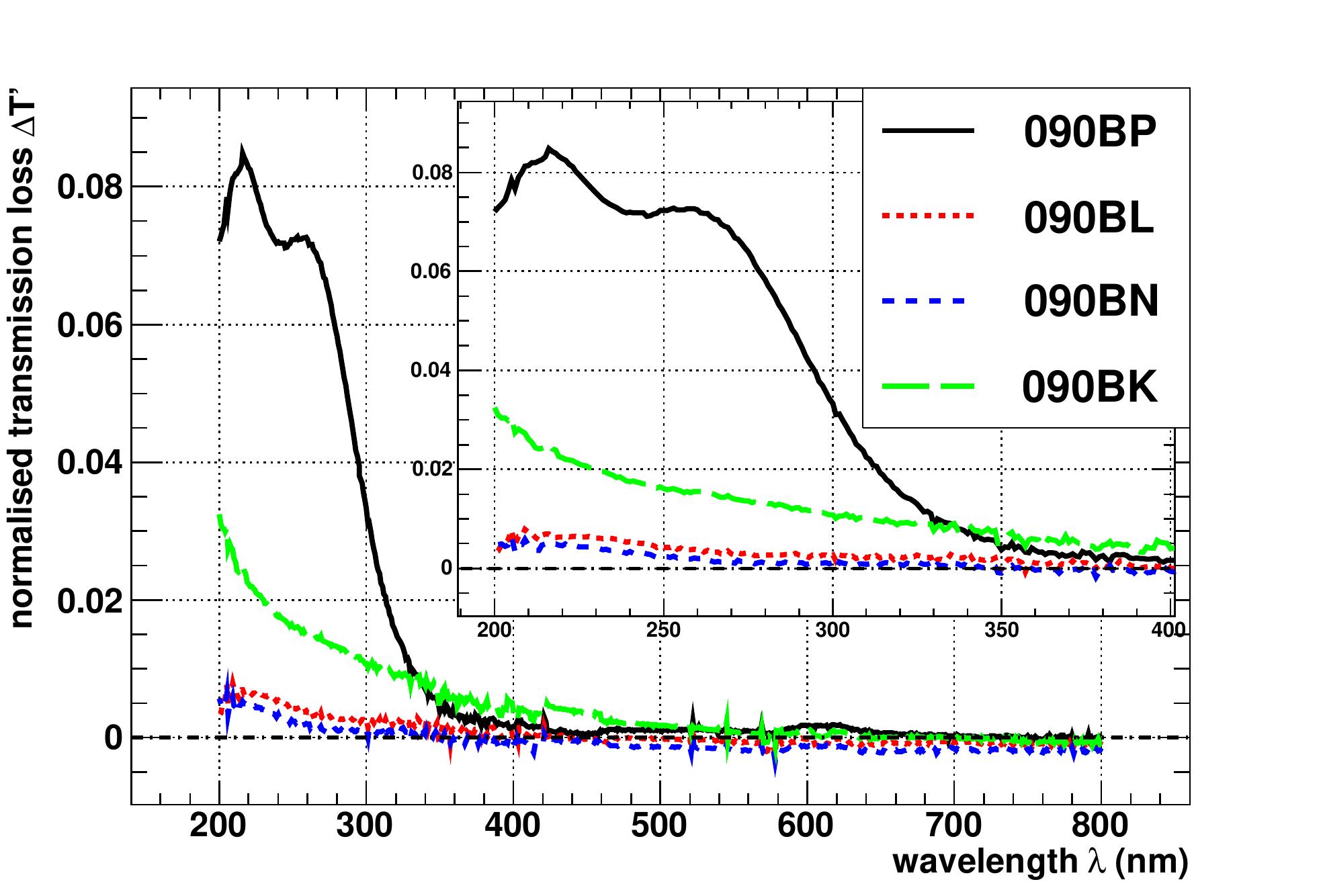}
	\caption{Normalized transmission loss $\Delta T'$ for Suprasil~2A samples (see Tab.~\ref{tab:SuprasilSamplesAndhydrogenContent}) as a function of wavelength after an accumulated dose of 100~krad. Inset shows most affected blue-UV region. Two absorption bands at wavelengths of 210~nm and 260~nm are clearly visible for the most hydrogen-depleted sample (090BP).}
	\label{fig:Suprasil2A_dT_100krad}
\end{figure}

\begin{figure}[!h]
	\centering
		\includegraphics[width=0.95\linewidth]{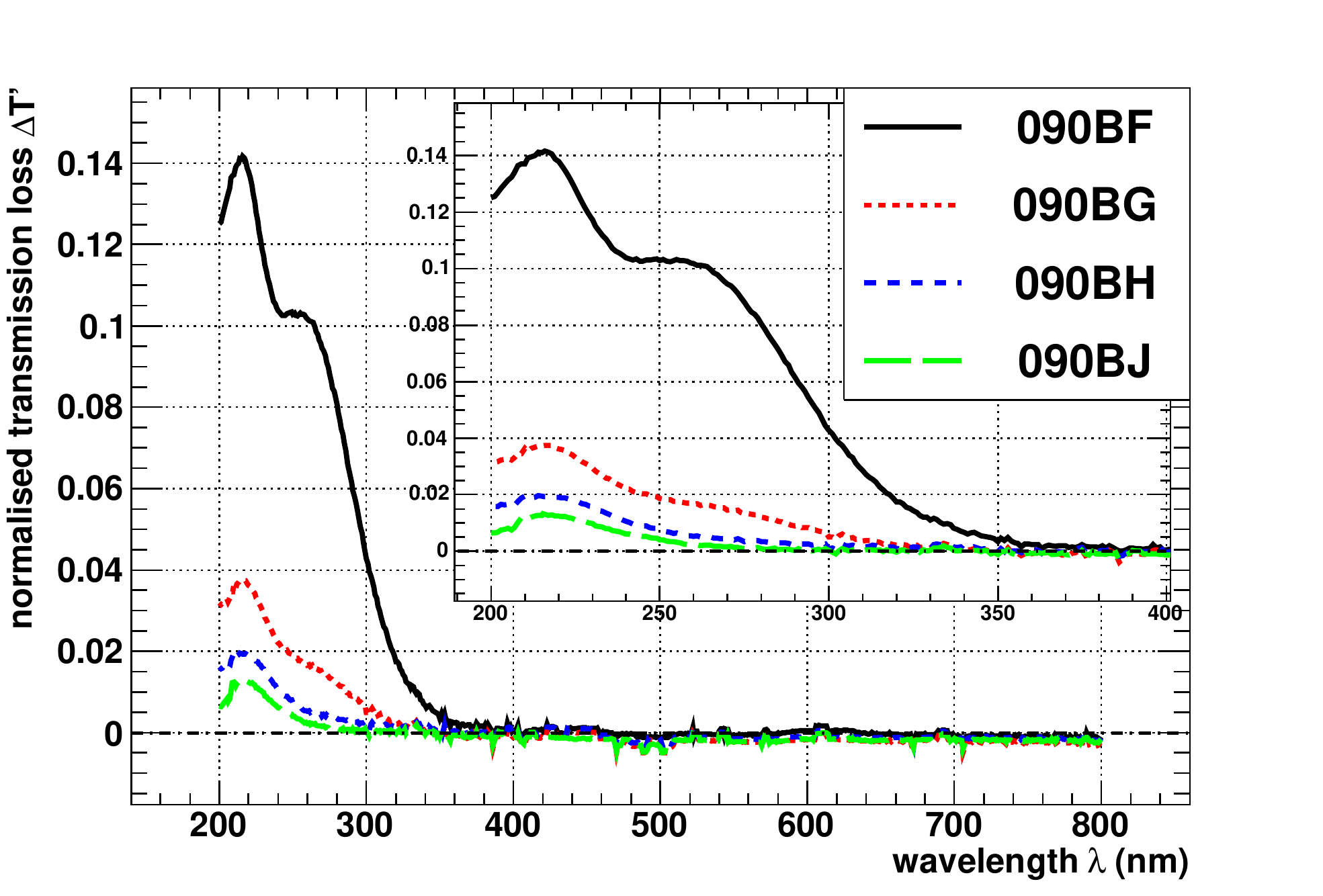}
	\caption{Normalized transmission loss $\Delta T'$ for Suprasil~311 samples (see Tab.~\ref{tab:SuprasilSamplesAndhydrogenContent}) as a function of wavelength after an accumulated dose of 100~krad. Inset shows most affected blue-UV region. Two absorption bands at wavelengths of 210~nm and 260~nm are clearly visible for the most hydrogen-depleted sample (090BF).}
	\label{fig:Suprasil311_dT_100krad}
\end{figure}

The normalized transmission difference $\Delta T'$ is computed according to Eqn.~\ref{eq:trans_loss}. Samples of both types with low hydrogen content exhibit enhanced radiation damage with peaks at the expected wavelengths of 210~nm and 260~nm. The magnitude of the degradation decreases with increasing hydrogen levels for both Suprasil types (see Fig.~\ref{fig:Suprasil2A_dT_100krad} and Fig.~\ref{fig:Suprasil311_dT_100krad}). Nevertheless, the Suprasil~2A sample with the highest hydrogen content (090BK) shows an increased degradation but with a different spectral shape compared to the hydrogen-depleted sample of the same type. This, however, is attributed to cleaning residues as will be explained later. No significant effects on the transmission properties above 400~nm were observed for all samples regardless of the hydrogen level present. 

\begin{figure}[!h]
	\centering
	\includegraphics[width=0.95\linewidth]{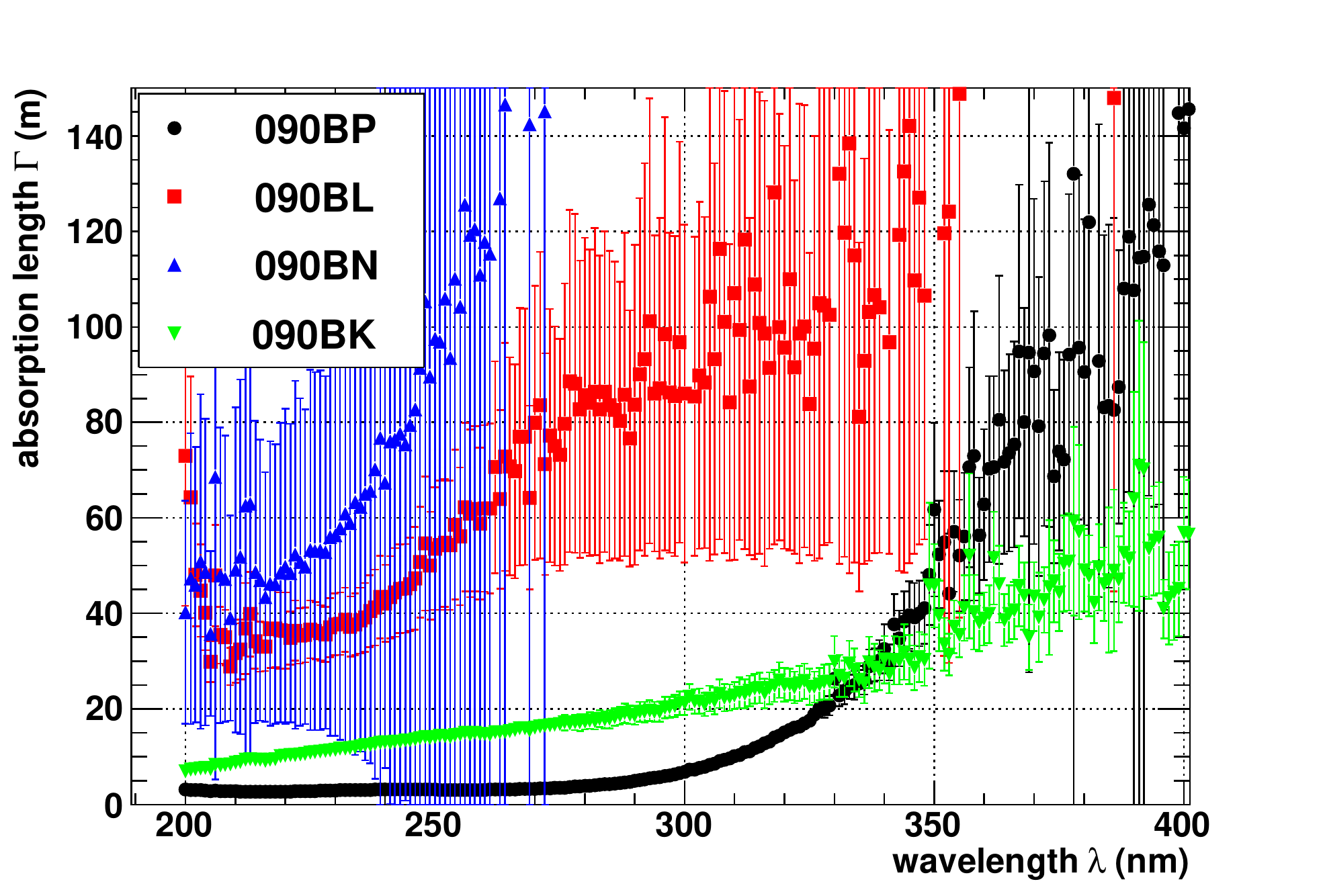}
	 	 \caption{Radiation induced absorption length $\Gamma$ for Suprasil~2A for different hydrogen levels (see Tab.~\ref{tab:SuprasilSamplesAndhydrogenContent}) as function of the wavelength.}
	\label{fig:Suprasil 2A_AbsLength_100krad}
\end{figure}

\begin{figure}[!h]
	\centering
	\includegraphics[width=0.95\linewidth]{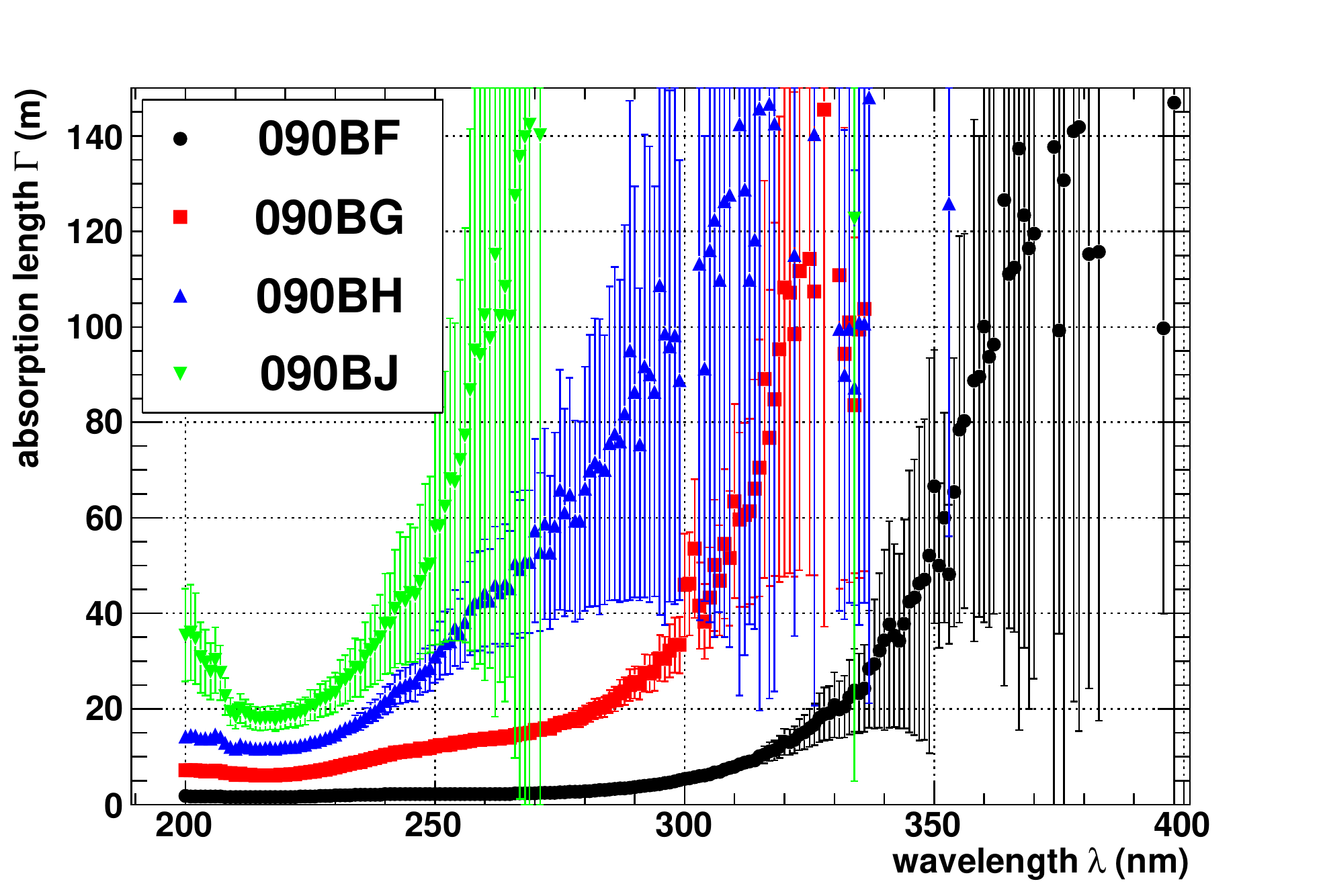}
	 	 \caption{Radiation induced absorption length $\Gamma$ for Suprasil~311 for different hydrogen levels (see Tab.~\ref{tab:SuprasilSamplesAndhydrogenContent}) as function of wavelength.}
	\label{fig:Suprasil 311_AbsLength_100krad}
\end{figure}

The corresponding absorption length $\Gamma$ due to radiation damage is given by
\begin{equation}
	\label{eq:abs_length}
	\Gamma = -\frac{L}{\ln(1-\Delta T')}\quad ,
\end{equation}
with $L$ being the sample length and $\Delta T'$ given by Eqn.~\ref{eq:trans_loss}. The absorption length $\Gamma$ drops to values as low as 2~m (see Fig.~\ref{fig:Suprasil 2A_AbsLength_100krad} and Fig.~\ref{fig:Suprasil 311_AbsLength_100krad}), which is comparable to the path lengths in the anticipated applications, at wavelengths below 400~nm.

\begin{figure}[h!]
	\centering	\includegraphics[width=0.95\linewidth]{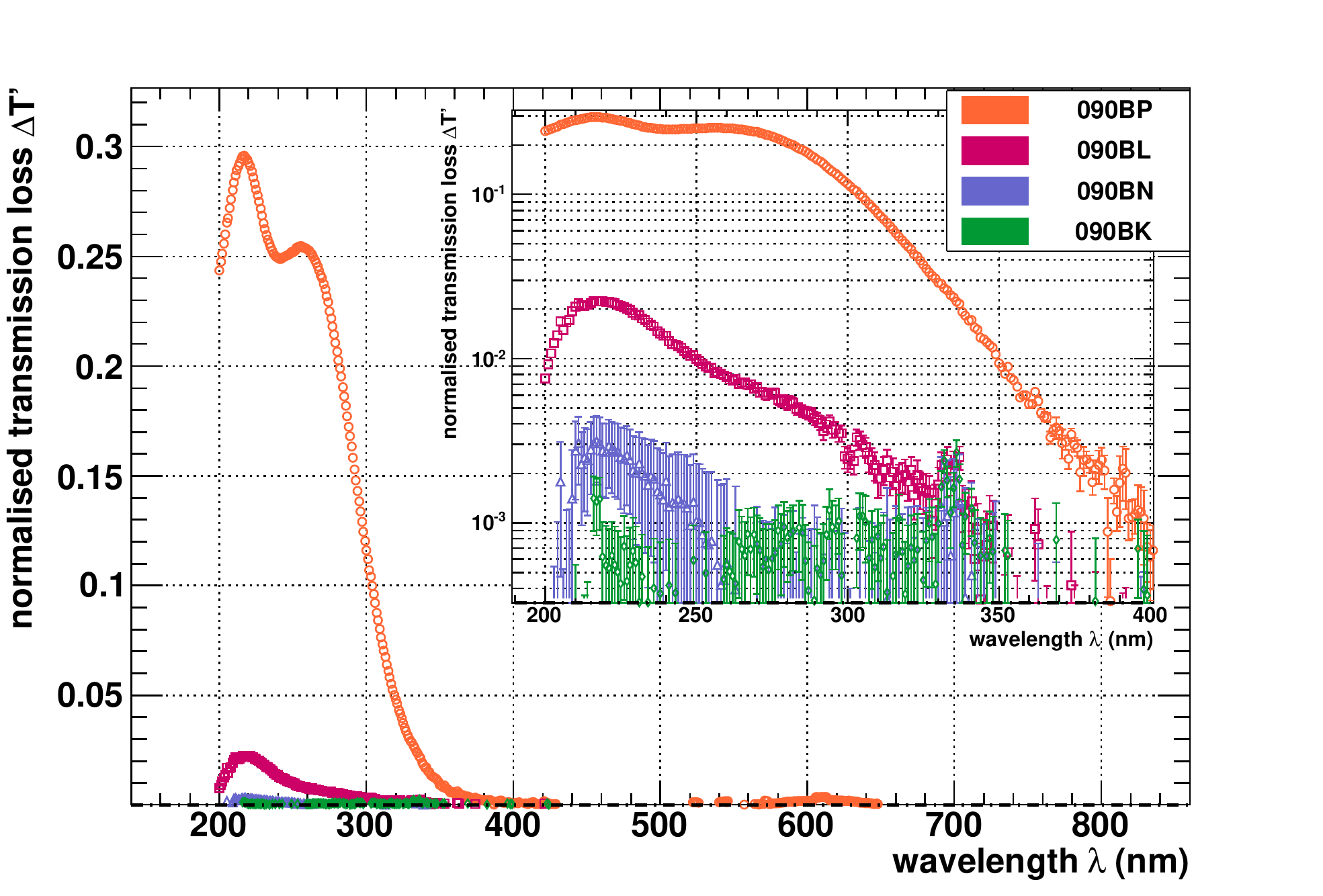}
	\caption{Normalized transmission loss $\Delta$T' of Suprasil~2A samples with different hydrogen content as a function of wavelength after 600~krad total integrated dose. The inset shows the the normalized transmission loss for wavelengths below 400~nm. For details of the different samples see Tab.~\ref{tab:SuprasilSamplesAndhydrogenContent}.}
	\label{fig:Suprasil2A_dT_600krad}
\end{figure}

\begin{figure}[!h]
	\centering	\includegraphics[width=0.95\linewidth]{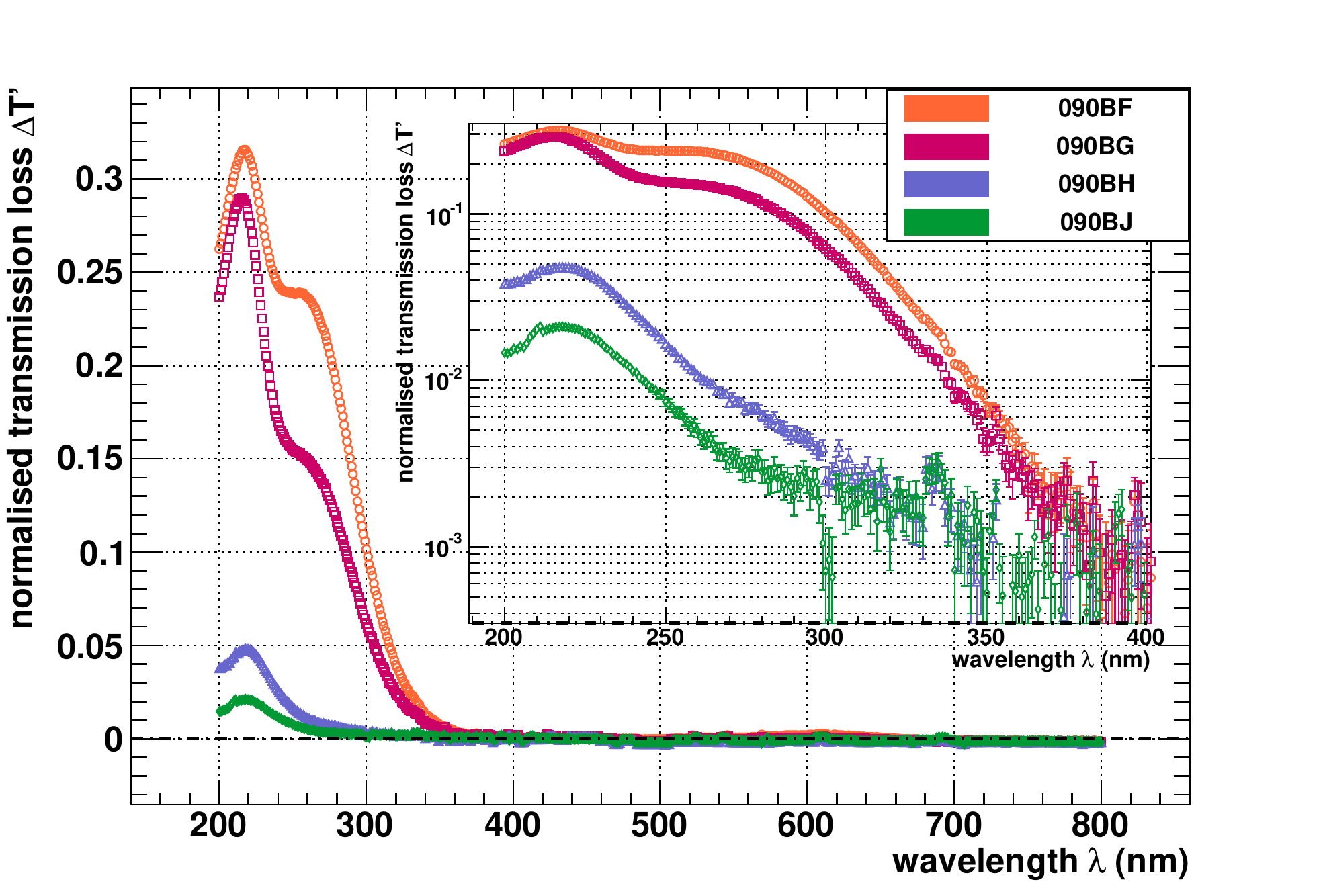}
	\caption{Normalized transmission loss $\Delta$T' of Suprasil~311 samples with different hydrogen content as a function of wavelength after 600~krad total integrated dose. The inset shows the the normalized transmission loss for wavelengths below 400~nm. For details of the different samples see Tab.~\ref{tab:SuprasilSamplesAndhydrogenContent}.}
	\label{fig:Suprasil311_dT_600krad}
\end{figure}

A further dose of 500~krad was applied to investigate the role of hydrogen and its consumption during irradiation. As known from the initial irradiation, the radiation damage is enhanced in hydrogen-depleted samples. This effect is observed, even more pronounced, after the second irradiation (see Fig.~\ref{fig:Suprasil2A_dT_600krad} and Fig.~\ref{fig:Suprasil311_dT_600krad}). Any radiation-induced damage is limited to wavelengths below 400~nm with two absorption lines around 210~nm and 260~nm. These absorption lines correspond to attached electrons on Si atoms, denoted by E' and to Non-Bridging Oxygen Hole (NBOH) defect centers~\cite{fused_silica:defects}.

\begin{figure}[!h]
	\centering
		\includegraphics[width=0.95\linewidth]{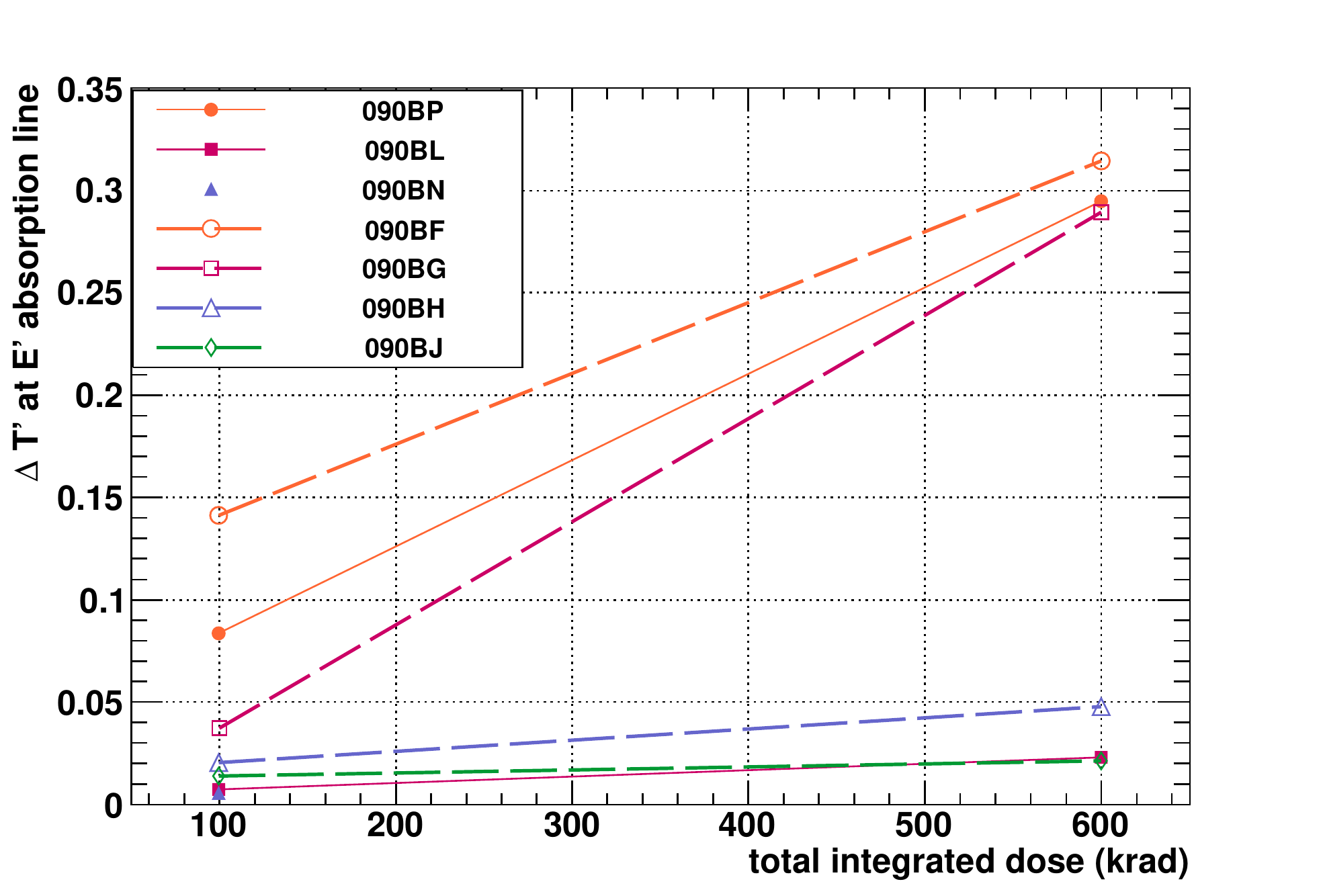}
		\caption{Peak normalized transmission loss $\Delta$T' at E' absorption line (210~nm) as a function of total integrated dose.}
	\label{fig:dT_at_E_vs_dose}
\end{figure}

The investigation of the radiation hardness of Heraeus Suprasil~2A and Suprasil~311 samples confirms that the existing defect models for UV laser applications also apply to ionizing radiation and shows clearly the influence interstitial hydrogen has on the level of damage. 
Furthermore, it was seen that the radiation hardness of off-the-shelf Suprasil~2A was superior to Suprasil~311 but with enhanced hydrogen levels both materials show similar properties. This emphasizes that not only the raw materials but also the production process of synthetic fused silica is relevant with respect to its radiation hardness. 

Comparing the optical transmission after 100~krad and 600~krad total integrated dose, two samples are noteworthy: 090BK (Suprasil~2A) and 090BG (Suprasil~311). 

Sample 090BK is the Suprasil~2A sample with the highest hydrogen concentration (see Tab.~\ref{tab:SuprasilSamplesAndhydrogenContent}). Optical characterization after 100~krad integrated dose showed exponentially decaying absorption properties up to 600~nm, however, no fused silica specific defect centers could be found. Moreover, after the 600~krad dose no such feature is found, and no radiation damage is found at all. This leads to the conclusion that the observed absorption after the 100~krad dose is due to improper cleaning of the sample prior to the optical characterization. 

Sample 090BG is the Suprasil~311 sample with the second lowest hydrogen concentration (see Tab.~\ref{tab:SuprasilSamplesAndhydrogenContent}). While radiation damage after the initial 100~krad dose was small, it increased dramatically after 600~krad total integrated dose.

The rise in transmission loss was investigated for the E' defect center at 210~nm. Two distinct trends in the rise, depending on the hydrogen concentration, are visible in Fig.~\ref{fig:dT_at_E_vs_dose}. Samples with lower hydrogen content show an increased optical absorption in the blue-UV regime with two distinct absorption bands visible for the most damaged samples. These bands can be associated with the E' and NBOH damage centers already known to exist for fused silica. 
A clear correlation between hydrogen concentration within a sample and the resulting optical transmission loss could be established.

\begin{figure}[!h]
	\centering	
	\includegraphics[width=0.9\linewidth]{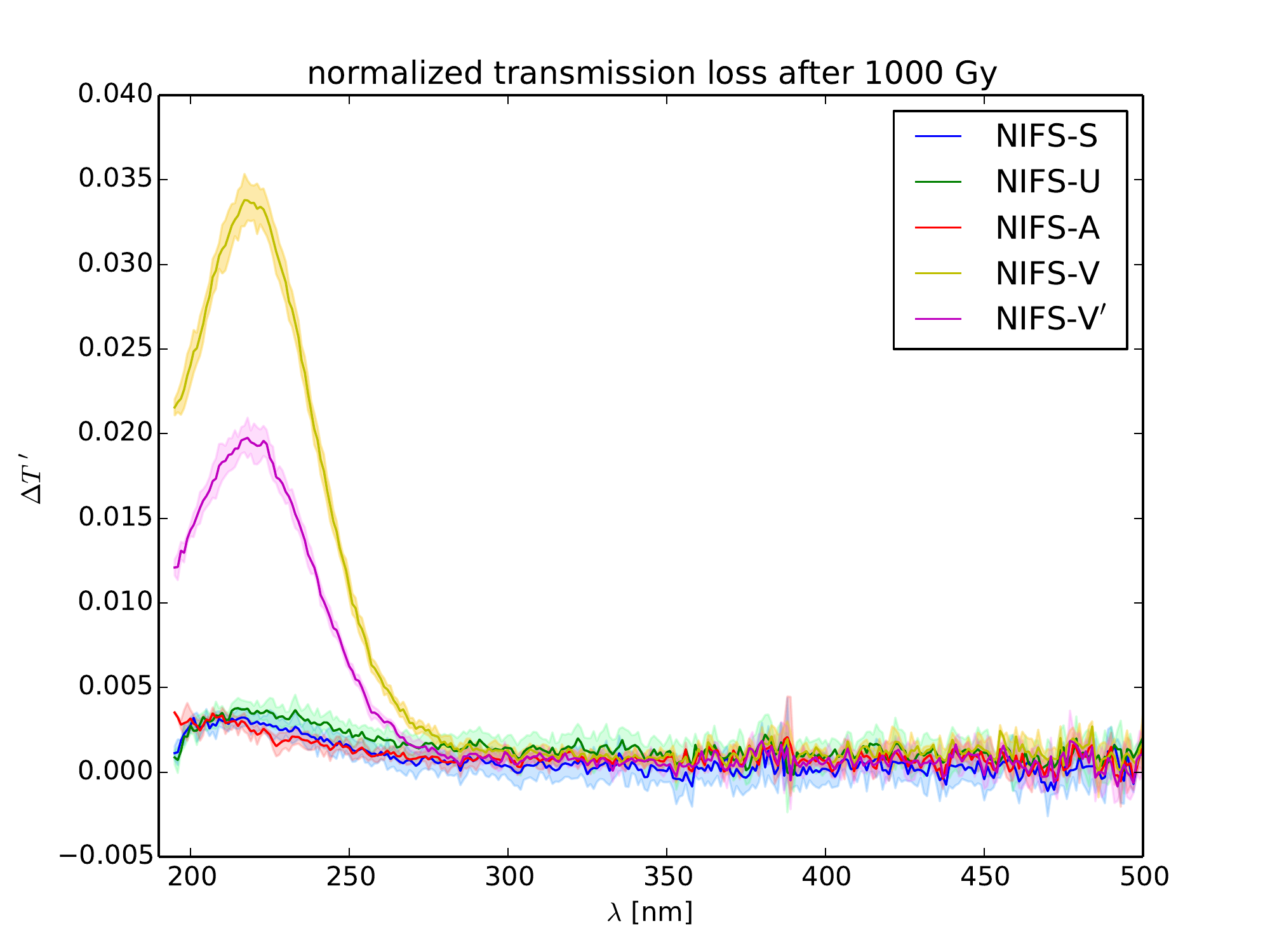}
	\caption{Normalized transmission loss for all NIFS-samples at wavelengths between 195~nm and 500~nm. One sees a broad absorption band around 220~nm for the NIFS-V series. The other samples show a very small difference in the same region.}
	\label{fig:Nikon_dT_100krad}
\end{figure}

The $\gamma$-ray irradiation study was extended to include another vendor, Nikon, which provided dry and wet synthetic fused silica samples (see Tab.~\ref{Tab:fused_silica-types} and \cite{panda:radiation_hardness_nikon}). All samples were irradiated with an integrated dose of 100~krad. The normalized transmission loss $\Delta T'$ was computed according to Eqn.~\ref{eq:trans_loss}. Figure~\ref{fig:Nikon_dT_100krad} shows the normalized transmission loss for the different NIFS grades. As expected transmission loss due to the induced radiation is visible starting at around 290~nm and peaking at 220~nm indicating damage of the silica network only. In agreement with the previous studies, the dry type (NIFS-V) shows a significant deviation.

\begin{figure}[!h]
	\centering	
	\includegraphics[width=0.9\linewidth]{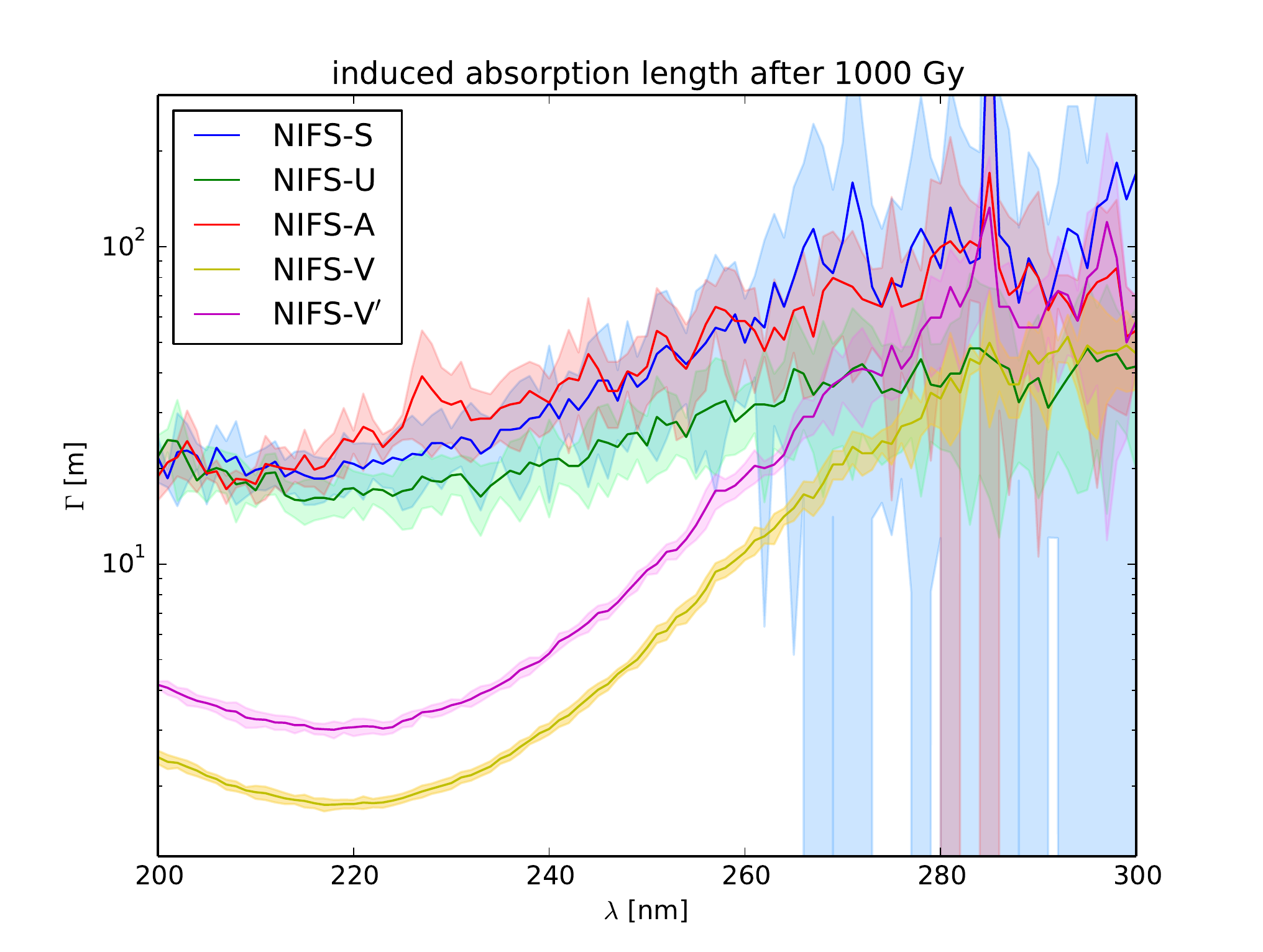}
	\caption{Radiation induced absorption length for all NIFS-samples. One clearly sees the inferior behavior of the NIFS-V samples.}
	\label{fig:Nikon_AbsLength_100krad}
\end{figure}

The corresponding absorption length $\Gamma$ is computed according to Eqn.~\ref{eq:abs_length}. Whereas for NIFS-S,-U, and -A the absorption lengths are above 10~m, the NIFS-V samples lay between 2 and 3~m (see Fig.~\ref{fig:Nikon_AbsLength_100krad}).

Based on experience from the BaBar DIRC and our R\&D, synthetic fused silica was chosen as radiator material for the \panda Barrel DIRC. 
Our own irradiation tests corroborated these findings and additionally established the 
crucial role interstitial hydrogen plays in preventing radiation damage. 
The damage models developed for UV lithography also apply to ionizing radiation, 
confirming the hydrogen consumption at higher doses.

The expected integrated dose for the Barrel DIRC over the \panda lifetime, see 
Fig.~\ref{fig:radmap_dirc}, is well below the doses applied in the irradiation tests mentioned above. Since at \panda the wavelengths of the photons are cut off below about 300~nm due to the glue joints between the radiator pieces and the lens material, the irradiation dose induced reduction of the radiation length has no impact on the \panda Barrel DIRC design.
Thus synthetic fused silica of sufficient grade and hydrogen content, available from several vendors, meets the detector requirements.

\subsubsection*{Optical Tests on the Radiators from Different Vendors}
\label{subsec:qualtest-radiator}
The  optical and mechanical quality of the DIRC radiators is of critical 
importance for the PID performance of the detector, since imperfections 
influence the photon yield and the single photon Cherenkov angle resolution (SPR).
Depending on the polar angle of the charged particle track, Cherenkov photons are 
internally reflected up to 400~times before exiting the bar.
The probability of photon loss during total internal reflection is determined by the surface roughness
and possible sub-surface damage, created in the fabrication process. A transport efficiency of 90\%
requires a radiator surface to be polished at the level of 10~\AA~or better. To maintain the magnitude of the Cherenkov angle
during the reflections, the bar surfaces have to be parallel and the squareness has to be better than 0.25~mrad. The combination of these tight optical and mechanical requirements makes the production of DIRC radiators challenging for the optical industry. 
Two primary fabrication methods, abrasive and pitch polishing, are available to 
fabricate DIRC radiator bars and plates. 
Pitch polishing was used for the radiator bar and plates for 
the BaBar DIRC and the Belle~II TOP.
This method achieved the required surface roughness and angular specifications
and produced sharp corners.
Several vendors propose different types of abrasive polishing and 
\panda will need to evaluate possible resulting sub-surface damage effects.
The use of wider bars or plates, resulting in a smaller total number of surfaces to be polished for the DIRC detector, is an attractive and cost-saving solution. During the \panda Barrel DIRC prototyping program
a total of about 30~bars and plates were produced by eight manufacturers
(Aperture Optical Sciences~\cite{aos},
InSync~\cite{insync},
Heraeus~\cite{Heraeus},
LZOS~\cite{lzos},
Nikon~\cite{nikon},
Schott Lithotec~\cite{schott-lithotec},
Zeiss~\cite{zeiss}, and
Zygo~\cite{zygo}), using different fabrication
techniques. The quality of these prototypes was tested in three separate experimental setups.
\begin{figure}[h!]
\begin{center}
\includegraphics[width=0.9\columnwidth]{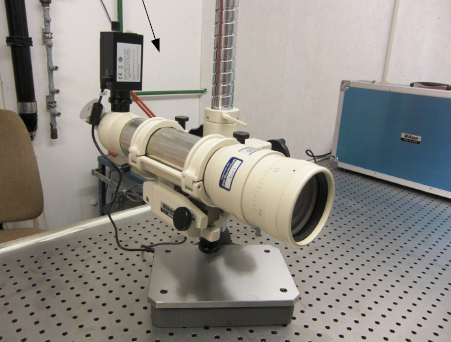}
\caption{The autocollimator (Nikon~6D) in the optical lab at GSI.}
\label{fig:proto-autocol}
\end{center}
\end{figure}

\subsubsection*{Parallelism and Squareness}
\label{subsec:qual-squareness}
\begin{figure}[h!]
\begin{center}
\includegraphics[width=0.9\columnwidth]{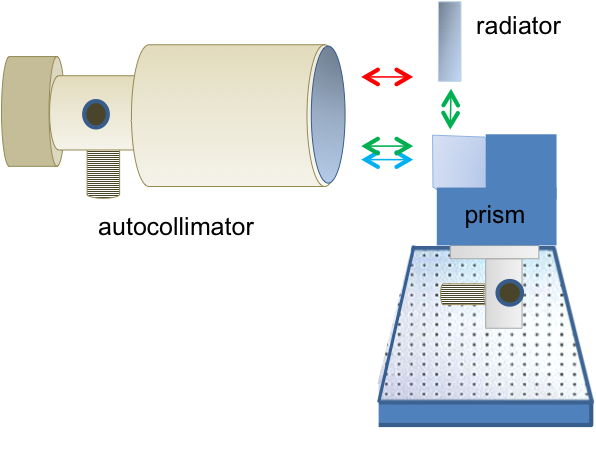}
\par
\vspace*{1cm}
\includegraphics[width=0.65\columnwidth]{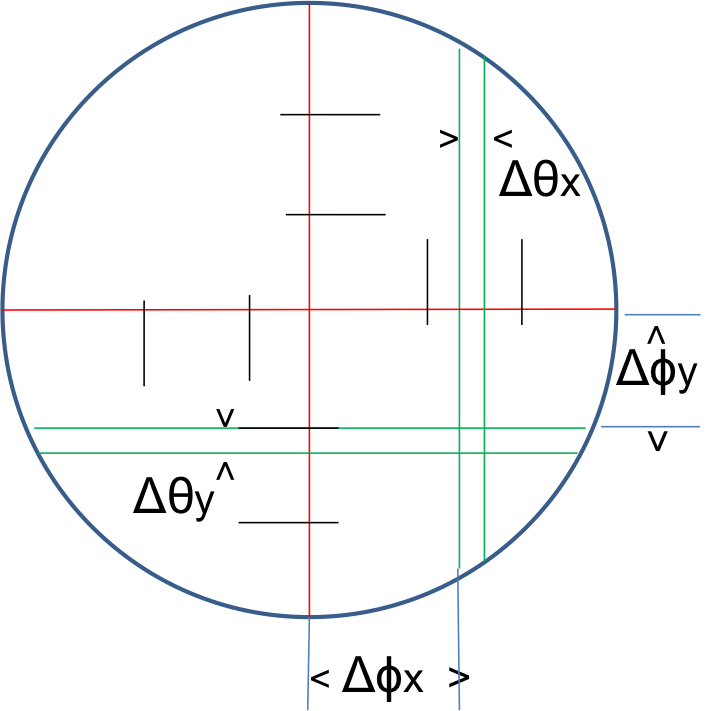}
\caption{Schematic illustration of the setup used to measure the squareness and the 
parallelism of the radiators.
\textit{Top:} The autocollimator (left) projects a reticle on the front 
surface of the radiator bar (red arrow) and the pentaprism (blue arrow). 
The reflections are aligned to each other and on the reticle in the autocollimator 
by moving the table underneath the pentaprism. 
The deviation of the reflection from the side of the radiator (green arrow) is 
then read off from the scale of the reticle.
\textit{Bottom:}
The deviation ($\Delta \varphi$) in x- and y-direction of the reflected reticle 
(green lines) from the reticle in the collimator (red line) is due to a non-squareness 
of the radiator. 
The splitting of the green lines shows a non-parallelism of the radiators. 
}
\label{fig:proto-autocol-schematic}
\end{center}
\end{figure}

Two setups exist at GSI to determine the parallelism and squareness of the bar surfaces.
One setup uses a laser, which is reflected from the bar sides.
The location of the reflected laser image is compared at a distance of more than 10~m 
for different orientations of the bar. 
The angular precision achieved was better than 0.1~mrad, sufficient to test the 
squareness of the radiator bars.
The second setup uses a Nikon 6D autocollimator (see Fig.~\ref{fig:proto-autocol}) 
and has, with 0.5~arcsecs (0.002~mrad), a much better accuracy. 
The distances ($\Delta \varphi_x$ and $\Delta \varphi_y$)
between the reticle in the ocular of the autocollimator, which is aligned 
to the front surface of the radiator, and the reticle image after reflection 
from the surface of the radiator via the pentaprism 
are determined without contact with the bar surfaces
(see Fig.~\ref{fig:proto-autocol-schematic}).
The deviation from the perfect parallelism $\Delta \beta$ can be determined from 
\begin{equation}
\Delta \beta = \frac{\Delta\theta}{n},
\end{equation}
where $\Delta\theta$ is the difference in the reflection from both surfaces 
and $n$ is the refractive index of the radiator material. 
The squareness of the faces of the radiator can be directly read off since the divisions 
in the reticle of the autocollimator are given in units of arcmin.

\begin{table}[h!]
\setlength{\tabcolsep}{6pt} 
\renewcommand{\arraystretch}{1.5} 
\caption{Measurement of the deviation from squareness of the radiator bar produced by Zeiss. 
The narrow sides (S) are 17~mm wide, the faces (F) 33~mm. E1 and E2 are the two ends of the bar.}
  \ \
\label{Tab:radiator-squareness}
\begin{center}
\begin{tabular}[]{@{\extracolsep{\fill}}ccc}
\hline
Angle & \multicolumn{2}{c}{$\Delta \varphi$} \\ 
between & [arcsec]  & [mrad]  \\
\hline 
S1/F1 & --15.0 $\pm$1 & 0.073 $\pm$ 0.005 \\
F1/S2 & 7.5 $\pm$1 &  0.036 $\pm$ 0.005 \\
S2/F2 & --1.5 $\pm$1 & 0.007 $\pm$ 0.005 \\
F2/S1 & 15.5 $\pm$1& 0.075 $\pm$ 0.005 \\\hline
E1/F1 & --25.5 $\pm$1& --0.124 $\pm$ 0.005 \\
E1/S2 & 5.3 $\pm$1& 0.026 $\pm$ 0.005 \\
E1/F2 & --14.0 $\pm$1& --0.068 $\pm$ 0.005 \\
E1/S1 & --27.0 $\pm$1& 0.131 $\pm$ 0.005 \\\hline
E2/F2 & --55 $\pm$1& --0.267 $\pm$ 0.005 \\
E2/S2 & --49.5 $\pm$1& --0.240 $\pm$ 0.005 \\
E2/F1 & 10.5 $\pm$1& 0.051 $\pm$ 0.005 \\
E2/S1 & --19.75 $\pm$1& --0.096 $\pm$ 0.005 \\

\hline
\end{tabular}
\end{center} 
\end{table}

\begin{table}[h!]
\setlength{\tabcolsep}{6pt} 
\renewcommand{\arraystretch}{1.5} 
\caption{Measurement of the paralelism ($\Delta \beta$) of the radiator bar radiator bar produced by Zeiss. 
The narrow sides (S) are 17~mm wide, the faces (F) 33~mm. E1 and E2 are the two ends of the bar.}
  \ \
\label{Tab:radiator-parallelism}
\begin{center}
\begin{tabular}[]{@{\extracolsep{\fill}}ccc}
\hline
Sides & \multicolumn{2}{c}{$\Delta \beta$} \\ 
 & [arcsec] & [mrad]  \\
\hline 
S1/S2 & 4.8 $\pm$1 & 0.023 $\pm$ 0.005 \\
F1/F2 & 6.1 $\pm$1 &  0.030 $\pm$ 0.005 \\
E1/E2 & 39 $\pm$1 &  0.128 $\pm$ 0.005 \\
\hline
\end{tabular}
\end{center} 
\end{table}

Results of measurements of the squareness and parallelism of a prototype 
radiator bar, produced by Zeiss, are shown in 
Tab.~\ref{Tab:radiator-squareness} and Tab.~\ref{Tab:radiator-parallelism}.
Only small deviations from the ideal bar shape are observed and the 
obtained values confirm that the bar fulfills the requirements.

\subsubsection*{Surface Roughness and Bulk Absorption}
\label{subsec:qual-surface}

\begin{figure}[htb]
\begin{center}
\includegraphics[width=0.98\columnwidth]{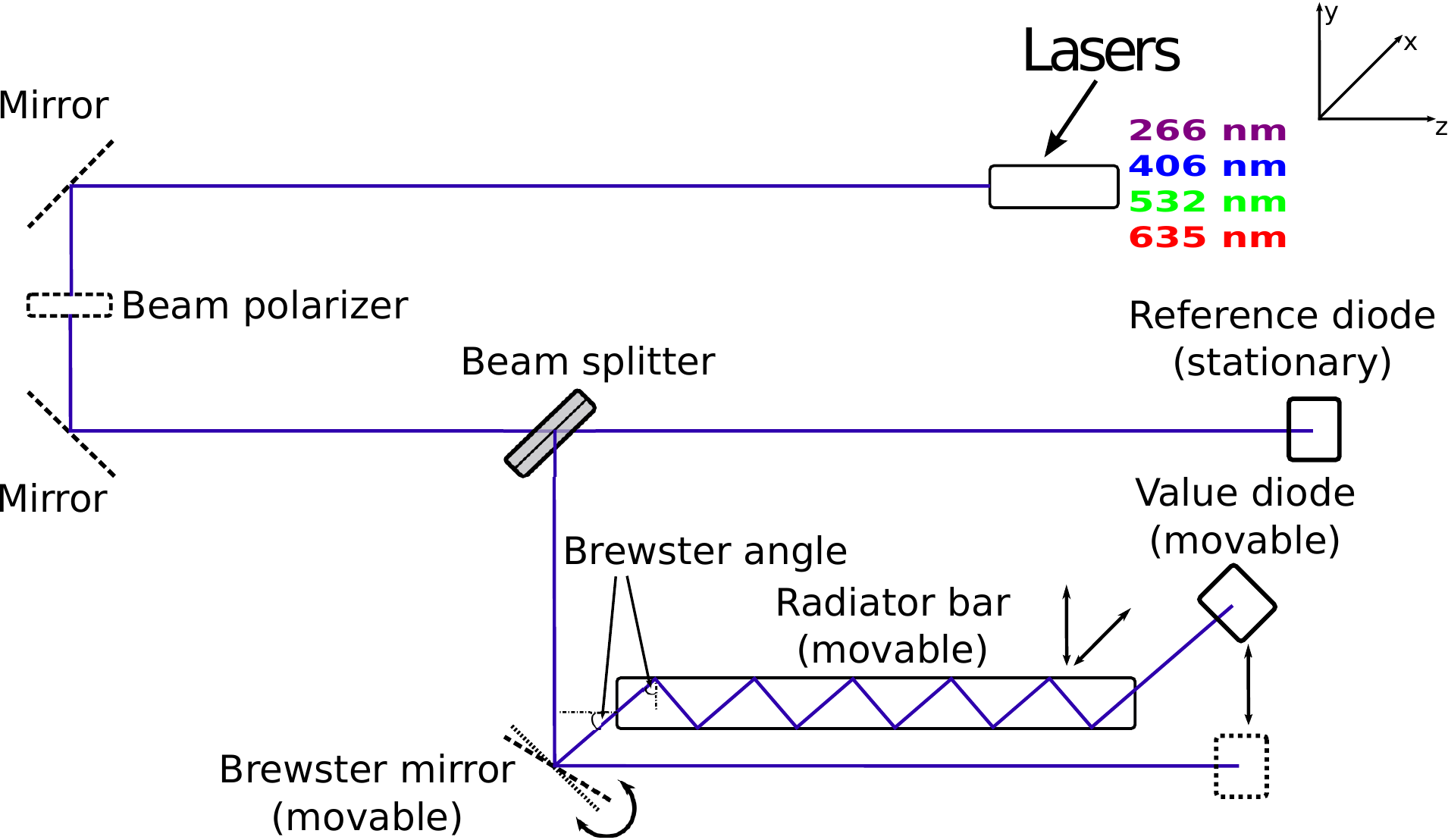}
\par
\vspace*{1cm}
\includegraphics[width=0.85\columnwidth]{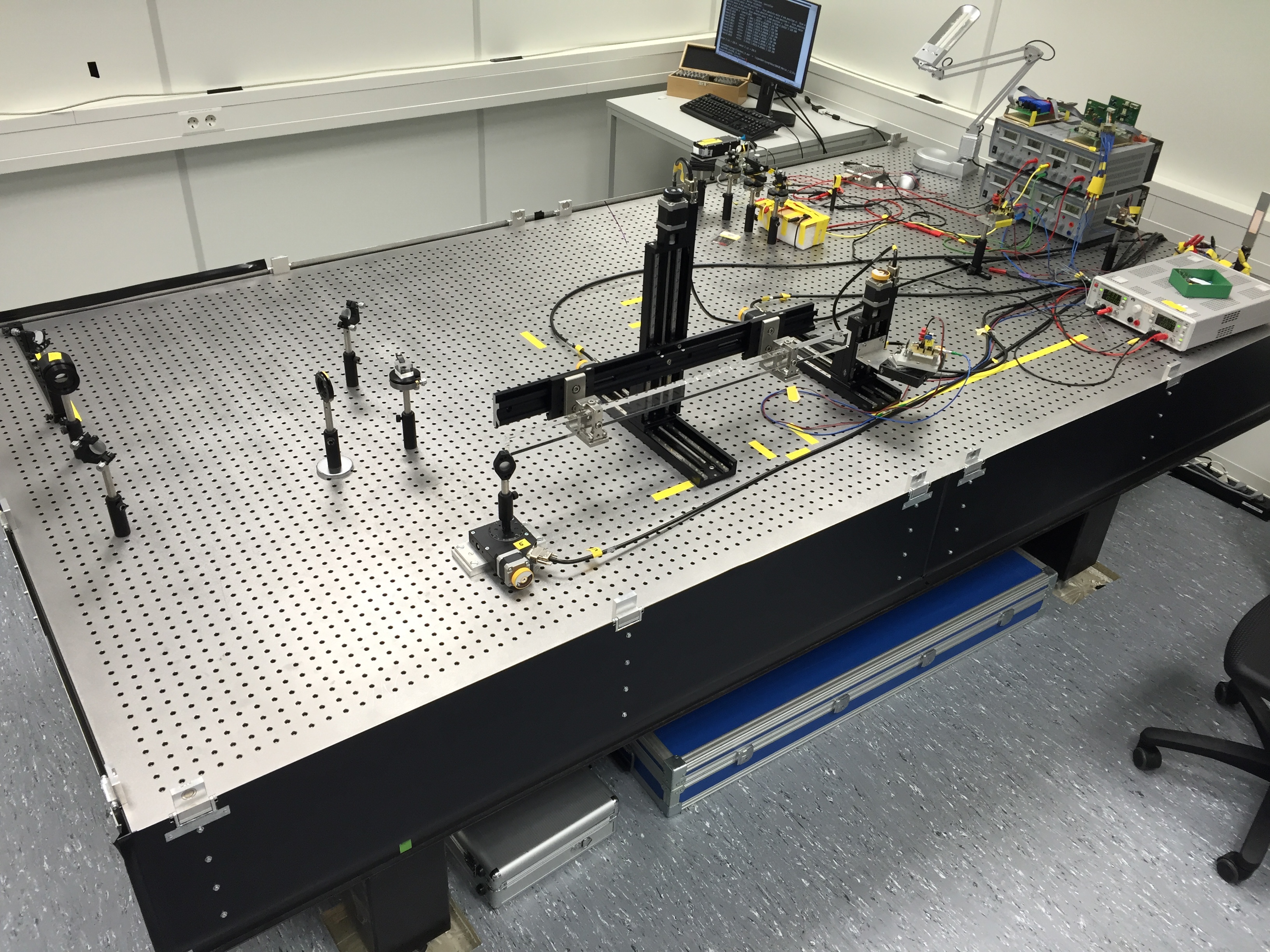}
\caption{
Schematic (top) and photo (bottom) of the setup used for the optical quality 
assurance of the prototype radiator bars for the Barrel DIRC. 
Two photodiodes are used: 
The ``reference'' diode, which is stationary and monitors the laser intensity, 
and the ``value'' diode, which measures the beam after it exits the bar.
}
\label{fig:current_setup2016}
\end{center}
\end{figure}

The optical quality of the \panda Barrel DIRC prototype radiator bars 
is evaluated using the setup shown in Fig.~\ref{fig:current_setup2016}.
The system is based on the method developed for the BaBar DIRC~\cite{babar:dirc}
and uses motion-controlled step motors and polarized laser beams with 
four different wavelengths to determine the coefficient of total 
internal reflection and the bulk attenuation of the radiators in a 
dark, temperature-stabilized room.

The transmitted intensity~$T$ is scanned for each bar using an array of 
laser entry points on the front bar surface, typically with a grid spacing of 
1--2~mm to accumulate several hundred measurements.
Possible laser intensity fluctuations are calibrated out using a (reference)
diode.
The mean value and RMS of $T$ are then extracted using a Gaussian fit to the 
data points. 

For the bulk transmission measurement the laser beam traverses the radiator 
bar parallel to the long bar axis. 
Since a part of the laser beam gets reflected from the bar ends, 
the intensity values have to be corrected for Fresnel losses. 

The reflection coefficient is measured by coupling the laser beam into 
the bar at Brewster angle to minimize the reflections on the end surfaces.
The laser then gets internally reflected from the side or face surfaces 
of the radiator, up to 50 times for a 1200~mm long bar, before it hits 
the (value) photodiode.

The coefficient of total internal reflection $R$ can then be calculated 
for each laser wavelength as:
\begin{equation}
T = R^{N} \cdot exp{\left(\frac{l}{\Lambda}\right)}\cdot(1-F)^{2} ,
\label{eq:attlength}
\end{equation}

where $\Lambda$ is the attenuation length, $N$ the number of internal reflections 
in the radiator bar, $l$ the length of the radiator, and $F$ the Fresnel correction.

\begin{table*}[tbh]
\setlength{\tabcolsep}{6pt} 
\renewcommand{\arraystretch}{1.5} 

  \caption{Bulk transmission, reflection coefficient and surface roughness for a test measurement with the InSync bar.}

  \ \
  
\label{Tab:radiator-absorption-reflectivity}
\begin{tabular*}{1.0\textwidth}[]{@{\extracolsep{\fill}}ccccc}
\hline  
Wavelength &  Bulk transmission & $\#$ faces & Reflection coefficient & Surface roughness \\ 

[nm] & [1/m] &  &  & [\AA] \\ 
\hline  
406 & $0.994 \pm 3.2 \cdot 10^{-4}$ & 49 & $0.99984 \pm 1.6 \cdot 10^{-5}$ & $4.9 \pm 1.3$ \\
532 & $0.997 \pm 2.7 \cdot 10^{-4}$ & 49 & $0.99991 \pm 1.4 \cdot 10^{-5}$ & $4.7 \pm 1.3$ \\
635 & $0.9994 \pm 8.0 \cdot 10^{-5}$ & 49 & $0.99996 \pm 1.5 \cdot 10^{-5}$ & $3.7 \pm 3.0$ \\
\hline
\end{tabular*} 
\end{table*}

\begin{figure}[h]
\begin{center}
\includegraphics[width=0.99\columnwidth]{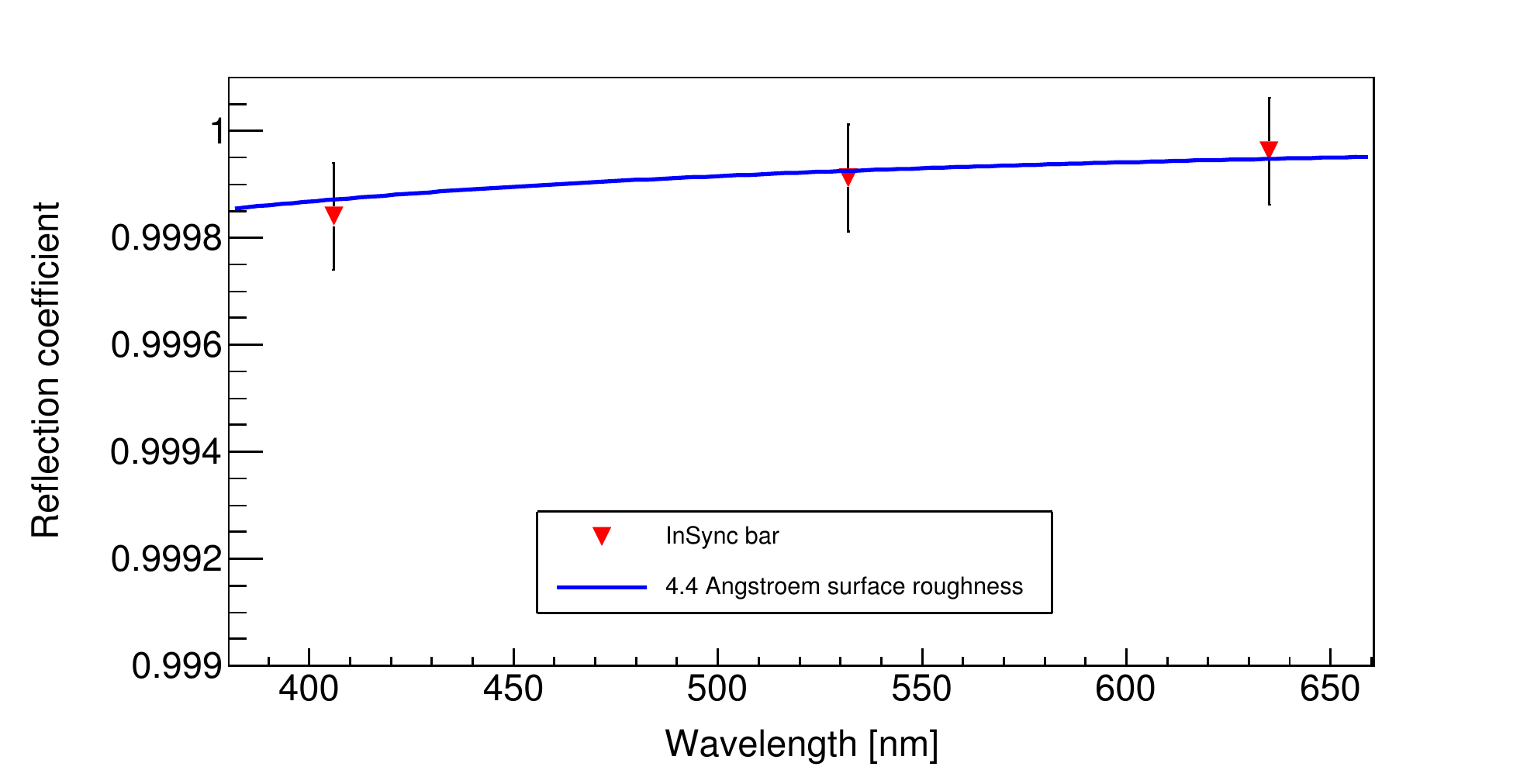}
\caption{Coefficient of total internal reflection as a function of the
laser wavelength for a prototype radiator bar produced by InSync.
The blue line is a fit to the data points based on scalar scattering theory. 
The error bars are determined from Gaussian fits to the results obtained for
all bar entry points during one scan.
}
\label{fig:bar_meas}
\end{center}
\end{figure}

The coefficient of total internal reflection $R$ can be related to the 
surface roughness via the scalar scattering theory~\cite{opt-surface}:

\begin{equation}
R = 1 - \left(\frac{n \cdot cos(\Theta_{Brewster}) \cdot H \cdot 4\pi}{\lambda}\right)^{2} ,
\label{eq:refelctivity}
\end{equation}
where $H$ represents the surface roughness and $\lambda$ the wavelength of the laser.

Figure~\ref{fig:bar_meas} shows the results from a measurement of a prototype 
bar fabricated by the company \mbox{InSync}.
The bar was produced to the specifications defined for the BaBar DIRC counter.
It has a length of 1200.04~mm, a width of 34.93~mm, and a thickness of 17.12~mm.
The calculated reflection coefficients, the corresponding surface roughness 
and bulk absorption are shown in Tab.~\ref{Tab:radiator-absorption-reflectivity}.
The surface roughness values are in good agreement with the interferometric 
measurement performed by InSync, which reported a surface roughness of 4--5~\AA~RMS.

The setup was recently upgraded to accommodate longer bars, up to 2.5~m in length, 
and wide plates. 
Furthermore, a UV laser ($\lambda = 266$~nm) was added to provide an additional data 
point for the bulk transmission and reflection coefficient measurements.
This increases the sensitivity to subsurface damage, improves measurement accuracy, 
and allows a detailed comparison of the techniques used to produce the radiator bars
and plates.

\subsection{Focusing with Lenses}
\label{subsec:lenses}

The original design of the \panda Barrel DIRC was guided by the successful 
BaBar DIRC detector~\cite{Adam05}. This rather conservative approach used 
almost the same cross section of the radiator bars but a much smaller expansion volume. 
Due to the large dimensions of the expansion volume of the BaBar DIRC detector 
(depth of the EV was 1100~mm), pinhole focusing could be used. 
The expansion volume cannot be made much smaller with the same radiator bar 
cross section and the same focusing method without the Cherenkov image 
becoming blurred. 
Lenses or mirrors as focusing elements are needed to provide the desired 
Cherenkov angle resolution.
Space limitations within the \panda detector favor lenses.
The optimum type of lens depends on the radiator type (narrow bar or 
wide plate) and the shape of the expansion volume.

\begin{figure}[bt]
\centering 
\includegraphics[width=.45\textwidth]{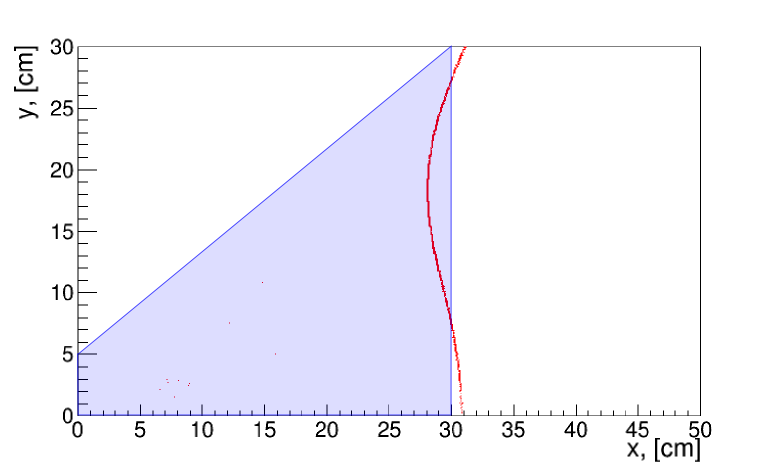}
\caption{\label{fig:lens_geant4} The 
focal plane of the spherical three-component lens as red curve simulated by the Geant4 software 
package. 
The blue shaded area depicts the expansion volume. } 
\end{figure}

The development of a lens system with a focal plane that matches the photon 
detector surface shape and maintains a consistently high photon yield for 
the entire \panda Barrel DIRC phase space was a significant challenge.
Conventional optics employ glass/air interfaces for refraction. 
However, the transition from a focusing convex fused silica surface to 
air traps many photons with steep incident angles by internal reflection 
in the fused silica (see Fig.~\ref{opt_lens}). 
\begin{figure}[tb]
\centering 
\includegraphics[width=.19\textwidth]{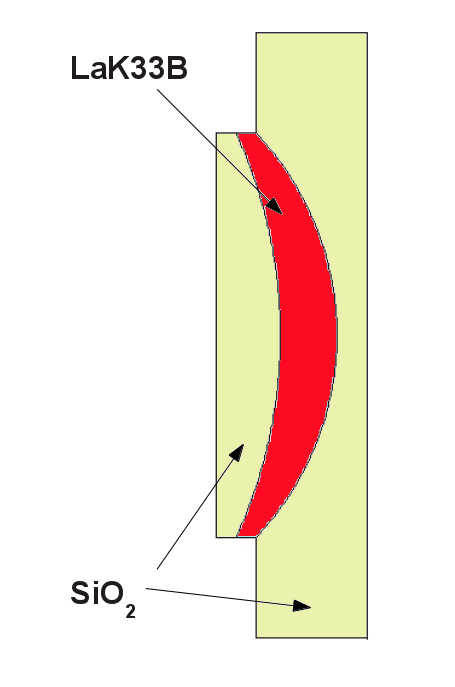}
\includegraphics[width=.26\textwidth]{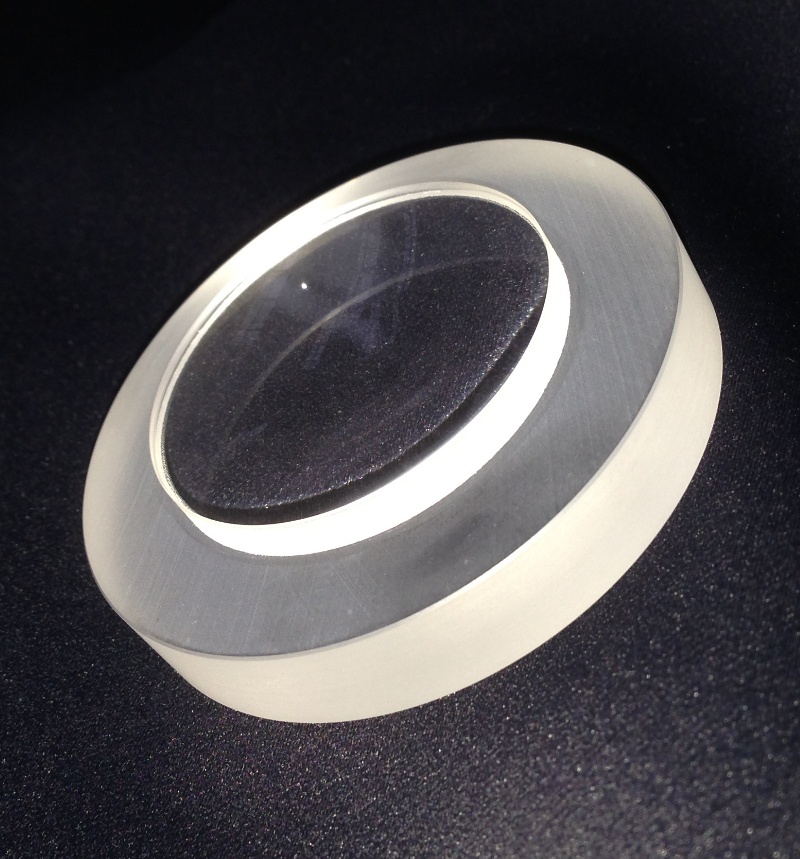}
\caption{\label{fig:L3Slens} The spherical three-component lens (left),
the photo of the prototype (right). The radiator will be attached to the left side.}
\end{figure}
Therefore, a lens that includes a material with a refractive index larger than 
fused silica was chosen. 
An early version of this lens used a single focusing surface and is described in 
Ref.~\cite{RICH13}. 
However, the Petzval condition~\cite{Petzval} has to be met to achieve a flat 
focal plane, which requires more than one refracting surface, as shown in 
Fig.~\ref{fig:L3Slens}.
This lens system is a single lens that consists of three parts: 
The two fused silica parts are for coupling the lens to the radiator bar 
and the expansion volume. 
The middle part, made from lanthanum crown glass (LaK33B), has two surfaces
with different curvatures. The left one in the schematic drawing in 
Fig.~\ref{fig:L3Slens} is a defocusing surface, the right one is a focusing surface. 
This LaK33B material was chosen due to a high refractive index 
of  \mbox{n$ = 1.786$} and a good transmission of  \mbox{T$ = 0.954$} for 
a \mbox{$10$~mm}-thick sample at a wavelength of \mbox{$\lambda = 380$~nm}. The 
lens was designed with the Zemax optical software~\cite{zemax} and cross-checked with the Geant 
simulation package. The results are shown in Fig.~\ref{fig:lens_geant4}.
 
A prototype of such a high-refractive index compound lens (see Fig.~\ref{fig:L3Slens}, right) has 
been built by a glass company~\cite{befort} and was tested in a Barrel DIRC prototype at CERN in 2015 
(see Sec.~\ref{sec:cern2015}). 

Radiation hardness of the lanthanum crown glass was initially a concern.
Therefore, a measurement of the radiation hardness of the 3-layer prototype 
lens and a sample of NLaK33B is underway~\cite{Kalicy15}.
Initial results, using an X-ray source~\cite{Kalicy17}, demonstrate 
that the NLaK33B material significantly exceeds the radiation hardness requirement 
for the \panda Barrel DIRC.

\subsection{Expansion Volume}
\label{subsec:ev}
The shape of the expansion volume was the subject of intensive simulation
studies~\cite{MP-MPatsyuk-PHD-THESIS-C}. 
Its length and opening angle determine the size of the photon readout area and, 
thus, the number of required photon sensors. 
The outcome of the optimization is an EV geometry comprising 16 compact prisms 
made of synthetic fused silica.
Each is coupled to a bar box and has a depth of 30~cm, a width of 16~cm, and 
an opening angle of 33$^\circ$.
The small size minimizes the cost of the sensors and readout electronics
while maintaining the required Cherenkov angle resolution.
Although the additional reflections on the prism sides complicate the hit 
pattern and the reconstruction of the Cherenkov angle, simulations (see Sec.~\ref{cha:simulation}) 
and test beam data (see Sec.~\ref{ch:performance}) have shown that these reflection do not cause 
problems for the PID performance.

\begin{figure}[h!]
\begin{center}
\includegraphics[width=0.85\columnwidth]{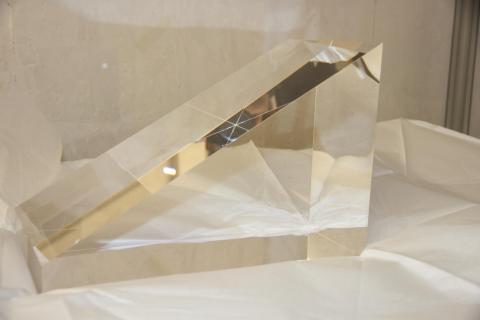}
\caption{One of two solid fused silica expansion volume prototypes produced for the
Barrel DIRC system tests with particle beams.
This prism has an opening angle of 30$^\circ$, a width of 170mm, and a depth of 300mm.}
\label{fig:EV}
\end{center}
\end{figure}

\section{Photon Sensors} \label{sec:sensors}

\subsection{Requirements} \label{subsection:sensors:requirements}

The reconstruction of the Cherenkov image requires two spatial coordinates or one spatial coordinate plus a time measurement. Additional measured variables can be used to over-determine the system and thus improve the detector performance and suppress background. Designs using two spatial coordinates plus a timing measurement in their reconstruction are known as 3D DIRC systems. The Barrel DIRC for \panda in its baseline design, using geometrical reconstruction, primarily relies on the two spatial coordinates with a precision of a few~mm. 
A reasonably precise time information is used to aid the reconstruction of the Cherenkov pattern and for background suppression. 
  
The magnetic field of the \panda target spectrometer (TS) solenoid puts severe design constraints 
on the photon readout. The sensor has to work in a magnetic field of $\approx$~1~T.
Since the image planes will be rather compact and because the average antiproton-proton  
interaction rate for  \panda at full luminosity will be $2\cdot10^7$~s$^{-1}$ 
(see Sec.~\ref{sec:pandaexp2}), 
the expected high single photon density at the sensor surface 
(see Eqn.~\ref{eqn:pixelrate}) requires a very high rate stability and a 
long lifetime of the counters in terms of integrated anode charge. 
By measuring the time-of-propagation of the Cherenkov photons from their creation 
point to the sensor surface,  ambiguities in the photon path can be resolved, 
improving the $\pi$/K separation. 
For optimum performance a time resolution of 100~ps ($\sigma$) or better 
is desirable for the geometric reconstruction and becomes a requirement
when the time-based imaging method is used for 
the narrow bars or wide plates. 
Finally, each particle traversing the radiators produces only a few detected Cherenkov 
photons. 
As a consequence, the photon sensors have to be of very high quality in terms of 
gain, quantum-, collection- and geometrical efficiency and feature a low dark count rate.

The requirements on the timing, rate capability, magnetic field tolerance, and active area ratio
are met by multi-anode Microchannel Plate PMTs (MCP-PMTs). 
The current design of the Barrel DIRC is based on MCP-PMTs with a size of of about 
60~mm$\times$60mm and a $8\times 8$ anode layout, using two microchannel plates of 10~$\mu$m 
pore diameter in a chevron configuration. 

\subsubsection*{Magnetic Field}
\begin{figure}[tb]
\centering 
\includegraphics[width=.48\textwidth]{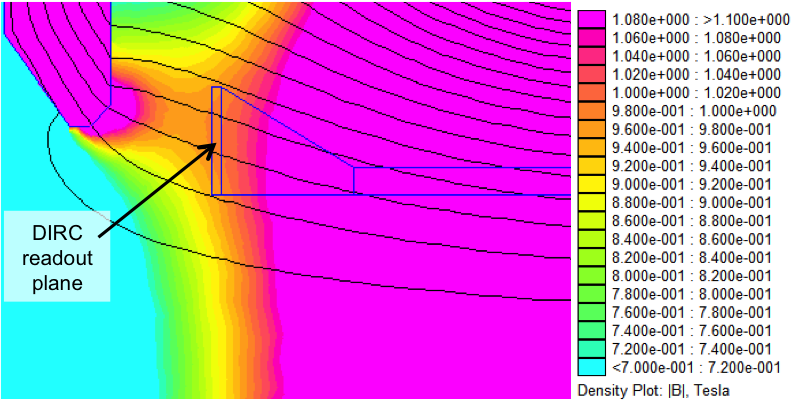}
\caption{\label{fig:BField} Magnetic field map in the readout area of the \panda Barrel DIRC.}
\end{figure}

The compact design of the \panda target spectrometer requires the photon detection 
system and initial digitization stages to be located inside the return yoke of the 
solenoid. 
As shown in Fig.~\ref{fig:BField}, the photon detection system of the Barrel DIRC 
is exposed to a magnetic field of about 1~T. 
The available construction space allows a moderate optimization of the sensor plane 
orientation relative to the direction of the magnetic field and the field lines are
expected to be perpendicular to the sensor front surface to within 15$^\circ$ or less.
The compact design and required large geometrical fill factor do not allow the 
installation of magnetic shielding. 
A suitable photon detection system should therefore work inside a magnetic field 
of up to 1.5 T (allowing for a safety margin of 50\% in the prediction of the 
magnetic flux).

\subsubsection*{Radiation Hardness}
The \panda experiment with hadronic interactions at high luminosity will produce
a large integrated radiation dose that the detectors have to withstand. 
However, as \panda is a fixed target experiment, most of the particles will be produced 
in the forward direction due to the Lorentz boost. 
The photon sensors, placed upstream with respect to the interaction point, are less affected.
An integrated radiation dose corresponding to $10^{11}$ neutrons/cm$^2$ 
is expected in this area over the lifetime of \panda (see Sec~\ref{sec:RadHardnessMatthias} 
and Fig.~\ref{fig:radmap_dirc} within), 
consisting mainly of neutrons and some electromagnetic background. 
This dose is not an issue for any commercially available MCP-PMT.

\subsubsection*{Area and Pixel Size}

The photons are focused with a lens onto the sensor plane at the back of the 
30~cm-deep EV, which will be equipped with 11 two-inch PMTs. 
The pixel size needed is in the order of 6$\times$6~mm$^{2}$, which matches 
MCP-PMTs like the PHOTONIS Planacon XP85112 or the Hamamatsu Prototype R13266,
as well as MAPMTs like the Hamamatsu H12700, which all feature 8$\times$8 anode arrays.
without a significant redesign of the focusing optics.

In principle, asymmetric pixels could be used to reduce the number 
of electronic channels. 
By combining neighboring pixels in a 2-inch MCP-PMT one could create a 
4$\times$8 pixel array with an effective pixel size of about 
6$\times$12~mm$^2$ and thereby reduce the number of readout channels 
by 50\,\%.
Simulation studies have shown~\cite{MP-MPatsyuk-PHD-THESIS-C} that this
configuration would not deteriorate the Cherenkov angle resolution per
photon significantly.
However, doubling the Cherenkov hit probability per pixel would also double 
the photon loss due to the dead time of readout electronics.
Therefore, the baseline design does not include asymmetric pixels.

\subsubsection*{Time Resolution}
The design of the Barrel DIRC requires a time resolution of 100~ps or better
for optimum performance of the reconstruction. 

For the geometrical reconstruction a good time resolution is needed to suppress 
combinatorial background from reflections inside the prism. 
Furthermore, photon timing better than 200~ps will allow, at least in principle,
a mitigation of the effect of chromatic dispersion~\cite{Benitez08-C}, which
would further improve the Cherenkov angle resolution of the Barrel DIRC.

The PID performance of the time-based imaging algorithm for the wide plate  
deteriorates if the time resolution per photon get worse than 100~ps.

\subsubsection*{Spectral Range}
Cherenkov photons are produced on spectrum as a function of $1/\lambda^2$, 
where $\lambda$ is the wavelength of the emitted Cherenkov photons.
The radiator material is transparent for visible and ultraviolet light 
and does not limit the spectral sensitivity of candidate photon 
detection systems. 
The spectral range, however, will be restricted by the wavelength-dependent
attenuation length of the optical materials used, in particular by the Epotek
glue (see Fig.~\ref{fig:atenuationlength}), to wavelengths larger than 290~nm.

\subsubsection*{Rate}\label{subsubsection:rate}
The design value of the average interaction rate at \panda is 20~MHz at 
full luminosity. 
The average multiplicity of tracks in the barrel region is 2 and
most of those particles will be above Cherenkov threshold.
On average about 50 Cherenkov photons will be detected per particle.
Assuming that the photons are isotropically distributed over all 
available pixels in a prism one can derive the photon hit 
rate for the individual readout pixels $R_{pixel}$ as follows:
\begin{equation}
\label{eqn:pixelrate}
\begin{split}
&R_{pixel}=\\
&\frac{2 \times 10^{7}  \frac{\text{events}}{\text{s}} \times 2  \frac{\text{tracks}}{\text{event}} \times 50  \frac{\text{photons}}{\text{track}}}{ 16  \text{ modules} \times 11\frac{\text{MCP-PMTs}}{\text{module}} \times 64\frac{\text{pixels}}{\text{MCP-PMT}}}\\
&~~~~~~~~\approx 180~\text{kHz}
\end{split}
\end{equation}

By using a safety margin of 10\% due to possible cross-talk between pixels 
and backsplash particles from the electromagnetic calorimeter, 
the average count rate is estimated to be $R_{pixel}\approx$~200~kHz.

\subsubsection*{Lifetime}

The expected accumulated anode charge can be calculated from the average 
pixel count rate by integrating over the 10 year \panda lifetime, taking 
the average luminosity over one machine cycle of the HESR 
(see Sec.~\ref{chap:panda}) into account.
Assuming a gain of 10$^{6}$ for the MCP-PMTs and 50\% duty cycle for \panda, 
simulations predict that the integrated anode charge over 10 years will 
accumulate to about 5~C/cm$^{2}$ for the Barrel DIRC. 


\subsection{Photon Devices}

For the detection of the Cherenkov photons MAPMTs, MCP-PMTs and SiPMs were evaluated. 
Their advantages and disadvantages are discussed below.

\subsubsection*{Multi-Anode Photomultiplier Tubes}

Multi anode dynode Photo Multiplier Tubes (MAPMTs) use a segmented anode and dynode structure to provide a correspondence between the position of the photon when entering the cathode and the readout pixel. E.g., the Hamamatsu H13700 series provides 256 3x3~mm$^2$ pixels combined with a conventional photocathode. The active area is 49x49~mm$^2$ and the MAPMT provides an excellent active area fraction of 89\%. The noise characteristics are very good as well. The PMT can be operated without pre-amplification. The quantum efficiency of the H9500 is given as 24\% at 420~nm. The spectral response is 300 to 650 nm. The main drawback lies in the sensitivity of these devices to magnetic fields. The gain drops rapidly even in small to moderate magnetic fields. The mechanical design and compactness of the \panda detector prevents the installation of effective magnetic shielding. Additionally, the pixel-to-pixel uniformity of MAPMTs shows large deviations. The typical uniformity is quoted as 1:3. Last but not least, the transit time spread of these and other MAPMTs is with $\approx$~0.3~ns too large for precise timing measurements.

\subsubsection*{Silicon Photomultipliers}

Several new developments in photon detection for future detectors concentrate on Silicon Photomultipliers (SiPMs). Conventional silicon photomultipliers consist of an array of avalanche photodiodes which are operated in Geiger-mode. Each of these photodiodes is able to detect single photons. When a photon crosses the depletion layer within one of the photodiodes, it can trigger an electrical avalanche discharge. If more than one diode in a silicon photomultiplier is triggered at the same time by several photons, the charges sum up and produce an electrical pulse with a charge proportional to the amount of detected photons.

This novel development of semiconductor photon sensors capable of detecting extremely low light levels provides a highly efficient, compact, easily customizable and magnetic field resistant alternative to the more conventional photon detection solutions like PMTs. Meanwhile there are many manufacturers offering a wide range of different SiPM models. An additional attractive feature of these devices is the possibility to integrate part of the read-out electronics into the design.

However, operating a photon detection system for an imaging Cherenkov counter requires the detection of single photons. This poses an inherent difficulty for semiconductor devices as thermal noise is indistinguishable from a signal generated by a true single photon hit. Although in recent SiPM models the original noise rates of MHz/mm$^{2}$ at room temperature came down to the level of about 100~kHz/mm$^{2}$ by the usage of better substrates to reduce afterpulsing effects and by applying grooves between the pixels to prevent optical crosstalk, SiPMs are still not a serious sensor alternative for the Barrel DIRC.
The thermal noise could be reduced to a tolerable level for single photon detection only by cooling the SiPM to a temperature of far below -20$^{\circ}$~C. This is not a viable option at \panda.
In addition, e.g. in the Hamamatsu MPPC (S10362-11 series) with a very high photon detection efficiency (PDE), an extremely high temperature sensitivity of the gain was found. This would have to be considered for detector applications, i.e. temperature stabilization is necessary.

Another major issue is the radiation dose exposure of the SiPMs in the \panda experiment. The radiation damage in the silicon substrate increases the bulk leakage current and hence the dark current, leading to more noise in the SiPMs making single photon detection even less practicable.

\subsubsection*{Microchannel Plate Photo Multiplier Tubes}

Microchannel Plate Photo Multiplier Tubes (MCP-PMTs) are the ideal sensors for applications where a low noise and sub-100~ps single photon detection is required inside a high magnetic field. They are available as multi-anode devices and provide a good active area ratio while still being rather compact in size. However, until recently the major drawback of MCP-PMTs has been serious aging issues. Ions in the residual gas produced by the electron avalanche are accelerated towards the photo cathode (PC) which gets damaged from this permanent bombardment. As a consequence the quantum efficiency (QE) drops while the integrated anode charge increases. Until recently (anno 2011) the rate conditions in \panda were far beyond the reach of any commercially available MCP-PMT where the QE had dropped by more than a half after typically $<$200 mC/cm$^{2}$, while for the Barrel DIRC up to 5 C/cm$^{2}$ are expected over the lifetime of \panda.

Our comparative measurements of the lifetime of MCP-PMTs (see Sec.~\ref{lifetime}) show clearly the enormous improvements of the most recent devices. The countermeasures against aging taken by the different manufacturers led to an increase of the lifetime by almost two orders of magnitude. The most important observation is the fact that ALD (atomic layer deposition) coated tubes show the best QE behavior. The $>$5 C/cm$^{2}$ integrated anode charge collected for the PHOTONIS XP85112 MCP-PMTs without a reduction of the QE make these devices a promising sensor candidate for the Barrel DIRC. The newly developed Hamamatsu 2$\times$2 inch$^{2}$ R13266 MCP-PMTs with ALD coating may also be an interesting option and are currently under investigation. Many more details about the lifetime issues are described in the following chapter.

\subsection{Evaluation of MCP-PMTs}\label{subsec:test_mcp}
\label{ch:mcp_pmts}

\begin{table*}[htb]
\setlength{\tabcolsep}{6pt} 
\renewcommand{\arraystretch}{1.5} 
\caption{Characteristics of the investigated lifetime-enhanced MCP-PMTs.}
\ \

\label{Tab:char}
{\small\begin{tabular*}{1.0\textwidth}[]{@{\extracolsep{\fill}}llllll}
\hline Manufacturer & BINP & PHOTONIS & \multicolumn{3}{c}{Hamamatsu}\\ 
\hline Type &  & XP85112 & R10754X-M16 & R10754X-M16M & R13266-M64 \\
Counter ID & \#1359/\#3548 & 1223/1332/1393 & JT0117 & KT0001/KT0002 & JS0022 \\
Pore diameter ($\mu$m) & 7 & 10 & 10 & 10 & 10 \\ 
Number of anodes & 1 & 8$\times$8 & 4$\times$4 & 4$\times$4 & 8$\times$8 \\ 
Active area (mm$^{2}$) & 9$^{2}$ $\pi$ & 53$\times$53 & 22$\times$22  & 22$\times$22 & 53$\times$53 \\ 
Total area (mm$^{2}$) & 15.5$^{2}$ $\pi$ & 59$\times$59 & 27.5$\times$27.5 & 27.5$\times$27.5 & 61$\times$61 \\ 
Geom. efficiency (\%) & 36 & 81 & 61 & 61 & 75 \\ 
\hline
\multirow{3}{*}{Comments}
 & better vacuum;	& better vacuum;	& film between				& ALD surfaces;		& ALD surfaces; \\ 
 & e-scrubbing; 	& 1-/1-/2-layer 	& 1$^{st}$\&2$^{nd}$ MCP 	& film between 		& film in front of\\ 
 & new PC			& ALD surfaces 		& 							& 1$^\mathrm{st}$\&2$^\mathrm{nd}$ MCP & 1$^\mathrm{st}$ MCP\\ 
\hline
\end{tabular*} 
}
\end{table*}

\subsubsection*{Measurement Setup and Investigated Types}

We have investigated the properties of many types of MCP-PMTs: circular-shaped single anode tubes from the Budker Institute of Nuclear Physics (BINP) in Novosibirsk, various square-shaped 2$\times$2 inch$^{2}$ 8$\times$8 pixel Planacon MCP-PMTs with different layouts from PHOTONIS, and several of the newly developed 1$\times$1 inch$^{2}$ array R10754X with four strips or 16 pads from Hamamatsu. Very recently, Hamamatsu has presented a larger square-shaped 2$\times$2 inch$^{2}$ prototype MCP-PMT R13266 with 8$\times$8 pixels, which is currently under investigation for possible usage in the Barrel DIRC. The technical characteristics of some of the investigated sensors are listed in Tab.~\ref{Tab:char}. 

The sensors were illuminated with a PiLas \cite{pilasC} laser which produces fast light pulses of 14~ps width ($\sigma$) at a wavelength of 372~nm; its maximum repetition rate is 1 MHz. The light is guided through a system of glass fibers, attenuated to the single photon level by neutral density filters and then focused onto the surface of the MCP-PMT with a system of micro lenses, which allows light spots from a few tens of $\mu$m to several cm in diameter. With the smaller spot sizes and an XY-scanner the gain and crosstalk behavior of the multi-pixel MCP-PMTs were investigated as a function of the surface position in steps of about 0.5 mm. For measurements of the rate capability typically a large laser spot was used.

Measurements of gain and time resolution as a function of the magnitude and the direction of a magnetic field were performed at a dipole magnet at the Forschungszentrum J\"ulich in Germany, which delivers a homogeneous field of up to 2.2~T over a pole shoe gap of 6 cm height. Usually the MCP-PMT signals were passively split after a 200-fold amplifier (Ortec FTA820A, 350 MHz bandwidth). One signal was directly fed into an ADC, while the other was discriminated (Philips Scientific 705) to determine the time delay between the MCP-PMT anode signal and the reference signal of the laser control unit. CAMAC and VME data acquisition systems were used to record the anode charge and the time delay for the signals of each pixel.

The most precise time resolution measurements were made with a LeCroy WavePro7300A with 3~GHz bandwidth and 20~GS/s sampling rate. This oscilloscope allows the determination of time resolutions at the few pico-second level.

\subsubsection*{Characteristics}

\paragraph*{Dark Count Rate}
\label{dark count}
$\;$\\

Each charged track will create a few thousand Cherenkov photons. After many reflections and other losses along the radiators and taking into account the Quantum Efficiency (QE) of the photon sensors only several tens of these photons will actually be detected. Therefore, it is important to use sensors with a moderately low dark count rate. From our measurements we find that at a gain of 10$^6$ and a threshold of 0.5 photo electrons the typical dark count rate for most of the tested MCP-PMTs is below 1 kHz/cm$^2$. These numbers are sufficient for the Barrel DIRC. Only the new BINP MCP-PMT with a modified photo cathode shows a dark count rate of more than 100 kHz/cm$^2$, while the Hamamatsu R10754 and R10754X show a significantly lower rate of $\sim$100 Hz/cm$^2$. We also observed that often the main fraction of the dark count rate comes from rather localized spots in the MCP-PMT indicating that most anode pixels have a very low dark count rate of only a few Hz.

\paragraph*{Gain inside Magnetic Field}
\label{magnetic field}
$\;$\\

\begin{figure*}[htb]
\begin{center}
\begin{minipage}{0.8\textwidth}
	\includegraphics[width=1.0\textwidth]{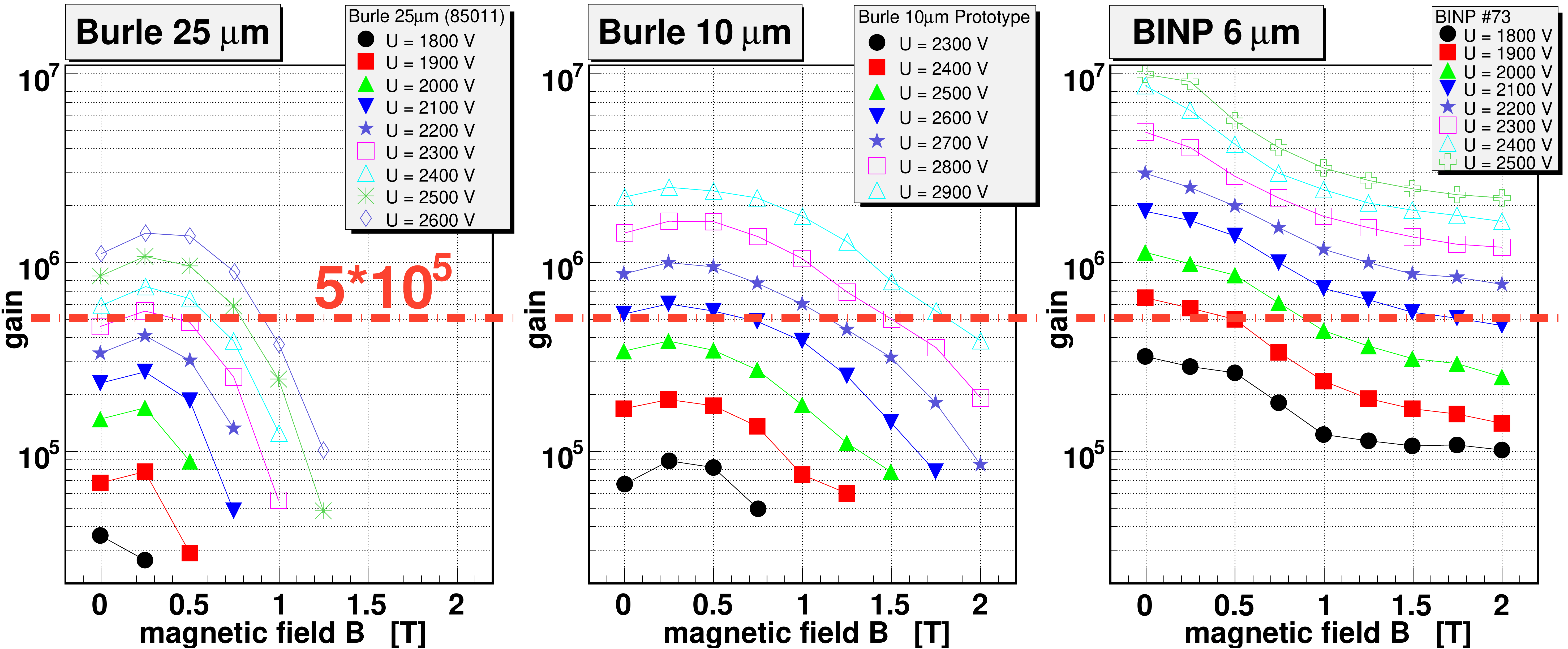}
\end{minipage}
\end{center}
\caption{Gain as a function of the magnetic field for different high-voltage
settings. Compared are MCP-PMTs of Burle-PHOTONIS with 25 $\mu m$
pore diameter (left), a PHOTONIS prototype with 10 $\mu m$ (middle)
and a BINP device with 6 $\mu m$ pore diameter (right). The minimum
gain of 5 x 10$^{5}$ for an efficient single photon detection is indicated
by the dash-dotted line.}
\label{fig:gain_vs_B1}
\end{figure*}

The behavior of the gain as a function of the magnetic field is shown
in Fig.~\ref{fig:gain_vs_B1} for different high-voltage settings of
three MCP-PMTs with different pore sizes. Clearly, the maximum gain reachable
with the MCP-PMT depends on the pore diameter. The 25 $\mu m$
device reaches just above 10$^{6}$ while with the MCP-PMT with
6 $\mu m$ pore size a gain of almost 10$^{7}$ is possible. These
results are compatible with earlier measurements~\cite{akatsu}.

\begin{figure}[htb]
\begin{center}
\begin{minipage}{0.49\textwidth}
	\includegraphics[width=1.0\textwidth]{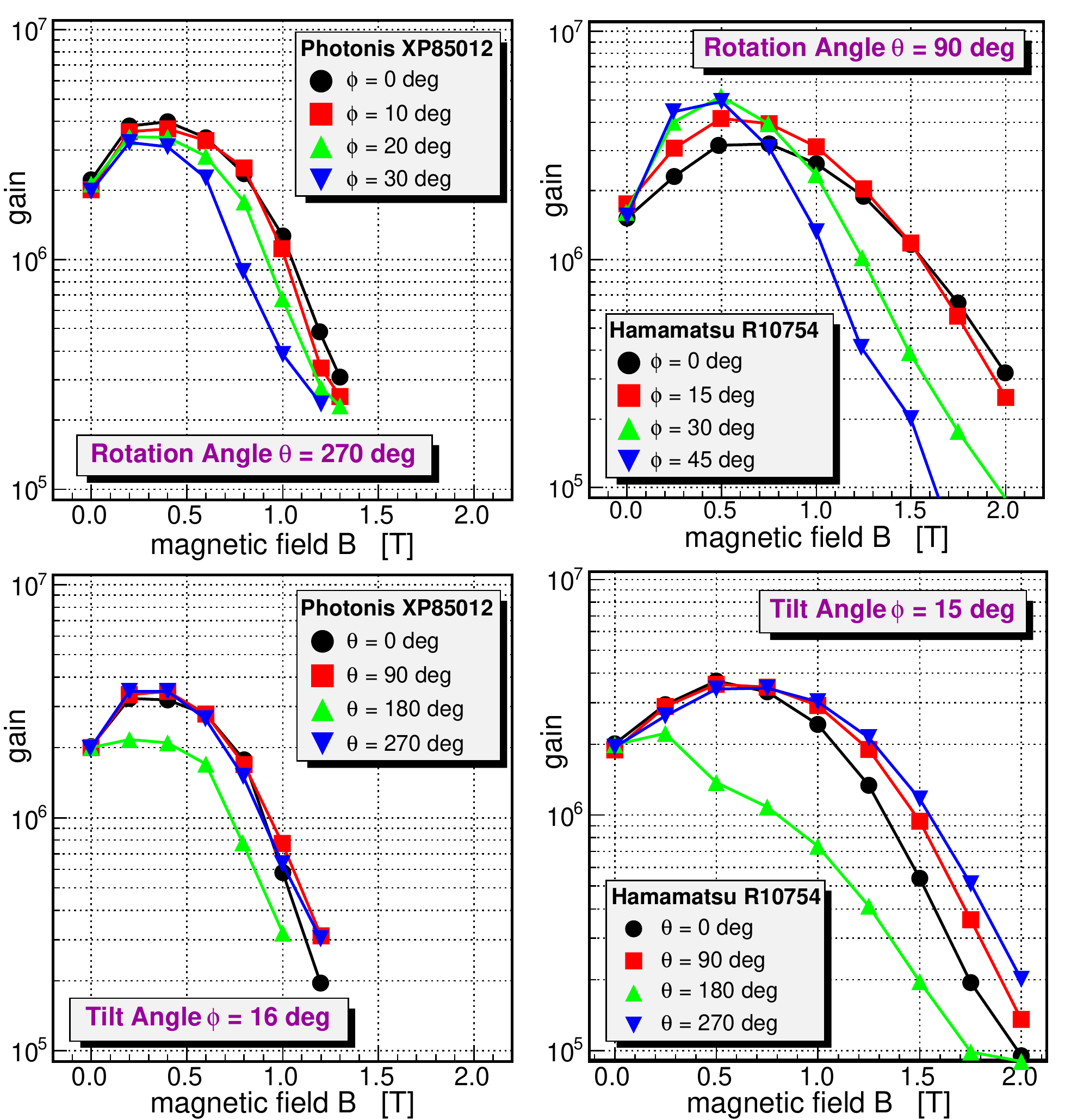}
\end{minipage}
\end{center}
\caption{Gain as a function of the magnetic field direction for the PHOTONIS XP85012 (left column) and the Hamamatsu R10754-00-L4 (right column). In the upper row the dependence on the tilt angle $\phi$ is shown, in the lower row that on the rotation angle $\theta$.}
\label{fig:gain_vs_B2}
\end{figure}

The dash-dotted line indicates the minimum gain of about 5$\cdot$10$^{5}$, which is still acceptable for an efficient single photon detection. From the plots it is obvious that the gain of the 25~$\mu m$ version
of the PHOTONIS Planacon XP85012 reaches this limit only at large high-voltage
settings. Since the gain collapses completely just above 1~T this device
does not meet the requirements for the Barrel DIRC. The PHOTONIS Planacon XP85112
with a smaller pore diameter of 10 $\mu m$ exhibits a larger gain and it is still
operable in the 2~T field of the \panda solenoid. Efficient single
photon detection appears possible up to at least 1.75~T, a high
voltage setting close to the recommended maximum for this device is needed though.
The best gain performance in a high magnetic field is observed for the
BINP MCP-PMT with 6 $\mu m$ pore diameter. The \panda gain limit
for single photon detection is reached at moderate operation voltages
even in a 2~T field.

Usually the gain reaches a maximum at $\sim$0.5~T and drops at higher fields. At a pore size of 25~$\mu$m the gain totally collapses just above 1~T, which can be attributed to the Larmor radius of the avalanche electrons at this field. Therefore, to efficiently detect single photons up to 1.5~T, as required in the Barrel DIRC, a pore size of $\le$10 $\mu$m is needed \cite{AL1}.

For the BINP MCP-PMT (see \cite{AL1}), the PHOTONIS XP85012, and the Hamamatsu R10754-00-L4 measurements of the gain dependence on the orientation of the PMT axis with respect to the field direction were also performed. The results for the two latter devices are displayed in Fig.~\ref{fig:gain_vs_B2}. In the upper row the gain dependence on the tilt angle $\phi$ between the PMT axis and the field direction is shown: this demonstrates that up to $\phi \approx 20^{\circ}$ no significant gain change is observed, while at larger angles the gain at higher field values starts to drop rapidly. Still, even at moderate tilt angles MCP-PMTs can be used for an efficient single photon detection in high magnetic fields. This is important for the PMT orientation in the Barrel DIRC and an enormous advantage compared to standard dynode-based PMTs.

In the lower row of Fig.~\ref{fig:gain_vs_B2} the gain behavior at different rotation angles $\theta$ of the PMT around the field axis and at a tilt angle $\phi \approx 15^{\circ}$ is shown: there is a significantly different slope at $\theta = 180^{\circ}$, when the capillaries of one of the two MCP layers point exactly along the field direction. At all other measured rotation angles the gain follows roughly the same slope.

\paragraph*{Time Resolution}
\label{time resolution}
$\;$\\

\begin{figure}[htb]
\begin{center}
\begin{minipage}{0.49\textwidth}
	\includegraphics[width=1.0\textwidth]{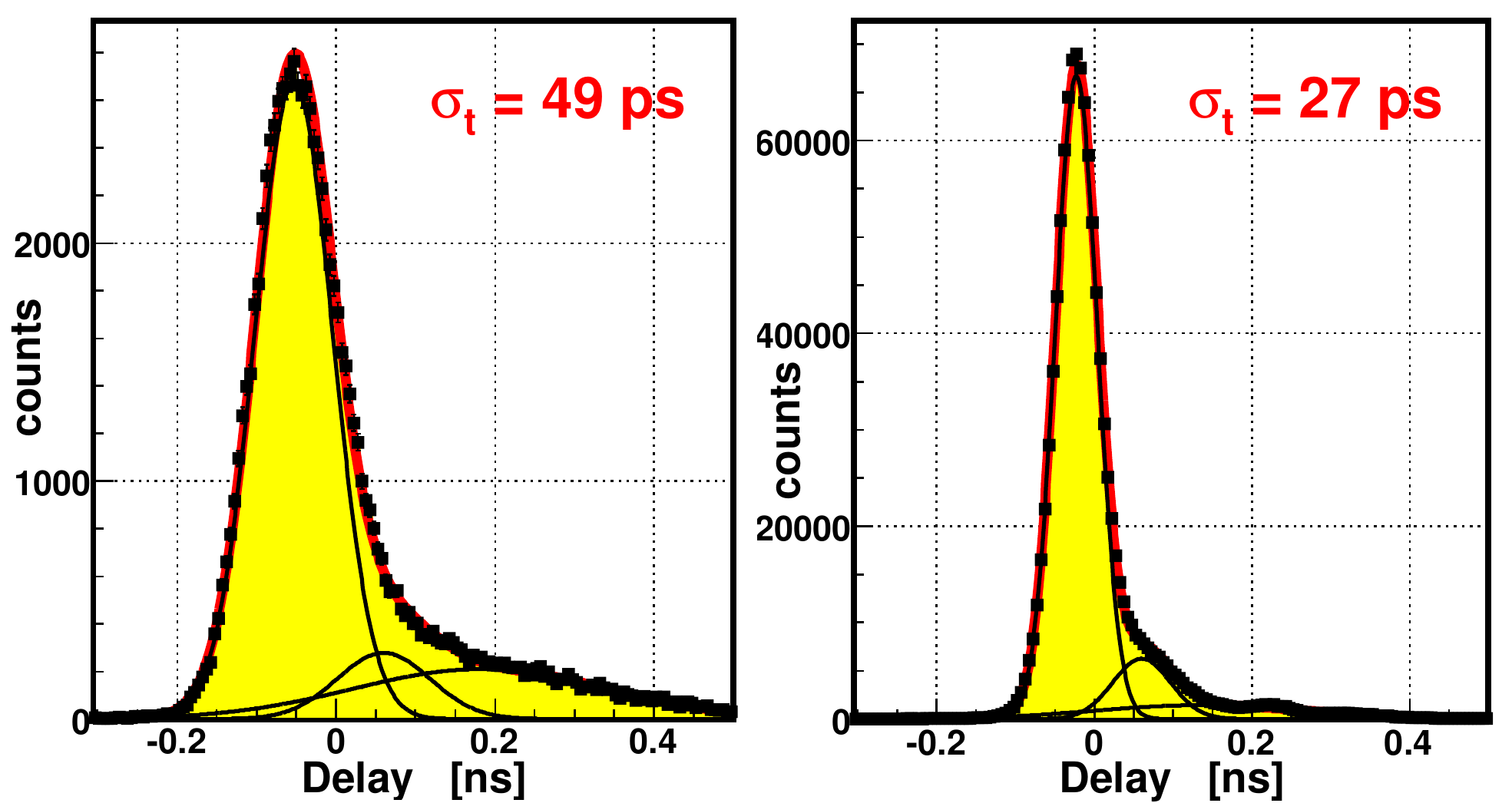}
\end{minipage}
\end{center}
\caption{Single photon time resolution for the PHOTONIS MCP-PMT
with 25 $\mu$m (left) and the BINP device with 6 $\mu$m (left) pore
diameter measured with a 3~GHz / 20 Gs oscilloscope. A LeCroy 821
leading edge discriminator and an Ortec VT120A amplifier were used.}
\label{fig:timeres}
\end{figure}

In Fig.~\ref{fig:timeres} the time resolution measured for the PHOTONIS MCP-PMT XP85012 with
25~$\mu m$ pores is compared to that of the BINP MCP-PMT with 6~$\mu m$ pores.
For the latter a resolution of 27~ps was obtained. This result still contains 
contributions from the finite time resolution of the electronics devices, the 
input channels of the oscilloscope, and in particular of the laser pulses. 
These resolutions were measured independently to be about 5-6~ps/channel for 
the oscilloscope channels and the same for the electronics devices used. 
The PiLas laser contributes 14~ps. Unfolding these contributions results in 
a net transit time resolution for single photons of $\sigma_{t}$ $\approx$~20~ps for the BINP MCP-PMT.

The distribution of the measured time resolutions \cite{AL1,AL2} usually consists of a narrow peak ($\sigma_{t}$) and a tail to one side which originates mainly from photo electrons backscattered at the MCP entrance. This behavior was seen for all investigated MCP-PMTs, though with different fractions. As listed in Tab.~\ref{Tab:tres}, the width of the peak was always $\le$50 ps, with the best resolutions of 27~ps and 23~ps (at 10$^6$ gain and after $\times$200 amplification of the MCP-PMT anode signal) for the BINP MCP-PMT with 6~$\mu$m pore diameter, respectively, for the Hamamatsu Prototype with 10~$\mu$m pore diameter. All measured time resolutions are without any correction for the resolutions of the used electronics modules and the laser pulse width.

The RMS width of the time distribution depends strongly on the height and extension of the tail. This can be partly controlled by building MCP-PMTs with a narrow gap between the PC and the first MCP, which reduces the amount of backscattered photo electrons reaching a MCP pore. In general it appears that all MCP-PMTs had a time resolution better than 60~ps.

\begin{table*}[htb]
\setlength{\tabcolsep}{6pt} 
\renewcommand{\arraystretch}{1.5} 
\caption{Single photon time resolutions of many investigated types of MCP-PMTs}
\ \

\label{Tab:tres}
\begin{tabular*}{0.98\textwidth}[]{@{\extracolsep{\fill}}llcc}
\hline  Manufacturer & Type & Pores [$\mu m$] & $\sigma_{t}$ [ps] \\
\hline  BINP & \#73 & 6 & 27 \\
\hline  \multirow{4}{*}{PHOTONIS} & XP85112 & 10 & 41 \\
                                        & XP85011   & 25 & 49 \\
                                        & XP85013   & 25 & 51 \\
                                        & XP85012   & 25 & 37 \\
\hline  \multirow{3}{*}{Hamamatsu}  & R10754-00-L4 (1'' $\times$ 1'') & 10 & 32 \\
									& R10754X-01-M16 (1'' $\times$ 1'') & 10 & 33 \\
									& Prototype R13266 (2'' $\times$ 2'') & 10 & 23 \\
\hline 
\end{tabular*} 
\end{table*}

The time resolutions were also measured as a function of the magnitude of the magnetic field, with no significant deterioration at higher fields being observed.

\paragraph*{Gain Homogeneity and Crosstalk}
\label{crosstalk}
$\;$\\

\begin{figure}[htb]
\begin{center}
\begin{minipage}{0.49\textwidth}
	\includegraphics[width=1.0\textwidth]{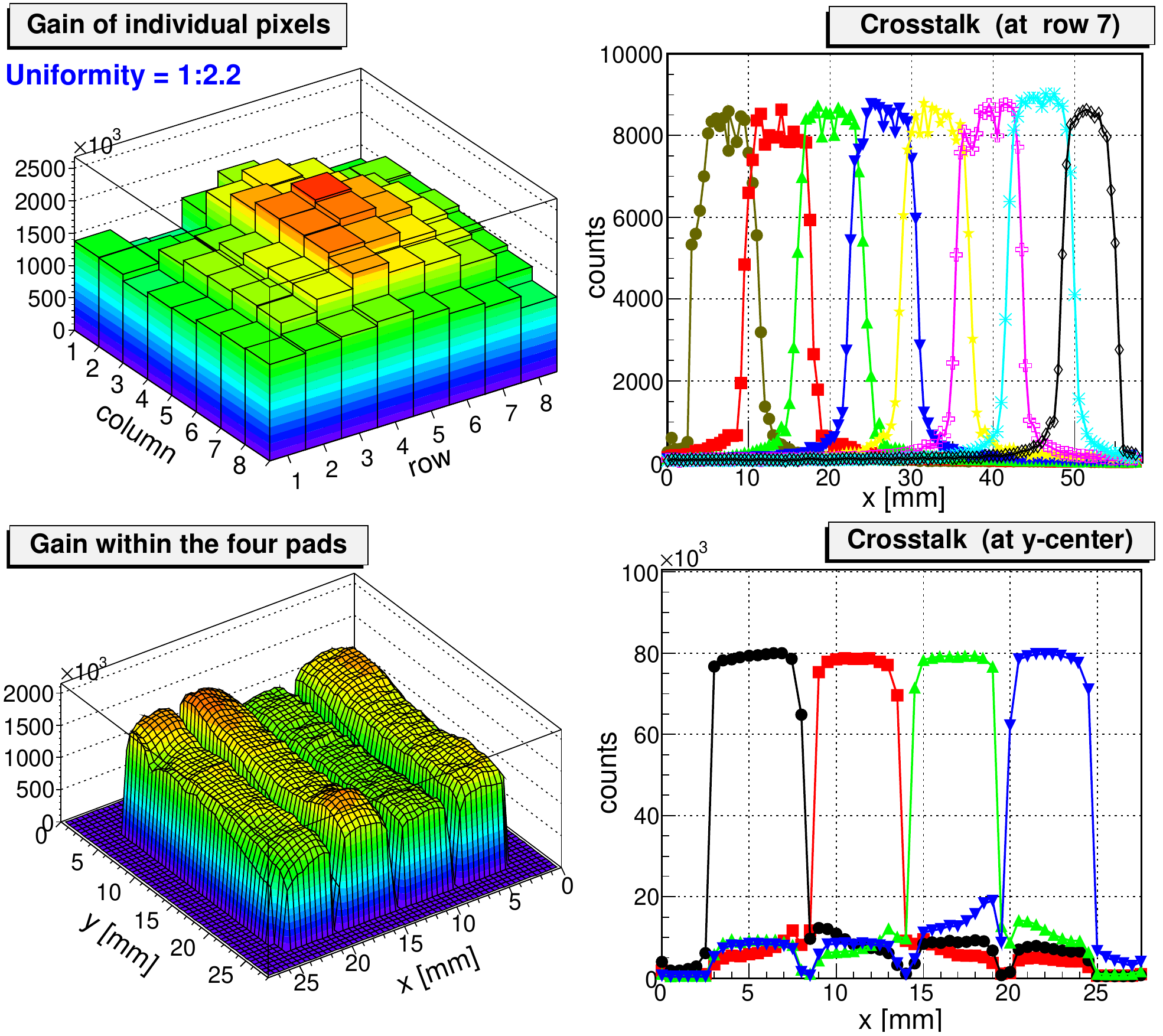}
\end{minipage}
\end{center}
\caption{Gain (left) and crosstalk behavior along one row of pixels (right) for the PHOTONIS XP85012 (upper) and the Hamamatsu R10754-00-L4 (lower).}
\label{fig:CountsCrosstalk}
\end{figure}

The response of the multi-anode MCP-PMTs was investigated with XY-scans across the active surface. The gain of the different pixels in a device can vary by a factor 3 to 5 as measured in some Hamamatsu and PHOTONIS tubes \cite{AL2}. The standard 25~$\mu$m pore MCP-PMTs of the latter manufacturer show typical gain variations up to a factor 2 across the 64 pixels, as plotted in Fig.~\ref{fig:CountsCrosstalk} (upper left) for the XP85012. The lowest gains are usually observed for the edge pixels and especially at the corners. The Hamamatsu R10754-00-L4 even shows significant gain inhomogeneities within one pad (Fig.~\ref{fig:CountsCrosstalk}, lower left), with measured fluctuations sometimes exceeding a factor 2. Currently, the new lifetime-enhanced MCP-PMT prototypes show somewhat larger gain fluctuations, which is expected to improve when the final tubes will be produced.

A lower gain may cause a reduced detection efficiency of the pixel. In fig \ref{fig:CountsCrosstalk} (right column) the number of counts of each pixel in a row is shown, when the active surface of the MCP-PMT was illuminated in steps of 0.5 mm along the x-coordinate (or column) while the y-position (or row) was kept constant.

These plots also show the crosstalk among the anode pixels. For the PHOTONIS XP85012 crosstalk is mainly visible at the transition to the adjacent pixels, most likely caused by charge sharing at the anode and by backscattered photo electrons at the MCP entrance, while pixels further away are hardly affected. In contrast, for the Hamamatsu R10754-00-L4 a significant response of all other pixels is observed when a certain pad is illuminated; even pixels far from the light spot can fire. Further investigations with the latter MCP-PMT showed that most of the crosstalk is of electronic nature and can be eliminated to a large extent by a modified construction of the tube: e.g., the second MCP layer is split into separate sectors each of the size of the adjacent anode pad \cite{inami}. With the latest lifetime-enhanced Hamamatsu R10754X tubes this electronics effect is solved and these devices show an even better crosstalk behavior than those from PHOTONIS.

\paragraph*{Gain Stability at High Rates}
\label{rate stability}
$\;$\\

\begin{figure}[htb]
\begin{center}
\begin{minipage}{0.49\textwidth}
\includegraphics[width=1.0\textwidth]{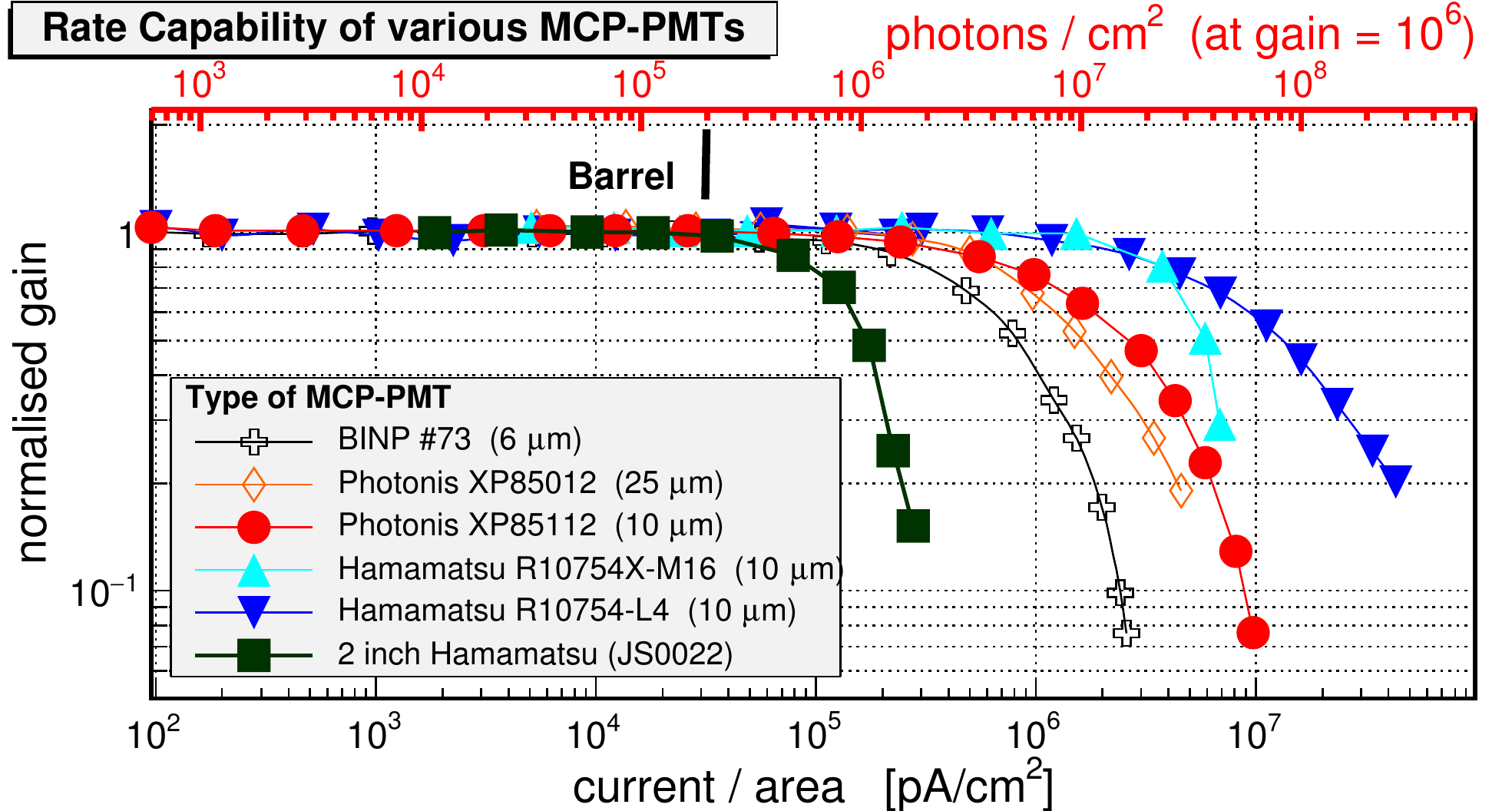}
\end{minipage}
\end{center}
\caption{Rate capability of various models of MCP-PMTs: the relative gain is plotted versus the anode current. At the upper axis the translation into a rate for single photons assuming a constant gain of 10$^6$ is given. The expected rate of detected photons for the Barrel DIRC is also indicated.}
\label{fig:CountsRates}
\end{figure}

The rate capability of MCP-PMTs is one of the most critical issues in high rate experiments like \panda. The expected photon density at the readout (anode) plane (after QE) is $\sim$200 kHz/pixel for the Barrel DIRC. At these photon rates the current in the high resistive material of the MCP capillaries may not flow off fast enough, which causes charge saturation effects. The result of this is a rapidly decreasing gain as seen in Fig.~\ref{fig:CountsRates} where the normalized gain is plotted versus the anode current. Assuming a certain gain of the tube (e.g., 10$^6$ in the figure) this current can be translated into a single photon density which is given at the upper axis.

The gain of older MCP-PMTs started dropping already at photon densities well below 1 MHz/cm$^2$ (e.g. PHOTONIS XP85011 in Fig.~\ref{fig:CountsRates}). However, little to practically no gain loss up to $\sim$2 MHz/cm$^2$ single photons is observed for the new PHOTONIS XP85112, while the Hamamatsu R10754 and R10754X are even capable of withstanding rates $>$5 MHz/cm$^2$ without a gain reduction. Although the new two-inch prototype MCP-PMT R13266 of Hamamatsu shows a significantly lower rate capability this new model would still qualify for the Barrel DIRC.

\subsubsection*{Aging and Lifetime}
\label{lifetime}

Aging of an MCP-PMT usually manifests itself in a reduction of its gain, its dark count rate and in particular its quantum efficiency (QE) when the integrated anode charge accumulates. While a lower dark count rate is desirable and the reduced gain can to some extend be compensated by a higher PMT voltage, the diminishing QE may lead to an unusable tube. The main cause of the QE drop appears to be feedback ions from the rest gas, especially heavy products like lead, which impinge on the photo cathode (PC) and damage it. It has also been speculated that neutral rest gas molecules like oxygen and carbon dioxide may pollute the PC surface and change its work function \cite{jinno}.

\paragraph*{Methods for Lifetime Improvement}
$\;$\\

An obvious way of reducing the amount of rest gas in the tube is to bake the microchannel plates to outgas the glass material and desorb the surfaces. Additionally, the vacuum inside the MCP-PMT is improved and the manufacturers often apply electron scrubbing to clean and polish the MCP surfaces. Besides these approaches the three main manufacturers of MCP-PMTs apply the following techniques to extend the QE lifetime:

\begin{itemize}

\item  In their latest MCP-PMT models the Budker Institute of Nuclear Physics (BINP) in Novosibirsk applies a special treatment to the bi-alkali PC which is baked in a vapor of caesium and antimony. This seems to increase the PC's hardness against feedback ions, but significantly increases the dark count rate of the tube \cite{barnyakov}. 

\item  A new and innovative approach is pursued by PHOTONIS. The surfaces and pores of the MCPs are coated with a very thin layer of secondary electron emissive material by applying an atomic layer deposition (ALD) technique \cite{arradiance,lappd1,lappd2}. This layer is expected to significantly reduce the outgassing of the MCP material.

\item  Hamamatsu first tried to eliminate the ion back flow from the anode side of the MCP-PMT by putting a thin protection layer of aluminum (film) between the two MCPs. In addition, potential gaps between the MCPs and the metal walls of the tube's frame were sealed with ceramic elements to hinder neutral atoms and molecules from the rear part of the MCP-PMT in reaching the PC \cite{jinno}. In their most recent MCP-PMTs Hamamatsu also applies the ALD technique, often combined with a film in front of or between the MCPs.

\end{itemize}

\begin{figure}[htb]
\centering
\includegraphics[width=.49\textwidth]{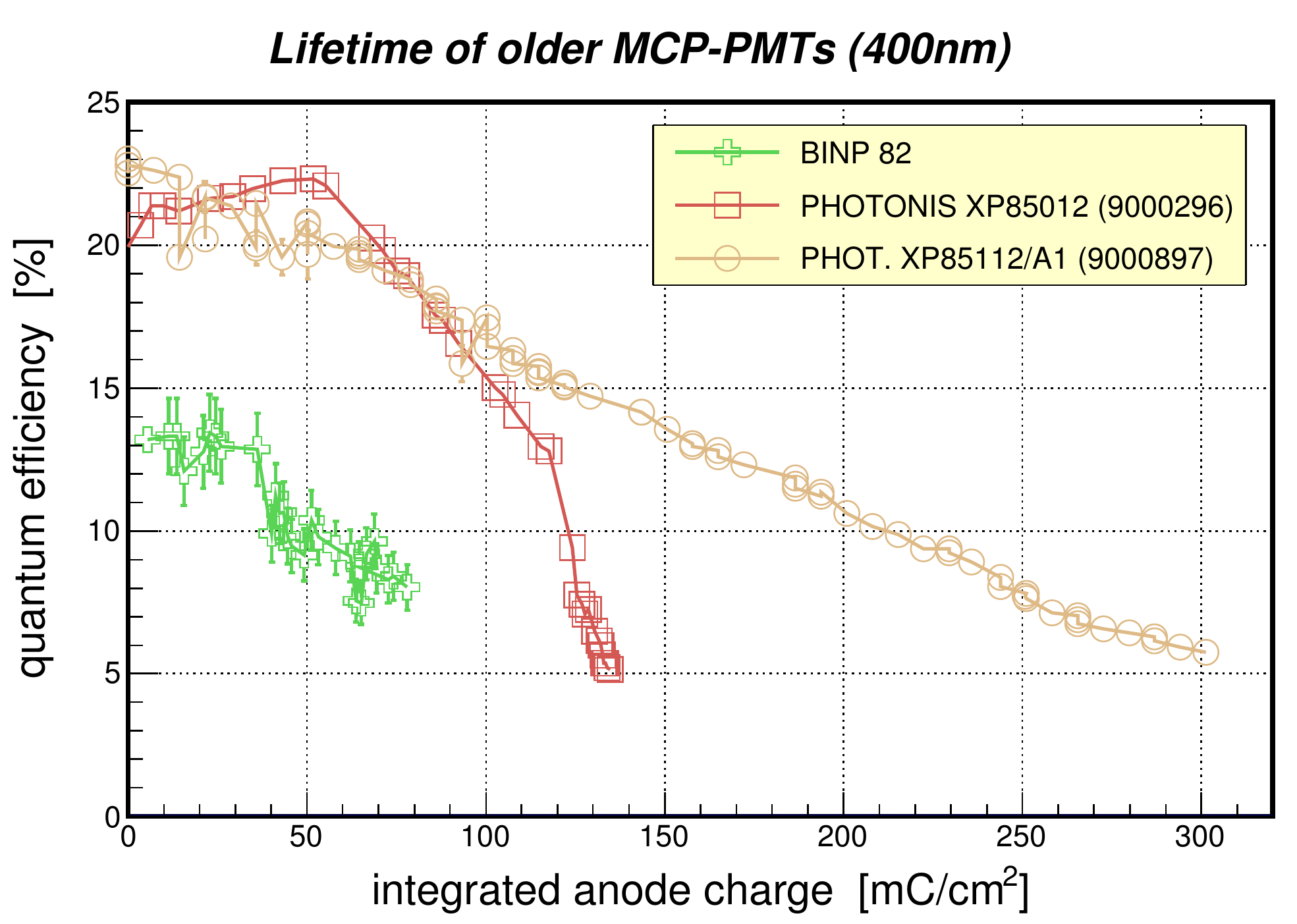}
\caption{Comparison of our aging measurements with not yet lifetime-enhanced MCP-PMTs: quantum efficiency as a function of the integrated anode charge at 400 nm.}
\label{fig:QEcomp_old}
\end{figure}

In the recent years we have measured the lifetime of several MCP-PMTs of the three manufacturers mentioned above. The first tubes from BINP (\#82) and PHOTONIS (XP85012-9000298 and XP85112-9000897) were still without the above-listed improvements (see Fig.~\ref{fig:QEcomp_old}). A list of the characteristics of the lifetime-enhanced MCP-PMTs discussed in this report are given in Tab.~\ref{Tab:char}.

\paragraph*{Setup of Lifetime Measurements}
$\;$\\

Until recently only few quantitative results on the lifetime of MCP-PMTs were available \cite{nagoya, barny}. Moreover, these were obtained in very different environments and therefore difficult to compare. The standard way of measuring the lifetime of an MCP-PMT is to determine the gain and especially the QE as a function of the integrated anode charge. If the QE has dropped by a certain percentage (e.g. 50\%) of its original value the sensor is presumed unusable. The \panda experiment is expected to run for at least 10 years at a 50\% duty cycle. Assuming the average antiproton-proton annihilation rate of 20 MHz and a sensor gain of 10$^{6}$, simulations show an integrated anode charge of 5 C/cm$^{2}$ expected for the MCP-PMTs of the Barrel DIRC.

The lifetimes shown in Fig.~\ref{fig:QEcomp_old}, which we determined for the first MCP-PMTs of BINP (\#82) and PHOTONIS (XP85012-9000298 and XP85112-9000897), were by far not sufficient for \panda. The QE had dropped by $>$50\% after only $\approx$200 mC/cm$^{2}$ integrated anode charge~\cite{alex, fred}.

The setup of our lifetime measurements is described in these publications~\cite{alex, AL4}. The MCP-PMTs were permanently and simultaneously illuminated with a blue (460 nm) LED at a rate comparable to that expected at the image plane of the Barrel DIRC, first 270 kHz and later 1 MHz to accelerate the measurement. The entire photo cathode of the MCP-PMT was homogeneously illuminated with near-parallel light. At the entrance window the light was attenuated to a level of $\sim$1 photon/cm$^2$ per pulse; at a gain of 7$\cdot$10$^5$ this corresponds to an integrated anode charge of $\sim$3.5 mC/cm$^2$/day ($\sim$14 mC/cm$^2$/day at the higher rate). The stability of the LED was controlled by measuring the current of a photodiode placed close to the MCP-PMTs. The MCP-PMTs' responses were continuously monitored by recording the pulse heights with a DAQ system at a highly prescaled rate. In irregular time intervals (a few days at the beginning, a few weeks later) the Q.E. of the photo cathode of each illuminated MCP-PMT was determined over a 300-800~nm wavelength band. The setup for the Q.E. measurements \cite{herold} consisted of a stable halogen lamp, a monochromator with 1~nm resolution and a calibrated reference diode (Hamamatsu S6337-01).

For each MCP-PMT, and in intervals of a few months, the photo current across the whole PC surface was measured in small steps of 0.5 mm at a wavelength of 372~nm to identify the regions where the QE degradation possibly starts.

\paragraph*{Results of Lifetime Measurements}
$\;$\\

\begin{figure}[htbp]
\centering
\includegraphics[width=.49\textwidth]{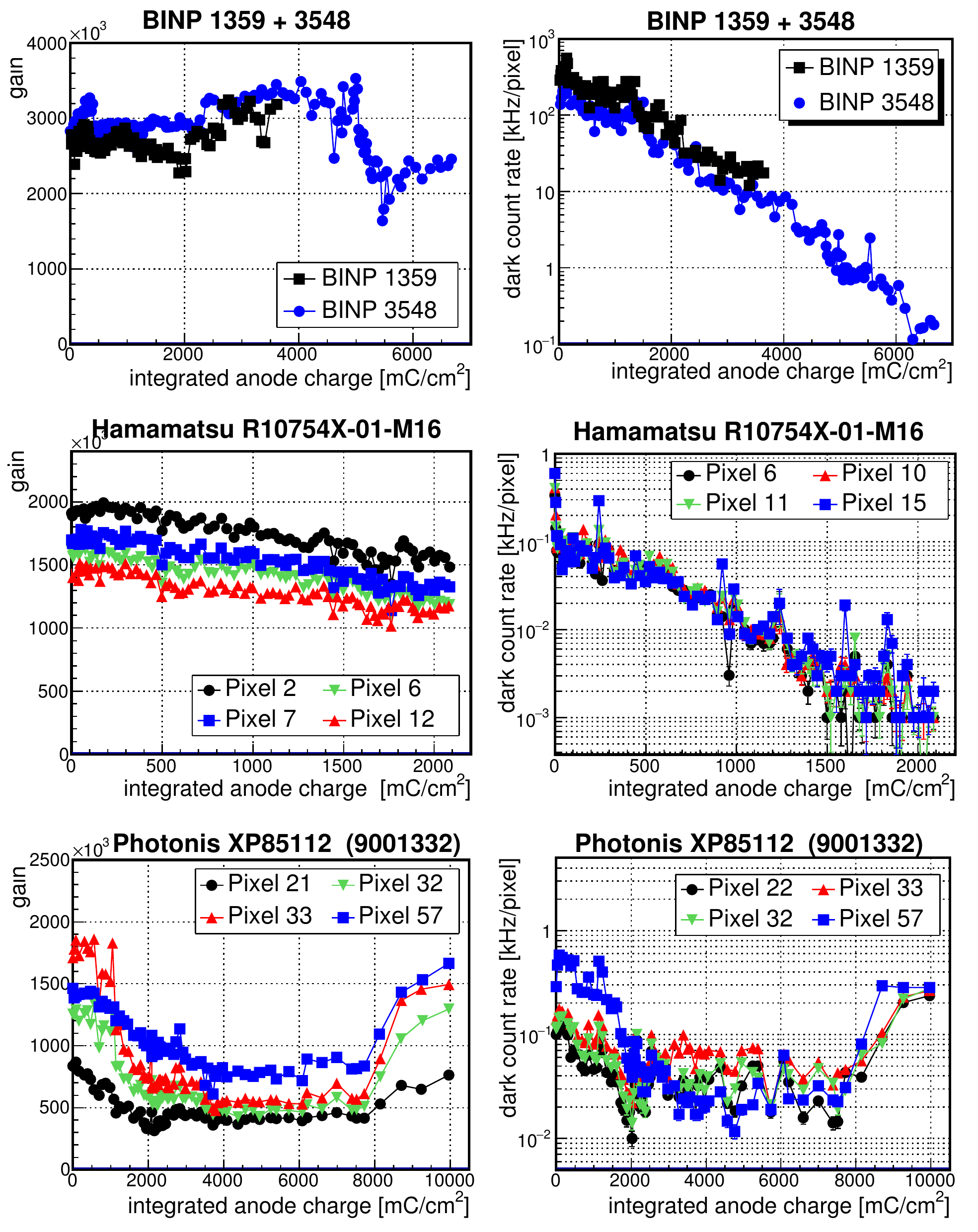}
\caption{Gain (left) and dark count rate (right) as a function of the integrated anode charge for selected MCP-PMTs.}
\label{fig:gainDcC}
\end{figure}

Important quantities for Cherenkov detectors are the gain and dark count rate of the used sensors. The gain has to be high enough for an efficient single photon detection and the dark count rate should be low since the photon yield per track is usually rather moderate. These quantities were measured as a function of the integrated anode charge as shown in Fig.~\ref{fig:gainDcC}. We observe that the gain changes are only moderate for most of the pixels of the displayed sensors and can easily be compensated for by increasing the tube voltage. On the other hand the dark count rate may drop by more than two orders of magnitude for the BINP and Hamamatsu MCP-PMTs. This finding indicates a change of the PC's work function during the illumination of the sensor. The PHOTONIS XP85112 does not show these massive changes in the dark count rate.

\begin{figure}[htb]
\centering
\includegraphics[width=.49\textwidth]{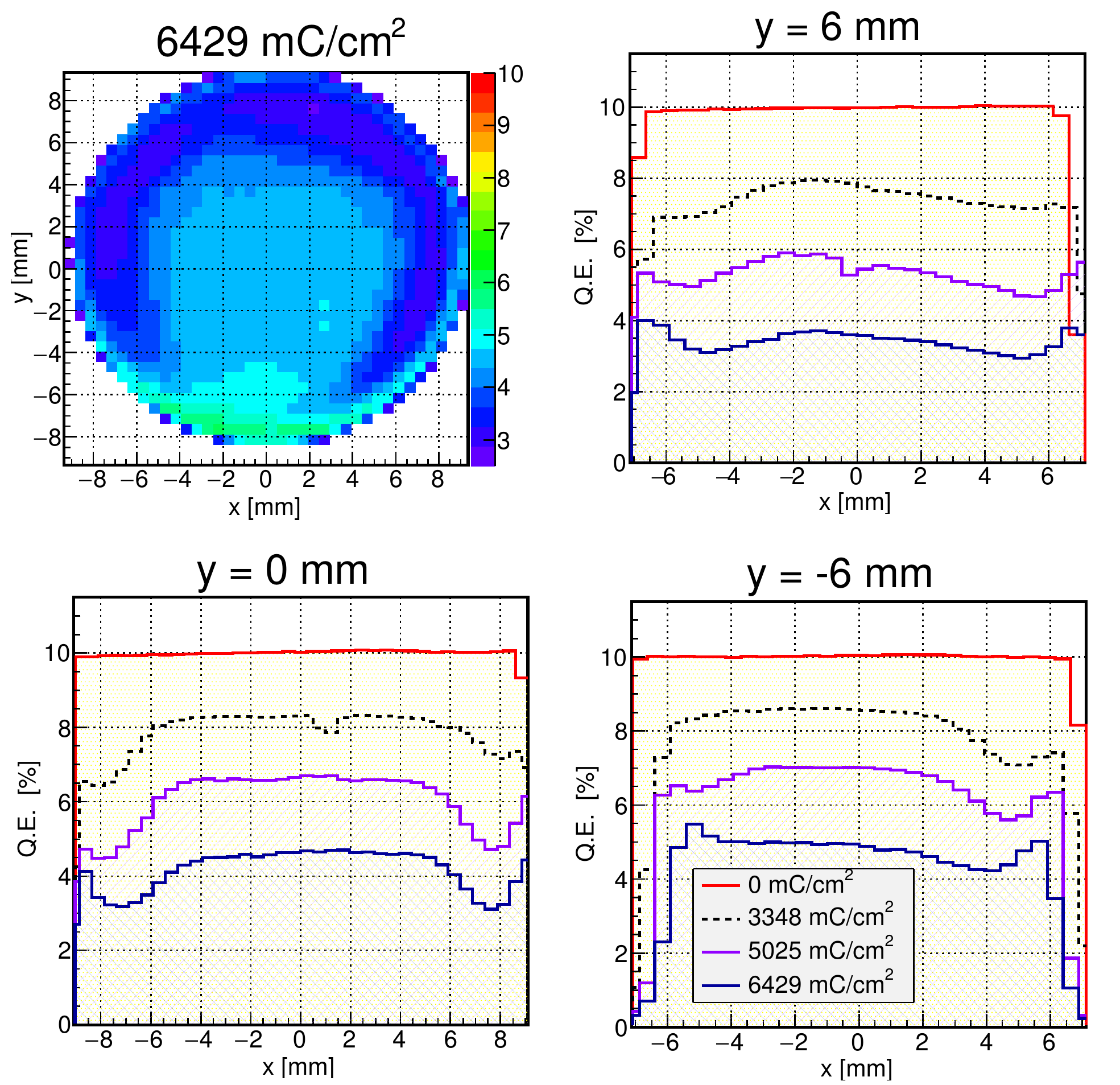}
\caption{QE at 372~nm as a function of the PC surface for the BINP \#3548 MCP-PMT with an active area of 18 mm diameter. Upper left: two-dimensional QE chart (in \% [color level]); other plots: QE x-projections at different y-positions and anode charges.}
\label{fig:QE2d_BINP}
\end{figure}

\begin{figure}[htb]
\centering
\includegraphics[width=.49\textwidth]{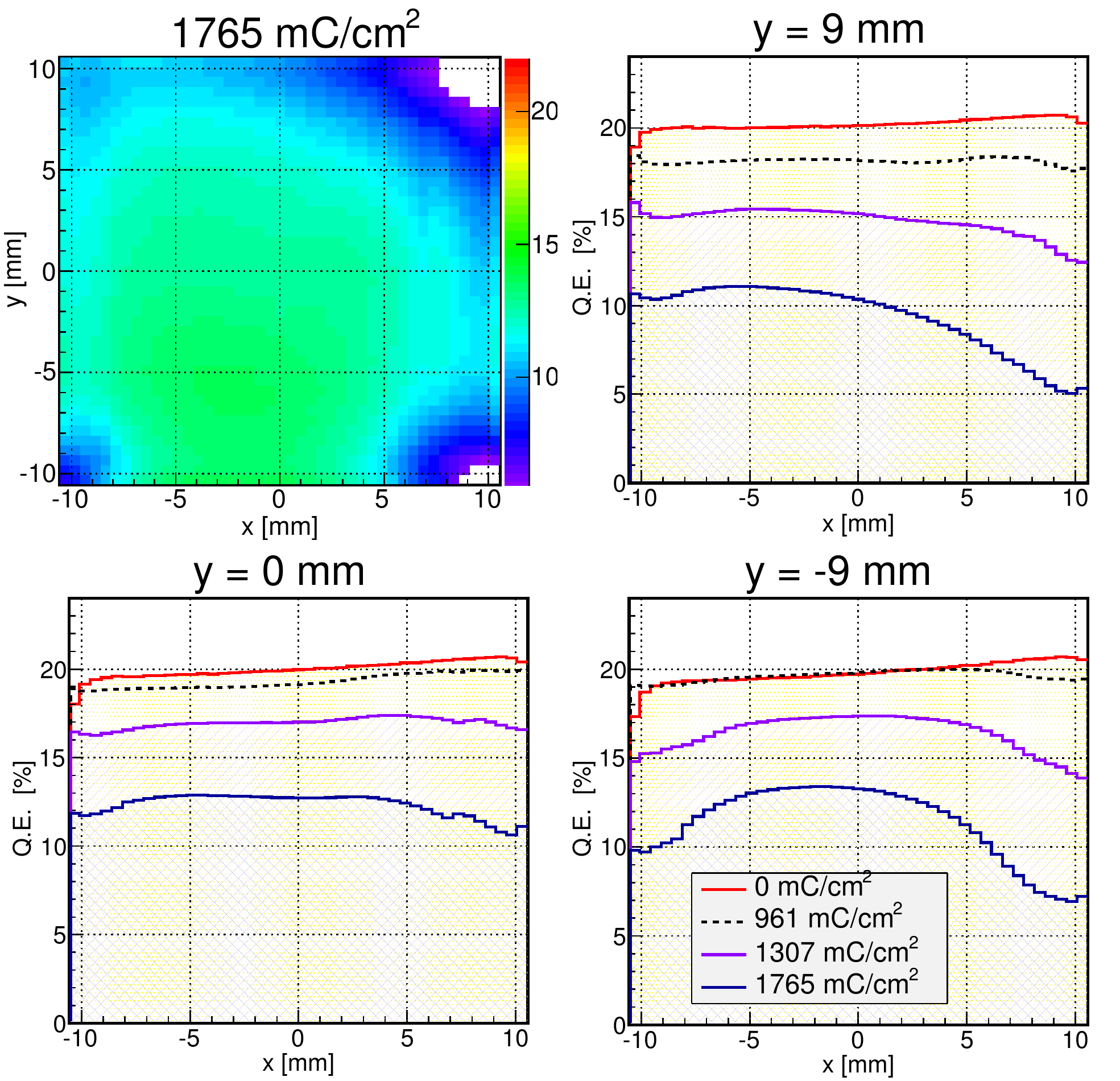}
\caption{QE at 372~nm as a function of the PC surface for the Hamamatsu R10754X-01-M16 (JT0117) MCP-PMT with an active area of 22 $\times$ 22 mm$^{2}$. The four plots display the same properties as in Fig.~\ref{fig:QE2d_BINP}.}
\label{fig:QE2d_Ham}
\end{figure}

\begin{figure}[htb]
\centering
\includegraphics[width=.49\textwidth]{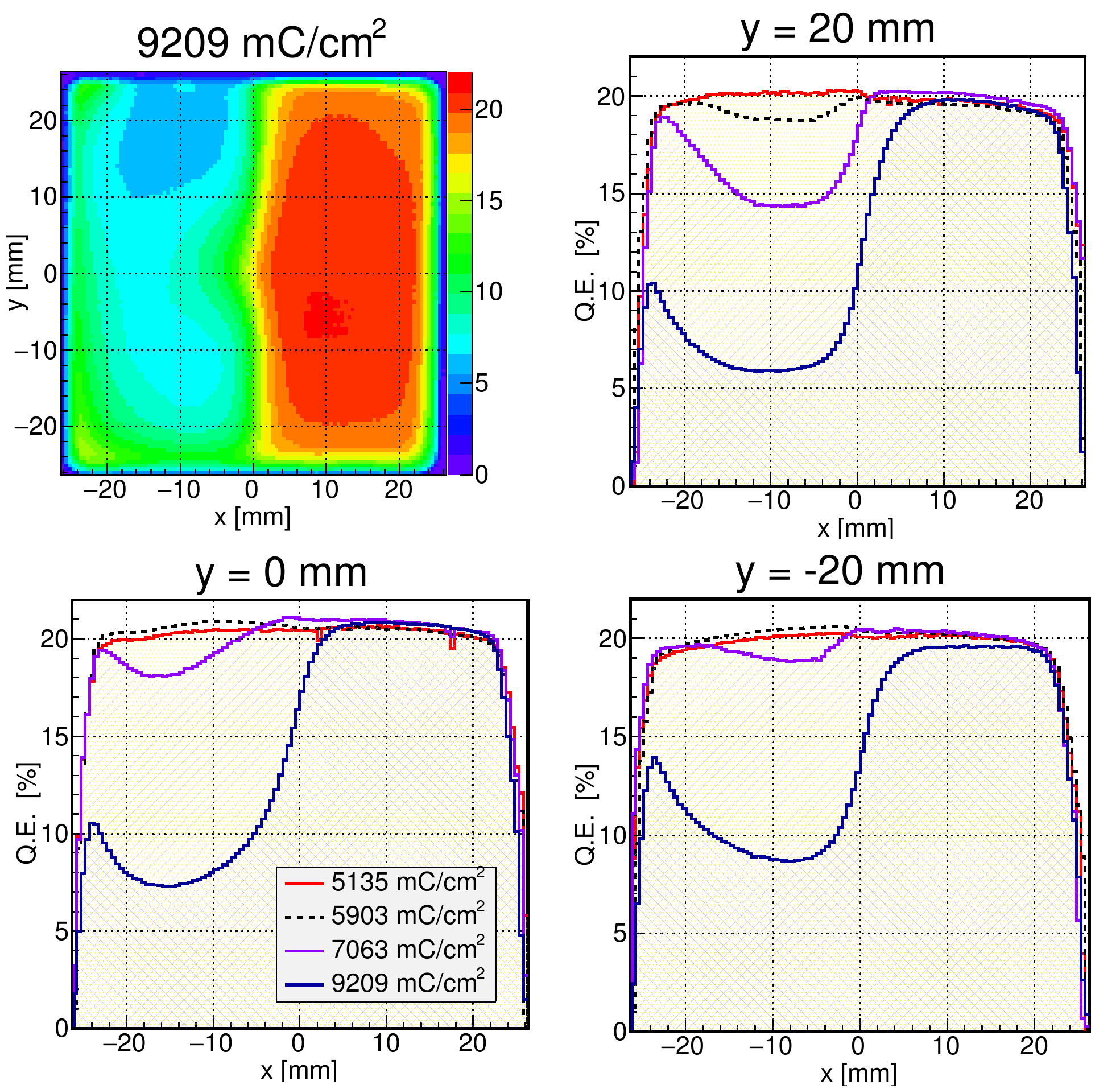}
\caption{QE at 372~nm as a function of the PC surface for the PHOTONIS XP85112 (9001223) MCP-PMT with an active area of 53 $\times$ 53 mm$^{2}$. The four plots display the same properties as in Fig.~\ref{fig:QE2d_BINP}.}
\label{fig:QE2d_Phot1223}
\end{figure}

\begin{figure}[htb]
\centering
\includegraphics[width=.49\textwidth]{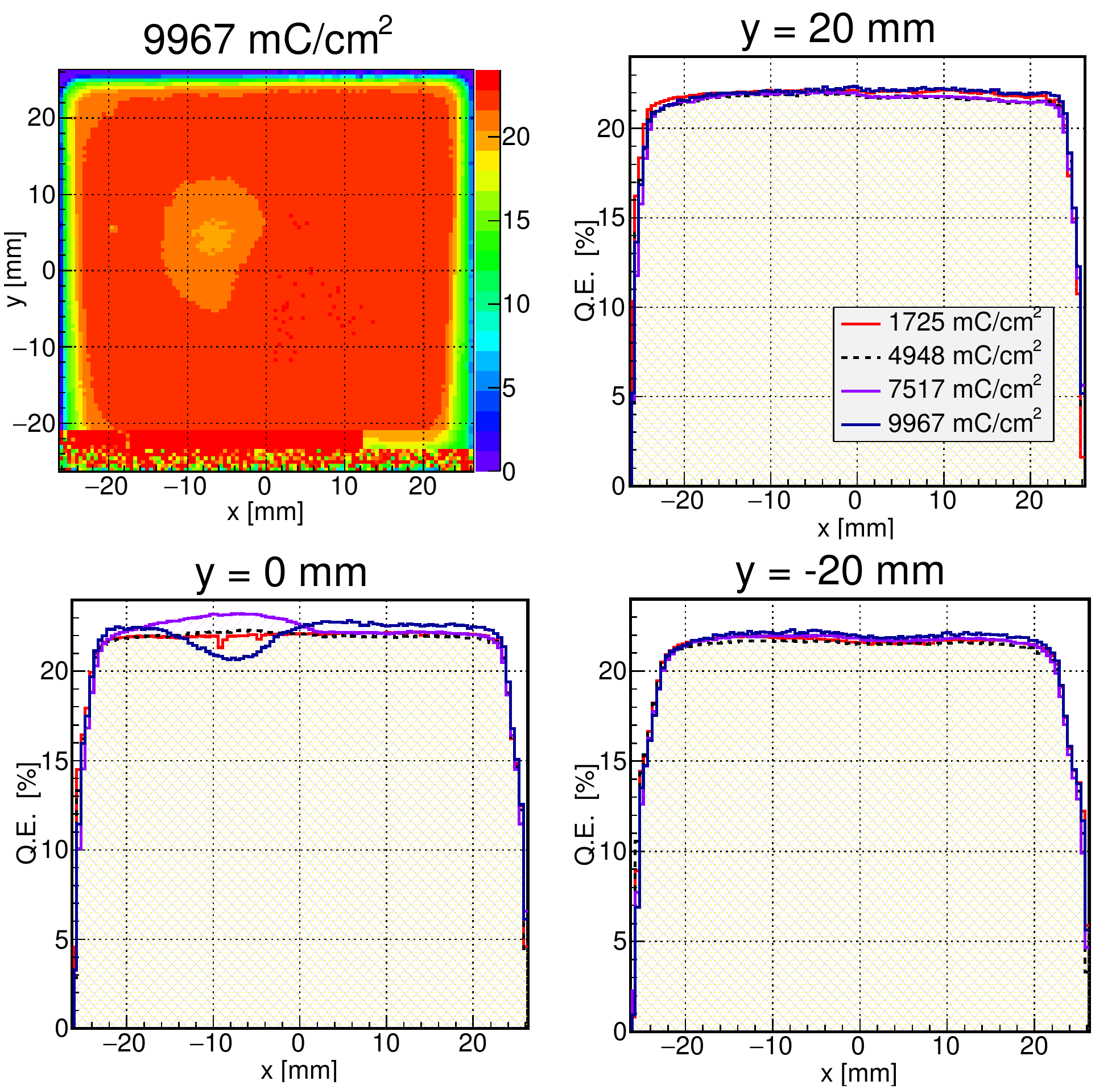}
\caption{QE at 372~nm as a function of the PC surface for the PHOTONIS XP85112 (9001332) MCP-PMT with an active area of 53 $\times$ 53 mm$^{2}$. The four plots display the same properties as in Fig.~\ref{fig:QE2d_BINP}.}
\label{fig:QE2d_Phot1332}
\end{figure}

The results of the QE scans across the PC surface are displayed in Fig.~\ref{fig:QE2d_BINP} to Fig.~\ref{fig:QE2d_Phot1332} for four MCP-PMTs. The upper left plot always shows a QE chart of the full PC surface with the integrated anode charge accumulated at the time of writing this document. For a better judgment of the magnitude of the observed QE changes three projections along the x-axis at different positions of y are plotted for each MCP-PMT. The histograms in these plots correspond to different anode charges, from the beginning of the illumination or the time when no QE degradation was observed yet to the highest anode charge. It is obvious that the MCP-PMTs from BINP (\#3548, Fig.~\ref{fig:QE2d_BINP}) and Hamamatsu (R10754X, Fig.~\ref{fig:QE2d_Ham}) show clear QE damages after >1 C/cm$^{2}$. From the QE chart and its projections it appears that the QE degradation starts at the corners (R10754X) or at the rim (\#3548) of the sensor. With progressing illumination the QE drop extends more and more to the inner regions of the PC. After an anode charge of 5025 mC/cm$^{2}$ and 1765 mC/cm$^{2}$ for the BINP and Hamamatsu MCP-PMT, respectively, the QE has dropped by more than 50\% of its original value in certain regions. The situation is different for the ALD-coated PHOTONIS XP85112 (Figs.~\ref{fig:QE2d_Phot1223} and \ref{fig:QE2d_Phot1332}), where basically no QE degradations up to $>$5 C/cm$^{2}$ is visible. Beyond this charge the sensor 9001223 shows the development of some QE damage at the upper left rim, but still at a tolerable level. Starting from $>$6 C/cm$^{2}$ a clear step emerges around x~=~0~mm. This stems from the fact that the right half of the PC (x~$>$~0~mm) of the sensor was covered during the illumination process. The sensor 9001332, which is of the same series as the 9001223 and whose PC was also covered on the right side (x~$>$~0~mm) during the illumination process, shows a small QE degradation only after close to 10 C/cm$^{2}$ anode charge. This is twice the charge needed for the Barrel DIRC. An effect of a slightly rising QE on the left side of the PC (x~$<$~0~mm) at increasing anode charges, as observed for the 9001332 at $\sim$7.5 C/cm$^{2}$ was already observed with another MCP-PMT from PHOTONIS \cite{alex} and usually indicates that the QE will soon begin to drop.

\begin{figure}[htb]
\centering
\includegraphics[width=.49\textwidth]{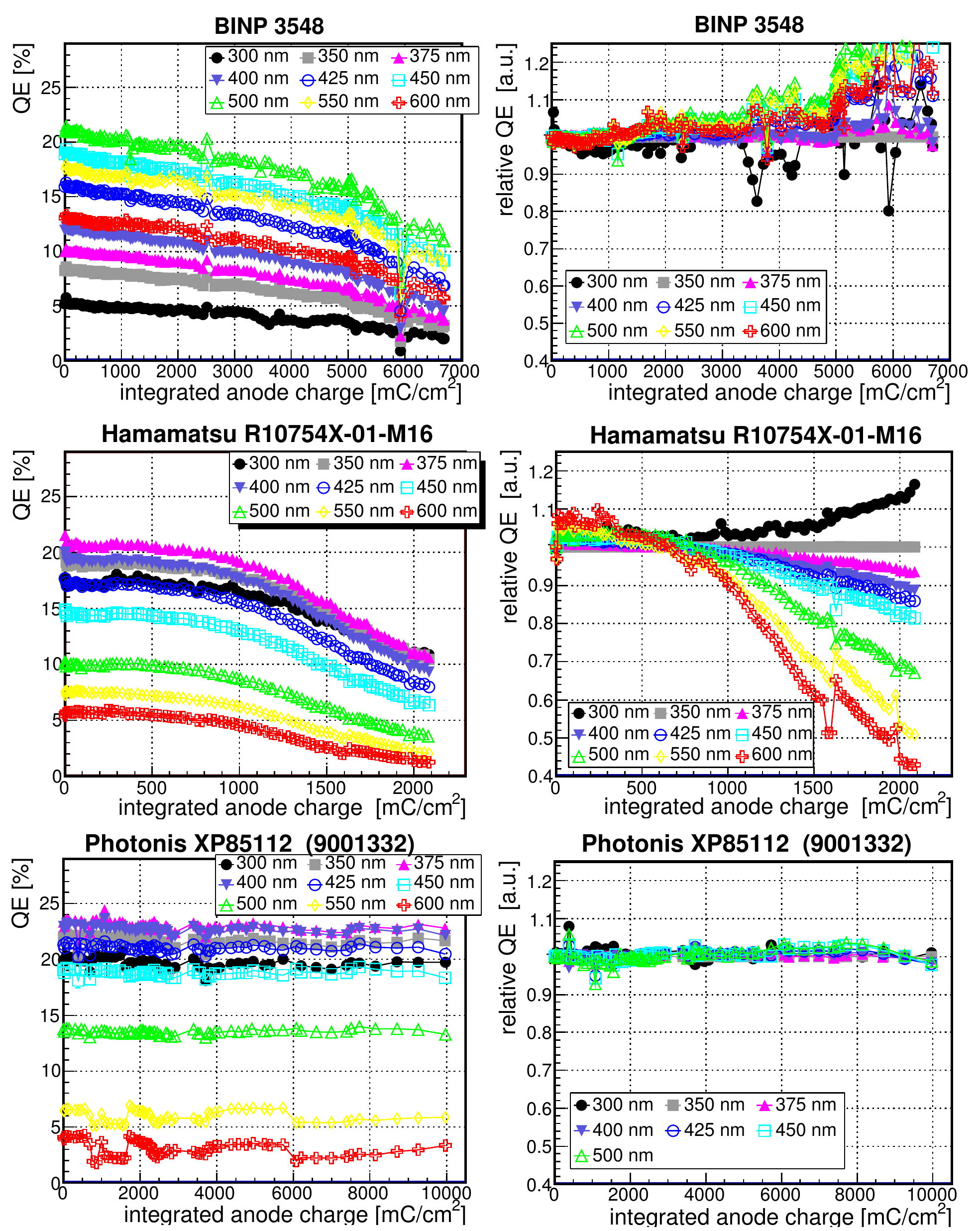}
\caption{QE (absolute and relative to 350 nm) as a function of the integrated anode charge and for different wavelengths.}
\label{fig:QEnm}
\end{figure}

It was reported earlier \cite{nagoya, alex} that the QE degrades faster for red than for blue light. To study the observed wavelength dependence we have measured the spectral QE as a function of the integrated anode charge for all investigated new MCP-PMTs. The results for different wavelengths are displayed in Fig.~\ref{fig:QEnm} for representative samples of MCP-PMTs treated with different techniques to reduce aging. It is obvious from the plots that the MCP-PMTs of the three manufacturers behave differently. While the QE of the Hamamatsu R10754X with a film as ion barrier starts dropping significantly beyond $\sim$1 C/cm$^{2}$ the QE of the BINP \#3548 with its modified PC shows a constant decline up to almost 7 C/cm$^{2}$ while the PHOTONIS XP85112 (9001332) shows practically no QE degradation up to 10 C/cm$^{2}$. A clear spectral dependence of the QE drop is only seen in the R10754X which could point to a change in the work function of the PC, possibly due to rest gas atoms and molecules adsorbed at the PC surface. The displayed BINP 3548 and PHOTONIS 9001332 MCP-PMTs do not exhibit a clear QE dependence upon the wavelength, while the PHOTONIS 9001223 (not displayed, see \cite{fred2}) definitely shows the beginning of a QE wavelength dependence immediately after the QE starts dropping at $>$6 C/cm$^{2}$.

\begin{figure*}[htb]
\centering
\includegraphics[width=.9\textwidth]{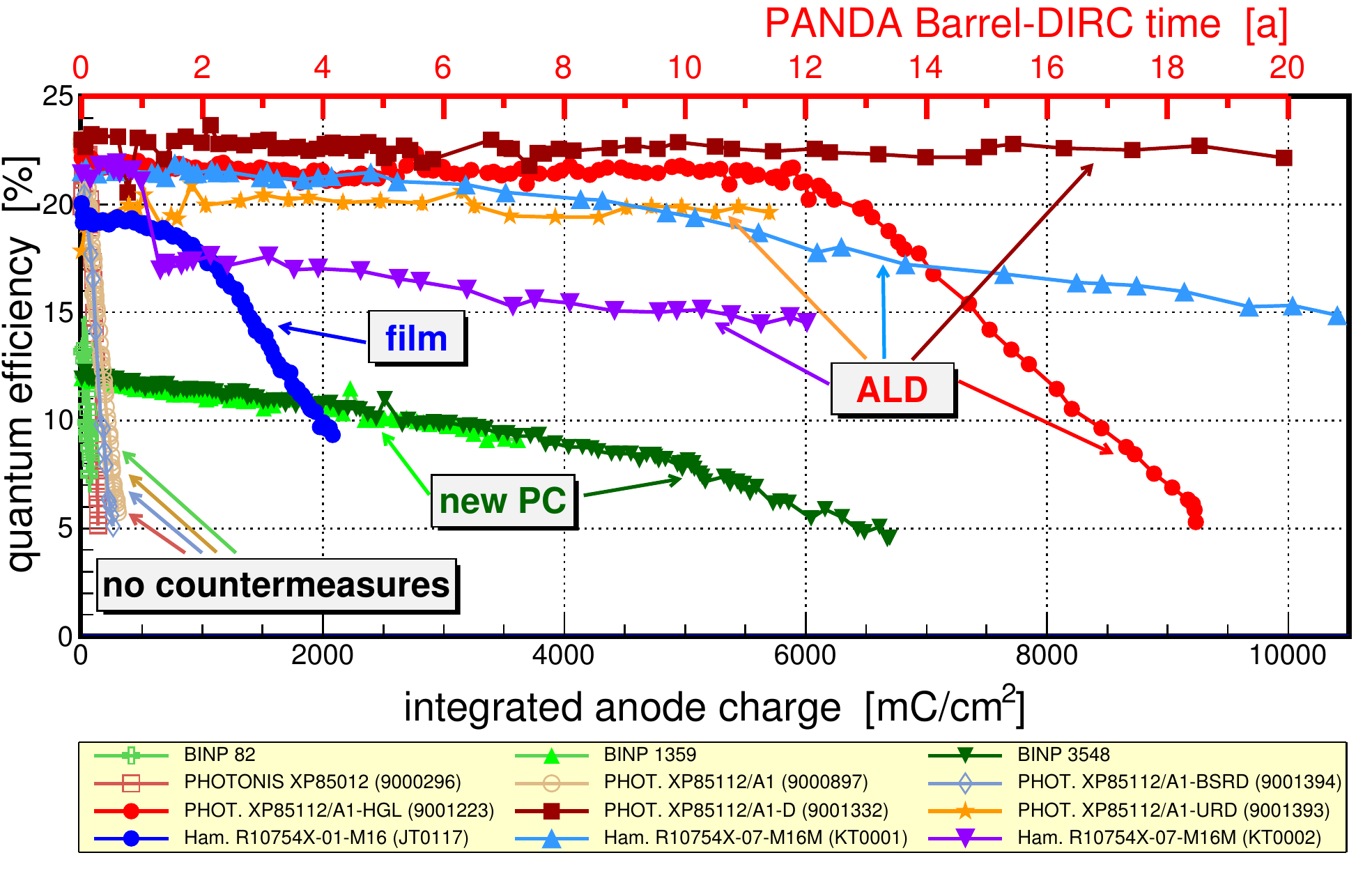}
\caption{Comparison of our MCP-PMT measurements: quantum efficiency as a function of the integrated anode charge at 400 nm.}
\label{fig:QEcomp}
\end{figure*}

Finally, in Fig.~\ref{fig:QEcomp} the QE at 400~nm is compared for all investigated MCP-PMTs. Clearly, the older MCP-PMTs (open symbols at the very left, see also Fig.~\ref{fig:QEcomp_old}) show a fast declining QE which drops below 50\% after $<$200 mC/cm$^{2}$. The situation is very different for the new lifetime-enhanced tubes. The QE of the Hamamatsu R10754X-01-M16 with a protection film is exhausted at $<$2 C/cm$^{2}$, while for the new ALD coated devices (R10754-07-M16M) the QE remains stable up to $>6$ C/cm$^{2}$ accumulated anode charge. The QE of the two BINP MCP-PMTs (\#1359 and \#3548) is continuously diminishing up to $\sim$3.5 C/cm$^{2}$ and $\sim$7 C/cm$^{2}$, respectively. All three new ALD coated PHOTONIS MCP-PMTs show a stable QE up to $\sim$6 C/cm$^{2}$. While for the 9001223 the QE starts dropping beyond 6~C/cm$^{2}$ the QE of the identically constructed 9001332 is still basically unaffected at $\sim$10 C/cm$^{2}$ integrated anode charge. The PHOTONIS 9001393 has a different design with two ALD layers, but also for this MCP-PMT the QE is stable up to at least 6 C/cm$^{2}$. The integrated anode charge of all ALD coated MCP-PMTs corresponds to at least 12 years of running the Barrel DIRC at the highest \panda luminosity.

\subsection{Conclusions}\label{subsec:concl}

Our intensive search for suitable photon sensors for the Barrel DIRC leads to the conclusion that MCP-PMTs are the most appropriate candidates. The tubes with 10 $\mu m$ pore size fulfill the requirements in terms of magnetic field immunity, time resolution, dark count rate, and gain stability at high photon rates. The recently developed techniques to prevent the photo cathodes from aging led to a ``quantum jump'' in the lifetime of these devices. Especially by coating of the MCPs with an ALD technique the lifetime of MCP-PMTs can be extended to $>$6 C/cm$^{2}$, which corresponds to $>$12 years running of the Barrel DIRC at the highest \panda luminosity. Further studies with modified MCP-PMTs in the attempt to extend the lifetime even more are currently ongoing.


\section{Front-End Electronics}
\label{sec:electronics}
\newcommand*{\dirich}{DiRICH\xspace}
\newcommand*{\padiwa}{PADIWA\xspace}
\newcommand*{\sodanet}{SODANET\xspace}
\newcommand*{\trbnet}{TRBnet\xspace}
The front-end electronics (FEE) is the interface between the photon sensors of the \panda Barrel DIRC and the DAQ. It provides the first step of the signal processing. 
Efficient hit detection, while maintaining excellent timing, is the main task of the FEE, even at the anticipated high interaction rates and in the context of the trigger-less \panda DAQ architecture. The stringent demand on a timing precision of $\le$100~ps, 
as required by the reconstruction methods, necessitates some post-processing of the signals, such as signal walk corrections. Therefore, in addition to the timing information, the signal amplitude or charge needs to be measured as well. However, due to the high interaction rate, the amount of data per hit needs to be kept reasonable.

While analog waveform sampling, as implemented for the Belle~II TOP counter~\cite{top_fee_tipp}, provides exceptional capabilities to extract the relevant signal features, a major drawback of this type of FEE is the requirement of an external trigger and the associated dead time for the readout. 

Thus a design solely based on discriminators is proposed for the \panda Barrel DIRC FEE. Their fast processing logic can provide high timing resolution with self-triggering capability. The charge measurement in such a design is achieved by measuring the Time-over-Threshold (TOT) and can be implemented by encoding the TOT of the original pulse in the digital output signal. The low power consumption and low cost of such a design are further important advantages.

\subsection{Time-over-Threshold Signal Properties}
\begin{figure}[!h]
	\centering
		\includegraphics[width=\columnwidth]{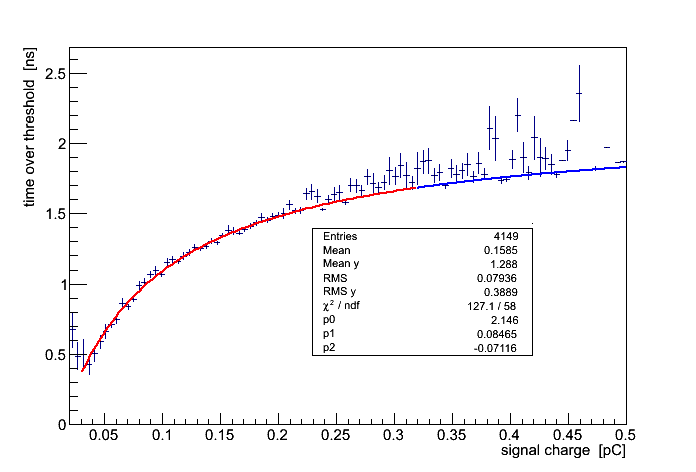}
	\caption{Time-over-Threshold (TOT) as function of measured signal charge for single photo-electron signals of MCP-PMTs. The red line is a fit of Eqn.~\ref{eq:tot_vs_q} to the data points. The blue curve is an extrapolation of the fit function to larger charge values which were not included in the fit.}
	\label{fig:50TotVsChargeFit}
\end{figure}

The TOT-based approach to walk correction of MCP-PMT signals has been studied and validated. Initial studies of MCP-PMT single photo-electron signals~\cite{MC-MCardinali-PHD-THESIS} show the expected non-linear correlations between signal charge and TOT (see Fig.~\ref{fig:50TotVsChargeFit}). The correlation can be described by
\begin{equation}
	\mathrm{TOT}(T) = b \cdot \left( 1 - \frac{2bT}{Q} \right), 
	\label{eq:tot_vs_q}
\end{equation}
where TOT is the Time-over-Threshold, $b$ is the true signal width, $Q$ is the signal charge, and $T$ is the applied threshold level.

\begin{figure}[!h]
 \centering
 \subfloat[Time-walk of single photo-electron signal as function of Time-over-Threshold (TOT). Photo-electrons back-scattered within the MCP-PMT are indicated by the black circle.]{
 \includegraphics[width=0.9\columnwidth]{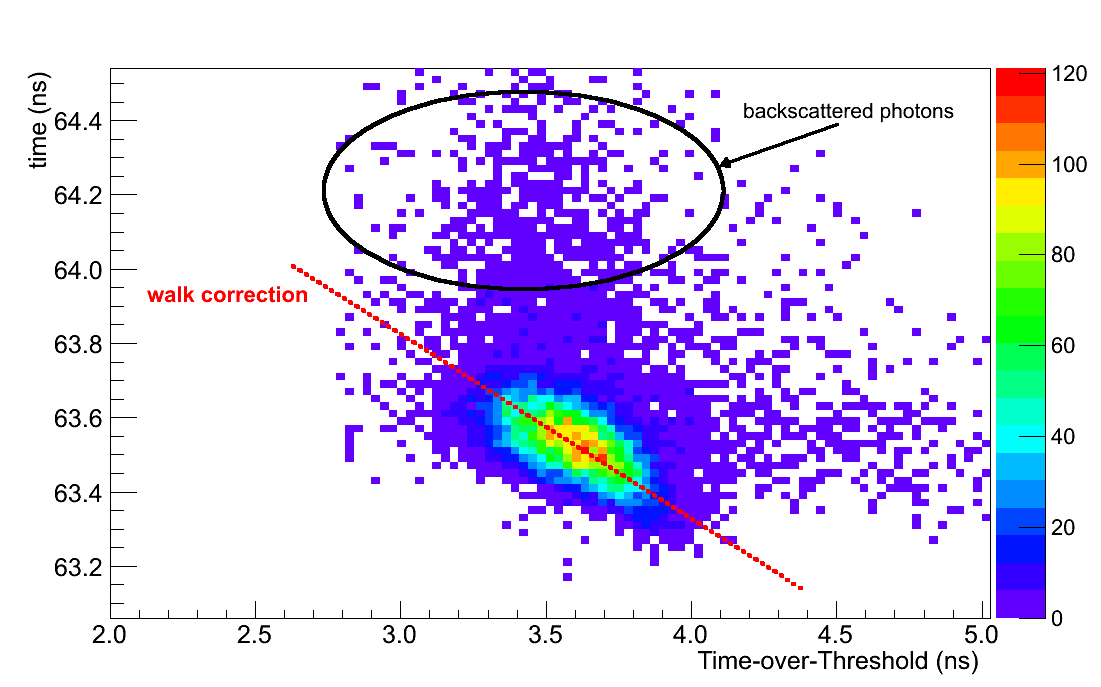}\label{fig:mcp_walk_tot}}
 \hfil
 \subfloat[Front-End Electronics (FEE) timing measured after applying a linear time-walk correction. The peak is fitted with a Gaussian function. The tail to larger values is caused by electrons back-scattered within the MCP-PMT.]{	
	\includegraphics[width=0.9\columnwidth]{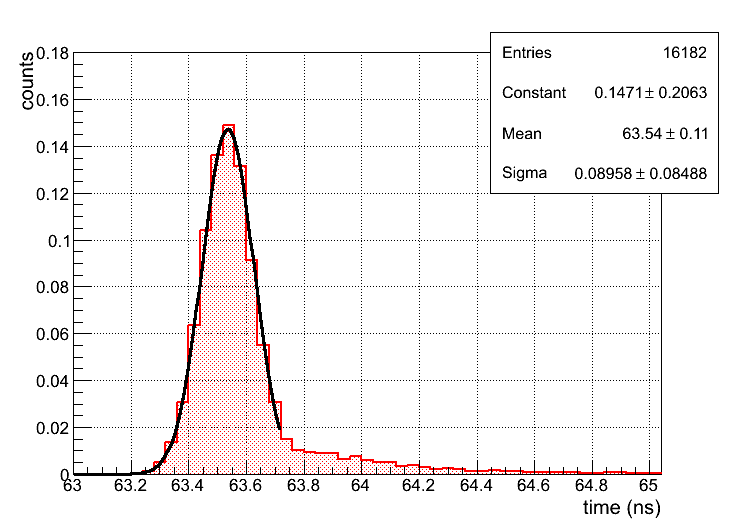}\label{fig:fee_timing}}
 \caption{Time-walk correction of MCP-PMT signals based on Time-over-Threshold (TOT) method. \cite{MC-MCardinali-PHD-THESIS}.}
\end{figure}

However, due to the, approximately, triangular MCP-PMT signal shape, the time walk is expected to exhibit a linear dependence on the TOT. Laboratory studies illuminating single pixels of an MCP-PMT with a fast laser pulser show indeed this kind of behavior (see Fig.~\ref{fig:mcp_walk_tot}). A linear time-walk correction has been successfully applied and a time resolution below 100~ps has been achieved (see Fig.~\ref{fig:fee_timing})~\cite{cardinali_rich, cardinali_tipp}.

\subsection{Requirements}
The FEE of the Barrel DIRC has to register single photon signals with high efficiency and precisely record their arrival time and the signal TOT. The hardware should be capable of processing signals from a large number of channels and must operate in an environment with a large magnetic field of 1-2~T. The radiation dose from ionizing and neutral particles is moderate (see Fig.~\ref{fig:radmap_dirc}). As derived in Sec.~\ref{subsubsection:rate} the maximum hit rate requirement is 180~kHz per pixel, respectively, 200~kHz per pixel with a 10\% safety margin included. This poses a challenge to the digitization stage of the analog signals as well as to the data transmission technology which has to handle and merge these channels.  

A special challenge arises from the fact that \panda will operate in trigger-less mode to ensure high flexibility for physics event selection. This means that the entire DAQ has to run continuously, transmit and pre-process data, and extract hit patterns to be able to provide PID information for online event filtering. A very important requirement for event building and reconstruction is the synchronization of the clocks of all sub-detectors. This is provided by the \sodanet~framework~\cite{sodanet-ref}.


The FEE has to provide calibration, monitoring and slow control functionality, e.g. setting thresholds or monitoring temperatures. Last, but not least, it has to cope with the tight volume constraints and low power consumption demands. 

\subsection{FPGA-Based Readout}
\label{subsec:fpgareadout}

\begin{figure}
    \centering
    \subfloat[The TRB3 is equipped with a central FPGA and four peripheral FPGAs near the corners.]
    {\includegraphics[width=.45\textwidth]{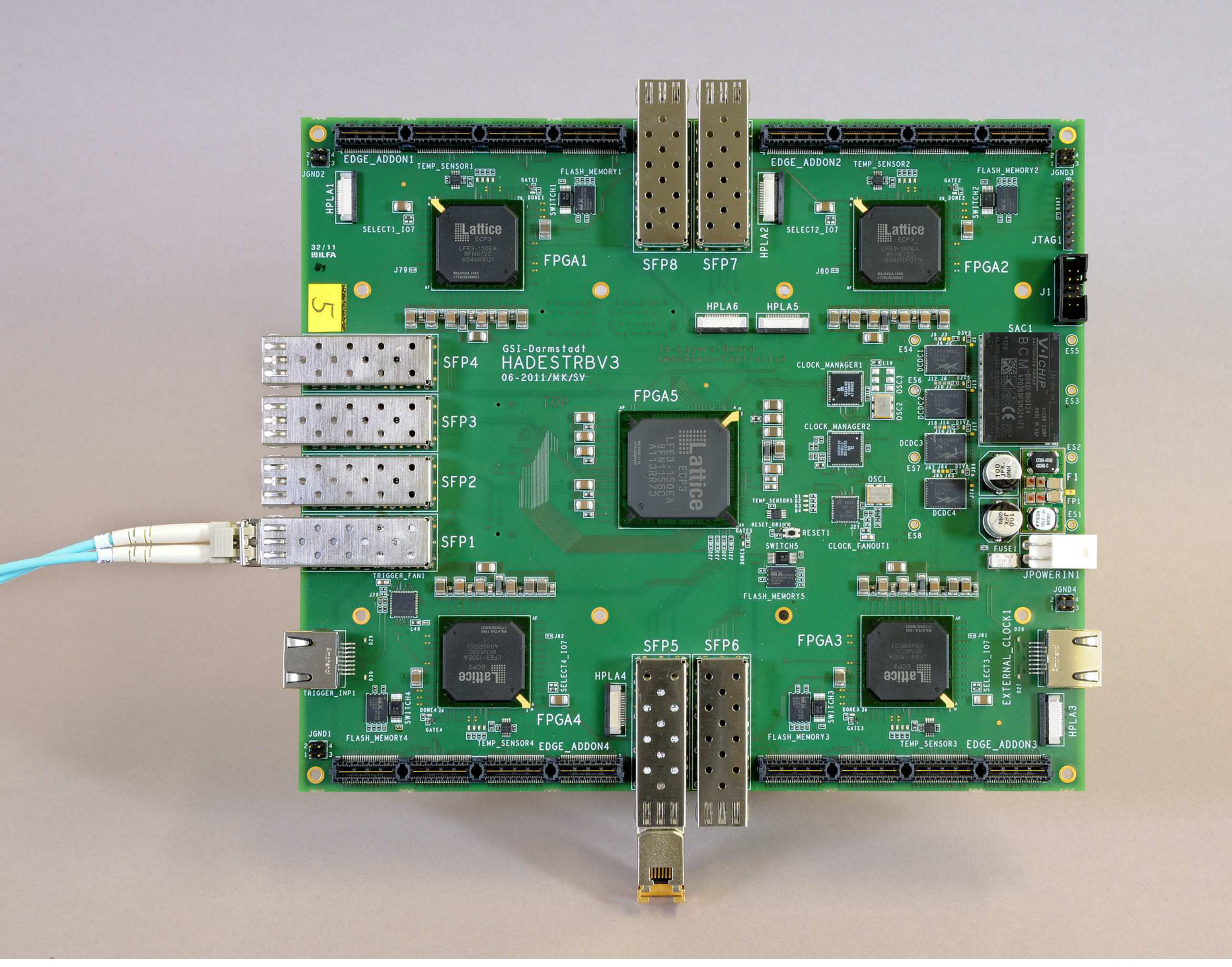}\label{fig:trb3}}\hfill
    \subfloat[The PADIWA front-end boards have connectors that plug in directly to the photon sensors.]
    {\includegraphics[width=.45\textwidth]{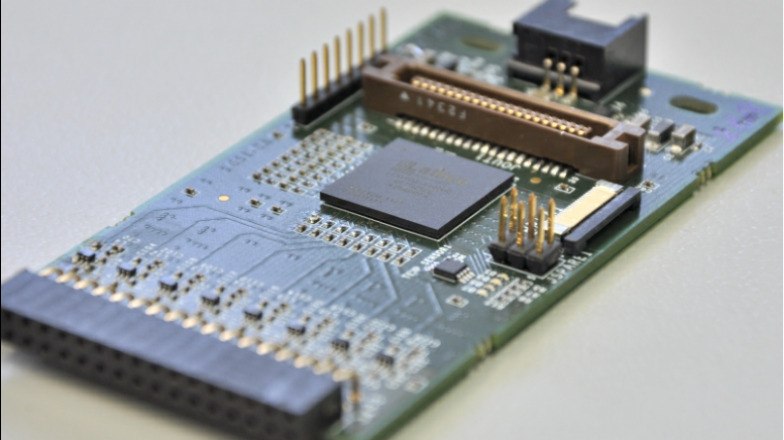}\label{fig:padiwa}}
    \caption{\panda FEE with TRB3 readout electronics~\cite{neiserC}~(a) and PADIWA frontend boards~(b).}
\end{figure}

The aforementioned requirements have led to the use of FEE designs based on FPGAs. The {\sl Time-to-Digital Converter Readout Board} (TRB3)~\cite{neiserC, comp-trb3-jinstC} is an advanced version of the {\sl Trigger Readout Board} (TRB2)~\cite{trb} that was originally developed for the {\sl High-Acceptance Dielectron Spectrometer} (HADES) experiment. Contrary to the earlier version, the {\sl time-to-digital converters} (TDCs) of the TRB3 are implemented in FPGA logic rather than in {\sl application-specific integrated circuit} (ASIC) hardware. Therefore, the TRB3 is equipped with four peripheral FPGAs, as shown in Fig.\autoref{fig:trb3}.
Each of them can be configured to provide $64$ TDC channels, resulting in $256$ TDC channels for one board. The TDC uses the FPGA's clock signal as coarse counter. The 200~MHz clock is equivalent to a time binning of 5~ns. The fine time is measured with the {\sl tapped delay line  method}~\cite{trb3-jinst}. It utilizes the propagation time of a start signal within a chain of $1$-bit full-adders. The hit signal flips one output after another from $1$ to $0$ within the chain until the TDC clock pulse latches the chain. This method relies on dedicated calibration of each TDC channel to mitigate non-linearity effects due to varying delay lengths between the full-adders of a chain.
The time precision achieved is better than $\approx$~10~ps~RMS for all the channels \cite{comp-trb3-jinstC}. The TDCs are capable of detecting multiple hits, with a maximum hit rate of 16.7~MHz. However, this is limited to bursts of $63$ hits. The TDCs expect {\sl low-voltage differential signaling} (LVDS) pulses on their inputs. The leading edge of the LVDS pulses can be measured with high precision, which is very important to determine the time-of-propagation of the Cherenkov photons in the \panda Barrel DIRC. The TOT measurement can be used to reduce the timing jitter.

The FPGA in the center of the board (see Fig.~\autoref{fig:trb3}) collects the event data from the peripheral FPGAs and combines them into packets for transmission via optical links. The current version of the readout board supports up to eight optical gigabit Ethernet links with a transfer rate of up to 3.2~Gbit/s for data transmission and slow control. In addition, the links can be configured to distribute a trigger signal with low latency in a setup with multiple readout boards, and thus capable of interfacing to \sodanet. Remote configuration, control and monitoring of the boards is based on the \trbnet framework~\cite{trbnet}.

This hardware has been successfully deployed in numerous laboratory and test-beam setups. The current FPGAs of the ECP3 family of Lattice semiconductor are cost efficient and very flexible in terms of development and redesign processes which is an important advantage.

Dedicated FEE boards equipped with discriminators -- called {\sl \panda DIRC Wasa} (\padiwa), Fig.~\autoref{fig:padiwa} -- were designed to digitize the analog pulses of MCP-PMTs. They have $16$ separate channels that generate fast LVDS signals where the pulse width depends on the signal TOT. The front-end boards are directly plugged on the output pins of the photon sensors in order to minimize noise pick-up by long wires in the analog signal chain. Twisted pair cables, carrying the digital (LVDS) signals, are used to connect the \padiwa cards with the TRB3. This combination was used for instrumenting the \panda Barrel DIRC prototypes.

\subsection{Radiation Hardness of the FPGAs}
\label{subsec:FEE-RadiationHardness}

The Barrel DIRC readout will be located within the \panda TS, which requires the hardware to function in harsh radiation environments. Even though the FEE is reachable for maintenance by opening the rear doors of the solenoid magnet, which is possible also in the in-beam position, the necessity of intervention has to be kept as small as possible. Thus the electronics and material have to be chosen adequately. The radiation environment in the Barrel DIRC has the potential to cause malfunctions in the FPGAs used for the readout. Ionized particles can cause two types of effects:
\begin{description}
    \item[Single event upsets]
        A {\sl single event upset} (SEU) occurs when an ionizing particle
        deposits its energy within the semiconductor, which leads to a 
        transient pulse in logic or support circuitry, or leads to a 
        bit flip by electric charge reallocation within a memory cell or register.
        The minimal energy which is needed for such a single event upset
        is called {\sl linear energy transfer threshold} (LET$_\textrm{th}$).
        This threshold depends on the amount of charge stored in a memory cell.
        Greater capacitance within the cells improves the radiation hardness,
        while the power consumption and the circuit times become worse. Single
        event upsets can lead to malfunctions in the instruction code registers
        as well as in the data registers. Single event upsets are soft errors.
    \item[Single event latchups] A {\sl single event latchup} (SEL) is the
        creation of a low-resistive path between the connections of a parasitic
        circuit element in a semiconductor. A resulting current can destroy the
        device by overheating. Single event latchups are hard errors. The reasons for this fatal error are known and eliminated in FPGAs since many years \cite{fpga_radhard}.
\end{description}
Single event upsets are the dominant factors which limit the application of FPGAs (but also of ASICs) in environments with high radiation exposure like \panda\@. Common but cost-intensive techniques to mitigate single event upset effects are {\sl triple module redundancy} and {\sl bit stream repair techniques}~\cite{fullerC}.

For the first method the logic is implemented redundantly within three independent blocks. Single event upsets can be detected as soon as one of the
blocks differs from the others. The failure can be fixed by reloading the affected block. It would be challenging or even impossible to implement a tapped delay line method with triple mode redundancy due to the different delays between the chains. Hence, it probably cannot be used for a TDC\@.

Bit stream repair techniques make use of the possibility to read out and reload the configuration of an FPGA without interrupting operation. One possibility is to reload the FPGA configuration frequently in order to correct potential single event upsets in the instruction logic without detecting them. Another alternative is to read out the registers and reload parts which are corrupted by single event upsets. The ECP3 FPGAs, mounted on the TRB3 boards, are equipped with a soft error detection on board. This feature is disabled at the moment and will be investigated in future tests. However, the disadvantage in both cases is that errors in the data registers are not recognized. Corrupted data can not be excluded.

In an alternative third method, which is implemented in the Lattice-FPGAs the FPGA is continuously comparing the data in the flash and the SRAm of the FPGA. If a difference is detected, the FPGA can be restarted occasionally. The flash will be reloaded to the SRAM. The boot process of the FPGA on the \dirich takes only in the order of $10^{-3}$~seconds.

Radiation hardness tests with the TRB3 FPGAs were performed by the CBM/HADES collaboration \cite{M.Traxler-privC-radH}, where 5$\times$10$^6$~ions/s impinged on a target resulting in an interaction rate of about 1\%. Each event had on average 200 charged reaction particles distributed over the detection surface of 3~m$^2$ or 30000~cm$^2$. The applied rate $R$ of charged particles passing each FPGA can be calculated as: 
\begin{equation}
R= \frac{5\times10^6 \cdot 0.01  \cdot 200}{3\times10^4} cm^{-2}s^{-1}= 333 cm^{-2}s^{-1}.
\end{equation}
Since the systematic error of these estimations is big and since the size of the FPGAs is 1~cm$^2$ we utilize $R=R_{FPGA}\approx300~FPGA^{-1}s^{-1}$.

Under these conditions one single event upset (SEU) happened every 5~hours among the 500~FPGA subjected to this irradiation, i.e. $N_{error}^{measured}$=0.2 relevant errors measured. 

In only about 1\% of all cases a SEU produces a relevant and visible error. Thus, with a neutron cross section of 2$\times$10$^{-14}$cm$^2$/bit (Xilinx ug116, 90~nm Virtex 4) and with a 4~Mbit-memory/FPGA
the number of estimated errors per hour $N_{error}^{estimated}$ is
 \begin{equation}
\begin{split}
N_{error}^{estimated}= 300cm^{-2}s^{-1} \cdot 500 \cdot 2\times10^{-14}cm^2 &\\
\cdot 0.01 \cdot 4\times10^6 \cdot 3600s = 0.43 
\end{split}
\end{equation}
Thus there are $\approx0.4$ relevant errors caused by SEUs per hour, which is roughly compatible with the measured value.
The error of these calculations is very big. The state of the art FPGAs do not have anymore a 90~nm technology. 
The ECP3 on the TRB~3 has 65~nm, the ECP5UM, which sits on the \dirich, 40~nm. However, it is not clear if the size is an advantage. 
A transistor being smaller is expected to be hit more rarely, yet the impact might be more severe then.

The development in the framework of the \dirich project continues, including the investigation of mitigation techniques.

\subsection{The \dirich System}

The successor of the \padiwa/TRB3 solution for the FEE is the \dirich, which is a cooperation of the \panda DIRC, CBM RICH, and HADES RICH groups. The goal of this development is to increase the level of integration and to avoid, as much as possible, the use of cables. Those can act as antennas that introduce noise into the system and take a lot of space in the setup. In the \dirich configuration the readout card connected to the photon sensor carries the discriminator and the TDC as well. All basic concepts have been tested and are validated. Optimization was performed leading to a prototype.

\begin{figure}[htb]
\centering
\includegraphics[width=.49\textwidth]{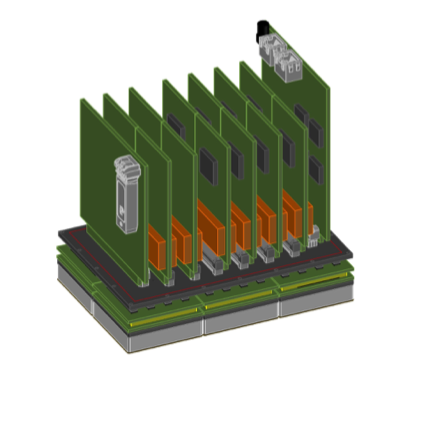}
\caption{Schematic of a  single MCP-PMT readout module for 6 MCP-PMTs on a common back plane: 12x \dirich FPGA-TDC front-end modules, DCM (left), and PM (right).}
\label{fig:DiRICH-setup}
\end{figure}

The front-end part of the HADES RICH readout chain is shown in Fig.~\ref{fig:DiRICH-setup}. It consists of modules capable of reading out 6~MAPMTs each, which are plugged into a common PMT carrier PCB. Each such module will be mounted on an aluminum frame structure, where the PCB provides mechanical fixation for the PMTs. The \panda Barrel DIRC will require a few modifications to adjust the footprint to the MCP-PMT layout on the expansion volume.

The PMT carrier PCB also serves as back plane for the FEE, which is plugged in directly from the backside. The back plane provides all necessary data-, power- and clock interconnects between the readout modules, minimizing the amount of cable connections to the modules.

\begin{figure}[htb]
\centering
\includegraphics[width=.49\textwidth]{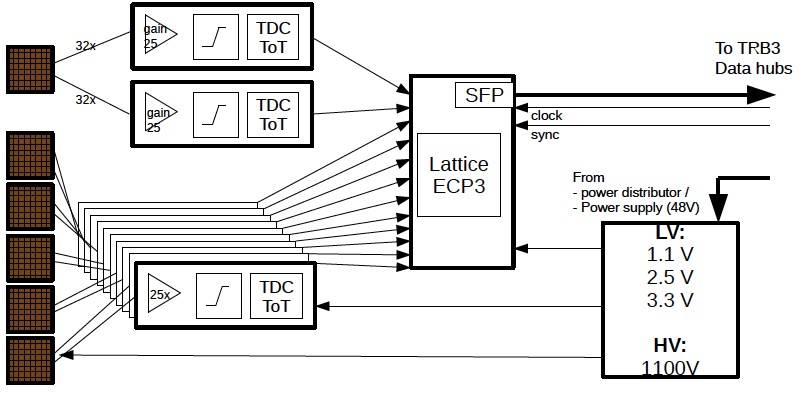}
\caption{\dirich read out concept}
\label{fig:DiRICH-readout}
\end{figure}

The FEE, that is plugged directly into the PCB back plane, comprises three different types of boards for (a)  analog signal handling and digitization, (b) data handling, and (c) power supply/distribution. The \dirich concept is shown in Fig.~\ref{fig:DiRICH-readout}.

The \dirich FPGA-TDC board provides an analog preamplifier, discriminator and TDC for 32 individual input channels. Each MCP-PMT (64 channels) is read out via 2 such boards. The analog inputs are galvanically decoupled from the MCP-PMT using SMD wide-band transformers for each channel, to avoid unwanted ground loops.

The characteristics of the single-ended analog MCP-PMT signal are a width of $\approx$~2~ns FWHM and a mean amplitude for single-photon signals of 8~mV on 50~Ohm and 0.1~pC with a large variation in signal amplitude.
Before discrimination, the analog MCP-PMT signal is amplified by a factor of ($\approx$~25) using a $\approx$~4~GHz-band-width transistor amplifier stage. Signal discrimination is implemented using the input comparators of the LVDS line receivers of the Lattice ECP5UM FPGA. The reference threshold voltage is generated individually for each channel using pulse width modulation with subsequent filtering by the FPGA itself.
The same FPGA hosts 32+1 FPGA-TDC channels re-using the design of TRB3 (tapped delay line approach, 200~MHz coarse counter), digitizing the leading-  and trailing edge of the discriminated analog signal to measure both signal arrival time and Time-over-Threshold, which is used for amplitude measurement.

Each time-stamp (leading and trailing edge) is decoded as a single data word (4~byte) using the \trbnet data format, and the data from all 32~channels are sent out via a common 2~Gbit/s serial link utilizing the \trbnet protocol routed through the back plane. A readout logic with matched readout window is used to implement a quasi self-triggered readout scheme.

The main purpose of the Data Combiner Module (DCM) is to combine the data from all 12~\dirich FPGA-TDC cards, 
mounted on a 6-PMT readout module, and to transfer the data via a single output link. 
It is based on a Lattice ECP3 FPGA, which is connected via individual 2~Gbit/s LVDS SERDES 
(Serializer-Deserializer) links to each of the  FPGA-TDC cards via the back plane. Data, in \trbnet data 
format, are sent out using a 13th 2~Gbit/s SERDES, connected to an optical Small Form-factor Pluggable
(SFP) transceiver on the board. Without hardware modification, the link speed of the output 
link can be increased to 2.4~Gbit/s (by increasing the basic clock from 200~MHz to 240~MHz). 
A further development of the DCM provides faster output link capability e.g. 2$\times$5~Gbit/s output 
links by utilizing new FPGAs, such as the Lattice ECP5UM5G FPGA, or the Kintex FPGA.

In case radiation hard links are necessary, instead of the SFPs we consider using radiation hard 
link technology developed at CERN for the LHC \cite{RadHard-FEE-CERN}.
The link consists of a radiation tolerant ASIC (GBTX) \cite{RadHard-GBTX-CERN} and opto-electronic 
components (Versatile link) \cite{RadHard-VL-CERN}.
This technology can be used to implement multipurpose high speed (up to 5~Gbit/s user bandwidth) 
bidirectional optical links, operable in radiation levels of 100~Mrad (1~MGy) \cite{RadHard-FEE-Test}.
Due to the modular structure of the DiRICH, a DCM board with radiation hard link components could 
be accommodated without changes to the other FEE boards. 

The DCM accepts external clock- and trigger/synchronization signals, and thus is capable to connect to \sodanet, which are distributed via LVDS fan out chips to each \dirich via individual clock/trigger lines located on the back plane.

The DCM also implements slow-control functionality, controlled via the \trbnet protocol on the output link, and can power-off/reboot individual cards via Power-enable lines on the back plane of the module.

The Power Module (PM) provides all Low Voltage (LV) DC power rails of 1.1V/1.2V/2.5V/3.3V from external cable connections to the back plane for distribution to the individual cards on the modules.
The PM provides active voltage measurement of the externally provided supply lines, allowing for a coarse regulation of the supply voltages on a remote Power Supply Board (PSB). It also provides current monitoring for each supply line.
There are fairly large current requirements on the low voltage lines with $\approx$~14A on the 1.1V-line and 3.5A on the
 1.2V-line, which supply all 12~\dirich modules on one 3x2 MCP-PMT module. Therefore, we consider the use of on-board DC/DC converters and a single 48~V supply to the PM, to reduce copper requirements for the supply lines.
Use of these DC/DC converters is optional, and the PM allows to bridge these and revert to the individual external LV supply lines.

In addition, the PM can serve as HV interface, distributing the HV supply 
via a common HV supply line on the back plane to each of the 6~PMTs. A special SAMTEC ERM8 back plane connector is used to allow for safe HV connection to the back plane.
 
\subsection*{Cooling of the FEE}

The highly-integrated FEE design produces a significant amount of heat in the compact readout unit.
The components of the highly-integrated \dirich system use a power of approximately 500~W for the entire Barrel DIRC 
while the solution based on PADIWA cards and TRB3 boards would use approximately 2~kW power.
The heat generated by the FEE will be extracted by a water cooling system. Space for the required supply lines are included in the mechanical design of the 
readout unit (see Sec.~\ref{sec:mech-readout}).

\section{Data Acquisition}

\subsection*{Data Rate Estimate}

The data rate estimates are based on the single photon hit rate per readout pixel induced by 20~MHz pbar-p collisions. A rate of 200~kHz/pixel is expected from the photon detector (see Sec.~\ref{subsection:sensors:requirements}). 

For the estimation of the data rate one has to take into account the present \trbnet decoding format, with 4 bytes per single time-stamp. A hit consists of a leading and a trailing edge time-stamp. In addition, some data overhead must be included, e.g. for synchronization messages (4~bytes) in the \trbnet data stream. Thus, 12~bytes per pixel hit are considered for further data rate estimates.

A single \dirich module with 32 channels and a hit rate of 200~kHz/pixel, produces a data rate of 200~kHz $\times$ 32 $\times$ 12~byte $\approx$~77~MB/s. It is connected via a 2~Gbit/s serial link to the DCM, capable of a maximum of 150~MB/s effective data rate.
The selected MCP-PMT has 64 channels, thus requiring 2 \dirich modules producing a data rate of 155~MB/s. Each of the 16 sectors contains eleven MCP-PMTs so that the data rate per sector is about 1.7~GB/s and the total Barrel DIRC data rate amounts to 27~GB/s.

An upgraded DCM with two 5-Gbit/s links would provide adequate throughput to cope with the expected data rate. Furthermore, the original PMT-backplane PCB design can be adapted to carry only up to four MCP-PMTs (620~MB/s) thus avoiding to overload the data link.

\begin{table*}[htb]
\setlength{\tabcolsep}{6pt} 
\renewcommand{\arraystretch}{1.5} 
\caption{Estimation of the Barrel DIRC data rate and corresponding cables links to the DAQ system.}
\ \

\label{Tab:data-rate}
{\begin{tabular*}{0.95\textwidth}[]{@{\extracolsep{\fill}}ll}
\hline
Average number of photon hits/event&	50 hits/event\\
Average photon hit rate per pixel &	200 kHz\\
Number of readout channels	& 11264 \\
Bytes/hit (leading+trailing / TOT)&	8 data + 4 overhead = 12 Bytes/hit\\
Data rate per MCP-PMT & 155~MB/s \\
Data rate per sector (11 MCP-PMTs) & 1.7~GB/s \\
Total data rate \panda Barrel DIRC &	27~GB/s\\
Number of fiber links from FEE to DPB &	96   (4.8 Gbit/s)\\
Number of 100 Gbit links from DPB to Compute Nodes&	4 \\
\hline
\end{tabular*} 
}
\end{table*}

Each of the Barrel DIRC sectors would then require six optical fiber output links bringing the total number of links to 96.
These links will be connected to the Data Processing Boards (DPB), where the data is aggregated to a smaller number of links to the compute nodes 
of the event builder. Proper distribution of input links to the DPBs will allow for a fairly homogeneous link utilization of 100~Gbit/s output links. 
Therefore, four 100~Gbit/s links to the compute nodes (see Tab.\ref{tab:M-AG-cable_cross_section}) will be sufficient to 
handle the estimated data rates, including a safety margin. The number of output links per DPB can be fairly flexibly adapted to adjust for increasing 
data rate requirements since these are well accessible outside the detector volume. A summary of the data rates can be found in Tab.~\ref{Tab:data-rate}.

\subsection*{DAQ System}
\begin{figure}[h]
	\centering
		\includegraphics[width=0.8\linewidth]{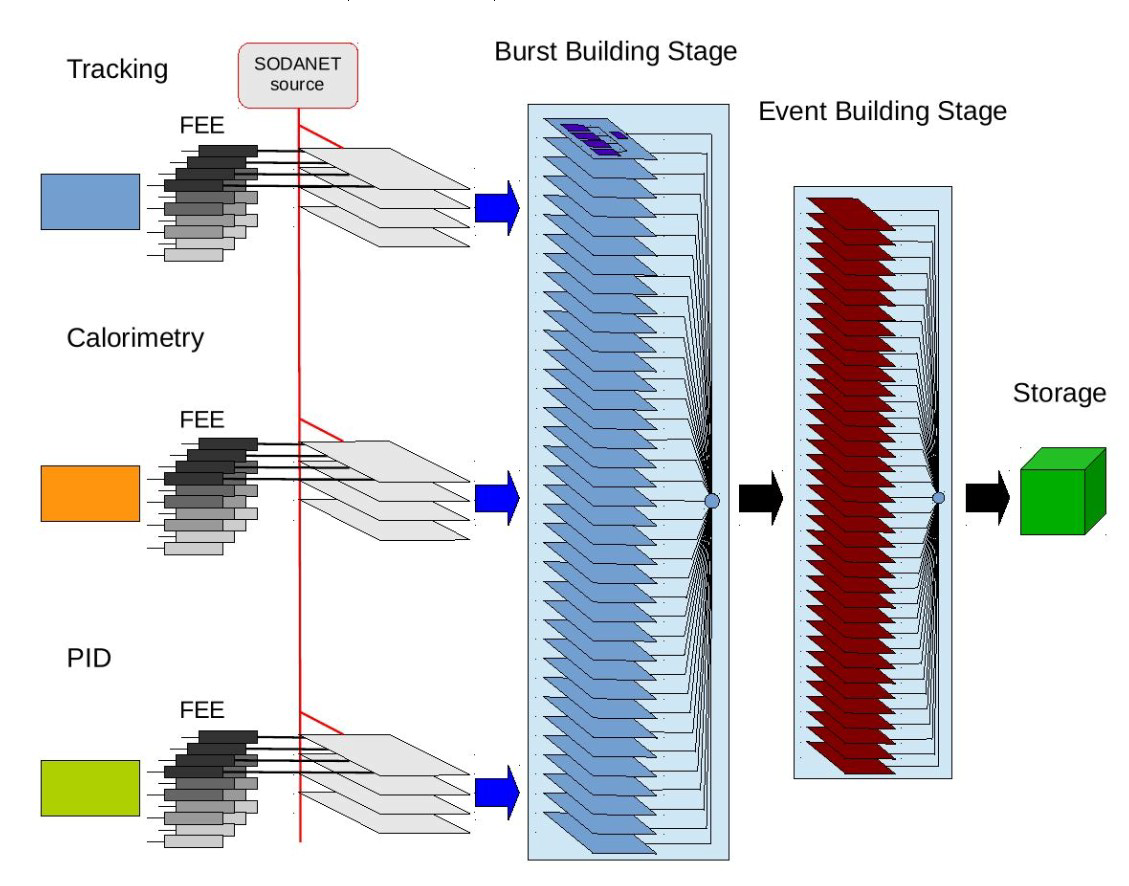}
	\caption{Schematic data flow of the \panda DAQ system and connection to \sodanet \cite{talk:daq}.}
	\label{fig:panda_daq_scheme}
\end{figure}

The \panda DAQ architecture (see Sec.~\ref{sec:daqbrief}) relies on precise time-stamping of detector information already in the FEE. Data from the numerous FEE modules are aggregated into fewer Data Concentrator (DC) modules. Based on the time-stamps the detector information is assembled into events. This is facilitated by exploiting the HESR beam structure (see Sec.~\ref{sec:hesr}) and grouping events into bursts (see also Fig.~\ref{fig:panda_daq_scheme}).
The reconstructed events are filtered online in a computing farm, allowing to simultaneously pursue a variety of  physics topics. The events passing the online filters are committed to storage.

Precise time-stamp generation and distribution within the \panda detector is provided by \sodanet (Synchronization Of Data Acquisition NETwork)~\cite{soda}. The final FEE has to be compatible with the \sodanet protocol. However, the \dirich uses a similar system already so that switching to \sodanet can be achieved by adapting the firmware and does not require adding new functionality.

Another important requirement imposed by the \panda DAQ architecture is the data reduction in the FEE by feature extraction and zero suppression. The Barrel DIRC, however, does not require zero suppression since it has a clear signal characteristic due to the MCP-PMT properties (see Sec.~\ref{ch:mcp_pmts}). Furthermore, data reduction is already implemented through the TOT implementation.

Processing of the recorded hits for the online event reconstruction requires the application of walk correction and time offset parameters that are stored in a calibration database which is part of the DAQ system~\cite{panda_tpr}. Subsequently the pattern recognition and PID can be performed. However, input from other detector systems is necessary for this step, most importantly the reconstructed particle track. The PID information is then combined with other detector information for event selection.

\section{Detector Control, Monitoring and Calibration} \label{sec:calibration}

\subsection{Detector Control System}\label{subsec:dcs}
The Detector Control System (DCS) provides control and monitoring for each detector system in \panda.

These Slow Controls (SC) of each \panda sub-detector are planned to be monitored and controlled from a common supervisory software application based on the Experimental Physics and Industrial Control System (EPICS)~\cite{EPICS}. 
 The quantities controlled and monitored to ensure safe and optimal operation of the Barrel DIRC are summarized in Tab.~\ref{Tab:dcs_parameters}.

\begin{table*}[htb]
\setlength{\tabcolsep}{6pt} 
\renewcommand{\arraystretch}{1.5} 
\caption{Barrel DIRC DCS parameters}
\ \

\label{Tab:dcs_parameters}
{\begin{tabular*}{0.95\textwidth}[]{@{\extracolsep{\fill}}lllll}
\hline
Component   & Monitor Item       & Location         & Number              & Alarm type     \\
\hline
\multirow{4}{*}{Electronics} & Temperature        & On FEE boards & 352 (2/board stack) & Software alarm \\
            & Low Voltage        & Power supply  & 176 (1/MCP-PMT)     & Software alarm \\
            & Current            & Near detector & 176 (1/MCP-PMT)     & Software alarm \\
            & Temperature        & On DCM boards & 32 (2/sector)       & HV interlock   \\ 
\hline        
\multirow{6}{*}{MCP-PMT}     & High Voltage       & Power supply  & 176                 & Software alarm \\
            & Current            & Near detector & 176                 & HV interlock \\
            & Bkgd rate / PMT    & Counting room & 176                 & HV interlock   \\
            & Bkgd rate / Sector & Counting room & 11                  & HV interlock   \\   
            & Integrated charge  & Counting room & 176                 & Software alarm \\
            & Laser hit rate     & Counting room & 176                 & Software alarm \\
\hline 
\multirow{2}{*}{Radiator}    & Temperature        & Radiator box  & 96 (2/bar)      & Software alarm \\
            &                    &               & or 32 (2/plate) & Software alarm \\
\hline 
\multirow{4}{*}{N$_2$ gas}      & Flow rate          & Inlet \& Outlet  & 64 (1/sector)       & Software alarm \\
            & Pressure           & Inlet \& Outlet  & 64 (1/sector)       & Software alarm \\
            & Temperature        & Inlet \& Outlet  & 64 (1/sector)       & Software alarm \\
            & Dew point          & Inlet \& Outlet  & 64 (1/sector)       & Software alarm \\
\hline 
\multirow{2}{*}{Laser}       & Intensity          & Far from detector & 1                   & Software alarm \\
            & Temperature        & Far from detector & 1                   & Software alarm \\
\hline
\end{tabular*} 
}
\end{table*}

All the DCS hardware for the Barrel DIRC is based on components-off-the-shelf 
(COTS) modules and industrial equipment.

 The individual HV, needed for each of the 176 MCP-PMTs, can be 
provided by commercially available multichannel power supply modules, hosted in crates located outside the beam area.
 Readily available HV modules featuring voltage setting resolutions of a few mV with ripple less than 10~mV peak-to-peak and current measurement resolutions
of a few nA, are well-qualified for the MCP-PMTs.
 Industrial crate controllers capable of using EPICS are also commercially available by various vendors such as iseg, CAEN and others. Such systems have been used successfully during several test beam campaigns. 
 
 Low voltage power supplies regulating the power to the FEE need to be located 
 as close as possible to the detector in racks  dedicated to the Barrel DIRC. 
A multichannel and modular approach, similar to the HV, is foreseen 
for the control and monitoring of the voltages, currents and
on-board temperatures for all FEE boards. 
The values of these low voltage levels will be in the range of 1-48 V and currents 
 can reach up to several Amperes.  
 While the actual values will be known upon completion of the \dirich FEE, 
commercially available low voltage power supplies are already being 
investigated, as well as  the feasibility to re-use in-house (at GSI) built 
power supplies in case the industrial equipment is not capable of standing the conditions of the 
\panda detector environment regarding the magnetic field and the radiation dose.
 
Environmental parameters, such as temperature and humidity, will be monitored by standard commercial devices at different locations inside the Barrel DIRC volume. 

To ensure that the bars are maintained in a low-moisture environment, dry nitrogen gas from liquid nitrogen boil-off will flow through each box. The gas will be monitored for humidity and filtered through a molecular sieve and mechanical filters to remove particulates. A part of the input N$_2$ gas leaks from the bar boxes and keeps also the bar box slots in the mechanical support structure free of condensation.

The EPICS-based \panda DCS features a supervisory layer where the Barrel DIRC-specific 
implementation will provide control, monitoring and archival functions for all parameters, 
including automated actions upon warnings and alarms.

\subsection{Laser Monitoring System}\label{subsec:monitoring}

The performance of the photon sensors and readout electronics 
will be evaluated by a Laser Monitoring System (LMS), based
on a laser pulser, such as the PiLas~\cite{pilasC} PiL040,
which produces 405~nm photon pulses with a trigger jitter of 
less than 30~ps.
The laser pulser, beam splitter, and a calibrated photon detector 
for monitoring the pulser intensity, will be located outside the \panda 
detector area, in a temperature controlled dark box. 
The light will be distributed by optical fibers to the 16 sectors
and coupled via diffusers into each prism to illuminate the entire 
readout plane.
Measurements of the photon hit time and time-over-threshold 
provide a calibration of the individual channel time delays and 
gain values.

The laser pulser will be operated at low intensity with per-pixel
hit probabilities below 10\%, corresponding to the single photon mode.
With a tunable trigger rate of up to 1~MHz, dedicated calibration runs 
are expected to take less than one minute.
A similar system, using the PiLas PiL040SM unit, was successfully
used for the prototype calibration during several test beam campaigns 
at GSI and CERN.

\subsection{Calibration and Alignment}

\subsubsection*{Time Calibration}
\label{subsec:caltime}

Time offsets between pixels, due to cable length and pixel-to-pixel
differences inside the photon sensors, have to be removed to achieve the time
resolution required for the Barrel DIRC.
The LMS will provide channel-by-channel T$_0$ values, which are then
stored in a database to properly calibrate the photon arrival times of all pixels.

\subsubsection*{Optical Calibration and Alignment}

The reconstruction of the Cherenkov angle from the hit pattern on the MCP-PMT 
array relies on the correct relative position and orientation of all optical elements 
and of the photon sensors and their pixels. 
The exact locations will be determined during the Barrel DIRC installation using
a laser survey system and, if required, a coordinate measuring machine.
The effect of misalignment between the DIRC and the tracking detectors and between DIRC 
components, like the bar/plate and the prism, can be corrected for using beam data.

\subsubsection*{In-Beam Calibration and Alignment}
\label{subsec:calonline}

After installation in \panda the Barrel DIRC alignment can be verified
using beam data.
Samples of muons, pions, kaons, and protons, identified either
by other \panda subdetectors or via kinematic fits, are available to 
calibrate the Barrel DIRC measurement of the Cherenkov angle. 

Muons can be identified by the muon chambers and provide a clean 
source of $\beta = 1$ particles. 
Decays from pair production of $\phi\phi$, $\Lambda\overline{\Lambda}$ or $K^{0}_{S}{K}^{0}_{S}$ 
can be used to obtain, after a few weeks of data taking, clean samples of pions, kaons and 
protons according to the decays:
\begin{eqnarray*}
\overline{p}p &\rightarrow &\phi \phi \rightarrow K^{+}K^{-}K^{+}K^{-}\\
\overline{p}p &\rightarrow &\Lambda\overline{\Lambda} \rightarrow \overline{p} \pi^{+}  p \pi^{-} \\
\overline{p}p &\rightarrow &K^{0}_{S}{K}^{0}_{S} \rightarrow \pi^{+}\pi^{-}\pi^{+}\pi^{-}
\end{eqnarray*}

The geometric reconstruction is then used to determine the Cherenkov
angle per photon and for each track and sensor pixel.
Any deviation of the measured Cherenkov angle in each of the calibration 
samples from the expected Cherenkov angle is then used to build 
a correction function or multi-dimensional lookup-table in the 
configuration database
to remove the effect of residual misalignments on the Barrel DIRC performance.
This is similar to the procedure used by the BaBar DIRC counter, where a
10\% improvement of the Cherenkov angle resolution was achieved
by using a per-photon correction of the Cherenkov angle, calculated
from a muon calibration sample~\cite{babar:nim-update}

\begin{figure}[h]
	\centering
		\includegraphics[width=0.8\linewidth]{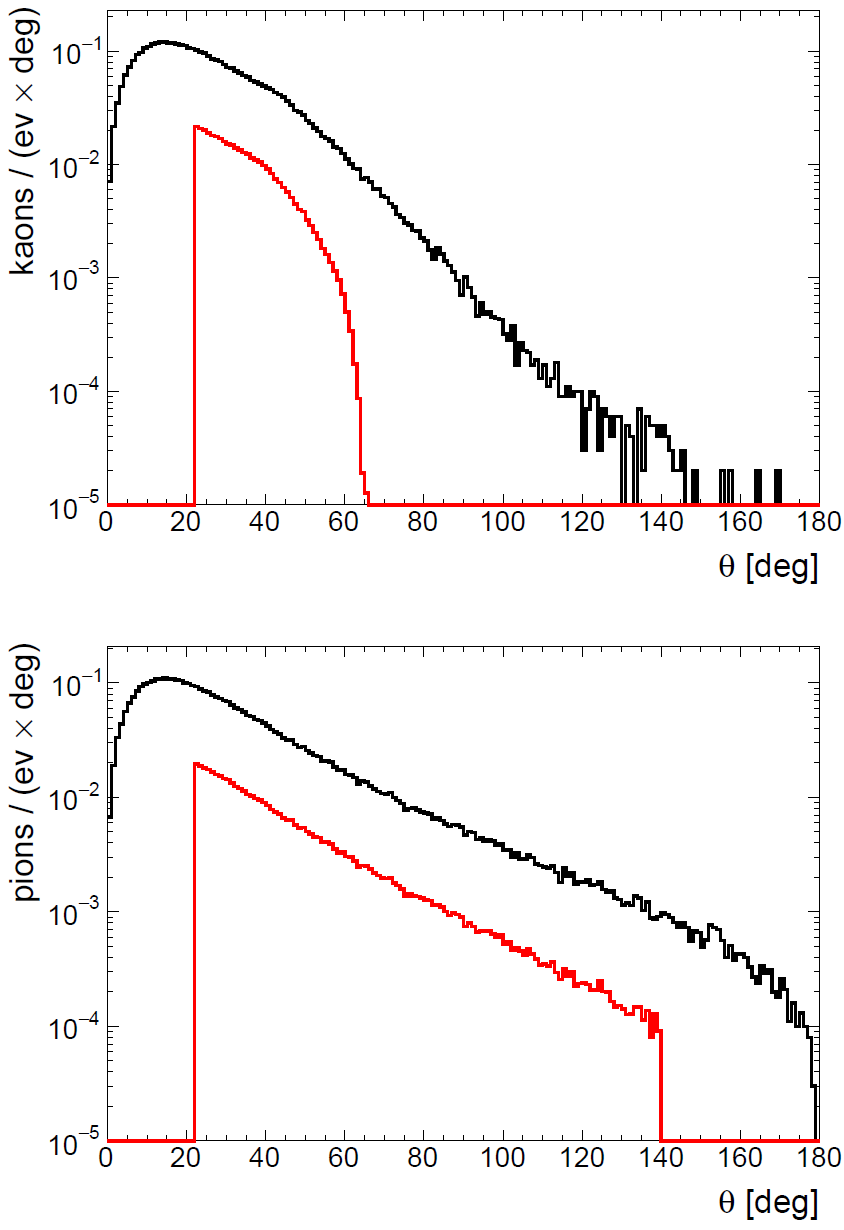}
	\caption{Polar angle distribution of kaons from $\overline{p}p \rightarrow \phi \phi$ 
	reactions (top) and pions from $\overline{p}p \rightarrow K^{0}_{S}{K}^{0}_{S}$ reactions 
	(bottom) at a beam momentum of 10~GeV/c. 
	The black curves are the distributions of 100\,000 events produced, the red curves are 
	the distributions of the detected particles.}
\label{fig:allignement-reactions}.
\end{figure}

Figure~\ref{fig:allignement-reactions} shows the polar angle coverage 
for 100\,000 calibration events for an antiproton beam with 10~GeV/c 
momentum: Kaons from $\overline{p}p \rightarrow \phi \phi$
are shown at the top, pions from $\overline{p}p \rightarrow K^{0}_{S}{K}^{0}_{S}$
at the bottom.
The generated distributions are shown in black, the distribution of 
charged pions and kaons well within the Barrel DIRC acceptance 
(transverse momentum $p_t > 100$~MeV/c, momentum $p > 300$~MeV/c for 
pions and $p > 700$~MeV/c for kaons) are shown in red.
 
Charged kaons are detected in the important forward region of the Barrel 
DIRC, from its lower angular limit 22$^{\circ}$ up to about 60$^{\circ}$. 
Pions are detected over the entire range, up to about 140$^{\circ}$.

Even taking into account that the initial luminosity of \panda is expected
to be a factor 10 below design, the 100\,000 calibration events shown
in Fig.~\ref{fig:allignement-reactions} can be collected within a few 
hours of data taking.

\section{Quality Assurance}
\label{sec:QA}

The \panda Barrel DIRC requires the production of many radiator bars 
or plates, optical elements such as lenses and prisms, photon sensors, 
and front-end boards. 
The high performance of the DIRC detector imposes very strict requirements 
on the quality of the different components. 
To ensure an efficient production, quality assurance procedures have been 
defined. 
The equipment and facilities for a semi-automated measurement of the component 
properties, the associated software tools, as well as the methods and facilities 
for the quality assurance tests, are described in the following sections.

\subsection{Quality Requirements}
\label{subsec:Q-Requirements}

The following requirements have to be met by the individual components to qualify for the \panda Barrel DIRC:

\textbf{Optical Elements}
\begin{itemize}
\item  Cherenkov Radiators
\begin{itemize}
\item  The surface roughness is 10~\AA~RMS or better for the large surfaces 
and 25~\AA~RMS or better for the ends of the bar.
\item  The squareness must not exceed a value of 0.25~mrad for side-to-face angles and the squareness of the side-to-end and face-to-end angles must not exceed 0.5~mrad.
\item The total thickness variation must not exceed a value of 25~$\mu$m.
\item The length of the radiators is 1200$^{\mathrm{+0}}_{\mathrm{-1}}$~mm and the 
thickness is 17$^{\mathrm{+0}}_\mathrm{{-0.5}}$~mm.
In the baseline design the width of the narrow bar is 53$^{\mathrm{+0}}_{\mathrm{-0.5}}$~mm
and the width of the wide plate in the design option is 160$^{\mathrm{+0}}_\mathrm{{-0.5}}$~mm.
\end{itemize}

\item Focusing Lenses
\begin{itemize}
\item The focal length (in synthetic fused silica) is 300\,mm $\pm$ 5\,mm.
\end{itemize}
\item Expansion Volumes
\begin{itemize}
\item The length is 300\,mm $\pm$ 1\,mm and the width is 160$^{\mathrm{+0}}_\mathrm{{-1}}$~mm.
\item The opening angle is $33^{\circ} \pm 1^{\circ}$.
\end{itemize}
\end{itemize}

\textbf{Photon Sensors}

\begin{itemize}
\item  Lifetime-enhanced MCP-PMTs ($>$5 C/cm$^{2}$ integrated anode charge) with a 10~$\mu$m pore 
diameter.
\item Anode layout with 8$\times$8 pixels of about 6$\times$6~mm$^{2}$ size with $\ge$80\% active area coverage.
\item  Quartz or Sapphire entrance window.
\item  $\geq$22\% peak QE and $\pm$0.5\% QE uniformity across the photo cathode surface.
\item  $\leq$10 kHz/cm$^{2}$ average dark count rate.
\item  $>$1 MHz/cm$^{2}$ rate stability of gain.
\item  $>$10$^{6}$ gain and a gain variation of less than a factor 2 between the anode pixels.
\item  Low to moderate cross talk between anode pixels.
\end{itemize}

\textbf{Front-End Electronics}

 \begin{itemize}
  \item Noise level below single photon signal level.
  \item $<$1\% dead channel count.
  \item Validated slow control communication capability.
 \end{itemize}

\subsection{Quality Assurance for the Radiators}
\label{subsec:QA-Radiators}

Although striae or inclusions are not expected to be an issue for the 
\panda Barrel DIRC, validation of the optical homogeneity will be 
part of the quality assurance protocol for the raw material.
A visual inspection will identify bubbles or inclusions and a laser
will be used to detect possible striae or layers with varying index
of refraction.
  
The setups described in Sec.~\ref{subsec:qualtest-radiator}, which were 
built to qualify the radiator prototypes from different vendors, will
also be used for the quality assurance (QA).
Radiator properties to be monitored during mass production include 
the bulk attenuation, surface roughness, subsurface damage, squareness 
and parallelism, flatness, and the sharpness of the edges.
The primary responsibility for QA will rest with the manufacturer. 
They will produce a QA report confirming compliance with the specifications 
and provide measurements of the dimensions, flatness, squareness and roughness 
of the surfaces. 
After delivery the radiators will be visually inspected for defects and
the need for post-shipment cleaning will be assessed. 
If required, radiators will be cleaned using the methods used for the 
BaBar DIRC before being placed into individual holders and stored 
under a HEPA filter.

The manufacturer's QA results will be cross-checked for each radiator 
using the setups and procedures described in Sec.~\ref{subsec:qualtest-radiator}.
The QA measurements foreseen at GSI are:
\begin{itemize}

\item Visual evaluation of inclusions in the radiator, scratches or chips.

\item Determination of bulk absorption length and coefficient of total internal reflection
for at least three laser wavelengths.

\item Determination of the squareness and parallelism.

\end{itemize}

\subsection{Quality Assurance for the Lenses, Mirrors, and Expansion Volume}
\label{subsec:QA-Lens}
The lenses will undergo a visual inspection for scratches and inclusions. 
Afterwards the focal length will be measured. 
The mirrors are off-the-shelf products and will also be visually inspected for defects. 
The EV prisms, made from synthetic fused silica, will be inspected visually in the same way. 
In addition, the dimension and form tolerance of each individual prism will be verified. 

\subsection{Quality Assurance for the Bar and Prism Boxes}
\label{subsec:QA-Barbox}
The radiators as well as the prisms will be housed in boxes from Carbon-Fiber-Reinforced 
Polymer (CFRP) for support, protection from the environment, and light tightness 
(see Sec.~\ref{cha:mech}). 

Tests foreseen at GSI are:
\begin{itemize}

\item After delivery, the parts of the boxes will be inspected visually for damage.
\item The dimensions and shape of each box will be measured.
\item Each box will be assembled and tested for light and gas tightness 
and, if necessary, cleaned prior to transfer to the cleanroom.
\end{itemize}

\subsection{Quality Assurance for the Photon Sensors}
\label{subsec:QA_MCP-PMT}
The QA measurements for the MCP-PMTs will be done at Erlangen. 
This requires the test of about 200~two-inch MCP-PMTs after the manufacturer 
has started the mass production. To be able to perform these tests efficiently efforts are 
ongoing to build a new semi-automated setup.

It is foreseen to measure the most important parameters of each MCP-PMT 
requiring only a few steps:

\begin{itemize}

\item  First, a Quantum Efficiency (QE) wavelength scan will be performed, followed by a QE position scan at one wavelength across the PC surface. This will qualify the peak QE and the QE uniformity. The procedure requires the measurements of low currents and is well established with the equipment already in use.

\item  In a second scan, the gain, the cross-talk between the anode pixels, the time resolution, the dark count rate, and the afterpulsing behavior will be measured simultaneously as a function of the position using a pico-second laser pulser.
This can be achieved with an untriggered DAQ which takes a certain amount of data at each scan position on a
grid with a 1~mm spacing. It is foreseen to do these measurements with the GSI TRB system using modified PADIWA front-end boards to allow for an accurate measurement of the signal charge. 

\item  The rate stability of the MCP-PMTs has to be measured separately. 
This is foreseen for selected tubes.

\item  To ensure a lifetime of $>$10 years an accelerated aging test will be done for 1--2  MCP-PMTs of each production batch. The main setup for the illumination and the measurement protocol already exists.

\end{itemize}
The QA measurements will be performed in parallel with the MCP-PMT fabrication 
and the position dependent parameters of each MCP-PMT will be stored in a 
database to be included in the detector simulations.

\subsection{Quality Assurance for the Front-End Electronics}
\label{subsec:QA-FEE}
An important aspect of testing the front-end electronics is the proper 
communication with the DCS, which is essential for any further checks. 
The assembled units will be characterized using analog signals comparable 
to signals from single photons. These signals can be supplied by a fast signal generator or an attached MCP-PMT illuminated by a fast laser pulser. This test allows for establishing noise levels present in the front-end electronics, which have an adverse impact on the detection efficiency, and detecting faulty channels either 
``hot'' or ``cold''. Depending on these results individual cards or the entire unit can be accepted for installation.

\putbib[./literature/lit_components]
\end{bibunit}

%% file: performance-validation/performance-validation.tex
\chapter{Performance}
\label{ch:performance}

Following the detailed detector simulation studies and measurements on 
test benches, many of the design concepts and components were
studied with a $\pi^-$ beam in cave~C at GSI and in a secondary 
hadron/lepton beam at the T9 beam line area of the CERN proton synchrotron.
Starting with a proof-of-principle prototype in 2008 and evolving to the large 
``vertical slice'' system prototype in 2015 and 2016, the simulation, reconstruction, 
detector resolution, and PID performance were validated during several 
test beam campaigns.

\begin{bibunit}[unsrt]

\section{Prototype Evolution}
\label{sec:proto_evo}

\textbf{Proof-of-Principle (2008)}

\begin{figure}[h!]
\centering
\includegraphics[width=0.95\columnwidth]{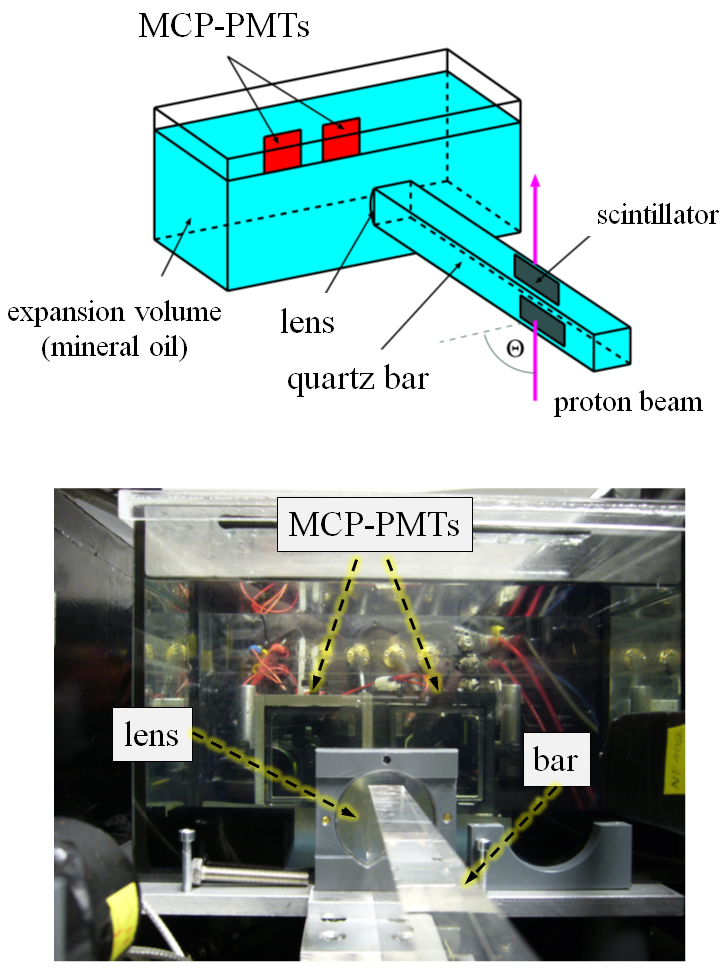}
\caption{Schematic (top) and photograph (bottom) of the 2008 prototype during 
the test beam campaign at GSI.}
\label{fig:proto_2008}
\end{figure}

The first \panda Barrel DIRC prototype is shown in Fig.~\ref{fig:proto_2008}. 
It was placed into proton beam at GSI with 2.0~GeV energy to serve as proof-of-principle for
observing a Cherenkov ring image by imaging with a focusing lens.
A fused silica spherical lens was attached to a fused silica radiator with 
refractive index matching liquid.
An air gap separated the lens from an acrylic glass container, filled with 
mineral oil.

Two 64-channel Microchannel Plate Photomultiplier Tubes (MCP-PMTs) were used to detect the 
Cherenkov photons and the TRB boards (version 2) with TOF add-on
front-end~\cite{trb2-pe} cards were used as readout.

Both sensors observed a Cherenkov signal, consistent with the pattern expected 
from simulation.\\

\textbf{Behavior of Cherenkov Hit Pattern (2009)}

\begin{figure}[h!]
\centering
\includegraphics[width=0.9\columnwidth]{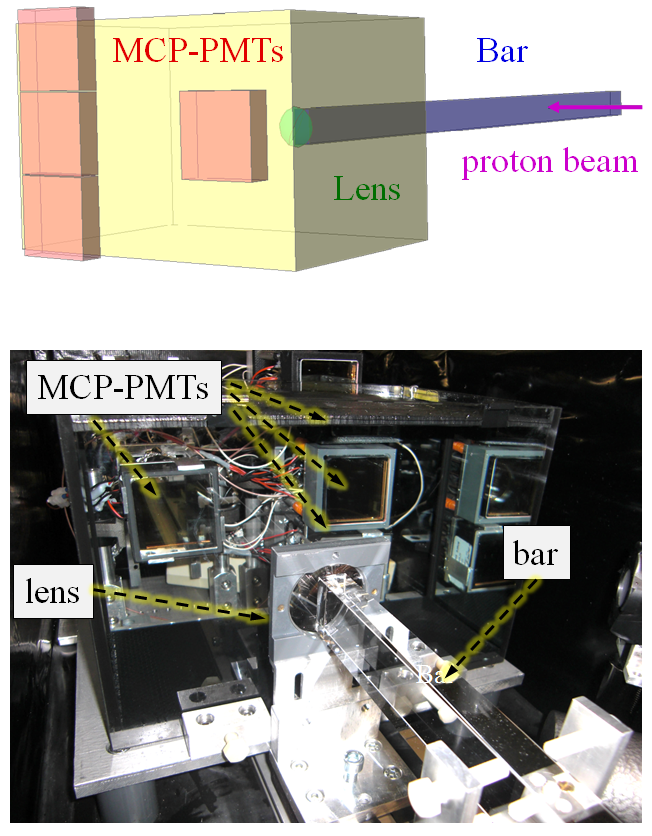}
\caption{Schematic (top) and photograph (bottom) of the 2009 prototype during 
the test beam campaign at GSI.}
\label{fig:proto_2009}
\end{figure}

\begin{figure}[h!]
\centering
\includegraphics[width=0.9\columnwidth]{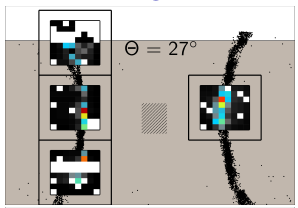}
\caption{Back view of the Cherenkov hit pattern recorded with the 2009 prototype 
in a 2~GeV proton beam at a polar track angle of $27^\circ$. 
The hashed rectangle at the center indicates the position of the radiator bar with 
respect to readout plane, the large gray rectangle represents the fill level of the used mineral oil 
(Marcol 82~\cite{marcol}) inside the EV. 
The black dots represent the expected Cherenkov rings from simulation.}
\label{fig:proto_2009_chrings27theta}
\end{figure}

In the next version of the prototype, shown in Fig.~\ref{fig:proto_2009},
the size of the acrylic container expansion volume (EV) was increased and the number 
of pixels and readout channels was doubled~\cite{RHoler-PHD-THESIS-pe}.
Preamplifiers were attached directly to the MCP-PMTs anode pins and the readout was done with
the TRB (version 2) and TOF add-ons.
The observed ring image is shown in Fig.~\ref{fig:proto_2009_chrings27theta} for a 
polar angle of $27^\circ$ between the bar and the 2~GeV energy proton beam.
The areas marked in white correspond to dead or inefficient electronics channels.
The shape of the pattern agreed well with simulation and both ring segments moved
across the detector plane when the polar angle was changed, as expected for 
Cherenkov photons. \\

\textbf{Study of Sensors and Lenses (2011)}

\begin{figure}[h!]
\centering
\includegraphics[width=0.8\columnwidth]{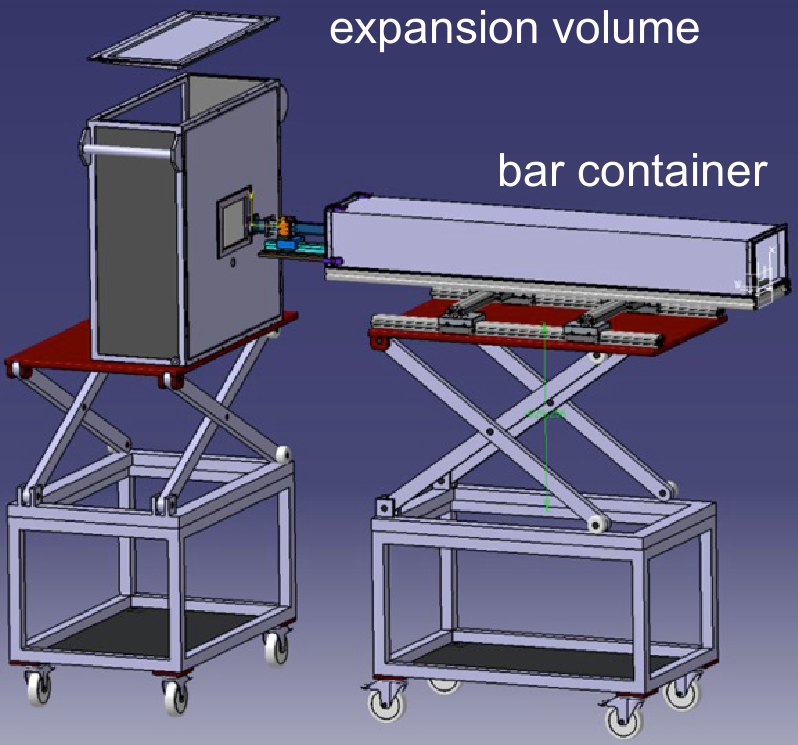}
\caption{Schematic view of the 2011 prototype used at GSI and CERN.}
\label{fig:2011_proto_cern}
\end{figure}

Two test beam campaigns were performed in 2011, one at GSI and one at CERN.
The main goal was to test different types of photon sensors and focusing 
lenses with Cherenkov light~\cite{GKalicy-PHD-THESIS-pe}.
The prototype was upgraded significantly compared to 2009 by increasing the 
size of the expansion volume and by making the prototype modular, allowing for
easy exchange of components, see Fig.~\ref{fig:2011_proto_cern}.
The expansion volume was a large aluminum box, filled with Marcol 82 mineral 
oil, with a glass entrance window to attach the bar and a large 
(80\,cm$\times$80\,cm) glass window at the back of oil tank for the photon
detectors.
The sensors were placed into plastic holders.
The holders were supported by aluminum masks, each custom-made for a specific 
range of polar angles, which also made the prototype light-tight.
Up to 11 sensors could be placed against the glass window at any one time.
An optical grease was used for the  coupling of the bar, lens, glass windows,
and sensors.
The list of photon detectors tested included Multianode Photomultiplier Tubes 
(MaPMTs) and MCP-PMTs with different sizes and anode configurations from 
two vendors, and an array of Silicon Photomultipliers (SiPM).
The SiPM array suffered from an unacceptable level of dark noise, in spite of
cooling the array with a Peltier element.
Although both the MaPMTs and the MCP-PMTs worked well during the beam test,
future prototypes used only MCP-PMTs since the MaPMTs are not an
option for the Barrel DIRC due the magnetic field in \panda
(see Sec.~\ref{sec:sensors}).

\begin{figure}[h!]
\centering
\includegraphics[width=0.75\columnwidth]{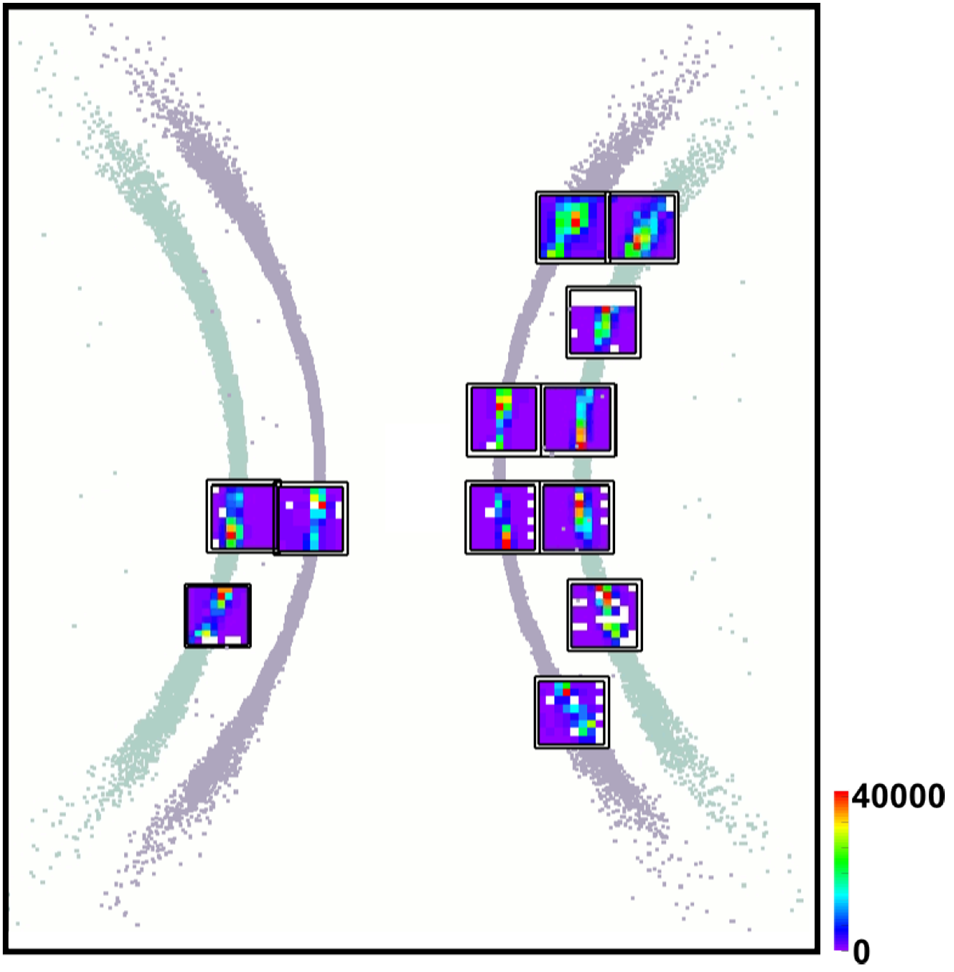}
\caption{Composite back view of the Cherenkov hit patterns for the 2011 prototype
for two runs with different sensor coverage.
The inner pair of rings segments corresponds to a polar angle of 120.2$^\circ{}$ 
and the outer pair to 109.6$^\circ{}$. 
The expected hit locations from Geant simulation are shown as dots.}
\label{fig:2011_chrings}
\end{figure}

Examples of the ring images obtained with this prototype for two different polar angles
are shown in Fig.~\ref{fig:2011_chrings} as a composite occupancy distribution.
The pixels marked in white correspond to dead or inefficient electronics channels,
primarily due to defective preamplifiers.
The observed pattern is in good agreement with the expectation from simulation 
for the two polar angles.

The photon yield was found to be a factor 3 lower than expected, primarily due to the 
poor transmission of the optical grease Rhodorsil Paste 7~\cite{silitech-pe},
used to couple the sensors to the glass window, which was replaced by the 
Eljen EJ-550 grease~\cite{eljen-pe} in future beam tests.

The beam test at the CERN PS T9 area resulted in the first measurement of the 
single photon Cherenkov angle resolution (SPR) for the \panda Barrel DIRC
with a value of SPR~$\approx 11$~mrad, consistent with the design goal and the
SPR value achieved by the BaBar DIRC. \\

\textbf{Fused Silica Prism Expansion Volume, Simulation Validation (2012)}

\begin{figure}[hbt]
	\centering
	\includegraphics[width=0.95\columnwidth]{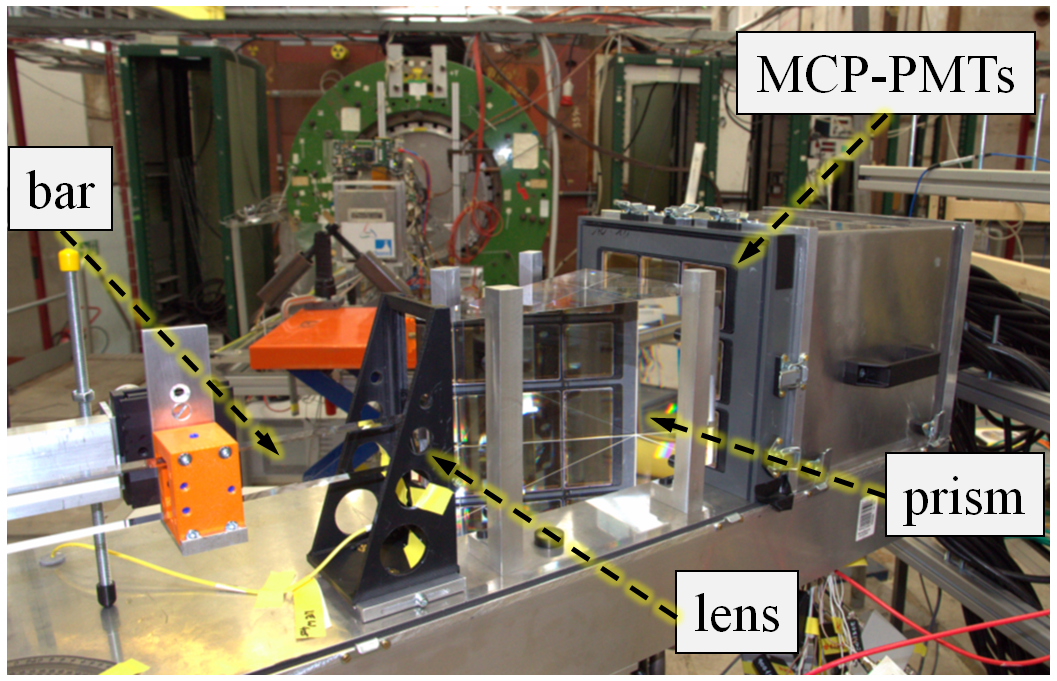}
	\caption{Photograph of the Barrel DIRC prototype at CERN in 2012. 
	}
	\label{fig:2012_proto_cern}
\end{figure}

\begin{figure*}[tb]
	\centering
	\includegraphics[width=0.95\textwidth]{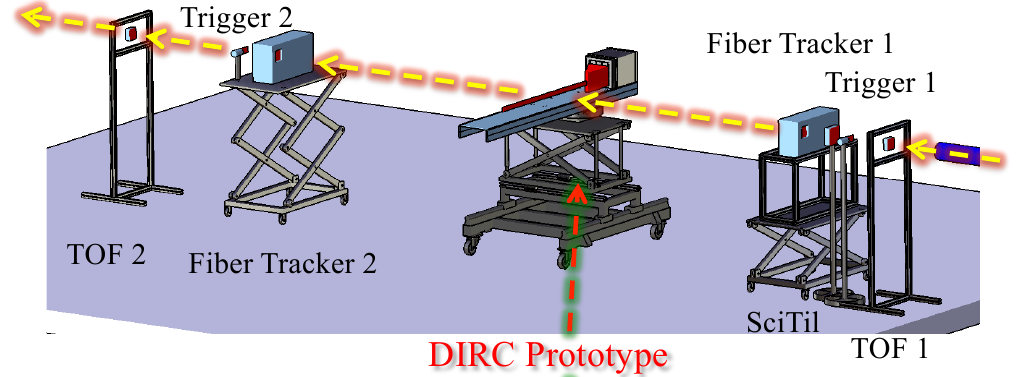}
	\caption{Detectors in the beam line at the CERN test beam campaign in 2012}
	\label{fig:2012_beamline_CERN}
\end{figure*}

For the next major prototype update the focus shifted to the compact expansion volume
geometry~\cite{GKalicy-PHD-THESIS-pe,MZuehlsdorf-PHD-THESIS-pe}.
A solid fused silica prism with a depth of 300~mm and a top opening angle of 30$^\circ{}$
was fabricated by industry and equipped with a a 3$\times$3 array of PHOTONIS Planacon 
MCP-PMTs, coupled to the large readout face.
The prototype setup during the 2012 beam test at the CERN PS (see Fig.~\ref{fig:2012_beamline_CERN}
and Fig.~\ref{fig:2012_proto_cern})
offered the first experience with the fused silica prism EV and with 2-layer compound
lenses (either spherical or cylindrical).
The bar, lens, prism, and MCP-PMTs were optically coupled using Eljen EJ-550
optical grease.

\begin{figure}[h!]
\centering
\includegraphics[width=0.85\columnwidth]{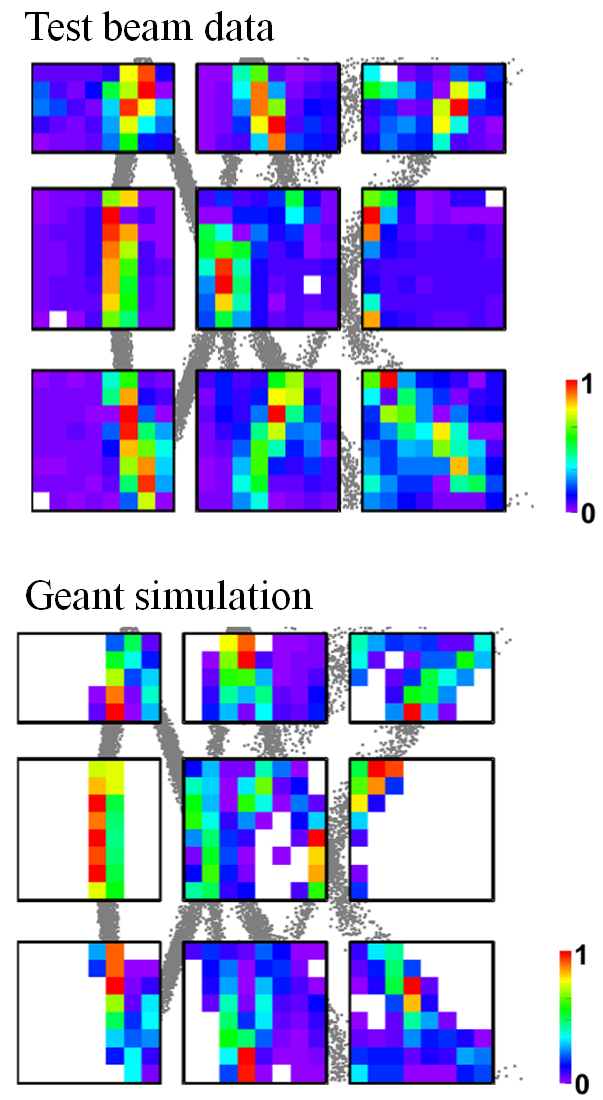}
\caption{Cherenkov hit pattern from experimental data (top) and simulation (bottom)
for the 2012 prototype. 
The normalized hit probability is shown for a spherical lens with air gap, a polar angle of 124$^\circ$ 
and a beam momentum of 10~GeV/c.
}
\label{fig:rings_2012}
\end{figure}

Figure~\ref{fig:rings_2012} shows the hit patterns for experimental data and for simulation, 
which look very different from the ring images of previous prototypes with oil tanks. 
The Cherenkov rings are now folded by side reflections inside the prism, which lead 
to overlapping ring segments.
The geometrical reconstruction method was used to determine the SPR and photon yield for
several configurations with different focusing options.
An important result was that the 2-layer spherical lens provided a photon yield of more than
17 signal photons per track for all polar angles, compared to traditional lenses that
operate with air gaps, which suffer from unacceptable photon loss for polar angles
around 90$^\circ{}$.
Furthermore, the simulation described the data very well and the geometrical reconstruction 
approach was successful in dealing with the additional ambiguities and backgrounds produced 
by the reflections inside the prism.
The configuration with the multi-layer lens and the fused silica prism became the
default for future test beam campaigns. \\

\textbf{Wide Plate as Radiator, New Readout Electronics (2014)}

\begin{figure}[h!]
\centering
\includegraphics[width=0.85\columnwidth]{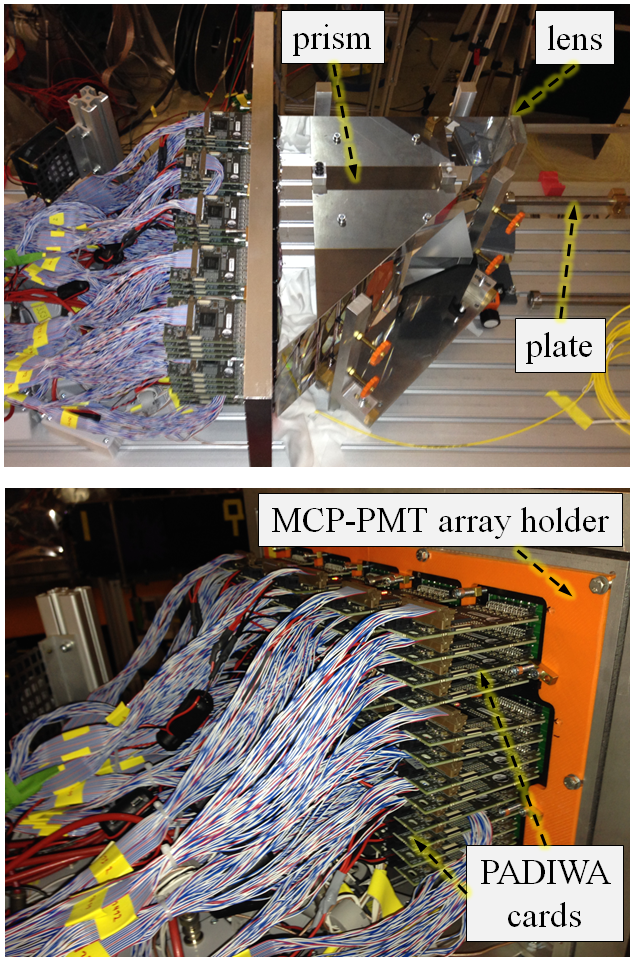}
\caption{Photographs of the readout end of the prototype in 2014
showing the optics (top) and the readout electronics (bottom).
}
\label{fig:2014_gsi_photos}
\end{figure}

In 2014 the prototype was modified to accommodate a wide fused silica plate as radiator
and a larger prism (with depth of 300~mm and a top angle of 
45$^\circ{}$)~\cite{MZuehlsdorf-PHD-THESIS-pe}.
A 1.7~GeV/c pion beam was used at GSI in the summer of 2014 to gain the first
experience with the wide radiator plate instead of a narrow bar.
The readout end of the prototype is shown in Fig.~\ref{fig:2014_gsi_photos}.
The array of 3$\times$5 PHOTONIS Planacon MCP-PMTs~\cite{photonis-pe} was coupled to the back of
the prism and the plate was either coupled directly to the front of the prism
or via a 2-layer cylindrical lens.
All optical components were coupled using Eljen EJ-550 optical grease.
The readout system was updated to the TRB version 3 in combination 
with the PADIWA amplifier and discriminator front-end cards~\cite{comp-trb3-jinst-pe}, 
which was mounted directly on the MCP-PMT backplane.

The primary goal was to study the performance of the plate and the new electronics.
The larger prism offered a potential performance improvement due to fewer 
reflections inside the prism and due to the larger sensor area with more pixels,
which separated photon paths to different pixels more clearly.

\begin{figure}[h!]
\centering
\includegraphics[width=0.95\columnwidth]{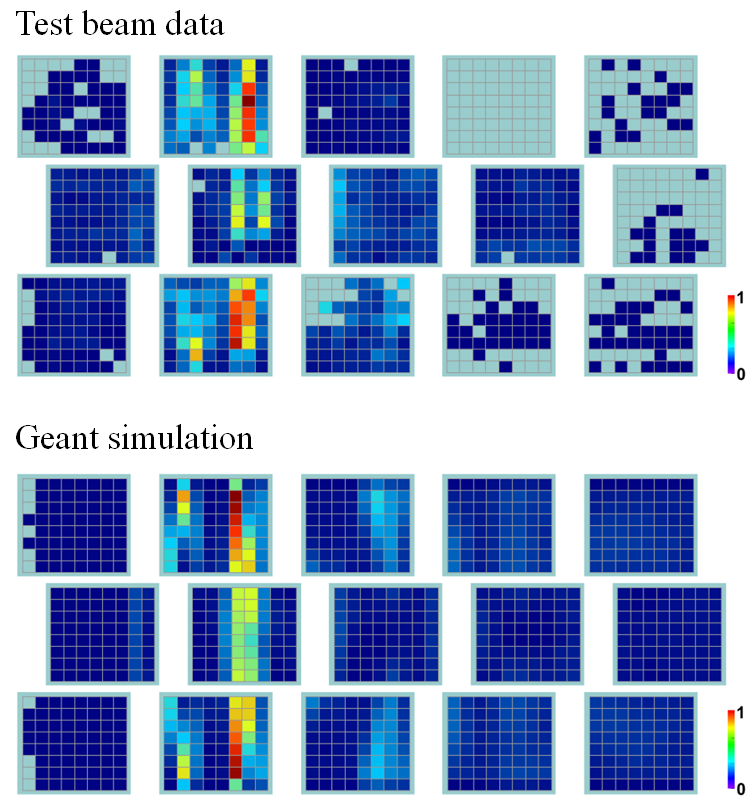}
\caption{Example for the normalized hit pattern recorded in 2014 
for the wide plate with the 2-layer cylindrical lens in
experimental data (top) and the corresponding pattern from simulation (bottom).
}
\label{fig:rings_2014}
\end{figure}

The high noise level in the GSI experimental area and the low thresholds on the PADIWA cards,
required to efficiently detect the MCP-PMT signals, caused oscillations within the readout
electronics.
This made it necessary to deactivate groups of channels with the highest sensitivity to
noise, which explains the gaps in the Cherenkov ring image for the data, shown and compared
to simulation in Fig.~\ref{fig:rings_2014}.
The experience with the noise-induced oscillations led to modifications of the PADIWA cards
after the beam time.
A low-pass filter was added to reduce the impact of high-frequency noise.

The second major complication experienced during the 2014 beam time was the rather large
beam spot size and divergence of the pion beam.
This effect caused the hit pattern to be smeared out
and ultimately meant that no quantitative measurements could be performed with the plate
geometry and the larger prism.
Data was also taken with the narrow bar configuration but the large beam divergence made it
impossible to determine the Cherenkov angle resolution.
This made another beam test at the CERN PS necessary in order to validate the PID performance 
of the narrow bar and to verify the time-based imaging approach for the plate.

\section{Prototype Test at CERN in 2015 - PID Validation of the Narrow Bar Design and the Wide Plate Design}

\label{sec:cern2015}

\begin{figure}[h!]
\centering
\includegraphics[width=.95\columnwidth]{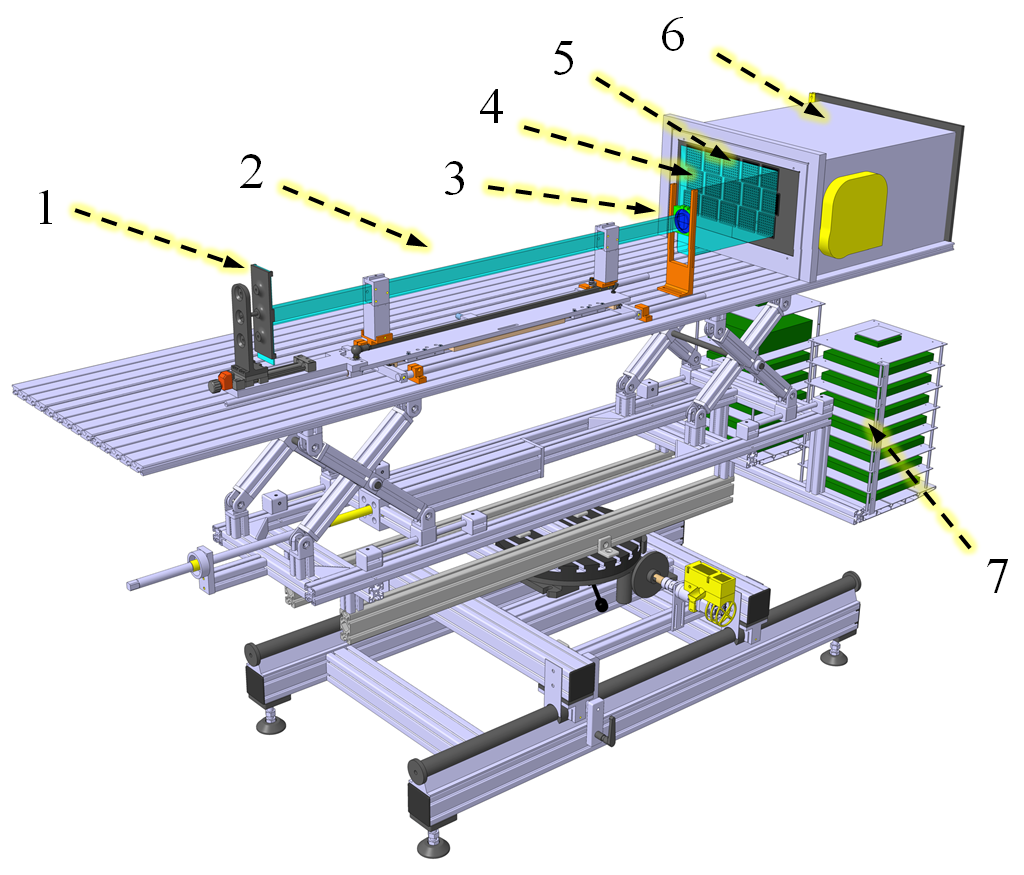}
\caption{Schematic of the prototype used at CERN in 2015, with 
	1: flat mirror, 2: radiator plate, 
	3: lens, 4: expansion volume, 
	5: array of 5$\times$3 MCP-PMTs, 6: readout unit, and 7: TRB stack.
	}
\label{fig:proto-cern-2015}
\end{figure}

\begin{figure}[htb]
\centering
\includegraphics[width=.95\columnwidth]{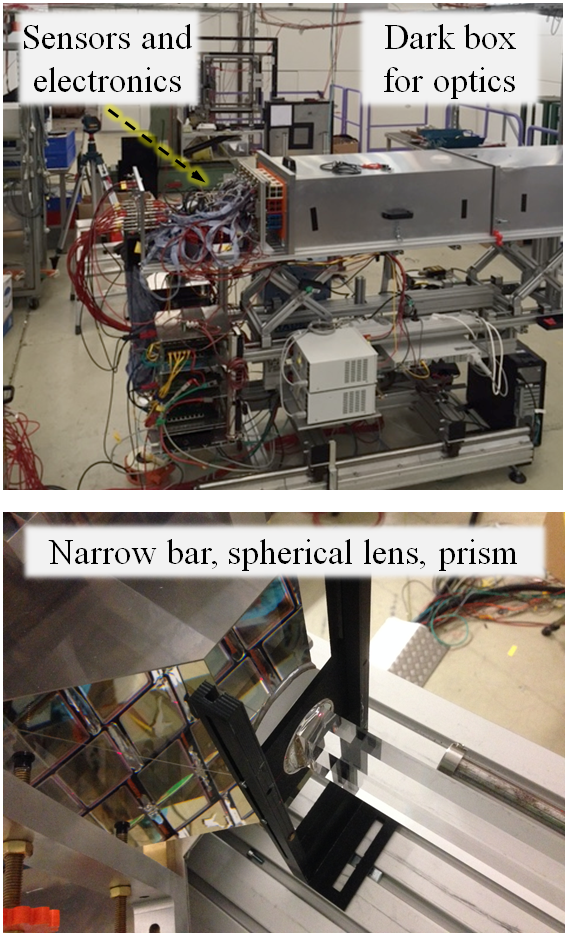}
\caption{Photograph of the 2015 prototype in the T9 beam line (top), close-up
of the 3-layer spherical lens between the narrow bar and the prism 
(bottom).}
\label{fig:proto-cern2015-photos}
\end{figure}

The goal of the test beam campaign at the CERN PS in 2015 was the validation of the
PID performance of the baseline design and of the wide plate.
The prototype, shown in Fig.~\ref{fig:proto-cern-2015} and 
Fig.~\ref{fig:proto-cern2015-photos}, comprised the essential 
elements of a ``vertical slice'' Barrel DIRC prototype: 
A narrow fused silica bar (17.1 $\times$ 35.9 $\times$ 1200.0~mm$^3$) 
or a wide fused silica plate (17.1 $\times$ 174.8 $\times$ 1224.9~mm$^3$), 
coupled on one end to a flat mirror, on the other end to a focusing lens, 
the fused silica prism as EV (with a depth of 300~mm and a top angle of 45$^\circ{}$), 
the array of MCP-PMTs, and the updated readout electronics.
The selection of lenses included 2- and 3-layer spherical and cylindrical lenses, 
with or without anti-reflective coating, as well as spherical lenses with air gaps.
The prototype support frame could be translated manually and rotated remotely 
relative to the beam, making it possible to scan the equivalent of the \panda Barrel DIRC 
phase space.

\begin{figure*}[thb]
\centering
\includegraphics[width=.9\textwidth]{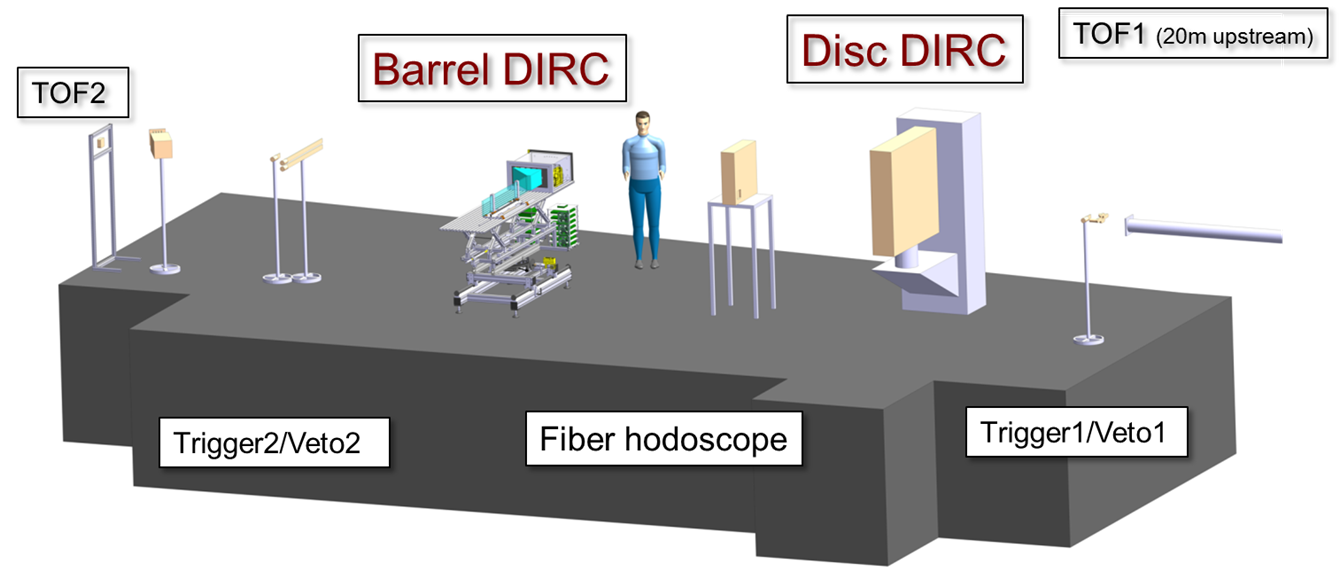}
\caption{Detector setup during the prototype test in the T9 beam line at CERN in 2015 (not to scale).
}
\label{fig:EXP-CERN2015}
\end{figure*}

The experimental setup used for the evaluation of the \panda Disc and Barrel DIRC 
prototypes during the beam times at CERN in May/June and July of 2015 is shown 
schematically in Fig.~\ref{fig:EXP-CERN2015}.
The momentum of the secondary lepton/hadron beam could be set to values between 
1.5 and 10~GeV/c with magnet settings available for positive and negative beam
polarity in steps of 0.5~GeV/c and 1~GeV/c.
The beam focus could be adjusted to either a small beam spot size near one of the two 
DIRC prototypes or to a parallel beam configuration. 
The polar angle between Barrel DIRC radiator and beam was determined using a 
precision scale and monitored using a camera. 
The vertical and horizontal position of the beam on the bar/plate was 
measured using scales.
A line laser system was used to align the prototype relative to the beam line.

\begin{figure}[thp]
\centering
\includegraphics[width=.95\columnwidth]{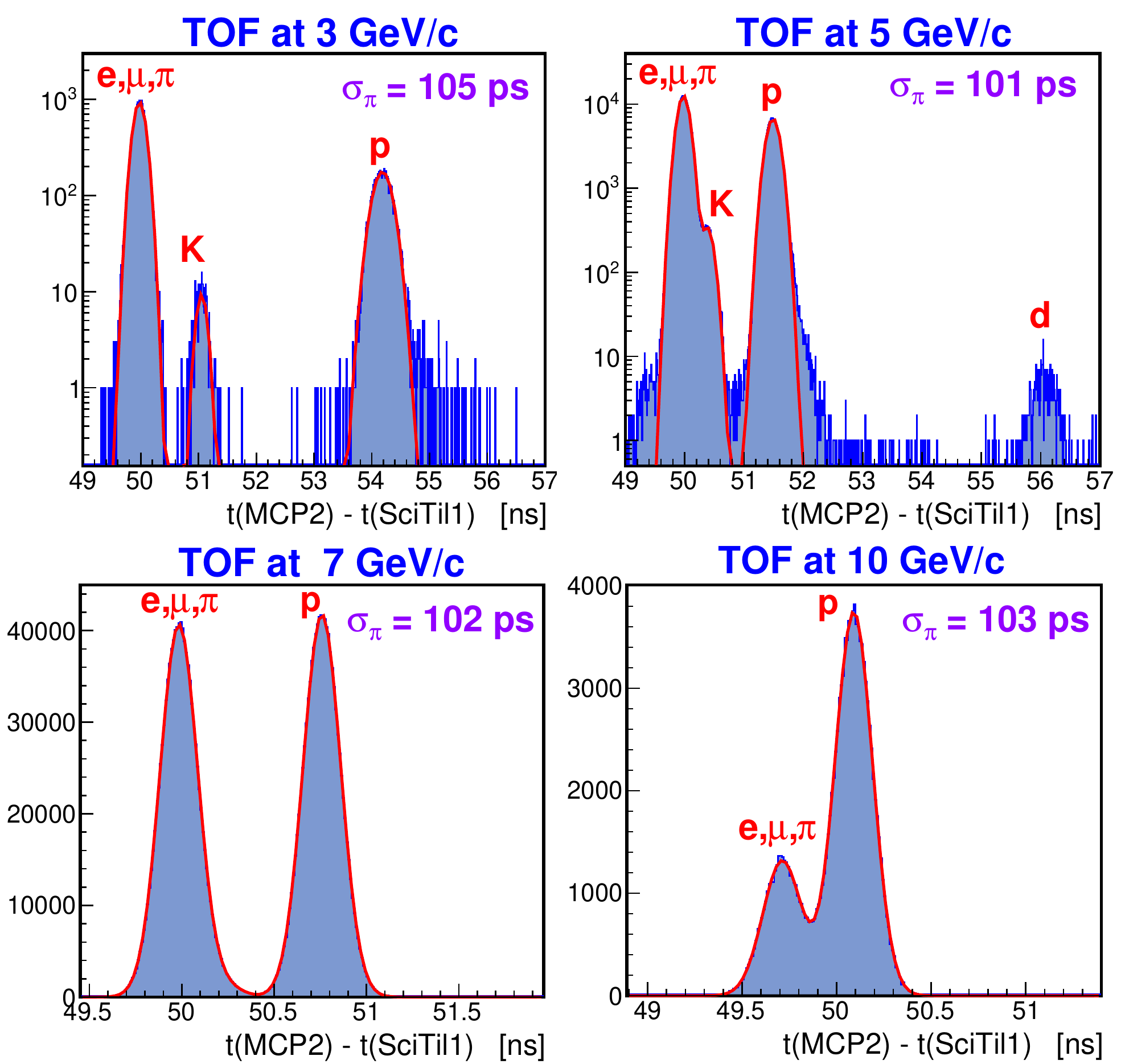}
\caption{$\pi/p$ tagging provided by the time-of-flight system.}
\label{fig:EXP-CERN2015-tof}
\end{figure}

Beam instrumentation included two scintillators with 40~mm diameter to define the 
trigger for the DAQ (Trigger1/2 in Fig.~\ref{fig:EXP-CERN2015}) and two veto counters
(Veto 1/2), sensitive to off-axis beam background.
A scintillating fiber hodoscope provided position information between the Disc and Barrel
DIRC prototypes.
A very fast time-of-flight (TOF) system~\cite{tof-jinst-pe}, positioned directly in 
the beam, was used for $\pi/p$ tagging. 
Each station (TOF1 and TOF2) consisted of a combination of a fast scintillating tile 
(SciTil) counter read out by silicon photomultipliers (SiPMs) and a PMMA radiator 
read out by an MCP-PMT. 
The first TOF station was placed into a gap between two magnets of the T9~beam line, 
about 24~m  in front of the Barrel DIRC prototype, the second station 5~m behind. 
The large distance of 29~m, in combination with the time resolution of 50--80~ps per 
TOF station, provided clean $\pi/p$ tagging at 7~GeV/c momentum and beyond, as 
can be seen in Fig.~\ref{fig:EXP-CERN2015-tof}.

The modular construction of the prototype made it possible to quickly exchange the radiator 
types and the lenses.  
The radiator bar or plate was either coupled directly to the synthetic fused silica prism, 
or a spherical or a cylindrical lens was placed between radiator and prism. 

The readout side of the prism was covered by 15 PHOTONIS Planacon MCP-PMTs XP85012/A1-Q, 
held in place by a 3D-printed 3$\times$5 matrix with offsets between the MCP-PMTs
designed for optimum Cherenkov ring coverage across all angles.

The latest generation of PADIWA front-end cards, modified with the low-pass filter for protection
from noise, was attached to the anode pins of the 15 Planacon MCP-PMTs.
The readout of about 1500 electronics channels was performed with a stack of the TRB version~3
boards~\cite{neiser-pe}.
 
During 34 days of data taking a total of some $5 \times 10^8$ triggers were recorded for a wide 
range of particle angles and momenta, similar to the expected \panda phase space, in 
different optical configurations.
The polar angle between the particle beam and the bar was varied between 20$^\circ$ 
and 155$^\circ$ and the intersection point between beam and bar was adjusted to values
between 6~cm and 93~cm from the readout end of the bar.

The T9 beam was predominantly composed of electrons, muons, pions, and protons.
Since the direct measurement of $\pi/K$ separation was not possible the PID
performance was evaluated for $\pi/p$ at 7~GeV/c instead.
At this momentum the Cherenkov angle separation of pions and protons (8.1~mrad) 
is approximately equivalent to the pion/kaon separation at 3.5~GeV/c (8.5~mrad), 
the 3~s.d. separation performance goal of the Barrel DIRC designs.
For systematic studies data was taken with beam momenta between 2~GeV/c and 
10~GeV/c.

The timing calibration was provided by two laser pulsers: a PiLas PiL040SM~\cite{pilas} 
with 405~nm wavelength and a trigger jitter of 27~ps and a PicoQuant PDL 800-D~\cite{picoquant} 
with a wavelength of 660~nm and a trigger jitter of 80~ps.
The laser pulsers were coupled into optical fibers, 
which were routed into the dark box covering the bar/plate and connected to an opal glass 
diffuser to illuminate the entire MCP-PMT plane.
Laser calibration runs were performed daily and after each configuration change.

\subsection{Simulation of the Prototype}
\label{sec:simulation-performance}

The simulation of the prototype was an important element of the 2015
beam tests, both during the preparation phase, when it was used to determine 
the optimum layout of the MCP-PMTs on the focal plane and the proper
location of the laser pulser fiber for calibration, and for the data analysis, 
to create the look-up tables (LUT) for the geometrical reconstruction.
A standalone Geant4 simulation was developed for the beam test, incorporating many
elements of the \panda Geant simulation, described in Sec.~\ref{sec:reco-geo-bars},
using the same material property tables and physics processes.
The beam detectors (TOF, Veto, Trigger) and the properties of the beam were 
included in the simulation.
For each detector configuration the detailed geometry of each optical
element, such as the relative orientation of the bar/plate relative to the
lens and the prism, were adjusted to the values measured during the detector
access periods.

\begin{figure}[h]
  \centering
  \includegraphics[width=.99\columnwidth]{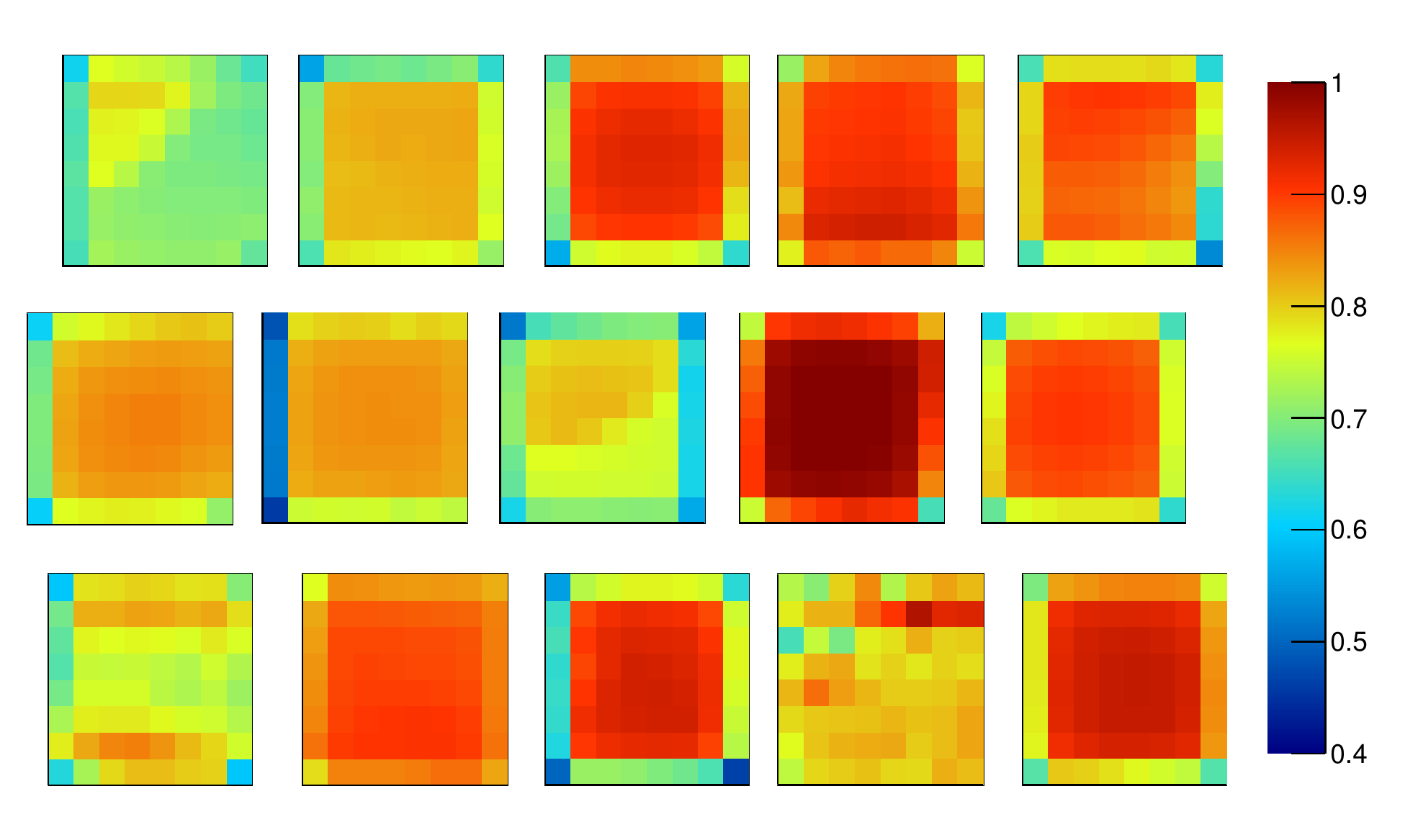}
  \caption{Map of the relative quantum efficiency (QE) of the MCP-PMTs used during 
  the prototype test at CERN in 2015, as implemented in simulations. 
  The absolute values of the QE the MCP-PMTs are determined as the product of 
  this map with the wavelength-dependent QE from Fig.~\ref{fig:qe_wave}.
  }
  \label{fig:qe_pix}
\end{figure}

\begin{figure}[h]
  \centering
  \includegraphics[width=.99\columnwidth]{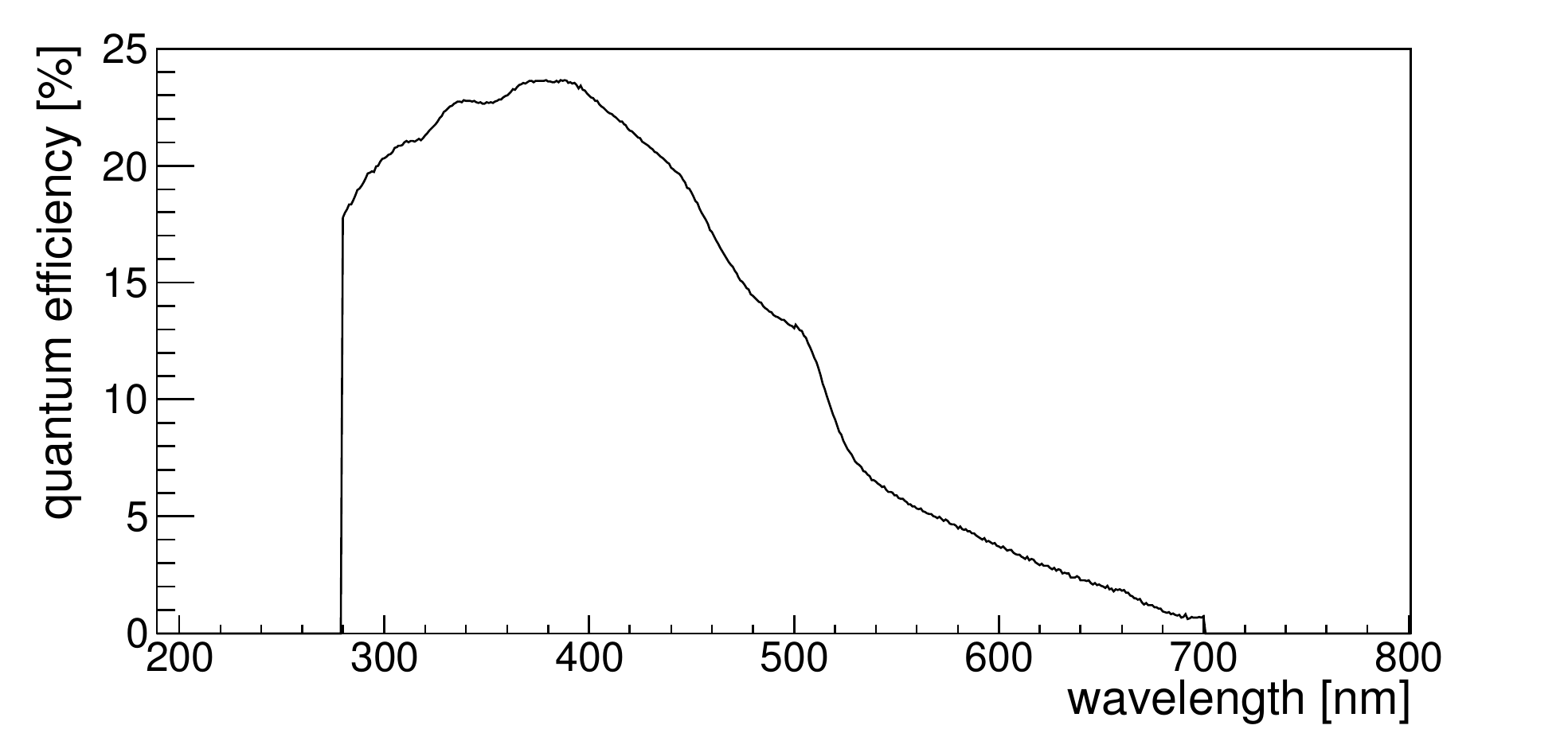}
  \caption{Wavelength dependence of the quantum efficiency of a Planacon XP85012/A1-Q MCP-PMT
  	in the Geant simulation.}
  \label{fig:qe_wave}
\end{figure}

In addition, the detailed properties of the specific Planacon MCP-PMTs used 
during this beam test were added to the simulation.
This included the observed charge sharing, dark count rate, collection efficiency, 
and the quantum efficiency (QE).
Each unit was scanned for QE and gain uniformity with a 372~nm laser pulser using
the setup described in Sec.~\ref{subsec:QA_MCP-PMT}.
The individual 2D maps of the QE were normalized to a reference and implemented
as relative QE maps in simulation, shown in the layout used during the beam test
in Fig.~\ref{fig:qe_pix}. 
The absolute value of the QE was taken from a scan of the QE as function of the
photon  wavelength, shown in Fig.~\ref{fig:qe_wave}, and multiplied with the 
relative QE maps to simulate the QE response of the MCP-PMT array.

The single photon timing resolution of the combination of the MCP-PMTs and the readout electronics
was initially set to 100~ps for simulation to reflect the expected electronics performance.
However, during the beam test a significantly worse resolution was observed and the 
time resolution in simulation was changed to 200~ps to better match the data.

\begin{figure}[h]
  \centering
  \includegraphics[width=.99\columnwidth]{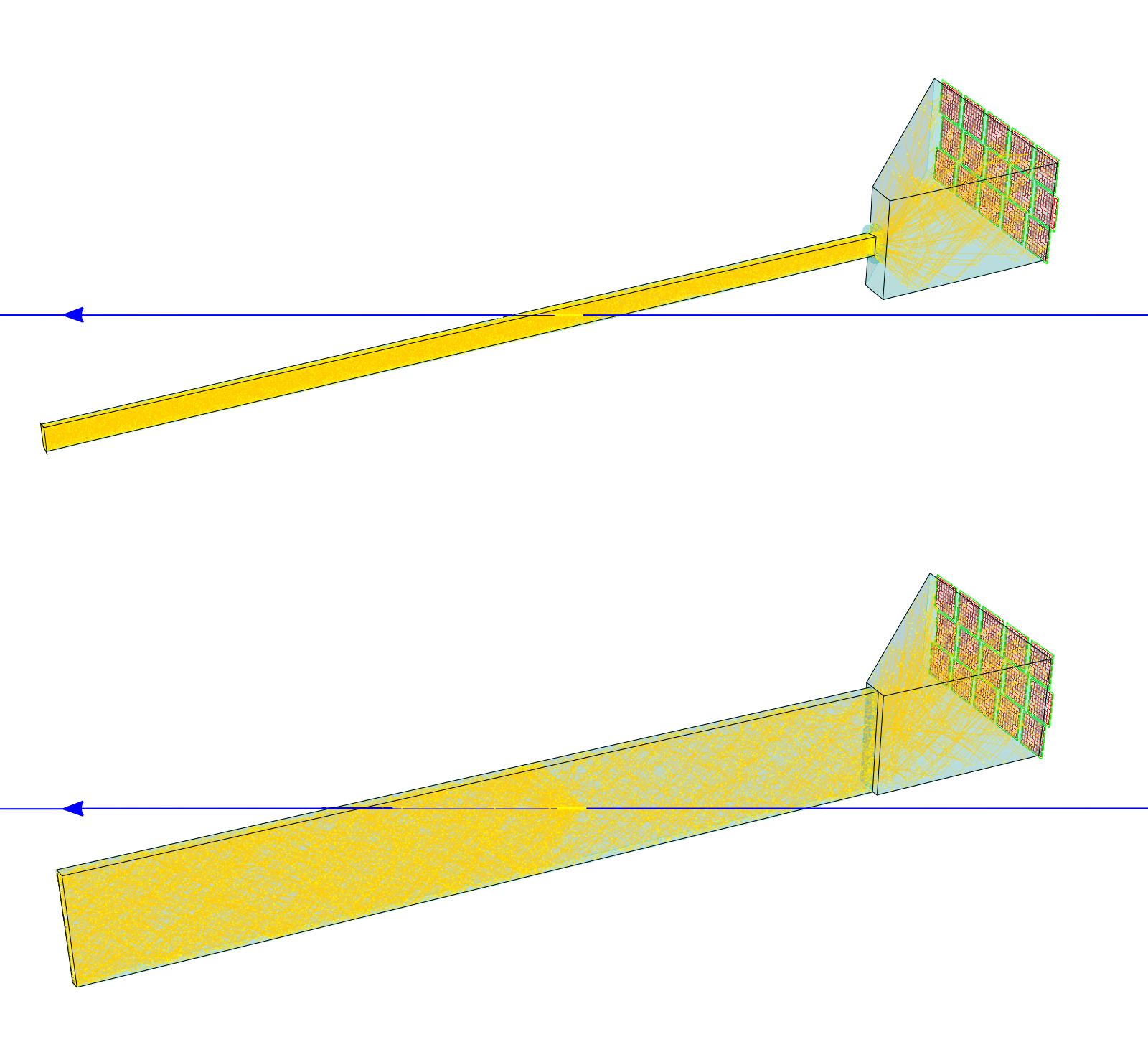}
  \caption{Example of the Geant simulation for a bar (top) and a plate (bottom) geometry.
    Pions with 7~GeV/c momentum and $25^{\circ}$ polar angle traverse the bar from right to left.
   }
  \label{fig:prtdirc_sim}
\end{figure}

\begin{figure}[h]
  \centering
  \includegraphics[width=.99\columnwidth]{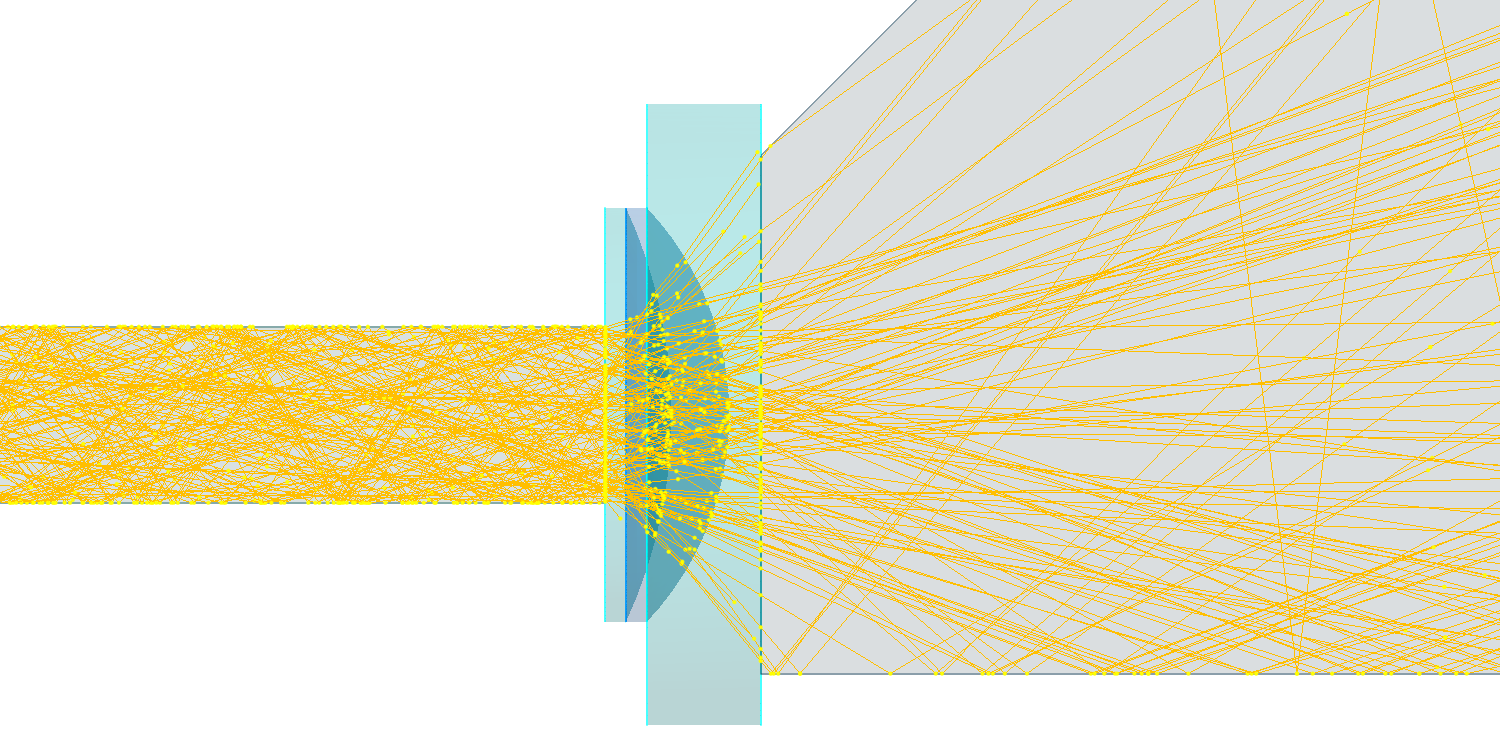}
  \caption{Close-up of the region of the 3-layer spherical lens in simulation. 
  The orange lines represent the Cherenkov photons originating from one $\pi^{+}$ with 7~GeV/c momentum 
  and 25$^{\circ}$ polar angle.}
  \label{fig:prtdirc_zoom}
\end{figure}

Figure~\ref{fig:prtdirc_sim} shows the event display of one simulated pion with 7~GeV/c momentum
and $25^{\circ}$ incident polar angle with respect to the radiator. 
The configurations with a narrow bar (top) and a wide plate (bottom) radiators are shown. 
Figure~\ref{fig:prtdirc_zoom} shows a close-up of the simulated event display with
the paths of the Cherenkov photons in the bar, 3-layer spherical lens, and the prism.

\subsection{Data Analysis}
\label{sec:prt_analysis}

\begin{figure}[h]
  \centering
  \includegraphics[width=.99\columnwidth]{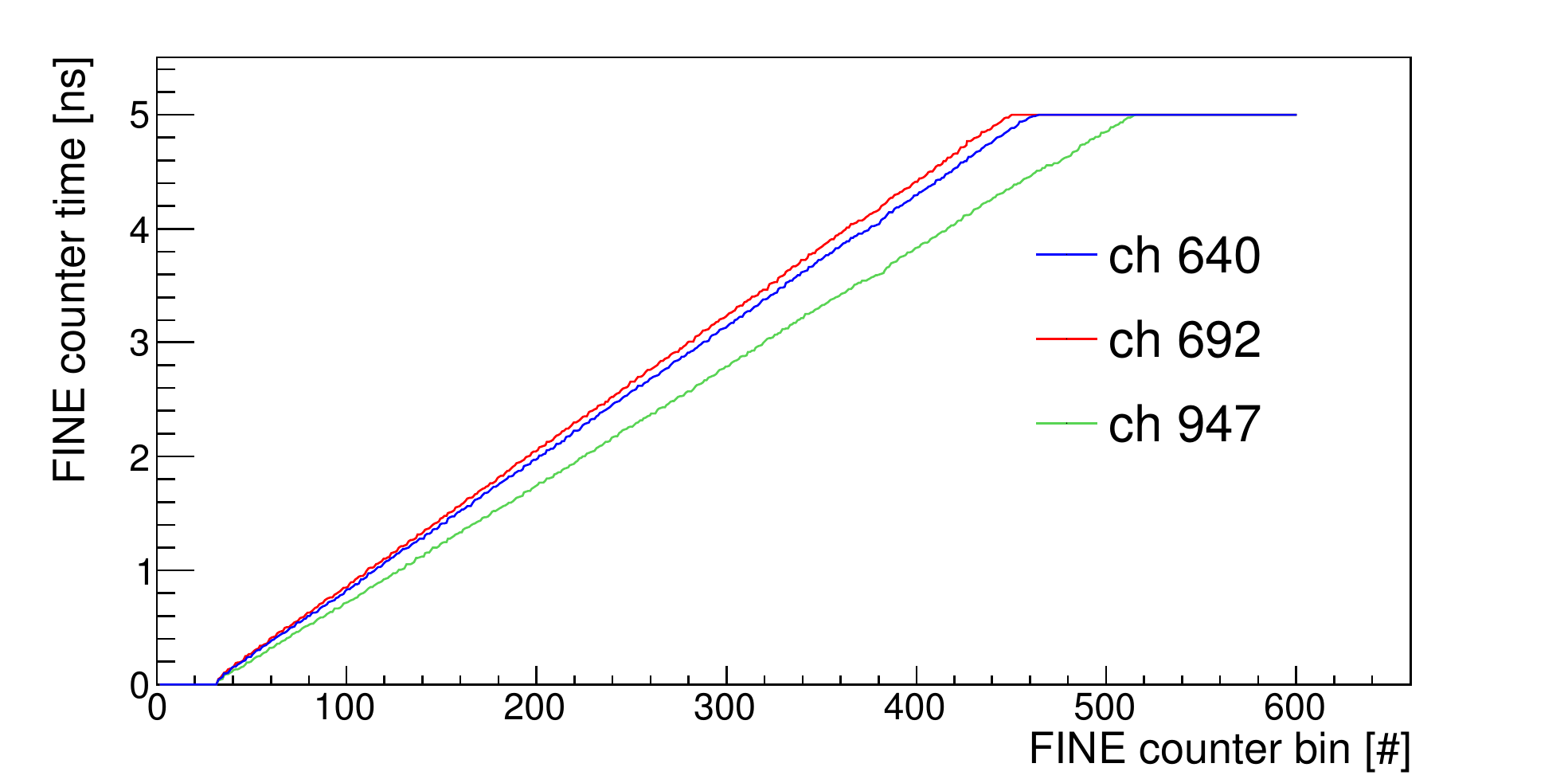}
  \caption{Example of the FINE counter conversion curves from TRB3 calibration for three channels.}
  \label{fig:finetime}
\end{figure}

The data from the prototype are stored in the list mode data format of the HADES 
data acquisition system protocol (TrbNet)~\cite{trbnet-pe} and converted offline 
into the CERN ROOT data format~\cite{root} for analysis.
The multi-hit TDCs on the TRBs record the time information for every channel with 
one or multiple signals above the discriminator threshold.
The most important information stored in the analysis file are times of leading and 
trailing edge of the detected signals. 
Each time is stored in three variables:
The EPOCH counter (with a range of 45.8~min), 
the COARSE counter (10.24~$\mu$s),
and the FINE counter with a range of approximately 5~ns. 
The differential nonlinearity of the FINE counter varies channel-by-channel and 
has to be calibrated using dedicated high-statistics calibration runs taken with 
the fast (8~ps RMS) internal pulser of the TRB.
Figure~\ref{fig:finetime} shows the result of the calibration for three channels, 
the curves that are used to convert the bin number to a fine time.

\begin{figure}[h]
  \centering
  \includegraphics[width=.99\columnwidth]{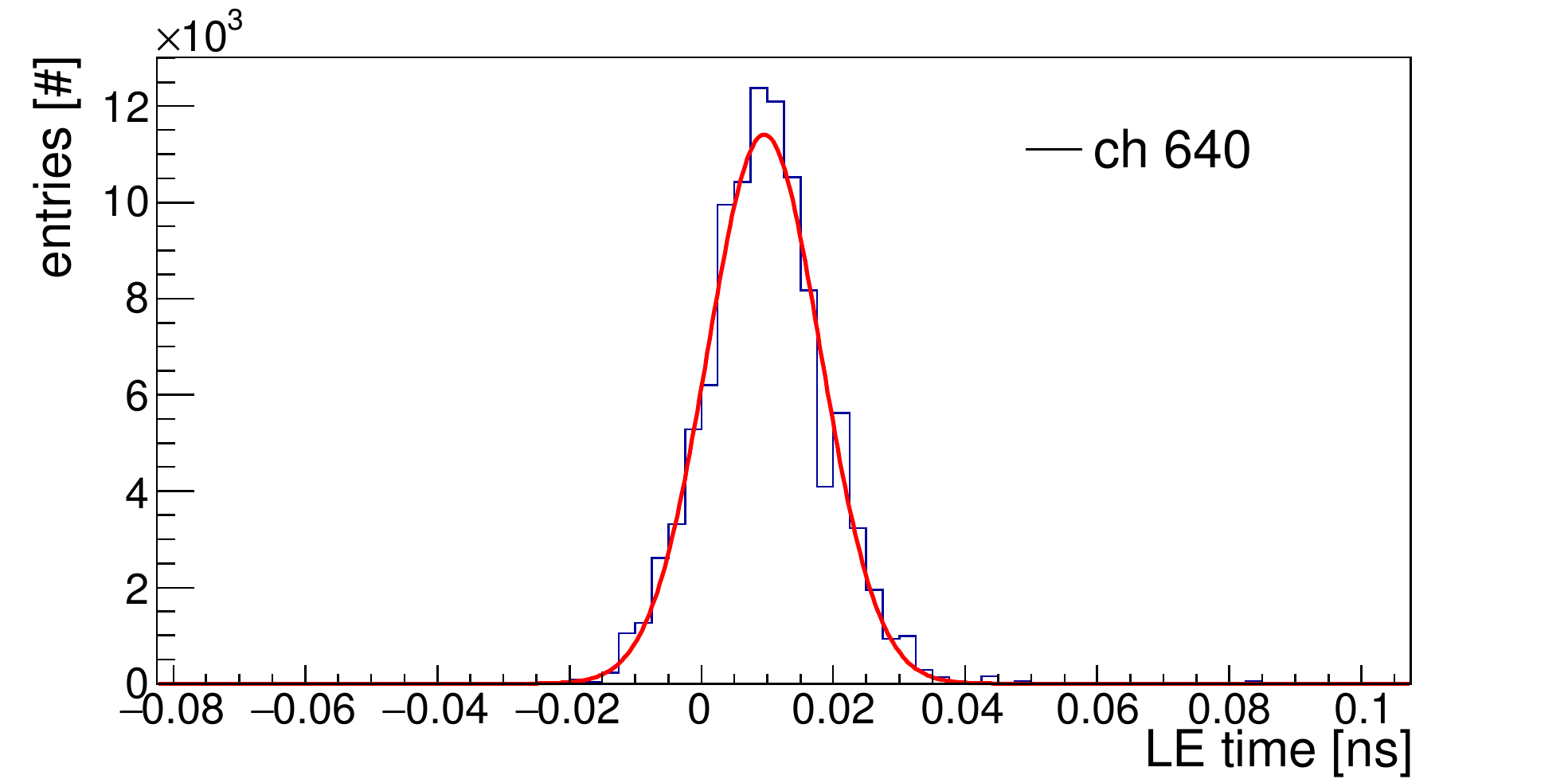}
  \caption{Example of the internal (leading edge, LE) time distribution for channel 640. The fit with a Gaussian yields
    a resolution of 8.5~ps.}
  \label{fig:trb_fit}
\end{figure}

The FINE time calibration curves were monitored closely during the test beam campaign
and found to be very stable so that only one calibration file is used for the entire 
beam test data.
The internal time resolution achieved in the 2015 data varies by channel, TDC, and TRB,
and covers the range from 7--20~ps RMS. 
Figure~\ref{fig:trb_fit} shows the example of the resolution for the channel~640.

\begin{figure}[h]
  \centering
  \includegraphics[width=.99\columnwidth]{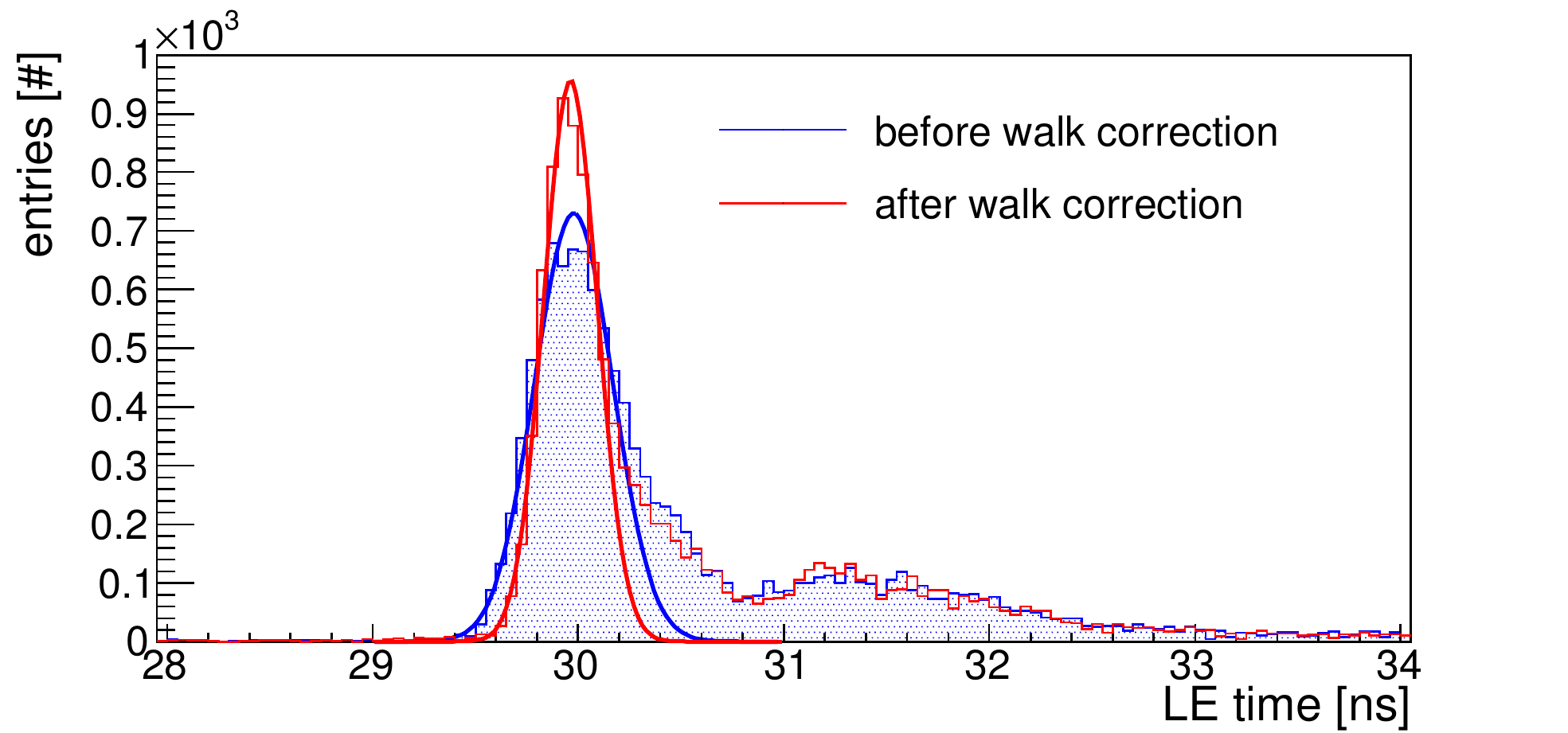}
  \caption{Example of the signal (leading edge, LE) time distribution of PicoQuant laser pulser
  calibration data for a typical Barrel DIRC pixel before time walk correction (blue) and after (red).
  The Gaussian fit in a narrow range around the peak results in resolutions
  of 190~ps and 140~ps, respectively.}
  \label{fig:pico_ex}
\end{figure}

As explained in Sec.~\ref{sec:electronics}, the time-over-threshold (TOT) information 
can be used to correct the data for time walk.
The TOT information is stored as the leading edge and trailing edge time of a signal.
The two times are recorded by the same TDC channel after the leading edge signal 
is delayed by about 30~ns. 
The exact value of the delay varies for each channel and, therefore, has to
be calibrated using the internal TRB3 pulser. 
Figure~\ref{fig:pico_ex} shows the leading edge time resolution for channel 640 
for data taken with the PicoQuant laser pulser.
The two histograms and corresponding fits to the data show the results before
and after a time walk correction using the TOT measurement, demonstrating the 
significant improvement from the TOT information.

Pixel-to-pixel time offsets due to differences in cable lengths and internal delays
on the readout cards were corrected using data recorded with the PiLas and 
PicoQuant laser pulser.
For each pixel the photon arrival time spectrum in laser data was fitted and the
mean values stored in a database.
These time constants were determined for each prototype configuration and subtracted 
from the leading time values to align all pixels in time space.
Finally, the event time offset was subtracted using simulation to facilitate
comparison of the experimental data to simulation.


The DAQ was started by a signal from the Trigger1 counter. 
Events were required to have signals close to the expected time in 
the Trigger1, Trigger2, TOF1, and TOF2 counters to ensure a well-defined
beam spot and a valid $\pi/p$ tag from the TOF system.

MCP-PMT signals (``hits'') were selected in a time window of $\pm$40~ns relative to 
the Trigger1 time.
Channels with excessive electronics noise above about 1~MHz and one
defective PADIWA card were masked and that same mask was applied to the simulation.

\begin{figure}[h]
	\centering
	\includegraphics[width=.99\columnwidth]{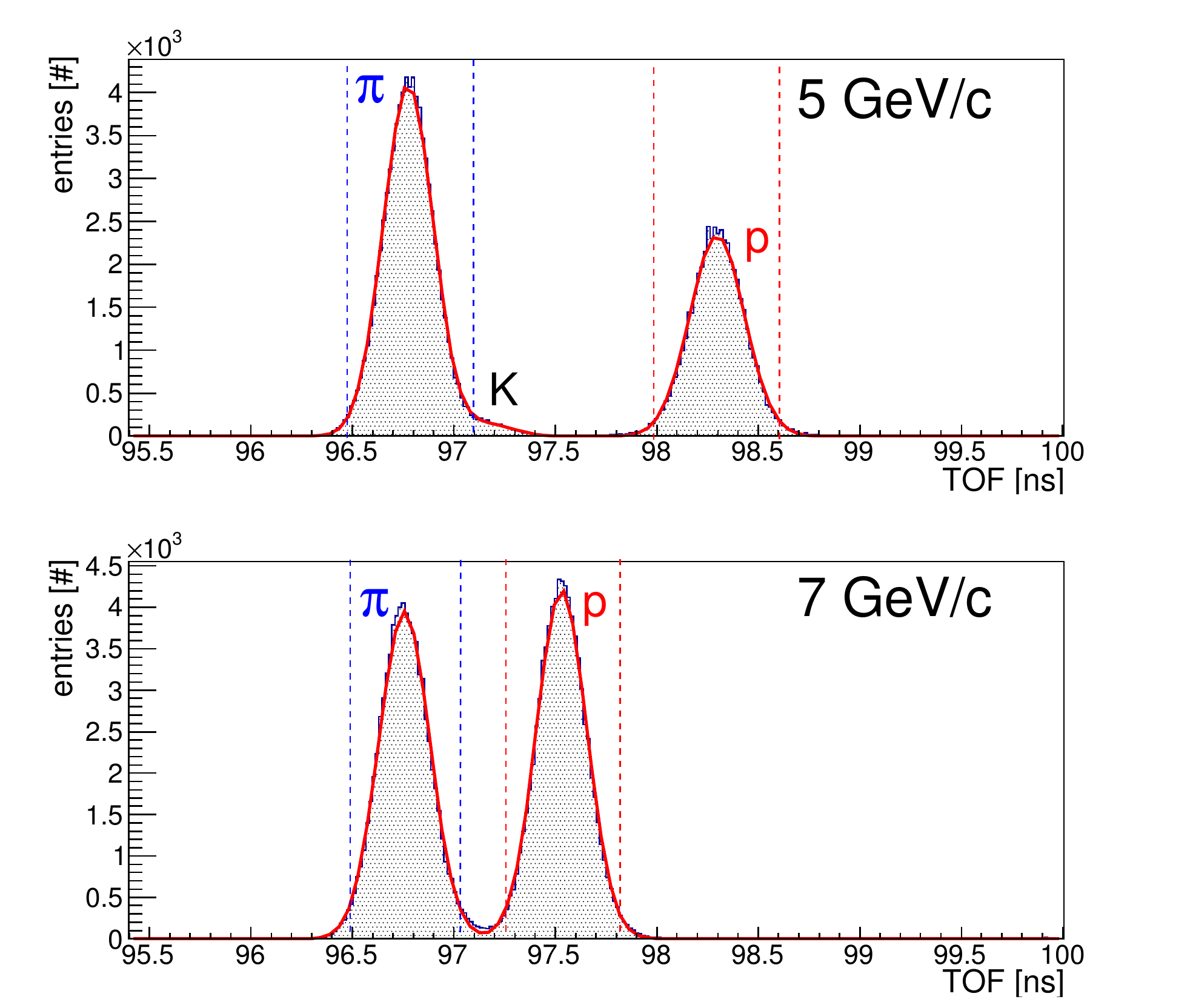}
	\caption{Time difference between the two TOF stations, separated by 29~m flight
		distance, for beam momenta of 5~GeV/c (top) and 7~GeV/c (bottom).
		The dashed lines indicate the two selection windows.
	}
	\label{fig:tof57}
\end{figure}

\begin{figure}[b]
  \centering
   \raggedright{Test beam data, pion tag}
   \hspace*{-2mm}\includegraphics[width=1.04\columnwidth]{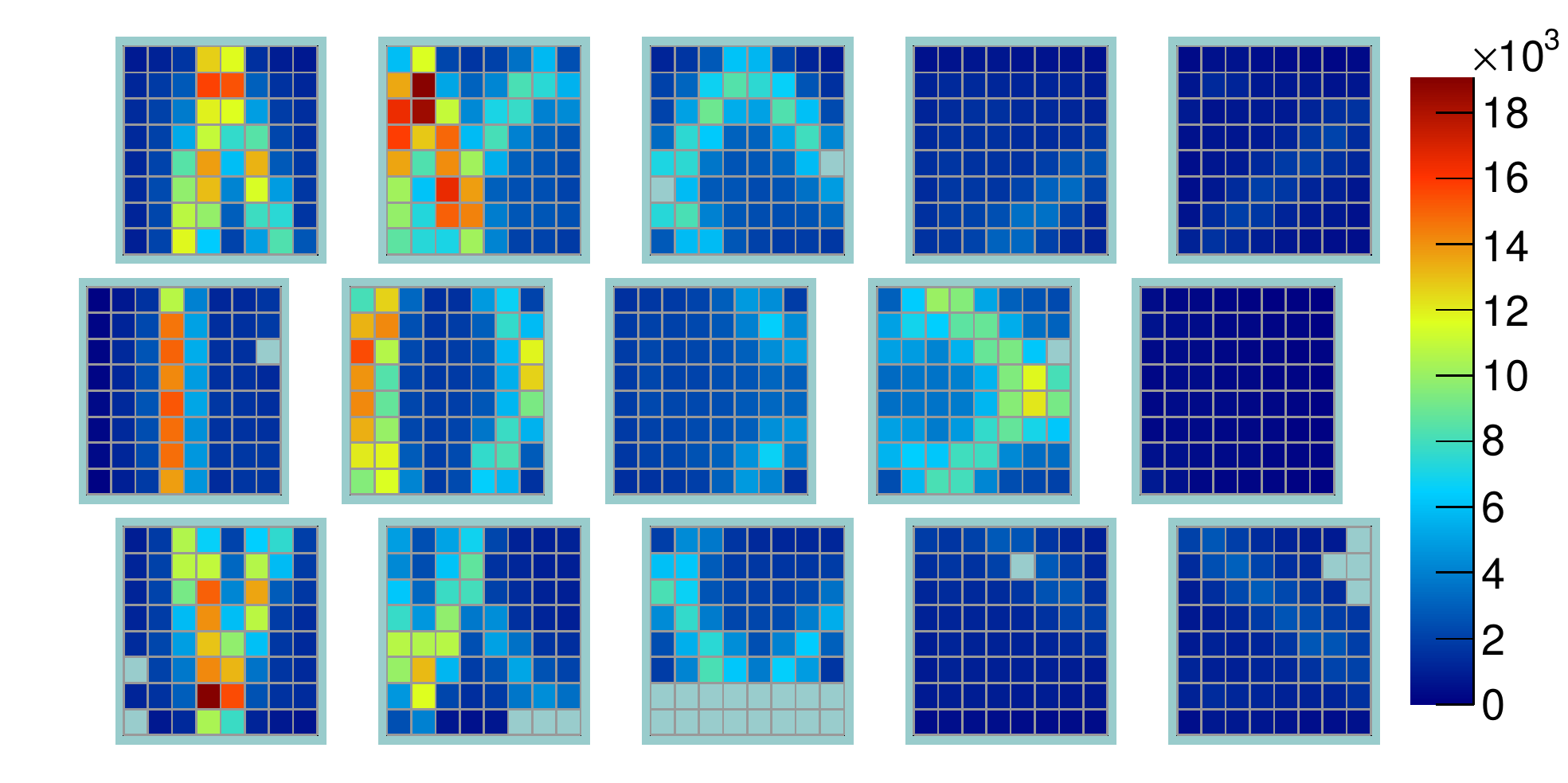}
   \raggedright{Test beam data, proton tag}
   \hspace*{-2mm}\includegraphics[width=1.04\columnwidth]{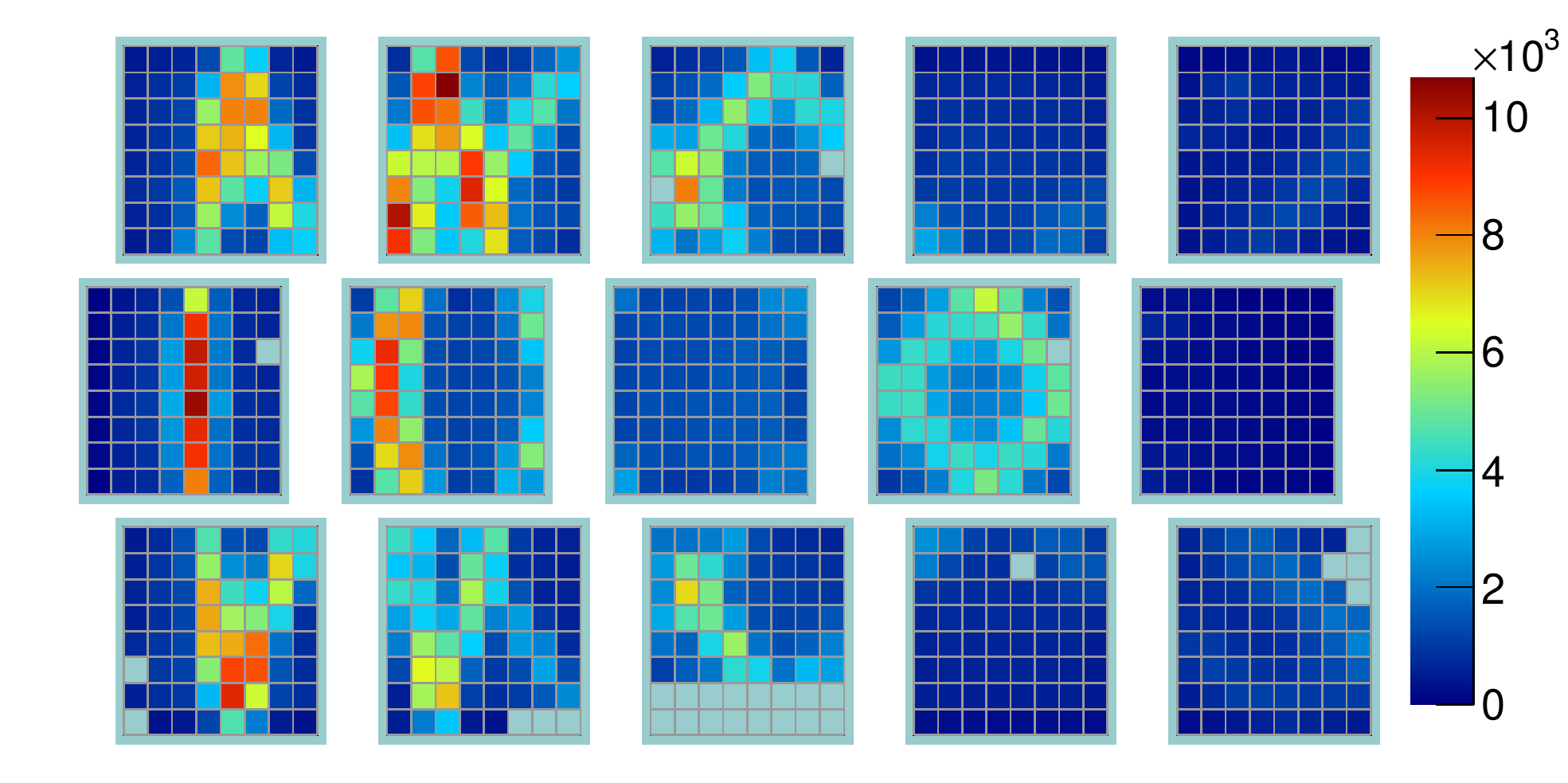}
   \raggedright{Geant simulation, protons}
   \hspace*{-2mm}\includegraphics[width=1.04\columnwidth]{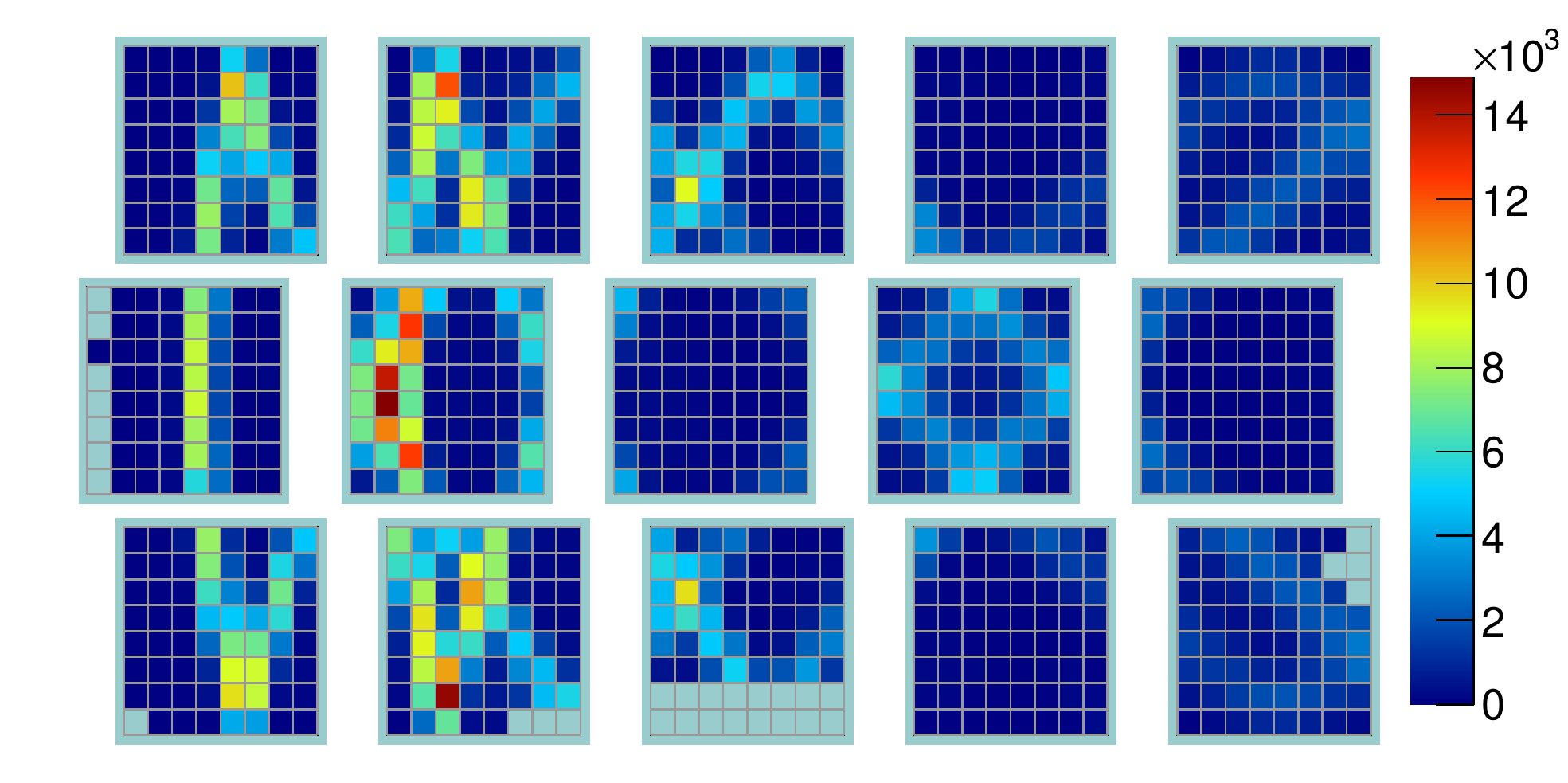}
   \caption{
    Accumulated hit pattern for the 2015 prototype, shown as number of signals per MCP-PMT pixel,
    for the narrow bar with a spherical 3-layer lens and a 5~GeV/c beam with a polar angle of 55$^\circ{}$.
    Experimental data for a pion tag (top) and proton tag (middle) are compared to
    Geant simulation for a proton beam (bottom).
	}
    \label{fig:hit_pattern-bar}
\end{figure}


The time difference measured by the two TOF stations was used to tag an
event as pion or as proton.
The time distributions, shown in Fig.~\ref{fig:tof57} for a beam momentum of 5~GeV/c (top)
and 7~GeV/c (bottom), were fitted with Gaussian functions near the pion and proton peak.
The $\pm 2\sigma$ window around the peak positions was used for selection, indicated
by the dashed lines.
For momenta up to 7~GeV/c this $\pi/p$ tag was very efficient with 
negligible mis-identification.

An example of the hit patterns is shown in Fig.~\ref{fig:hit_pattern-bar} 
for the configuration with the narrow bar and the 3-layer spherical lens. 
The beam momentum was 5~GeV/c and the polar angle between bar and beam was 55$^{\circ}$.
The top and middle figure are for the tagged pions and protons in the
experimental data, respectively.
The complex folded Cherenkov ring image is visibly shifted horizontally by about one column between
the pion and proton tag, due to the smaller Cherenkov angle for protons at 5~GeV/c.
The simulated hit pattern for protons is shown in Fig.~\ref{fig:hit_pattern-bar} (bottom)
and agrees very well with the experimental data, although there was more background
in the beam data than in the simulation.
For higher momenta, when the Cherenkov angle difference for pions and protons 
is smaller, the hit patterns become more difficult to distinguish by eye.

\begin{figure}[h]
  \centering
  \includegraphics[width=.98\columnwidth]{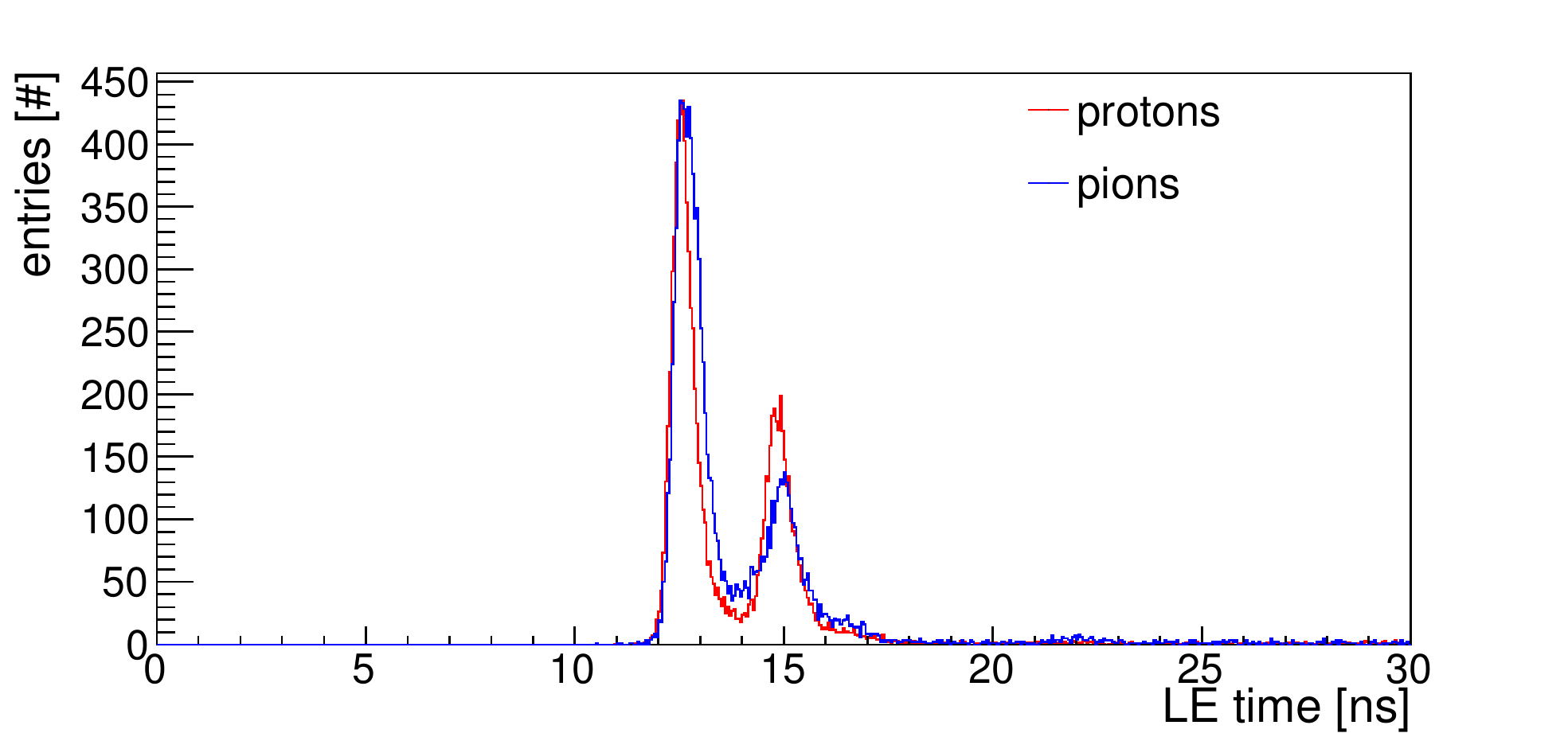}
  \caption{Example of the corrected leading edge time for one typical pixel for a 
    beam with 7~GeV/c momentum and $55^{\circ}$ polar angle.
    Data from the proton tag are shown in red, from the pion tag in blue.
    }
  \label{fig:letime}
\end{figure}

In addition to the difference in the spatial hit pattern there is also an important 
difference in the arrival time of the Cherenkov photons, which forms the basis
of the time-based imaging algorithm, explained in Sec.~\ref{sec:reco-plate}. 
Figure~\ref{fig:letime} shows the corrected leading edge time of the 
detected photons from events tagged as protons (red) and pions (blue) with 7~GeV/c beam momentum 
and $55^{\circ}$ polar angle for the configuration with the narrow bar and the 
3-layer spherical lens.
Since the hit times were corrected for time-of-flight of the beam particles,
the time spectrum corresponds to the time-of-propagation of the Cherenkov photons 
from the emission to the detection.
The multiple peaks in the distribution are due to different paths in the bar 
and prism leading to the same pixel.
A small but significant shift can be seen between the signals from pions and 
protons due to the difference in the Cherenkov angle (about 8.1~mrad).

Two reconstruction methods were used to determine the figures of merit 
(SPR and photon yield) and 
to evaluate the PID performance of different prototype configurations.

The look-up-tables (LUT) for the geometrical reconstruction (Sec.~\ref{sec:reco-geo-bars}) 
were created using the standalone Geant simulation of the prototype.
The single photon Cherenkov angle $\theta_{C}$ was calculated by combining
the beam direction vector with all possible photon directions from the LUT 
for pixels with a hit. 
For a given hit each possible Cherenkov angle value is called an ambiguity.

Two additional selection criteria were applied to reduce the ambiguity 
background in the Cherenkov angle spectra in the beam data.
The size of the beam spot on the bar/plate was reduced with a tight cut
on the beam spot in the fiber hodoscope and in the downstream TOF 
station TOF2.
This narrows the beam profile to a width of about 10~mm, reducing the 
effect of the beam divergence.

The second selection is applied during the geometrical reconstruction. 
Once the photon direction vector is determined, the path of the photon inside
the bar and the prism can be calculated. 
Assuming the group velocity of a photon with 380~nm wavelength (the average 
wavelength of photons detected in this prototype), the expected photon 
propagation time can be calculated and subtracted from the measured 
photon hit time. 

\begin{figure}[h]
  \centering
  \includegraphics[width=1\columnwidth]{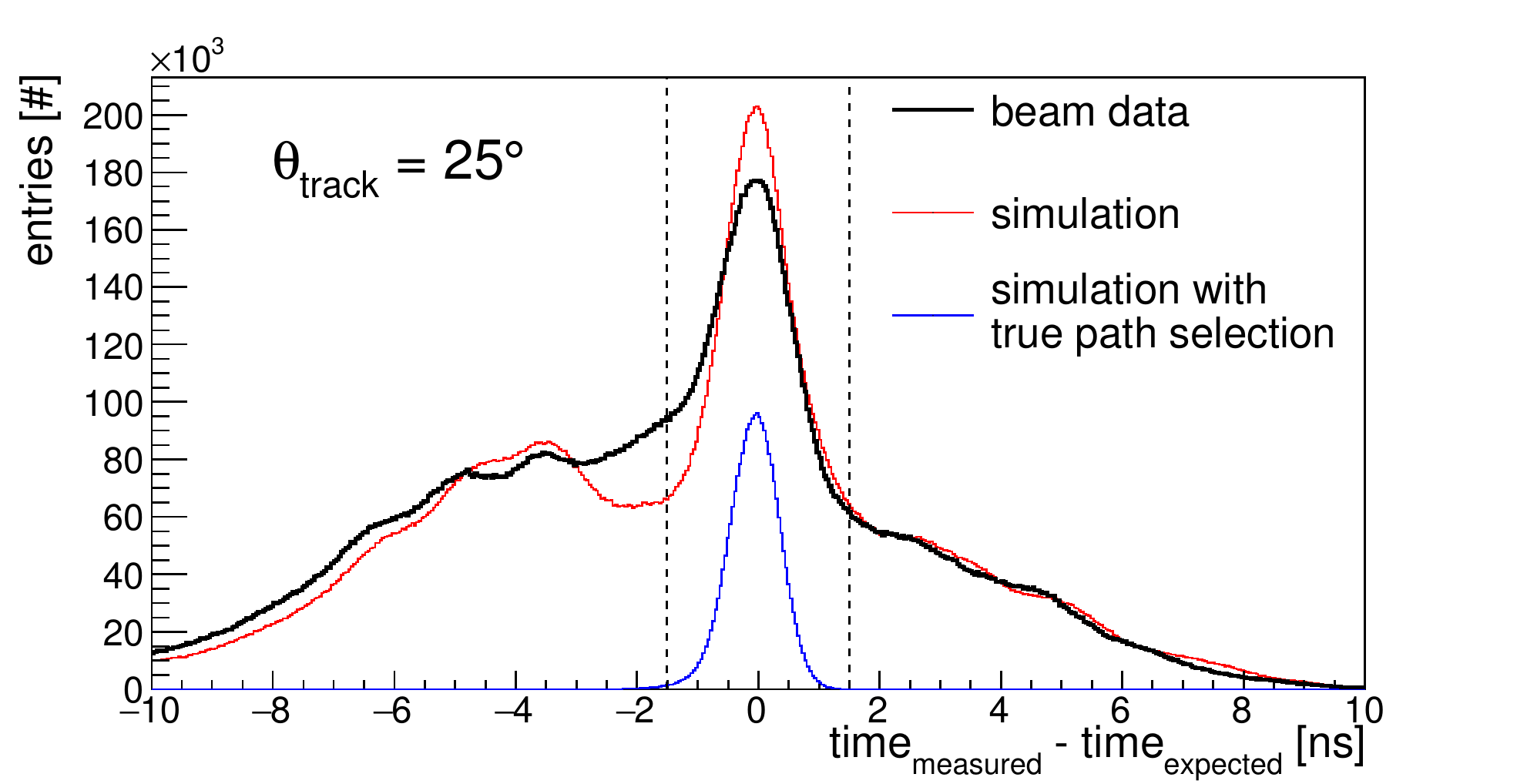}
  \includegraphics[width=1\columnwidth]{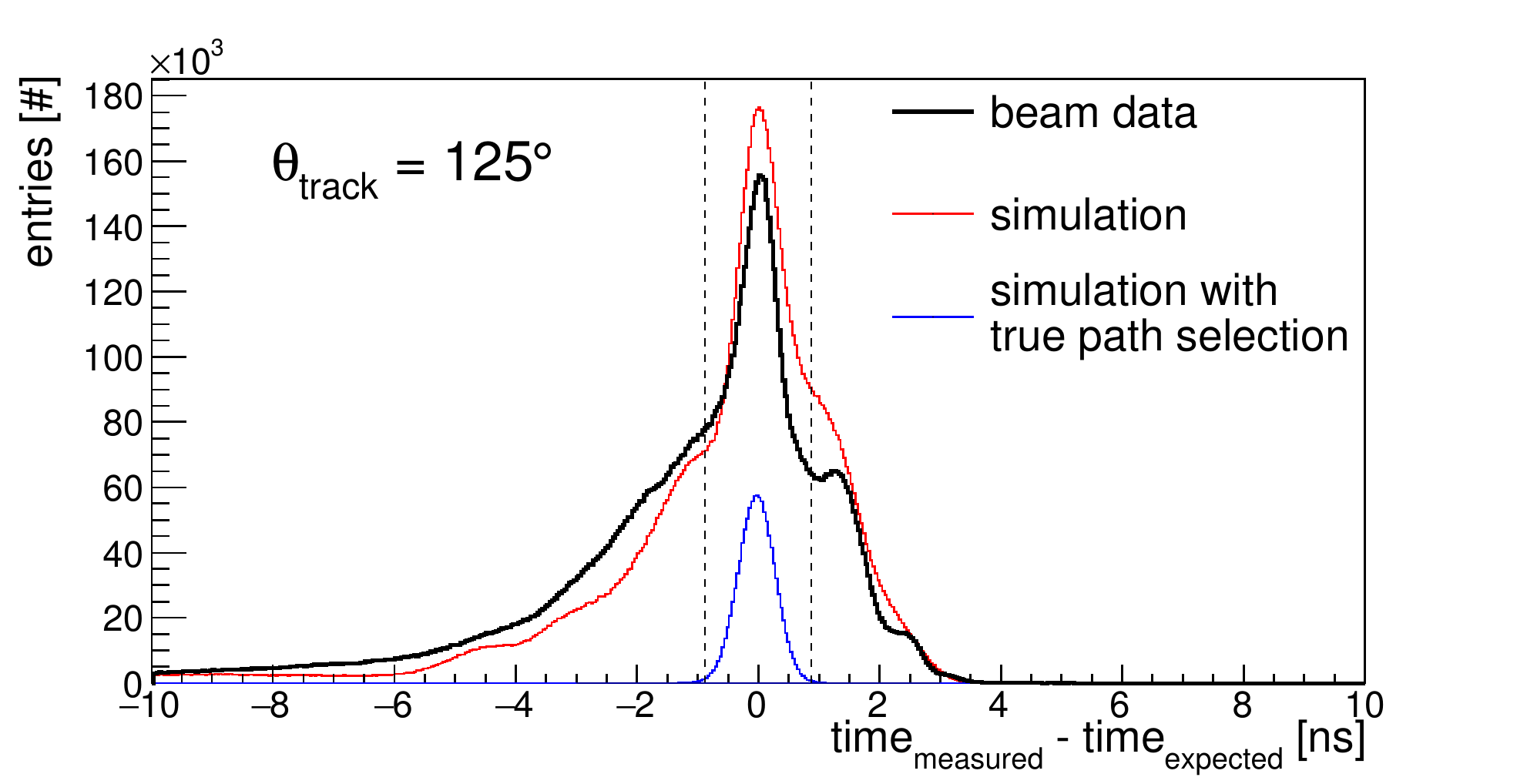}
  \caption{Difference between the measured and expected arrival time of 
    Cherenkov photons for the narrow bar and the 3-layer spherical lens.
	The beam momentum was 7~GeV/c and the polar angle was either $25^{\circ}$
    (top) or $125^{\circ}$ (bottom).
    Proton-tagged beam data are compared to simulation.
  }
  \label{fig:timediff}
\end{figure}

Figure~\ref{fig:timediff} shows the difference between measured and expected 
hit time, $\Delta t$, for the configuration with a narrow bar and the 3-layer spherical lens,
a beam momentum of 7~GeV/c and a polar angle of either $25^{\circ}$
(top) or $125^{\circ}$ (bottom). 
The red line shows the simulation result in comparison to the beam data with
a proton tag (black line).
The shape of both distributions, dominated by the ambiguity background, agrees 
reasonably well. 
As for the Cherenkov angle distribution, the correct photon propagation paths 
create a peak around $\Delta t \approx 0$  while ambiguous paths form a complex 
background.
This is seen clearly in simulation when the reconstruction is performed only
for the correct photon path (blue line).
The distribution for $25^{\circ}$ is significantly wider than for $125^{\circ}$
due to the longer photon path in the bar and prism, leading to a much larger
chromatic dispersion in the photon propagation time.
The selection on $\Delta t$ is then defined as, for example, $\vert \Delta t \vert \le 1.5$~ns
for $25^{\circ}$ and $\vert \Delta t \vert \le 0.8$~ns for $125^{\circ}$,
as indicated by the vertical dashed lines.
This cut reduces the number of ambiguities per photon significantly.

\begin{figure}[h]
	\centering
	\includegraphics[width=1\columnwidth]{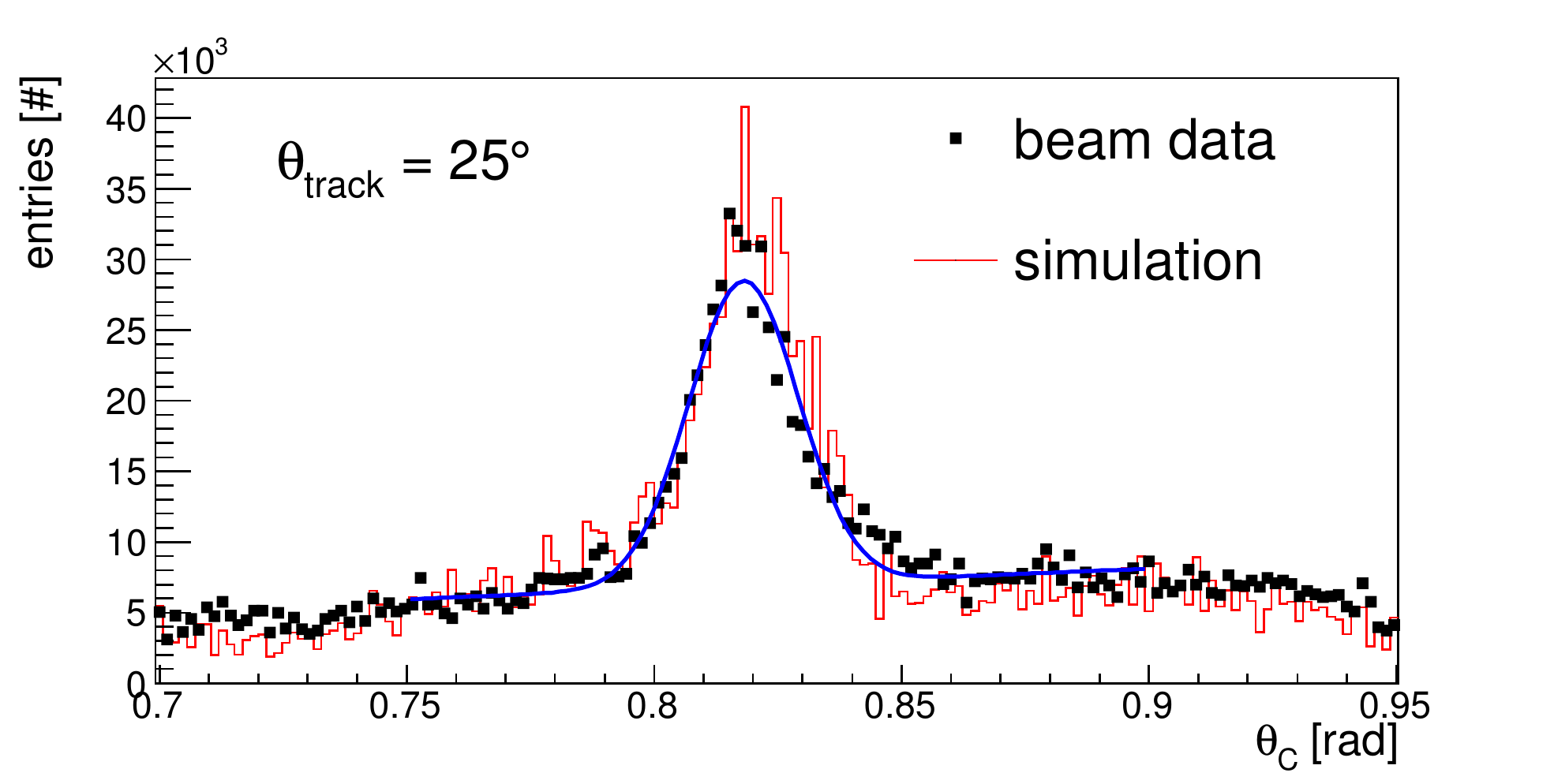}
	\includegraphics[width=1\columnwidth]{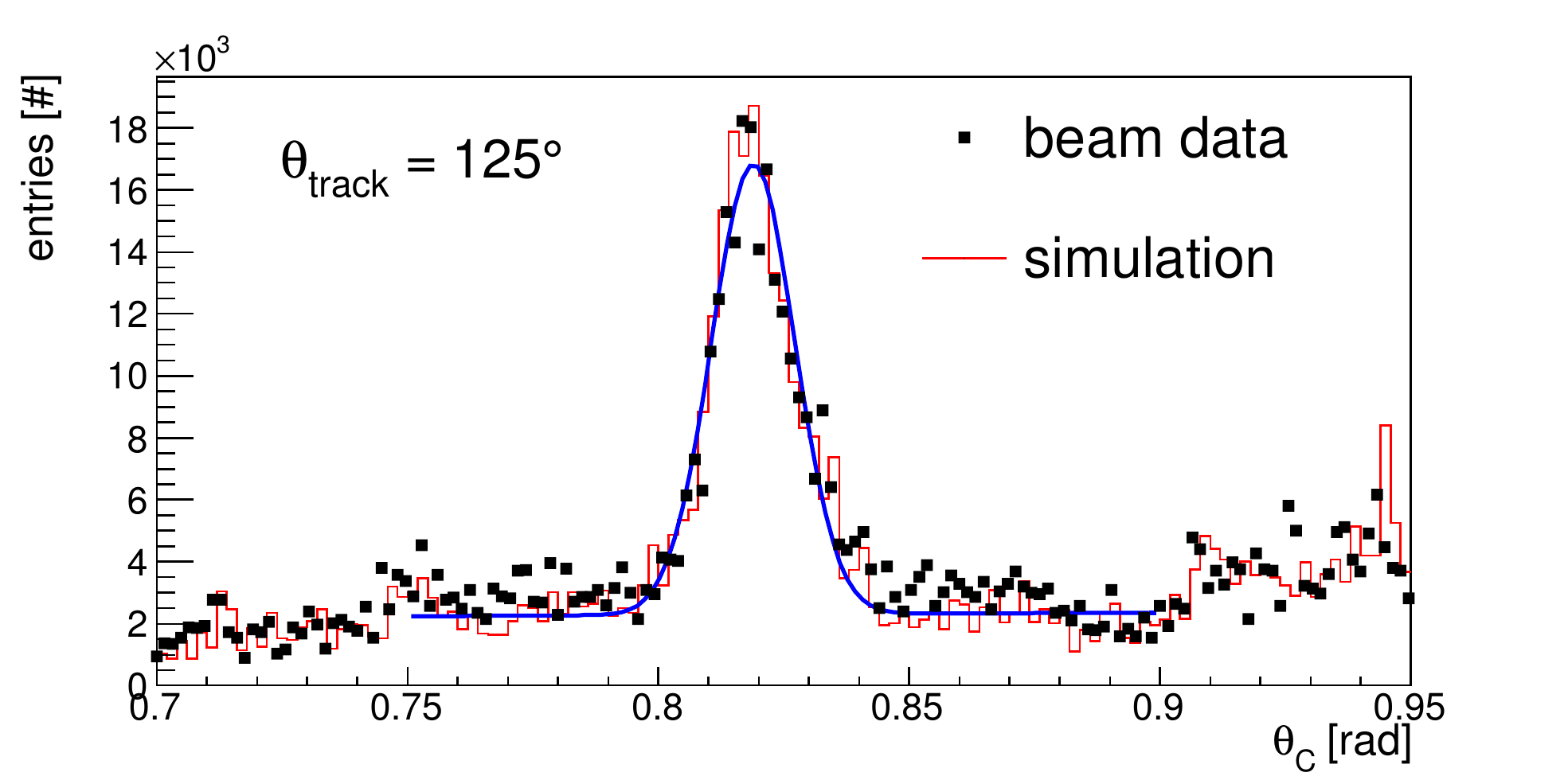}
	\caption{Single photon Cherenkov angle distribution for 5000 events in data (proton
		tag) and simulation (protons) for the narrow bar and the 3-layer spherical lens.
		The beam momentum was 7~GeV/c and the polar angle was either $25^{\circ}$
		(top) or $125^{\circ}$ (bottom).
	} 
	\label{fig:spr_ca}
\end{figure}

Figure~\ref{fig:spr_ca} shows $\theta_{C}$ after all selection cuts 
for the configuration with a narrow bar and the 3-layer spherical lens.
The beam momentum was 7~GeV/c and the polar angle was either $25^{\circ}$
(top) or $125^{\circ}$ (bottom).
The beam data for 5000 proton-tagged events (points) is compared to 
simulated protons (line).
A fit of a Gaussian plus a linear background to the beam data distributions
yields SPR values of 11~mrad at $25^{\circ}$ and 8~mrad at $125^{\circ}$.
The simulation describes the properties of the experimental data well, 
both in the signal region and in the area of the combinatorial background.

\begin{figure}[h]
	\centering
	\includegraphics[width=1\columnwidth]{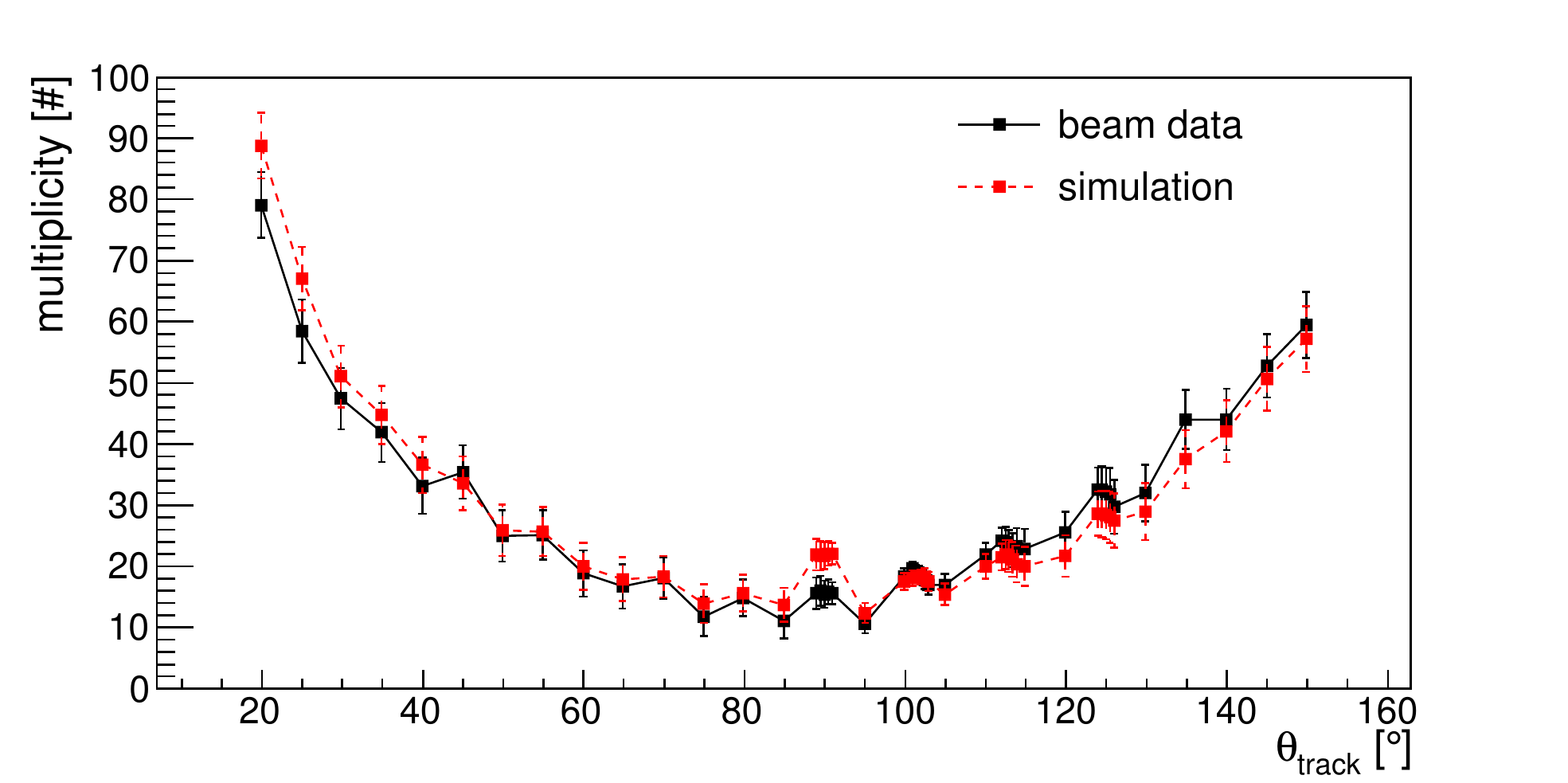}
	\includegraphics[width=1\columnwidth]{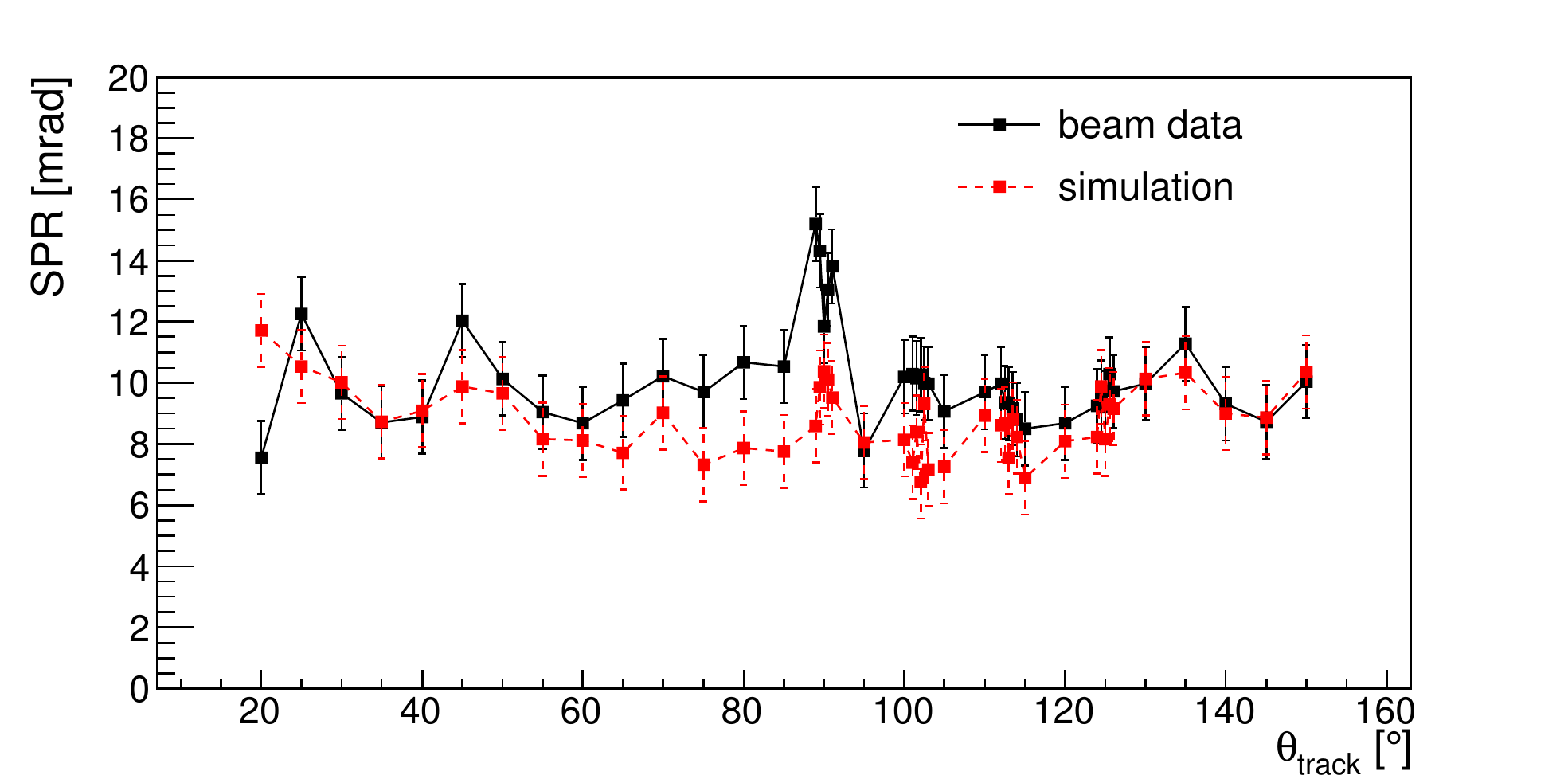}
	\caption{Photon yield (top) and SPR (bottom) as a function of the track polar angle 
		for the narrow bar and the 3-layer spherical lens for tagged protons at
		7~GeV/c beam momentum in data (black) and Geant simulation (red). 
		The error bars correspond to the RMS of the distribution in each bin. }
	\label{fig:nph-spr}
\end{figure}

To eliminate the contribution of the ambiguity background on the determination
of the photon yield, only photons with at least one ambiguity in a $\pm 3\sigma$
window around the expected value of the Cherenkov angle are counted.
The reconstructed photon yield as a function of the track polar angle is shown 
in Fig.~\ref{fig:nph-spr}\,(top) for the configuration with the narrow bar radiator and 
the 3-layer spherical lens.

The number of Cherenkov photons from the beam data (black) ranges from 12 to 80 
and is in agreement with simulations (red). 
The distribution has a peak near perpendicular incidence at $90^{\circ}$ where 
the entire Cherenkov cone is totally internally reflected.
The yield drops for smaller and larger polar angles as part of the ring escapes 
the bar until it rises again as the length of the particle track in the bar increases.

For polar angles below 40$^\circ{}$ the simulation overestimates the photon yield 
by about 10\%, which may be an indication of an incorrect mirror reflectivity 
value used in the simulation.
Around $90^{\circ}$ the photon yield in the beam data is more than 30\% below the
simulation.
This difference is most likely due to the way the MCP-PMTs were selected for this
beam test.
The newer units with the higher gain were placed into the columns on the left side 
of the prism (as viewed from behind) to get the best efficiency for the angles 
below 60$^\circ{}$, which is the most difficult range in terms of PID in \panda.
The older units were placed in the right column, where they only affect the ring
image of polar angles in the range of 80--100$^\circ{}$, a less demanding 
region for \panda.
Since the modified PADIWA cards shaped and attenuated the signal more than expected,
it is likely that this loss in amplitude affected mainly the older MCP-PMTs, 
and, thus, the polar angles around 90$^\circ{}$. 

It should be noted that both the measured and simulated photon yield of the 
prototype are lower than the yield expected for the baseline design in \panda 
due to the larger gaps between MCP-PMTs in the prototype setup.
Furthermore, the prototype photon yield numbers include a contribution 
from charge sharing between MCP-PMT anode pads. 
This effect is included in the prototype simulation, based on measurements
performed for the PHOTONIS Planacon MCP-PMTs, and estimated to contribute on 
average about 15\% to the reported photon yield.

The single photon Cherenkov angle resolution for the same data set 
(narrow bar and the 3-layer spherical lens for tagged protons at 7~GeV/c momentum) 
is shown in Fig.~\ref{fig:nph-spr}\,(bottom). 
The beam data and simulation are consistent within the RMS of the distributions 
for the forward and backward angles.
Again, the beam data performance is somewhat worse than simulation for 
polar angles around 90$^\circ{}$ and the less efficient MCP-PMT/PADIWA
coverage is contributing to this effect as well.
The poorer timing resolution and the additional background in the data particularly 
affects this polar angle region because the shape of the combinatorial background
is especially complicated due to many overlapping ambiguities.
This makes the $\Delta t$ selection less efficient, the fits to the $\theta_C$ distributions 
less stable and the width larger.

\begin{figure}[tbh]
  \centering
  \includegraphics[width=.99\columnwidth]{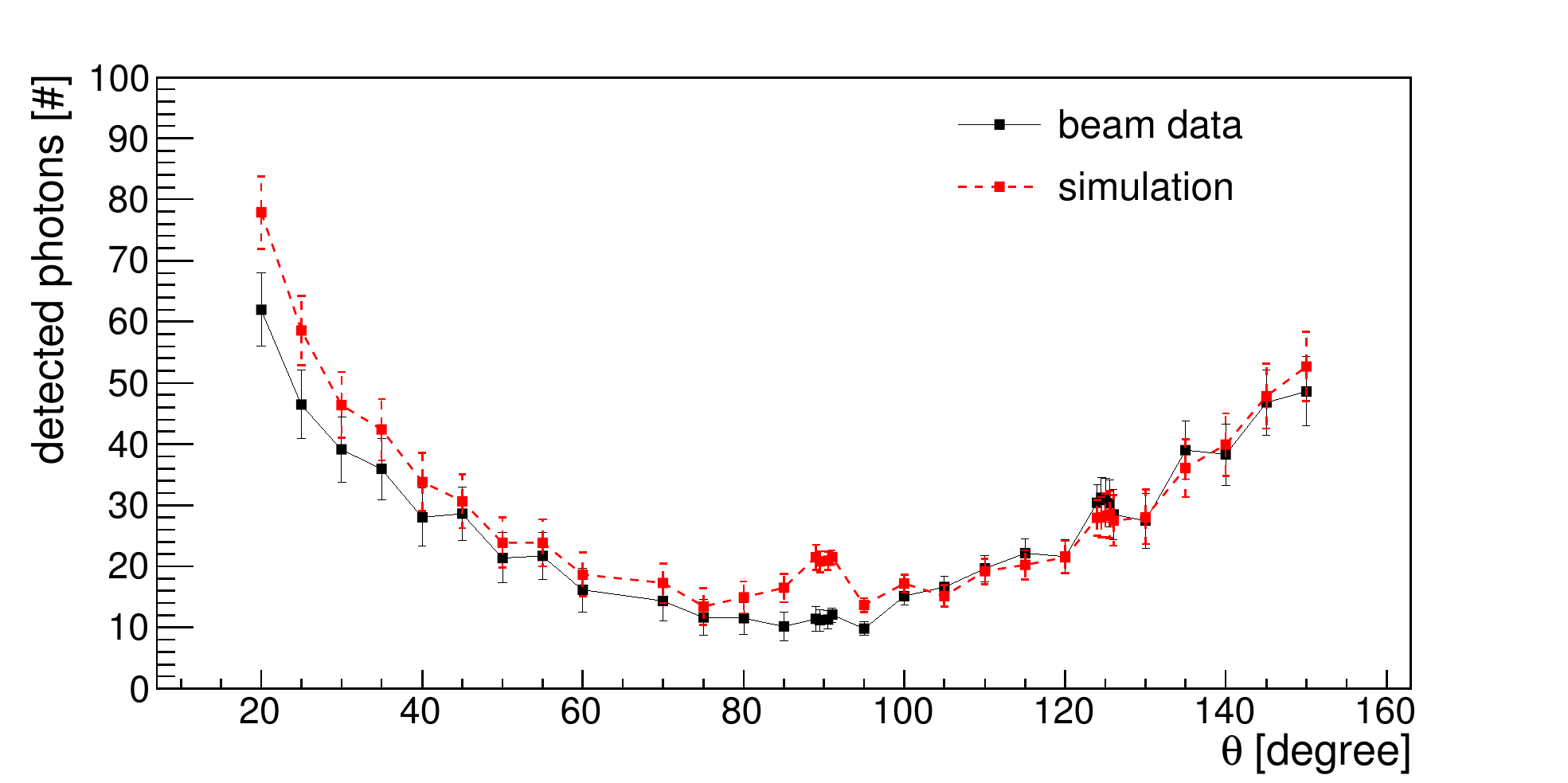}
  \includegraphics[width=.99\columnwidth]{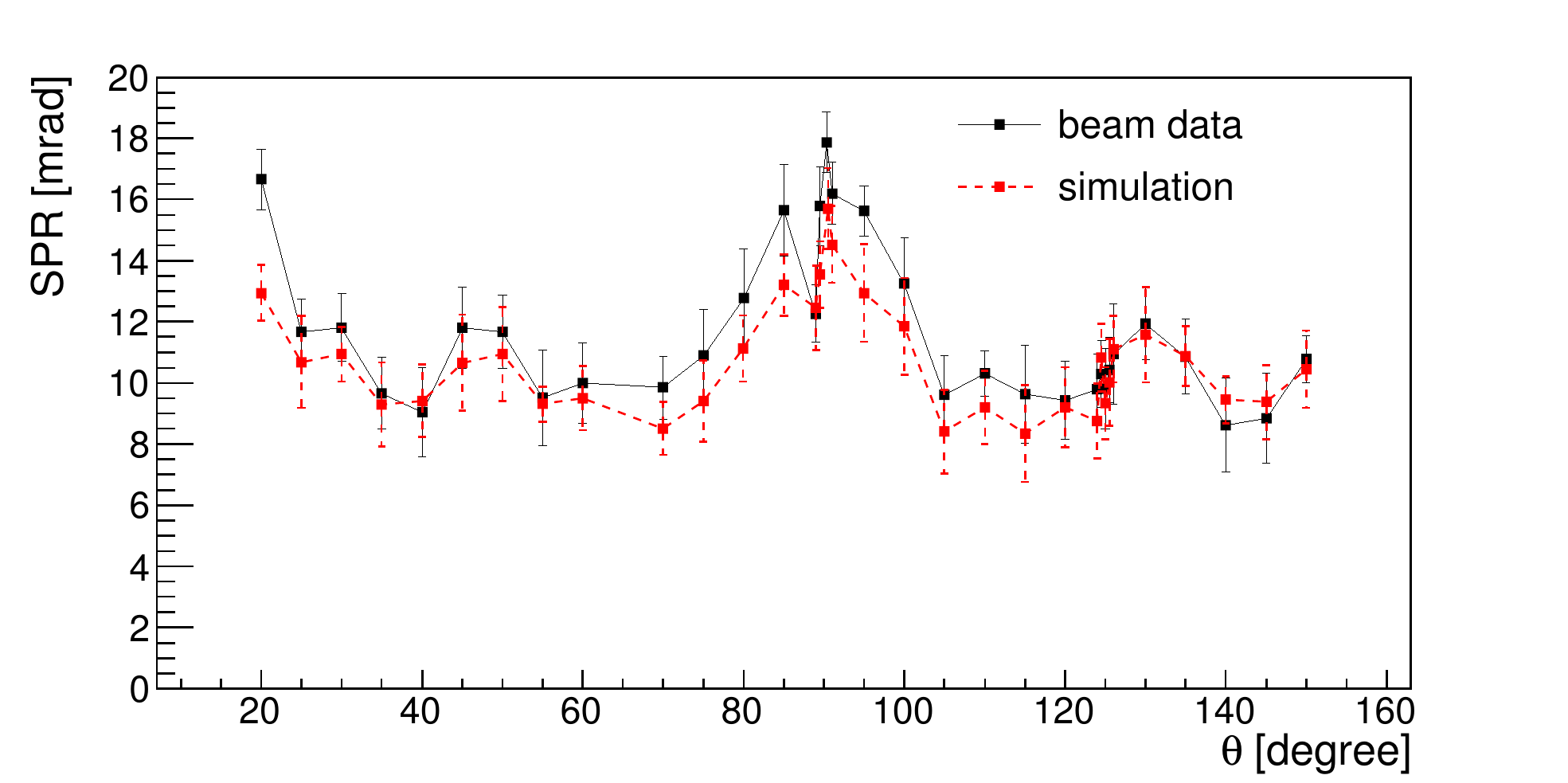}
  \caption{
      Photon yield (top) and SPR (bottom) as a function of the track polar angle for the narrow bar 
      and the 2-layer spherical lens for tagged protons with 7~GeV/c momentum in data 
      (black) and Geant simulation (red). 
      The error bars correspond to the RMS of the distribution in each bin. 
    }
  \label{fig:photons_sim-data7gev}
\end{figure}

The photon yield and SPR for the narrow bar and a 2-layer spherical lens are compared
for the 7~GeV/c beam to simulation in Fig.~\ref{fig:photons_sim-data7gev}.
The overall performance is worse than for the 3-layer spherical lens, as expected.
The differences between the simulation and the experimental data are similar to the
spherical lens configuration.

\begin{figure}[tbh]
  \centering
  \includegraphics[width=.99\columnwidth]{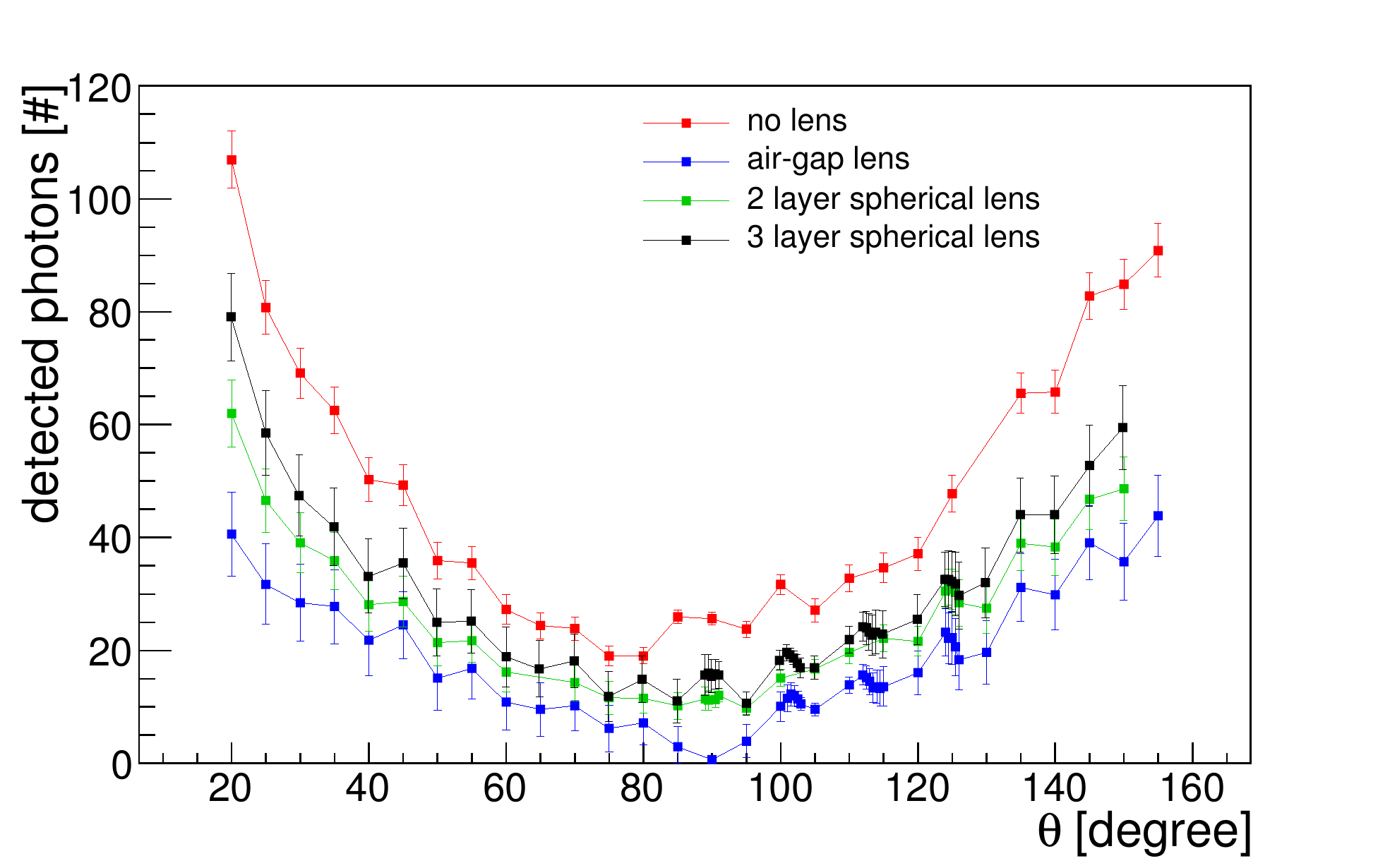}
  \includegraphics[width=.99\columnwidth]{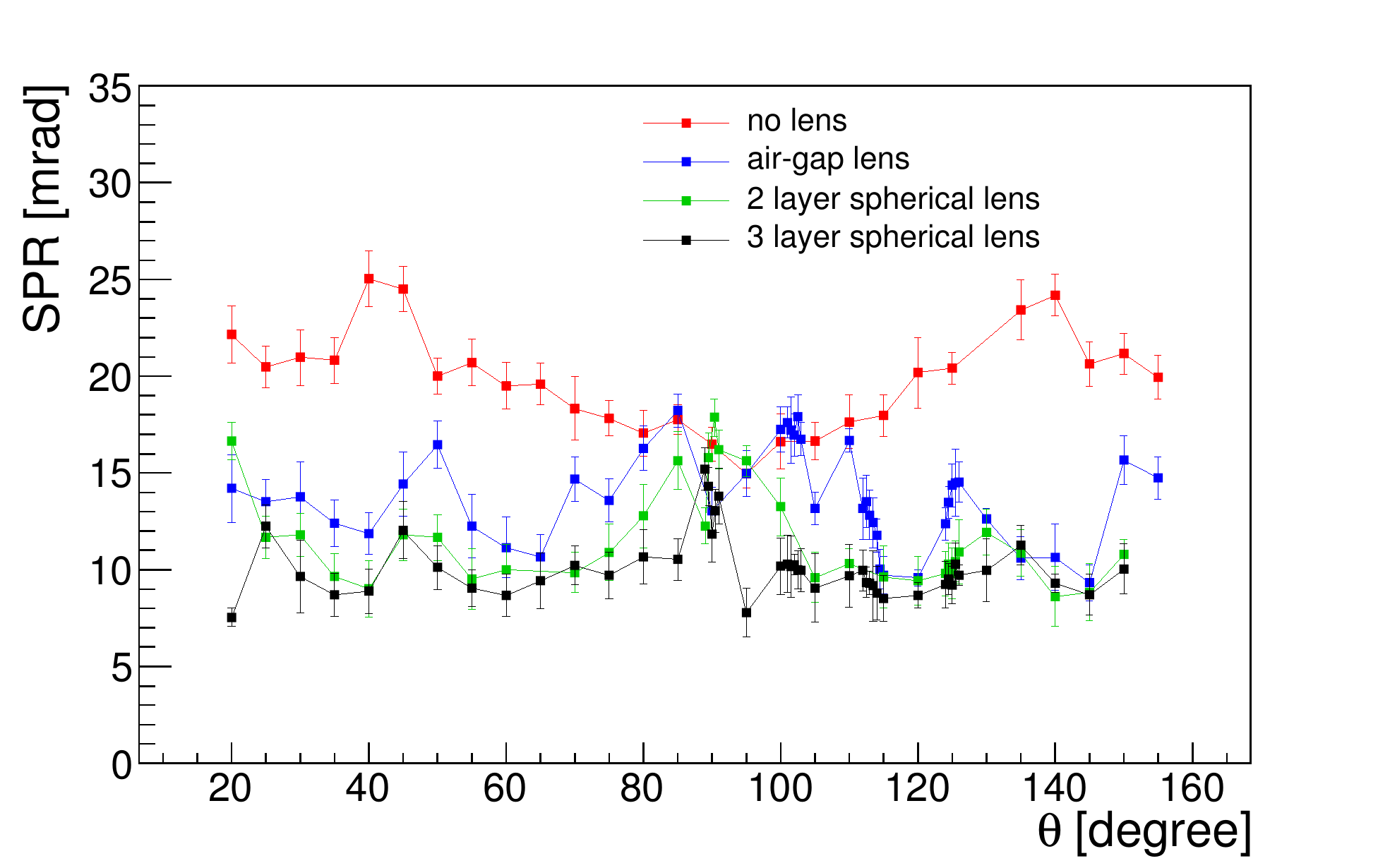}
     \caption{
         Photon yield (top) and SPR (bottom) as a function of the track polar angle for the narrow bar 
         and various focusing options for tagged protons with a momentum of 7~GeV/c.
         The error bars correspond to the RMS of the distribution in each bin. 
		}
  \label{fig:nph-spr_data7gev}
\end{figure}

Figure~\ref{fig:nph-spr_data7gev}\,(top) compares the photon yield measured for the narrow
bar and various focusing options with tagged protons at 7~GeV/c momentum.
The best photon yield is achieved when the bar is coupled directly to the prism.
The two multi-layer lens configurations show a lower yield by at least 40\%, primarily
due to losses from reflections at the unpolished sides of the lens, 
but still perform well with a yield of 10 photons or more at all angles.
An unacceptable photon loss is observed for the spherical lens with an air gap,
in particular for polar angles around $90^{\circ}$.

The single photon Cherenkov angle resolution is shown in Fig.~\ref{fig:nph-spr_data7gev}\,(bottom)
for tagged protons at 7~GeV/c for the narrow bar and various focusing options.
Although the photon yield is highest for the configuration without focusing, the SPR
is by far the worst (red line). The 3-layer spherical lens provides the best resolution.

These findings for the figures of merit are in good agreement with the simulation 
design study, presented in Sec.~\ref{sec:sim-design-options}, and confirm the choice 
of the baseline design configuration.

\subsection{PID Performance of the Narrow Bar Design}
\label{sec:prt_performance}

Both the geometrical and the time-based imaging reconstruction were applied to the
geometry with the narrow bar and the 3-layer spherical lens to determine the
$\pi/p$ separation at 7~GeV/c, the equivalent of the $\pi/K$ separation at 3.5~GeV/c.

In the geometrical reconstruction the Cherenkov angle of the track is determined
by fitting for each track the single photon Cherenkov angle distribution for all
hits and ambiguities to a Gaussian plus a straight line.
The width of the difference between this measured Cherenkov angle and the
expected Cherenkov angle is defined as the Cherenkov angle resolution per track 
$\sigma_{C,track}$.
This approach is similar to the ``track maximum likelihood fit'' method used for the
BaBar DIRC~\cite{babar-dirc-pe}.

\begin{figure}[tbh]
  \centering
  \includegraphics[width=.98\columnwidth]{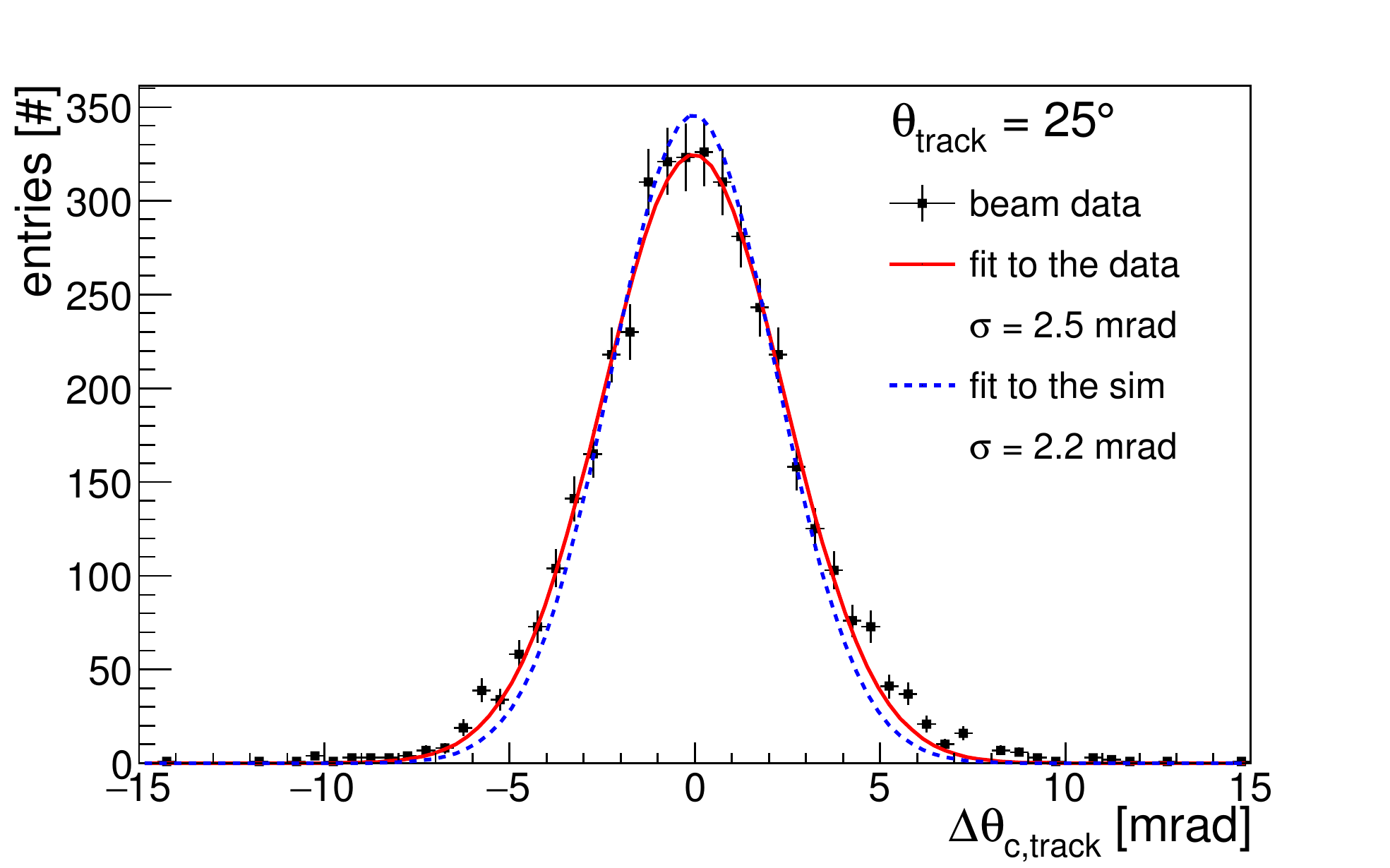}
  \includegraphics[width=.98\columnwidth]{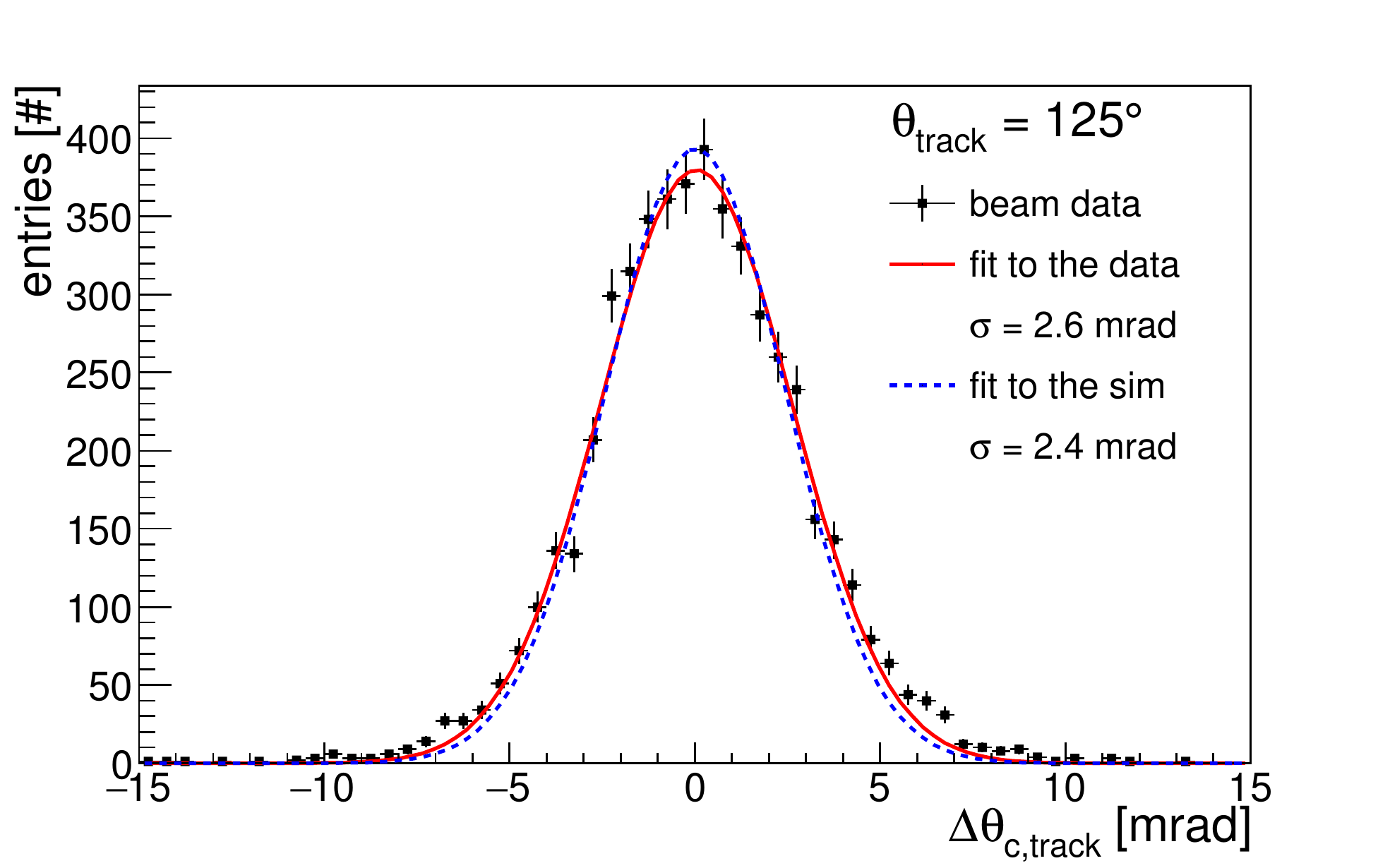}
  \caption{Resolution of the reconstructed Cherenkov angle per track for proton tags (data)
  and protons (simulation) with a momentum of 7~GeV/c, a polar angle of $25^{\circ}$ (top) 
  and $125^{\circ}$ (bottom).
  The narrow bar with the 3-layer spherical lens is used.}
  \label{fig:trackres}
\end{figure}

The track-by-track Cherenkov angle fit was performed for 5000 proton-tagged events
in the beam data and is compared to 5000 simulated protons in Fig.~\ref{fig:trackres}.
The distributions are shown for the narrow bar with the 3-layer spherical lens,
a momentum of 7~GeV/c, a polar angle of $25^{\circ}$ at the top and $125^{\circ}$ 
at the bottom.

A Gaussian function (red line) was used to fit the distributions in the range of 
$\pm 4$~mrad, resulting in the Cherenkov angle per track resolution 
of $\sigma_{C,track}=2.5\pm 0.2$~mrad for the $25^{\circ}$ polar angle and 
and $\sigma_{C,track}=2.6\pm 0.2$~mrad for the $125^{\circ}$ polar angle. 
Both beam data distributions are in reasonable agreement with the simulations 
(dashed blue line) which, however, overestimated the resolution by about 9\%.

With these values of the track Cherenkov angle resolution the $\pi/p$ separation
at 7~GeV/c can be calculated as the Cherenkov angle difference of pions and
protons at this momentum ($\Delta(\theta_C)=8.1$~mrad), divided by $\sigma_{C,track}$.
This results in a 3.3~s.d. separation value at $25^{\circ}$ and 3.1~s.d 
for $125^{\circ}$.
The corresponding values for the $\pi/K$ separation power at 3.5~GeV/c
($\Delta(\theta_C)=8.5$~mrad) are 3.5~s.d. separation at $25^{\circ}$ 
and 3.3~s.d for $125^{\circ}$.

The second approach to utilizing the geometrical reconstruction results for 
PID evaluation is to perform a direct track-by-track particle hypothesis 
likelihood test.
Instead of fitting for the Cherenkov angle, as done in the first method
discussed above, only the likelihood is calculated for the single photon 
Cherenkov angle distribution for a track to originate from a pion or a proton.

\begin{figure}[tbh]
  \centering
  \includegraphics[width=.99\columnwidth]{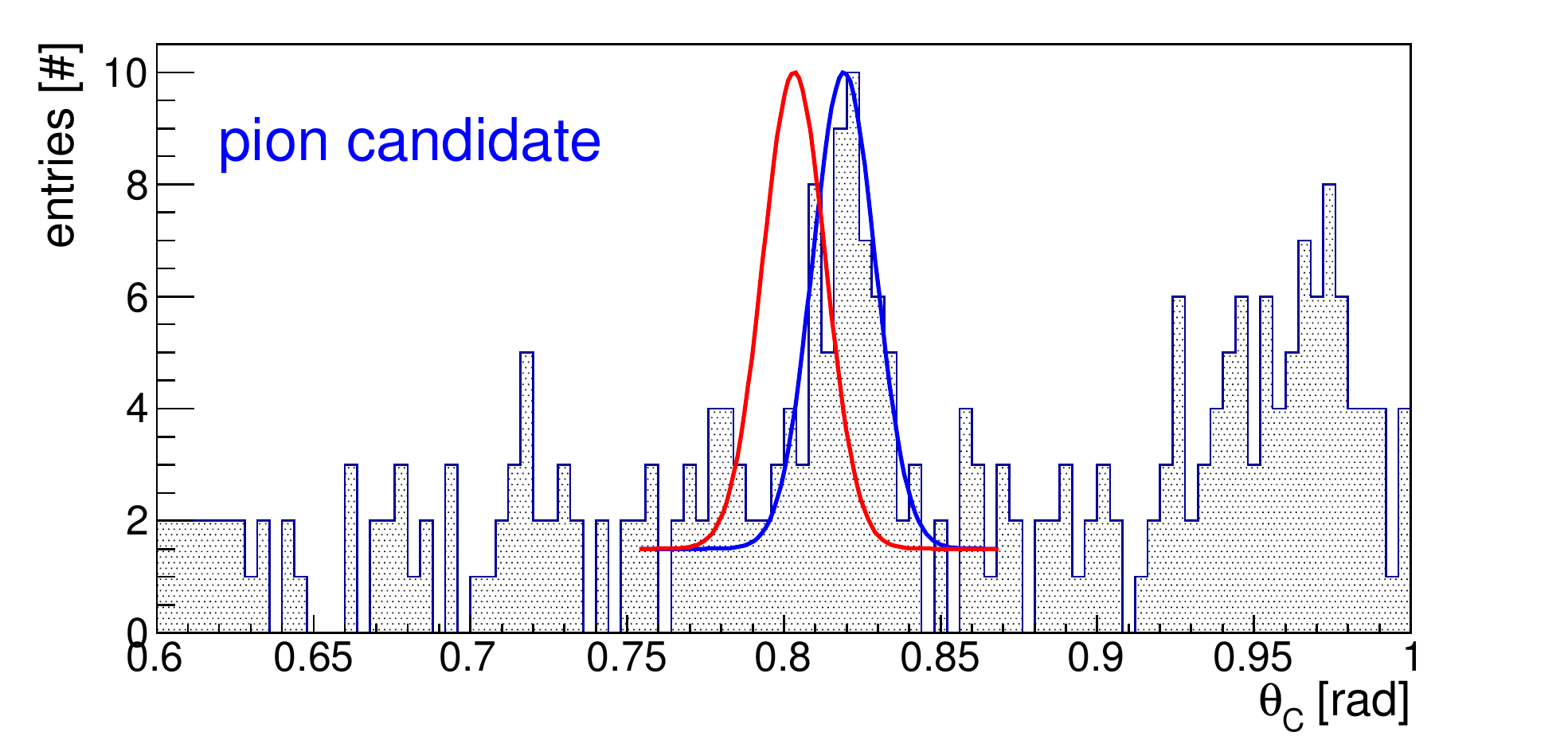}
  \includegraphics[width=.99\columnwidth]{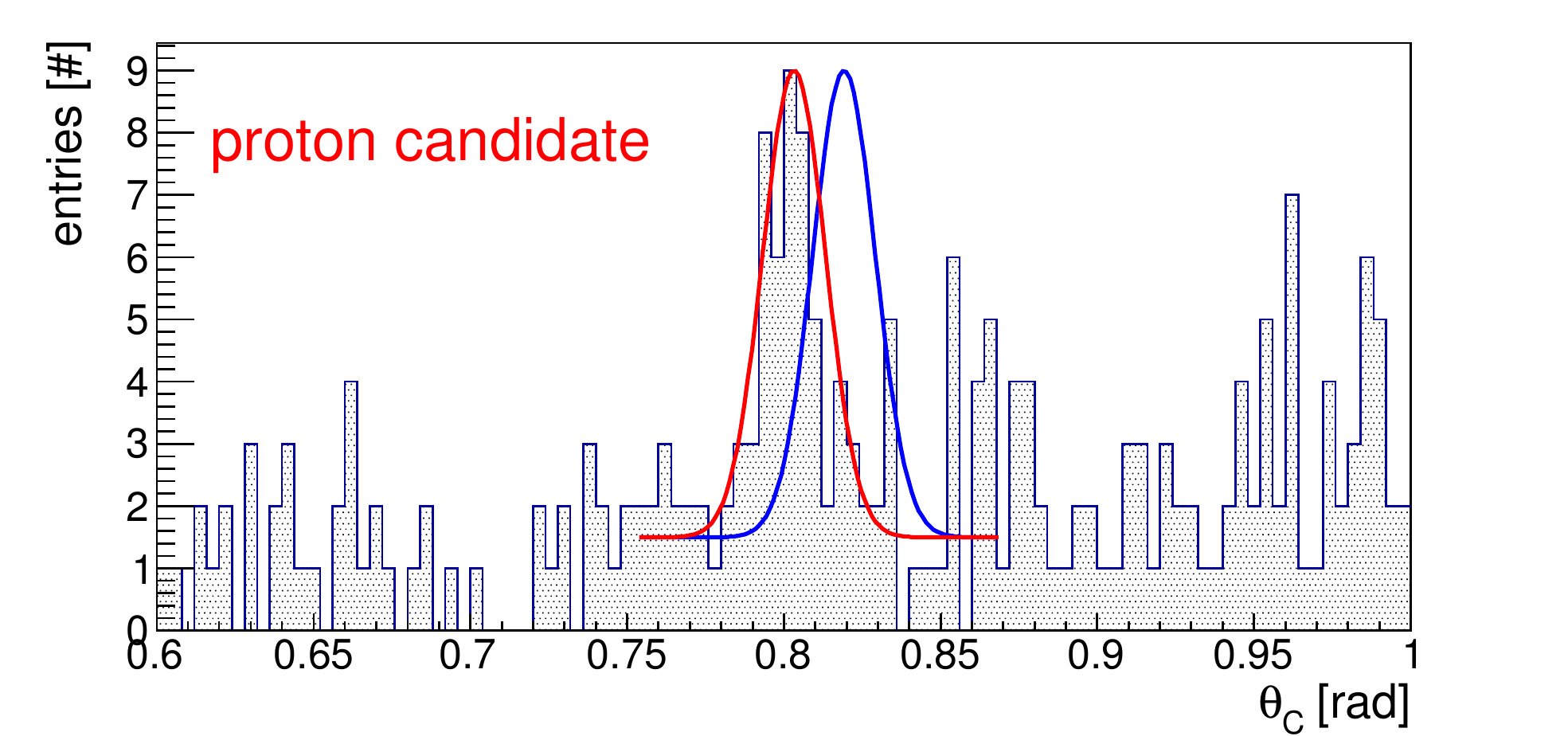}
  \caption{
  Examples of the single photon Cherenkov angle distributions for single TOF-tagged
  beam data events.
  The red and blue lines indicate the expected distribution for the proton and pion mass 
  hypotheses.
  The distributions are for the narrow bar with the 3-layer spherical lens, a  
  beam momentum of 5~GeV/c with a $25^{\circ}$ polar angle.
  }
\label{fig:lh_geomreco}
\end{figure}

For each event the single photon Cherenkov angle distribution for all hits and 
ambiguities is compared to a Gaussian plus a linear background, 
where the mean value of the Gaussian is fixed to the expected Cherenkov angles 
for either pions or protons, respectively, and the Gaussian width is fixed to the 
expected SPR for that polar angle.
Figure~\ref{fig:lh_geomreco} shows the examples of the single photon Cherenkov 
angle distribution for two TOF-tagged beam data events taken at 5~GeV/c 
momentum and $25^{\circ}$ polar angle using the narrow bar and the 3-layer spherical lens. 
The red and blue lines indicate the expected distribution for the proton and pion mass 
hypothesis. 
The upper distribution is in agreement with the pion hypothesis, providing larger 
likelihood value for pions comparing to protons. 
The opposite is seen for the bottom distribution for the proton candidate.

\begin{figure}[tbh]
  \centering
  \includegraphics[width=.99\columnwidth]{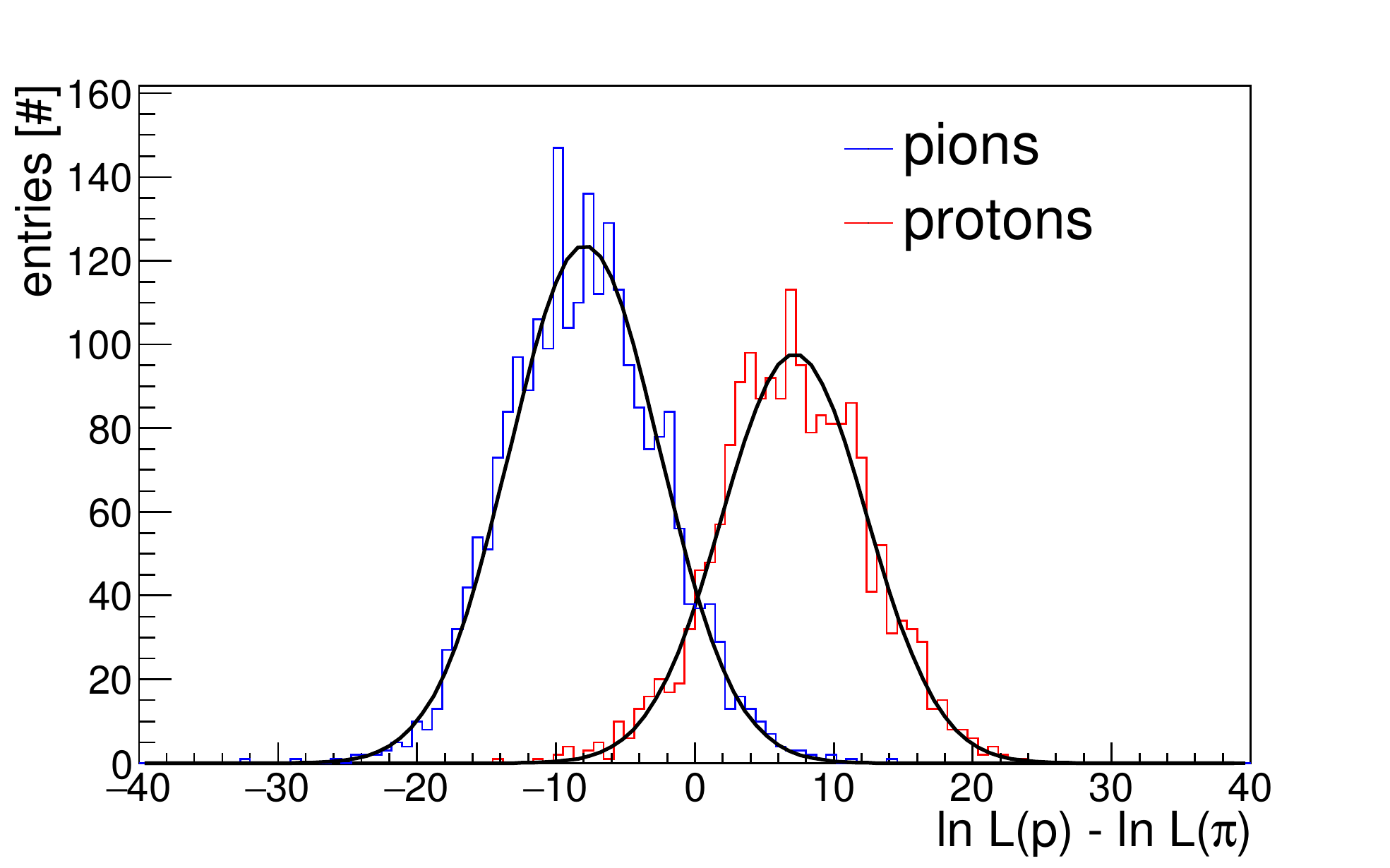}
  \caption{
  Proton-pion log-likelihood difference distributions for proton-tagged (red) and 
  pion-tagged (blue) beam events as result of geometrical reconstruction. 
  The distributions are for the narrow bar with the 3-layer spherical lens, a 
  beam with 7~GeV/c momentum and $25^{\circ}$ polar angle. The $\pi/p$ separation
  power from the Gaussian fits is 2.9~s.d.
    }
\label{fig:lh_geomreco_7_25}
\end{figure}

The result of the track-by-track unbinned likelihood calculation of the $\pi/p$ 
hypothesis tests for the beam data taken at 7~GeV/c and $25^{\circ}$ polar angle
with the narrow bar and the 3-layer spherical lens is shown in Fig.~\ref{fig:lh_geomreco_7_25}.
The separation power in this case is defined by Eq.~\ref{eq:seppower} and gives
2.9 standard deviations. 
This value is slightly lower than the separation power deduced from the track 
Cherenkov angle resolution. 
The most likely cause of this difference is the influence of non-Gaussian tails,
which is ignored in the track Cherenkov angle approach but visible in the
per-track hypothesis test.

The third approach to evaluate the PID performance of the baseline design is 
applying the time-based imaging method.
The $\pi/p$ hypothesis test is performed for each event on the leading edge 
time distribution of each pixel with a hit.
The probability density functions (PDFs) are taken from a statistically independent
beam data sample with the exact same detector configuration and beam condition,
separated by the TOF tags.
An example is to take from one run only even event numbers to calculate the PDFs and
perform the likelihood test only on the odd event numbers.

\begin{figure}[tbh]
  \centering
  \includegraphics[width=.99\columnwidth]{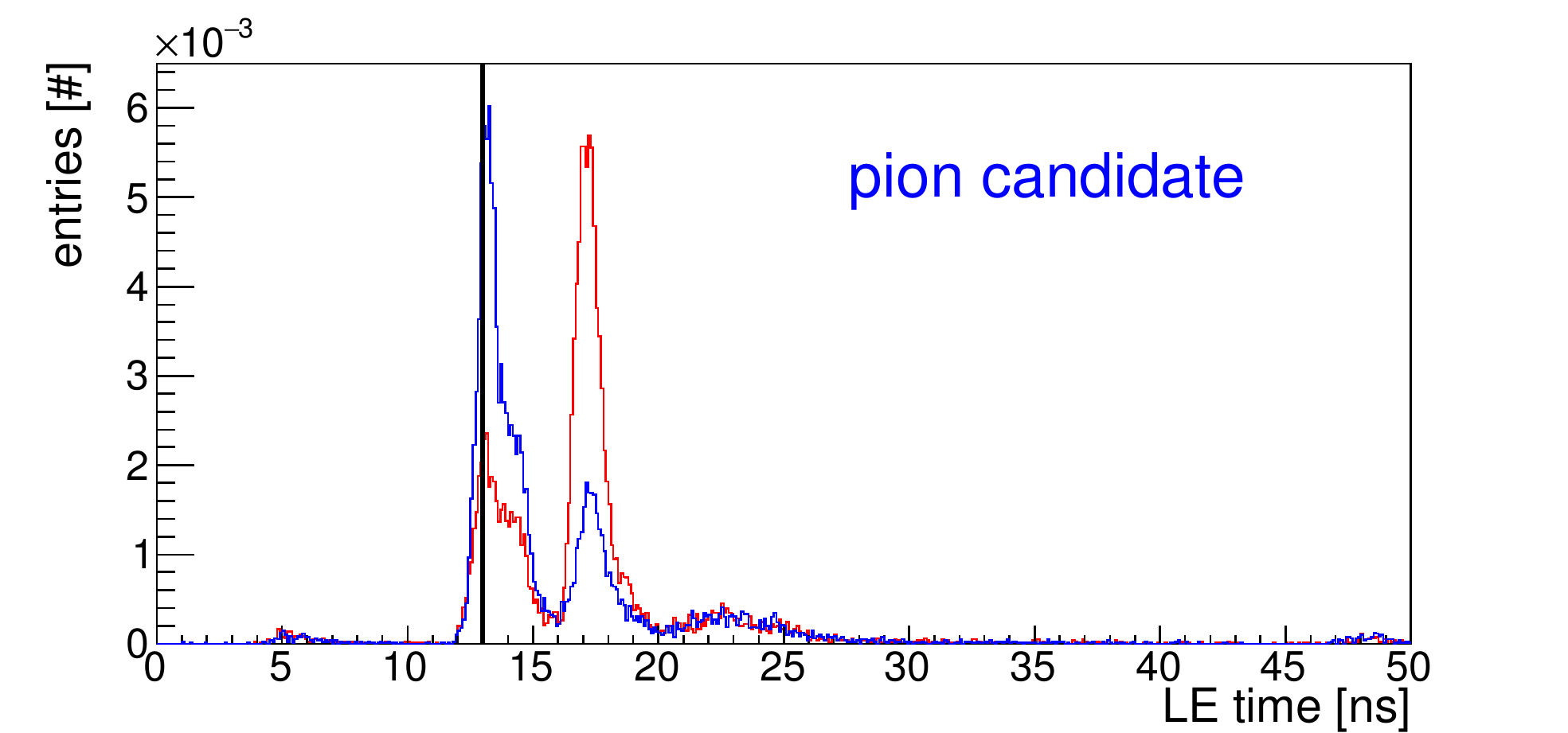}
  \includegraphics[width=.99\columnwidth]{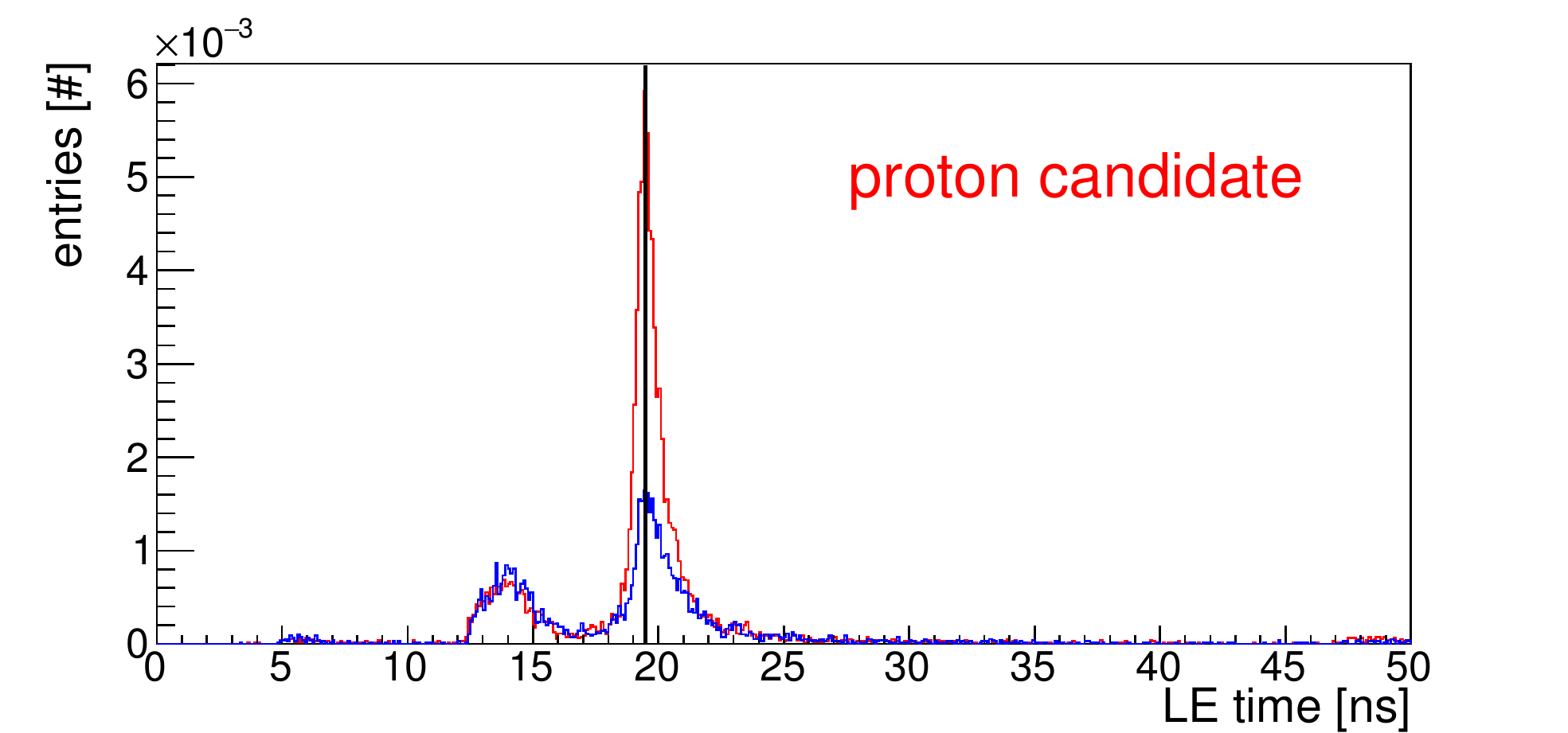}
  \caption{
  Probability density functions for protons (red) and pions (blue) determined from 
    the beam data at 7~GeV/c and a polar angle of $25^{\circ}$ for one example pixel
    for two different events. 
    The distributions are for the narrow bar and the 3-layer spherical lens.
        The vertical lines indicate the observed hit times for a track tagged as pion
         (top) and as proton (bottom).
    }
\label{fig:lh_tireco-bar}
\end{figure}

Figure~\ref{fig:lh_tireco-bar} shows examples of the time-based imaging PDFs 
determined from data for proton-tags (red) and pion-tags (blue) at 7~GeV/c
and $25^{\circ}$ polar angle for the narrow bar and the 3-layer spherical lens. 
The vertical line shows the hit times recorded by the example pixel in these
two events, one a good pion candidate, one a good proton candidate.

\begin{figure}[bth]
  \centering
  \includegraphics[width=.99\columnwidth]{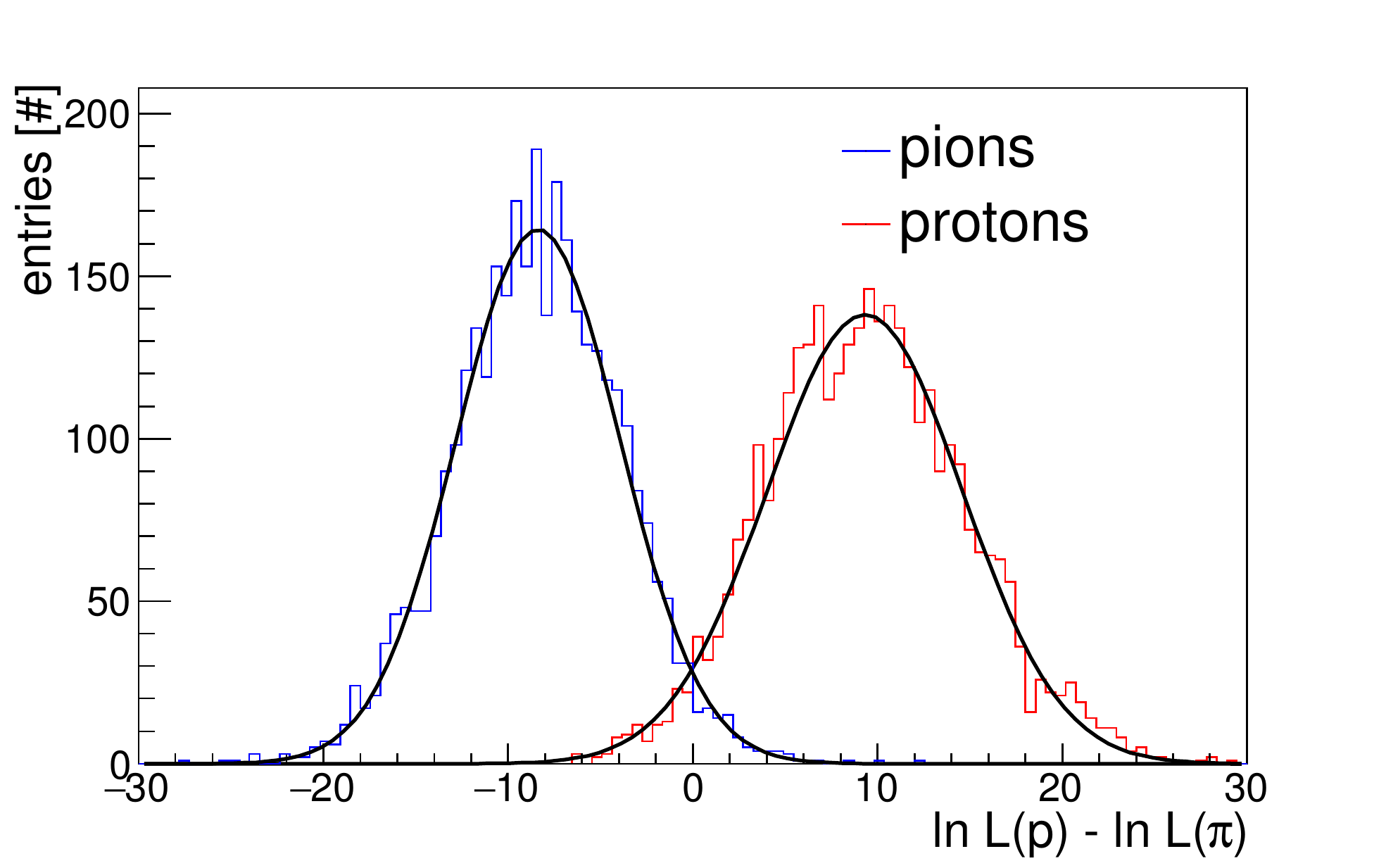}
  \caption{
    Proton-pion log-likelihood difference distributions for proton-tagged (red) and 
    pion-tagged (blue) beam events as result of the time-based imaging reconstruction. 
    The distributions are for the narrow bar with the 3-layer spherical lens, a 
      beam with 7~GeV/c momentum and $25^{\circ}$ polar angle.  The $\pi/p$ separation
    power from the Gaussian fits is 3.6 standard deviations.
    }
\label{fig:lh_tirecoS-bar}
\end{figure}

The resulting proton-pion log-likelihood difference distributions are 
shown in Fig.~\ref{fig:lh_tirecoS-bar} for the beam data taken at 7~GeV/c and 
$25^{\circ}$ polar angle with the narrow bar and the 3-layer spherical lens.
The separation power determined from the Gaussian fits is 3.6 standard deviations. 

\begin{figure}[t]
  \centering
  \includegraphics[width=.98\columnwidth]{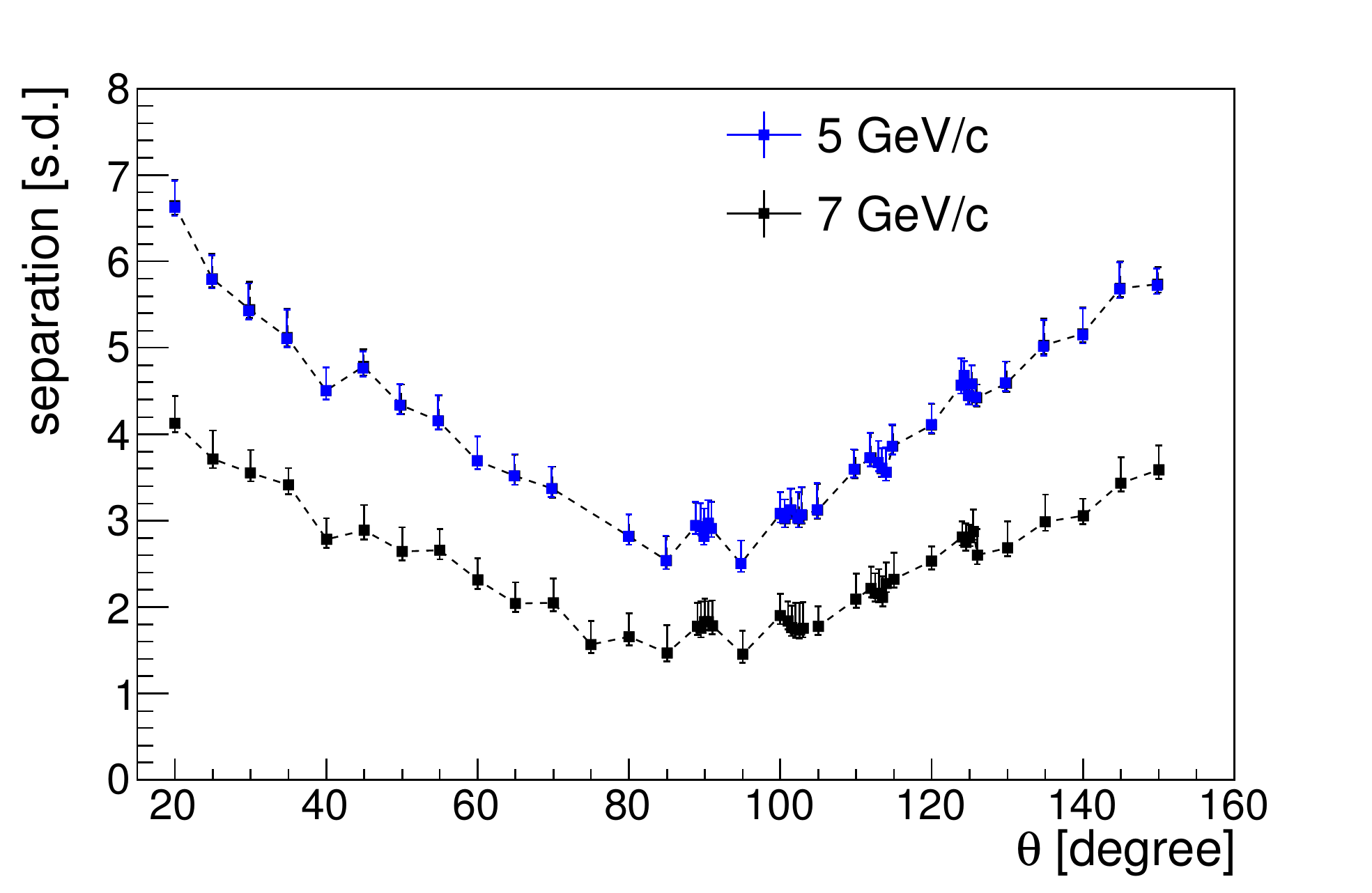}
  \caption{
  Proton/pion separation power from time-based imaging in the beam data as the function 
  of the polar angle at 5~GeV/c (blue) and 
  7~GeV/c (black) momentum for the narrow bar with the 3-layer spherical lens. 
  }
  \label{fig:prt_ti_bar}
\end{figure}

The results of applying the time-based imaging reconstruction to two sets of data, 
polar angle scans with 5~GeV/c and 7~GeV/c momentum, are summarized in 
Fig.~\ref{fig:prt_ti_bar}.
The separation power from the proton-pion log-likelihood difference distributions 
for proton-tagged and pion-tagged beam events is shown as a function of the polar
angle for the narrow bar and the 3-layer spherical lens.

The value of the separation power is proportional to the number of detected photons.
Therefore, the distributions roughly follow the typical shape of the photon yield
in the Barrel DIRC. 
The errors include symmetrical and asymmetrical parts. 
The symmetrical error corresponds to the fit errors of the likelihood distributions,
whereas the asymmetrical error reflects the quality of the PDFs.
The simulation suggests that at least 50k events should be used to calculate the
PDFs and that the log-likelihood difference becomes systematically smaller when
fewer events are used.
In the data only about 30k tagged events were available for this study, causing 
a systematic underestimation of the separation power.

The separation power for the time-based imaging reconstruction is the best 
of the three methods tested. 
In spite of the timing difficulties it exceeds 4~s.d. proton/pion separation power
for the most difficult region for the
Barrel DIRC PID in \panda, for high-momentum tracks at forward angles.

It is important to note in this context, that the PID requirement for the 
\panda Barrel DIRC is strongly momentum-dependent due to the asymmetric 
kaon phase space (see Sec.~\ref{cha:design-goals}).
The 3~s.d. $\pi/K$ separation only has to be achieved for polar angles less
than 35$^\circ{}$.
For 45$^\circ{}$ the maximum momentum for the 3~s.d. performance already drops
to 2.5~GeV/c.
For the beam test this means that the achieved $\pi/p$ separation at 
5~GeV/c and 7~GeV/c translates into a $\pi/K$ separation that is better than 
required for the entire kaon phase space.

The performance would, presumably, be even better, if the timing resolution obtained
during the beam test would not have been a factor 2-3 worse than the 100~ps goal.
However, the design with the narrow bar and the spherical lens is robust against 
timing deterioration and delivers excellent PID performance for both the geometrical 
and the time-based imaging reconstruction methods, meeting or exceeding the PID 
requirements for \panda.

\subsection{PID Performance of the Wide Plate Design}
\label{sec:prt_performance_plate}

\begin{figure}[b]
  \raggedright{Test beam data, pion tag}
  \hspace*{-2mm}\includegraphics[width=1.04\columnwidth]{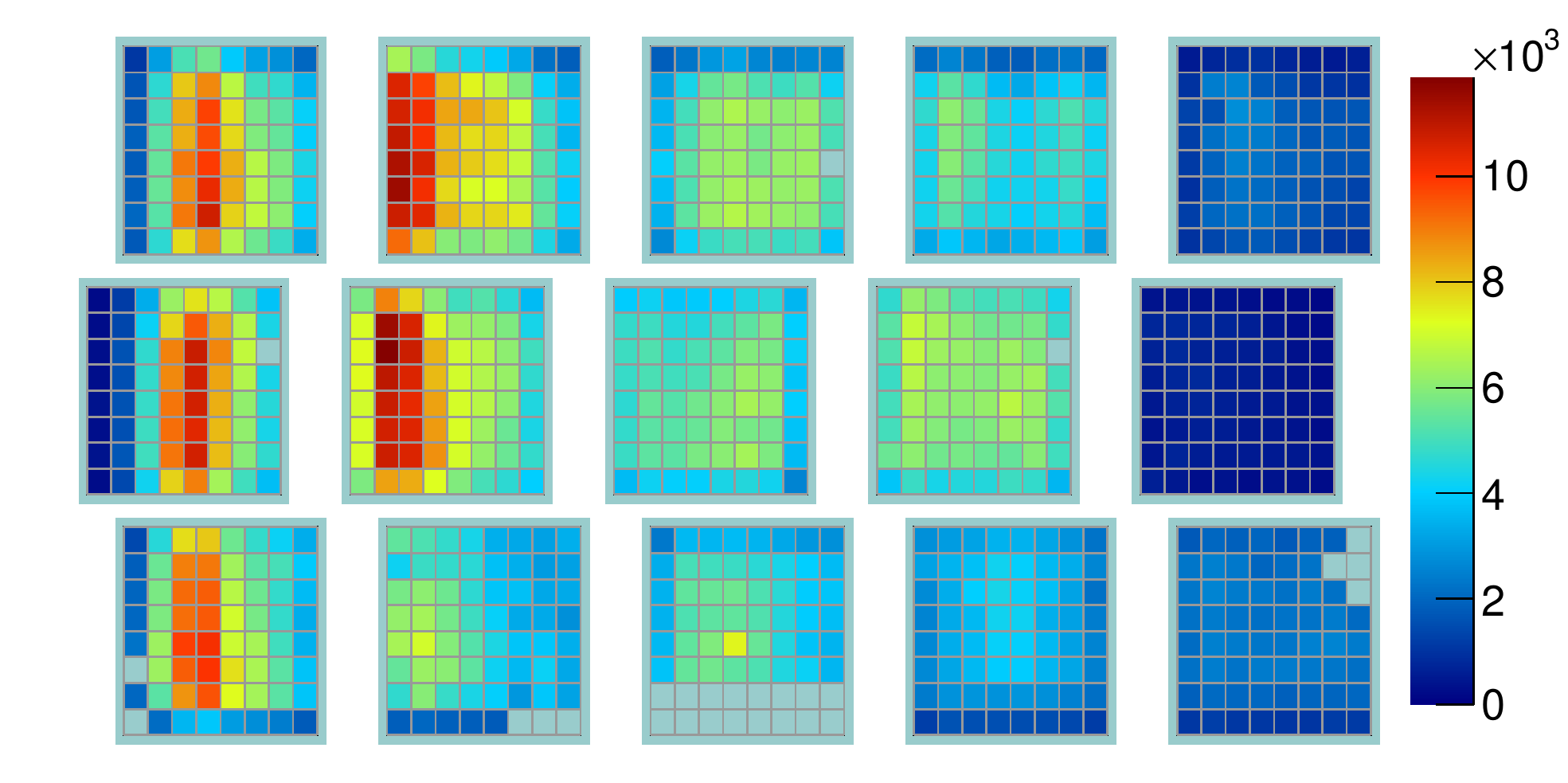}
  \raggedright{Test beam data, proton tag}
  \hspace*{-2mm}\includegraphics[width=1.04\columnwidth]{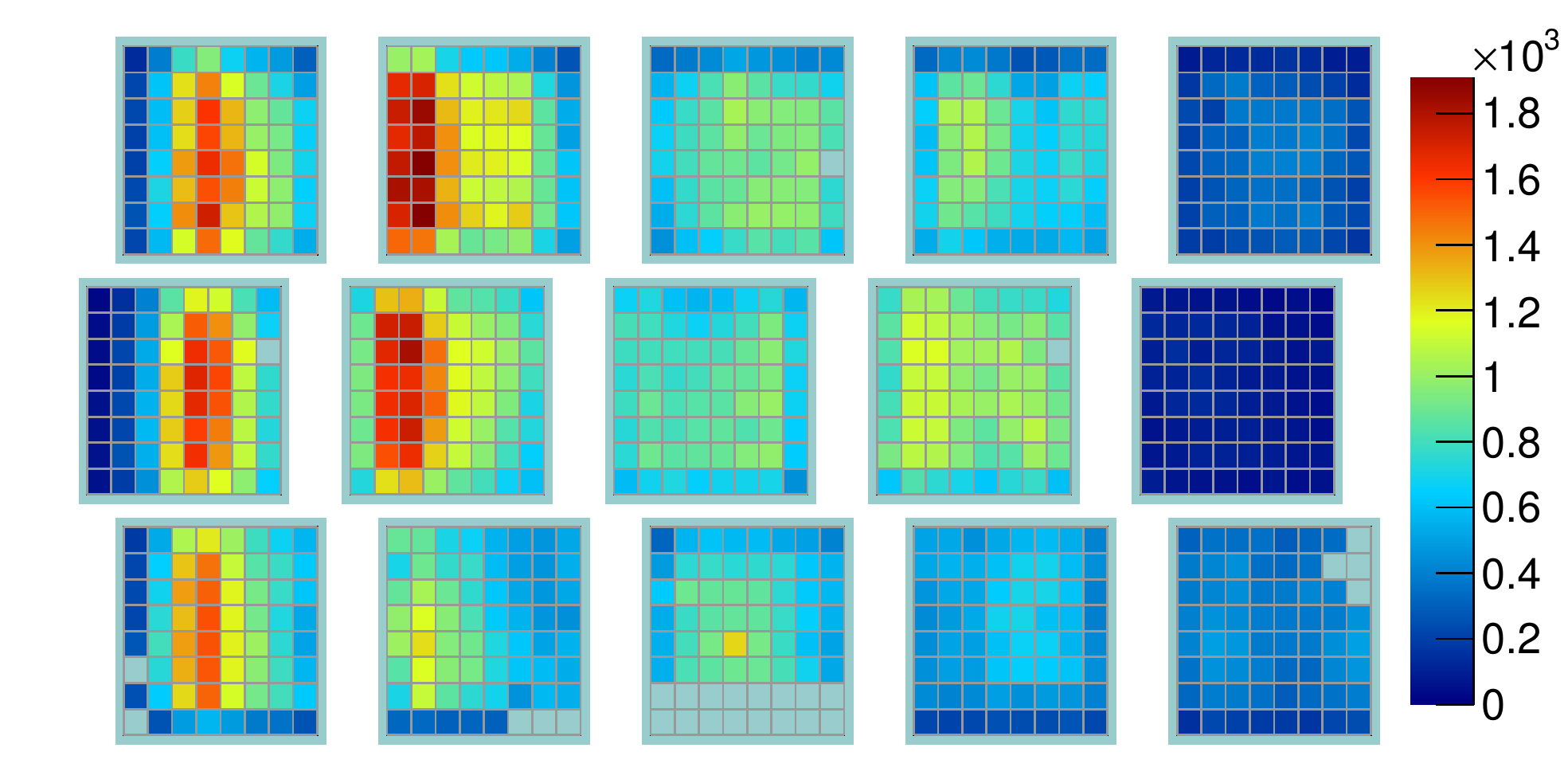}
  \raggedright{Geant simulation, protons}
  \hspace*{-2mm}\includegraphics[width=1.04\columnwidth]{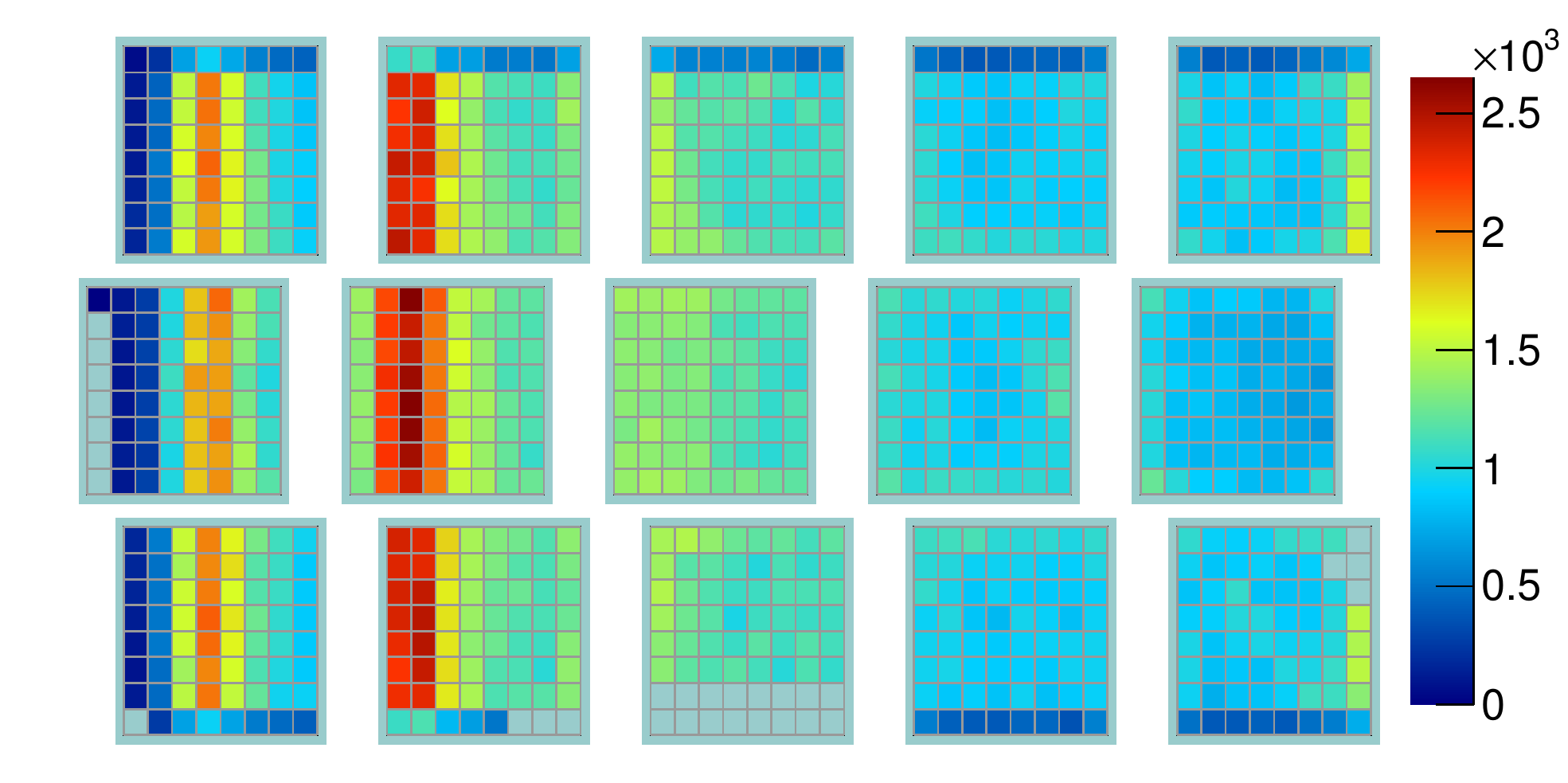}
  \caption{
    Accumulated hit pattern for the 2015 prototype, shown as number of signals per MCP-PMT pixel,
    for the wide plate without focusing and a 7~GeV/c beam with a polar angle of 25$^\circ{}$.
    Experimental data for a pion tag (top) and proton tag (middle) are compared to
    Geant simulation for a proton beam (bottom).
  }
  \label{fig:hit_pattern-plate}
\end{figure}

The PID performance of the wide plate was evaluated with a 2-layer cylindrical 
lens and without any lens for various polar angles and beam momenta.
Figure~\ref{fig:hit_pattern-plate} shows the hit pattern for the wide plate 
without a focusing lens at 7~GeV/c momentum and a polar angle of 25$^{\circ}$
for tagged pions and protons.
The pion sample (top) appears visually rather similar to the proton sample, which is 
in reasonable agreement with the simulation.

\begin{figure}[tbh]
  \centering
  \includegraphics[width=.99\columnwidth]{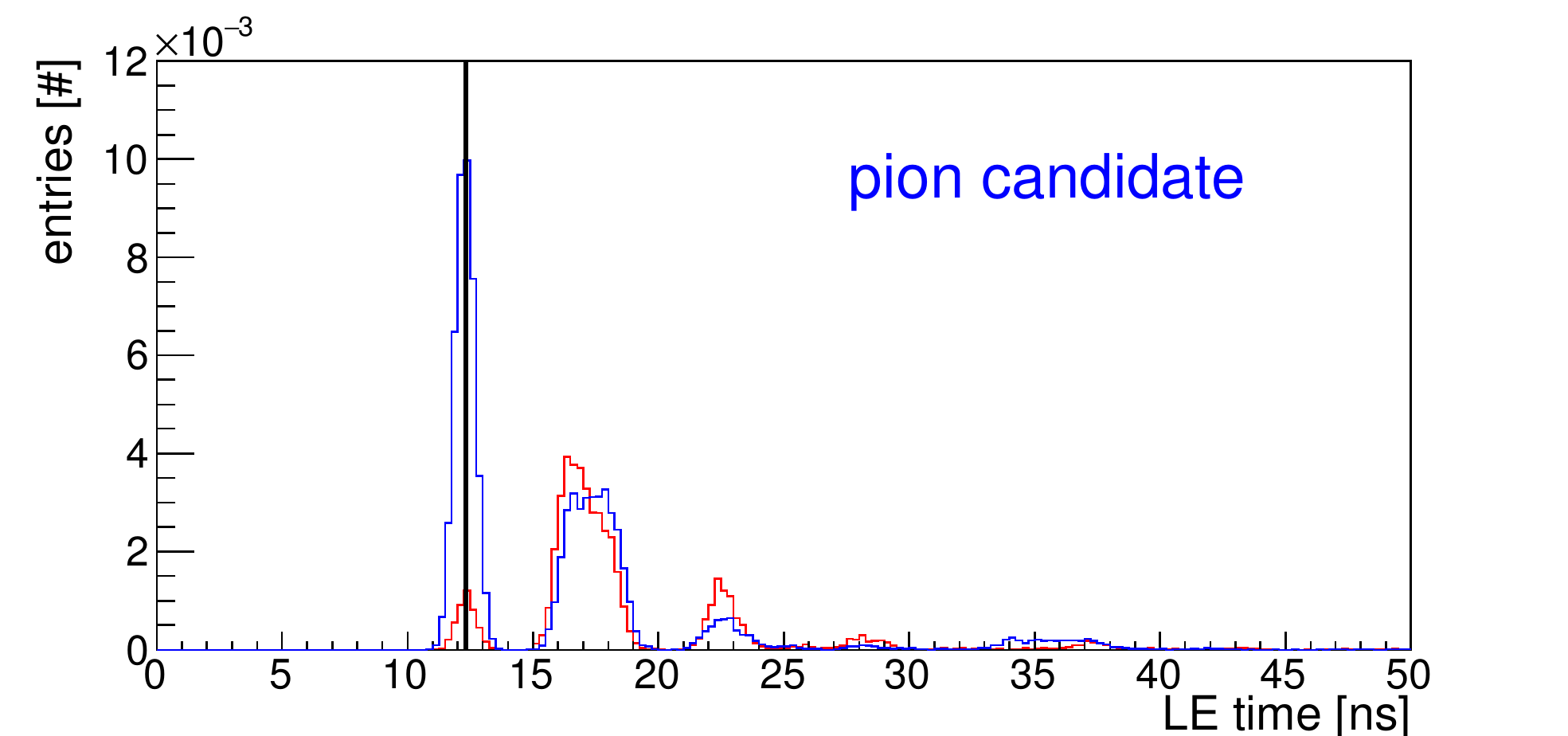}
  \includegraphics[width=.99\columnwidth]{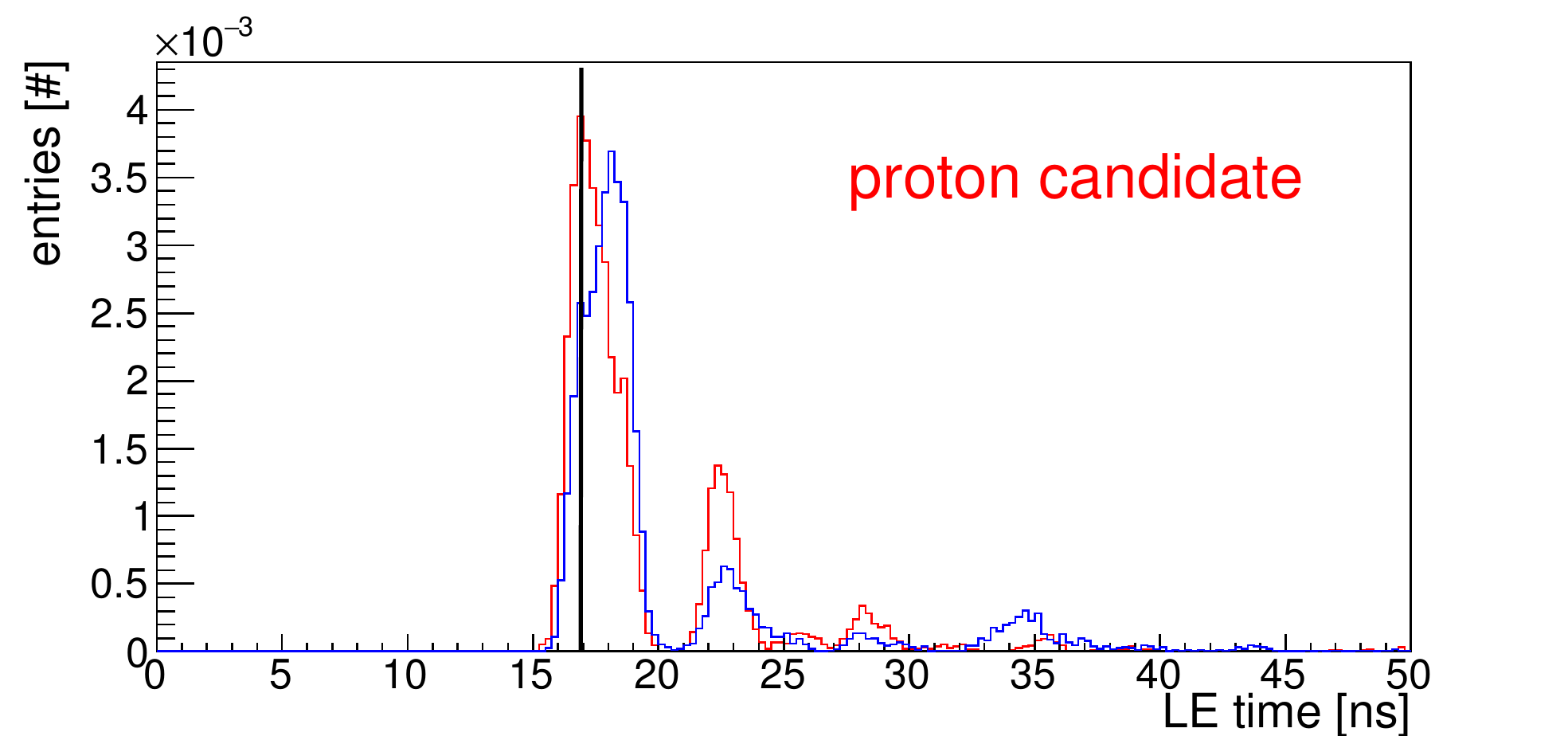}
  \caption{
  Probability density functions for protons (red) and pions (blue) determined from 
    the beam data at 7~GeV/c and a polar angle of $25^{\circ}$ for one example pixel
    for two different events. 
    The distributions are for the wide plate without focusing.
    The vertical lines indicate the observed hit times.
    }
\label{fig:lh_tireco-plate}
\end{figure}

Figure~\ref{fig:lh_tireco-plate} shows single-pixel examples of the time-based imaging 
PDFs for the plate without focusing, determined from data for events with a
proton-tag (red) and a pion-tag (blue) at 7~GeV/c and $25^{\circ}$ polar angle.
The hit times in the pixel, shown as vertical lines, are in good agreement
with the pion hypothesis in the upper figure, and with the proton hypothesis 
in the lower figure.

\begin{figure}[bth]
  \centering
  \includegraphics[width=.99\columnwidth]{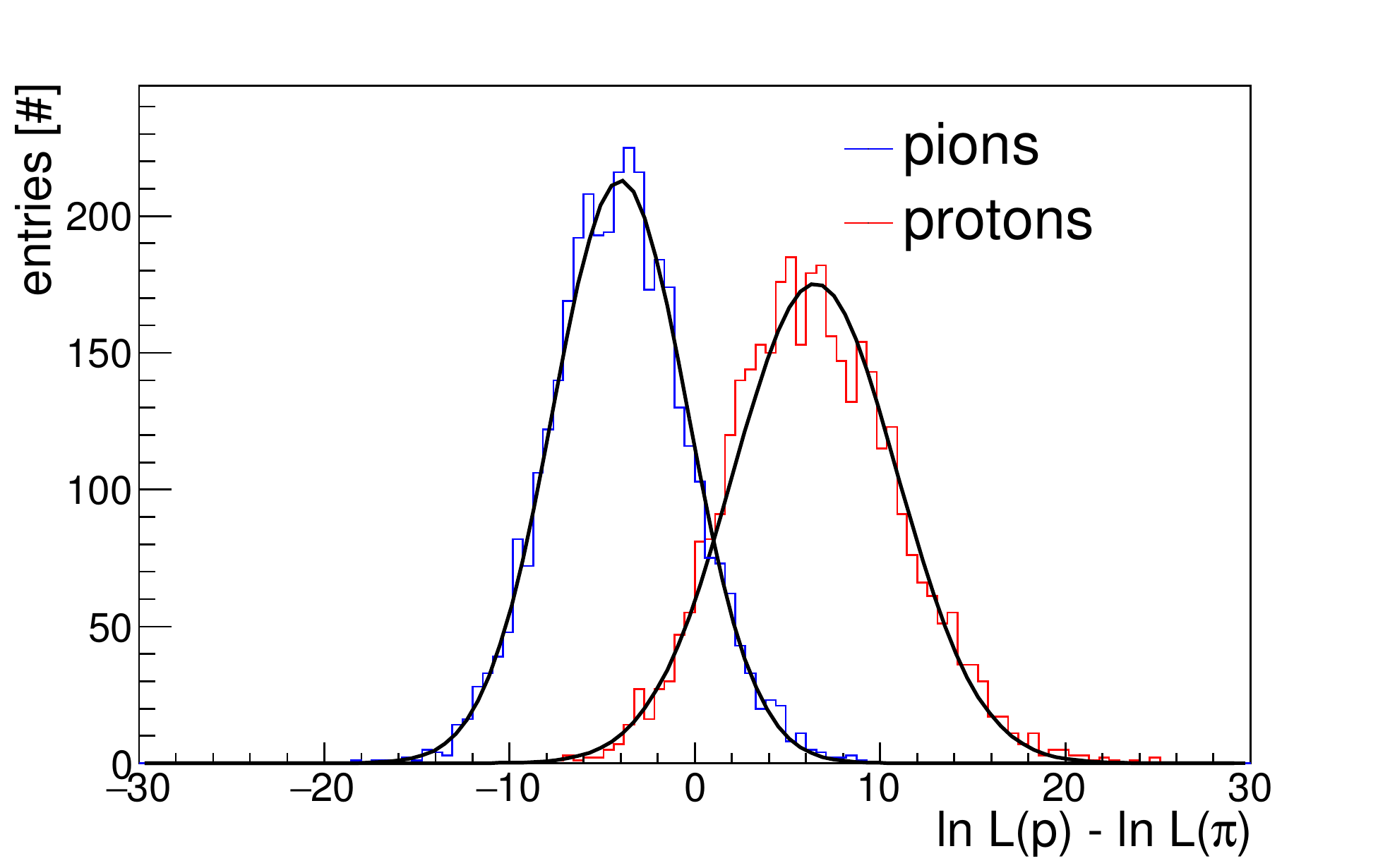}
  \caption{
    Proton-pion log-likelihood difference distributions for proton-tagged (red) and 
    pion-tagged (blue) beam events as result of the time-based imaging reconstruction. 
    The distributions are for the wide plate without focusing, a 
    beam with a 7~GeV/c momentum and a $25^{\circ}$ polar angle. 
    The $\pi/p$ separation power extracted from the Gaussian fits is 2.6 standard deviations.
    }
\label{fig:lh_tirecoS-plate}
\end{figure}

The result of the unbinned likelihood calculation for the plate without 
focusing at 7~GeV/c and $25^{\circ}$ polar angle is shown in 
Fig.~\ref{fig:lh_tirecoS-plate}.
The proton/pion separation power in this case is 2.6 standard deviations and
does not yet quite meet the PID goals for the \panda Barrel DIRC.

\begin{figure}[tbh]
  \centering
  \includegraphics[width=.98\columnwidth]{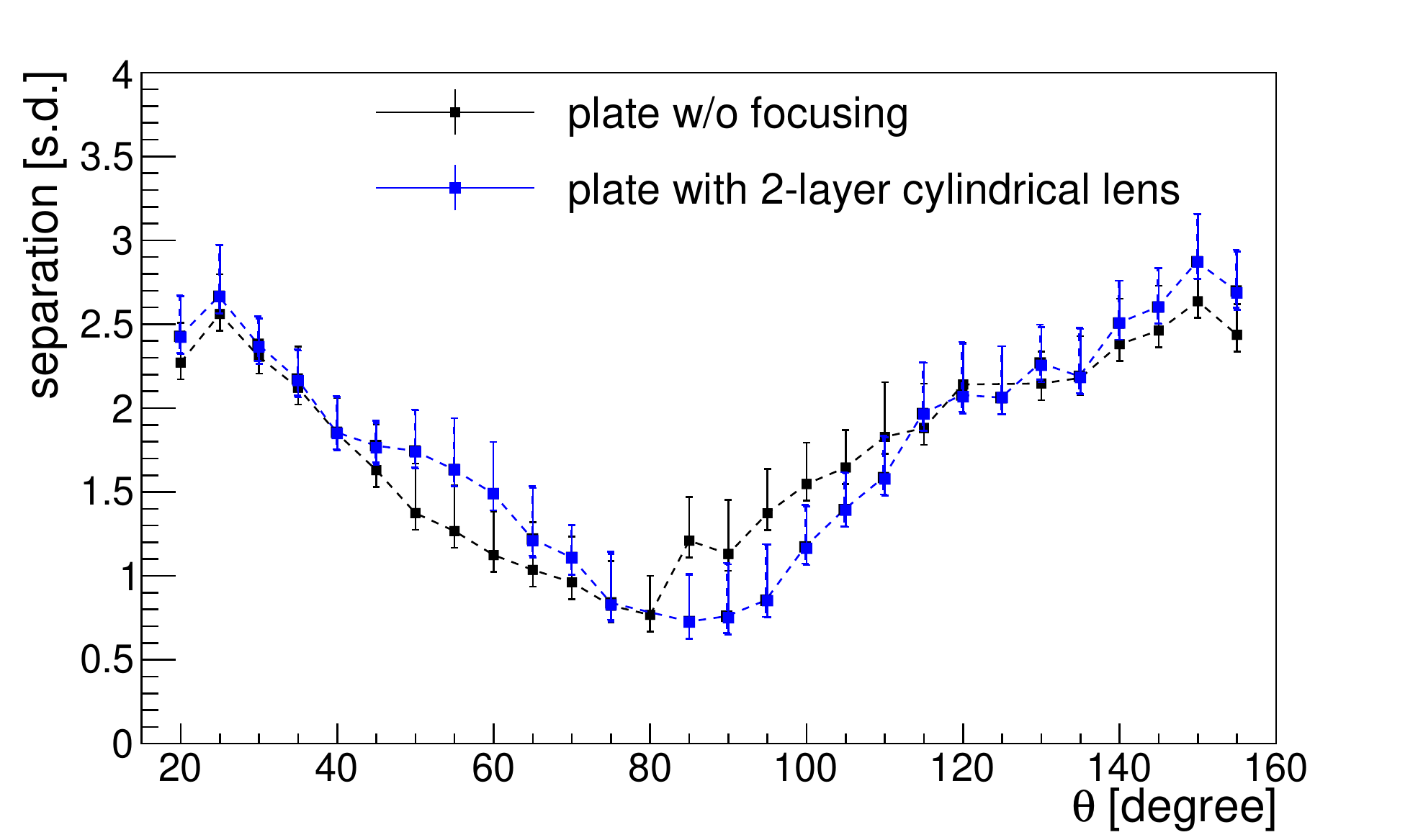}
  \caption{
  Proton/pion separation power from time-based imaging as the function of the polar angle at 
  7~GeV/c (black) momentum in the beam data for the wide plate without focusing (black) and 
  with the 2-layer cylindrical lens (blue).
  }
  \label{fig:prt_plate_comp}
\end{figure}

The separation power in the 7~GeV/c beam data for the plate without focusing is compared
as a function of polar angle to the plate with the 2-layer cylindrical lens in Fig.~\ref{fig:prt_plate_comp}.
For the steep forward and backward angles the performance is slightly better with the lens
while the design without a lens shows slightly better separation for polar angles between
80$^\circ{}$ and 110$^\circ{}$, probably because the larger photon loss due to reflection 
inside the lens lowers the photon yield.

\begin{figure}[tbh]
  \centering
  \includegraphics[width=.98\columnwidth]{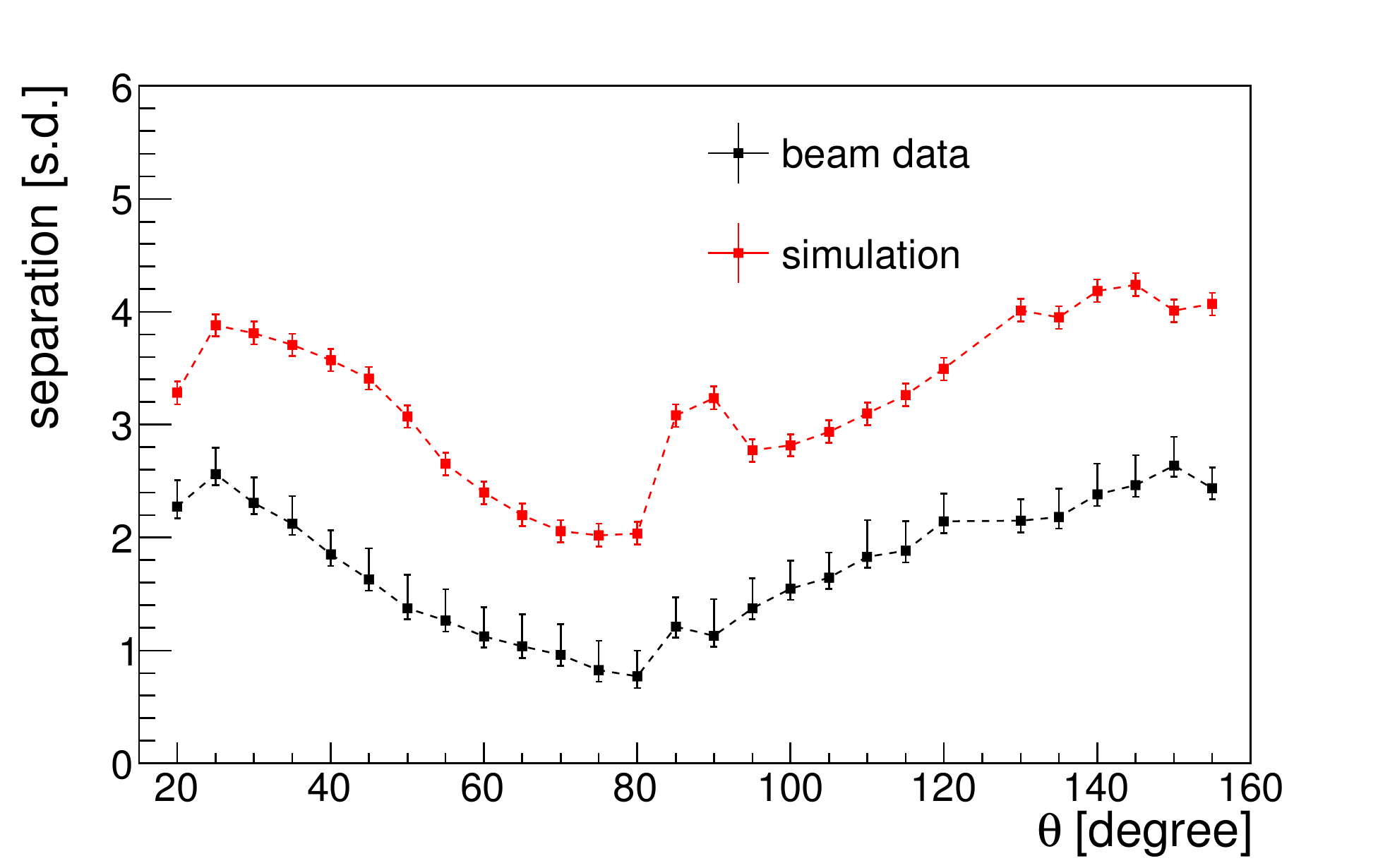}
  \caption{
  Proton/pion separation power from time-based imaging as the function of the polar angle at 
  7~GeV/c momentum for the wide plate without focusing in the beam data (black)
  and the simulation (red), assuming 100~ps time resolution and a 3~mm RMS beam spot (red).
  }
  \label{fig:prt_ti_plate}
\end{figure}

Figure~\ref{fig:prt_ti_plate} shows this proton/pion separation power of the wide plate 
at 7~GeV/c momentum as a function of the polar angle. 
It is compared to the simulation, which used the design timing resolution of 100~ps and a
beam spot size of 3~mm RMS.

The proton/pion separation power for the wide plate without lens does not reach the 
3~s.d. goal of the \panda Barrel DIRC PID.
This is predominantly caused by the timing resolution, which was a factor 2--3 worse than
expected.
The limited size of the data sample used to generate the timing PDFs and the photon 
detection efficiency loss on the older, less efficient MCP-PMTs, also caused lower
separation power values.
The drop in the separation power for steep forward and backward angles is caused by the
size of the beam spot.

The Geant simulation with an assumed time resolution of 100~ps, which overestimates the 
performance of the design for all polar angles (since the resolution in data was considerably
worse) shows that the wide plate, with improved timing of about 100~ps, should in 
fact be able to deliver the $\pi/K$ performance required for \panda PID.

\begin{figure}[tbh]
  \centering
  \includegraphics[width=.98\columnwidth]{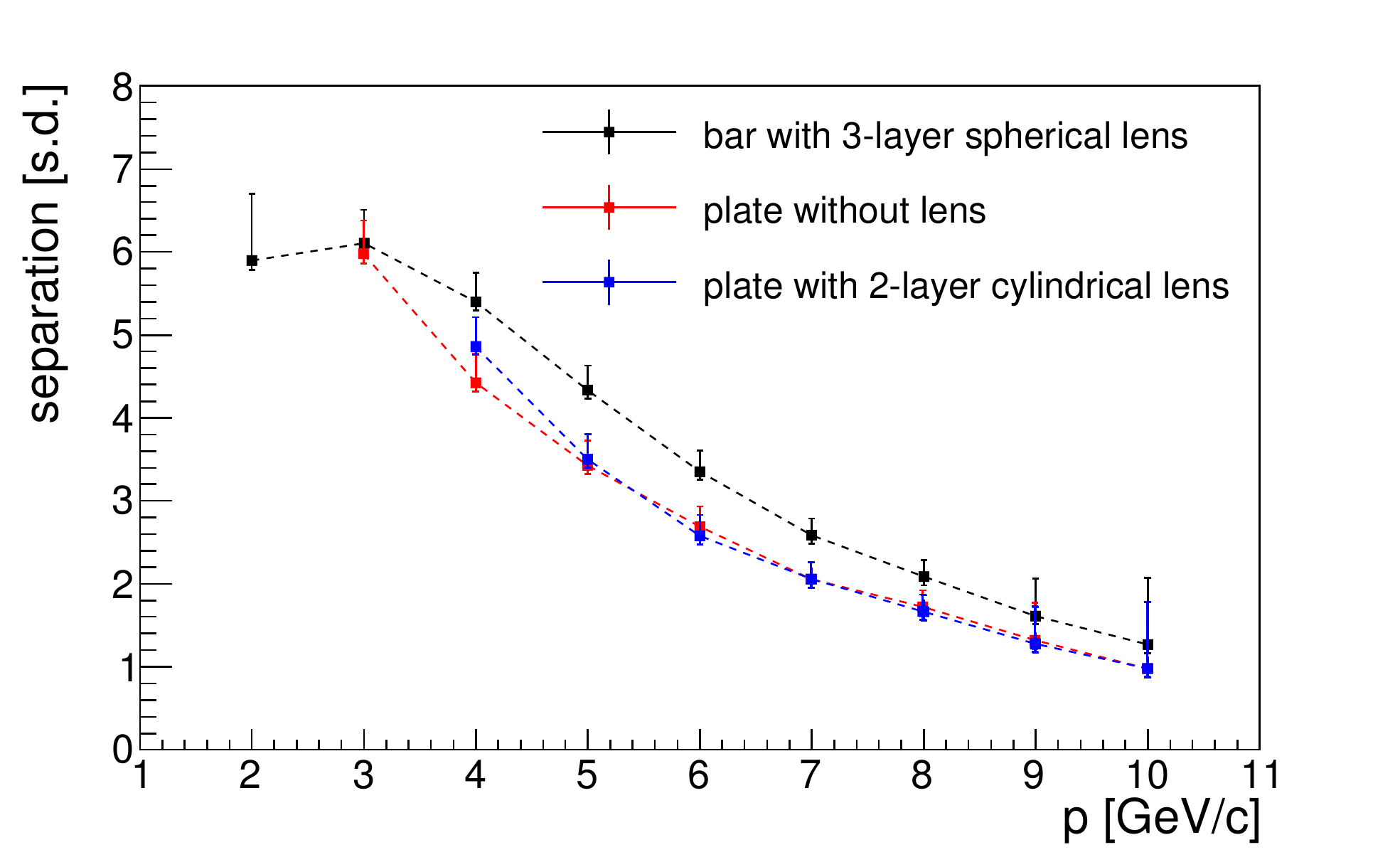}
  \caption{
  Proton/pion separation power as the function of the momentum for
    beam data with $125^{\circ}$ polar angle and different geometry configurations.}
  \label{fig:prt_ti_mom}
\end{figure}

Finally, in Fig.~\ref{fig:prt_ti_mom}, the proton/pion separation power for the 
narrow bar with the 3-layer lens is compared to the performance obtained by the
wide plate without focusing and with a 2-layer cylindrical lens as a function
of the beam momentum for a polar angle of $125^{\circ}$.

\subsection{Conclusion of the 2015 Prototype Test}
\label{sec:conclusion_2015}

The design with the narrow bar and the spherical lens is found to meet or exceed
the PID requirements for \panda.
It is robust against timing deterioration and delivers excellent $\pi/K$ separation
for both the geometrical and the time-based imaging reconstruction methods. 

The geometry with the wide plate and the 2-layer cylindrical lens performs significantly
worse than the narrow bar geometry and does not quite reach the \panda PID goals.

\section{Prototype Test at CERN in 2016 - PID Validation of the Wide Plate Design}

The 2015 prototype test demonstrated that the figures of merit and the $\pi/K$ 
separation power of the geometry based on narrow bars exceeded the \panda PID 
requirements for the entire pion and kaon phase space.
The performance of the wide plate, however, fell short of reaching the 3~s.d. 
$\pi/K$ separation goal.
An additional beam test campaign was performed at CERN in October/November 2016 
to validate the PID performance of the wide plate after improving several key 
aspects of the prototype configuration.

\subsection{Prototype Improvements Prior to Beam Test}

A detailed comparison of the experimental data from the 2015 beam test to the 
Geant prototype simulation identified several issues with the data quality which 
directly affected the plate PID performance.

\textbf{Readout Electronics Timing Precision}

The timing precision of the PADIWA/TRB readout chain during the beam tests in 2015 was 
found to be a factor 2-4 worse than the 100~ps goal.
Since the time-based imaging performance depends strongly on the timing precision, 
the readout electronics needed to be improved.

The capacitance in the input low-pass filter of the PADIWA was reduced from 48~pF to 10~pF,
significantly reducing the effect of signal shaping, thus improving the timing precision.
The gain of the preamplifier on the PADIWA was increased from a value of about 7-10 
to a value of 20--25, improving
both the timing precision and the hit detection efficiency of the readout.

\begin{figure}[htb]
	\includegraphics[width=.98\columnwidth]{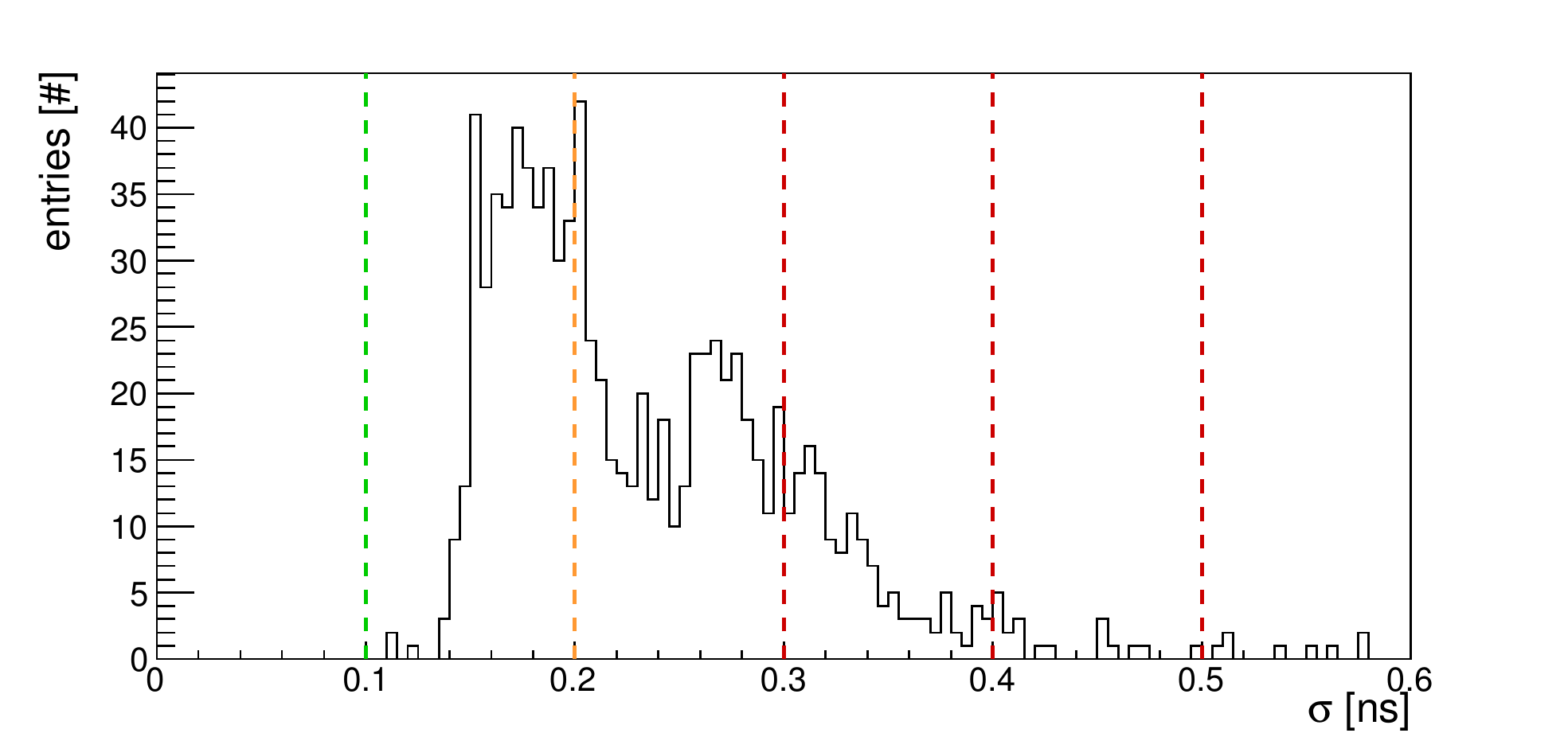}
	\includegraphics[width=.98\columnwidth]{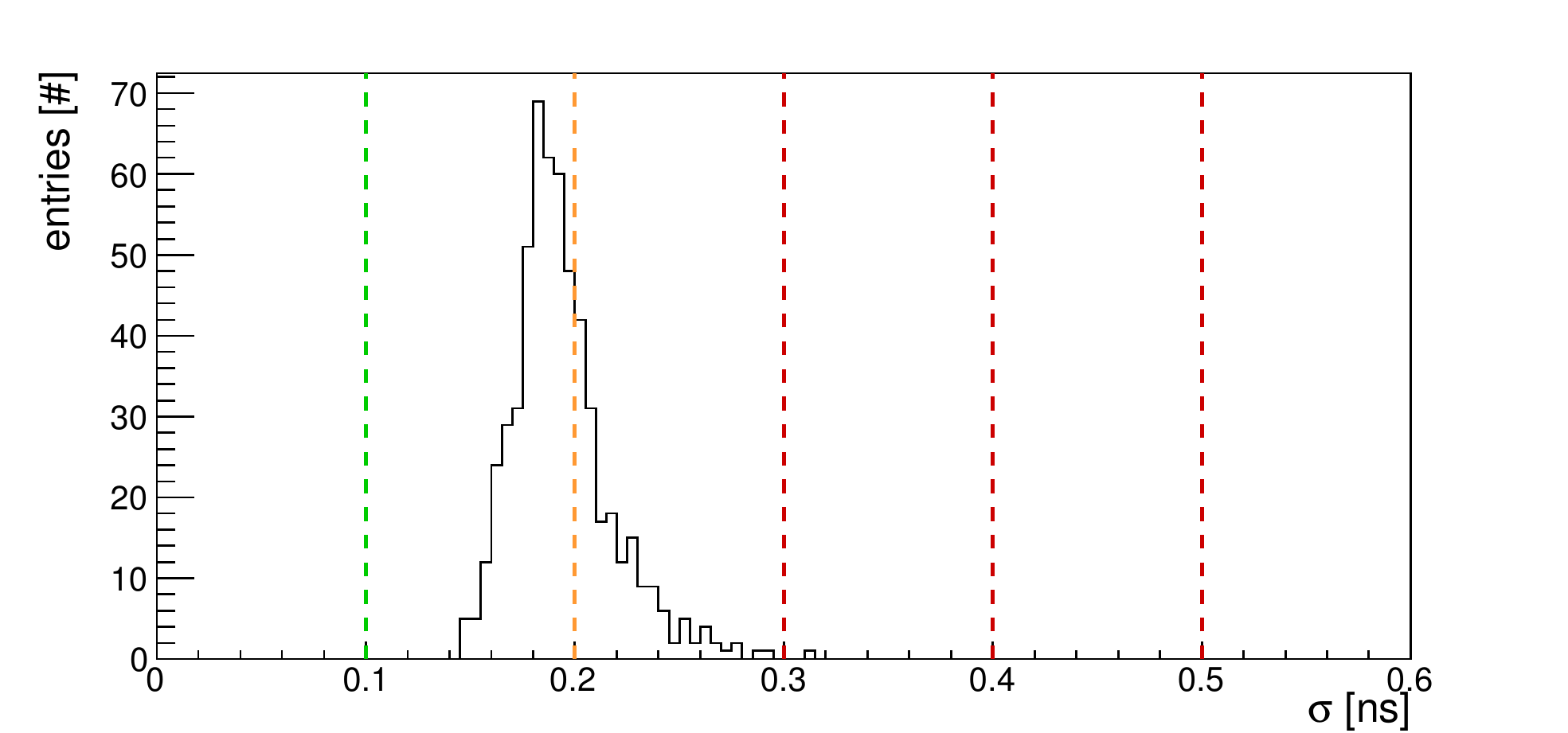}
	
	\caption{
		Timing precision per channel from PiLas laser pulser data for the prototype readout configuration
		in 2015 (top) and in 2016 (bottom).
	}
	\label{fig:time_res_pilas_20152016}
\end{figure}

Figure~\ref{fig:time_res_pilas_20152016} compares the timing precision per channel
in PiLas laser pulser data, as observed in 2015, to the performance obtained in 2016, 
after the modifications to the readout electronics.
A clear improvement is visible and, although the timing precision is still a factor
of 1.5--2.5 worse than the nominal 100~ps goal, the large tail above 300~ps in the 
2015 timing precision distribution was successfully removed.

\textbf{MCP-PMT Sensor Quality}

The Cherenkov angle resolution and photon yield obtained in 2015 for the narrow bar with 
the 3-layer spherical lens for polar angles around 90$^{\circ}$ were considerably worse 
than expected from the prototype Geant simulation.
The most plausible explanation was that at these polar angles the photons were primarily
detected by the older MCP-PMT models, placed on the right side of the MCP-PMT array.
Those MCP-PMTs had a lower gain and were less uniform in gain and quantum efficiency
than the newer models, placed on the left side.

Therefore, a smaller prism with a top angle of 30$^{\circ}$ was used in 2016 (the depth 
is still 300~mm), which reduced the size of the MCP-PMT array from $3 \times 5$ to $3 \times 3$,
so that only the newer, higher-quality units were used in 2016.

\begin{figure*}[thb]
	\centering
	\includegraphics[width=.8\textwidth]{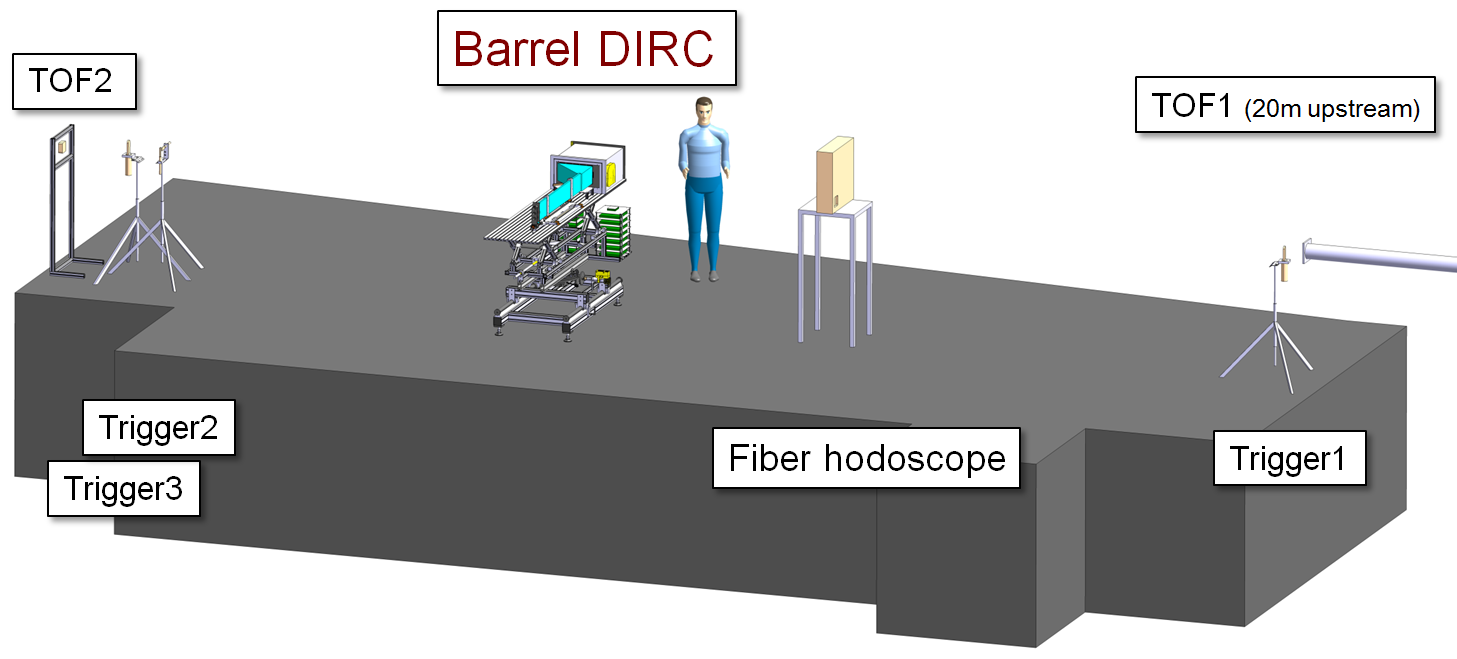}
	\caption{Detector setup during the prototype test in the T9 beam line at CERN in 2016 (not to scale).
	}
	\label{fig:EXP-CERN2016}
\end{figure*}

\textbf{Event Statistics}

The size of the event sample available for creating the timing probability density 
functions (PDFs) in each pixel and for testing the PID performance was found to 
have a major systematic impact on the result.
Simulation demonstrated that the $\pi/p$ separation power increases steadily with the 
number of events in the analysis and only approaches the high-statistics limit after 
typically 50\,000--100\,000 selected events were used.
For tight cuts on the beam instrumentation detectors, in particular the fiber hodoscope,
the event selection efficiency in data can be below 1\%.
Therefore, larger statistics samples of at least $10^7$ triggers per configuration
were taken in 2016, whenever possible.

\textbf{Beam Spot Size and Divergence}

To minimize divergence the beam, was configured in the ``parallel beam'' focus mode, 
which created a beam spot of about 40--50~mm diameter on the plate.
This will cause large photon propagation time differences inside the plate, in particular
for steep forward incidence angles.
An additional trigger counter was added to the beam instrumentation in 2016 to make much
tighter cuts on the beam spot size possible.

\subsection{Prototype Configuration in 2016}

\begin{figure}[b]
	\centering
	\includegraphics[width=.98\columnwidth]{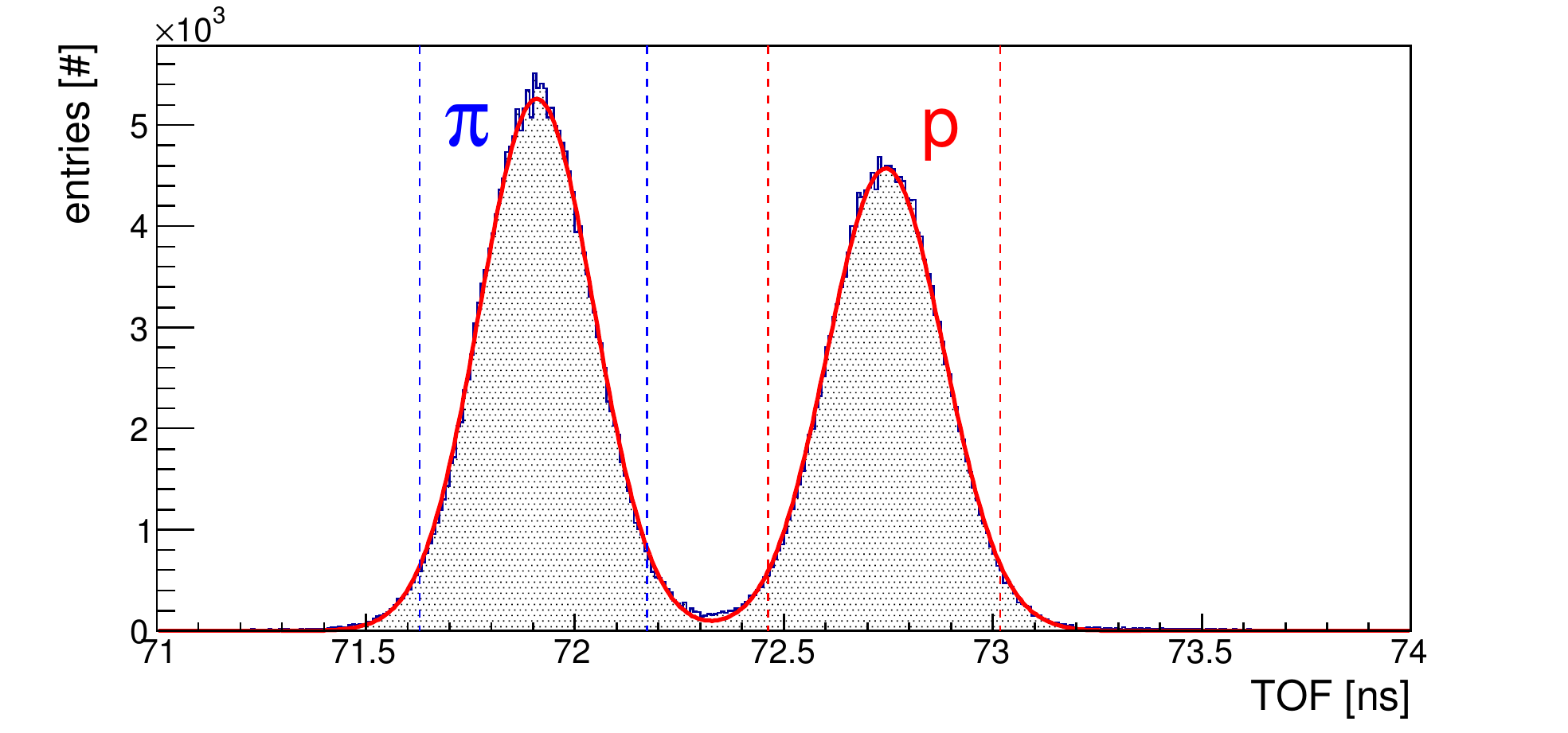}
	
	\caption{Time difference between the two TOF stations, separated by 29~m flight
		distance, for a beam momentum of 7~GeV/c.
		The dashed lines indicate the $\pi$ and $p$ selection windows.
	}
	\label{fig:tof57-2016}
\end{figure}

The beam line configuration at the CERN PS/T9 area in 2016 is shown in Fig.~\ref{fig:EXP-CERN2016}.
Beam instrumentation included two scintillators with 40~mm diameter to define the 
trigger for the DAQ (Trigger1/2 in Fig.~\ref{fig:EXP-CERN2016}) and a smaller 
scintillator finger with a width of 8~mm (Trigger3) to constrain the beam spot.
A scintillating fiber hodoscope provided position information upstream from the Barrel
DIRC prototype.
The same time-of-flight (TOF) system was used as in 2015.
The two TOF stations were again separated by a distance of about 29~m, leading to
clean $\pi/p$ tagging at 7~GeV/c momentum, as can be seen in Fig.~\ref{fig:tof57-2016}.

\begin{figure}[h]
	\centering
	
	\includegraphics[width=.95\columnwidth]{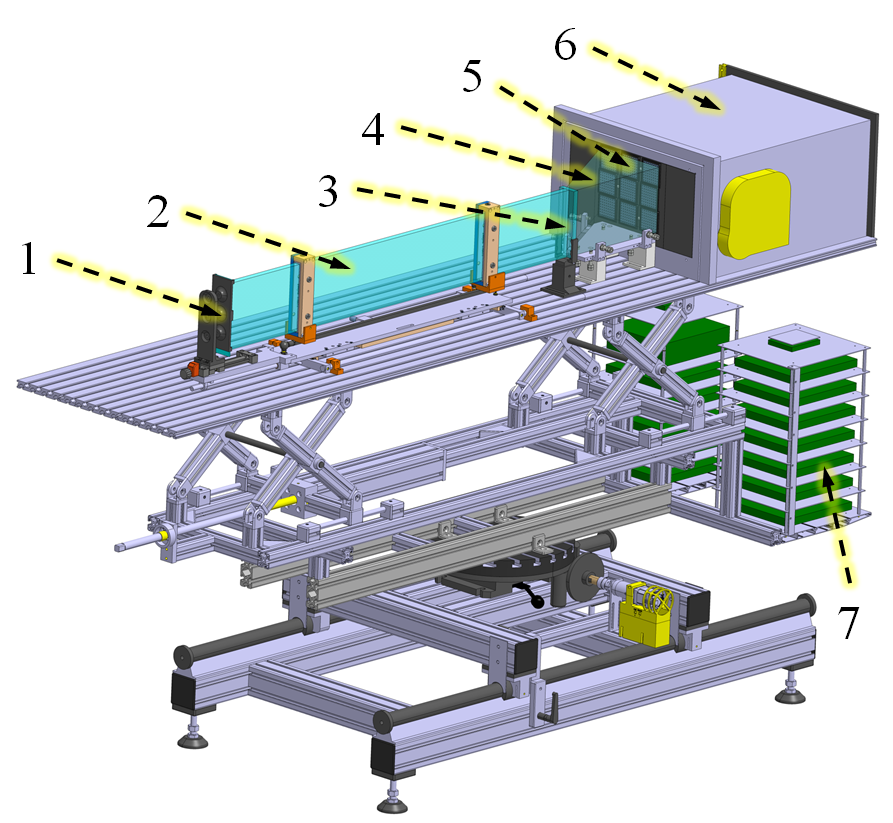}
	
	\caption{Schematic of the prototype used at CERN in 2016, with 1: flat mirror, 2: radiator plate, 
		3: lens, 4: expansion volume, 
		5: array of 3$\times$3 MCP-PMTs, 6: readout unit, and 7: TRB stack.}
	\label{fig:proto-cern-2016}
\end{figure}

\begin{figure}[h]
	\centering
	\includegraphics[width=.85\columnwidth]{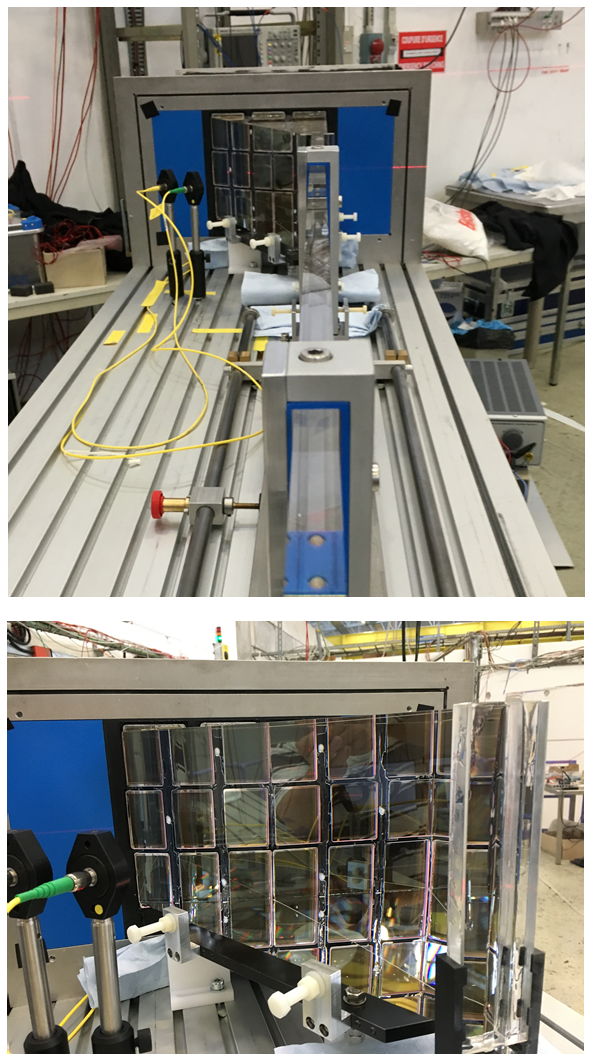}
	\caption{Photographs of the 2016 prototype in the T9 beam line: view along the
		length of the plate (top) and close-up of the coupling of the prism to the
		2-layer cylindrical lens and to the $3 \times 3$ MCP-PMT array (bottom).}
	\label{fig:proto-cern2016-photos}
\end{figure}

The prototype, shown in Fig.~\ref{fig:proto-cern-2016} and Fig.~\ref{fig:proto-cern2016-photos}, 
comprised a wide fused silica plate (17.1 $\times$ 174.8 $\times$ 1224.9~mm$^3$), 
coupled on one end to a flat mirror, on the other end to the 2-layer cylindrical lens, 
the fused silica prism as EV (with a depth of 300~mm and a top angle of 30$^\circ{}$), 
the 3$\times$3 array of PHOTONIS Planacon XP85012 MCP-PMTs, and the modified readout 
electronics.
The prototype support frame could be translated manually and rotated remotely 
relative to the beam, making it possible to perform a scan of a number of 
polar angle/momentum points.

\subsection{Results of the 2016 Prototype Test}

The calibration and simulation of the prototype data, as well as the data analysis,
was very similar to the procedure described in detail in Sect.~\ref{ch:performance}.
About $4.9 \times 10^8$ triggers were recorded using the mixed hadron beam at the CERN PS/T9 
beam line with the hadron-enriched target (H3).
Since the primary goal of this beam test was the validation of the PID performance
of the wide plate, in particular the $\pi/K$ separation towards the upper momentum range
in \panda, most of those triggers, approximately 340M, were taken with the beam momentum 
of 7~GeV/c.
The $\pi/p$ Cherenkov angle difference at this momentum (8.1~mrad) is close to the $\pi/K$ 
Cherenkov angle difference at 3.5~GeV/c (8.5~mrad).

In addition, high-statistics runs were taken several times per day with 
the internal electronics pulser to monitor the TDC calibration and with
the PiLas picosecond laser pulser to determine the time offsets between the pixels.
Time walk effects in the time measurements of the TOF stations and the prototype
MCP-PMTs were corrected using time-over-threshold information.

The event selection was based on the coincidence of the three trigger counters,
a clean $\pi$ or $p$ tag from the TOF system, and a selection of fibers in the
hodoscope.
Depending on the beam momentum and polar angle, the selection efficiency was
typically between 0.5\% and 1\%.

MCP-PMT signals (``hits'') were selected in a time window of $\pm$40~ns relative to 
the Trigger1 time.

\begin{figure}[htb]
	\raggedright{Test beam data, proton tag}
	\vspace*{-6mm}
	\begin{center}
          \hspace*{-12mm}
	  \includegraphics[width=1.2\columnwidth]{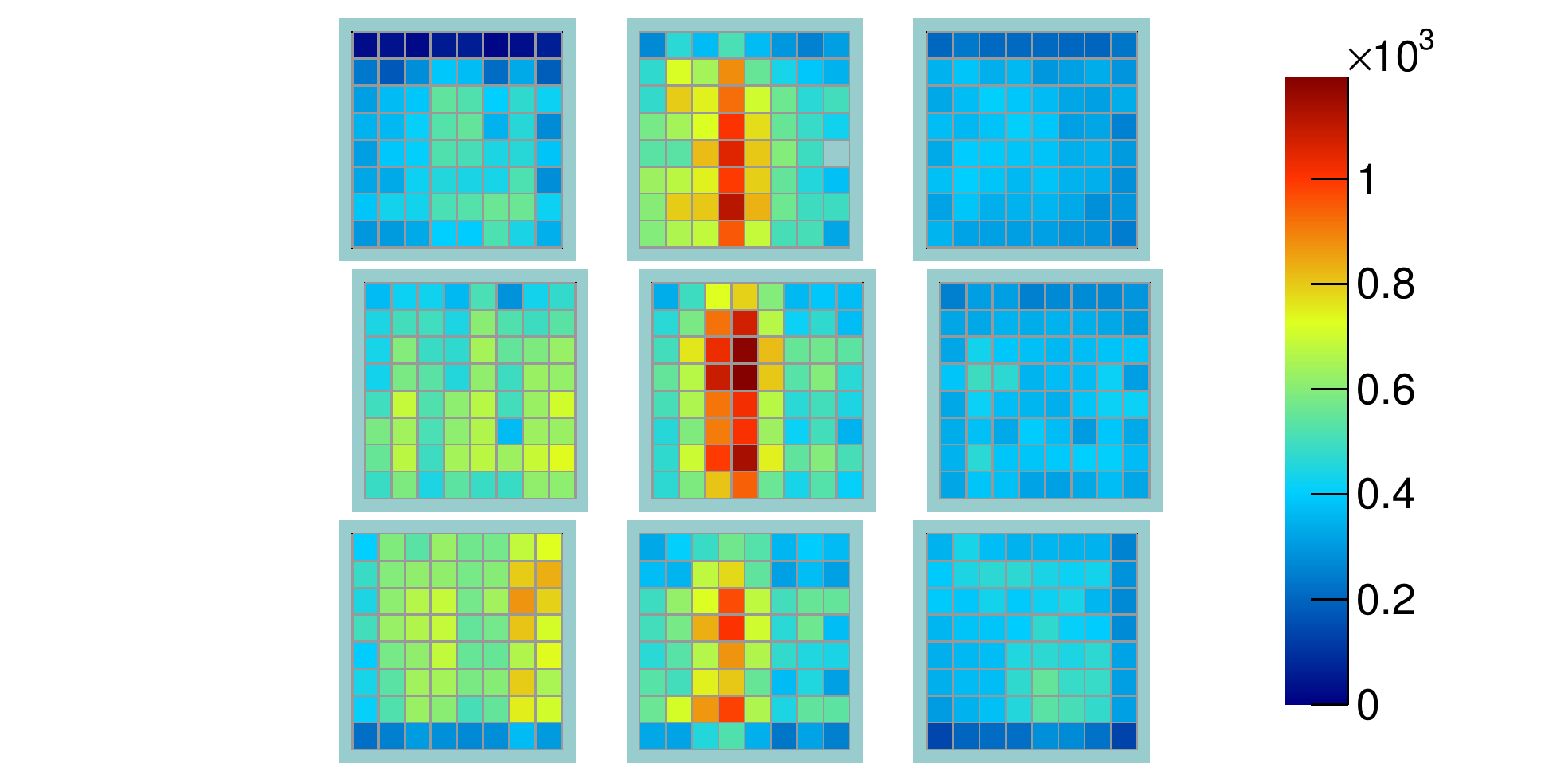}
	  \raggedright{Geant simulation, protons}   
          \hspace*{-12mm}
	  \includegraphics[width=1.2\columnwidth]{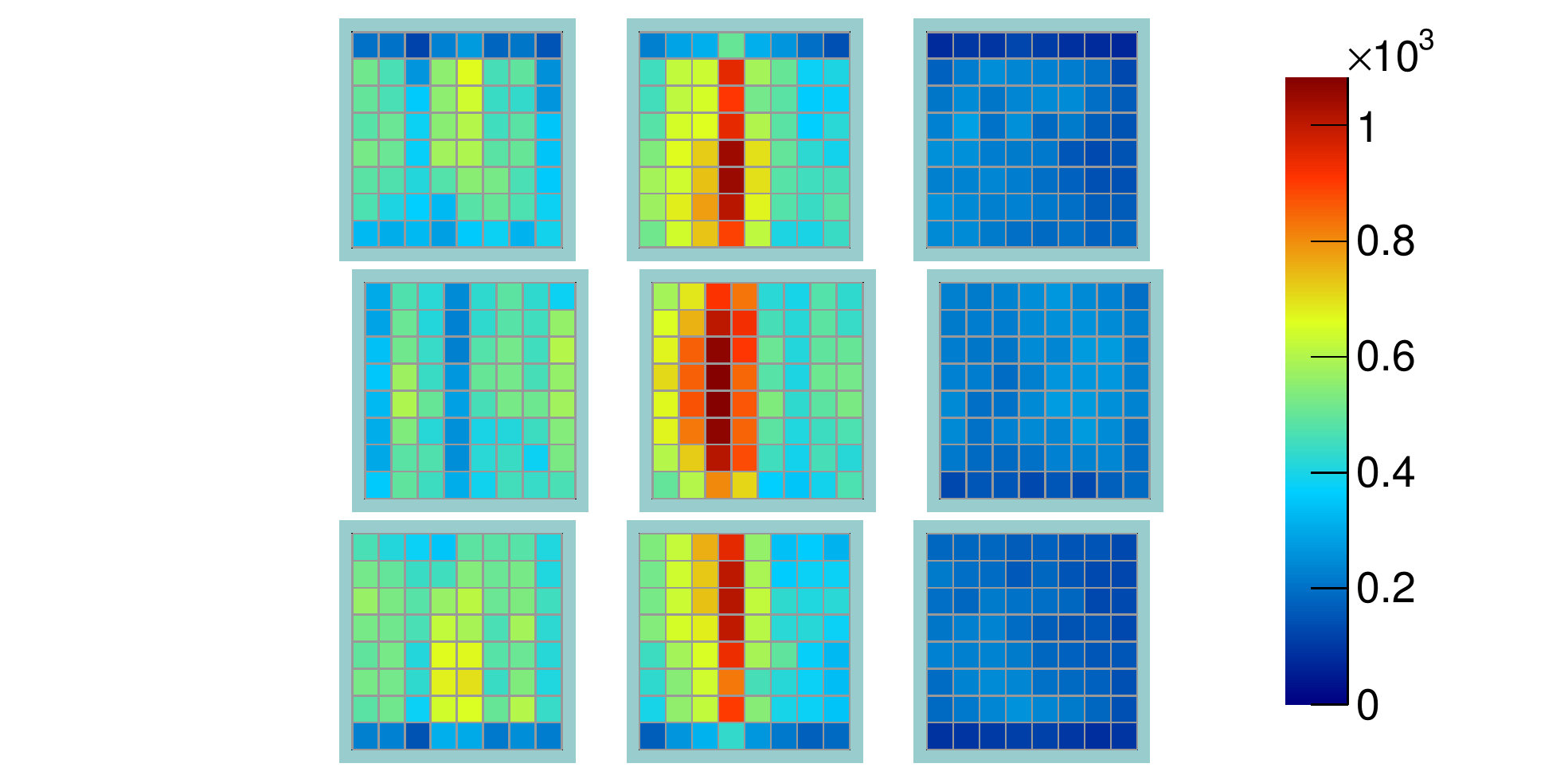}
	\end{center}
	\caption{
		Accumulated hit pattern for the 2016 prototype, shown as number of signals per MCP-PMT pixel,
		for the wide plate with a cylindrical 2-layer lens and a 7~GeV/c beam with a polar angle of 25$^\circ{}$.
		Experimental data for a proton tag (top) are compared to
		the Geant prototype simulation for a proton beam (bottom).
	}
	\label{fig:hit_pattern-plate-2016}
\end{figure}

Figure~\ref{fig:hit_pattern-plate-2016} shows the hit pattern for the wide plate 
with the cylindrical focusing lens at 7~GeV/c momentum and a polar angle of 
25$^{\circ}$ for tagged protons and the prototype simulation for a proton beam.
The simulation is in reasonable agreement with the data.

\begin{figure}[tbhp]
	\centering
	\includegraphics[width=1\columnwidth]{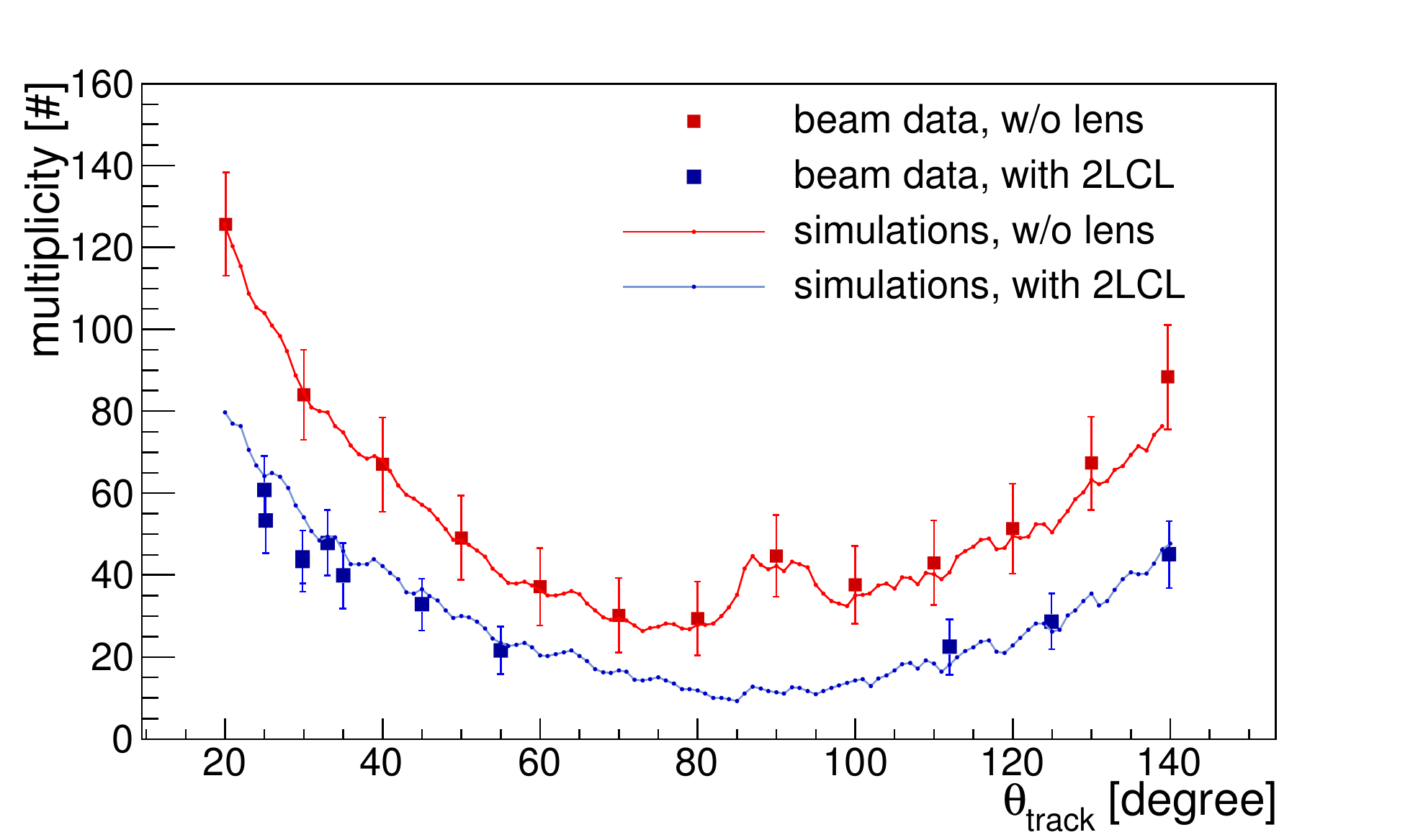}
	\caption{Photon yield as a function of the track polar angle for the wide plate
		without lens (red) and with the 2-layer cylindrical lens (``2LCL'', blue) for tagged 
		protons at 7~GeV/c beam momentum in data (points) and Geant prototype simulation (lines). 
	}
	\label{fig:nph-2016}
\end{figure}

The reconstructed photon yield as a function of the track polar angle is shown 
in Fig.~\ref{fig:nph-2016} for the configuration with the wide radiator plate,
with and without the 2-layer cylindrical lens.

The geometric reconstruction method is used to calculate the expected photon
propagation time in the plate, lens, and prism for each pixel.
Although this algorithm does not deliver precise results for the wide plate,
the calculated value can be used to put a loose cut of $\pm 5$~ns 
on the difference between the measured and expected hit to further reduce the 
background.

\begin{figure}[h]
	\centering
	\includegraphics[width=0.8\columnwidth]{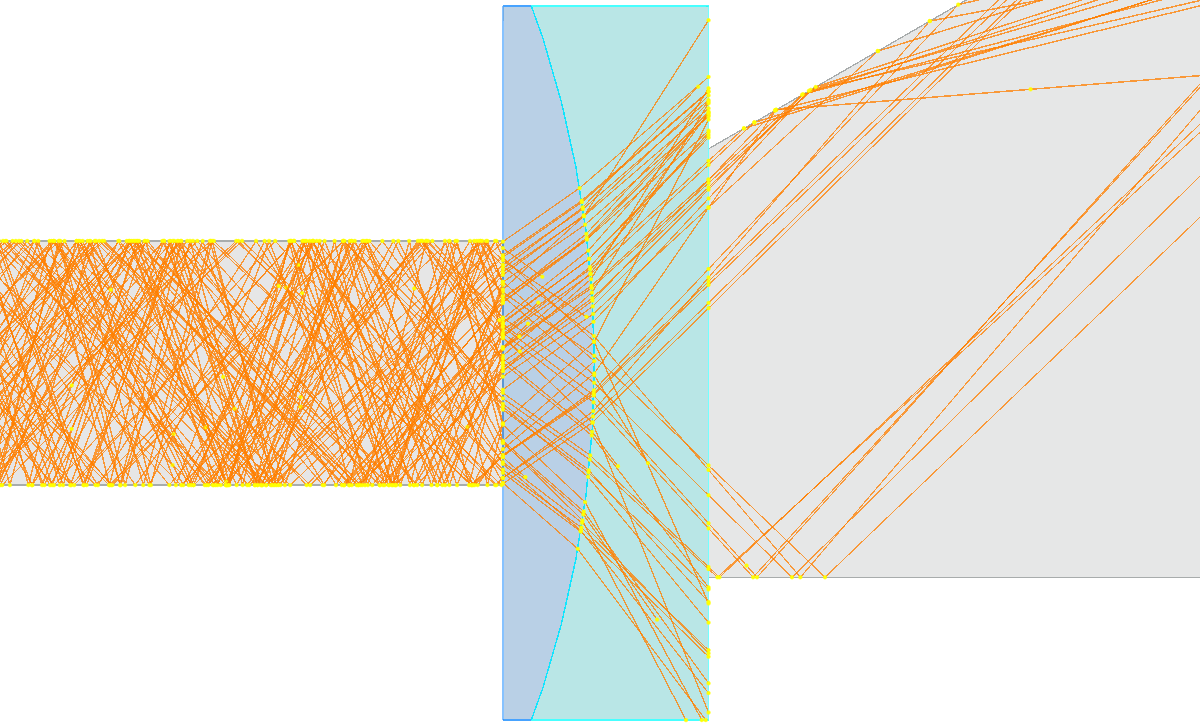}  
	\caption{Close-up of the region of the 2-layer cylindrical lens in simulation
		for the 2016 configuration. 
		The orange lines represent the Cherenkov photons originating from one $\pi^{+}$ with 7~GeV/c momentum 
		and 90$^{\circ}$ polar angle.}
	\label{fig:lens-2016}
\end{figure}

The simulation describes the experimental data well, with remaining differences
of up to 10\%.
The photon yield for the 2-layer cylindrical lens is, as expected, substantially 
lower than the yield for the plate coupled directly to the prism.
While most of this difference is due to the loss of photons inside the lens,
a significant fraction of the photons are lost at the interface of the lens
and the prism.
This loss is caused by a size mismatch of the lens and the smaller prism
used in 2016, illustrated for the prototype simulation in Fig.~\ref{fig:lens-2016}.
Steep internal photon angles are particularly affected and have a significant
probability to miss the entrance into the prism, while the steep forward beam
angles, which are of particular interest during this beam test, are mostly
unaffected by this size mismatch.

The time-based imaging method was used to determine the PID performance of the wide plate,
in particular the $\pi/p$ separation power.
The probability density functions (PDFs) were determined from statistically independent
beam data samples with the exact same detector configuration and beam condition,
selected by using the TOF tags.

\begin{figure}[bth]
	\centering
	\includegraphics[width=.99\columnwidth]{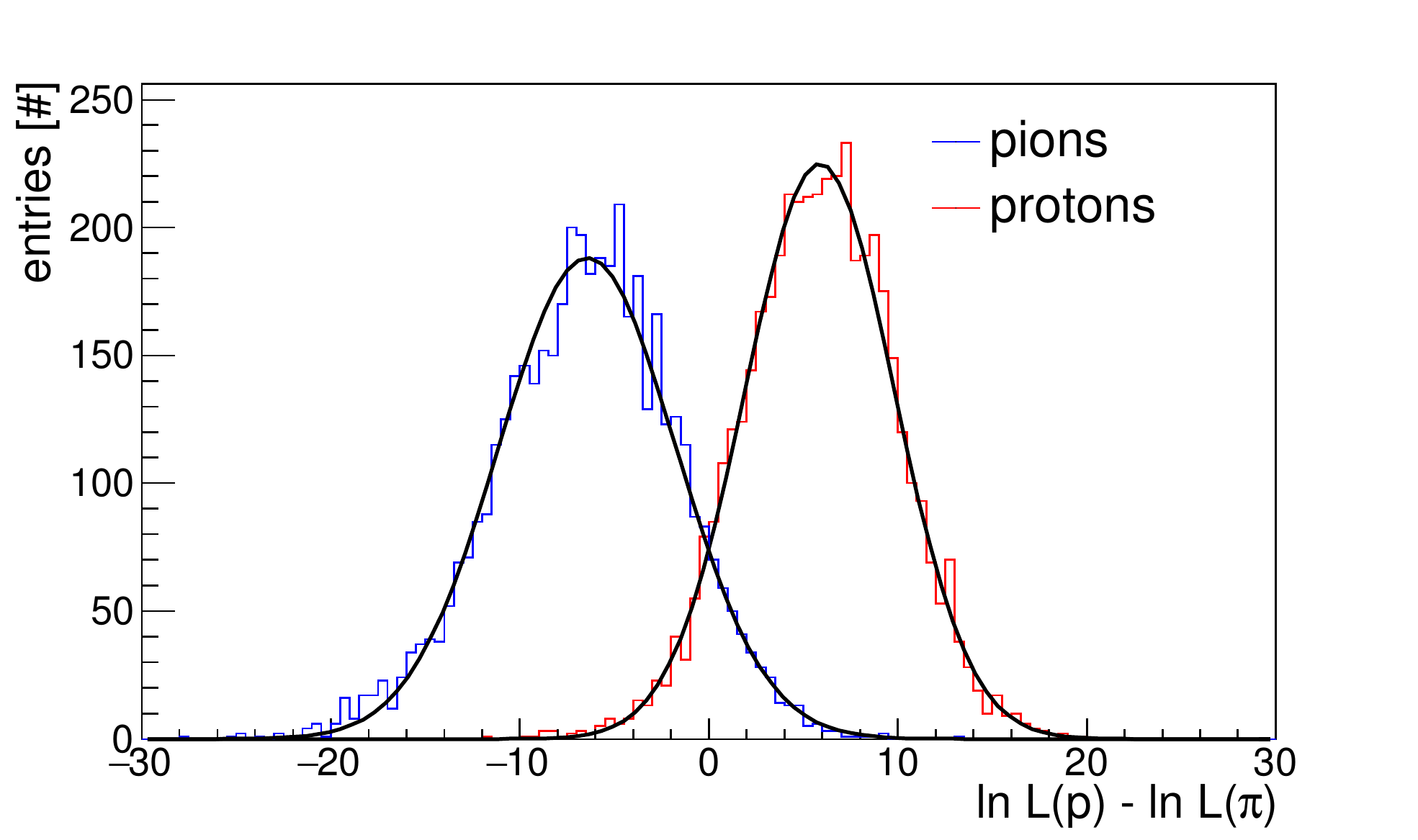}
	\includegraphics[width=.99\columnwidth]{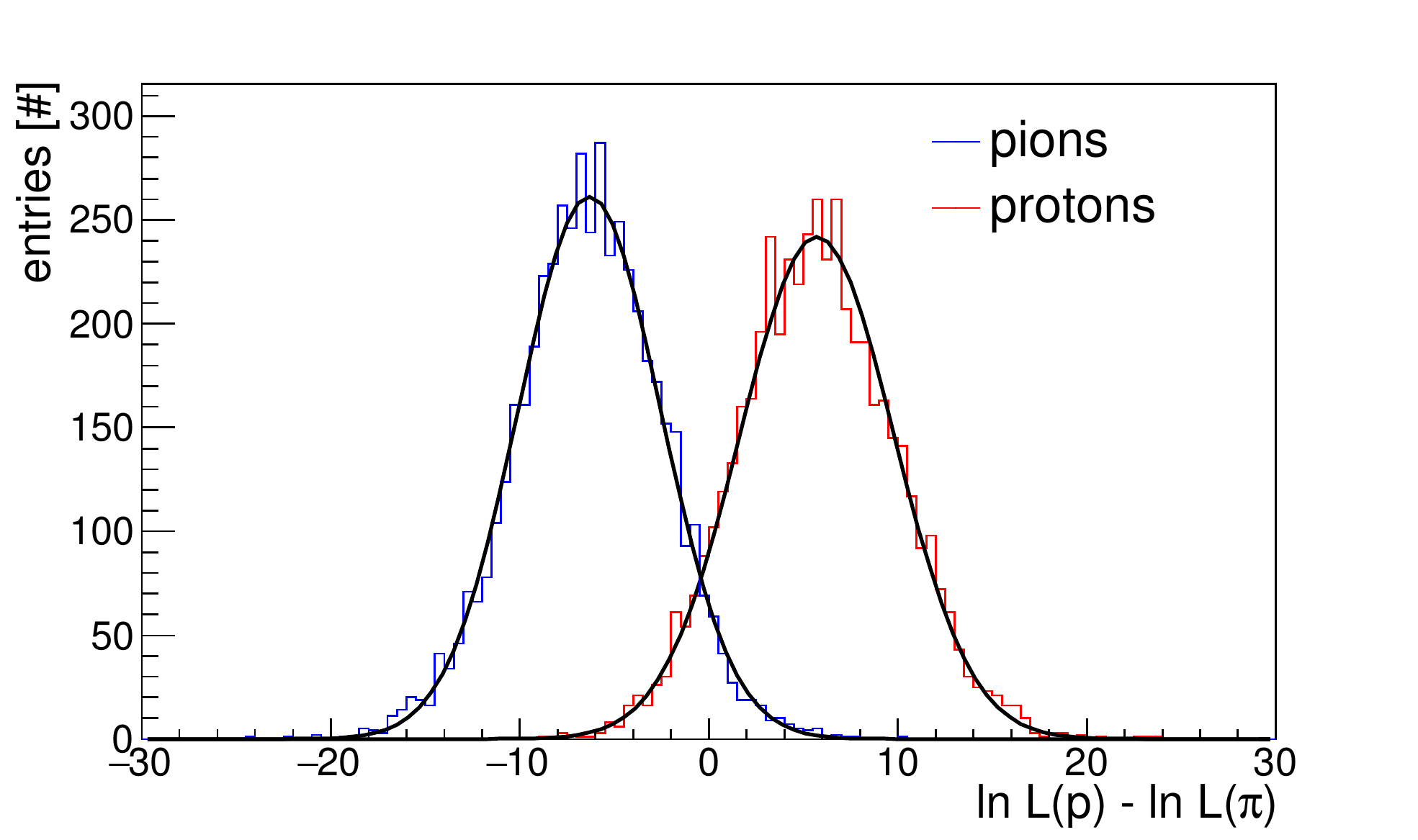}
	\caption{
		Proton-pion log-likelihood difference distributions for proton-tagged (red) and 
		pion-tagged (blue) beam events from 2016 as a result of the time-based imaging 
		method. 
		The distributions are for the wide plate without focusing (top) 
		and with a cylindrical 2-layer lens (bottom), a 
		beam with 7~GeV/c momentum and $25^{\circ}$ polar angle.  
		The separation power values from the Gaussian fits are 2.8 standard deviations 
		(s.d.) without focusing and 3.1~s.d. with focusing.
	}
	\label{fig:lh_tirecoS-plate2016}
\end{figure}

The result of the unbinned likelihood calculation for the plate with 
and without focusing at 7~GeV/c momentum and $25^{\circ}$ polar angle is shown in 
Fig.~\ref{fig:lh_tirecoS-plate2016}.
The observed $\pi/p$ separation power is $2.8^{+0.4}_{-0.2}$~standard deviations
(s.d.) for the plate without focusing.
For the plate with the 2-layer cylindrical lens the $\pi/p$ separation is
$3.1^{+0.1}_{-0.1}$\,s.d.,
in good agreement with the prototype simulation, which predicts a 
$3.3^{+0.1}_{-0.1}$\,s.d. separation value.
This is a clear improvement compared to the plate results of the 2015 
test beam campaign.

\begin{figure}[bth]
	\centering
	\includegraphics[width=.99\columnwidth]{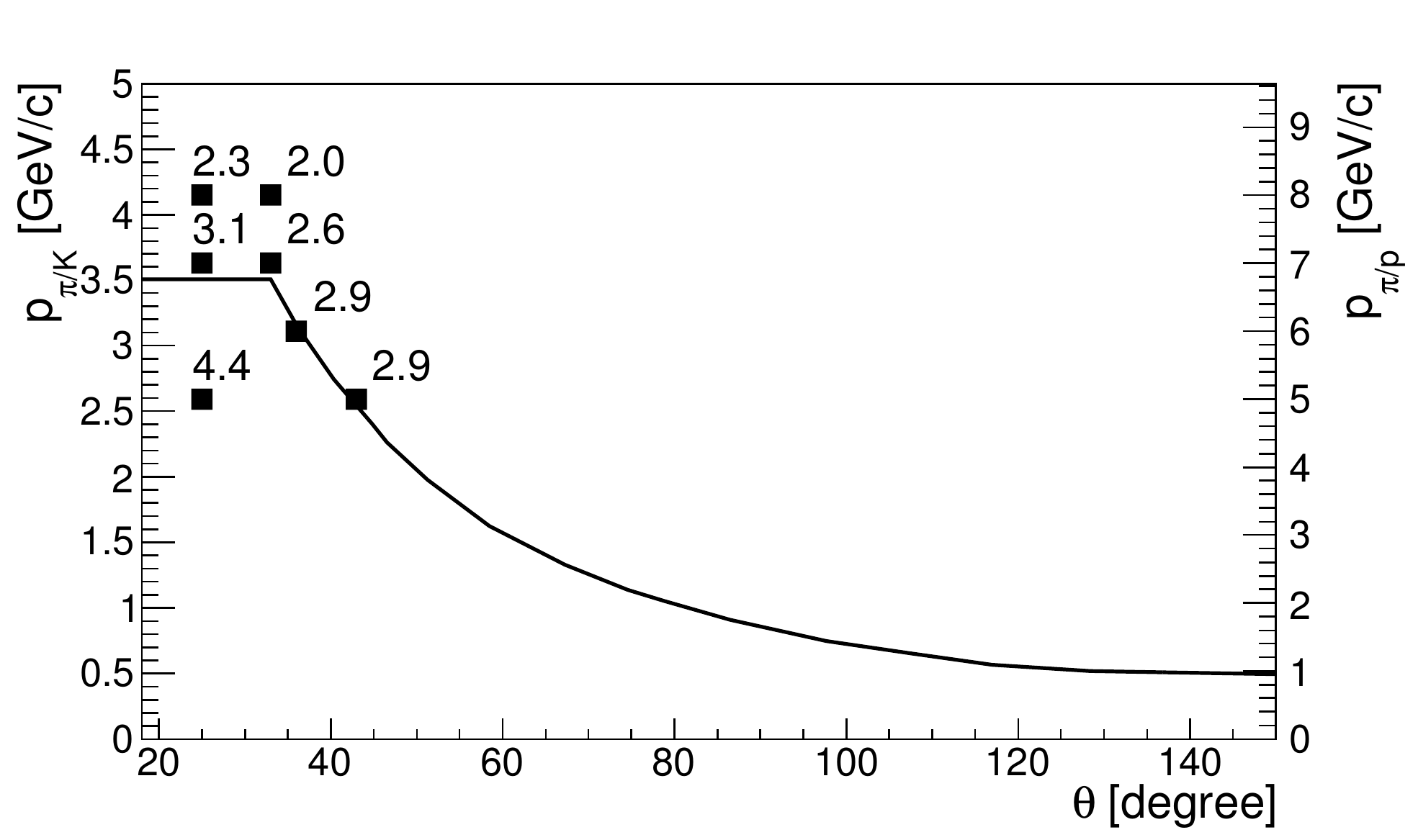}
	\caption{
		Proton-pion separation power as a function of momentum and polar angle for 
		the wide plate with a cylindrical 2-layer lens in 2016.
		The p$_{\pi/p}$ momentum denotes the beam momentum during the beam test while
		p$_{\pi/K}$ is the pion/kaon momentum where the $\pi/K$ Cherenkov angle difference
		is the same as the $\pi/p$ Cherenkov angle difference at p$_{\pi/p}$.
		The area below the black line corresponds to the final-state phase space 
		for charged kaons from various benchmark channels (see Section~\ref{cha:design}) 
		where the 3~s.d. separation goal has to be reached.
		The error of the separation power values varies between $^{+0.1}_{-0.1}$\,s.d. and
		$^{+0.4}_{-0.2}$\,s.d., depending on the available event statistics.
	}
	\label{fig:lh_phasespace-plate2016}
\end{figure}

Figure~\ref{fig:lh_phasespace-plate2016} shows the $\pi/p$ separation power 
from the 2016 data for the wide plate with the 2-layer cylindrical lens for 
various points in the \panda Barrel DIRC phase space.
The black line indicates the boundary of the expected final state kaon
phase space in \panda (see Sec.~\ref{cha:design-goals}), where the goal
for the Barrel DIRC is defined as at least 3~s.d. $\pi/K$ separation.

The $\pi/p$ separation power for all points near or inside the expected 
kaon phase space region is close to the 3~s.d. goal.
Even at 7~GeV/c, the observed $\pi/p$ separation is $3.1^{+0.1}_{-0.1}$~s.d. 
at 25$^\circ{}$ polar angle and $2.6^{+0.3}_{-0.1}$~s.d. at 33$^\circ{}$.
The errors are dominated by the systematics, in particular the asymmetric
error associated with the event statistics.

To predict the performance of the design with the wide plate in \panda based on the
2016 beam test, one needs to consider several unavoidable performance limitations of 
the prototype setup compared to the \panda Barrel DIRC configuration:
\begin{itemize}
	\item The cylindrical lens used during the beam test was made of only two layers, 
	which caused the focal plan to be much less flat than expected for the 3-layer 
	cylindrical lens in \panda.
	Furthermore, the thicker NLaK layer in the lens and the size mismatch between  
	lens and prism caused photon losses.
	\item The timing precision with the readout based on the PADIWAs and TRBs was
	on average about a factor 2 worse than the precision expected for the DiRICH 
	electronics.
	\item The MCP-PMTs used in 2016 are older models with a lower peak quantum efficiency (QE)
	and larger non-uniformity of the gain and QE than the MCP-PMTs that will be used in \panda.
	\item The prism had a smaller opening angle ($30^\circ{}$ instead of $33^\circ{}$)
	and the imaging plane was covered with an array of $3 \times 3$ MCP-PMTs with
	wider gaps between the MCP-PMTS which caused additional photon loss compared
	to the $3 \times 3 + 2$ array design for \panda.
	\item Eljen EJ-550 optical grease was used for the coupling between the 
	plate, lens, and prism, as well as between the prism and MCP-PMTs.
	The optical property of these connections was significantly worse than the
	silicone cookie coupling method and Epotek glue are expected to provide in \panda.
	\item Since the prototype data was taken without a magnetic field, charge-sharing
	between the anode pads of the MCP-PMTs caused about 15-20\% additional hits in 
	neighboring pixels on the sensors, which led to a deterioration of the spatial 
	resolution. Charge sharing will not be a factor in \panda due to the $\approx$~1~T 
	magnetic field in the region where the MCP-PMTs will be placed.
	\item The measurement of the beam position on the plate in 2016 was improved compared 
	to 2015 but still considerably worse than the expected angle and position resolution 
	of the \panda tracking system.
\end{itemize}

\begin{table*}[hbt]
	\centering
	\begingroup
	\setlength{\tabcolsep}{6pt} 
	\renewcommand{\arraystretch}{1.5} 
	\caption{Table of the $\pi/p$ separation power observed in 2016 for the 
		wide plate with the cylindrical lens at different polar angles, compared 
		to the expectation from the prototype simulation and the corresponding 
		expected $\pi/p$ and $\pi/K$ separation power for different simulation 
		configurations for the prototype and the \panda Barrel DIRC
		(see text).}
	\vspace*{3mm}
	\label{tab:sep-power-2016}
	{\small
		\begin{tabular}{|c|c|c|c|c|c|}
			\hline
			& \multicolumn{3}{l|}{$\pi/p$ Separation at 7~GeV/c [s.d.]} 
			& \multicolumn{2}{l|}{$\pi/K$ Separation at 3.5~GeV/c [s.d.]} \\\cline{2-6}
			\multirow{3}{*}{Polar Angle [$^\circ{}$]} & \multirow{3}{*}{Measurement} 
			& \multicolumn{4}{c|}{Geant Simulation Configuration} \\  \cline{3-6}
			& & \multicolumn{2}{c|}{Prototype} & \multicolumn{2}{c|}{\panda} \\\cline{3-6}
			& & 2016 & Final Optics & Full Detector & Timing $\sigma_t = 200$~ps\\\hline
			
			25  &	3.1 $^{+0.1}_{-0.1}$ &	3.3 $\pm$ 0.1 & 4.3 $\pm$ 0.1 &	6.6 $\pm$ 0.1 & 5.7 $\pm$ 0.1  \\
			33  &	2.6 $^{+0.3}_{-0.1}$ &	3.1 $\pm$ 0.1 & 3.9 $\pm$ 0.1 &	5.4 $\pm$ 0.1 & 4.8 $\pm$ 0.1  \\
			112 &	1.8 $^{+0.4}_{-0.1}$ &	1.9 $\pm$ 0.1 & 2.8 $\pm$ 0.1 &	4.1 $\pm$ 0.1 & 2.6 $\pm$ 0.1  \\
			125 &	2.3 $^{+0.3}_{-0.1}$ &	2.4 $\pm$ 0.1 & 3.0 $\pm$ 0.1 &	4.7 $\pm$ 0.1 & 3.1 $\pm$ 0.1  \\ \hline
			
		\end{tabular}
	}
	\endgroup
\end{table*}

Given these limitations, the observed $\pi/p$ separation has to be extrapolated
to the expected PID performance of the wide plate in the \panda Barrel DIRC using 
the tuned detailed Geant simulation.

Table~\ref{tab:sep-power-2016} compares the observed $\pi/p$ separation power 
from the 2016 beam test to the expectation from the prototype simulation and 
to the expected performance for several other simulation configurations,
including the full \panda Barrel DIRC setup.

The prototype simulation describes the $\pi/p$ separation power for the 2016 data 
within the errors.
Since the simulation, as previously shown, agrees similarly well with the other 
relevant observables, such as the hit pattern, timing precision, and photon yield, 
the use of the simulation to evaluate the expected performance in \panda is justified.

For a momentum of 7~GeV/c and a polar angle of $25^\circ{}$ the prototype simulation 
predicts a $\pi/p$ separation power of (3.3 $\pm$ 0.1)~s.d., which corresponds to a 
$\pi/K$ separation power of (6.6 $\pm$ 0.1)~s.d. at 3.5~GeV/c and $25^\circ{}$ in \panda.
The predicted performance exceeds the PID requirement for the \panda Barrel 
DIRC for high-momentum particles across the entire final state kaon phase space.

The extrapolation from 3.3~s.d. for the prototype simulation to 6.6~s.d. for 
the \panda simulation is a rather large step.
To understand the performance drivers in more detail, the Geant simulation was used 
to study additional configurations, summarized in Table~\ref{tab:sep-power-2016}.

The 2016 prototype simulation was modified by replacing the 2-layer cylindrical 
lens with the 3-layer cylindrical lens, the $30^\circ{}$ prism with the full-size 
$33^\circ{}$ prism, and the array of 9 MCP-PMTs with an array of 11 MCP-PMTs.
Other important parameters, like the QE and timing precision, were left unchanged.
The outcome is listed for different polar angle values in column 4 in 
Table~\ref{tab:sep-power-2016} under the header ``Final Optics.''
For 7~GeV/c and $25^\circ{}$ the $\pi/p$ separation power improves from 3.3~s.d. in 
the 2016 simulation to 4.3~s.d. in the Final Optics configuration.

Next, the \panda simulation, which includes all the expected properties of the 
Barrel DIRC components, in particular the DiRICH timing precision and the higher
QE of the next-generation MCP-PMTs, was modified to simulate the effect of a
timing precision deterioration from 100~ps to 200~ps. 
The outcome is shown in column 6 under the header ``Timing $\sigma_t = 200$~ps.''
For 3.5~GeV/c and $25^\circ{}$ the $\pi/K$ separation power goes from 6.6~s.d. in 
the default simulation to 5.7~s.d. for the worse timing precision.

\subsection{Conclusion of the 2016 Prototype Test}
\label{sec:conclusion_2016}

The 2016 beam test showed that the Barrel DIRC design with the wide plate and 
cylindrical lens can be expected to meet or exceed the \panda PID requirements.
The simulation demonstrated that optical components and MCP-PMTs of high quality, 
as well as the excellent timing precision of the DiRICH readout, are of critical 
importance to reach the PID design goal for the full kaon final state phase space.

\section{Design Decision}

The prototype tests in 2015 and 2016 successfully validated the PID performance 
of both radiator geometries, the narrow bar with the spherical lens, and the wide plate
with the cylindrical lens.
At 7~GeV/c momentum and $25^{\circ}$ polar angle the $\pi/p$ separation power was 
3.6~s.d. for the narrow bar and 3.1~s.d. for the wide plate.
Similarly good results were obtained for other polar angles and momenta.

Provided that the expected technical characteristics of the MCP-PMTs, lenses, 
and readout electonics, and are achieved, the observed $\pi/p$ separation power 
values at or above 3~s.d. extrapolate to $\pi/K$ separation powers of 4--7~s.d. 
in the full \panda Barrel DIRC simulation for the entire final state kaon phase 
space in \panda,

Since both radiator geometries are capable of meeting the PID requirements, 
other factors have to be taken into account to decide which geometry 
is selected as baseline design for the \panda Barrel DIRC.

The wide plates have the advantage that fewer pieces have to be produced
by industry. 
As discussed in Sec.~\ref{sec:cost}, the total cost of the \panda Barrel DIRC
with wide plates as radiators is, therefore, expected to be about 15\% lower 
than the design with narrow bars.

The design with narrow bars, on the other hand, has several performance advantages.

The PID performance of the narrow bars is superior to the wide plate
for most of the \panda Barrel DIRC phase space.
The $\pi/p$ separation power measured for the narrow bar with the 3-layer
spherical lens in 2015 was 3.6~s.d., compared to the value of 3.1~s.d.
observed for the wide plate with the 2-layer cylindrical lens in 2016.

The PID information from the geometrical reconstruction algorithm, explained 
in Sec.~\ref{sec:reco-geo-bars}, is only meaningful for narrow bars.
This method provides a proven and reliable alternative to the time-based 
imaging, which still carries the technical risk that the method for 
calculating the probability density 
functions analytically has yet to be developed and validated with real data.
Since the geometrical reconstruction is primarily based on the measurement 
of the spatial coordinate, this approach can provide PID even when the timing
precision is much worse than expected.
The 2015 beam test demonstrated that the $\pi/p$ separation power from the geometrical
reconstruction (3.3~s.d.) was only slightly worse than the result of the
time-based imaging (3.6~s.d.).

\begin{figure}[bth]
	\centering
	\includegraphics[width=.99\columnwidth]{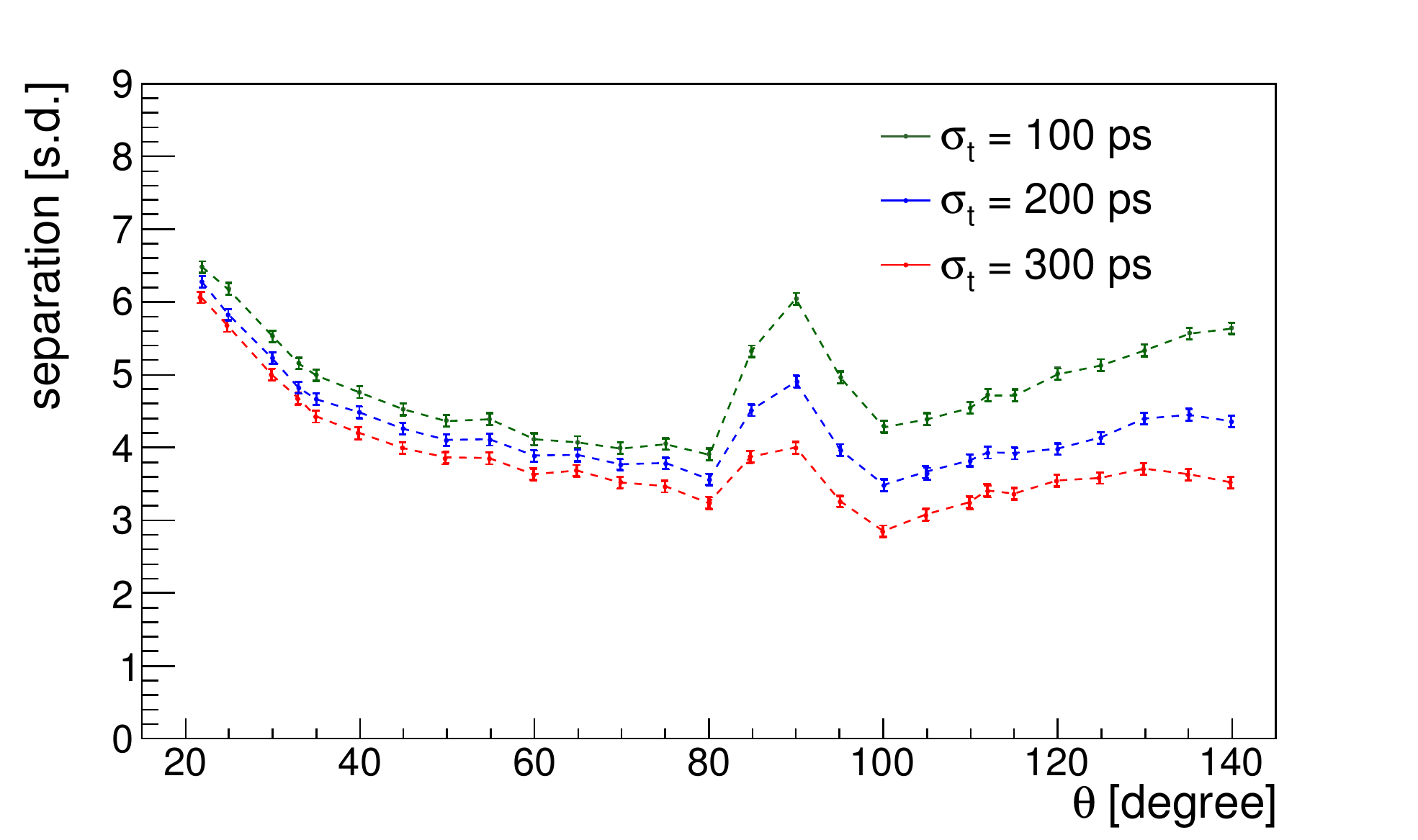}
	\includegraphics[width=.99\columnwidth]{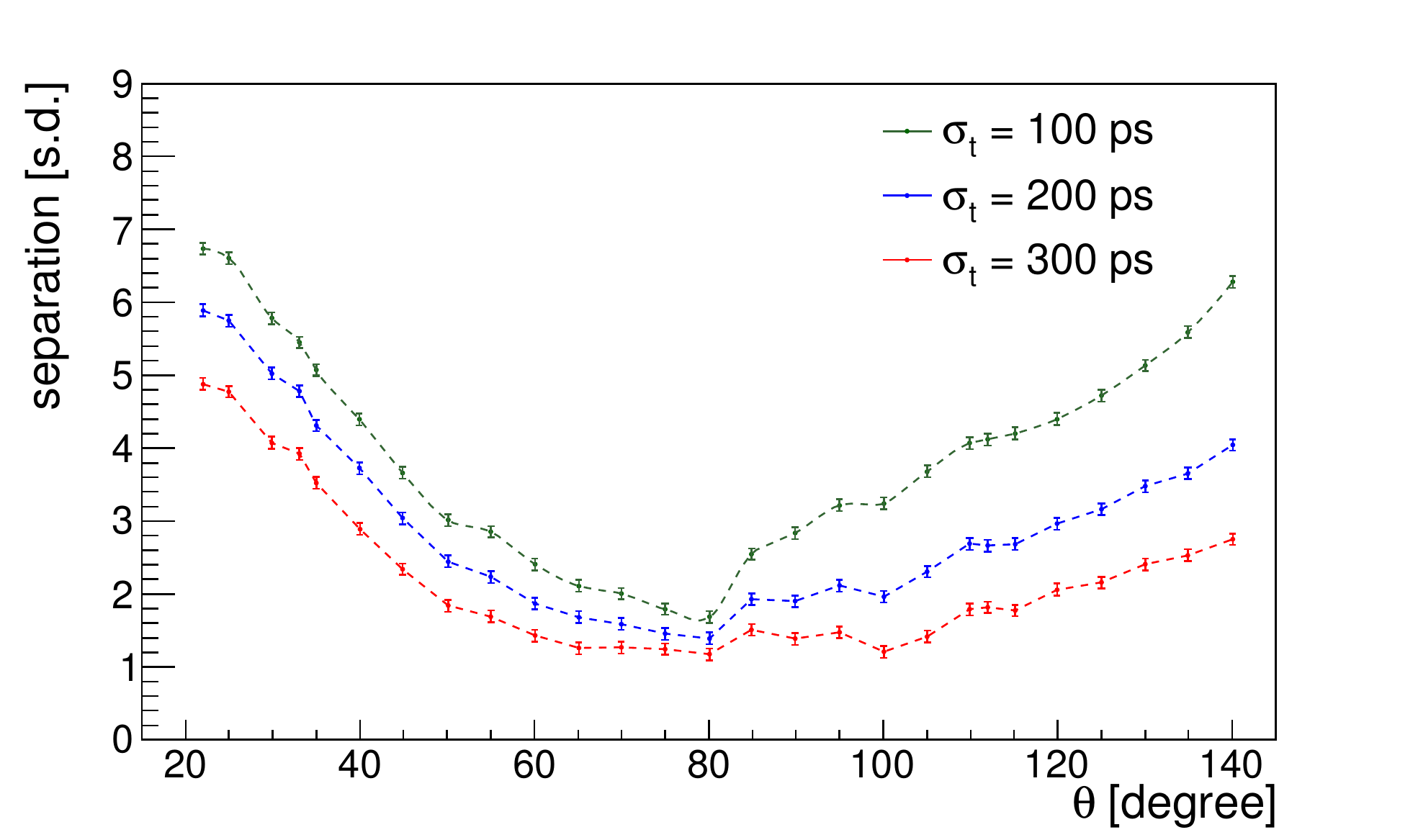}
	\caption{
		Pion/kaon separation power at 3.5~GeV/c momentum for the \panda Barrel
		DIRC simulation as the function of the polar angle 
		for the narrow bar with the 3-layer spherical lens (top) and 
		for the wide plate with the 3-layer cylindrical lens (bottom).
		The time-based imaging method is used and different values of the 
		timing precision, $\sigma_t$, are assumed in the simulation.
	}
	\label{fig:pik_sep_bar_plate_timeres}
\end{figure}

Furthermore, the geometrical reconstruction provides the ability to separate 
Cherenkov photons from background using the difference between the measured 
and expected photon propagation time in the radiator bar. 
As described in Sec.~\ref{sec:reco-geo-bars}, this information is essential in
dealing with pile-up effects at high interaction rates in \panda, and cannot
be calculated with sufficient accuracy for the design with wide plates.
It should also be noted that this algorithm can be used to determine 
the event time from the Barrel DIRC data with very little dependence on other 
\panda subdetectors.

Another advantage of the narrow bar geometry is the finer radiator segmentation 
in the azimuth angle.
The impact of multiple tracks hitting one radiator and the challenge of separating 
Cherenkov photons from these tracks was discussed in Section~\ref{cha:simulation}. 
Due to the width of the radiators, this probability is a factor 3 smaller
for the narrow bar geometry, making this design less sensitive to multi-track
effects.

Figure~\ref{fig:pik_sep_bar_plate_timeres} shows the $\pi/K$ separation power 
at 3.5~GeV/c momentum as the function of the polar angle and the timing
precision in the \panda Barrel DIRC simulation for the design using the narrow 
bar with the 3-layer spherical lens (top) and for the wide plate with the 3-layer 
cylindrical lens (bottom).
The design with the wide plate shows a significantly stronger dependence on the timing
precision than the narrow bars.
Especially in the forward direction, for polar angles below $40^\circ{}$,
where the pion and kaon momenta are the highest, the performance of the 
wide plate deteriorates quickly when the timing precisions of 100~ps is
not reached, while the performance of the design with narrow bars remains 
mostly unchanged.

Therefore, the \panda Barrel DIRC design with narrow bars provides a larger 
margin for error and can be expected to perform significantly better during 
the first \panda physics run due to the dependence of the wide plate geometry 
on excellent timing.

Due to these key performance advantages, the geometry with the narrow 
bars and the 3-layer spherical lens was selected as the baseline design 
for the \panda Barrel DIRC.

\putbib[./literature/lit_performance-validation]
\end{bibunit}

%% file: mechanics/mechanics.tex
\begin{bibunit}[unsrt]
\chapter{Mechanical Design and Integration}
\label{cha:mech}

The mechanical design of the \panda Barrel DIRC has to meet the following requirements:

\begin{itemize}
\item Use of non-magnetic and radiation-hard materials.
\item Ability to remove components for maintenance (sensors and electronics).
\item Option to install or remove bar boxes without interference with other \panda systems.
\item Secure and precise assembly and alignment.
\item Protection against mechanical instabilities or damage due to thermal expansion.
\item Placement of sensitive fused silica bars or plates and prisms in hermetically sealed containers.
\item Minimal material budget and radiation length in the acceptance region of the electromagnetic calorimeter.
\item Low construction cost.
\end{itemize}

These conditions have to be met within the tight spatial environment of the complex 
\panda target spectrometer (TS). 

The mechanical support structure for the Barrel DIRC bar boxes must also serve as 
the support of the SciTil detector, located in close proximity at a slightly larger radius.
The mechanical design has to provide the possibility to detach the entire readout unit, comprising 
the expansion volumes, electronics and sensors, from the \panda detector and the 
radiator barrel for access to the inner detectors. 
To simplify installation, each module should be mounted on rails to slide into individual slots 
in the support structure.

\section{Design Approach} \label{sec:mechdesign}

\begin{figure*}[htb]
\begin{center}
\includegraphics[width=0.7\textwidth]{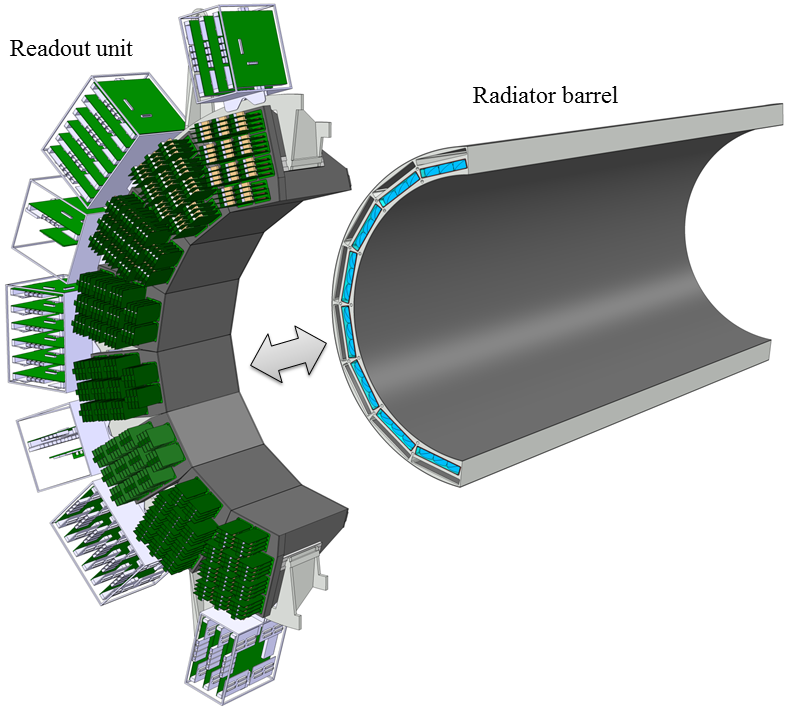}
\end{center}
\vspace*{-4mm}
\caption{Mechanical design of the two main parts of the \panda Barrel DIRC - 
half-section view: Readout unit and radiator barrel.}
\label{fig:M-AG-design_concept}

\end{figure*}

The \panda Barrel DIRC (Fig.~\ref{fig:M-AG-design_concept}) consists of two main parts: 
The radiator barrel, which contains the radiator bars or plates inside the bar boxes and 
also serves as support for the SciTil detector, and the readout unit, which includes 
the prism expansion volumes, photon sensors, and Front-End-Electronics (FEE). 
The design is modular and allows the installation or removal of each individual 
sealed container holding the optical components.
This is possible during scheduled shutdowns without significant interference with 
other \panda subdetectors.
The relative alignment between the radiator barrel and the readout unit is ensured
by alignment pins and bushings.
All major mechanical components are expected to be built from aluminum alloy and 
Carbon--Fiber–-Reinforced Polymer (CFRP) to minimize the material budget and weight 
and to maximize the stiffness.

The dimensions of these mechanical structures are shown in Tab.~\ref{tab:dim-barrel-dirc}.

\begin{table}[h]
	\centering
\setlength{\tabcolsep}{6pt} 
\renewcommand{\arraystretch}{1.5} 
	\caption{Dimensions of the \panda Barrel DIRC mechanical structures.}
	\ \ 
	\label{tab:dim-barrel-dirc}
		\begin{tabular}{ccrl}
		\hline
	Part & Property & \multicolumn{2}{c}{Value} \\
		\hline
		\multirow{4}{*}{Barrel} & Int. radius & 448 & mm \\
		                    & Ext. radius &     538 & mm \\
		                    & Tot. weight &   $\approx400$ & kg \\
		                    & z position &     -1190 to +1270 & mm \\
		                    & $\Delta$ z &     2460 & mm \\
		\hline
		\multirow{4}{*}{Readout} & Int. radius & 448 &  mm \\
                                    & Ext. radius &  1080 & mm \\
		                    & Tot. weight &   $\approx500$ & kg \\
		                    & z position &     -1710 to -1190 & mm \\
		                    & $\Delta$ z &     520 & mm \\
		\hline
		\end{tabular}
\end{table}

\subsection{Radiator Barrel}

\begin{figure}[htb]
	\begin{center}
		\includegraphics[width=1\columnwidth]{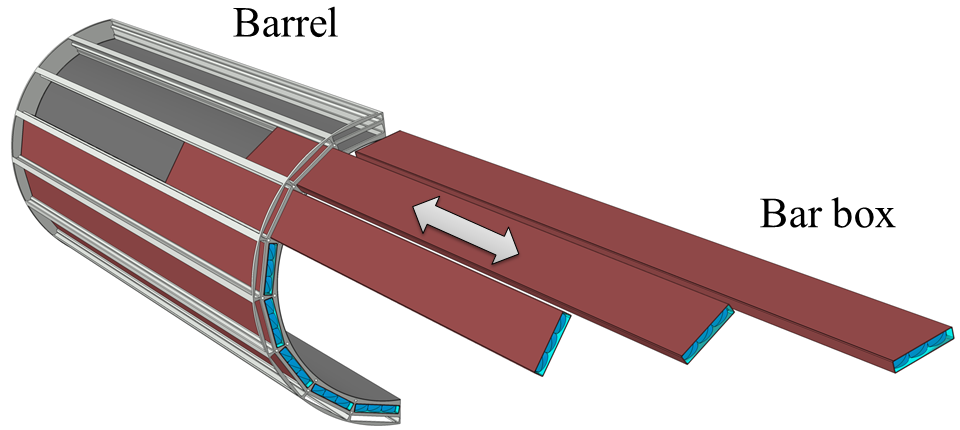}
	\end{center}
	\vspace*{-4mm}
	\caption{Mechanical design concept of the radiator barrel - half-section view. 
	Modular design to install or remove single bar boxes.}
	\label{fig:M-AG-barrel_overview}
\end{figure}

\begin{figure*}[htb]
	\begin{center}
		\includegraphics[width=0.8\textwidth]{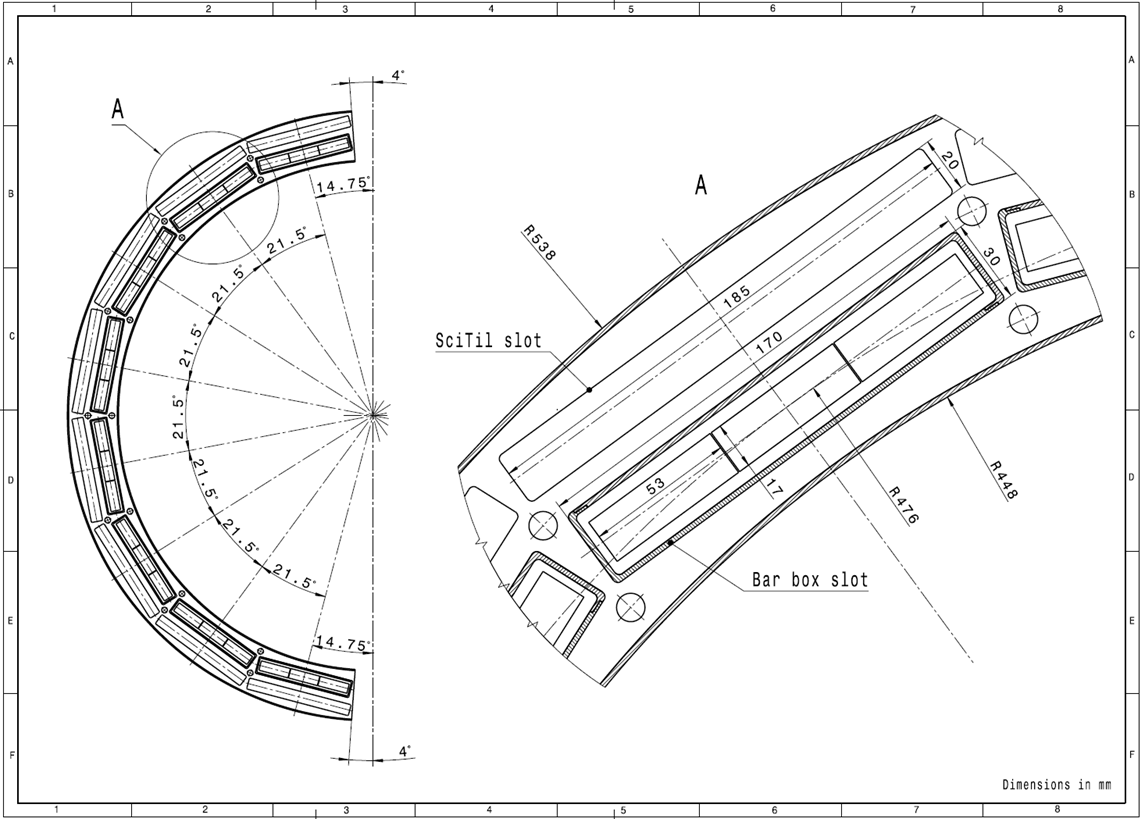}
	\end{center}
	\vspace*{-4mm}
	\caption{Sketch of a cross-section of the radiator barrel with a zoom
	into the area covered by one bar box.}
	\label{fig:M-AG-slots_drawing}
\end{figure*}

The mechanical design concept for the barrel part is based largely on the 
BaBar DIRC detector design approach~\cite{babar-dirc-nim-mec}. 
The support structure holds 16 bar boxes filled with radiator bars or plates,
as shown in Fig.~\ref{fig:M-AG-barrel_overview} and, in more detail 
with dimensions, in Fig.~\ref{fig:M-AG-slots_drawing}.

\begin{figure*}[htb]
	\begin{center}
		\includegraphics[width=0.8\textwidth]{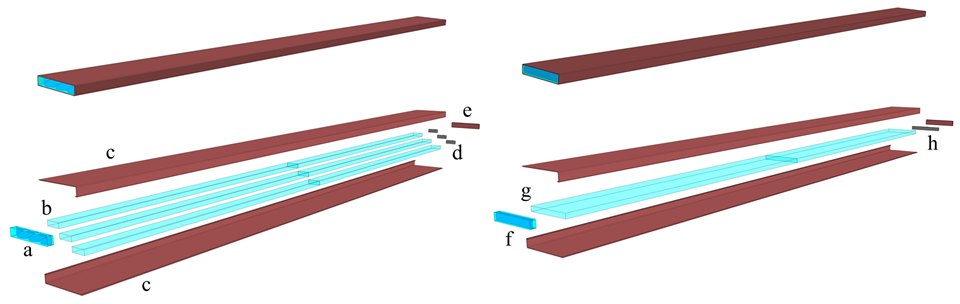}
	\end{center}
	\vspace*{-4mm}
	\caption{Bar box - exploded view - for the geometries with narrow bars (left) and 
	with a wide plate (right). 
	(a) Block of spherical lenses, (b) radiator bars, (c) bar box shells,  
	(d) spring-loaded mirrors, (e) bar box end cap, 
	(f) cylindrical lens, (g) radiator plate, (h) spring-loaded mirror.}
	\label{fig:M-AG-bar_box_expl}
\end{figure*}

Each bar box contains three radiator bars (or one plate), produced 
by gluing two shorter radiator pieces end-to-end. 
The main components of one bar box are shown in Fig.~\ref{fig:M-AG-bar_box_expl}.

The upstream end of the bar box is defined by the focusing lens system which forms the 
optical connection to the prism expansion volume. 
On the downstream end a flat mirror is attached to every radiator, bar or plate, 
perpendicular to the long axis of the radiator. 
The mirrors are spring-loaded to account for small differences in the bar lengths and  
to protect the glue joints against movement along the long axis of the radiator during 
transport.
To avoid photon loss and to prevent potential damage from physical contact each radiator is 
placed on small fixed buttons made from nylon or PEEK.
Similar buttons define the space between the radiators and the side and top covers of the bar box.
The buttons opposite the direction of the gravitational load will be spring-loaded 
to maintain a constant force.
The narrow bars are optically isolated from their neighbors by a $\approx$100\,$\mu$m air gap, 
enforced by two custom aluminum foil spacers or capton shims per bar.

The bar boxes are kept under a constant purge from boil-off dry nitrogen
to maintain a clean and dry environment and avoid possible contamination from 
outgassing of the glue and other materials used in the construction.

\begin{figure*}[htb]
	\begin{center}
		\includegraphics[width=0.7\textwidth]{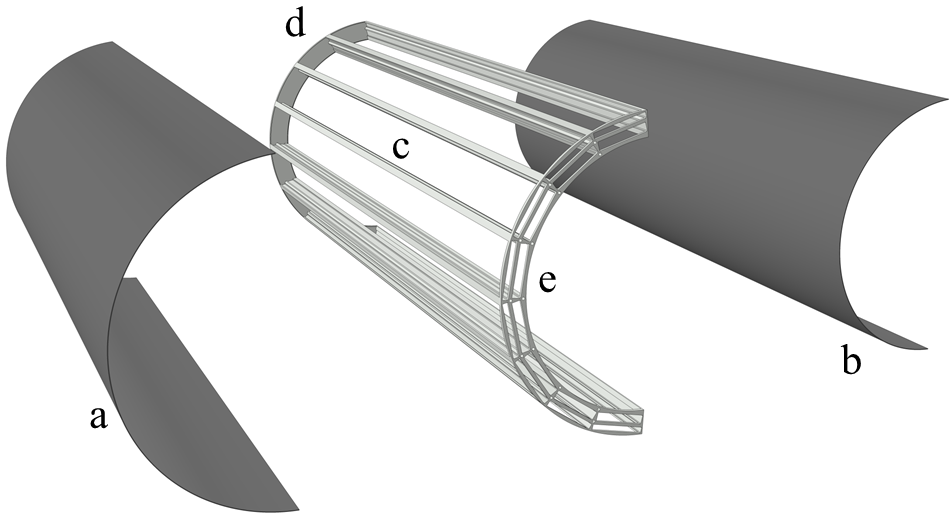}
	\end{center}
	\vspace*{-4mm}
	\caption{Support structure of the barrel - exploded view of a half section: (a) outer sheet, 
	(b) inner sheet, (c) rails, (d) downstream half-ring, and (e) upstream half-ring.}
	\label{fig:M-AG-support_exploded}
\end{figure*}

The support structure of the barrel is a hollow cylindrical frame made of two halves. 
Each half consists of rails held by two half-rings at the ends (Fig.~\ref{fig:M-AG-support_exploded}).
The nitrogen supply lines are integrated into the rail profiles. 
The whole structure is surrounded by thin inner and outer sheets to achieve a high stiffness. 
The upstream half-ring includes precision-machined rails for the precise and repeatable 
positioning of the bar boxes (Fig.~\ref{fig:M-AG-slots_drawing}).

\begin{figure}[htb]
	\begin{center}
		\includegraphics[width=1\columnwidth]{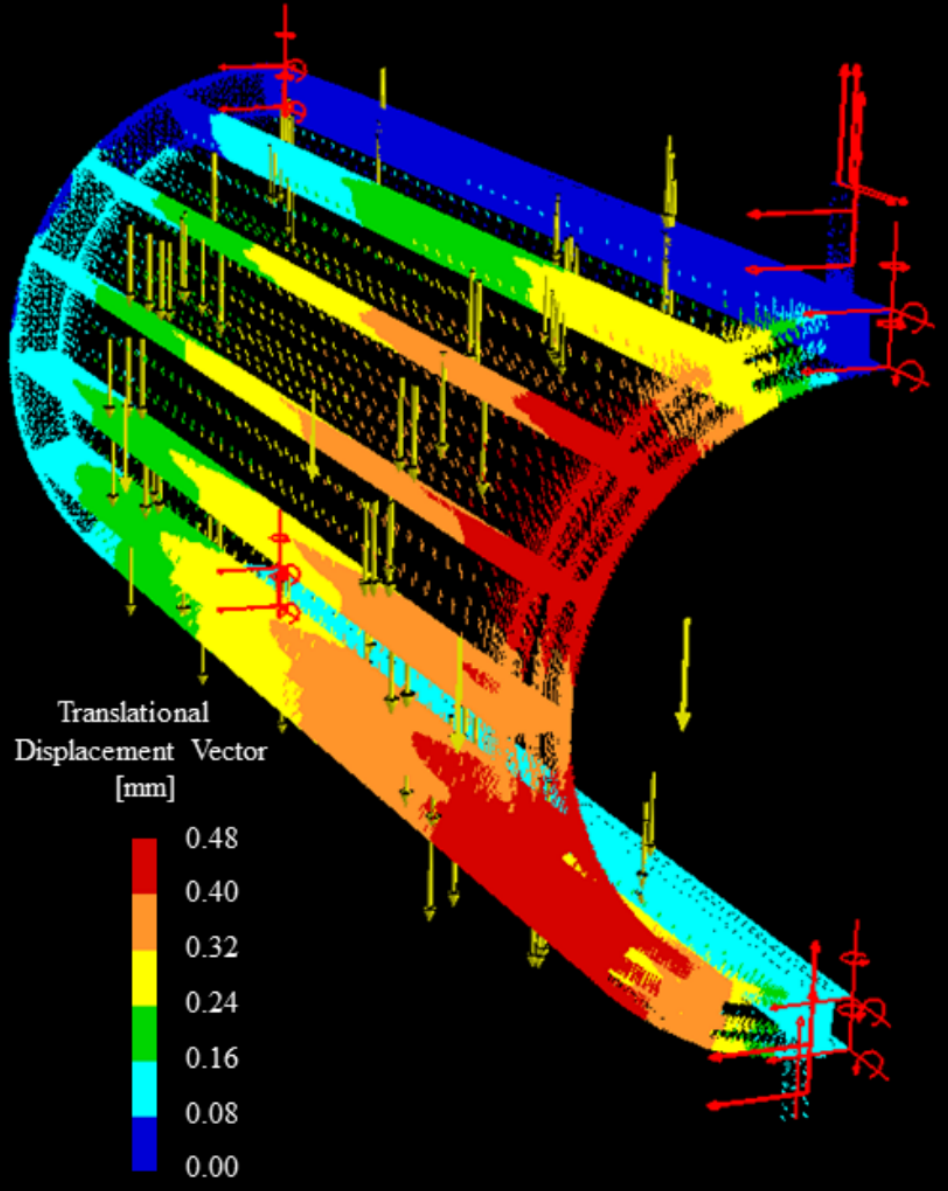}
	\end{center}
	\vspace*{-4mm}
	\caption{Support structure of the barrel - half-section. 
	FEM analysis showing the translational displacement vectors. 
	Materials: aluminum alloy (rails and half-rings), CFRP (inner and outer sheet). 
	Loads: weight of fully equipped bar boxes and own weight of support structure.}
	\label{fig:M-AG-ex_FEM_dis}
\end{figure}

\begin{figure}[htb]
	\begin{center}
		\includegraphics[width=1\columnwidth]{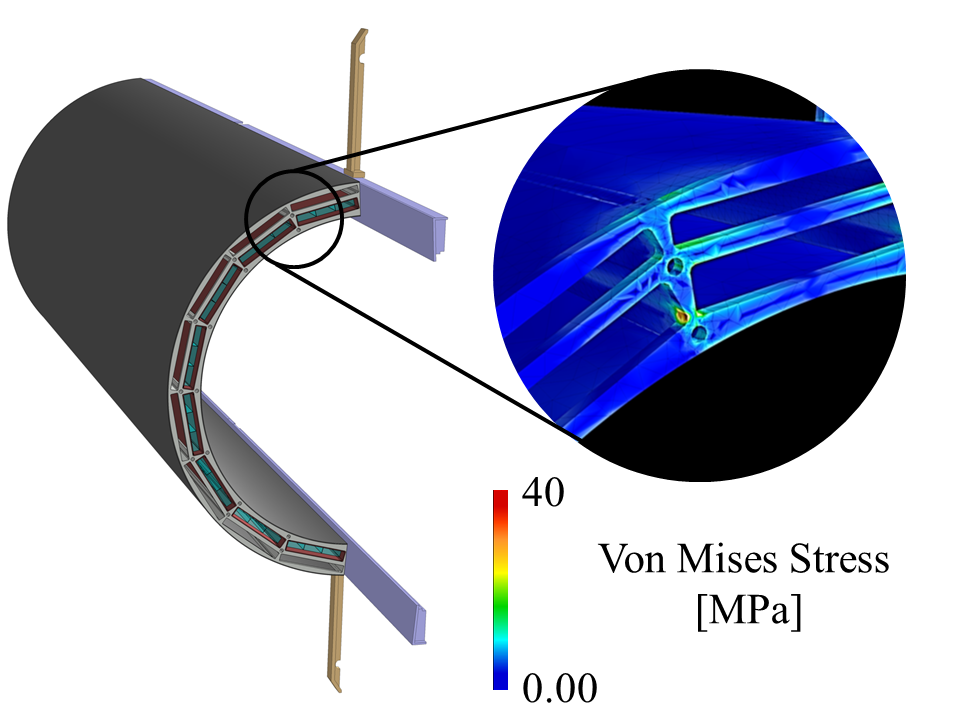}
	\end{center}
	\vspace*{-4mm}
	\caption{FEM analysis of the radiator barrel support frame showing the stress 
	distribution (von Mises) in the corners of the slots. 
	}
	\label{fig:M-AG-ex_FEM_peaks}
\end{figure}

The design goal is to limit the maximum displacement to less than 0.5~mm at any point.
A the analysis of the support structure (Fig.~\ref{fig:M-AG-ex_FEM_dis})
using the finite elements method (FEM)
shows that this goal is reached with the current design. 
The stress levels are moderate with uncritical stress peaks, far below the permissible 
elastic limit of a typical aluminum alloy, in the corners of the slots 
(Fig.~\ref{fig:M-AG-ex_FEM_peaks}). 

The active area of the synthetic fused silica radiator bars or plates covers about 85\% of 
the full azimuthal angle.
The loss in coverage is caused in equal parts by the $\pm 4^\circ{}$ gap at the top and 
bottom of the Barrel DIRC, due to the target beam pipe, and by the space between 
adjacent bar boxes, required for the rails and mechanical support structure.

\begin{figure}[htb]
	\begin{center}
		\includegraphics[width=1\columnwidth]{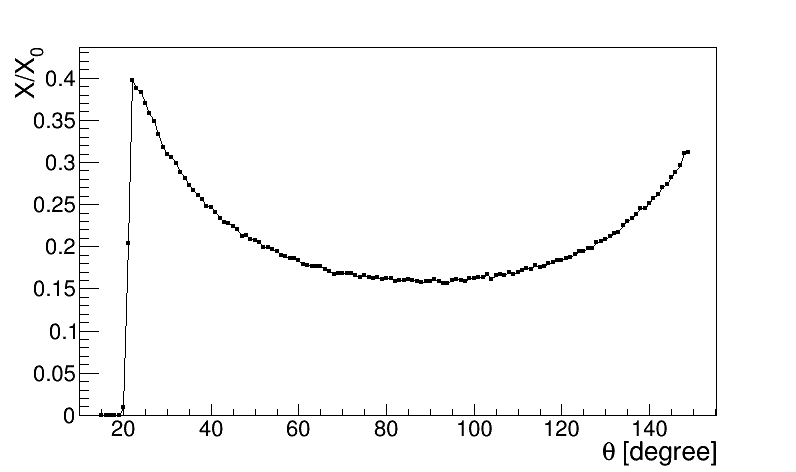}
	\end{center}
	\vspace*{-4mm}
	\caption{Material budget of the Barrel DIRC  as function
		of the polar angle in units of radiation length ($X_0$).
		The values are averaged over the azimuth angles of the Barrel DIRC acceptance and 
		determined from the full Geant detector simulation.}
	\label{fig:mec-radlength}
\end{figure}

The material of the mechanical components and the fused silica radiators adds up, on 
average, to about 16\% radiation length at normal incidence.
Due to the longer path lengths in the matrial this value increases to 40\% for steep forward 
angles, as shown in Fig.~\ref{fig:mec-radlength}.

\subsubsection*{SciTil Integration}
\begin{figure}[htb]
	\begin{center}
		\includegraphics[width=1\columnwidth]{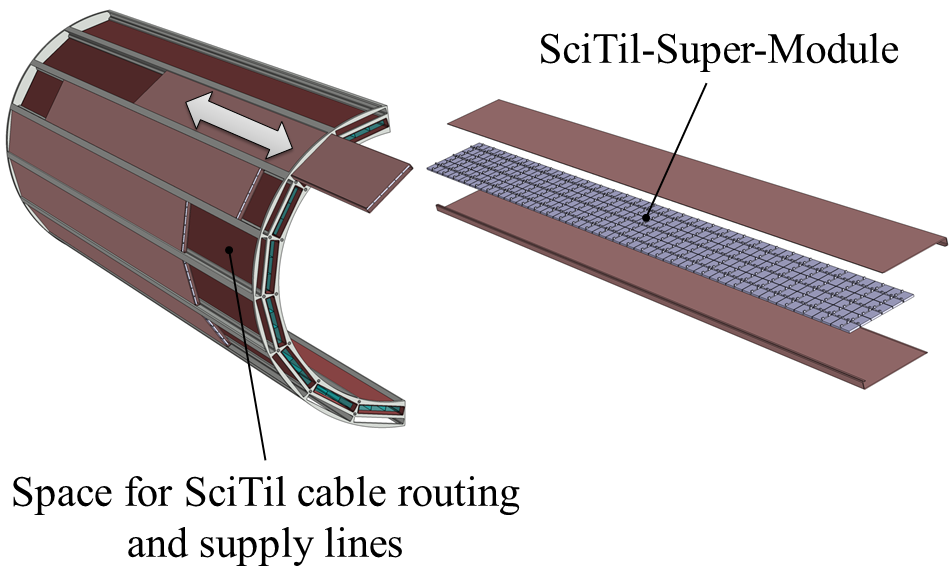}
	\end{center}
	\vspace*{-4mm}
	\caption{Position of SciTil Super Modules integrated in the radiator barrel.}
	\label{fig:M-AG-scitil_int}
\end{figure}

The Barrel DIRC support structure includes space for 16 SciTil boxes on the outer side 
with one SciTil super module per box (Fig.~\ref{fig:M-AG-scitil_int}). 
Upstream, in front of the super modules, appropriate space is reserved for supply lines,
cooling, and cable routing.
The modular design approach allows access to the SciTil boxes during shutdowns.

\subsection{Readout Unit}
\label{sec:mech-readout}
\begin{figure*}[htb]
	\begin{center}
		\includegraphics[width=0.78\textwidth]{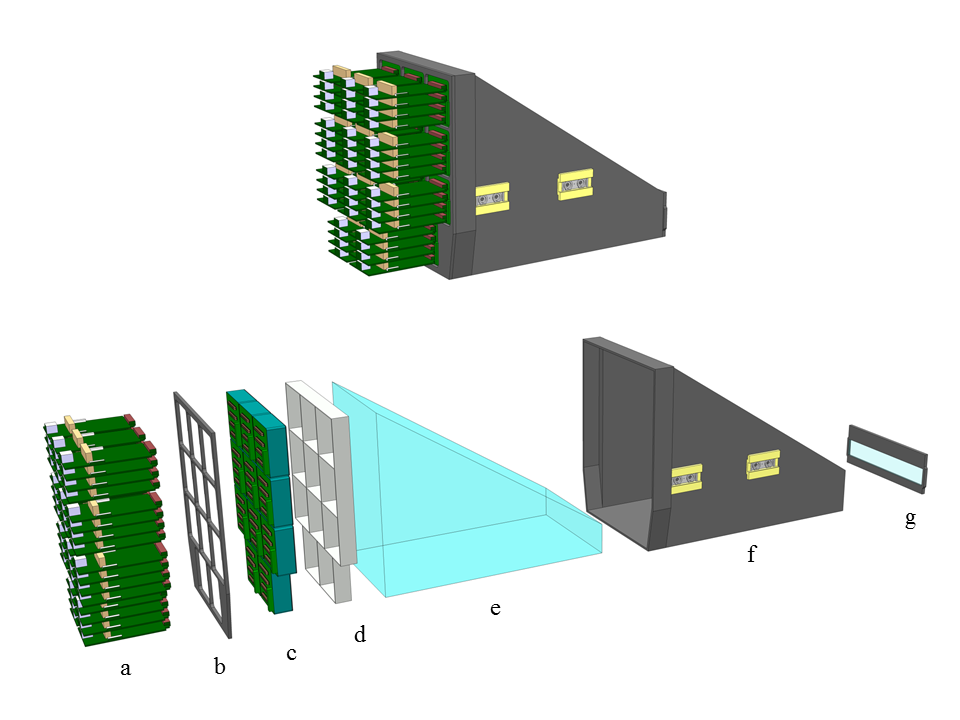}
	\end{center}
	\vspace*{-4mm}
	\caption{Components of the prism box - exploded view: (a) FEE, 
	(b) box flange, (c) MCP-PMT unit, (d) grid, (e) synthetic fused silica prism, 
	(f) box with linear unit, and (g) coupling flange.}
	\label{fig:M-AG-prism_box}
\end{figure*}

The design of the \panda Barrel DIRC expansion volume region is very different
from the BaBar DIRC. 
Instead of one large tank filled with ultra-pure water it is based on 16 optically 
isolated boxes with synthetic fused silica prisms.
Each prism box is light-tight, purged with boil-off dry nitrogen, contains
one prism and 11 MCP-PMTs (Fig.~\ref{fig:M-AG-prism_box}), 
and is coupled optically to the bar box by a coupling flange and
a silicone cookie, made, for example, from Momentive TSE3032 or RTV615~\cite{momentive3-mec} material (the latter is still to be proven radiation hard).

The light-seal on the upstream end of the prism box is provided by the prism box
flange and a grid which holds the MCP-PMTs in place and also provides the ability 
to access and exchange single MCP-PMTs in situ.
Optical coupling between the MCP-PMTs and the prism will be achieved by the same
silicone cookie material.

\begin{figure*}[htb]
	\begin{center}
		\includegraphics[width=0.7\textwidth]{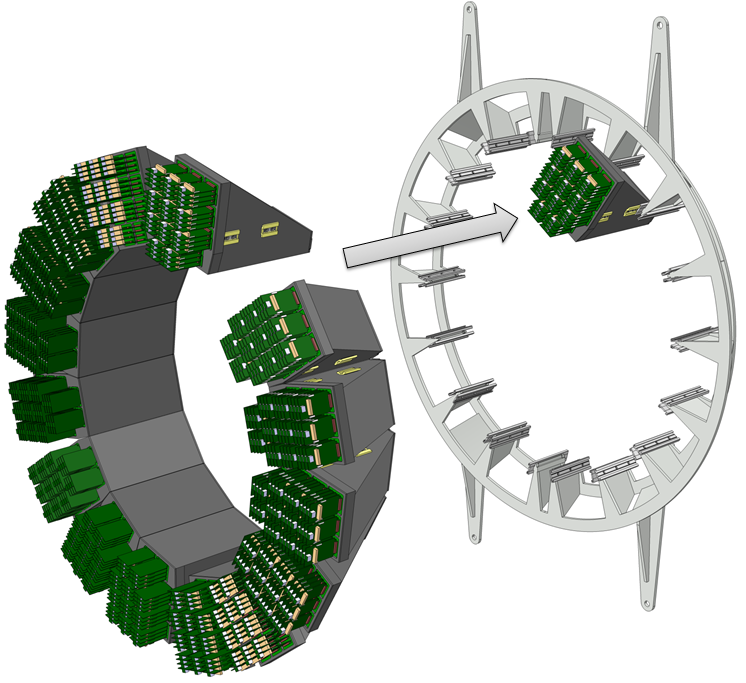}
	\end{center}
	\vspace*{-4mm}
	\caption{Suspension of 16 prism boxes inside the readout unit support ring. 
	Each box is aligned and positioned by precision linear slides in the axial direction.}
	\label{fig:M-AG-ev_box_frame}
\end{figure*}

The prisms are supported by small, round nylon or PEEK buttons to minimize the 
contact area and to avoid photon loss. 
The support structure of the prism boxes is based on a circular ring segment, attached 
on the cryostat by four arms. 
Each box is aligned and positioned on a precision linear slide 
(Fig.~\ref{fig:M-AG-ev_box_frame}).

\begin{figure}[htb]
	\begin{center}
		\includegraphics[width=0.9\columnwidth]{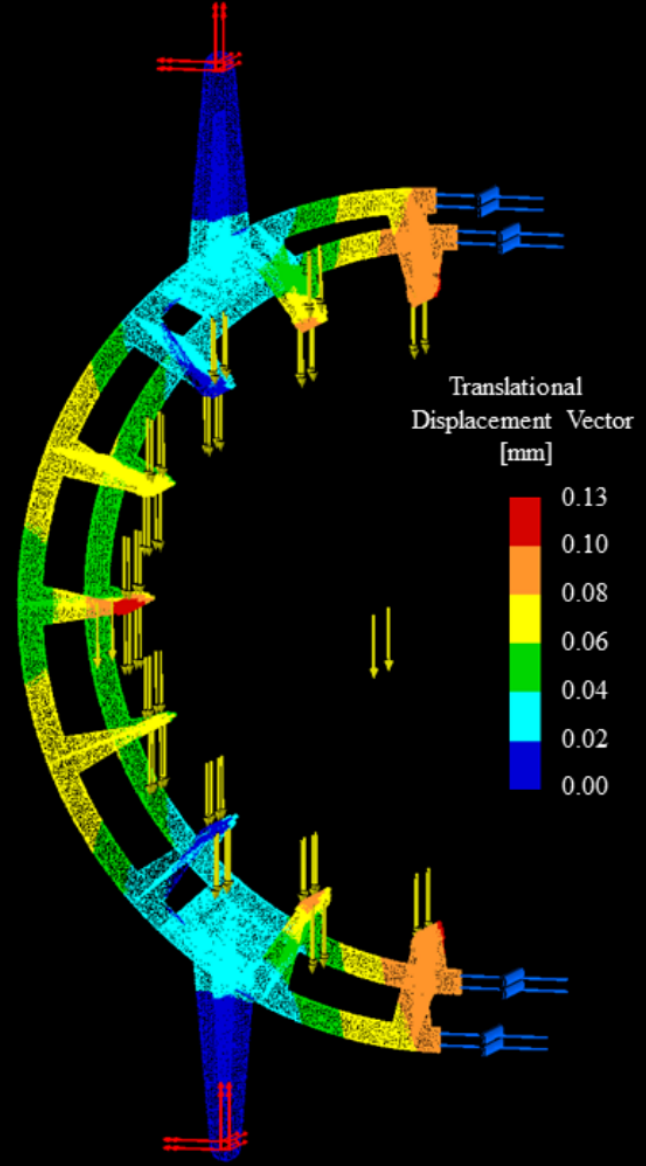}
	\end{center}
	\vspace*{-4mm}
	\caption{Support ring of the readout unit. 
	FEM analysis showing the translational displacement vector (analysis of half-ring 
	due to symmetry). Material: aluminum alloy, loads: weight of fully equipped 
	prism boxes and weight of aluminum ring structure.}
	\label{fig:M-AG-ev_frame}
\end{figure}

\begin{figure}[htb]
	\begin{center}
		\includegraphics[width=1\columnwidth]{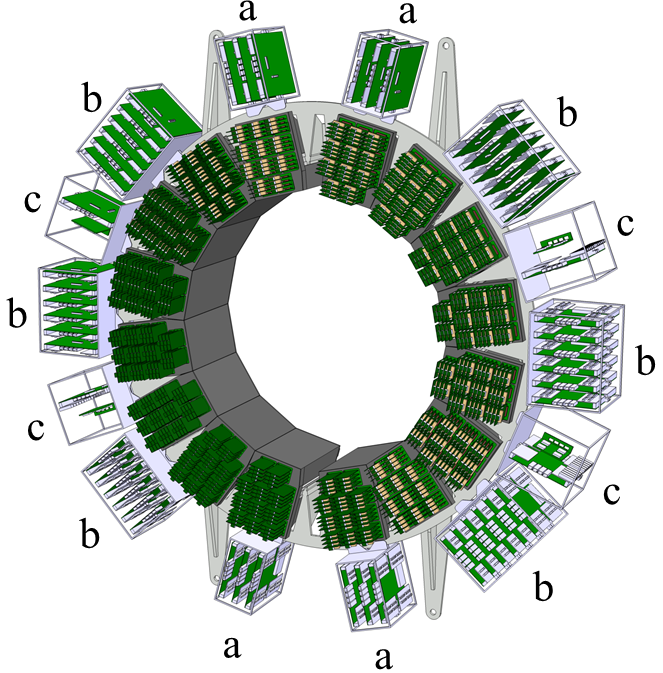}
	\end{center}
	\vspace*{-4mm}
	\caption{Readout unit including 16 prism boxes and readout electronics on top - 
	(a) sub-rack including TRBs for one box, (b) sub-rack including TRBs for two 
	boxes, (c) sub-rack including Central Trigger System and network switch.}
	\label{fig:M-AG-ev_com}
\end{figure}

Since the entire weight of the readout unit is supported by the cryostat, no weight 
force is transmitted to the radiator barrel in order to maintain the correct alignment.  
FEM simulations (Fig.~\ref{fig:M-AG-ev_frame}) were used to design the support ring, 
which is made of aluminum alloy and has a diameter of 1468~mm. 
In the outer region, at larger radii, there is sufficient space for mounting small racks 
with readout electronics as well as patch panels and to integrate supply lines for 
nitrogen purging and the FEE cooling system (Fig.~\ref{fig:M-AG-ev_com}). 
Due to the location on the far upstream end of the \panda TS
special efforts to minimize material or to create a homogeneous radiation 
length profile were not required.

\clearpage
\section{Integration into \panda}

\subsection{Neighboring Subdetectors} \label{sec:surdets}

\begin{figure}[htb]
	\begin{center}
		\includegraphics[width=1\columnwidth]{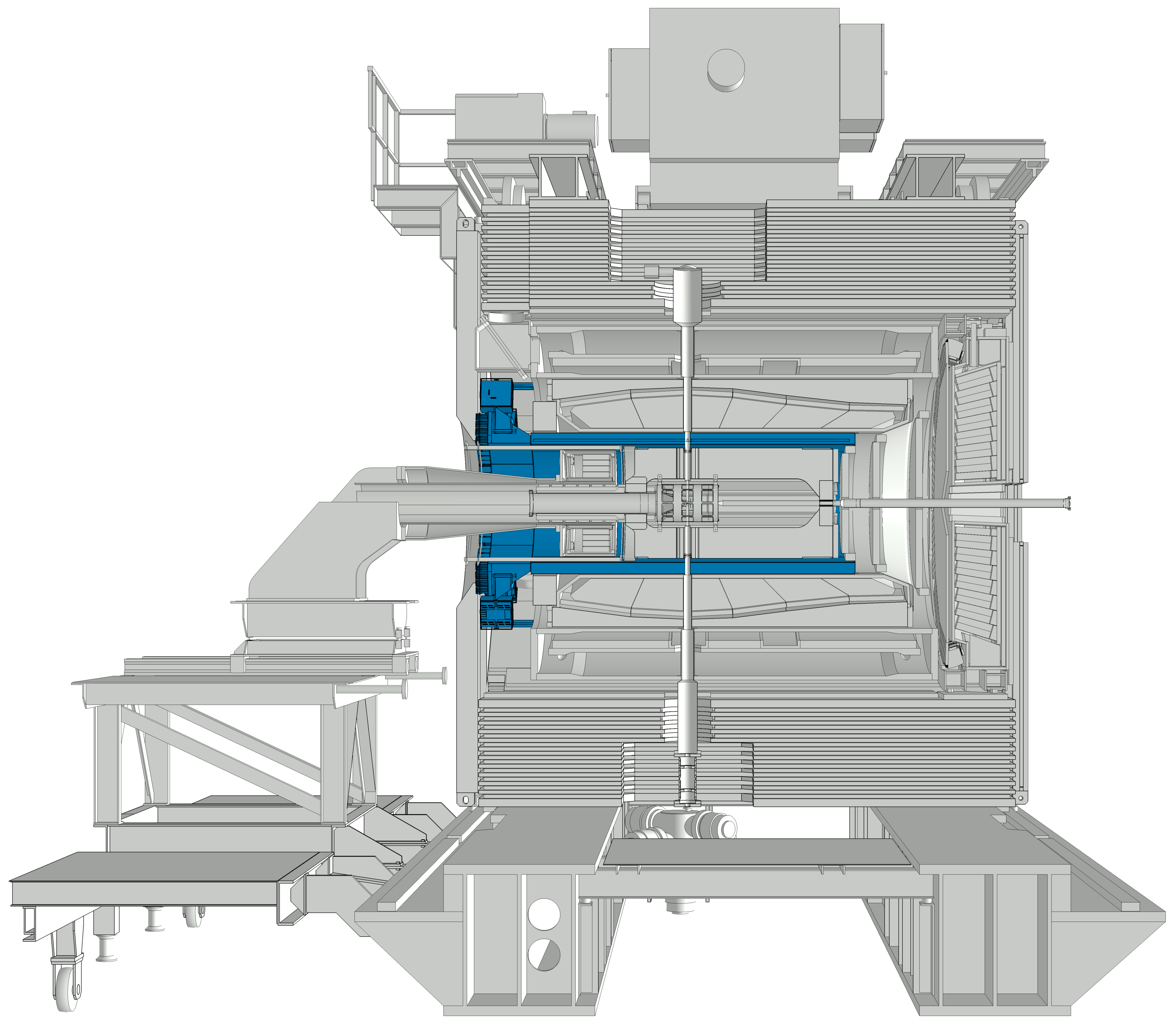}
	\end{center}
	\vspace*{-4mm}
	\caption{Cross section of the \panda Target Spectrometer with the Barrel DIRC marked in blue.
	The auxiliary platform used for detector installation is seen on the left.}
	\label{fig:M-AG_TS_DIRCs}
\end{figure}

The Barrel DIRC is located in the densely packed TS volume, as can be seen in 
Fig.~\ref{fig:M-AG_TS_DIRCs}, in close proximity to several \panda subdetectors. 
In the downstream part of the barrel, for polar angles between 22$^\circ{}$ and 140$^\circ{}$,
the Barrel DIRC shares a boundary with the central tracking detectors, 
which comprise the Straw Tube Tracker (STT) and the Micro Vertex Detector (MVD).
The Backward Endcap Electromagnetic Calorimeter (EMC) is located near the inner radius 
of the Barrel DIRC at larger polar angles.
In the forward direction the Barrel DIRC borders on the Gaseous Electron Multiplier 
(GEM) detector and the Barrel EMC surrounds the common 
Barrel DIRC/SciTil support structure.

To facilitate detector integration the groups representing all subdetectors 
have agreed on dimensions and volumes with an additional 4~mm clearance 
between neighboring volumes.

\subsection{Installation Procedure} \label{sec:mounting}

\begin{figure*}[htb]
	\begin{center}
		\includegraphics[width=0.8\textwidth]{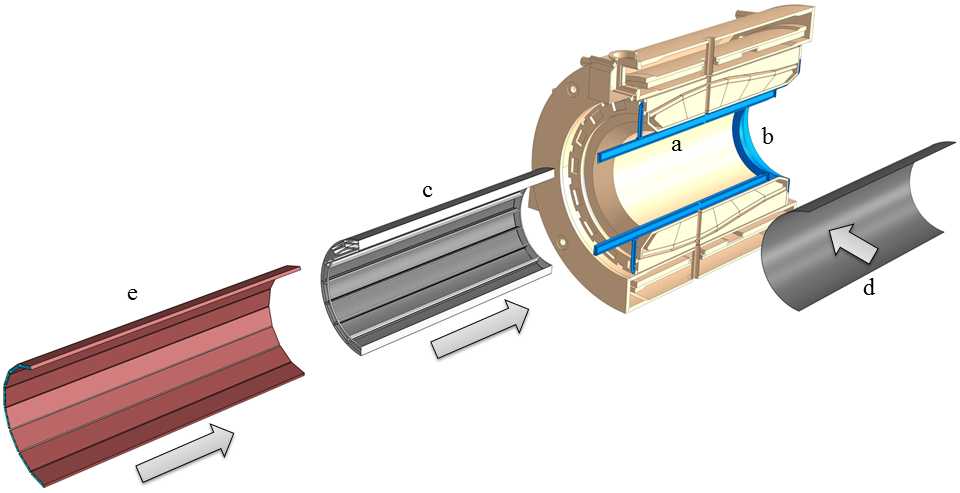}
	\end{center}
	\vspace*{-4mm}
	\caption{Installation procedure of the barrel - half section view: 
	(a) central tracker (CT) beams, (b) downstream CT beam support cone, (c) half-frame of the barrel, 
	outer sheet mounted, (d) inner sheet of the half-frame, (e) eight bar boxes.}
	\label{fig:M-AG-barrel_in}
\end{figure*}

The first step of the \panda Barrel DIRC installation procedure is to mount the 
two halves of the radiator barrel support onto the central tracker (CT) beams and
to connect the forward half-rings to the downstream cone that supports the CT beams
(Fig.~\ref{fig:M-AG-barrel_in}). 
The outer sheets of the frames are already attached at that point while  
the inner sheets are not to ensure access to the fastening 
points inside the support structure. 
To maintain appropriate clearances with respect to the barrel calorimeter
the two halves of the support structure will be transported on a trolley,
placed on the auxiliary platform behind the upstream end of the \panda 
detector. 
After the two half-barrels have been installed, the inner sheets of the frames can 
be mounted and the placement of the support structure is surveyed to verify
that the alignment is correct.
Next the bar boxes, connected to strong-backs and rotated into the appropriate 
angular orientation, are lifted by a fixture in place, ready to slide them into
their respective slots in the support rings.
A laser system is used to verify that the bar box is parallel to the rails
before each box is guided into its slot.
After completion of the bar box insertion a survey is performed to measure the
location of each box.

\begin{figure*}[htb]
	\begin{center}
		\includegraphics[width=0.8\textwidth]{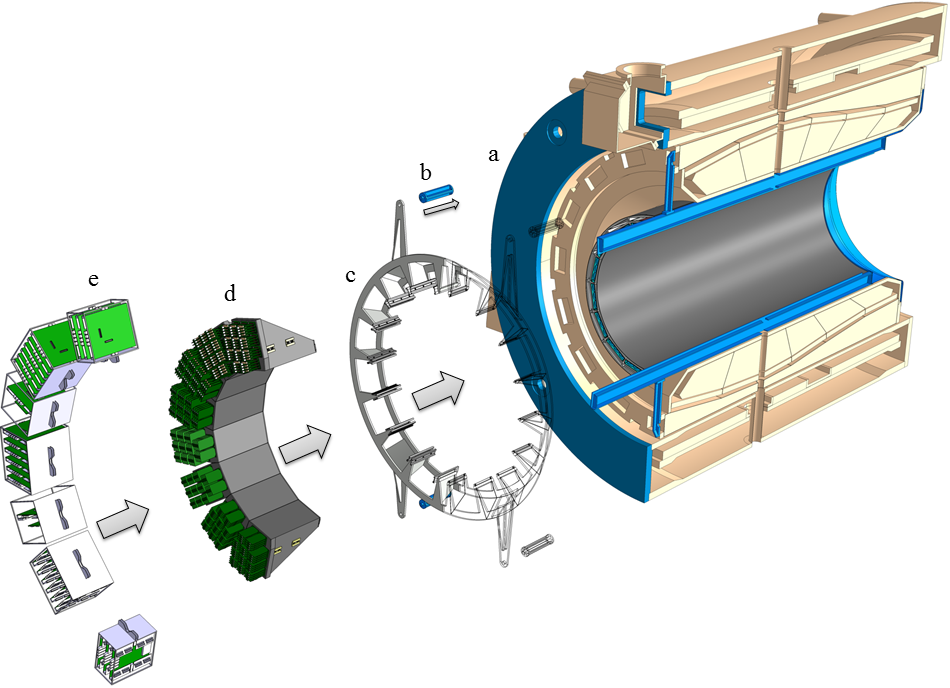}
	\end{center}
	\vspace*{-4mm}
	\caption{Installation procedure of the readout unit - half section view: (a) cryostat 
	upstream flange, (b) rigid spacers, (c) readout unit support ring, (d) eight prism boxes, 
	(e) readout electronics.}
	\label{fig:M-AG-ex_in}
\end{figure*}

The installation of the prisms follows a similar strategy.
A trolley transports the fixture into place where it is attached 
to the upstream flange of the cryostat (Fig.~\ref{fig:M-AG-ex_in}). 
After the support ring is aligned and fixed and the position surveyed, each 
prism slot can be equipped with one prism box, followed by the insertion of the
prisms, the attachment of the sensors and the readout electronics.

\section{Supply Lines and Cable Routing} \label{sec:cable}

\begin{figure*}[htb]
	\begin{center}
		\includegraphics[width=0.8\textwidth]{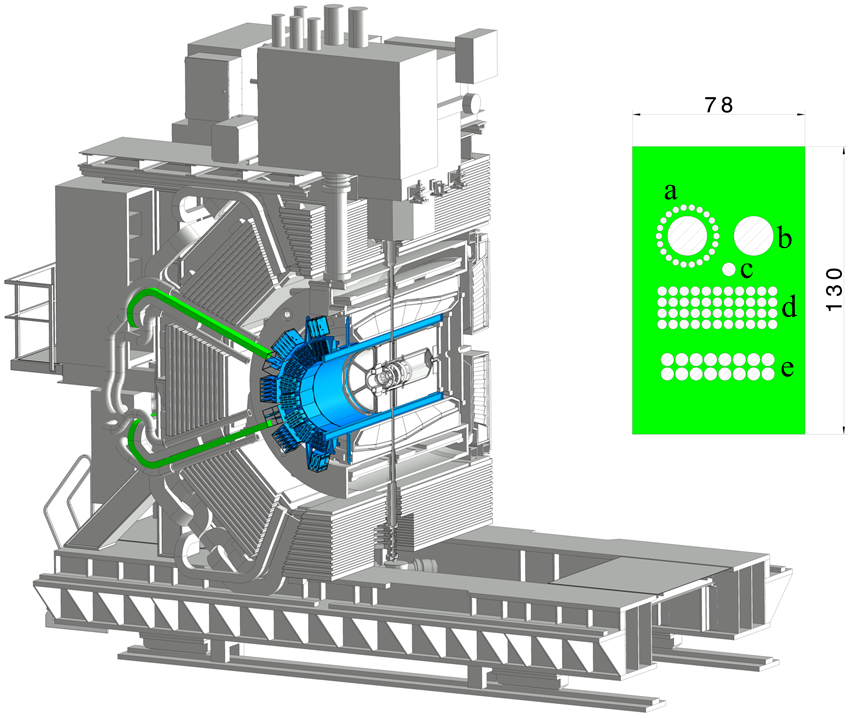}
	\end{center}
	\vspace*{-4mm}
	\caption{View of installed \panda Barrel DIRC - half-section view. 
	Cable ducts (marked in green) used in each quarter of the detector. 
	Cross section of one single cable duct - dimensions in mm: (a) LV, arranged around FEE cooling line, 
	(b) FEE cooling, (c) ethernet, (d) HV, and (e) nitrogen supply.}
	\label{fig:M-AG-cable_trays}
\end{figure*}

All electrical cables of the Barrel DIRC will be selected in compliance with 
the FAIR cable rules (fire safety, radiation resistance, bending radius, etc.). 
They are divided into four cable harnesses which are merged in each quarter 
of the readout unit support. 
Four cable ducts (Fig.~\ref{fig:M-AG-cable_trays}), integrated in and routed 
along the solenoid barrel, are used as the cable paths into the main supply 
chain and further to the service area in the \panda hall. 
The lines for the nitrogen flush system and the FEE cooling,
as well as the fibers for the laser pulser, are routed along 
the readout unit support ring. 
An overview of the present status of the supply lines, the cables and their cross 
sections is shown in Tab.~\ref{tab:M-AG-cable_cross_section}.

\begin{table*}[h]
	\centering
\begingroup
\setlength{\tabcolsep}{6pt} 
\renewcommand{\arraystretch}{1.5} 
	\caption{Table of the present status of the total number of cables and supply lines. 
	In the single cross section the insulations are included.}
	\ \
	\label{tab:M-AG-cable_cross_section}
	{\small
	\begin{tabular*}{1.0\textwidth}[]{@{\extracolsep{\fill}}lcccc}
	\hline
	Type&Connection&Number of units& \multicolumn{2}{c}{Cross section [mm$^2$]} \\
        &                   &                               & Single unit & Total  \\\hline
		HV cables & 176 MCP-PMTs & 176 coaxial cables &  15 &  2640 \\ 
		LV cables &  44 TRBs &  88 cables &  7 & 616 \\
		Readout cables &  4 TRB hubs & 4 ethernet cables &  30 & 120	\\
		\multirow{2}{*}{FEE cooling lines} &  FEE	&  4 inlet lines  &  250	&  1000	 \\ 
		                                   &  FEE	&  4 outlet lines &  250 &  1000	 \\
		\multirow{2}{*}{Nitrogen supply lines} & prism \& radiator boxes & 32 inlet lines &  29 & 928 \\ 
		                                       &  prism \& radiator boxes & 32 outlet lines &  29 & 928 \\ \hline
		
	\end{tabular*}
	}
\endgroup
\end{table*}

\section{Assembly Procedures} \label{sec:assembly}

The bar boxes will be assembled in a cleanroom, currently under construction at 
the Helmholtz-Institut Mainz. 
The design is very similar to the cleanroom used at SLAC for the assembly 
of the BaBar DIRC.
Large optical tables, covered by HEPA filters, will be used to inspect, 
qualify, clean, and glue the radiators, lenses, and mirrors and to 
place them into the bar boxes. 
The gluing, assembly, and storage will be based on the experience gained with the bar box 
assembly for the BaBar DIRC~\cite{babar-dirc-nim-mec} and the Belle~II TOP~\cite{belleII:TOP-mec}.
After assembly the completed bar boxes are placed in storage under a constant
nitrogen purge.
Procedures for the transport of the bar boxes to GSI/FAIR may be similar to the
method proposed for the transport of the BaBar DIRC bar boxes from SLAC to 
Jefferson Lab for the GlueX experiment~\cite{gluex-dirc-mec}. 
The outcome of that transport, planned for the spring/summer of 2017, will
be analyzed and necessary corrections applied.

The prism boxes will be assembled either in the same cleanroom in Mainz or
in the optical lab at GSI where a work table with HEPA filter coverage 
is available.

\section{Maintenance} \label{sec:maintenance}

An important goal of the mechanical concept is to use materials and components 
which enable a maintenance-free operation. 
Therefore, no scheduled maintenance, other than replacement of air filters, is 
currently foreseen. 
The performance of the sensors and readout electronics is monitored with 
an internal electronic pulser and a laser pulser system (see Sec.~\ref{subsec:monitoring}).
Should any of the readout cards or sensors require intervention, in situ access is
possible during a brief shutdown.
Any major intervention, like realignment or inspection of optical components,
can be realized while the \panda detector is in the parked position 
away from the beam line during the longer, scheduled shutdown periods.

\putbib[./literature/lit_mechanics]

\end{bibunit}

%% file: organization/organization.tex
\chapter{Project Management} \label{ch:orga} 
\begin{bibunit}[unsrt]

\section{Collaboration Structure} \label{sec:collabstruc}

The \panda Cherenkov group comprises physicists, engineers, and students from the Universities 
Erlangen-N\"urnberg, Giessen, Glasgow, and Mainz as well
as BINP Novosibirsk, GSI Darmstadt, JINR Dubna, and SMI Vienna. 
These institutions share the responsibilities for the Barrel DIRC, the Endcap Disc DIRC, 
the forward RICH detector, and the Barrel Time-of-Flight system.

The project management, design, and construction of the Barrel DIRC is currently concentrated at 
GSI and the primary responsibility for the tests of the photon detectors is at Erlangen University.
Otherwise the expertise on optical elements, electronics, software development and tests of prototypes 
with particle beams are shared within the whole Cherenkov group.
Specific Barrel DIRC work packages will be assigned to the groups in the upcoming MoU of \panda. 

List of institutions currently participating in the \panda Barrel DIRC R\&D and construction planning:
\begin{itemize}
\item GSI Helmholtzzentrum f\"ur Schwerionenforschung GmbH, Darmstadt, Germany
\item Friedrich Alexander Universit\"at Erlangen-N\"urnberg, Erlangen-N\"urnberg, Germany
\item II. Physikalisches Institut, Justus Liebig-Universit\"at Gie\ss{}en, Gie\ss{}en, Germany
\item Institut f\"ur Kernphysik, Johannes Gutenberg-Universit\"at Mainz, Mainz, Germany
\end{itemize}

\section{Schedule} \label{sec:schedule}

\begin{figure*}[htb]
\begin{center}
    \includegraphics[width=.95\textwidth]{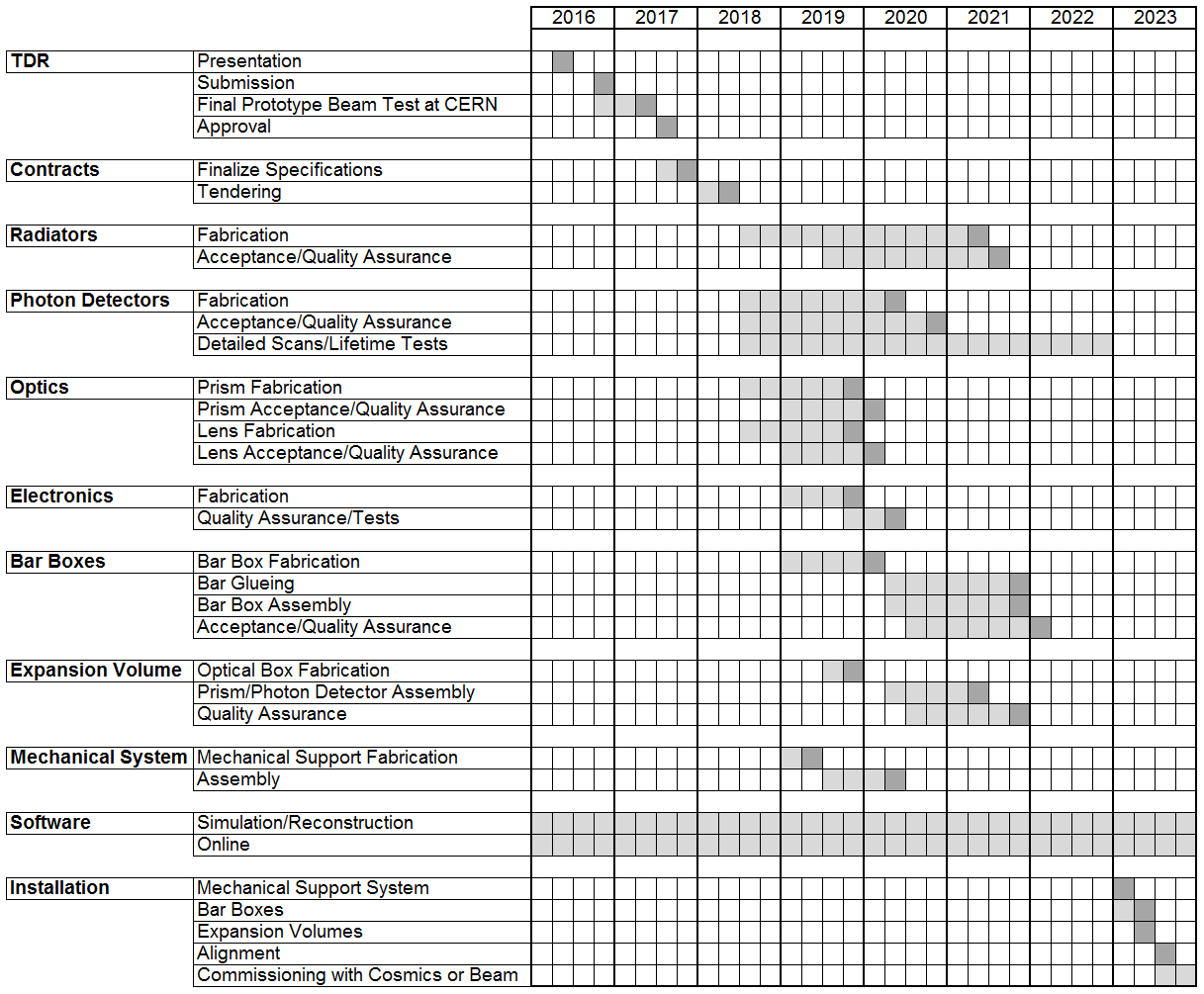}
\end{center}
\caption{Schedule for the \panda Barrel DIRC project from the presentation of the TDR in 2016 
through the installation and commission in 2023.
The time line for component production is based on estimates from industry. }
\label{fig:orgaschedule}
\end{figure*}

The Barrel DIRC project schedule through 2023 is shown in Fig.~\ref{fig:orgaschedule}. 
The project can be divided into six phases:
\begin{enumerate}
\item 2016: Submission of the TDR to FAIR.
\item 2017--2018: Finalize the specifications for radiators, photon detectors, optics, and electronics, call for tenders, 
establish fabrication contracts.
\item 2018--2021: Industrial fabrication of components.
\item 2021: Assembly of optical and mechanical components.
\item 2022--2023: Installation into \panda when hall is available and \panda detector is ready for installation.
\item 2023: Commissioning with cosmic rays and/or beam.
\end{enumerate}

The main milestones for the Barrel DIRC are:
\begin{itemize}
\item Approval of TDR, expected in Q3/2017.
\item Signed contracts for industrial fabrication of components, expected in Q2/2018.
\item Completion of photon sensor production, expected in Q2/2020.
\item Completion of radiator production, expected in Q2/2021.
\item Final assembly of bar boxes, expansion volume, and mechanical support, expected in Q4/2021.
\end{itemize}

The schedule for installation and commissioning depends on two additional milestones external to the Barrel DIRC project:
\begin{itemize}
\item Building milestone for availability of the \panda hall, currently projected for Q4/2021.
\item \panda detector ready for installation of Barrel DIRC mechanical support, currently projected for Q1/2023.
\end{itemize}

Figure~\ref{fig:orgaschedule} shows that the Barrel DIRC schedule is consistent with the external milestones.

\section{Cost} \label{sec:cost}

The estimated cost of the construction of the \panda Barrel DIRC is about 4.1~M\EUR{}
for the baseline design, using three narrow bars per sector, or about 3.6~M\EUR{}
for the design option, based on one wide plate per sector.
The cost of the two designs differs only in the fabrication cost for the radiators and lenses.

The dominant contribution to the construction cost are the fabrication of the fused silica radiators 
and the photon sensors.
The fabrication costs were calculated based on budgetary quotes obtained in the spring of 2016 
from several companies and include the production of 10\% additional units as spares..
Only businesses that are considered pre-qualified as potential vendors for the \panda Barrel DIRC 
production, based on demonstrated experience and/or the successful fabrication of pre-series 
prototypes, were considered.
This includes four companies in Europe, USA, and Japan for the radiator and prism production
and two companies in USA and Japan, plus, possibly, a third, European, company for the
photon detector production.

The cost estimates of the other components, such as mechanical elements, HL/LV, and DAQ  are
based on experience gained by other detector systems and information from experts.

The \panda Barrel DIRC is an in-kind contribution of Germany to the \panda experiment.
Funding for the construction is provided by the BMBF/Germany as part of the approved Projekt\-mittel\-antrag (PMA).
This in-kind contribution is valued at 2\,690\,000 \EUR{} (cost basis 2005). 
The PMA funds were transfered to GSI in 2012 and will become available once the TDR is accepted by FAIR. 
Applying standard cost escalation factors from 2005 to 2012 the 2.69~M\EUR{} in 2005~\EUR{} translate to
3.51~M\EUR{} in 2012~\EUR{}.

This means that the construction cost of the \panda Barrel DIRC exceeds the PMA budget 
by about 590~k\EUR{} or 17\% for the baseline design with three bars per sector and 
by about 50~k\EUR{} or 2\% if the design option with wide radiator plates is used instead.

It should be noted that, due to the likely production of bars, prisms, and sensors by companies
outside of Europe, exchange rate fluctuations add a significant uncertainty to the estimation
of the cost of the \panda Barrel DIRC system.
The budgetary quotes for the bars, prisms, and sensors were provided primarily in USD and a 
conversion rate of 1.12~USD per 1~\EUR{} (April 2016) was applied. 
Since the exchange rate at the time of the tender process cannot be predicted a cost 
estimate uncertainty of $\pm$15\% should be assumed.

This means that the design option with wide plates can be considered to be covered by the 
available PMA funds while the baseline design exceeds the PMA budget.

\section{Manpower} \label{sec:workpackages}

The manpower required and available is a mixture of staff, postdoctoral research associates, and PhD students 
as well as master and bachelor students, involved in R\&D, design, assembly, and testing. 
All major items for production are outsourced. 
The optical tests as well as the PMT testing are assumed to be done by two experienced physicists with assistance from students.

\section{Safety} \label{sec:safety}

The design and construction of the Barrel DIRC, including the infrastructure for its operation, will be performed according 
to the safety requirements of FAIR and the European and German safety rules. 
Detailed procedures for the assembly, installation, and operation of the Barrel DIRC will be provided to 
ensure personnel safety and the integrity of the Barrel DIRC components and avoid interference with other parts 
of the \panda experiment. 
There are no hazardous gases or flammable components in the Barrel DIRC design.
The primary hazards are mechanical and electrical.

\subsection*{Mechanics}

The strength of the Barrel DIRC support structures has been computed (FEM calculations) with physical models in the course 
of the design process and the required safety margins were applied. 
Additional forces during a quench of the super conducting magnet have been taken into account.

\subsection*{Electrical Equipment}

All electrical equipment in \panda will comply with the legally required safety code and concur to standards for 
large scientific installations following guidelines worked out at CERN to ensure the protection of all personnel 
working at or close to the components of the \panda system. 
Power supplies will have safe mountings independent of large mechanical loads. 
Hazardous voltage supplies and lines will be marked visibly and protected from damage by nearby forces. 
All supplies will be protected against over-current and over-voltage and have appropriate safety circuits and 
fuses against shorts.  
DC-DC converters have to be cooled to avoid overheating and the power supply cables will be dimensioned correctly 
to prevent overheating.  
All cabling and optical fiber connections will be executed with non-flammable halogen-free materials according 
to up-to-date standards. 
A safe grounding scheme will be employed throughout all electrical installations of the experiment. 
Smoke detectors will be mounted in all appropriate locations. 

\subsection*{Radiation Aspects}

Shielding, operation and maintenance of all \panda components will be planned according to European and German 
safety regulations to ensure the proper protection of all personnel. 
The access to the experimental equipment during beam operation will be prohibited and the access during maintenance 
periods will be cleared after radiation levels are below the allowed thresholds.

The Barrel DIRC equipment may become activated by low energy protons and neutrons leading to low-energy radioactivity 
of the activated nuclei. 
Therefore, all equipment has to be monitored for radiation before it is taken out of the controlled area.

\end{bibunit}